\documentclass{cernyrep} 
\pdfoutput=1
\usepackage{texnames}
\usepackage{tabularx}
\usepackage[T1]{fontenc}
\usepackage{subfigure}
\usepackage[bookmarks=false, colorlinks=true, linktoc=page, linkcolor=blue, citecolor=blue, urlcolor=blue]{hyperref}
\usepackage{import}
\newboolean{includetikzforreece}
\setboolean{includetikzforreece}{true}
\ifthenelse{\boolean{includetikzforreece}}{
  \usepackage{TIKZ/tikz}
  \usetikzlibrary{trees}
  \usetikzlibrary{decorations.pathmorphing}
  \usetikzlibrary{decorations.markings}
  \tikzset{
    photon/.style={decorate, decoration={snake}, draw=black},
    wino/.style={draw=redwine},    
    electron/.style={draw=black, postaction={decorate},
      decoration={markings,mark=at position .55 with {\arrow[draw=black]{>}}}},
    scalar/.style={draw=black, dashed,postaction={decorate},
      decoration={markings,mark=at position .55 with {\arrow[draw=black]{>}}}},
    gluon/.style={decorate, draw=black,
      decoration={coil,amplitude=4pt, segment length=5pt}}
  }
  \tikzstyle{blob}=[circle,
    thick,
    minimum size=0.4cm,
    draw=gray!80,
    fill=gray!60]   
  \tikzstyle{redblob}=[circle,
    thick,
    minimum size=0.4cm,
    draw=red!80,
    fill=red!60] 
}

\begin{document}
\begin{flushright}
CERN-TH-2016-111
\end{flushright}
\title{Physics at a 100 TeV $pp$ collider: beyond the Standard Model phenomena}
 
% Makes sure folders use master authors.tex file
\IfFileExists{../authors.tex}{\author{
Editors: \\
T.~Golling$^{1}$,\, 
M.~Hance$^{2}$,\, 
P.~Harris$^{3}$,\, 
M.L.~Mangano$^{4}$,\, 
M.~McCullough$^{4}$,\, 
F.~Moortgat$^{3}$, \,
P.~Schwaller$^{5}$,\, 
R.~Torre$^{6}$,\,
\\
Contributors:
\\
P.~Agrawal$^{7}$,\,
D.S.M.~Alves$^{8,9}$
% ...
\vspace*{2cm}
}

\institute{
$^{1}$ University of Geneva, Geneva, Switzerland \\
$^{2}$ University of California, Santa Cruz, Santa Cruz, CA, USA \\
$^{3}$ CERN, EP Department, CH-1211 Geneva, Switzerland \\
$^{4}$ CERN, TH Department, CH-1211 Geneva, Switzerland \\
$^{5}$ DESY, Notkestr. 85, D-22607 Hamburg, Germany \\
$^{6}$ Institut de Th\'eorie des Ph\'enom\`enes Physiques, EPFL, CH1015 Lausanne, Switzerland \\
$^{7}$ Department of Physics, Harvard University, Cambridge, MA 02138, USA\\
$^{8}$ Center for Cosmology and Particle Physics, Department of Physics, New York University, New York, NY 10003, USA\\
$^{9}$ Department of Physics, Princeton University, Princeton, NJ 08544, USA\\
$^{10}$ Department of Physics, University of Basel, Klingelbergstr.\ 82, CH-4056 Basel, Switzerland\\
$^{11}$ Max-Planck-Institut f\"ur Physik, F\"ohringer Ring 6, D-80805 M\"unchen, Germany\\
$^{12}$ Univ Lyon, Univ Lyon 1, ENS de Lyon, CNRS, Centre de Recherche Astrophysique de Lyon UMR5574, F-69230 Saint-Genis-Laval, France\\
$^{13}$ High Energy Physics Division, Argonne National Laboratory, Argonne IL, USA\\
$^{14}$ Theoretical Physics Division, Physical Research Laboratory, Ahmedabad-380009, India\\
$^{15}$ Institut f\"ur Theoretische Physik, Universit\"at Heidelberg, Germany \\
$^{16}$ Physik-Department T30d, Technische Univertit\"at M\"unchen,
James-Franck-Strasse 1, D-85748 Garching, Germany\\
$^{17}$ Max-Planck-Institut f\"{u}r Kernphysik, Saupfercheckweg 1, D-69117 Heidelberg, Germany\\
$^{18}$ Department of Physics, University of Notre Dame, IN, USA \\
$^{19}$ High Energy Physics Group, Blackett Lab., Imperial College, Prince Consort Road, London SW7 2AZ, UK\\
$^{20}$ PRISMA Cluster of Excellence and Mainz Institute for Theoretical Physics, Johannes Gutenberg
University, 55099 Mainz, Germany \\
$^{21}$ Department of Physics, Indian Institute of Technology, Kanpur-208016, India\\
$^{22}$ Department of Physics and Astronomy, University of Victoria, Victoria, BC V8P 5C2, Canada\\
$^{23}$ Perimeter Institute for Theoretical Physics, 31 Caroline St. N, Waterloo, Ontario, Canada\\
$^{24}$ Department of Physics, University of California, Davis, Davis, CA 95616, USA\\
$^{25}$ LPTHE, CNRS, UMR 7589, 4 Place Jussieu, F-75252, Paris, France \\
$^{26}$ Institute of Theoretical Science, University of Oregon, Eugene, OR 97403\\
$^{27}$ University of California, Santa Barbara, CA, USA \\
$^{28}$ Maryland Center for Fundamental Physics and Department of Physics,
University of Maryland, College Park, MD 20742, USA\\
$^{29}$ School of Natural Sciences, Institute for Advanced Study, Princeton, NJ 08540, USA\\
$^{30}$ Fysiska institutionen, Lunds universitet, Lund, Sweden\\
$^{31}$ LEPP, Department of Physics, Cornell University, Newman Laboratory, Ithaca, NY 14853, USA\\
$^{32}$ Department of Physics, University of Washington, Seattle WA, USA\\
$^{33}$ Theoretical Particle Physics and Cosmology Group, Dept. of Physics, King's College London, London WC2R 2LS, UK\\
$^{34}$ Michigan Center for Theoretical Physics (MCTP), Physics Department, University of Michigan, Ann Arbor, MI 48104 USA\\
$^{35}$ C.N. Yang Institute for Theoretical Physics, Stony Brook University, Stony Brook, NY 11794, USA\\
$^{36}$ Department of Physics, Brown University, Providence, RI 02912, USA\\
$^{37}$  New High Energy Theory Center, Department of Physics, Rutgers University, 136 Frelinghuisen Road, Piscataway, NJ 08854, USA\\
$^{38}$ Department of Physics and Astronomy, University of California, Irvine, CA 92697, USA\\
$^{39}$ Theoretical Physics Department, Fermilab, Batavia, IL 60510, USA\\
$^{40}$ Institute of Physics, University of Silesia, Uniwersytecka 4, 40-007 Katowice, Poland\\
$^{41}$ Department of Physics, University of Cincinnati, Cincinnati, Ohio 45221, USA\\
$^{42}$ Mitchell Physics Building, Texas A\&M, College Station, TX 77843, USA\\
$^{43}$ Northwestern University, Evanston IL, USA \\
$^{44}$ Pittsburgh Particle Physics, Astronomy, and Cosmology Center (Pitt-PACC)
Department of Physics and Astronomy, University of Pittsburgh, Pittsburgh, PA 15260, USA \\
$^{45}$ Korea Institute for Advanced Study (KIAS), Seoul 130-012, Korea \\
$^{46}$ Physics Department, Technion, Israel Institute of Technology, Haifa 32000, Israel\\
$^{47}$ SLAC National Accelerator Laboratory, Menlo Park, CA 94025, USA \\
$^{48}$ Institute for Particle Physics Phenomenology (IPPP), Department of Physics, Durham University, Durham, DH1 3LE, UK\\
$^{49}$ Department of Physics, University of Florida, Gainesville, FL 32611, USA\\
$^{50}$ Department of Physics, Duke University, Durham NC, USA\\
$^{51}$ Institute for Theoretical Particle Physics and Cosmology, RWTH Aachen University, D-52056
Aachen, Germany\\
$^{52}$ Physik-Institut, Universit\"at Z\"urich, CH-8057 Z\"urich, Switzerland\\
$^{53}$ Center for Theoretical Physics, University of California, Berkeley, CA 94720, USA\\
$^{54}$ IFT UAM-CSIC, Cantoblanco, 28049, Madrid, Spain \\
$^{55}$ Department of Physics and Astronomy, Michigan State University, East Lansing, MI 48824, USA\\
$^{56}$ Institute for Theoretical Physics, Albert Einstein Center, University of Bern, Sidlerstr. 5, CH-3012 Bern, Switzerland\\
$^{57}$ William I. Fine Theoretical Physics Institute, School of Physics and Astronomy, Univ. of Minnesota, Minneapolis, MN 55455, USA\\
$^{58}$ IFAE, Universitat Aut\`onoma de Barcelona, E-08193 Bellaterra, Barcelona, Spain\\
$^{59}$ Nikhef, Science Park Amsterdam, Netherlands \\
$^{60}$ Department of Physics and Astronomy, Johns Hopkins University, Baltimore, MD 21218, USA\\
$^{61}$ Dept of Physics,University of Illinois at Urbana-Champaign,1110 West Green Street Urbana, IL 61801, USA\\
$^{62}$ Ottawa-Carleton Institute for Physics, Carleton University,1125 Colonel By Drive, Ottawa, Ontario K1S 5B6, Canada\\
$^{63}$ Department of Physics, University of Alberta, Edmonton, AB T6G 2E1, Canada\\
$^{64}$ Maryland Center for Fundamental Physics, Department of Physics, University of Maryland, College Park, MD 20742, USA\\
$^{65}$ CP3-Origins and DIAS, University of Southern Denmark, Campusvej 55, 5230
Odense M, Denmark\\
$^{66}$ Ernest Orlando Lawrence Berkeley National Laboratory, University of California, Berkeley, CA 94720, USA\\
$^{67}$ Department of Physics, The University of Chicago, Chicago, IL 60637, USA\\
$^{68}$ Enrico Fermi Institute and Kavli Institute for Cosmological Physics, The University of Chicago, Chicago, IL 60637, USA\\
$^{69}$ Department of Physics, University at Buffalo, The State University of New York, Buffalo, NY 14260-1500, USA\\
$^{70}$ Center for Theoretical Physics, Massachusetts Institute of Technology, Cambridge, MA 02139, USA\\
}
}{}

\maketitle % this produces the title block

\begin{abstract}
This report summarises the physics opportunities in the search and
study of physics beyond the Standard Model at
a 100~TeV $pp$ collider.
\end{abstract}
 
\setcounter{tocdepth}{3}
\tableofcontents
 
\newpage 

\newcommand{\met}{\ensuremath{E\!\!\!\!/_T}}
\def\iab{\ensuremath{\mathrm{ab}^{-1}}}
\def\ifb{\ensuremath{\mathrm{fb}^{-1}}}
\def\ipb{\ensuremath{\mathrm{pb}^{-1}}}
\def \gsim{\mathrel{\vcenter
     {\hbox{$>$}\nointerlineskip\hbox{$\sim$}}}}
\def \lsim{\mathrel{\vcenter
     {\hbox{$<$}\nointerlineskip\hbox{$\sim$}}}}
\newcommand{\Ht}{\ensuremath{H_{T}}}
\newcommand{\pt}{\ensuremath{p_{T}}}
\newcommand{\ttbar}{\ensuremath{t\bar{t}}}

%
%=================================================================
\section{Foreword}
A 100~TeV $pp$ collider is under consideration, by the high-energy
physics community~\cite{benedikt,cepc_website},
as an important target for the future development
of our field, following the completion of the LHC and High-luminosity
LHC physics programmes. The physics opportunities and
motivations for such an ambitious
project were recently reviewed in~\cite{ArkaniHamed:2015vfh}. The
general considerations on the strengths and reach of very high energy
hadron colliders have been introduced long ago in the classic pre-SSC
review~\cite{Eichten:1984eu}, and a possible framework to establish the
luminosity goals of such accelerator was presented recently
in~\cite{Hinchliffe:2015qma}.

The present document is the result of an extensive study, carried out
as part of the Future Circular Collider (FCC) study towards a Conceptual
Design Report, which includes separate Chapters dedicated to Standard
Model physics~\cite{SM-report}, physics of the Higgs boson and EW symmetry
breaking~\cite{Higgs-report}, physics beyond the Standard
Model (this paper), physics of heavy ion
collisions~\cite{Dainese:2016gch} and physics with the FCC injector
complex~\cite{inj-report}. Studies on the physics programme of an
$e^+e^-$ collider (FCC-ee) and $ep$ collider (FCC-eh)
at the FCC facility are proceeding
in parallel, and preliminary results are documented
in~\cite{Gomez-Ceballos:2013zzn} (for FCC-ee) and
in~\cite{AbelleiraFernandez:2012cc} (for the LHeC precursor of FCC-eh).

\section{Introduction} %\footnote{Editors: all}} 
\label{sec:intro}

Experimental measurements at a 100 TeV collider would cover previously unexplored territory at energies never before reached in a laboratory environment.  Standard Model calculations will enable precise predictions for the phenomenology of the known particles and forces in this new frontier.  The comparison of observations against predictions will allow for the structure of the Standard Model to be tested at unprecedented energies and with unparalleled precision.  If observations and predictions agree within estimated uncertainties then this would provide a stunning confirmation of our present understanding of nature.  If, on the other hand, observations do not agree with theoretical predictions this would mark a breakdown of the Standard Model of particle physics and the rise of new physical processes.  In this way, amongst its many roles, a 100 TeV collider may discover new laws of nature.

In its significance for our understanding of nature, the discovery of new physics at the energies accessible to a 100 TeV collider would be unrivaled, but it would not necessarily be unheralded.  There are a number of reasons to believe that a new physical description of nature beyond the Standard Model may be required at these energies.  In this section we will summarise the landscape beyond the Standard Model accessible to a 100 TeV collider.

At the deepest level, one may discover new symmetries of spacetime at 100 TeV, for which a leading candidate is supersymmetry.  Supersymmetry as a new high energy symmetry of spacetime is theoretically motivated from a number of perspectives, covering dark matter, unification of the forces, and the electroweak hierarchy problem.  In Sec.~\ref{sec:SUSY} we will summarise these motivations in detail.  Sec.~\ref{sec:SUSY} will then go on to systematically consider the rich phenomenology of the various new particles predicted in supersymmetric theories, with an aim to connect this phenomenology with concrete supersymmetric scenarios to provide a clear context for the interpretation of measurements.

One hint of new physical effects that may be unearthed at a new energy scale come from presence of dark matter. It is now well established that dark matter is prevalent throughout the universe. To explain its large abundance, many different mechanisms of new effects beyond the Standard Model have been proposed. This spectra of models  extend form supersymmetric models to other exotic models that go beyond the basic precepts of supersymmetry. A remarkable aspect of dark matter models is that with a loose set of assumptions, many of these models give concrete predictions of the current dark matter abundance originating from the early universe. In many cases, coverage of the allowed parameter space can be obtained with a 100 TeV machine.  In Sec.~\ref{sec:DM}, we review the different classes of dark matter models and the characteristic searches both at a hadronic collider and beyond that drive the sensitivity to these models. 

In Sec.~\ref{sec:beyond} we will discuss the reach for a future 100 TeV collider on a variety of new physics signatures that are not typical of supersymmetric and Dark Matter models. Most of these signatures originate from the decay of heavy resonances to Standard Model particles, but also indirect probes of new physics based on the measurement of the production of Standard Model particles at high invariant masses will be discussed. To make quantitative the assessment of the 100 TeV collider reach for these signatures, different benchmark models will be considered, related to some of the main issues of the SM, among which the hierarchy problem and the origin of neutrino masses. The list of studies that we present are at a very preliminary level and the list itself is far from being complete. However, they constitute a solid starting point that allows to identify the main experimental issues associated to the different signatures and that are essential for the design of the future facilities, i.e.~both the collider and the detectors.

It is clear that the exploration of physics beyond the Standard Model using proton-proton collisions with a center-of-mass energy of 100 TeV bears unprecedented challenges.  Energy ranges with dynamic scales ranging from the sub~GeV to 10s of TeV  become a necessity to maximize the capability demanded to study different physical effects. High instantaneous luminosity will require exquisite techniques to mitigate pileup. Such techniques include standard techniques at the LHC such as charged particle vertexing studies, but also more advanced approaches such as neutral vertex association through fast timing or with hadron fragmentation structure~\cite{Cacciari:2014gra,Berta:2014eza,Bertolini:2014bba}. In order to contain the highest $\pt$ jets (of tens of TeV) fully in the hadronic calorimeter a depth of at least 12~nuclear interaction lengths ($\lambda_{I}$) is necessary. 

To identify highly boosted hadronically decaying top quarks, $W$, $Z$ and $H$ bosons with transverse momenta in the multi-TeV range a very fine hadronic calorimeter lateral segmentation in $\eta\times\phi$ of at least $0.025\times0.025$ is needed. This allows for the measurement of the boson jet substructure.  Even this segmentation might limit the identification or these objects for the highest accessible $\pt$ objects. Tracker or tracker+ECAL based jet substructure methods might offer a solution~\cite{Schaetzel:2013vka,Larkoski:2015yqa,Spannowsky:2015eba}.  However, the increasing presence of long-lived neutral hadrons at higher $\pt$ represents a difficult challenge in jet substructure that may have limitations~\cite{Bressler:2015uma}.  This will also affect the capability to measure the polarisation of these objects.  

Reconstructing leptonic decays of the top quark and $W$, $Z$ and $H$ bosons will be limited in the highly boosted regime: the small opening angles, e.g. between the lepton and the $b$-jet in a $t\to W(\to \ell\nu) b$ decay result in non-isolated leptons which are very hard to distinguish from e.g. a lepton from a $b$-hadron decay.

Another challenge is the momentum resolution for multi-TeV muons.  The size of the ATLAS detector~\cite{Aad:2008zzm}, for instance, is driven by the size of the muon system with the goal to measure the transverse momentum of muons with $\pt =1$~TeV with a 10\% uncertainty, resulting in a diameter of ATLAS of about 25~m.  Scaling this up to muons of $\pt =10$~TeV with a similar resolution pushes the size of the detector and of the magnetic field to unfeasible dimensions, and alternative strategies are needed and are currently under study. 

The tracking system will also be challenged to efficiently reconstruct multi-TeV objects.  Identification of $b$ jets or $\tau$ leptons with a $\pt$ well above 1 TeV is largely unexplored, even at the LHC.  The $b$-tagging performance of the current ATLAS and CMS detector deteriorate dramatically in the $\pt$ range between a few hundred GeV and 1 TeV.  The $\tau$ lepton identification and the decay components of the tau suffer from similar limitations of resolution in the tracker and calorimeters in the highly boosted regime. More generally, the high boost of Standard Model particles results in very collimated objects and makes high demands on tracking capabilities in very dense environments. Charge particle angular separation can currently go to a level roughly $0.01\times0.01$ in $\eta\times\phi$, which has an impact for b-quark jets at roughly~1~TeV. Perserving consistent performance for a 100 TeV detector would require separation in part of the detector at angles roughly 10 times smaller.  

Even as much of the discussion of the beyond the Standard Model (BSM) physics potential of a 100 TeV collider focuses on the high-\pt{} regime, reconstruction of low-\pt{} and displaced objects will also be critical to discover many new physics scenarios.   BSM models with weak couplings or compressed mass spectra may lead to low-\pt{} final states.  Such events may only be distinguished above backgrounds by tagging soft or displaced decay products, such as soft leptons, kinked or disappearing tracks, highly-displaced vertices, or even measuring the charged particle $dE/dx$.  Tracking and calorimetry must therefore be hermetic for prompt, high-\pt{} objects in addition to soft and/or displaced objects.

In addition to the challenges of resolving high \pt~objects, data rates and detector readout will have to be sufficiently fast to correspond with the high collision rate. The high collision rates and demands for high granularity will significantly increase the data rate coming out of the detector. Triggering of both low energy anomalous objects, such as disappearing tracks, and high \pt~objects demands more sophisticated high data volume readouts and high speed pattern recognition, especially under the onset of pileup.

While the design of a new detector poses interesting and difficult problems, many of the technologies currently being investigated for HL-LHC already go in the detection of improving the granularity and data rate of the detectors. Additionally, much of the interesting BSM physics requires high \pt objects, for which the demands on basic calorimetric resolution, tracking performance for simple objects such as quark or gluon jets or missing transverse energy can be met with existing technologies. Nevertheless, a clear need for more information is present.  

All the BSM physics benchmarks discussed in this section are used to identify the most relevant features needed by the new detectors. This should lead to a compromise between the feasibility of the desired detector and the coverage of the largest possible spectrum of new physics signatures. A realistic fast simulation of different detector configurations and a close collaboration between theorists and experimentalists is crucial in this phase of the study to assess the limitations and to study solutions for them.

%\label{sec:beyond}

\newpage 
\newcommand{\HT}{\Ht}
\newcommand{\chioz}{\ensuremath{\widetilde{\chi}_{1}^{0}}}
\newcommand{\LSP}{\chioz}
\newcommand{\Stop}{\ensuremath{\tilde{t}}}
\newcommand{\gluino}{\ensuremath{\widetilde{g}}}
\newcommand{\squark}{\ensuremath{\widetilde{q}}}
\newcommand{\TeV}{\ensuremath{\mathrm{TeV}}}
\newcommand{\GeV}{\ensuremath{\mathrm{TeV}}}
\newcommand{\meff}{\ensuremath{m_{eff}}}
\newcommand{\slepton}{\ensuremath{\tilde{\ell}}}

\section{Supersymmetry}
%\texorpdfstring{\footnote{Editors: McCullough, Hance}}{}}
\label{sec:SUSY}

%M.~Hance and M.Mc~Cullough; contributors:...

%Feel free to introduce subsections. Each subsection should be a
%separate file, to be included with an 
%\verb| \input{subsection}| statement.

%As labels for the subsections, try to use things like
%\verb|sec:SUSY_xxx|. Ditto for references to equations, figs:
%\verb|\label{eq:SUSY_xxx}|, etc

\subsection{Introduction}
\label{sec:susy_intro}
As the detailed theoretical study of quantum field theory progressed in the 1970's it slowly emerged, in various respects, that a new symmetry of spacetime was possible.  Under the known spacetime symmetries the fields we observe, such as scalars (spin-0), fermions (spin-$1/2$), and vectors (spin-1), all form different representations of the Lorentz group and they transform under the known spacetime symmetries in their own way, independently.  However, it was realised that under the new hypothetical spacetime symmetry these different representations may themselves transform into one another and would together be combined into a larger representation of a larger symmetry, a supersymmetry \cite{Volkov:1973ix,Wess:1974tw}.  These larger representations are called superfields, since they contain multiple component fields.  Perhaps the most economical superfield, known as a chiral superfield, contains a complex scalar boson and a Weyl fermion.  Supersymmetry imposes specific relations between the interactions and masses of these component fields.  If gravitational interactions are to respect this symmetry, which they must if the symmetry is realised at a fundamental level, then the supersymmetry manifests as a gravitational theory, known as supergravity \cite{Freedman:1976xh,Deser:1976eh}.  This theory contains, in addition to the spin-2 graviton, a spin-$3/2$ partner fermion known as the gravitino which again has interactions purely dictated by the supersymmetry.

What would the discovery of supersymmetry mean for our understanding of nature?  It would be nothing short of the discovery of an entirely new spacetime symmetry.  The discovery of any symmetry signifies a fundamental shift in our perspective on fundamental physics and such discoveries are rare.  Examples include the early $\text{U}(1)_{EM}$ symmetry of electromagnetism, the $\text{SU}(3)_{C}$ symmetry of QCD and more recently the embedding of electromagnetism into the full $\text{U}(1)_{Y} \times \text{SU}(2)_{W}$ symmetry of the electroweak sector, of which the W and Z bosons are the additional force carriers and the Higgs serves as the final cornerstone in the theory.  However, the last time a new spacetime symmetry was discovered harks back to the work of Einstein and the discovery of the diffeomorphism invariance of the laws of physics.  The discovery of supersymmetry would thus mark a monumental shift in how we perceive nature at its most fundamental level.

If supersymmetry is realised in nature, how could we tell?  As already stated, in supersymmetry every field is inextricably tied to its superpartner.  This means that if supersymmetry were an exact symmetry of nature every particle we know would have a superpartner: every boson would have a fermion partner, with the suffix `ino', every fermion a boson partner, with the prefix `s'.  Thus the Higgs boson would have a fermion `Higgsino' partner, the photon a `photino', and the gluon a `gluino'.  The electron would have a scalar `selectron' partner, the top quark a `stop squark' partner and so forth.  Exact supersymmetry would require that all superpartners have equal mass to their observed counterpart and hence practically all of the superpartners would have been discovered already.  However, it turns out that the fundamental, high-energy, supersymmetry may persist if the masses of the partners are split from their known counterparts by an amount known as the soft mass.  The term `soft' is used as this form of supersymmetry breaking preserves the supersymmetric features of the theory at high energies.  The soft mass may lift the mass of, for example, the selectron by an amount $\tilde{m}$ above the electron mass.  At energies $E\gg \tilde{m}$ the theory will still be supersymmetric and maintain all of its appealing features, although the partner will not be directly apparent at energies $E\ll \tilde{m}$, essentially hiding supersymmetry from low-energy observers.\footnote{Besides the mass terms, for the scalars there are also scalar trilinear couplings which mimic the matter Yukawa couplings.  These couplings are known as `A-terms' and, while they break supersymmetry, this breaking is soft.}

If one simply takes the Standard Model of particle physics and supersymmetrizes it, then in the simplest variant, known as the MSSM, the only complication beyond the adding of superpartners is that two separate Higgs doublets are required, for reasons related to anomaly cancellation.\footnote{In renormalizable models two Higgs doublets are also required as a result of holomorphicity of the superpotential.}  Each doublet obtains a vacuum expectation value.  The vev for the Higgs doublet coupling to the up-type fermions is $v_u$, and the one coupling to down-type fermions is $v_d$.  The ratio of these parameters arises frequently and is commonly known as $\tan \beta = v_u/v_d$.  Thus, all told, the number of fields is a mite greater than a simple doubling of fields.

The couplings of the additional superpartners to the known fields is predicted by supersymmetry itself.  Thus in many cases once the mass of the superpartners are chosen it is possible to predict the collider production rate for a specific superfield.\footnote{In some cases additional soft terms may also enter, such as scalar trilinear couplings, thus these may need to be chosen to make predictions for some sparticles.}  Because of this predictability, in this supersymmetry section the experimental prospects for discovering supersymmetry at a 100 TeV collider can be broken down for each particular superpartner.  In Section~\ref{sec:susy_xs} predictions for the production cross sections of different particles are presented, and following this search strategies and projections for stop squarks (Section~\ref{sec:SUSY_stop}), gluinos (Section~\ref{sec:susygluino}), the first two generations of squarks (Section~\ref{sec:susy_squarks}), and electroweakinos such as winos, binos, and higgsinos (Section~\ref{sec:susy_ewino}) are presented.  However, before considering the experimental sensitivity it is useful to explore further the theoretical aspects of supersymmetry, to inform our interpretation of the search projections with regard to expectations for well-motivated mass ranges.

The goal of this section is not to promote supersymmetry for any of its particular virtues, nor is it to use supersymmetry to provide a precise physics motivation for a 100 TeV collider, however it is pragmatic to consider the superpartner mass ranges suggested by certain theoretical perspectives to provide a reference point for experimental projections. For the sake of a broad discussion, the various superpartner soft masses may be broken down into three categories.  We may also use symmetries to understand their expected proximity to one another.  Since the Higgs mass has been measured we will keep it separate; however, although they are not strictly related to each other, all of the other scalar masses may be broadly described with a generic parameter $\tilde{m}_0$.  The symmetry broken by the scalar soft masses is supersymmetry, and it may be naturally small.  In isolation, the Higgsino mass $\mu$ pairs the up-type Higgsino $\widetilde{H}_u$ with the down-type Higgsino $\widetilde{H}_d$ to form a Dirac mass.  This parameter respects supersymmetry, but breaks a Peccei-Quinn symmetry of the MSSM, thus it can be naturally small and is not a priori connected with the scale of supersymmetry breaking.  Also the three gaugino masses are not strictly related to each other, however in many models they are not hierarchically different.  Thus, for the purposes of broad discussion, we may lump them into one parameter $\widetilde{M}_{1/2}$.  This parameter breaks supersymmetry and a continuous R-symmetry, thus it may also be naturally small.

With these considerations in mind it is clearly possible to have hierarchies amongst these parameters.  In particular, it is natural to have $\widetilde{M}_{1/2} \ll \tilde{m}_0$ and/or $\mu \ll \tilde{m}_0$, or it may be that they are all comparable $\widetilde{M}_{1/2} \sim \mu \sim \tilde{m}_0$.  We will now consider in detail some of the theoretical and phenomenological features of supersymmetry with a specific view towards motivating certain mass ranges for the different types of superpartner.

\subsubsection{Dark Matter}
When the SM is supersymmetrized some remarkable features arise.  The first is that if an additional $\mathcal{Z}_2$ global symmetry known as `R-parity' is imposed, to help avoid potentially phenomenologically unacceptable features such as fast proton decay, then the theory contains not one, but a number of fields that are compelling candidates for explaining the dark matter.  Most importantly, these fields have the required masses and couplings to satisfy the required ingredients for the so-called `WIMP Miracle', which naturally generates a dark matter abundance in the region of the observed abundance for stable weak-scale particles \cite{Ellis:1983ew,Goldberg:1983nd}.  The main candidates are the so-called `neutralinos', comprising the neutral Higgsinos, Wino, and Bino, which may all mix under electroweak symmetry breaking.  There are also the sneutrinos, which are a priori interesting dark matter candidates \cite{Hagelin:1984wv,Ibanez:1983kw,Falk:1994es}, although as the simplest incarnations are already in tension with direct detection searches we will not consider sneutrinos further here.  As the dark matter searches are covered in Sec.~\ref{sec:DM} we will not consider the dark matter candidates in any more detail, however it is worth keeping in mind throughout this section that the provision of good dark matter candidates remains a strong motivation for considering supersymmetric theories.  Neutralino dark matter thus motivates the mass range $\mu,\widetilde{M}_{1/2} \lesssim \mathcal{O}(\text{few TeV})$, otherwise it would not be possible to obtain the correct relic density and they would overclose the Universe.  This clearly points to a mass range that is within kinematic reach of a 100 TeV collider.

\subsubsection{Gauge Coupling Unification}
An unexpected surprise that arises whenever the Standard Model is supersymmetrized connects the behaviour of the Standard Model gauge couplings to a deep idea concerning the nature of the forces at extremely high energies.  When the superpartners are added, it was found that upon evolving the $\text{U}(1)_{Y}$, $\text{SU}(2)_{W}$, and $\text{SU}(3)_{C}$ gauge couplings up to high energies they appeared to unify at energies close to $E\sim 10^{16}$ GeV \cite{Dimopoulos:1981yj,Dimopoulos:1981zb}.  This is shown in Fig.~\ref{fig:unification}.  Of course, that two lines will cross is almost guaranteed, however three lines crossing almost at a point is strongly suggestive of a deeper structure.

%%%%%%%%%%%%%%%%%
\begin{figure}[tbp]\begin{center}
\includegraphics[width=0.6\textwidth]{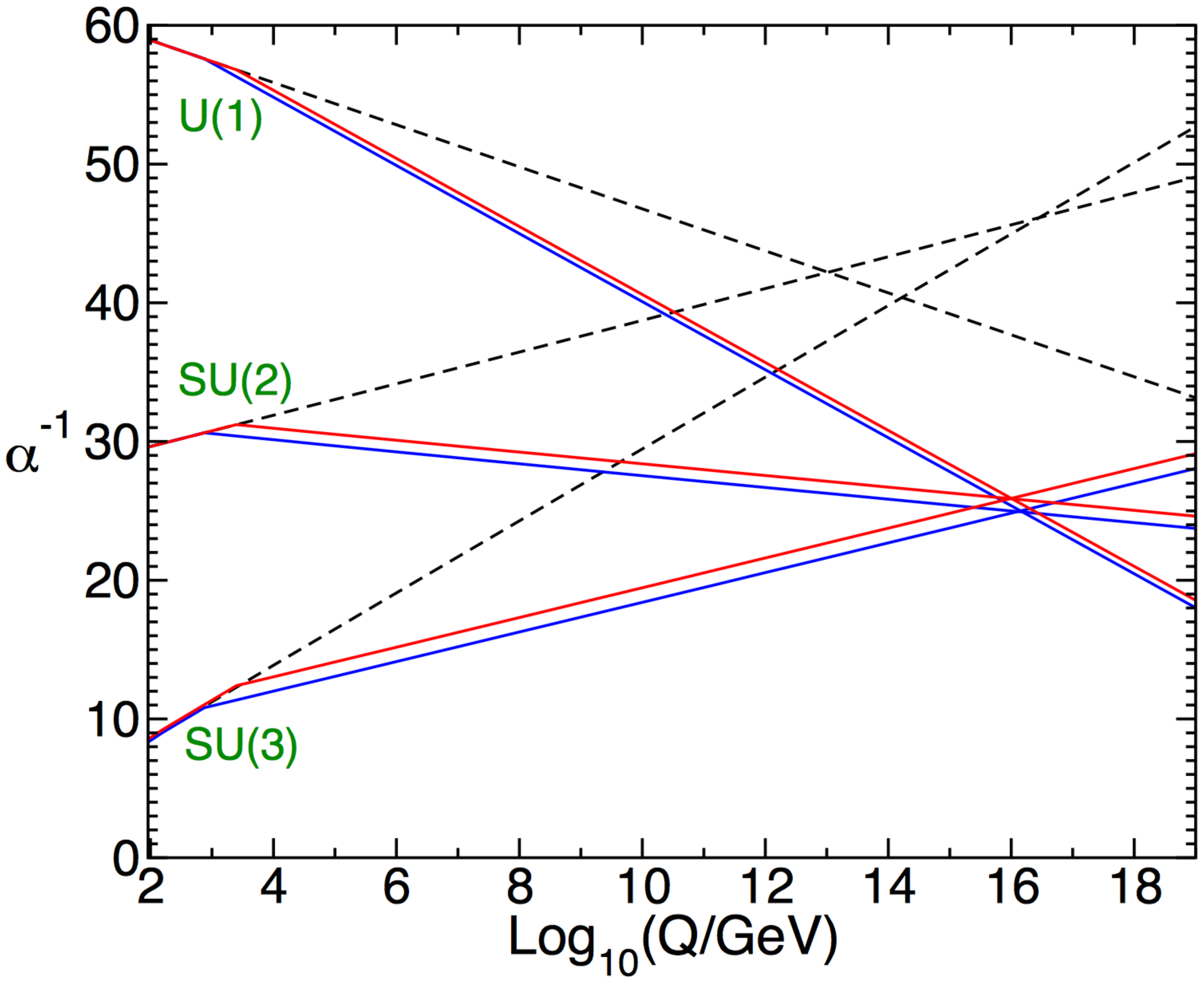}
\end{center}
\caption{Renormalization group evolution of gauge couplings up to high energies, taken from \cite{Martin:1997ns}.  The Standard Model gauge couplings are shown in dashed black and the gauge couplings with superpartners added, with masses in the range $0.75 \to 2.5$ TeV, are shown in red and blue.  Unification of the forces at high energies is clearly apparent in the supersymmetric case.}
\label{fig:unification}
\end{figure}%
%%%%%%%%%%%%%%%%%

Ever since the unification of the electroweak forces, it has been believed that further unification of all gauge forces, now including $\text{SU}(3)_{C}$, may occur at very high energies.  A variety of larger gauge groups into which they may unify have been proposed, however the simplest is arguably an $\text{SU}(5)$ gauge symmetry \cite{Georgi:1974sy}.\footnote{It is also possible that the gauge forces unify with gravity, in the context of String Theory, however we will not discuss this possibility here.}  It is deeply compelling that the Standard Model matter gauge representations neatly fall into multiplets of a larger symmetry, such as $\text{SU}(5)$, as this need not have been the case.  A key feature which must arise at the unification scale in such a theory is that the gauge couplings must themselves become equal.  Thus supersymmetric gauge coupling unification is strongly suggestive that supersymmetry may go hand-in-hand with the unification of the forces and, if discovered, the superpartners would provide a low energy echo of physics at extremely high energies.

When considering the role of the superpartners in supersymmetric unification one finds that some are more relevant than others.  The reason is that since the matter fermions of the Standard Model fill out complete unified representations, so must their partners, the squarks and the sleptons.  Thus although the masses of squarks and sleptons may change the scale at which unification occurs they do not significantly alter whether or not the couplings will unify, unless they are split by large mass differences themselves.  This means that the most important superpartners for gauge coupling unification are the fermions: the gauginos and the Higgsinos.

Studies of supersymmetric gauge coupling unification generally find that for successful unification it is necessary to have gauginos and higgsinos not too far from the weak scale.  If the gaugino and Higgsino mass parameters are taken equal, then unification requires $\mu,\widetilde{M}_{1/2} \lesssim \mathcal{O}(10 \text{ TeV})$ 
%at $1\sigma$, where $\sigma$ denotes the magnitude of estimated 
with some uncertainty due to unknown threshold corrections at the unification scale \cite{Arvanitaki:2012ps}.  The scalar soft masses, $\widetilde{m}_0$, may be arbitrarily heavy while preserving successful gauge coupling unification.  This realization led to the consideration of so-called `Split-Supersymmetry' theories \cite{Wells:2004di,ArkaniHamed:2004fb,Giudice:2004tc}, in which the main motivations for the mass spectrum are taken from gauge coupling unification and dark matter, as discussed previously.

The fact that, in addition to the gauge forces, also the matter particles are unified in representations of the unified gauge symmetry group, can imply relations between the Yukawa couplings of quarks and leptons at the unification scale \cite{Georgi:1974sy,Chanowitz:1977ye,Nanopoulos:1978hh,Georgi:1979df,Antusch:2009gu,Antusch:2013rxa}. To compare such predictions with the measured values of the fermion masses, one has to take into account the supersymmetric loop threshold corrections at the soft breaking mass scale \cite{Hempfling:1993kv,Hall:1993gn,Carena:1994bv,Blazek:1995nv,Antusch:2008tf,Antusch:2015nwi}, which depend on the masses of the superpartners. Including them in the analysis, and using the measured fermion masses and Higgs mass as constraints, unified theories are even capable of predicting the complete sparticle spectrum \cite{Antusch:2015nwi,Antusch:2016nak}. An example from a recent analysis is shown in Fig.\ \ref{fig:GUTs-sparticle-spectrum}. The superpartner masses are found to be $\lesssim  \mathcal{O}(5 \text{ TeV})$, testable at a 100 TeV pp collider.

\begin{figure}
\begin{center}
\includegraphics[scale=.45]{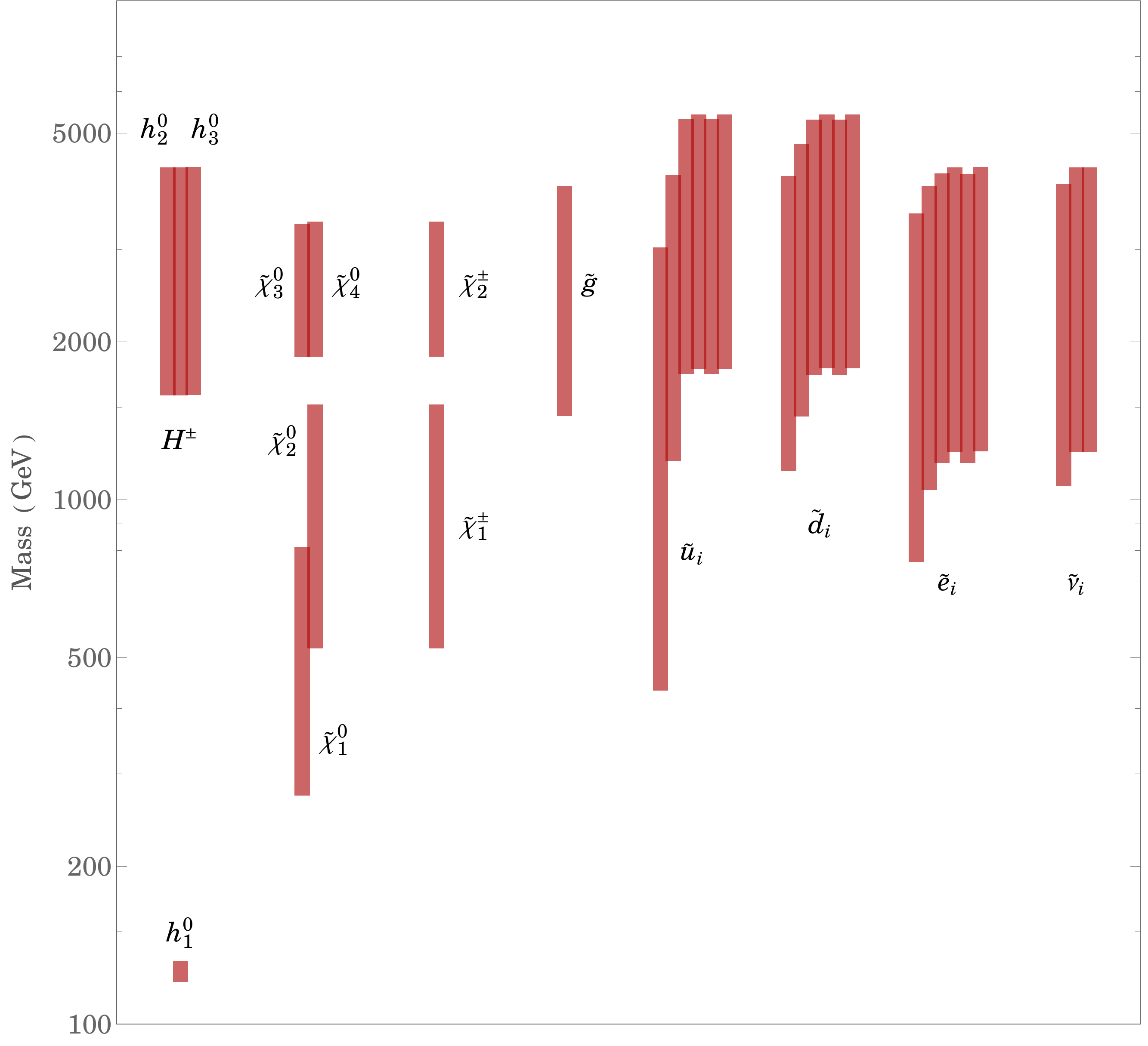}
\end{center}
\caption{Prediction for the superpartner and Higgs boson masses ($1\sigma$ HPD intervals) in classes of $\text{SU}(5)$ unified theories with the unification scale relations $\frac{y_\tau}{y_b}=-\frac{3}{2}$, $\frac{y_\mu}{y_s}=6$, $\frac{y_e}{y_d}=-\frac{1}{2}$ and universal (CMSSM) boundary conditions for the soft breaking parameters \cite{Antusch:2015nwi}.
}
\label{fig:GUTs-sparticle-spectrum}
\end{figure}

%To summarize, as with dark matter, gauge coupling unification motivates Higgsino and gaugino soft masses in the approximate range $\mu,\widetilde{M}_{1/2} \lesssim \mathcal{O}(\text{10's TeV})$, once again suggesting that much of the parameter space motivated by this consideration should be within reach of a 100 TeV collider.

To summarize, as with dark matter, gauge coupling unification and the unification of matter particles in representations of the unifying gauge symmetry group motivate  the existence of superpartners of the Standard Model particles with masses $\lesssim \mathcal{O}(\text{10 TeV})$, once again suggesting that much of the parameter space motivated by this consideration should be within reach of a 100 TeV collider.

\subsubsection{The Higgs Mass}
As is common in physics, when new symmetries are introduced to a theory, the predictive power often increases.  Because supersymmetry is softly broken, many new parameters associated with this breaking are introduced and certain aspects of the increased predictivity are lost.  However, some predictability beyond the SM remains and the Higgs boson mass is a prime example.

In the Standard Model, when the theory is written in the unbroken electroweak phase there are only two fundamental parameters in the scalar potential, the doublet mass $m_H$, and the quartic coupling $\lambda$.  In the broken electroweak vacuum this translates to two fundamental parameters, the Higgs vacuum expectation value $v=246$ GeV, and the Higgs scalar mass $m_h$.  Once these two parameters are set, all other terms, such as the Higgs self-couplings, are determined.  Supersymmetric theories take this one step further as supersymmetry relates the Higgs scalar potential quartic coupling to the electroweak gauge couplings in a fixed manner.  The story is complicated a little relative to the Standard Model by the two Higgs doublets required in supersymmetric theories, however since the quartic couplings in the scalar potential are no longer free parameters, once the vacuum expectation value is set $v = \sqrt{v_u^2 + v_d^2} = 246$ GeV, the Higgs mass is now also predicted by the theory.  At tree level, this prediction is
\begin{equation}
m_h = M_Z |\cos 2 \beta| ~~.
\end{equation}
Clearly for any value of $\beta$ this prediction is at odds with the observed value of $m_h \approx 125$ GeV and thus for consistency additional contributions to the Higgs doublet quartic terms are required.  Within the MSSM the only potential source is from radiative corrections at higher orders in perturbation theory.  The dominant corrections arise from loops of particles with the greatest coupling to the Higgs, the stop squarks \cite{Ellis:1990nz,Ellis:1991zd}.  If the soft mass splitting between the top-quark and stop squarks is large enough then radiative corrections which are sensitive to this supersymmetry breaking may spoil the supersymmetric prediction for the Higgs quartic couplings and allow for contributions that may bring the Higgs boson mass within the observed window.

%%%%%%%%%%%%%%%%%
\begin{figure}[tbp]\begin{center}
\includegraphics[width=0.4\textwidth]{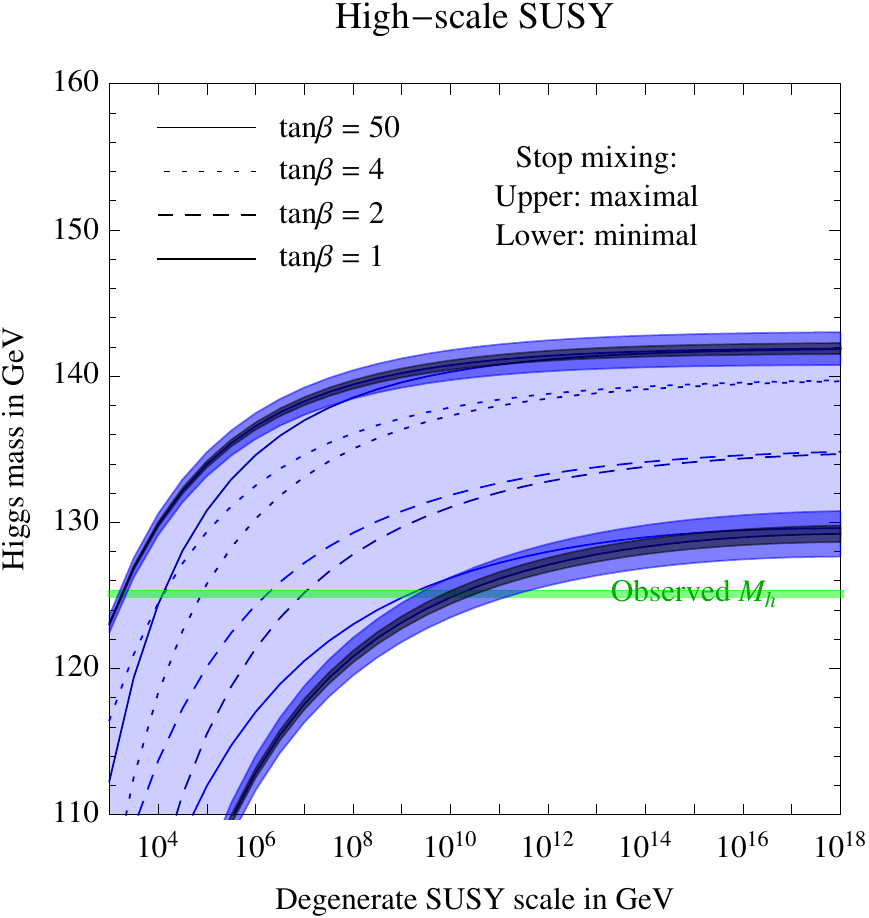} \qquad \includegraphics[width=0.4\textwidth]{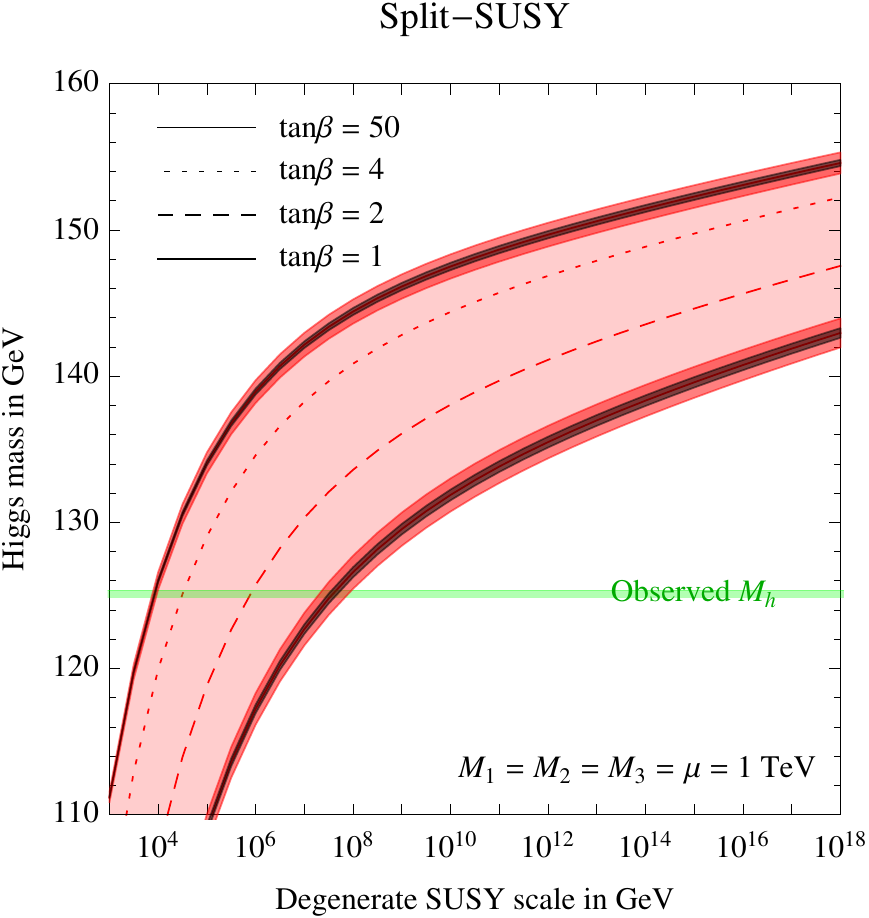}
\end{center}
\caption{Higgs mass predictions as a function of the supersymmetry breaking soft mass scale and the Higgs sector parameter $\tan \beta$, taken from \cite{Bagnaschi:2014rsa}.  In the High-Scale scenario all soft masses $\mu,\widetilde{M}_{1/2},\widetilde{m}_0$ are varied together, whereas in the Split SUSY scenario $\mu,\widetilde{M}_{1/2}$ are kept at 1 TeV and only the scalar soft masses $\widetilde{m}_0$ are varied.}
\label{fig:MSSMhiggsmass}
\end{figure}%
%%%%%%%%%%%%%%%%%

In Fig.~\ref{fig:MSSMhiggsmass} we show the expected soft mass parameter scales which reproduce the observed Higgs mass.  Clearly, within the MSSM the observed Higgs mass may be reproduced for scalar masses in the range $1 \text{ TeV} \lesssim \widetilde{m}_0 \lesssim 10^8 \text{ TeV}$.\footnote{In fact, if the soft scalar trilinear term $\tilde{A}_t$ is chosen so as to maximise the shift in the Higgs mass, the lightest stop squark could be as light as $\sim 500$ GeV \cite{Hall:2011aa}.}  Furthermore, if we consider the range $ \tan \beta > 4$, then scalar masses below $\mathcal{O}(\text{10's TeV})$ are required.  This is the first upper bound we have encountered for the scalar soft masses, resulting directly from the Higgs mass measurements.  Theoretically, this has given rise to a reduction in the allowed parameter space of supersymmetric theories and in the context of Split SUSY, where previously scalar masses could take almost any value, now the Higgs mass measurements have led to the so-called `Mini-Split' scenario \cite{Arvanitaki:2012ps,ArkaniHamed:2012gw}, where there is an upper bound on the value of the scalar soft masses.

There are variants of the MSSM in which the Higgs mass may also be raised above the MSSM tree-level prediction by utilizing additional effects deriving from couplings to new fields.  If the coupling is to new fields in the superpotential then such theories are typically variants of the NMSSM, in which the Higgs doublets couple to an additional gauge singlet.  Alternatively, the corrections may arise from coupling to new gauge fields, due to additional contributions to the quartic scalar potential predicted by supersymmetric gauge interactions.  Importantly, in these scenarios the additional enhancements of the Higgs mass only serve to reduce the required value of the radiative corrections, and hence the required value of the scalar soft masses.  Thus the required scalar soft mass values shown in Fig.~\ref{fig:MSSMhiggsmass} serve as an approximate upper limit for theories beyond the MSSM.

To summarize, the measurement of the Higgs mass has now provided information that is key to understanding the expected mass ranges of superpartners relevant to a 100 TeV collider, particularly for the stop squarks.  Although scalar masses may be as large as $\widetilde{m}_0 \sim 10^8$ TeV, for a broad range of parameter space, if it is the case that $\tan \beta > 4$ this upper bound is reduced significantly to $\widetilde{m}_0 \lesssim \mathcal{O}(10\text{'s TeV})$.   All told, the observed Higgs mass may in some cases already point towards scalar superpartners within the expected reach of a 100 TeV collider.

It should also be noted that an appealing feature of superysmmetric models is that electroweak symmetry breaking may be driven radiatively upon RG evolution from high to low scales \cite{Ibanez:1982fr,Inoue:1982pi,Ibanez:1982ee,Ellis:1983bp,AlvarezGaume:1983gj}.  This attractive feature may not be possible in all scenarios, including the Mini-Split models, depending on parameter choices.

\subsubsection{Naturalness and the Hierarchy Problem}
Finally, we arrive at a question that has been a driving force within fundamental physics research, and we find a supersymmetric answer to this question in which one of the most magical aspects of supersymmetry comes to the fore.  Briefly, before considering the hierarchy problem in detail, it is worthwhile to explain why this central feature of supersymmetric theories has been left to the end of this section.  The reason is twofold.  First, as we will see, a total supersymmetric resolution of the hierarchy problem looks increasingly under tension from LHC measurements, hence this motivation for supersymmetry is perhaps waning relative to the others, at least in its purest form, and this trend may continue with additional LHC data.  Secondly, this discussion was deliberately left until the end to reinforce the notion that it is not necessary to rely on naturalness arguments in order to discuss supersymmetry as a well-motivated new spacetime symmetry, or as an interesting phenomenological framework which may lie at the core of deep questions in fundamental physics concerning dark matter and the unification of the forces.

If the Standard Model of particle physics could be taken in isolation it would be a well-defined quantum field theory with the Higgs mass as a renormalized input parameter, which could in principle take any value desired.  However, this is not the case and the Standard Model must itself be viewed as a low energy effective description of some more fundamental theory at higher energy scales, as there are numerous reasons to expect new physics at energies $M_{\text{New}}$ far above the weak scale.  We will discuss a sample here.  The most obvious example is the theory which UV-completes QFT at the Planck scale $M_{\text{New}}=M_P \sim 10^{18}$ GeV to provide a consistent unification of quantum mechanics and general relativity, i.e. the theory of quantum gravity.  This may be preceded, at lower energies, by the grand unified theory of the gauge forces, at the scale $M_{\text{New}}=M_{GUT} \sim 10^{16}$ GeV.  This may be preceded by the Peccei-Quinn breaking scale $M_{\text{New}}=f_a \gtrsim 10^9$ GeV associated with the axion solution of the strong CP-problem, or by the right-handed neutrino mass scale $M_{\text{New}}=M_N \gtrsim 10^{11}$ GeV.  In any of these cases there should exist new fields with mass characterized by the relevant energy scale, coupled to the Higgs.  Even in the absence of new physics at these energies, hypercharge exhibits a Landau pole and becomes strongly coupled at very high energies, thus even within the SM there is reason to believe in the existence of new physics at extremely high energies, although realistically quantum gravitational effects will have entered before that scale, rendering an unambiguous discussion of this feature difficult.

The possibility of new physics at high energies is not a problem in itself, rather the problem is concerned with how the weak scale may be so far below $M_{\text{New}}$.  The reason is that even if we were to set the tree-level Higgs mass to a value hierarchically below $M_{\text{New}}$, this situation would not be stable at the quantum level.  Radiative corrections, most often depicted through one-loop diagrams, will in general give corrections to the Higgs mass, and hence the weak scale, of $\delta M_H \sim \mathcal{O} (M_{\text{New}})$.  One could choose to finely tune parameters such that all contributions contrive to cancel at low energies, leading to $M_H \ll M_{\text{New}}$ when all corrections are included.  However, in arguments elucidated by Wilson, t' Hooft \cite{Hooft:1979bh}, and specifically quoting Susskind \cite{Susskind:1978ms}, ``observable properties of a system should not depend sensitively on variations in the fundamental parameters''.  This is the core of the hierarchy problem: a finely-tuned scenario for the weak scale is unnatural, seemingly implausible, although still possible.

The supersymmetric solution to this problem is straightforward to sketch.  All fermions enjoy a chiral symmetry acting on their individual Weyl components.  A fermion mass, whether Dirac or Majorana, breaks this chiral symmetry.  This means that if the mass, and hence breaking of the chiral symmetry, is small then a fermion may remain naturally smaller than other mass scales in the theory and this will remain true at the quantum level.  In fact, we are already familiar with this in the Standard Model.  While we may wonder at the origin of the huge hierarchy between the electron mass and the tau mass, we do not puzzle over the quantum stability of this mass difference.  This lies at the core of the supersymmetric resolution of the hierarchy problem.  Supersymmetry ties the mass of a scalar field to the mass of its fermionic superpartner, and since supersymmetry does not break the chiral symmetry enjoyed by the fermion, and the chiral symmetry protects the fermion mass from large quantum corrections, so too must the mass of its scalar partner be protected, by proxy.

This means that in a supersymmetric theory it is perfectly natural for the mass of the individual components of a superfield to be hierarchically below other mass scales in the theory, even if two superfields with vastly separated masses are coupled to each other with $\mathcal{O}(1)$ couplings.  This is extraordinary and is quite at odds with naive intuition, which is what makes this property of supersymmetry so magical.  In practice it means that in a supersymmetric theory the weak scale could be comfortably below new physics at a scale $M_{\text{New}}$, even if this is identified with the Planck scale.  Thus supersymmetry may provide a concrete foundation for the Standard Model fields all the way up to the scale of quantum gravity.

Of course in nature supersymmetry must be broken and once the symmetry is broken at a scale $\tilde{m}$, which represents the soft mass scale, the Higgs mass is no longer protected from quantum corrections.  Thus supersymmetry is effective in protecting the Higgs mass all the way down from a high mass scale to the supersymmetry breaking scale $M_{\text{New}} \to \tilde{m}$, however from the soft mass scale down to the weak scale, $\tilde{m} \to m_h$ supersymmetry is no longer present.  This means that for a natural theory without tuning we must expect $\tilde{m} \sim m_h$, and conversely if $\tilde{m} \gg m_h$ there must be some fine tuning to realize the weak scale below the soft mass scale.  These qualitative arguments may be made quantitative.  A well motivated measure for the degree of tuning in the weak scale with respect to a given fundamental parameter in the theory, $a$, is \cite{Ellis:1986yg,Barbieri:1987fn}
\begin{equation}
\Delta [a] = \frac{\partial \log M_Z^2}{\partial \log a^2}  ~~.
\end{equation}
By minimising the weak scale potential at large $\tan \beta$ we find
\begin{equation}
M_Z^2 = -2 (m^2_{H_u} + |\mu|^2)  ~~,
\end{equation}
where $m^2_{H_u}$ is the soft mass for the up-type Higgs which includes all radiative corrections.  Let us consider the tree-level contribution from the $\mu$-term, along with the one-loop contributions from stop squarks and the winos, and the two-loop contribution from gluinos, which are given by
\begin{eqnarray}
\delta m^2_{H_u} (\tilde{t}) &=& -\frac{3 y_t^2}{4 \pi^2} m^2_{\tilde{t}} \log (\Lambda/m_{\tilde{t}}) \\
 \delta m^2_{H_u} (\widetilde{W}) &=&-\frac{3 g^2}{8 \pi^2} (m^2_{\widetilde{W}} +m^2_{\tilde{h}})\log (\Lambda/m_{\widetilde{W}}) \\
  \delta m^2_{\tilde{t}} &=& \frac{2 g_s^2}{3 \pi^2} m_{\tilde{g}}^2 \log (\Lambda/m_{\tilde{g}}) ~~,
\end{eqnarray}
where $\Lambda$ is a UV-cutoff at which the full UV-completion of supersymmetry kicks in, and the last term may be inserted into the first to obtain an estimate of the tuning from gluinos.  Conservatively taking $\Lambda=10$ TeV we arrive at the following expectations for a theory which is only tuned at the $10\%$ level \cite{Craig:2013cxa}:
\begin{equation}
\mu \lesssim 200 \text{ GeV} ~~,~~ m_{\tilde{t}} \lesssim 400 \text{ GeV} ~~,~~ m_{\widetilde{W}} \lesssim 1 \text{ TeV} ~~,~~ m_{\tilde{g}} \lesssim 800 \text{ GeV} ~~,~~ 
\end{equation}
This picture is clearly at odds with the stop mass values required to achieve the observed Higgs mass in the MSSM, shown in Fig.~\ref{fig:MSSMhiggsmass}.  However it may be that non-minimal structure beyond the MSSM lifts the Higgs mass without requiring large stop masses, thus this constraint is not too significant.  More importantly, current constraints on the Higgs boson couplings, which would typically be modified if the stop squarks were light, already place stringent constraints on light stop scenarios.  Furthermore, direct searches for stops and gluinos, already show that significant portions of this parameter space are in tension with LHC 8 TeV data (for a thorough overview see \cite{Craig:2013cxa}).  Finally, in many (but not all) concrete scenarios it is expected that the first two generation squarks should not be significantly heavier than the stop squarks and, as the production cross section is enhanced due to valence quarks in the initial state, constraints on first two generation squarks are very strong, indirectly placing strong constraints on the naturalness of many supersymmetric theories.

Where do these strong constraints leave the supersymmetric solution to the hierarchy problem?  As we are on the brink of a paradigm shift in our understanding of electroweak naturalness a number of possibilities are plausible.

It could be that the weak scale is meso-tuned, as in Mini-Split supersymmetry, and the \ae sthetic motivations for supersymmetry as a new spacetime symmetry are justified, whereas the naturalness arguments were misguided, to at least some degree, since supersymmetry does solve the big hierarchy problem and we are instead left with a relatively small tuning of the weak scale up to energies as high as $\mathcal{O} (10^8)$ TeV.  This scenario is in some sense quite successful.  A fundamental Higgs boson of mass $m_h \lesssim 135$ GeV is predicted, gauge coupling unification and successful dark matter candidates are realized, all at the cost of accepting some meso-tuning.  Although not necessarily guaranteed, the gauginos should be below mass scales of $\sim \mathcal{O}(\text{few TeV})$, mostly driven by the dark matter requirement.

Another possibility which has only recently been explored is that the Mini-Split spectrum is realized in nature, with all of the above successes, however the theory is not actually tuned due to a hidden dynamical mechanism which renders the hierarchy from the weak scale to the soft mass scale natural \cite{Batell:2015fma}.  This can be achieved by employing the cosmological relaxation mechanism of \cite{Graham:2015cka} in a supersymmetric context.  In this case both the \ae sthetic arguments for supersymmetry and the naturalness arguments for the weak scale were well founded, however the two may have manifested in an entirely unexpected manner, with a cocktail of symmetries and dynamics protecting the naturalness of the weak scale up to the highest energies.  As before, the gauginos should be below mass scales of $\sim \mathcal{O}(\text{10's TeV})$, however this expectation comes from the fact that a loop factor suppression between scalars and gauginos is expected in this model and in addition the scalars cannot be arbitrarily heavy due to the finite cutoff of the cosmological relaxation mechanism.

Alternatively, a reevaluation of the fine-tuning in the infrared may be required if a spectrum with heavy squarks is made natural due to correlations between soft mass UV-boundary conditions and the infrared value of the Higgs mass, as in `Focus Point' supersymmetry \cite{Feng:1999zg} or in the recently proposed `Radiatively-Driven' natural Supersymmetry \cite{Baer:2012up}.  In these cases gauginos, Higgsinos, and most likely also stop and sbottom squarks are expected to still be in the sub-$10$ TeV range.  The first two generation squarks may be somewhat heavier.
 
Finally, it is still possible that the weak scale is relatively natural due to supersymmetry, however the sparticles have evaded detection until now.  If this is the case it is likely the stop squarks are still relatively light, in the range of a few 100's of GeV, and the Higgs mass is raised by an additional tree-level term.  For the stop squarks to evade detection there are a number of possible scenarios.  We will discuss just a few here.  One is an example of a so-called `compressed' spectrum (see e.g.\ \cite{LeCompte:2011cn}), where the mass splitting between the stop and the stable neutralino is so small that the tell-tale missing energy signature carried away by the neutralino is diminished to the point of being unobservable.  Another possibility is `Stealth Supersymmetry' \cite{Fan:2011yu,Fan:2012jf}, where again the missing energy signatures are diminished, however in this case from sparticle decays passing through a hidden sector.  Yet another possibility is for R-parity violating decays of the superpartners \cite{Barbier:2004ez}, since in this case missing-energy signatures are removed and the collider searches must contend with larger backgrounds (see e.g.\ \cite{Csaki:2011ge} for models and collider phenomenology).  For a natural spectrum the first two generations of squarks must also have evaded detection.  One possibility is to raise their mass above experimental bounds, which is compatible with naturalness if they stay within an order of magnitude or so of the gluinos and stops \cite{Papucci:2011wy,Dimopoulos:1995mi,Cohen:1996vb}.  Dirac gauginos also offer opportunities for suppressing collider signatures, at no cost to the naturalness of the theory \cite{Fox:2002bu,Kribs:2012gx}, as Dirac gauginos may naturally be heavier than their Majorana counterparts.  This scenario allows not only for the suppression of gluino signatures at the LHC, but also suppresses the t-channel gluino exchange production of the first two generation squarks.

In summary, if we wish for supersymmetry to provide a comprehensive solution to the electroweak hierarchy problem, then the full cohort of sparticles should lie below $\mathcal{O}(\text{few TeV})$.  Otherwise we are forced into considering at least some fine tuning of the weak scale or alternatively the introduction of an additional mechanism, beyond supersymmetry, to enable a natural weak scale.

\subsubsection{Summary}
Having whetted our appetite with a variety of theoretical considerations we are now well placed to understand the connection between theoretically motivated mass ranges in supersymmetry and the experimental reach of a 100 TeV collider for supersymmetry.  A brief summary of the theory motivation for superpartner mass ranges is as follows.  
\begin{itemize}
\item  Supersymmetric dark matter leads us to expect electroweak fermions comprising some admixture of the bino, wino, and/or Higgsino, with mass below $\sim$ few TeV.  If the bino and wino masses are not hierarchically separated from the gluino mass then we may also expect gluinos below $\mathcal{O}(10\text{'s TeV})$.

\item  Expectations from gauge coupling unification and the unification of matter particles in representations of the unifying gauge symmetry group are similar, and motivate the existence of superpartners of the Standard Model particles with masses $\lesssim \mathcal{O}(\text{10 TeV})$. More detailed predictions for the superpartner masses are possible in the context of specific unified models.

%Expectations motivated by gauge coupling unification are similar, although less robust, with gauginos below $\mathcal{O}(10\text{'s TeV})$, and potentially also Higgsinos, although it is difficult to quantify the expected upper limit.

\item  The measured Higgs mass points towards scalar superpartners below $\sim 10^8$ TeV, and in many well-motivated cases the upper limit may be as low as $\mathcal{O}(10\text{'s TeV})$.  Strictly speaking this applies mostly to the stop squarks, however in many models they are within an order of magnitude of the first two generation squarks as well.
\item  Naturalness points towards superpartners that are as light as can be possible given current experimental constraints.  With stops ideally below $\sim$ 400 GeV, gluinos below $\sim$ few TeV, and first two generation squarks below $\sim$ few TeV.  Relaxing the naturalness criterion raises the masses at the price of increased fine tuning.  If the weak scale is natural then it is likely that supersymmetry has been hidden by an exotic scenario, that may require specialized techniques to dig the signal out from background.
\end{itemize}
Let us now consider the experimental prospects for supersymmetry at 100 TeV.  Numerous studies have already shown that a potential proton-proton collider operating at $\sqrt{s} = 100$ TeV greatly extends the kinematic reach for superpartners, into the many-TeV range \cite{Cohen:2013xda, Jung:2013zya, Low:2014cba, Cohen:2014hxa, Ellis:2014kla, Acharya:2014pua, Gori:2014oua, diCortona:2014yua, Berlin:2015aba, Beauchesne:2015jra}.
Previous studies have focused primarily on pair production of superpartners, both strongly-interacting \cite{Cohen:2013xda, Jung:2013zya, Cohen:2014hxa} and weakly-interacting \cite{Low:2014cba, Acharya:2014pua, Gori:2014oua, diCortona:2014yua, Berlin:2015aba}.  In Sec.~\ref{sec:susy_xs} we will consider the pair production cross sections for various superpartners at 100 TeV, including NLO corrections.  In Secs.~\ref{sec:SUSY_stop} to \ref{sec:susy_ewino} we will focus on searches specific to particular superpartners, often employing simplified models.  In Sec.~\ref{sec:SUSY_exotic} we discuss the exotic signatures of long-lived charged superpartners and in Sec.~\ref{sec:SUSY_indirect} potential indirect constraints on stop squarks from modifications of Higgs pair production.  In Sec.~\ref{sec:SUSY_model} potential measurements at a 100 TeV collider are interpreted in the context of two specific supersymmetric models, the `constrained MSSM' and Mini-Split Supersymmetry.  Sec.~\ref{sec:SUSY_post} discusses the next steps to be made after discovering supersymmetry at a 100 TeV collider.  Finally, in Sec.~\ref{sec:susy_summary} we summarize, focussing on a broad characterization of the expected reach of a 100 TeV collider and the potential implications for our understanding of supersymmetry.

\subsection{Cross Sections for Production of SUSY Particles}%\footnote{Eds: M.Kr\"amer, J.Lindert}}
\label{sec:susy_xs}
In this section we present reference cross sections for the production
of SUSY particles at 100\;TeV. 

We first focus on the pair production of squarks and gluinos, 
\begin{equation}
pp \to \tilde{q}\tilde{q}, \tilde{q}\tilde{q}^{*}, \tilde{q}\tilde{g}, \tilde{g}\tilde{g} + X,
\label{eq:prod}
\end{equation}
where the charge conjugated processes $pp \to \tilde{q}^*\tilde{q}^*$
etc.\ are included. We assume 10 mass-degenerate squark flavours, 
$\tilde{q} \in \{u_{L/R},d_{L/R},c_{L/R},s_{L/R},b_{L/R}\}$ and have
suppressed the chirality labels in Eq.(\ref{eq:prod}) and below. The production of
stops is treated separately, as the large Yukawa
coupling between top quarks, stops and Higgs fields gives rise
to potentially large mixing effects and mass splitting. Thus, for stop
production we consider the pair production of the lighter mass eigenstate, $\tilde{t}_1$,  
\begin{equation}
pp \to \tilde{t}_1 \tilde{t}_1^* + X. 
\end{equation} 

First, in Figure~\ref{fig:xsect_all_100} we show cross section
predictions for the various squark and gluino production processes,
assuming degenerate squark and gluino masses. For squark/gluino masses near 2\;TeV 
the inclusive cross section is of the order 100\;pb. The relative
size of the various production channels depends on the squark/gluino 
masses and is driven by the corresponding parton luminosities. The cross
sections include NLO SUSY-QCD corrections \cite{Beenakker:1996ch, Beenakker:1997ut} and the resummation of
threshold logarithms at next-to-leading logarithmic (NLL) accuracy
\cite{Beenakker:2009ha, Beenakker:2010nq}, as
described in Ref.\cite{Borschensky:2014cia}. 

\begin{figure}[t]
  \begin{center}
       	\includegraphics[width=11.5cm]{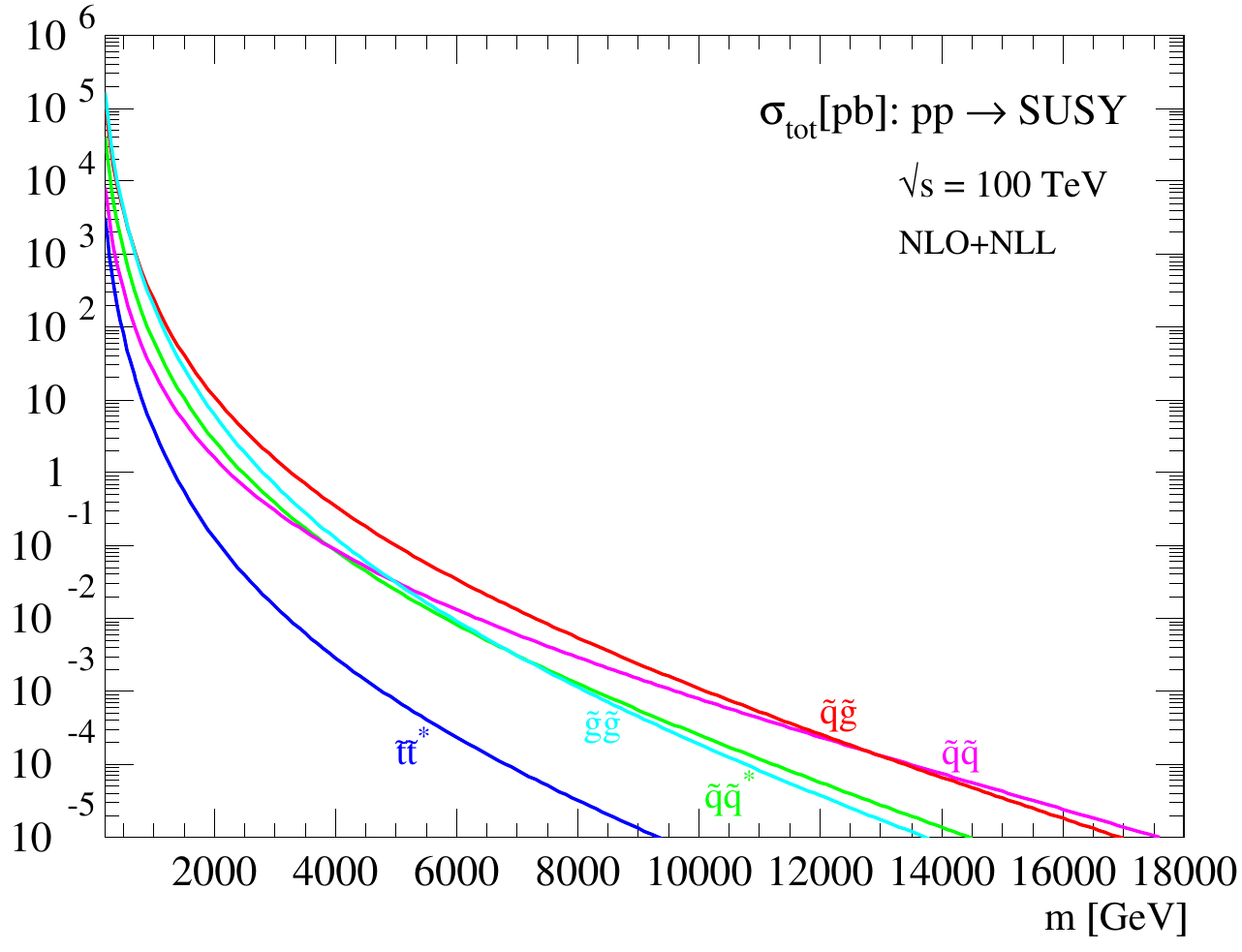}
  \end{center}
  \caption{NLO+NLL cross sections for squark and gluino
    pair-production, $pp \to \tilde{q}\tilde{q},
    \tilde{q}\tilde{q}^{*}, \tilde{q}\tilde{g},
    \tilde{g}\tilde{g},\tilde{t}_1 \tilde{t}_1^*  + X$, at $\sqrt{S} =
    100$\;TeV, as a function of the degenerate squark and gluino mass
    $m_{\tilde{q}}=m_{\tilde{g}}=m$. From Ref.~\cite{Borschensky:2014cia}.}
  \label{fig:xsect_all_100}
\end{figure}

We will now consider individual production processes in more detail,
starting with gluino pair production in a simplified model with the
squarks decoupled. In Fig.~\ref{fig:xsect_gluino100} we show the NLO+NLL
cross section, $pp \to \tilde{g}\tilde{g}$, including the theoretical
uncertainty from scale variation and the parton distribution
functions, as determined following the procedure described in
Ref.~\cite{Borschensky:2014cia}. The individual sources of the uncertainty are shown in
the lower plot for the mass range $1\;{\rm TeV} \le m_{\tilde{g}} \le
4\;{\rm TeV}$. Fig.~\ref{fig:xsect_squark100} shows the corresponding
results for 
squark-antisquark production, $pp \to \tilde{q}\tilde{q}^{*}$, with
gluinos decoupled. 

\begin{figure}[t]
  \begin{center}
       	\includegraphics[width=7.5cm]{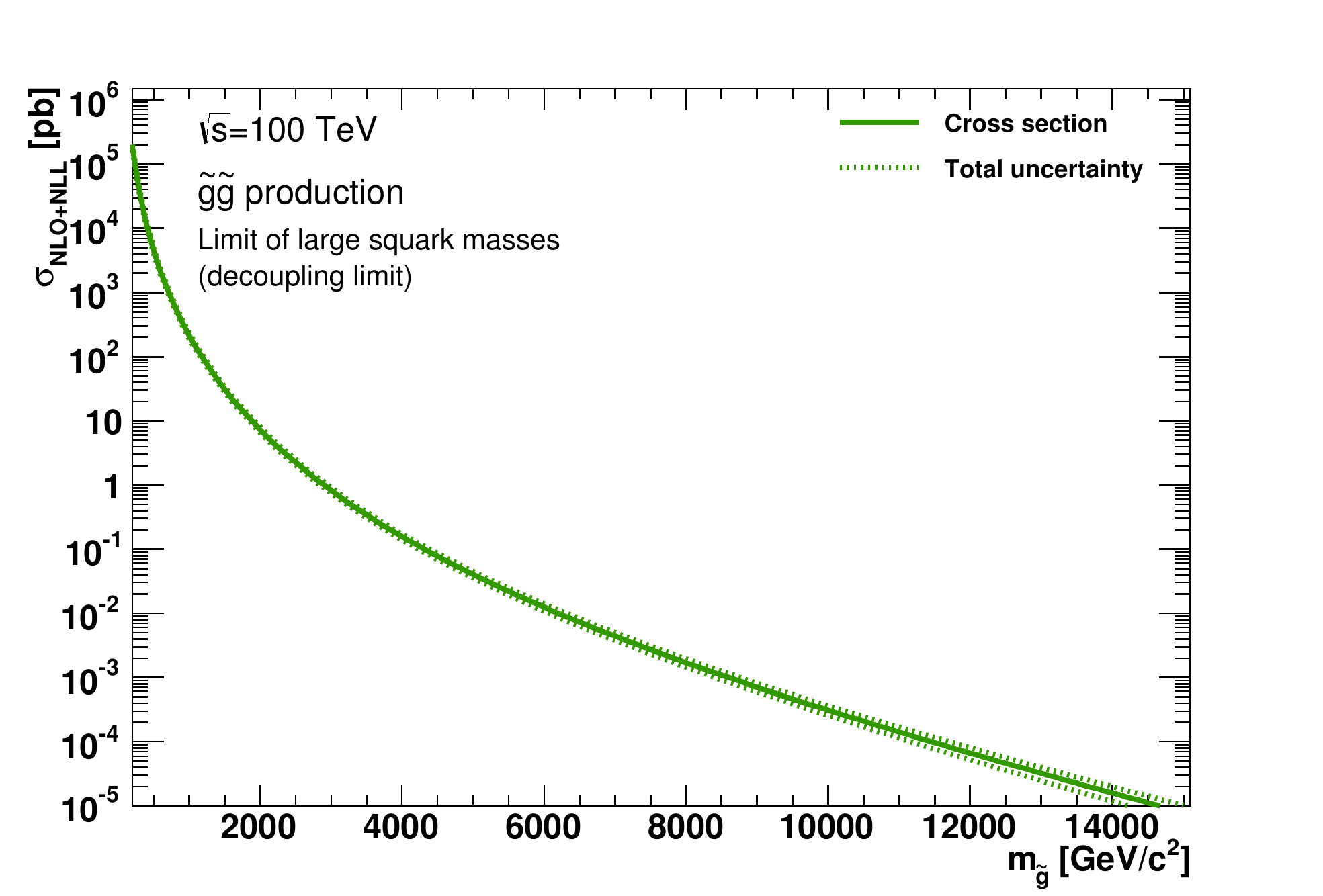}
       	\includegraphics[width=7.5cm]{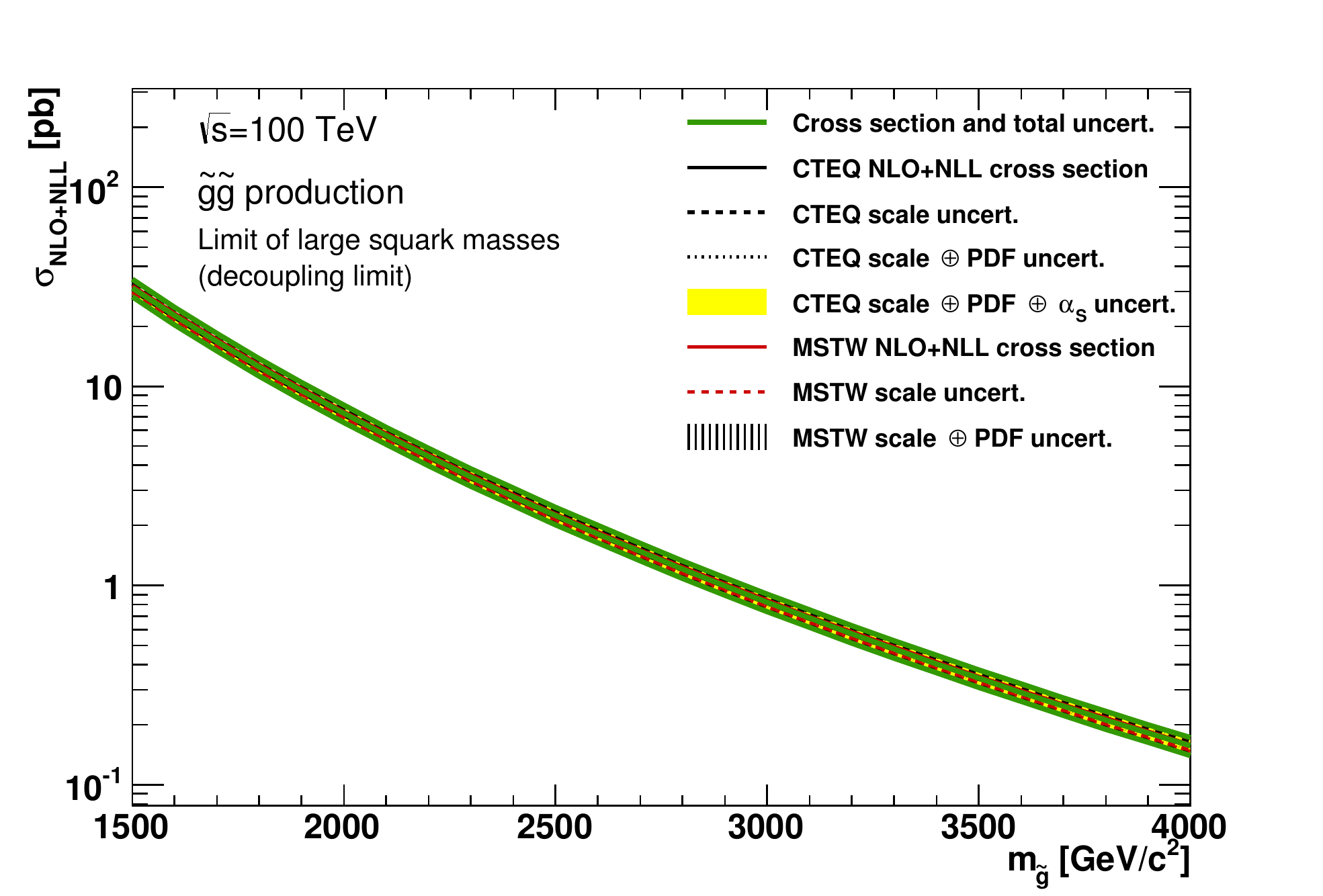}
  \end{center}
  \caption{NLO+NLL cross section for gluino
    pair-production, $pp \to 
    \tilde{g}\tilde{g} + X$, at $\sqrt{S} =
    100$\;TeV, as a function of the gluino mass with squarks 
    decoupled. 
The black (red) lines correspond to the cross section and scale
uncertainties predicted using the CTEQ6.6 \cite{Nadolsky:2008zw} (MSTW2008 \cite{Martin:2009iq}) pdf set. The yellow (dashed black) band corresponds to the total CTEQ6.6 (MSTW2008) uncertainty, as described in \cite{Borschensky:2014cia}.
The green lines show the final cross section and its total
uncertainty. From Ref.~\cite{Borschensky:2014cia}.}
\label{fig:xsect_gluino100}
\end{figure}

\begin{figure}[t]
  \begin{center}
       	\includegraphics[width=7.5cm]{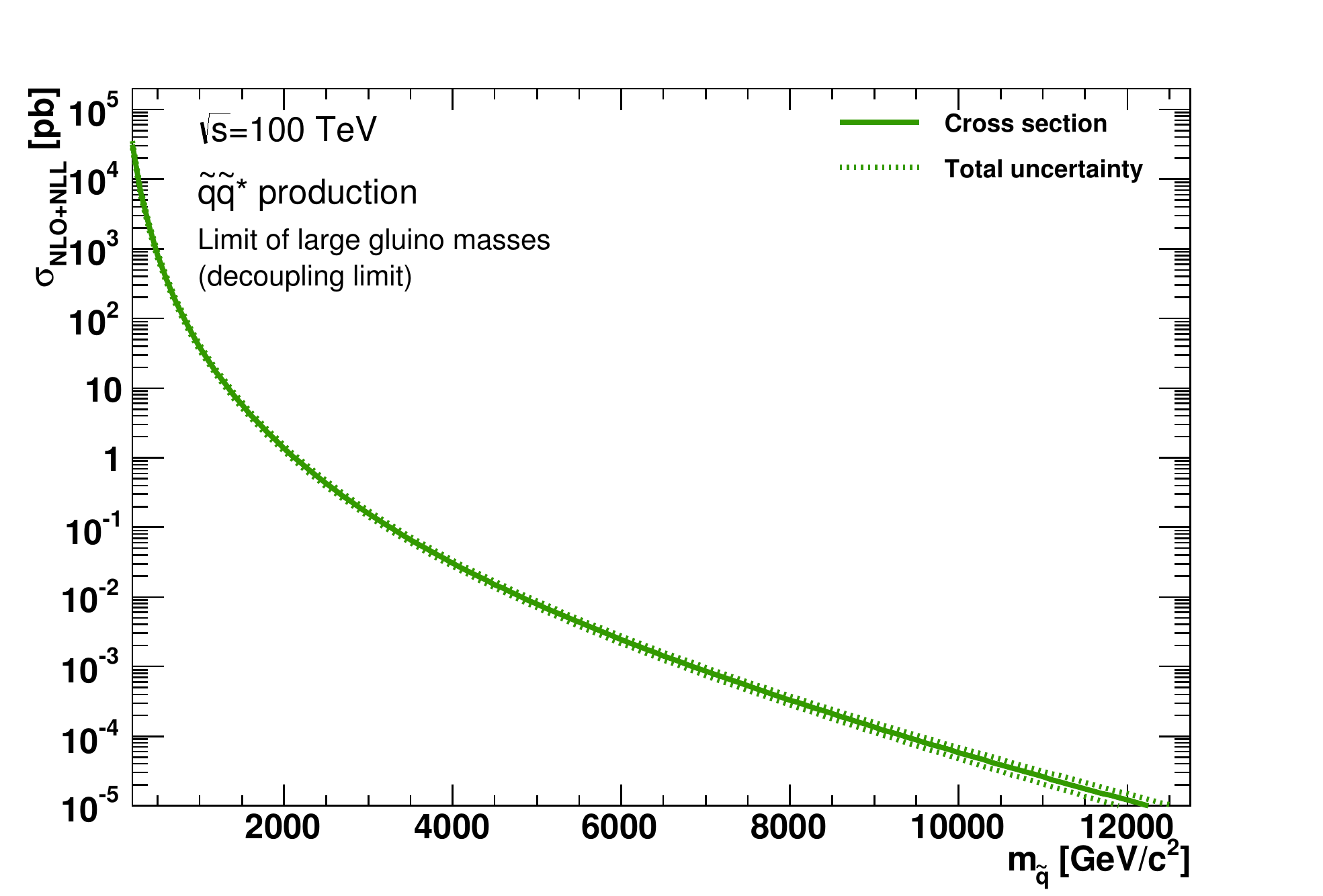}
       	\includegraphics[width=7.5cm]{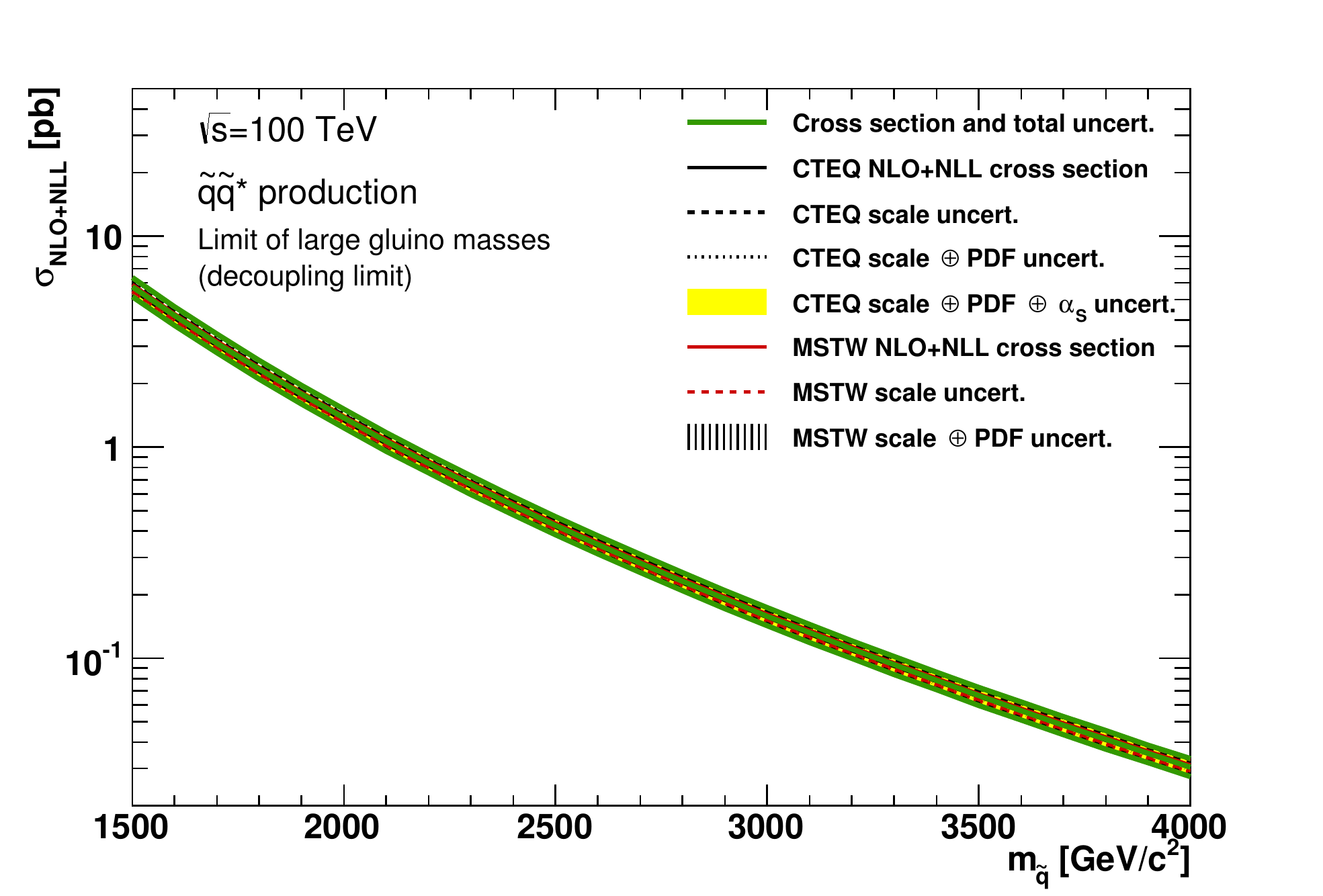}
  \end{center}
  \caption{NLO+NLL cross section for squark-antisquark
    production, $pp \to 
    \tilde{q}\tilde{q}^* + X$, at $\sqrt{S} =
    100$\;TeV, as a function of the squark mass with gluinos
    decoupled. 
The black (red) lines correspond to the cross section and scale
uncertainties predicted using the CTEQ6.6 \cite{Nadolsky:2008zw} (MSTW2008 \cite{Martin:2009iq}) pdf set. The yellow (dashed black) band corresponds to the total CTEQ6.6 (MSTW2008) uncertainty, as described in \cite{Borschensky:2014cia}.
The green lines show the final cross section and its total
uncertainty. From Ref.~\cite{Borschensky:2014cia}.}
  \label{fig:xsect_squark100}
\end{figure}

Finally, in Fig.~\ref{fig:xsect_stop100}  we show the cross section
for the pair production of the lighter stop mass eigenstate in a model
where all other sparticles are decoupled. Note that these cross
sections are approximately equal to the cross section for the lighter
sbottom mass eigenstate, 
assuming that the rest of the coloured SUSY spectrum is
decoupled.

\begin{figure}[t]
  \begin{center}
       	\includegraphics[width=7.5cm]{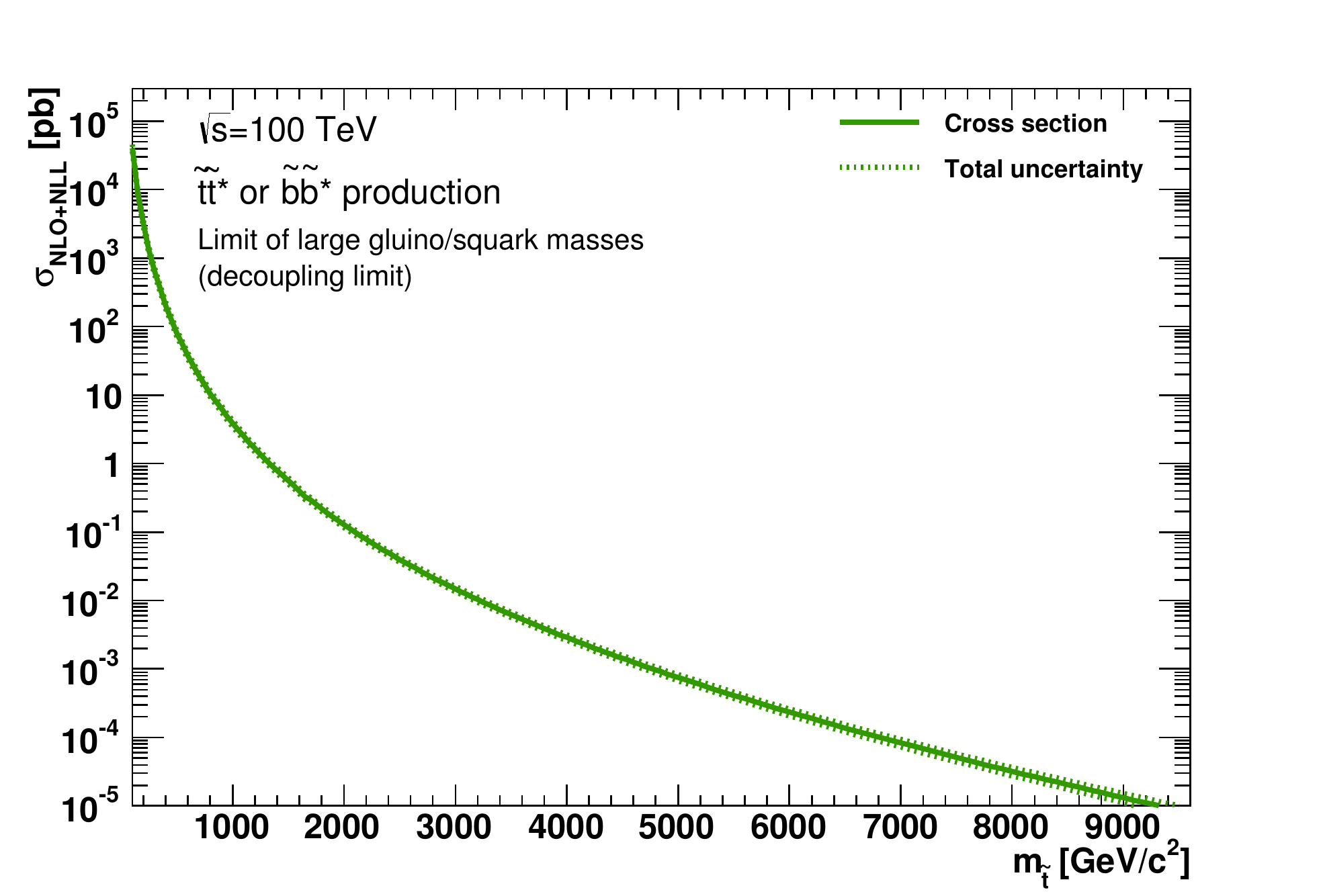}
       	\includegraphics[width=7.5cm]{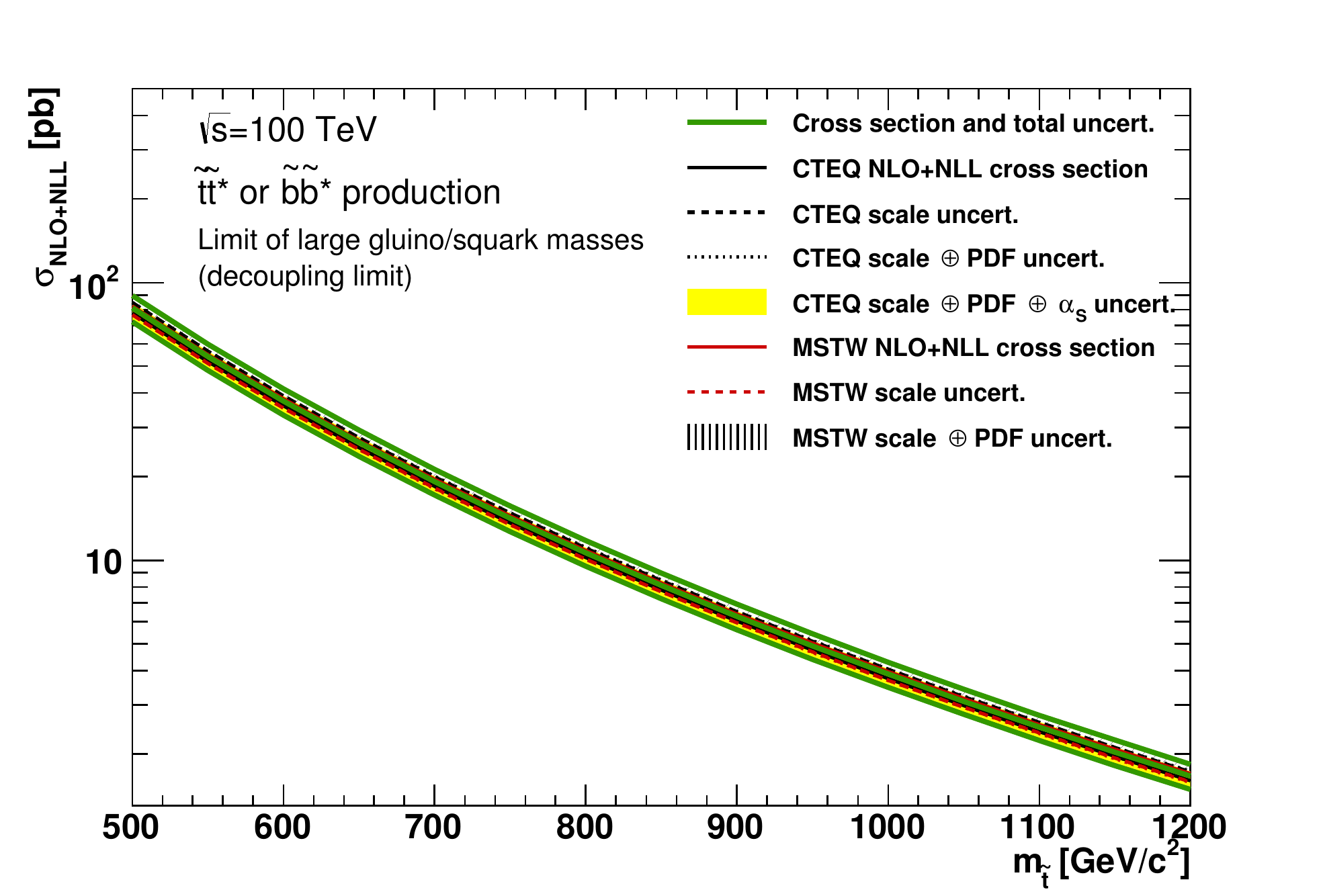}
  \end{center}
  \caption{NLO+NLL cross section for stop-antistop production, $pp \to 
    \tilde{t}_1\tilde{t}^*_1 + X$, at $\sqrt{S} =
    100$\;TeV, as a function of the stop mass with all other coloured
    sparticles     decoupled. The black (red) lines correspond to the cross section and scale
uncertainties predicted using the CTEQ6.6 \cite{Nadolsky:2008zw} (MSTW2008 \cite{Martin:2009iq}) pdf set. The yellow (dashed black) band corresponds to the total CTEQ6.6 (MSTW2008) uncertainty, as described in \cite{Borschensky:2014cia}.
The green lines show the final cross section and its total
uncertainty. From Ref.~\cite{Borschensky:2014cia}.}
  \label{fig:xsect_stop100}
\end{figure}

Besides higher-order QCD corrections, the production of squarks and gluinos 
receives Born-level~\cite{Bornhauser:2007bf} and higher-order electroweak (EW)
contributions~\cite{Hollik:2008yi,Hollik:2008vm,Mirabella:2009ap,Germer:2010vn,Germer:2011an,Germer:2014jpa,Hollik:2015lha}. 
These EW corrections are enhanced well above the TeV scale due to large logarithms of 
Sudakov type.  For $\sqrt{\hat s} \gg M_W$, NLO EW corrections can be 
at the level of several tens of percent of the LO cross section. In Fig.~\ref{fig:xsect_nloew} 
we illustrate the effect of such EW corrections for the case of squark-antisquark 
(left) and stop-antistop (right) production, where for squark-antisquark production 
we separate different chirality combinations (LL, RR, LR+RL). The production 
of left-handed squark-antisquark pairs receives NLO EW corrections with respect to the LO 
predictions of up to $-30\%$, while for the other production modes and for stop-antistop production
NLO EW corrections are smaller. These large NLO EW corrections are partly compensated 
(or even overcompensated) by the contribution from photon-induced production. However, these 
contributions are accompanied by very large intrinsic PDF uncertainties~\cite{Hollik:2015lha}, which may substantially 
alter the size of the electroweak corrections.
Overall, any precision study of SUSY particle production in the multi-TeV range should include 
higher-order EW corrections and photon-induced production.

\begin{figure}[htbp]
  \begin{center}
       	\includegraphics[width=.49\textwidth]{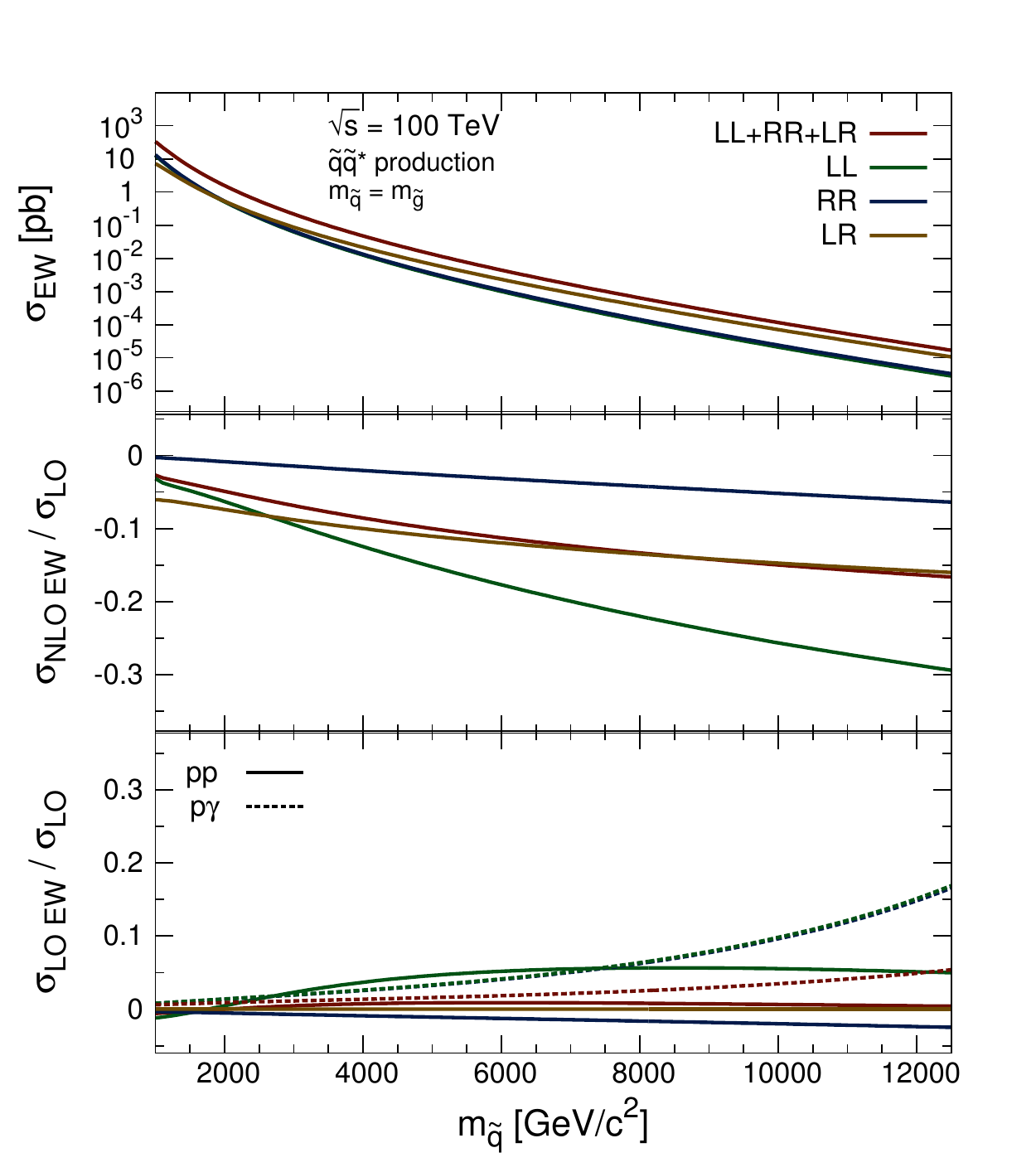}
       	\includegraphics[width=.49\textwidth]{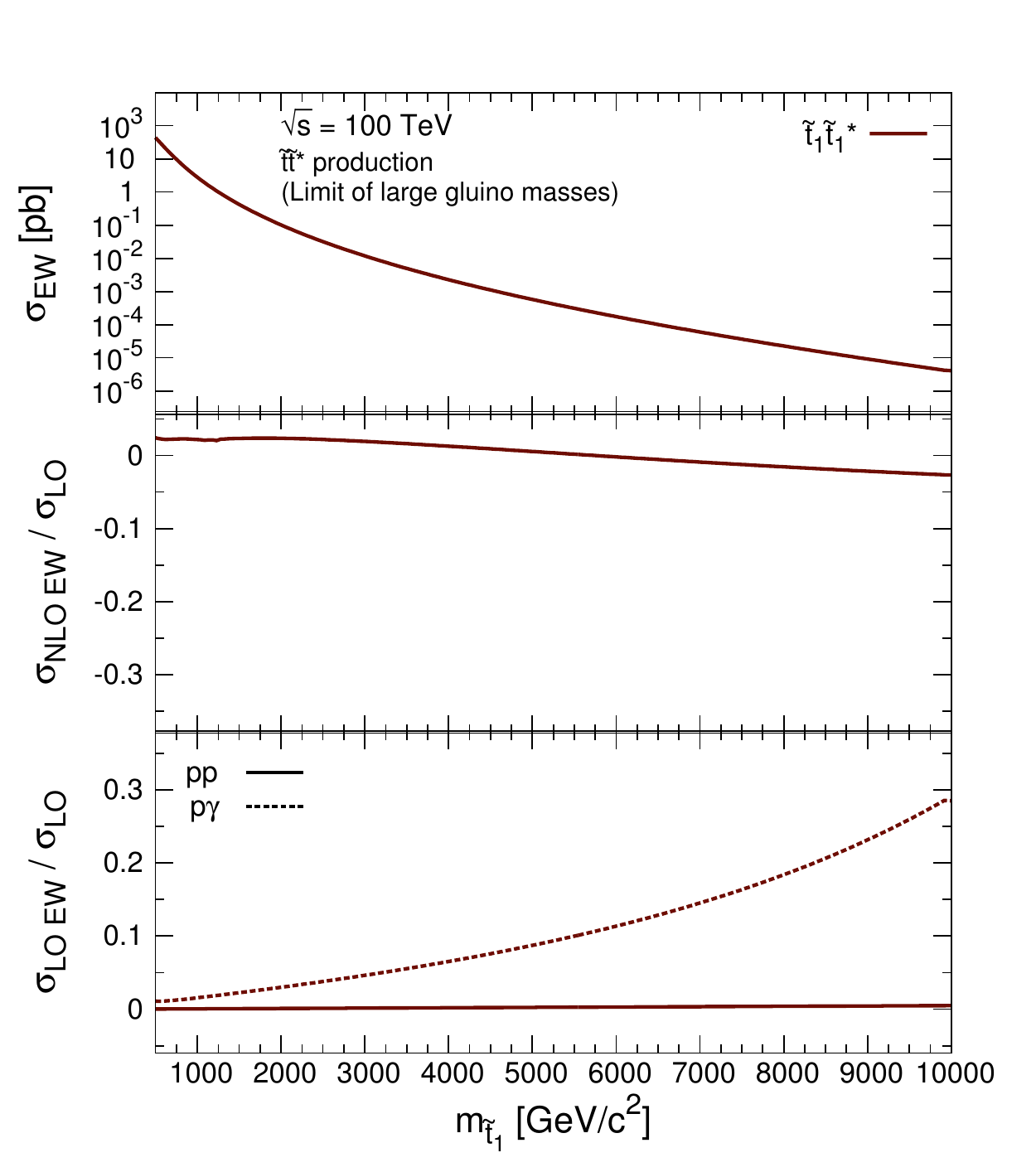}
  \end{center}
  \caption{Cross sections for squark-antisquark production, $pp \to \tilde{q}\tilde{q}^* + X$, and stop-antistop production, $pp \to \tilde{t}_1\tilde{t}^*_1 + X$, at $\sqrt{S} = 100$\;TeV including EW corrections $\sigma_{\mathrm{EW}}=\sigma_{\mathrm{LO}}+\sigma_{\mathrm{LO\,EW}}+\sigma_{\mathrm{NLO\, EW}}$, as a function of the produced squark/stop mass. In the case of squark--antisquark production all squarks and the gluino have the same mass $m_{\tilde q}=m_{\tilde g}$, while in the case of stop-antistop production the gluino is decoupled and all light-flavor squark masses are set to $m_{\tilde q}=5000$ GeV. All cross sections are obtained using NNPDF2.3QED \cite{Ball:2013hta}.}
  \label{fig:xsect_nloew}
\end{figure}

The associated production of neutralinos with squarks and 
gluinos provides a complementary probe of SUSY particle
production. In Fig.~\ref{fig:xsect_n1sq_100} we present the (leading-order) cross section for 
$pp \to \tilde{\chi}_1^0+ \tilde{q}$ in a simplified model
with degenerate squarks of the first two generations  
and a bino $\tilde{\chi}_1^0$. All other SUSY particles are
decoupled. The cross section for $pp \to \tilde{\chi}_1^0+ \tilde{g}$
is shown in Fig.~\ref{fig:xsect_n1go_100}. Again, we consider a pure
bino $\tilde{\chi}_1^0$, and set the gluino and the squarks of the
first two generations to a common mass. The cross sections have been
obtained with {\tt MadGraph5}~\cite{Alwall:2014hca}.

\begin{figure}[htp]
  \begin{center}
       	\includegraphics[bb=100 430 600 700, clip]{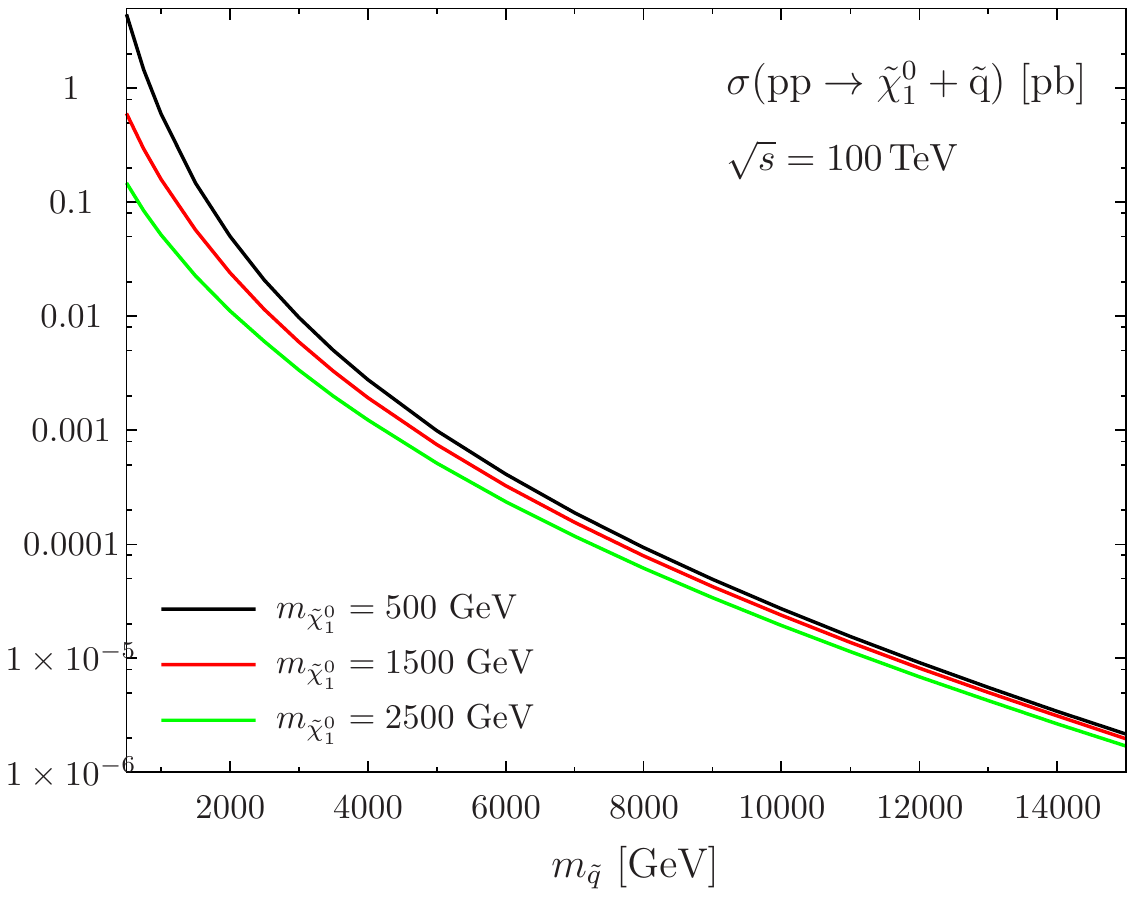}
  \end{center} 
  \caption{LO cross section for neutralino-squark associate 
    production, $pp \to \tilde{\chi}_1^0 + \tilde{q}$, at $\sqrt{s} =
    100$\;TeV, as a function of the squark mass for three different
    values of the neutralino mass. We assume a
    simplified model with degenerate masses for the squarks of the
    first two generations, a pure bino $\tilde{\chi}_1^0$, and all
    other SUSY particles decoupled.}
  \label{fig:xsect_n1sq_100}
\end{figure}

\begin{figure}[htp]
  \begin{center}
       	\includegraphics[bb=100 430 600 700, clip]{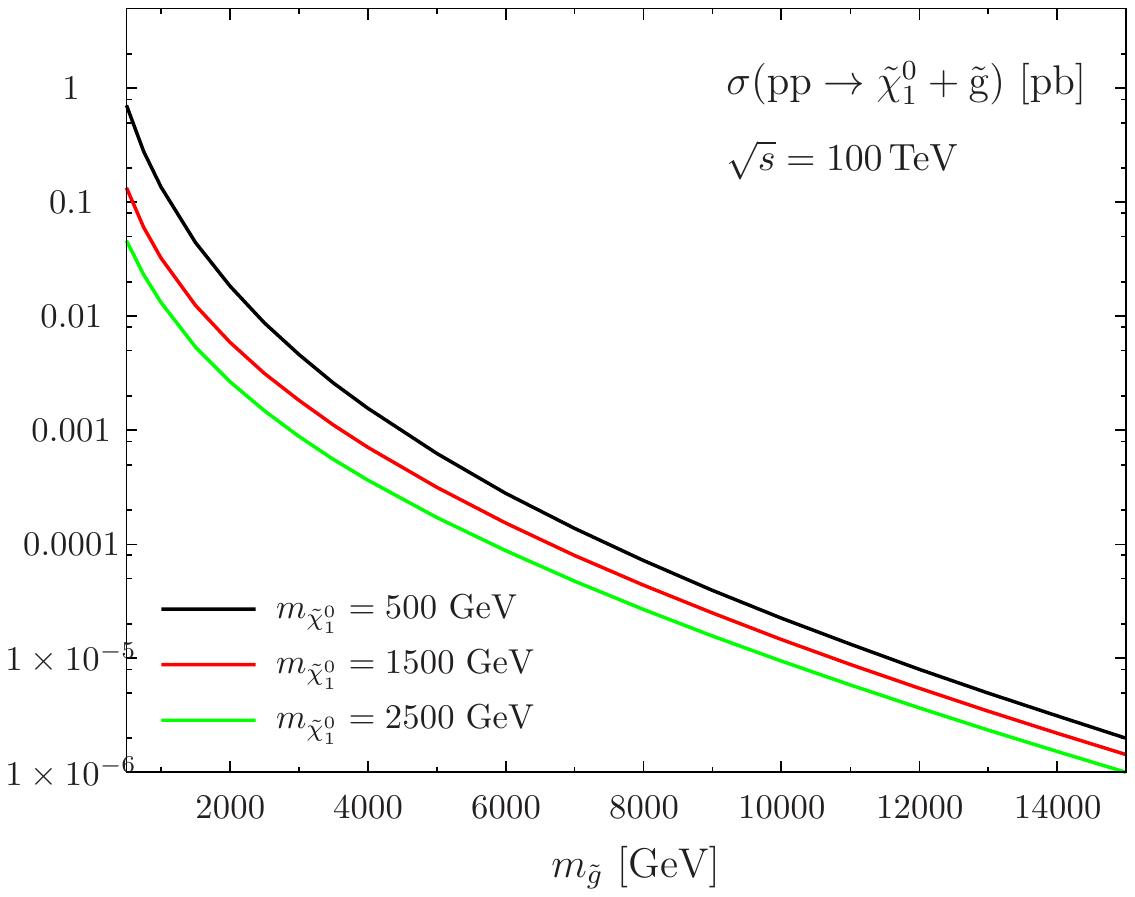}
  \end{center} 
  \caption{LO cross section for neutralino-gluino associate 
    production, $pp \to \tilde{\chi}_1^0 + \tilde{g}$, at $\sqrt{S} =
    100$\;TeV, as a function of the gluino mass  for three different
    values of the neutralino mass. We assume a
    simplified model with degenerate masses for the gluino and the squarks of the
    first two generations, a pure bino $\tilde{\chi}_1^0$, and all
    other SUSY particles decoupled.}
  \label{fig:xsect_n1go_100}
\end{figure}

\subsection{Stop Squarks}
\label{sec:SUSY_stop}
The largest radiative correction to the Higgs potential arises from top loops, thus the scalar partner of
the top (stop) is of critical importance for understanding if supersymmetry solves the hierarchy problem.  In this section, we will study the reach for stops at a future 100 TeV hadron collider.

Motivated by dark matter and proton decay, we consider $R$-parity to be a good symmetry and imagine a neutral
lightest supersymmetric particle (LSP) that is  stable on collider time scales.
We will refer to the LSP as a neutralino (\chioz),
but it could have quantum numbers which differ from the usual MSSM neutralinos.
Thus we take a simplified model which consists of a stop and a much lighter neutralino,
and, in this model, the decay $\Stop\rightarrow t \chioz$ occurs 100\% of the time.

\subsubsection{Leptonic Decays}

The LHC experiments have performed many searches for stops~\cite{Khachatryan:2015wza,Aad:2015pfx}, and such searches will be an important piece of the LHC and HL-LHC physics programs.
However, the kinematic regime accessible at $\sqrt{s}=100$~TeV is completely different from that of the LHC, and will require new search strategies.
From the cross sections shown in Section~\ref{sec:susy_xs} we see that a 100 TeV machine will easily produce multi-TeV stops, so we expect a large fraction of
the parameter space to contain multi-TeV top quarks.  

LHC techniques fail at higher energies precisely because the top quarks are highly boosted.
In Figure~\ref{fig:SUSY_stop_topboost} we show the top quark \pt{} and the average $\Delta R$ between its decay products as a function of the stop and neutralino masses.
In most of the parameter space accessible at $\sqrt{s}=100$~TeV the top decay products are contained in a cone of the same size of an LHC jet: $\Delta R \lesssim 0.5$.
In some cases the separation between them is even smaller than the size of LHC calorimeter cells.
For example, an $8$~TeV stop and a light neutralino give a large fraction
of top quarks with $p_T^t \approx 5$~TeV. This corresponds to a separation between the $W$ and the $b$ from the top decay of $\Delta R \approx 0.07$,
to be compared with a tower of the CMS hadronic calorimeter in the
barrel $\Delta \eta \times \Delta \phi \sim \mathcal{O}\left(0.1\times 0.1 \right)$~\cite{Chatrchyan:2008aa}.

%%%%%%%%%%%%%%%%%%%%%%%%%%%%%%%%%%%%%
\begin{figure}[ht]
  \raisebox{-0.048\height}
           {\includegraphics[trim=3mm 0 0 0,clip=true, width=.45 \textwidth]{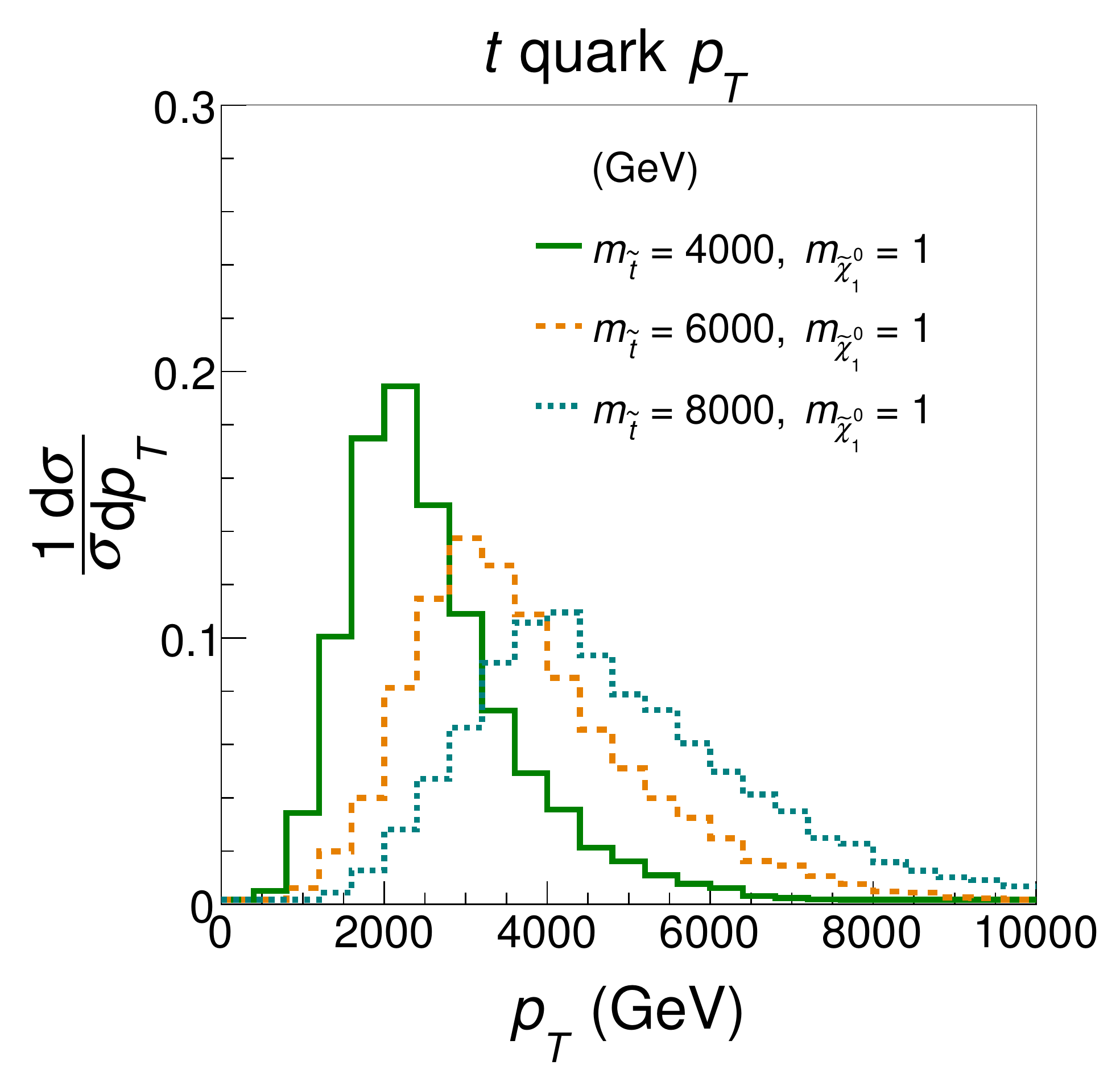}}
           \includegraphics[width=.407 \textwidth]{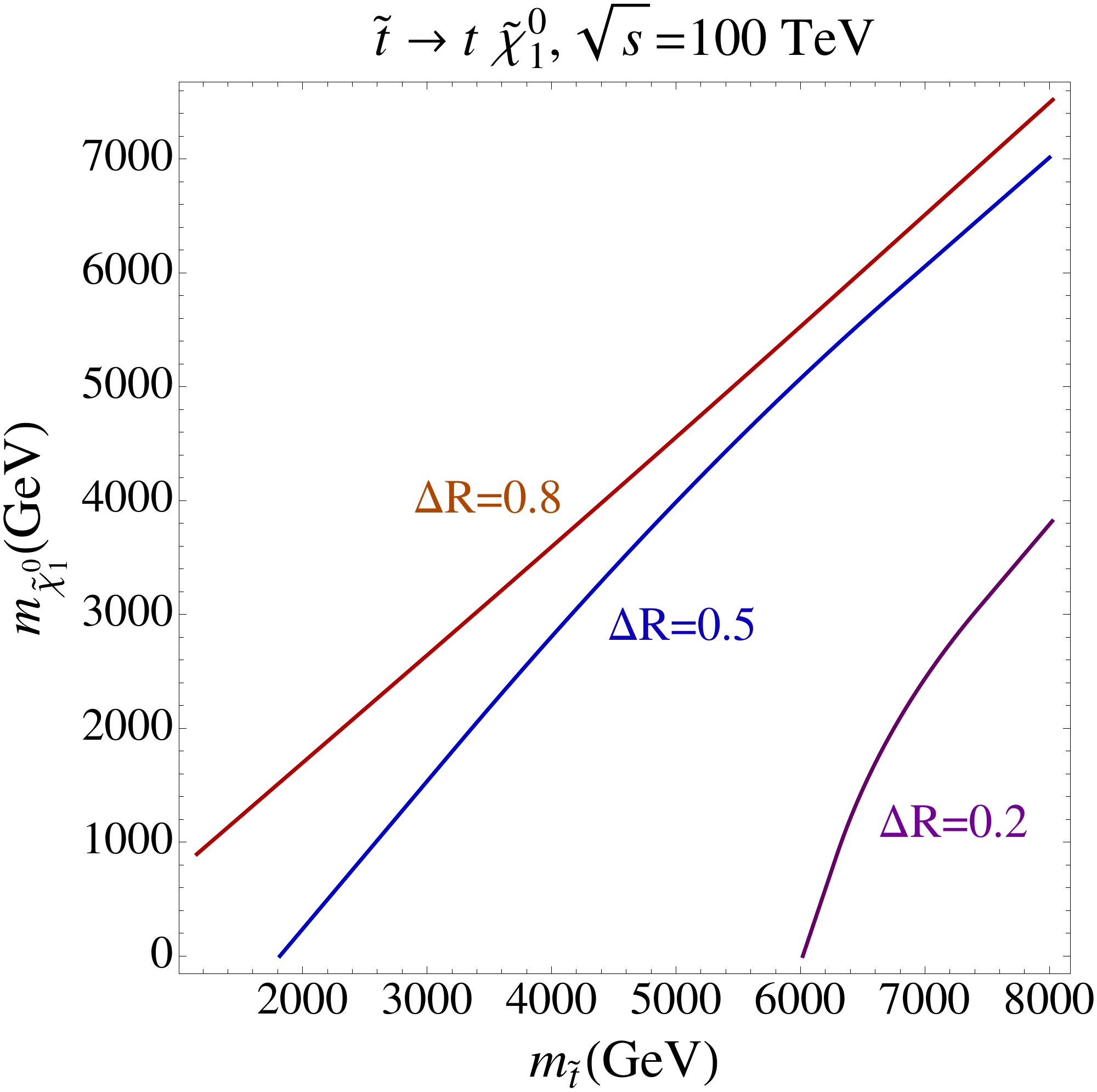}
           \caption{
             Left: The \pt{} distribution of the leading top quark for $m_{\Stop}=4,6,8$ TeV, setting $m_{\chioz} = 1\, {\rm GeV}$.
             Right: Average $\Delta R$ between $W$ and $b$ from top decays as a function of $m_{\Stop}$ and $m_{\chioz}$.}
           \label{fig:SUSY_stop_topboost}
\end{figure}
%%%%%%%%%%%%%%%%%%%%%%%%%%%%%%%%%%%%%%

It is clear that traditional LHC searches, which aim to reconstruct the top quark from its decay products, will be ineffective,
unless detector granularities improve considerably. The same is true for algorithms specifically designed to tag top quarks~\cite{Plehn:2010st, Kaplan:2008ie},
as was shown in Ref~\cite{Cohen:2014hxa}. Therefore we avoid relying on substructure techniques, and instead build our search around the requirement of a muon inside a jet.
This greatly reduces the SM backgrounds while making the analysis almost insensitive to the future detector design.
Similar techniques are already in use at hadron colliders to tag
$b$-jets~\cite{Acosta:2005zd, Abazov:2004bv, Abulencia:2005qa, Aaltonen:2007dm, Aaltonen:2009ad, Aaltonen:2010se, Aad:2009wy, ATLAS:2010afa, Chatrchyan:2012jua, ATLAS-CONF-2013-068}. 

The analysis was performed with the Snowmass background samples~\cite{Avetisyan:2013onh} for the $t\bar t$+jets, single $t$+jets, $t\bar tV$+jets, and $V$+jets ($V = W, Z$) background processes.
An $H_T$-binned QCD multijet sample was also produced, following the same prescription as the Snowmass samples.  Signal samples were produced unbinned in \Ht.  All samples were generated with
{\tt MadGraph5}~\cite{Alwall:2011uj} and showered with {\tt Pythia6}~\cite{Sjostrand:2006za}.  The detector simulation was implemented using {\tt Delphes}~\cite{deFavereau:2013fsa} with the
Snowmass combined detector card~\cite{Anderson:2013kxz}. The signal cross-section was computed at NLL + NLO in~\cite{Borschensky:2014cia}, consistent with the calculations presented in
Section~\ref{sec:susy_xs}.

Our selection requirements are (applied in the order in which they are listed):

\begin{enumerate}
\item At least two $\Delta R=0.5$ anti-$k_t$ jets~\cite{Cacciari:2008gp} with $|\eta|<2.5$ and $\pt >$ 1 TeV must be present in the event.
\item We require at least one muon with $p_{T}^{\mu}>$ 200 GeV inside a $\Delta R=0.5$ cone centered around the axis of one of the leading two jets.
\item Events in which at least one isolated lepton (either an electron or a muon) with $\pt >$~35~GeV and $|\eta|<2.5$ is present are rejected.
  Our isolation criterion requires that the total \pt{} of all particles within a $\Delta R < 0.5$ cone around the lepton be less than $10\%$ of its \pt.
\item $\Delta \phi_{\met J} > 1.0$, where $\Delta \phi_{\met J}$ is the minimum $|\Delta \phi|$ between missing energy (\met) and any jet in the event with $\pt>200$ GeV and $|\eta|<2.5$.
\item After the previous cuts are applied we define three signal regions: $\met >$ 3, 3.5 or 4 TeV. 
\end{enumerate}

This set of cuts is designed to optimize the stop mass reach for light neutralinos. As we approach the diagonal of the $m_{\Stop}-m_{\chioz}$ plane,
the top gets a smaller fraction of the initial energy, and its decay products become more separated.
In addition, the total visible energy and \met{} in the event are considerably reduced.
In this compressed region of parameter space the natural candidate to recover sensitivity is a dilepton
search~\cite{Cohen:2014hxa}. This leads us to consider also the signal region defined by the following set of requirements: 

\begin{enumerate}
\item At least two $\Delta R~=~0.5$ anti-$k_T$ jets with $|\eta|<2.5$ and $\pt>500$ GeV in the event.
\item The presence of two isolated leptons (either electrons or muons) with $p_T^{\ell}>35$ GeV is required. The isolation criterion is the same described for the boosted signal region. 
\item $\met > 2$ TeV.
\item $\Delta \phi_{\met J,\,\ell} > 1.0$, where $\Delta \phi_{\met J,\,\ell}$ is the minimum $|\Delta \phi|$ between \met{} and any jet with $\pt>200$~GeV and $|\eta|<2.5$, and any isolated lepton with $p_{T}^{\ell}>35$ GeV and $|\eta|<2.5$.
\end{enumerate}

The expected mass reach of the compressed and boosted searches is shown in Figure~\ref{fig:SUSY_stop_exclusion_discovery} for 3000~\ifb{} of integrated luminosity. We assume a $20\%$ systematic uncertainty on both signal and background.
Exclusion is defined at 95\% confidence level, and the significance for discovery is 5~$\sigma$.  Background and signal are modeled as Poisson distributions with Gaussian systematics.  Exclusion limits are computed using a modified Frequentist
procedure (${\rm CL}_{\rm s}$) computed using {\tt ROOSTATS}~\cite{Moneta:2010pm}.
We find that stops with masses of $\approx 5.5(8)$ TeV can be discovered (excluded) if the neutralino is massless.  In most of the parameter space we can exclude neutralino masses up to 2~TeV. In the compressed region we can discover stops up to 1.5 TeV. The impact of larger systematic uncertainties on both signal and background is discussed in~\cite{Cohen:2014hxa}.

Another conclusion of the study is that an integrated luminosity of 3000~\ifb{} does not saturate the potential of a 100 TeV collider.  A factor of 10 more in integrated luminosity would extend the discovery reach on the stop mass up to 8 TeV (for a massless neutralino) and the exclusion to 10 TeV~\cite{Cohen:2014hxa}.

%%%%%%%%%%%%%%%%%%%%%%%%%%%%%%%%%%%%%%%%%
\begin{figure}[ht]
  \includegraphics[width=.47 \textwidth]{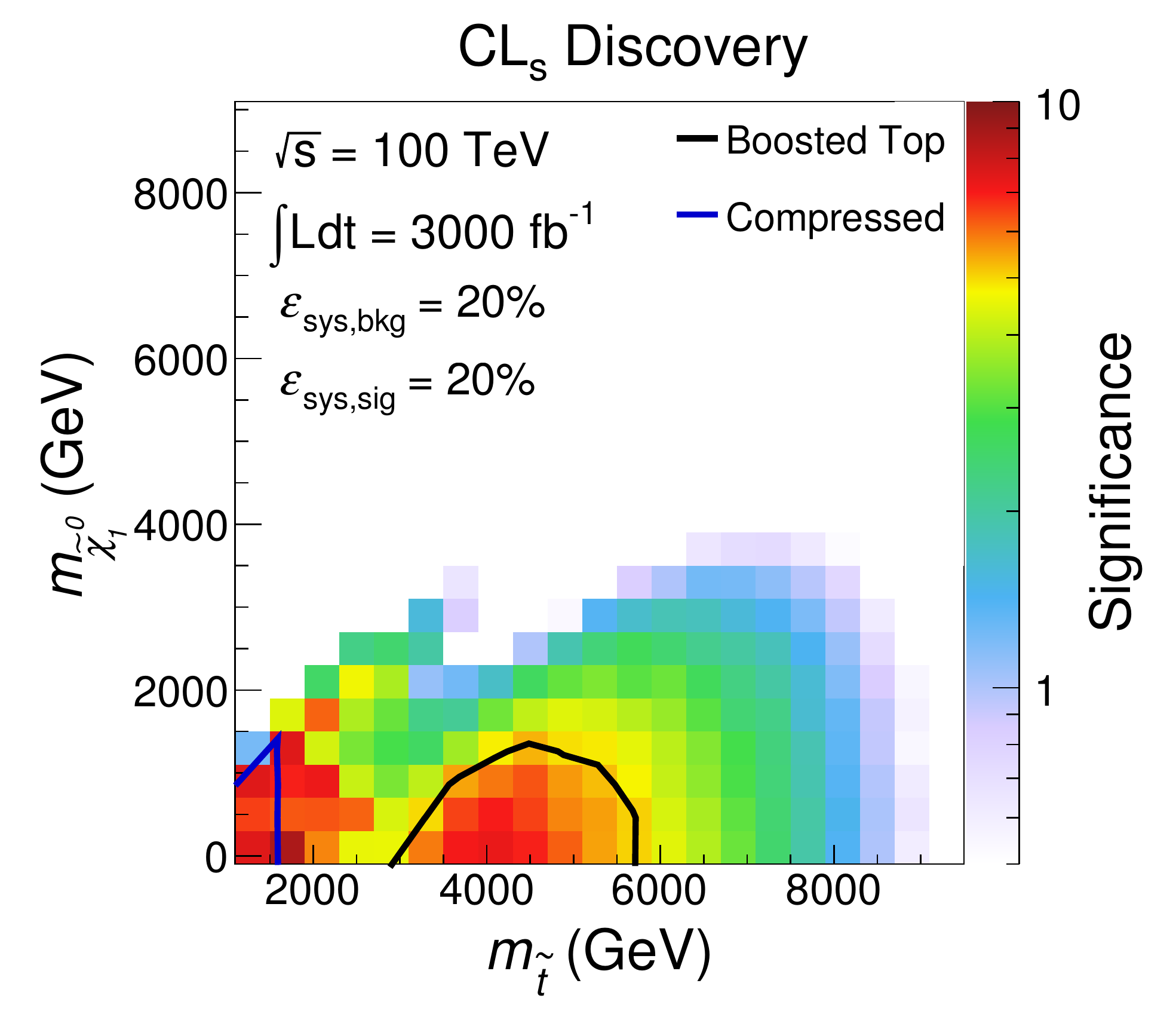}
  \includegraphics[width=.50 \textwidth]{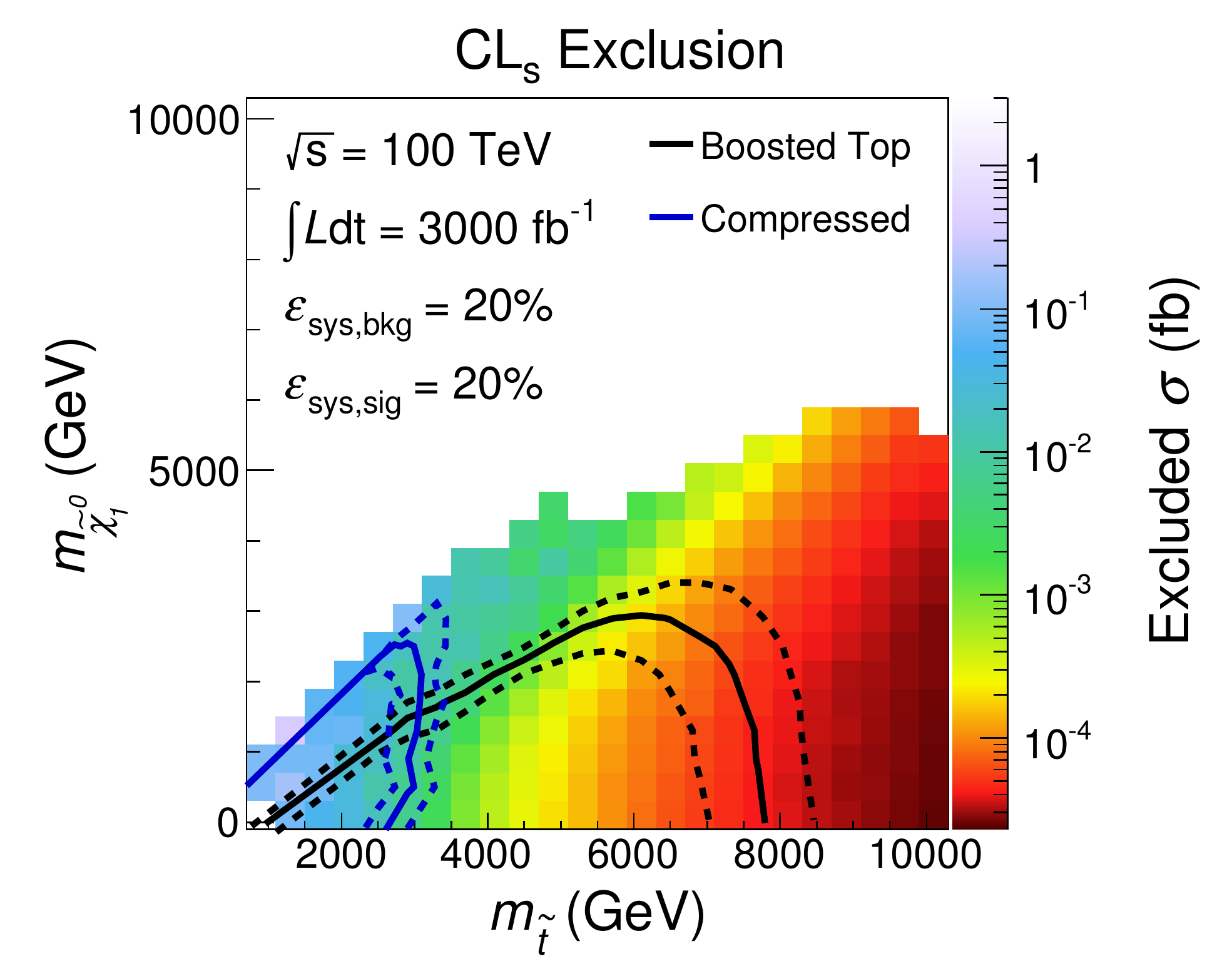}
  \caption{Left: Discovery potential and Right: Projected exclusion limits for 3000~\ifb{} of total integrated luminosity at $\sqrt{s}=100$~TeV. The solid lines show the expected discovery or exclusion obtained from the boosted top (black) and compressed spectra (blue) searches. In the boosted regime we use the \met{} cut that gives the strongest exclusion for each point in the plane. The dotted lines in the left panel show the $\pm 1\sigma$ uncertainty band around the expected exclusion.}
  \label{fig:SUSY_stop_exclusion_discovery}
\end{figure}
%%%%%%%%%%%%%%%%%%%%%%%%%%%%%%%%%%%%%%%%%

\subsubsection{Hadronic Decays}
In this section we consider fully hadronic decays using strategies inspired
by~\cite{Plehn:2010st,Plehn:2012pr,Kaplan:2012gd,Dutta:2012kx,Buckley:2013lpa,Stolarski:2013msa}. 
Experimental searches for this channel from the 8 TeV run of the LHC are reviewed in~\cite{Chatrchyan:2013sza,Aad:2015pfx}
with a current bound of about 700 GeV for very light neutralinos.
The limit weakens with increasing neutralino mass, disappearing completely for a neutralino heavier than about 300 GeV.   

The fully hadronic channel has two advantages over leptonic searches.
The first is that it has the largest branching fraction for the top decays.
The second is that it has no inherent missing energy from neutrinos, so all the missing energy comes from the neutralinos.
This allows many backgrounds to be reduced by vetoing events with leptons. 

Here we will present a very crude estimate of the reach at a future 100 TeV collider.
We will choose stringent cuts to get a very pure signal sample, and then compute the signal efficiency using literature and simplified parton level simulations.
The results, presented in Ref.~\cite{Stolarski:2013msa}, are summarized in Table~\ref{tab:SUSY_stop_results} for $\sqrt{s}=$100 TeV, and for other future collider scenarios.
Those results use tree-level cross sections given by MadGraph 5~\cite{Alwall:2011uj}, but for our 100 TeV study we use NLL+NLO results from~\cite{Borschensky:2014cia}.

\begin{table}[t]
  \centering
  \begin{tabular}{|c|c|c|c|c|}
    \hline
    Collider	&	Energy		&	Luminosity		&		Cross Section &  Mass   \\       \hline		\hline
    LHC8 	&	8 TeV		&	20.5 fb$^{-1}$			&  10 fb     &	650 GeV                \\ \hline
    LHC 	&	14 TeV			&	300 fb$^{-1}$			&  3.5 fb     &	1.0 TeV                \\
    HL LHC	&	14 TeV		&	3 ab$^{-1}$			&  1.1 fb   &	1.2 TeV    	\\
    HE LHC &    	33 TeV                         &       3 ab$^{-1}$                &   91 ab   &      3.0 TeV     \\	
    FCC-hh &    	100 TeV                         &       1 ab$^{-1}$                     &    200 ab     &   5.7 TeV     \\	
    \hline
  \end{tabular} \hspace{-0.138cm}\vline
  \vspace{0.3cm}
  \caption{The first line gives the current bound on stops from the LHC 8 TeV data~\cite{Chatrchyan:2013sza,Aad:2015pfx}. The remaining lines give the estimated 5$\sigma$ discovery reach in stop pair production cross section and mass for different future hadron collider runs (from~\cite{Stolarski:2013msa}).  At 100 TeV, NLL+NLO cross sections can be used to extend the reach.}
  \label{tab:SUSY_stop_results}
\end{table}%

%\subsubsubsection{Signal}
\label{sec:SUSY_stop_signal}

We use top tagging~\cite{Thaler:2008ju, Kaplan:2008ie,Almeida:2008yp,Plehn:2010st,Thaler:2011gf,Soper:2012pb,Schaetzel:2013vka} to distinguish signal from background.  %For other FCC-hh studies on top quark reconstruction see~\cite{?}, but note that
Since highly boosted top tagging may suffer from intrinsic limitations due to the nature of calorimeters~\cite{Bressler:2015uma}, the search presented here avoids specialized substructure variables and instead uses top-tagging techniques established at the LHC.   This is applied to stop searches in theory studies in~\cite{Plehn:2010st,Plehn:2012pr,Kaplan:2012gd,Dutta:2012kx,Buckley:2013lpa,Stolarski:2013msa}. Top tagging has been used by experiments at the LHC~\cite{Chatrchyan:2012ku,Aad:2012raa} in other types of searches, and from~\cite{Chatrchyan:2012ku} we take the efficiency of top tagging to be 50\% for tops with $p_T > 500$ GeV. From the same search we take the fake rate to be 5\% for the same $p_T$ range. There is very little data for $p_T > 800$ GeV, but we will use these efficiencies throughout out study, even at very high energy. The HPTTopTagger~\cite{Schaetzel:2013vka} study focuses on $p_T > 1$ TeV and finds somewhat lower tagging efficiency but also lower fake rates. 

Therefore, we make the following cuts taking the efficiency from the literature:
\begin{itemize}
\item Require both tops decay hadronically (46\%),
\item Require one $b$-tag (70\%)~\cite{CMS-PAS-BTV-11-001,ATLAS-CONF-2011-102},
\item Require both tops pass a top tagger (25\%).
\end{itemize}

We also simulate pair production of 6 TeV stops decaying to a nearly massless (1 GeV) neutralino at a 100 TeV machine. The simulation is done at parton level with MadGraph 5~\cite{Alwall:2011uj} and is used to compute the efficiency for the following two cuts:
\begin{itemize}
\item Require that both tops have $p_T > 500$ GeV (97\%),
\item Require missing transverse energy bigger than 4 TeV (38\%).
\end{itemize}

The first cut justifies the efficiency of the top tagger cut from above. The efficiency of the second cut is computed after the first cut is applied, and the total efficiency of all cuts is 3.0\%.

In order to estimate the size of the backgrounds, we use the same combination of cut efficiencies obtained from the literature and parton level Monte Carlo. Because of our requirement of  $b$ and top tags, the dominant backgrounds will be those with on-shell tops. In searches at the LHC, the dominant background is $\ttbar$ production where one of the tops decays to a hadronic $\tau$. At 100 TeV, however, this background is made negligibly small by the large missing energy cut. Therefore, the dominant background is $\ttbar Z$ where the $Z$ decays to neutrinos and is highly boosted to pass the missing energy cut. 

The production cross section at 100 TeV is 46 pb. Applying just the 4 TeV missing energy cut reduces the effective cross section to 130 ab. Applying the requirement of both tops having $p_T > 500$, as well as branching ratios and $b$ and top tags reduces the effective cross section to 1.4 ab, so with these hard cuts, even this potentially large background can be reduced to be essentially negligible until extremely high luminosity is reached. Other more exotic backgrounds such as four top and $\ttbar ZZ$ production are not considered here, but they are expected to be subdominant.

%\subsubsubsection{Estimate of Reach}
%\label{sec:SUSY_stop_reach}

%We now estimate the reach in terms of stop pair production cross section and stop mass.
Our results are summarized in Table~\ref{tab:SUSY_stop_results}.
We estimate the $\sigma$-significance as number of signal events divided by the square root of the number of background events.
This can be rewritten as a discovery of $N_\sigma$ being achieved with the following signal cross section
\begin{equation}
  \sigma_{s} = \frac{N_\sigma}{\varepsilon_s}\sqrt{\frac{\varepsilon_b \sigma_b}{L}}
  \label{eq:nsigma}
\end{equation}
where $\varepsilon_s$ is the signal efficiency computed in Section~\ref{sec:SUSY_stop_signal},
$\varepsilon_b \sigma_b$ is the effective background cross section,% computed in Section~\ref{sec:SUSY_stop_bg},
and $L$ is the integrated luminosity of the collider run.
Our cuts are such that the expected number of background events is $\mathcal{O}(1)$, so we need $\mathcal{O}(5)$ events for a 5$\sigma$ discovery.
In this regime, Eq.~(\ref{eq:nsigma}) is not strictly correct, but will suffice as a reasonable approximation here.
We find that at 100 TeV machine with 1 ab$^{-1}$ of luminosity can discover stops with pair production cross section of 200 ab.
Using leading order cross sections from Madgraph 5, this corresponds to a discovery reach of 5.7 TeV. 

%There is now an NLL+NLO computation of the production cross section available~\cite{Borschensky:2014cia}, and we show the results in Figure~\ref{fig:SUSY_stop_xsec}.
Using the cross sections in Section~\ref{sec:susy_xs} to estimate the mass reach leads to slightly stronger limits, albeit with higher-order corrections included for signal while background is still treated at leading order.
%This is not strictly consistent because leading order cross sections are used for the background, but a more precise computation is not available for the corner of the background phase space that we are interested in.
With these caveats, we can use Eq.~\ref{eq:nsigma} and Figure~\ref{fig:xsect_stop100} %\ref{fig:SUSY_stop_xsec}
to estimate the reach.  The final results are shown in Table~\ref{tab:NLOresults}. Going from leading order to the more precise calculation extends the reach by about 10\%. 

%% \begin{figure}
%%   \centerline{\includegraphics[width=.6\textwidth]{BSM_SUSY_figures/xs.pdf}}
%%   \caption{Stop pair production cross section at 100 TeV computed at NLL+NLO from~\cite{Borschensky:2014cia}. The red lines correspond to the 5$\sigma$ discovery reach at with integrated luminosity of 1 ab$^{-1}$ (top) and 30 ab$^{-1}$ (bottom).}
%%   \label{fig:SUSY_stop_xsec}
%% \end{figure}

\begin{table}[t]
  \centering
  \begin{tabular}{|c|c|c|c|}
    \hline
    Energy		&	Luminosity		&		Cross Section &  Mass   \\       \hline		\hline
    100 TeV                         &       1 ab$^{-1}$                &   200 ab   &      6.2 TeV     \\	
    100 TeV                         &       30 ab$^{-1}$                     &    36 ab     &   7.9 TeV     \\	
    \hline
  \end{tabular} \hspace{-0.138cm}\vline
  \vspace{0.3cm}
  \caption{Discovery reach at a 100 TeV collider using NLL+NLO cross sections for two different luminosity benchmarks. }
  \label{tab:NLOresults}
\end{table}%

The analysis here is a very naive estimate of the reach, and there many things that could be done to make it more precise including implementing top decays and hadronization as well as a realistic detector simulation.
One can also consider more sophisticated cuts which vary for the different stop masses. It would also be interesting to see what the top tagging efficiency and fake rates look like at even higher top momenta.
These and other issues are left for future study.

% cross sections:  (Borschensky et al) http://inspirehep.net/record/1306918
%  Stop coannilihation http://inspirehep.net/record/1291558

\subsection{Gluinos}
\label{sec:susygluino}
Gluinos are a critical component of supersymmetric theories.  With regard to the hierarchy problem they only enter the Higgs potential at two loops embedded within a stop loop. However, due to the large top Yukawa, and reasonably large QCD gauge coupling, these corrections can be large, thus gluinos are still important for understanding the role of supersymmetry in addressing the hierarchy problem.

In this section we study the reach of a 100 TeV collider for gluinos in the context of several simplified models.
The gluino and LSP will always be considered to be relatively light, sometimes along with light-flavor squarks.
The LSP is assumed to be stable, and all decays are assumed to be prompt.  
Depending on the spectrum of heavy scalar sparticle masses, gluino decays can be mediated by light-flavor squarks or by stops.
When the decays are mediated by light-flavor squarks, the gluino effectively
undergoes a three-body decay to two quarks and the invisible \LSP.  For decays mediated by stops, the gluino decays to $\ttbar+\chioz$.
In cases where light-flavor squarks are also accessible, the gluinos and
squarks can be produced in association with each other, leading to a third class of signatures.  The three signatures considered in this
section, along with the analysis strategies used to confront them, are shown below:
\renewcommand{\arraystretch}{1.5}
\setlength{\tabcolsep}{8pt}
\begin{center}
  \begin{tabular}{c|c|c}
    Simplified Model & Decay Channel & Search Strategy\\
    \hline
    Gluino-neutralino (light flavor) &  $\gluino \rightarrow q\,\overline{q}\,\chioz$ & jets+\met, mono-jet \\
    Gluino-neutralino (heavy flavor) &  $\gluino \rightarrow t\,\overline{t}\,\chioz$ & Same-sign dilepton\\
    \multirow{2}{*}{Gluino-squark with a massless neutralino} & $\gluino \rightarrow \big (q\,\overline{q}\,\chioz / q\, \squark^*\big)$; & \multirow{2}{*}{jets+\met} \\
                                                              & $\squark \rightarrow \big( q\,\chioz / q\,\gluino \big) $                 &  \\
  \end{tabular}
  \label{tab:susy_gluino_summary}
\end{center}

\subsubsection{Pair Production} % Tim Cohen
\label{sec:susy_gluinopairprod}
Models of split and mini-split SUSY can have scalar superpartner masses well above the masses of the
gauginos \cite{Wells:2004di, ArkaniHamed:2004fb, Giudice:2004tc, Arvanitaki:2012ps, ArkaniHamed:2012gw}.
In this case, the gluino (\gluino) and LSP (\chioz) are left as the only accessible superpartners at a $\sqrt{s}=100$ TeV collider.
However, the large cross section for gluino production, as shown in Section~\ref{sec:susy_xs},
makes this a likely discovery channel for SUSY at present and future colliders.  A full description of all analyses summarized here is available in Refs.~\cite{Cohen:2013xda, Cohen:2014hxa}.

Parton level events for all searches were generated using \texttt{Madgraph5} v1.5.10 \cite{Alwall:2011uj}.
All signals involve the pair production of SUSY particles and are matched using MLM matching up to 2 additional jets.
The $k_t$-ordered shower scheme with a matching scale of \texttt{qcut}=\texttt{xqcut}=$100\text{ GeV}$ was used.
We do not account for any possible inadequacies inherent in the current Monte Carlo technology,
\emph{e.g.} electroweak gauge bosons are not included in the shower.

The gluinos and squarks were treated as stable at the parton level.  These events were subsequently decayed and showered using
\texttt{Pythia6}~\cite{Sjostrand:2006za} and passed through the \texttt{Delphes} detector simulation
\cite{deFavereau:2013fsa} using the ``Snowmass" detector parameter card \cite{Anderson:2013kxz}.  Total production cross sections
were computed at NLO using a modified version of \texttt{Prospino} v2.1 \cite{Beenakker:1996ed, Beenakker:1996ch, Beenakker:1997ut}, and 
stop cross sections were computed at NLL using \cite{Borschensky:2014cia}, consistent with the results shown in Section~\ref{sec:susy_xs}.

\subsubsubsection{Gluino-neutralino with light flavor decays}
In a simplified gluino-neutralino model with decays to light flavor quarks, the gluino is the only kinematically accessible colored particle.
The squarks are completely decoupled and do not contribute to gluino production diagrams.
The gluino undergoes a prompt three-body decay through off-shell squarks,  $\gluino \rightarrow q\,\overline{q}\,\chioz$,
where $q$ is one of the light quarks and $\chioz$ is a neutralino LSP.
The only two relevant parameters are the gluino mass $m_{\gluino}$ and the neutralino mass $m_{\chioz}$.  

The background is dominated by $W/Z + \text{jets}$, with subdominant contributions from $t\,\overline{t}$ production.
Single top events and $W/Z$ events from vector boson fusion processes are negligible.
In all cases, there are decay modes which lead to multi-jet signatures.
The $\met$ can come from a variety of sources, such as neutrinos, jets/leptons that are lost down the beam pipe, and energy smearing effects. 

The first analysis used to confront such signals is inspired by an ATLAS upgrade study~\cite{ATL-PHYS-PUB-2013-002}.
After an event preselection, rectangular cuts on one or more variables are optimized at each point in parameter space to yield maximum signal significance.
Specifically, we simultaneously scan a two-dimensional set of cuts on $\met$\ and $H_T$, where $\met$\ is the magnitude of the missing transverse
momentum and $H_T$ is defined as the scalar sum of jet $p_T$.   Following a standard four-jet pre-selection, the following cuts are applied:

%\vspace{5pt}
%\textbf{\textsc{Search Strategy}: Simultaneous optimization over $H_T$~and $\met$}
%\vspace{-5pt}
\begin{itemize}
\item $\met/\sqrt{H_T} > 15$ GeV$^{1/2}$
\item The leading jet $p_T$ must satisfy $p_T^\text{leading} < 0.4\, H_T$
\item $\met > (\met)_\text{optimal}$
\item $H_T > (H_T)_\text{optimal}$
\end{itemize}

The discovery reach and limits for all several future collider scenarios in the full $m_{\gluino}$ versus $m_{\chioz}$ plane can be seen in Fig.~\ref{fig:GOGO_Comparison}.
%The $14~\TeV$ $300~\ifb$ limit with massless neutralinos is projected to be at a gluino mass of $2.3~\TeV$ (corresponding to 110 events),
%while the $14~\TeV$ $3000~\ifb$ limit is projected to be at $2.7~\TeV$ (corresponding to 175 events).
%The $14~\TeV$ LHC with $3000~\ifb$ could discover a gluino as heavy as $2.3\TeV$ if the neutralino is massless,
%while for $m_{\chioz}\gtrsim 500 \text{ GeV}$ the gluino mass reach rapidly diminishes.
%At a $33~\TeV$ machine with $3000~\ifb$, the limit with massless neutralinos is projected to be $5.8~\TeV$ (corresponding to 61 events).
%The $33~\TeV$ proton collider with $3000~\ifb$ could discover a gluino as heavy as $4.8\TeV$ if the neutralino is massless,
%while for $m_{\chioz}\gtrsim 1~\TeV$ the gluino mass reach rapidly diminishes.
For a $100~\TeV$ collider with $3000~\ifb$, the limit with massless neutralinos is projected to be $13.5~\TeV$ (corresponding to 60 events).
The $100~\TeV$ proton collider with $3000~\ifb$ could discover a gluino as heavy as $11~\TeV$ if the neutralino is massless,
while for $m_{\chioz}\gtrsim 1~\TeV$ the gluino mass reach rapidly diminishes.  

\begin{figure}[tbp]
  \centering
  \includegraphics[width=.48\columnwidth]{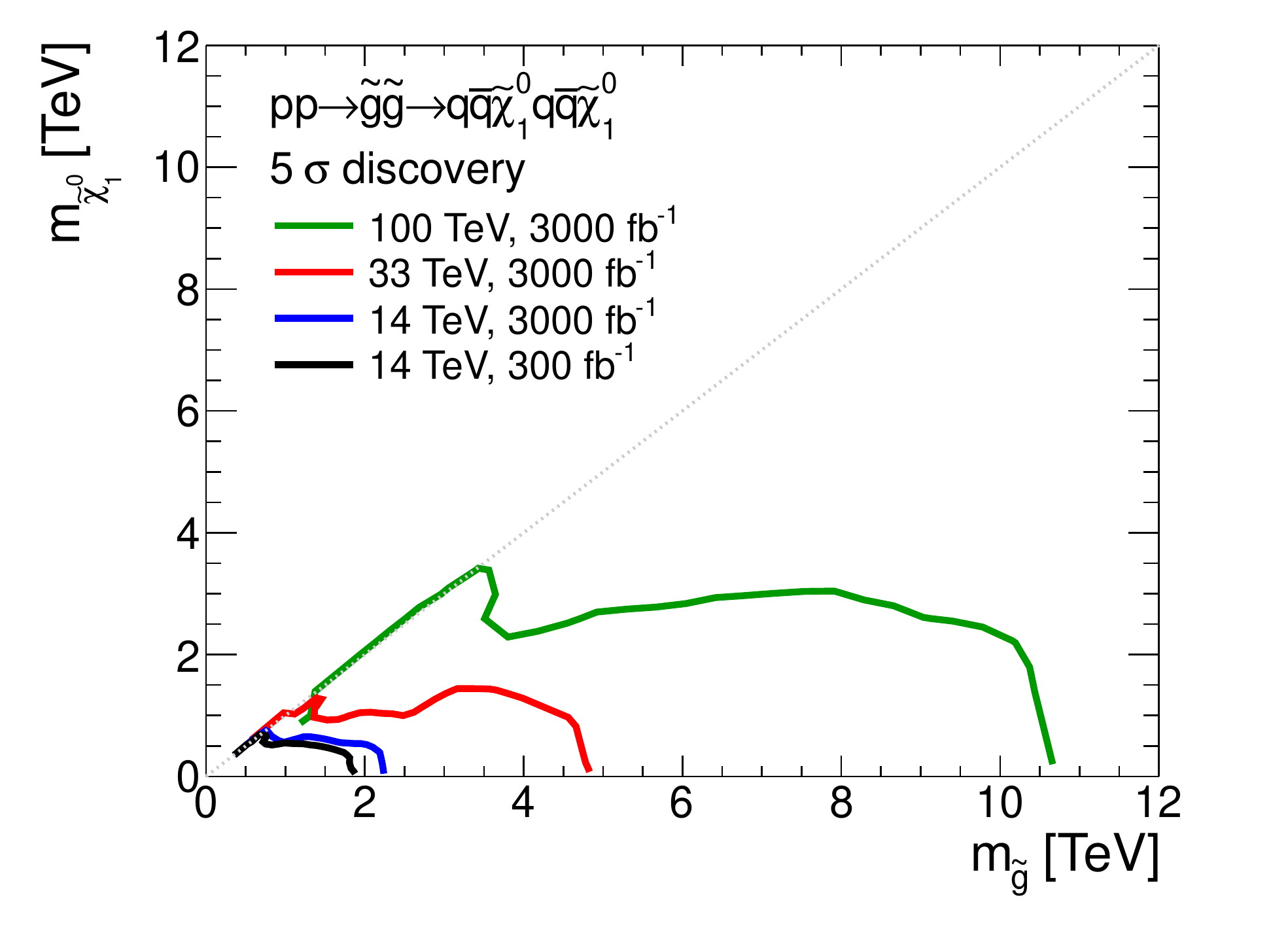}
  \includegraphics[width=.48\columnwidth]{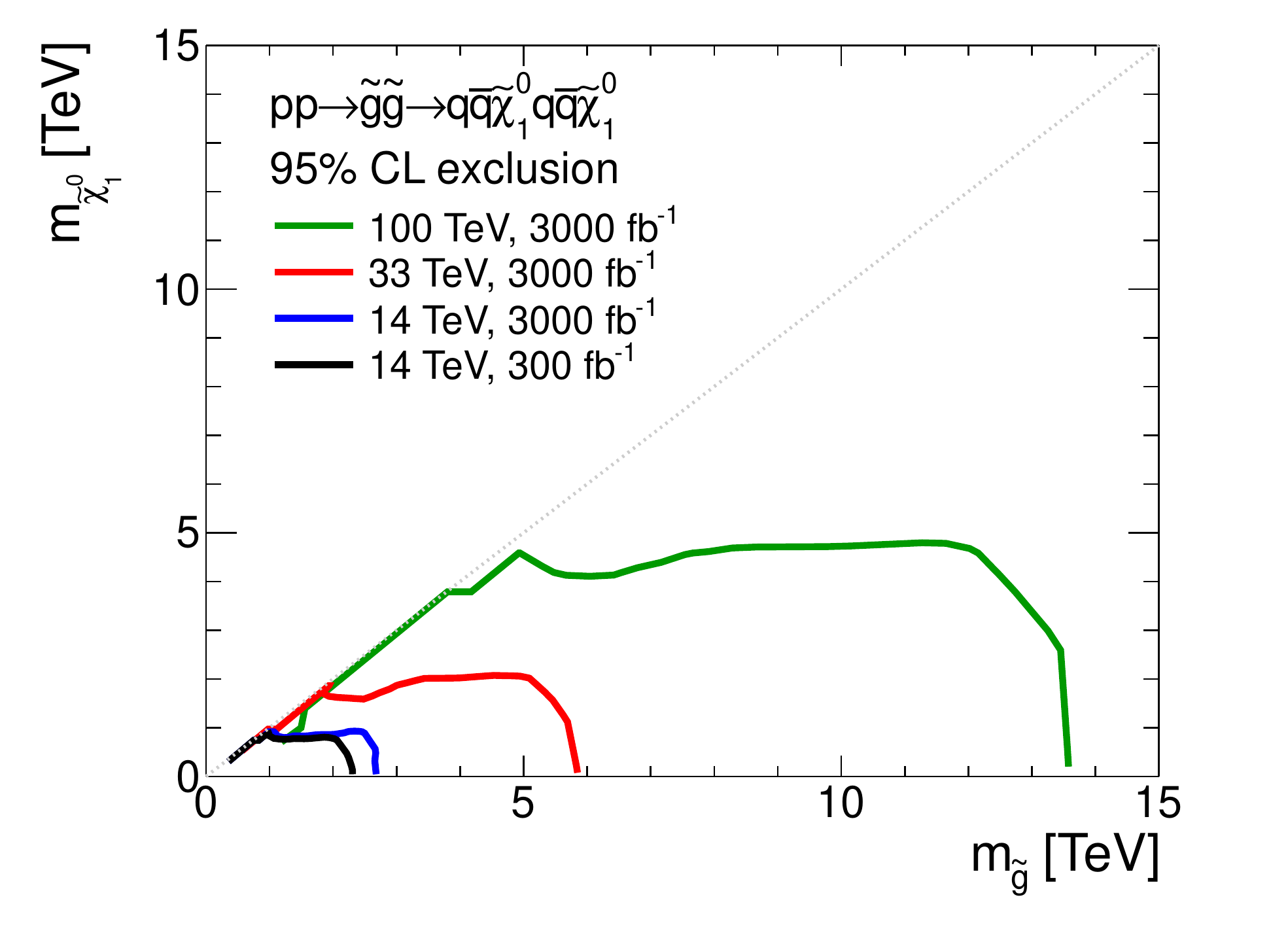}
  \caption{Results for the gluino-neutralino model with light flavor decays.
    The left [right] panel shows the $5\,\sigma$ discovery reach [$95\%$ CL exclusion] for the four collider scenarios studied here.
    A $20\%$ systematic uncertainty is assumed and pile-up is not included.}
  \label{fig:GOGO_Comparison}
\end{figure}

A separate analysis is used to target the compressed region of parameter space of this simplified model, where:
\begin{equation}
  m_{\gluino} - m_{\chioz} \equiv \Delta m \ll m_{\gluino}.
\end{equation}

For models with this spectrum, the search strategy of the previous section does not provide the optimal reach.
With compressed spectra the gluino decays only generate soft partons, thereby suppressing the \Ht{} signals and reducing the efficiency for passing the $4$ jet requirement.
A more effective strategy for compressed spectra searches relies instead on events with hard initial state radiation (ISR) jets to discriminate signal from background.  

The dominant background is the production of a $Z$ boson in association with jets, where the $Z$ boson decays into a pair of neutrinos ($Z \rightarrow \nu \nu$),
leading to events with jets and a significant amount of missing transverse energy. Subleading backgrounds are the production of a $W$ boson which decays
leptonically $\big(W \rightarrow \ell\, \nu\big)$ in association with jets, where the charged lepton is not reconstructed properly.
Finally, when considering events with a significant number of jets, $\ttbar$ production in the fully hadronic decay channel $\big (t\rightarrow b\,q\,q'\big)$ can be relevant.

In this study, we will apply two different search strategies that are optimized for this kinematic configuration and will choose the one
that leads to the most stringent bound on the production cross section for each point in parameter space.  Some of the cuts chosen below
are inspired by recent public results from ATLAS~\cite{ATLAS-CONF-2012-147} and CMS~\cite{CMS-PAS-EXO-12-048} on monojet searches.
Following a standard pre-selection, we first define a search strategy that selects events with a very hard leading jet
%% \vspace{5 pt}
%% \textbf{\textsc{Search Strategy 1}: Leading jet based selection}
%% \vspace{-5 pt}
\begin{itemize}
\item at most 2 jets
\item leading jet must have $p_T > (\text{leading jet } p_T)_\text{optimal}$ and $|\eta| < 2.0$
\item second jet is allowed if $\Delta \varphi(j_{2},\met) > 0.5$
\item $\met > \big(\met\big)_\text{optimal}$
\end{itemize}
where both $\big(\met\big)_\text{optimal}$ and $(\text{leading jet } p_T)_\text{optimal}$ are determined simultaneously by taking the values in the range $1 - 10$ TeV that yields the strongest exclusion.

The second search strategy targeting the compressed regime uses a $\met$-based selection with no jet veto:
%% \vspace{5 pt}
%% \textbf{\textsc{Search Strategy 2}: $\met$~based selection without jet veto}
%% \vspace{-5 pt}
\begin{itemize}
\item leading jet with $p_T > 110$ GeV and $|\eta| < 2.4$
\item $\met > (\met)_\text{optimal}$
\end{itemize}
with $\met$ varied in the range $(1,10)~\TeV$.  No requirement is placed on a maximum number of jets.
Note that for higher jet multiplicities the production of top quark pairs in the fully hadronic decay mode starts to dominate over $W/Z$ + jets production.

The discovery reach and limits for the compressed searches for all four collider scenarios in the full $m_{\gluino}$ versus $m_{\chioz}$ plane are shown in Fig.~\ref{fig:GOGO_Compressed_Comparison}.
%With $300$~\ifb{}, this search can exclude gluino masses of up to approximately $900$ GeV for a mass difference of $5$ GeV, with reduced reach for larger mass differences.
%The limits increase to around $1$ TeV with a factor of $10$ more data.  This improves the reach near the degenerate limit by roughly $200\text{ GeV}$ compared
%to the $H_T+\met$-based analysis; the $H_T+\met$-based searches do not begin to set stronger limits until $\Delta\gtrsim50\text{ GeV}$.  The discovery reach of
%this search is gluino masses up to 800 GeV near the degenerate limit. Unlike the exclusion reach, the discovery reach for this search is not a substantial
%improvement over the $H_T+\met$-based analysis, even in the degenerate limit.  This occurs because the signal efficiency using these searches is such that
%there are not enough events to reach $5\sigma$ confidence.  

%For a $33$ TeV proton collider with $3000$ fb$^{-1}$ of data, the exclusion reach for a mass difference of $5$ GeV covers gluino masses of up to approximately $1.8$ TeV,
%with reduced reach for larger mass differences.   For very small mass differences in the range of $5$ to $50$ GeV discoveries could be made for gluino masses up to $1.4$ TeV.
%This search improves the exclusion (discovery) reach near the degenerate limit by roughly $800\text{ GeV}$ ($400\text{ GeV}$) compared to the $H_T+\met$-based analysis;
%the $H_T+\met$-based searches do not begin to set stronger limits until $\Delta\gtrsim50\text{ GeV}$. 

For a $100$ TeV proton collider with $3000$ fb$^{-1}$ of data, the exclusion reach for a mass difference of $5$ GeV covers gluino masses of up to approximately $5.7$ TeV,
with reduced reach for larger mass differences.   For very small mass differences discoveries could be made for gluino masses up to $4.8$ TeV.
This search improves the exclusion (discovery) reach near the degenerate limit by roughly $1.7$ TeV ($1.3\TeV$) compared to the $H_T+\met$-based analysis;
the $H_T+\met$-based searches do not begin to set stronger limits until $\Delta\gtrsim500\text{ GeV}$.  

\begin{figure}[tbp]
  \centering
  \includegraphics[width=.48\columnwidth]{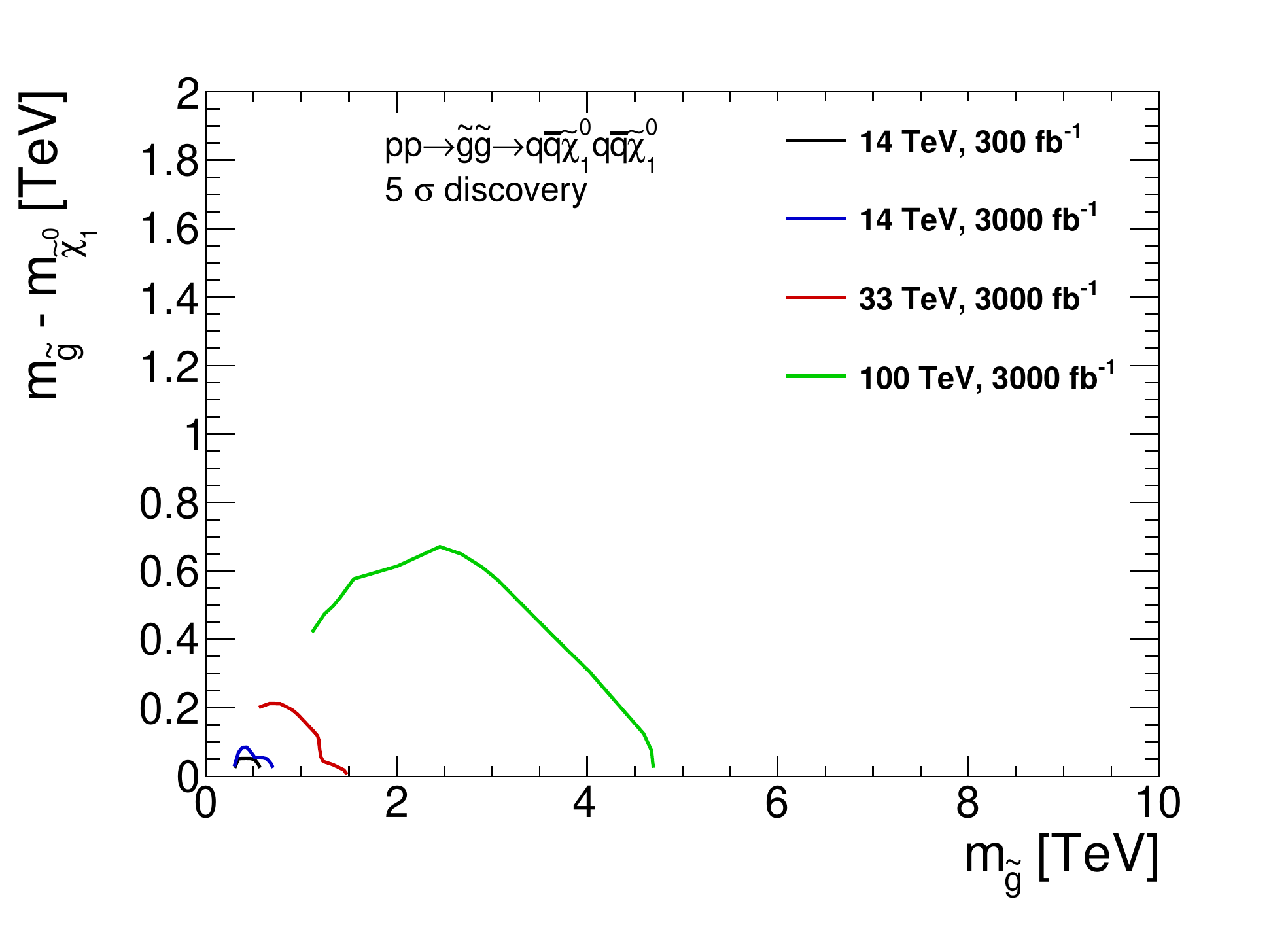}
  \includegraphics[width=.48\columnwidth]{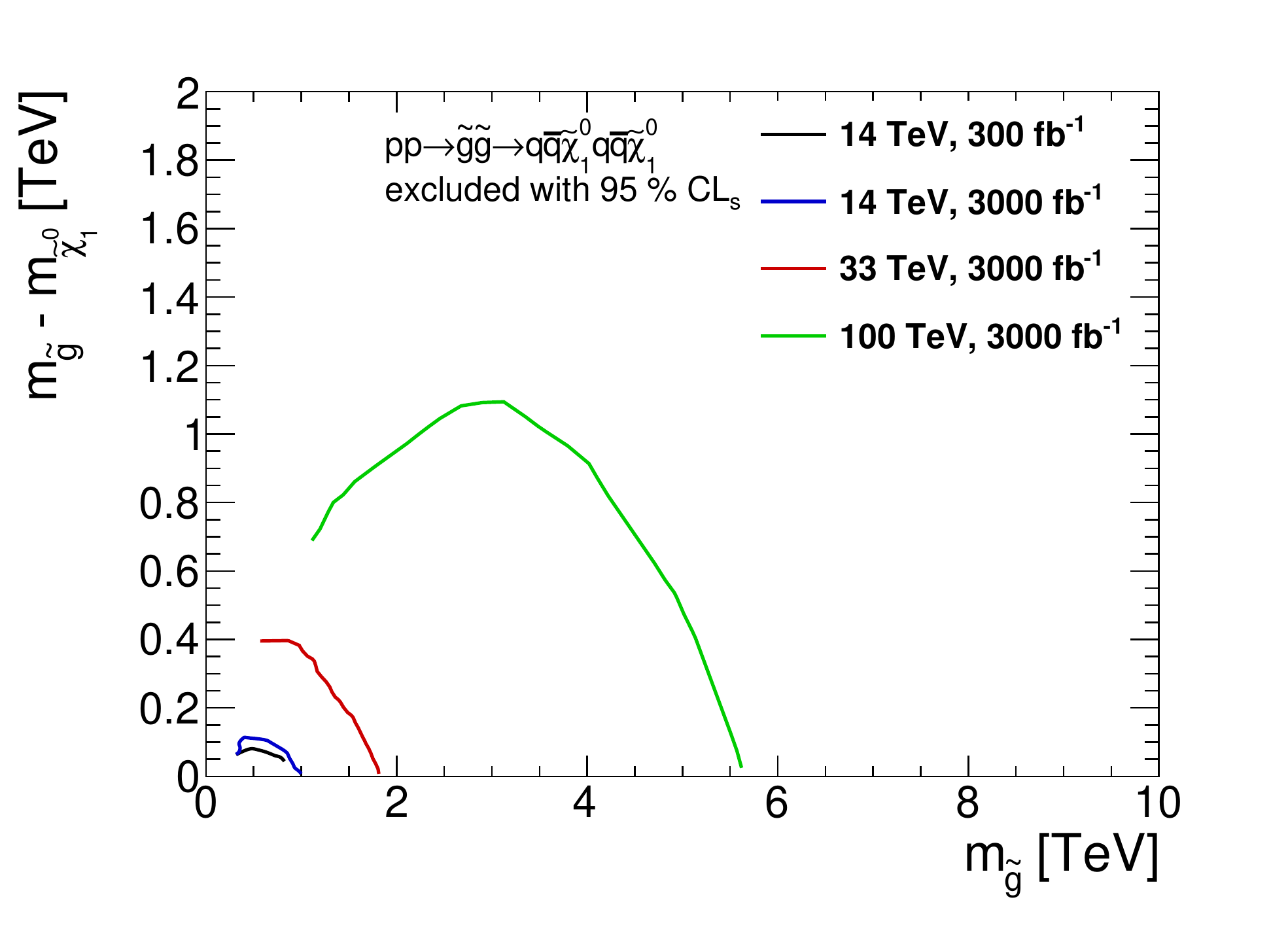}
  \caption{Results for the gluino-neutralino model with light flavor decays for the analyses that target the compressed region of parameter space.  The left [right] panel shows the $5\,\sigma$ discovery reach [$95\%$ CL exclusion] for the four collider scenarios studied here.  A $20\%$ systematic uncertainty is assumed and pile-up is not included.}
  \label{fig:GOGO_Compressed_Comparison}
\end{figure}

\subsubsubsection{Gluino-neutralino with heavy flavor decays}
In a gluino-neutralino model with decays to heavy flavour quarks, the gluino is the only kinematically accessible colored particle.  The squarks are completely decoupled and do not contribute to gluino production diagrams.
The gluino undergoes a prompt three-body decay through off-shell stops,  $\gluino \rightarrow t\,\overline{t}\,\chioz$, where $t$ is the top quark and $\chioz$ is a neutralino LSP.
The only two relevant parameters are the gluino mass $m_{\gluino}$ and the neutralino mass $m_{\chioz}$.  

The model produces two $t\,\overline{t}$ pairs along with considerable $\met$~(away from the compressed region of parameter space), and therefore provides an interesting benchmark scenario for searches involving
a combination of hadronic activity, leptonic signatures and $b$-tagging.  A search which requires same-sign di-leptons (SSDL) is one viable approach to eliminating the SM background since this final state is
highly suppressed in the SM.  A SSDL pair is required and any remaining leptons are not allowed to form a $Z$-boson, inspired by the CMS collaboration in~\cite{Chatrchyan:2012paa}.  We note that this was the
only channel explored in this scenario; it would be interesting to investigate how an all hadronic final state search would perform at the higher energy machines.

The analysis used to derive the results below requires an SSDL pair, which is very efficient at eliminating backgrounds. The dominant background is top pair production, where both tops decay leptonically (the di-leptonic channel).
There are subdominant backgrounds from $W\,b\,b$, which are accounted for by including the $BJ$ Snowmass particle container \cite{Avetisyan:2013onh}.
All backgrounds simulated for Snowmass are included and their rates are found to be negligible.
Since the SSDL requirement is very effective at suppressing backgrounds, only a mild cut on $\met$~is necessary to observe this model.
This implies that this search will also be very effective in the compressed regions of parameter space where $m_{\gluino} \simeq m_{\chioz}$.

After preselection, the following are used as discriminating variables.  Eight model points, three with very low LSP mass, three with medium LSP mass, and two with high LSP mass are used to define eight signal regions, which rely on some combination of the following cuts.

%% \vspace{5 pt}
%% \textbf{\textsc{Search Strategy}: Same-sign di-lepton based selection}
%% \vspace{-5 pt}
\begin{itemize}
\item Symmetric $M_{T2} > \big(\text{symmetric } M_{T2}\big)_{\text{optimal}}$
\item $p_T > \big(p_T\big)_\text{optimal}$ for the hardest lepton
\item $\met > \big(\met \big)_\text{optimal}$
\item $N_\text{jets} > \big(N_\text{jets} \big)_\text{optimal}$
\item $N_{b\text{-jets}} > \big(N_{b\text{-jets}} \big)_\text{optimal}$
\item $m_\text{eff} > \big(m_\text{eff} \big)_\text{optimal}$
\item $(H_T)_{\text{jets}}  >  \big( (H_T)_{\text{jets}} \big)_\text{optimal}$
\end{itemize}

Symmetric $M_{T2}$ is defined in the canonical way~\cite{Lester:1999tx, Barr:2003rg, Burns:2008va}, where the SSDL pair is used for the visible signal and the invisible particle test mass is assumed to be zero; \meff{} is defined as the scalar sum of the \pt{} of all visible objects and $\met$.

The results for the gluino-squark-neutralino model are given in Fig.~\ref{fig:GOGO_HeavyFlavor_Comparison}.  The $14~\TeV$ $300~\ifb$ limit is projected to be $1.9~\TeV$ (corresponding to 73 events),
and the $3000~\ifb$ limit is projected to be $2.4~\TeV$ (corresponding to 67 events).
The $14~\TeV$ LHC with $3000~\ifb$ could discover a gluino (with $\gluino\rightarrow t\,\overline{t}\,\chioz$) as heavy as $2.0~\TeV$ if the neutralino is massless.
The $33~\TeV$ $3000~\ifb$ limit is projected to be $4.0~\TeV$ (corresponding to 243 events).
A $33~\TeV$ proton collider with $3000~\ifb$ could discover a gluino (with $\gluino\rightarrow t\,\overline{t}\,\chioz$) as heavy as $3.4~\TeV$ if the neutralino is massless.
The $100~\TeV$ $3000~\ifb$ limit is projected to be $8.8~\TeV$ (corresponding to 224 events).
A $100~\TeV$ proton collider with $3000~\ifb$ could discover a gluino (with $\gluino\rightarrow t\,\overline{t}\,\chioz$) as heavy as $6.4~\TeV$ if the neutralino is massless.
Note that due to the relatively weak cuts that can be placed on $\met$, the SSDL signal is robust against models with almost degenerate gluino and neutralino. 

\begin{figure}[tbp]
  \centering
  \includegraphics[width=.48\columnwidth]{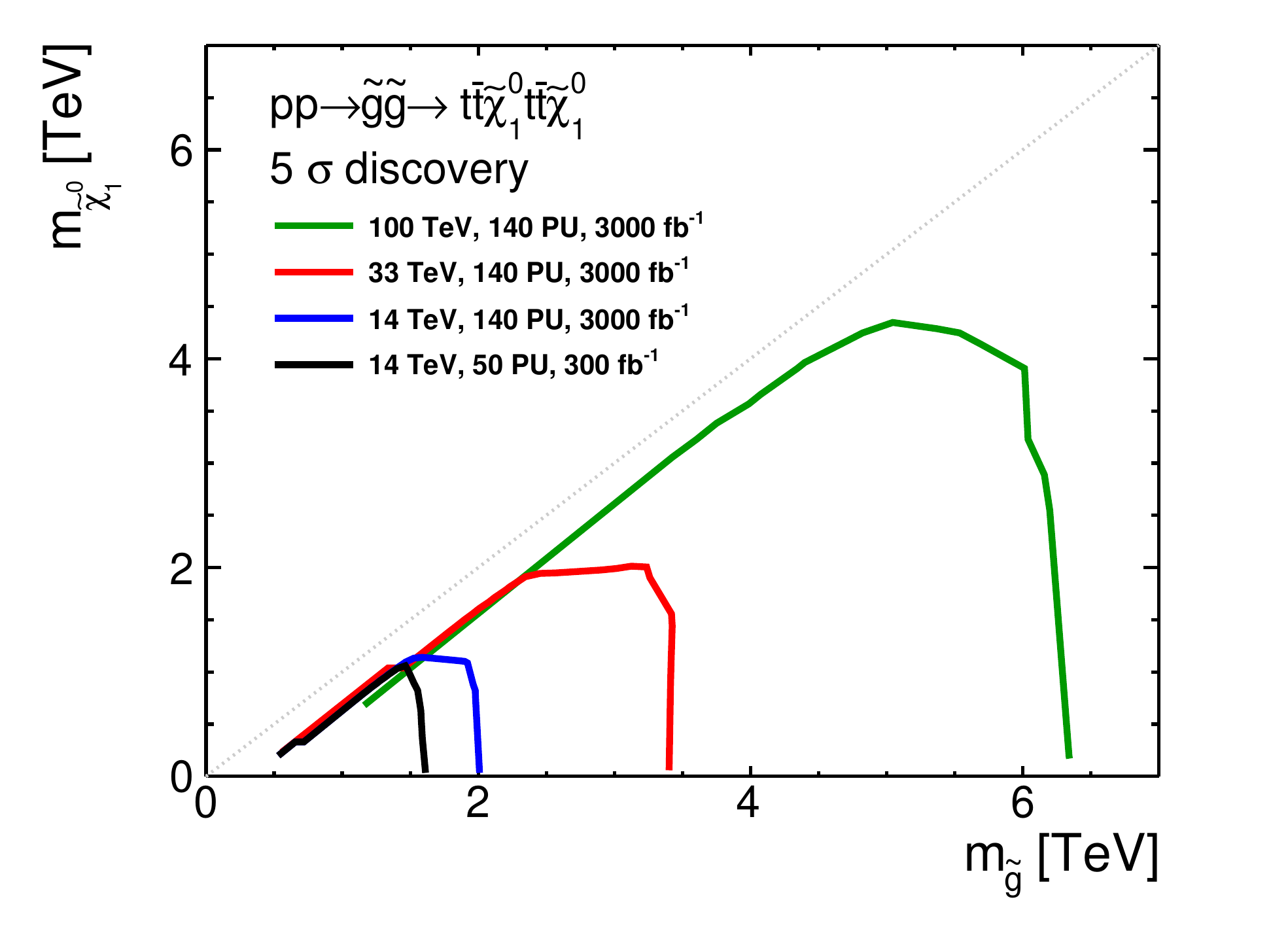}
  \includegraphics[width=.48\columnwidth]{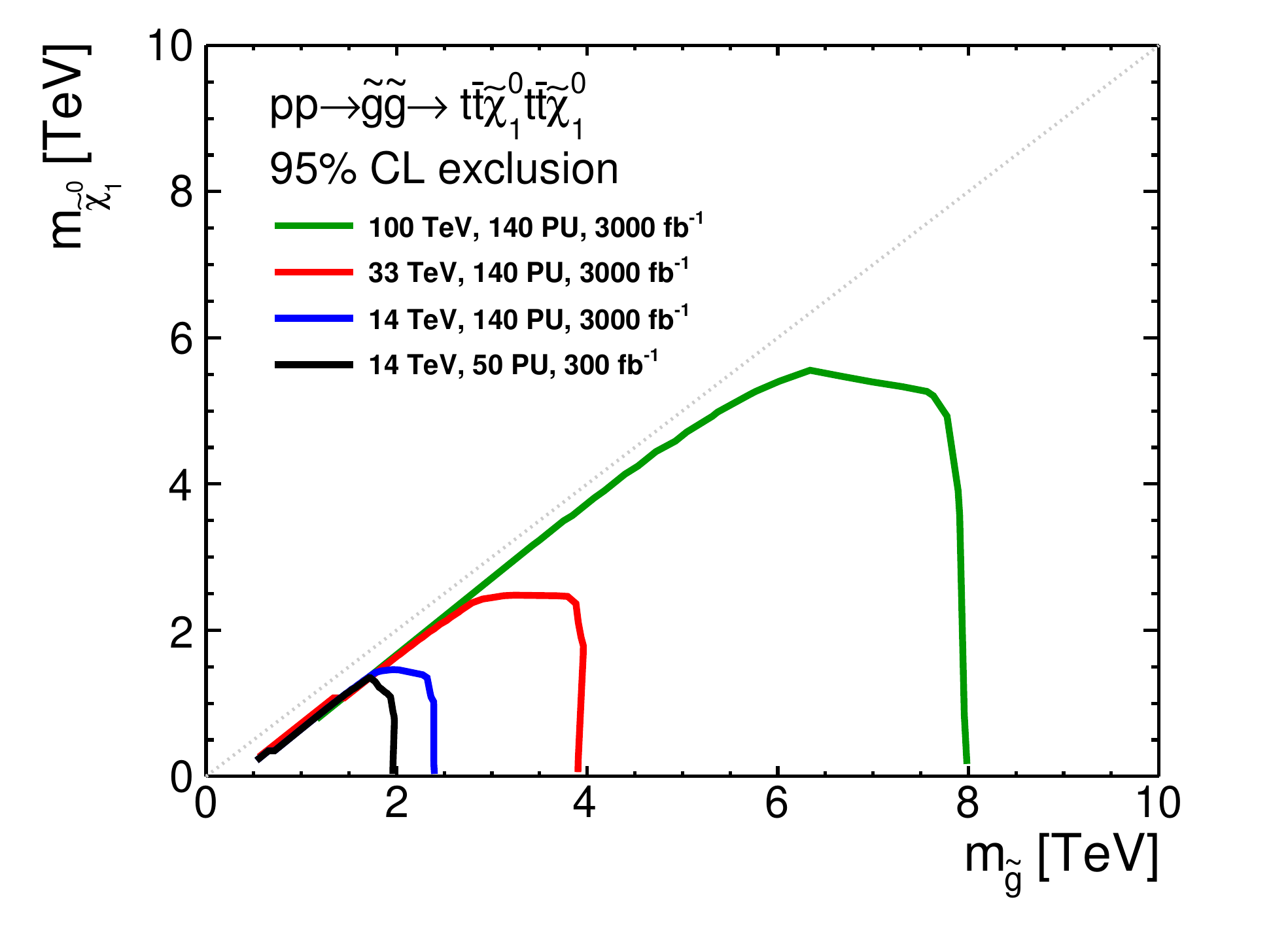}
  \caption{Results for the gluino-squark-neutralino model.  The neutralino mass is taken to be $1 \text{ GeV}$.
    The left [right] panel shows the $5\,\sigma$ discovery reach [$95\%$ CL exclusion] for the four collider scenarios studied here.
    A $20\%$ systematic uncertainty is assumed and pile-up is included.}
  \label{fig:GOGO_HeavyFlavor_Comparison}
\end{figure}

\subsubsection{Associated Production} % Cohen, Ellis, Zheng
\label{secsusy:associatedgluino}

In gluino-squark neutralino models, the gluino, the first and second generation squarks, and the LSP are all kinematically accessible.
The only relevant parameters are the squark mass $m_{\squark}$, which is taken to be universal for the first two generations, the gluino mass $m_{\gluino}$,
and the neutralino mass $m_{\chioz}$.  We consider two scenarios: one in which the neutralino is massless, and the gluino and squark have similar masses;
and another in which the neutralino is also light, but the squark mass is substantially above that of the gluino mass.

\subsubsubsection{Associated production \texorpdfstring{with $m_{\squark}\sim m_{\gluino}$}{of light gluinos and squarks}}

For this study we fix the neutralino mass $m_{\chioz}=1~\GeV$, which captures the relevant kinematics for
$m_{\gluino},m_{\squark}\gg m_{\chioz}$. The decay mode is chosen depending on the mass hierarchy.  

%For a full MSSM model, which in particular would imply a specific neutralino composition, there will in general be a non-zero branching ratio for the squark to decay to a neutralino and a quark when kinematically allowed. If the decay directly to a gluino is kinematically allowed however it will tend to dominate, and in this study for simplicity we assume that the squark is weakly coupled to the neutralino and decays to the gluino proceed with 100\% branching ratio when kinematically allowed. Likewise for $m_{\gluino} > m_{\squark}$, the branching ratio of the gluino to 3-body versus 2-body decays depends on the masses and coupling of the squarks to the neutralino, and we take the 2-body branching ratio to be 100\% in this region of parameter space.  To capture the transition region where $m_{\gluino} \simeq m_{\squark}$, parameter choices along the line $m_{\gluino} = m_{\squark}$ are included; the gluino decay is taken to be 3-body and the squarks are assumed to decay directly to the neutralino.

This model is a good proxy for comparing the power of searches that rely on the traditional jets and $\met$ style hadron collider search strategy
to discriminate against background.  The final state ranges from two to four (or more) hard jets from the decay (depending on the production channel) and missing energy.
The current preliminary limits on this model using $20$~\ifb{} of $8$ TeV data are $m_{\gluino} = 1750 \text{ GeV}$ and $m_{\squark} = 1600 \text{ GeV}$
(ATLAS \cite{ATLAS-CONF-2013-047}) assuming a massless neutralino.

Following an identical analysis strategy as for the gluino-neutralino model, described in Section~\ref{sec:susy_gluinopairprod},
the results for the gluino-squark-neutralino model are given in Fig.~\ref{fig:GOSQ_Comparison} and in Table~\ref{tab:susy_gosqresults}.
A 100 TeV collider can exclude up to 16 TeV in mass for $m_{\gluino}\sim m_{\squark}$.

\begin{table}[tbp]
  \centering
  \begin{tabular}{| c c | c c |}
    \hline
    $\sqrt{s}$     &$\int{\mathcal{L}\mathrm{d}t}$    &\multicolumn{2}{c|}{95\% CL Exclusion}\\ \relax
        [TeV]         &[fb$^{-1}$]                   &Mass Reach [TeV]     &$N$ produced\\
        \hline
        14             &300                      &2.8                  &155\\
        14             &3000                     &3.2                  &293\\
        \hline
        33             &3000                     &6.8                  &132\\
        \hline
        100            &3000                     &16                   &136\\
        \hline
  \end{tabular}
  \caption{95\% CL exclusion limits on associated gluino-squark production for various collider scenarios.  The last column indicates the total number of squark pairs produced at a squark mass at the 95\% CL exclusion limit for the given collider scenario.}
  \label{tab:susy_gosqresults}
\end{table}

\begin{figure}[tbp]
  \centering
  \includegraphics[width=.48\columnwidth]{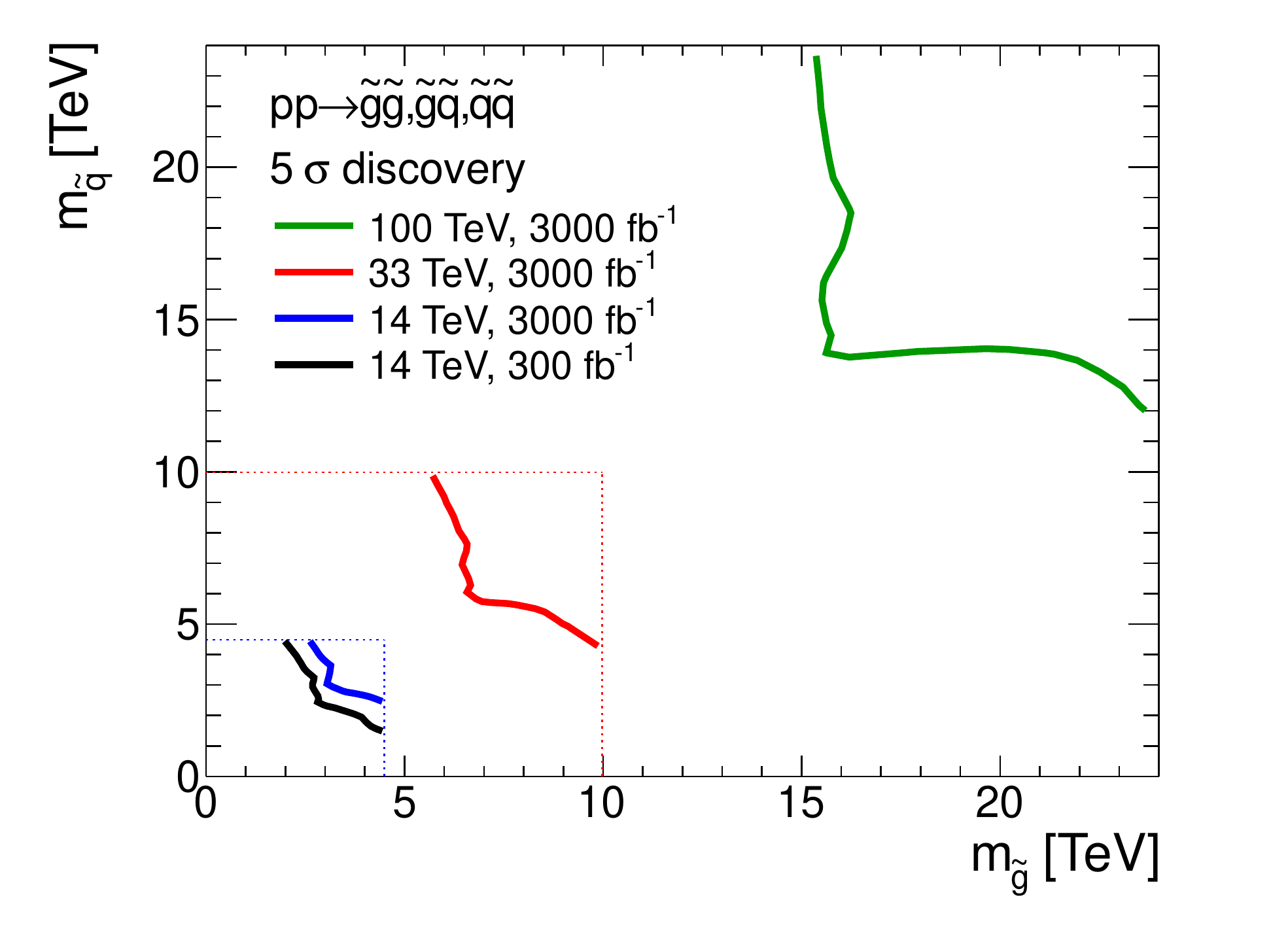}
  \includegraphics[width=.48\columnwidth]{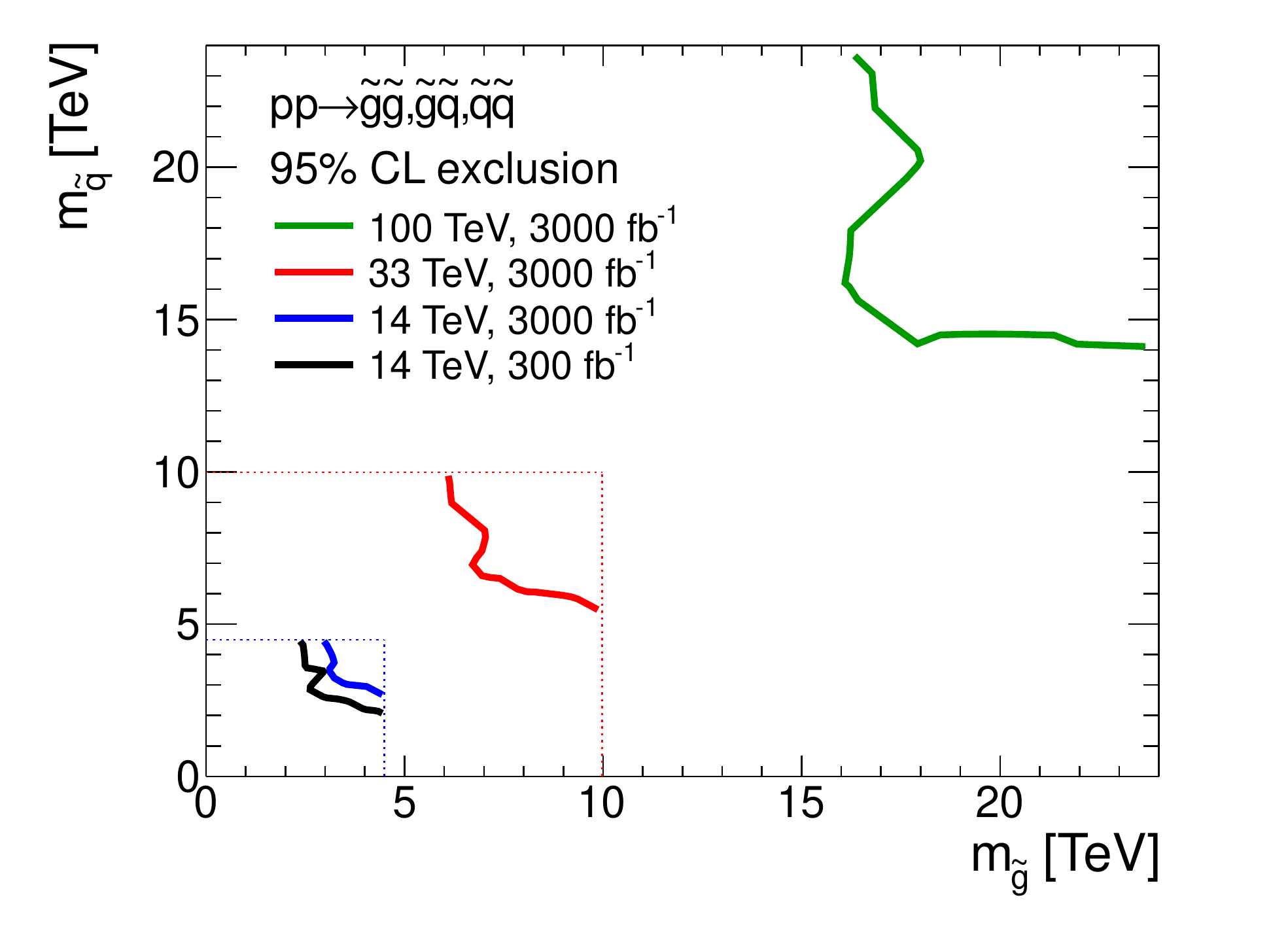}
  \caption{Results for the gluino-squark-neutralino model.  The neutralino mass is taken to be $1 \text{ GeV}$.  The left [right] panel shows the $5\,\sigma$ discovery reach [$95\%$ CL exclusion] for the four collider scenarios studied here.  A $20\%$ systematic uncertainty is assumed and pile-up is not included.}
  \label{fig:GOSQ_Comparison}
\end{figure}

\subsubsubsection{Associated production \texorpdfstring{with $m_{\squark}>m_{\gluino}$}{of light gluinos and heavy squarks}}
The gluino-squark-neutralino model in the previous section was probed in a region where $m_{\gluino}\sim m_{\squark}$.
In this section, we consider squark-gluino associated production in a region of parameter space in which the gluinos are relatively light,
while the squarks are heavier, but not completely decoupled.
This work is documented more completely in~\cite{Ellis:2015xba}, where we have analysed the prospects for squark-gaugino associated production at a 100 TeV collider.

Squark-gluino associated production is interesting because it has the potential to probe much higher squark masses than those reached in pair production.
Spectra with a hierarchy between the gluino and the first two generation squarks are predicted in many scenarios, such as anomaly-mediated SUSY breaking \cite{Randall:1998uk, Giudice:1998xp},
or in ``mini-split"-type models \cite{Wells:2003tf, Acharya:2007rc, Arvanitaki:2012ps}.

We consider two simplified models for squark-gluino associated production. In both, the particle content consists only of first and second generation squarks, gluino, and a Bino LSP ($\chioz = \tilde{B}$).
The two models correspond to different choices of the LSP mass:
\begin{itemize}
\item Non-compressed: $M_1 = 100$ GeV  (results in Fig. \ref{NonComp.FIG}) 
\item Compressed: $m_{\gluino} - m_{\chioz} =15$ GeV (results in Fig. \ref{Comp.FIG}) 
\end{itemize}
where we take the first and second generation squarks to be degenerate in mass, and decouple all other superpartners.
Our results are insensitive to the choice of $M_1 = 100$ GeV in the non-compressed spectra, as the LSP is effectively massless for $m_{\chioz} \ll m_{\gluino}$.
The compressed spectra are consistent with the gluino-neutralino dark matter (DM) coannihilation region \cite{Profumo:2004wk, Ellis:2015vaa}.

Events from squark-gluino associated production have distinctive event topologies, with a hard leading jet and significant $\met$.
Both arise primarily from the decay of the heavy squark, since the gluino is produced at relatively low \pt.
As in the gluino simplified models above, the dominant sources of background are top pair production and production of an SM boson + jets \cite{Cohen:2013xda}.
However, both of these backgrounds fall off rapidly both with increasing $\pt (j_1)$, $\met$, and $\met\sqrt{\Ht}$ (where $\Ht$ is the scalar sum of the jet transverse energies).
This can be seen for an example spectrum point in Fig. \ref{PT.FIG}.

\begin{figure}[tbp]
  \centering
  \includegraphics[width=0.6\textwidth]{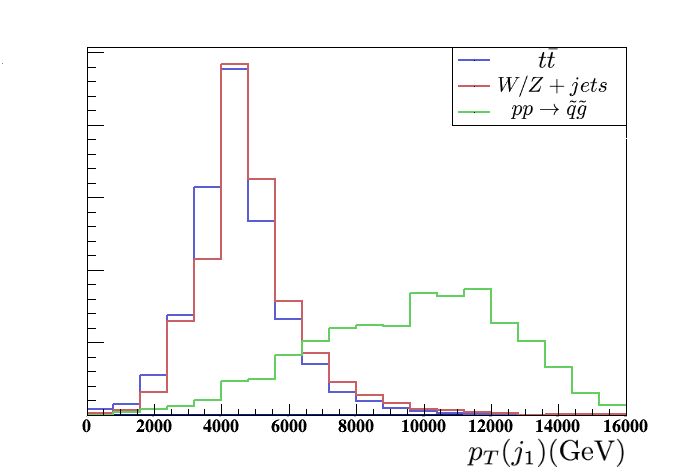}
  \caption{Example distribution of the leading jet $\pt$ for $p p \to \squark\gluino$, showing that the leading jet $\pt$ of the signal (green) is a good discriminatory variable.
    Shown here is the spectrum with $m_{\squark} \simeq 26$ TeV and $m_{\gluino} \simeq 4$ TeV. All events shown satisfy $\met > 2$ TeV.}
  \label{PT.FIG}
\end{figure}

The leading jet typically has a $p_T (j_1) \sim m_{\squark}/2$, while the decay of the squark into the LSP $\squark \rightarrow q \gluino \rightarrow 3\, q \chioz$ results in a highly boosted neutralino and
large $\met$. As such, heavy squark - light gluino associated production events have a striking collider signature with very low SM backgrounds.

We impose the following baseline cuts for both spectra:\begin{equation*} \Ht > 10 \,\,\TeV,\,\,\,\, \met /\sqrt{\Ht} > 20 \,\, \GeV^{1/2} . \end{equation*}
For the non-compressed spectra we impose the additional cut:\begin{equation*}8 \,\,\, \mathrm{jets \,\,\,with}\,\,\, \pt > 50\,\, (150) \,\, \mathrm{GeV} \,\end{equation*}
where the softer cut is optimized for heavier squarks and lighter gluinos, while the harder cut is optimized for lighter squarks and heavier gluinos.
We then scan over leading jet $\pt$ and $\met$ cuts in order to maximize the significance, $\sigma$:
\begin{equation}\label{sig}
  \sigma \equiv \frac{S}{\sqrt{1 + B + \lambda^2 B^2 + \gamma^2 S^2}}\,\,,
\end{equation}
where S (B) is the number of signal (background) events passing all cuts, and $\gamma$ $(\lambda)$ parameterize systematic uncertainties associated with signal (background) normalization.
We have verified that the optimal cuts render any ``background" from gluino pair production subdominant to the SM background.

Our results are shown in Figs.~\ref{NonComp.FIG} and \ref{Comp.FIG} for the non-compressed and compressed spectra, respectively.
We have assumed a conservative 3 ab$^{-1}$ integrated luminosity \cite{Hinchliffe:2015qma}.

%%%%%%%%%%%%%%%%%%%%%%%%%%%%%%%%%%%%%%%%%
\begin{figure}[tbp]
  \subfigure[Non-Compressed]{\includegraphics[width=.48 \textwidth]{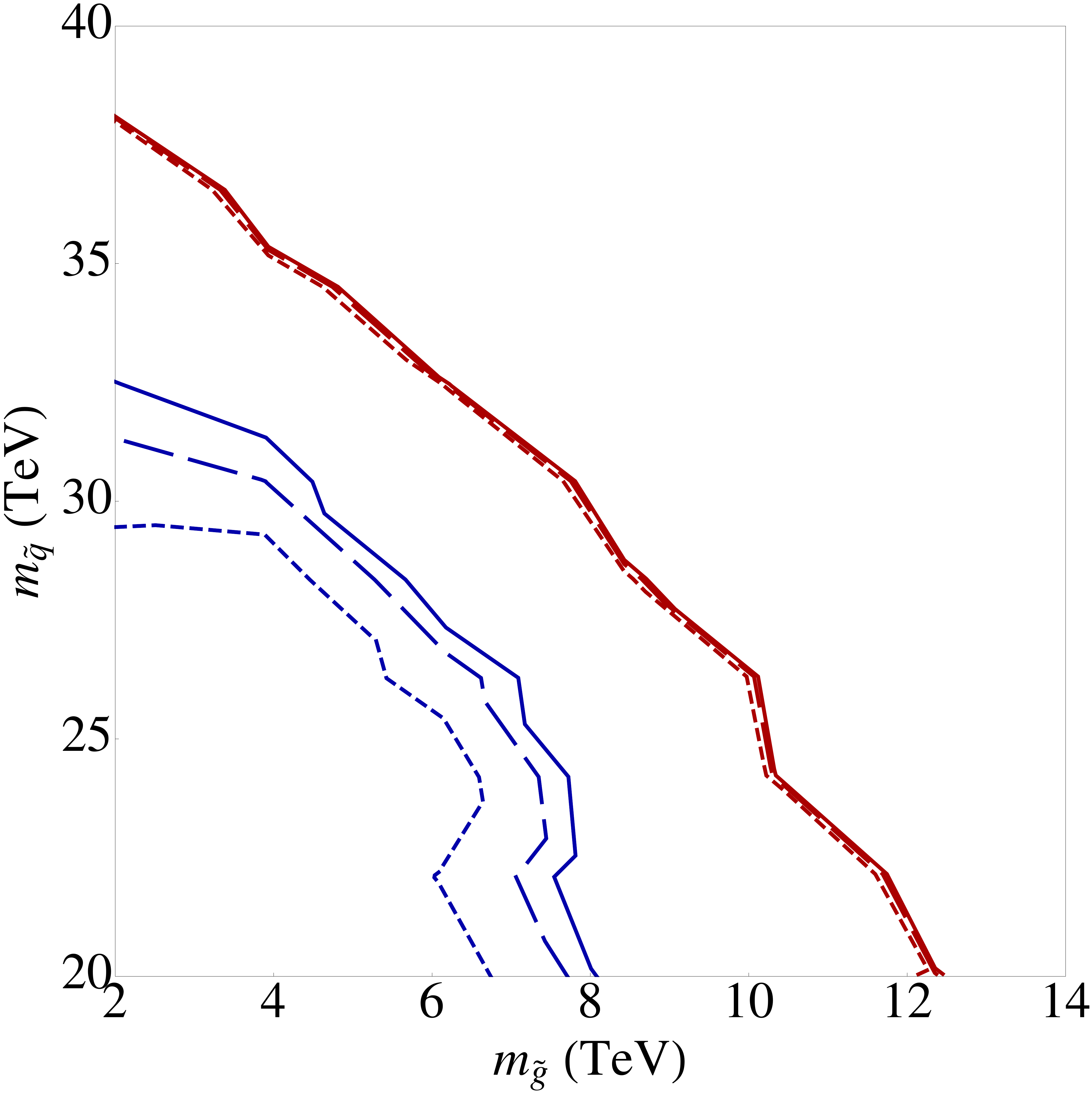}\label{NonComp.FIG}}
  \subfigure[Compressed]{\includegraphics[width=.48 \textwidth]{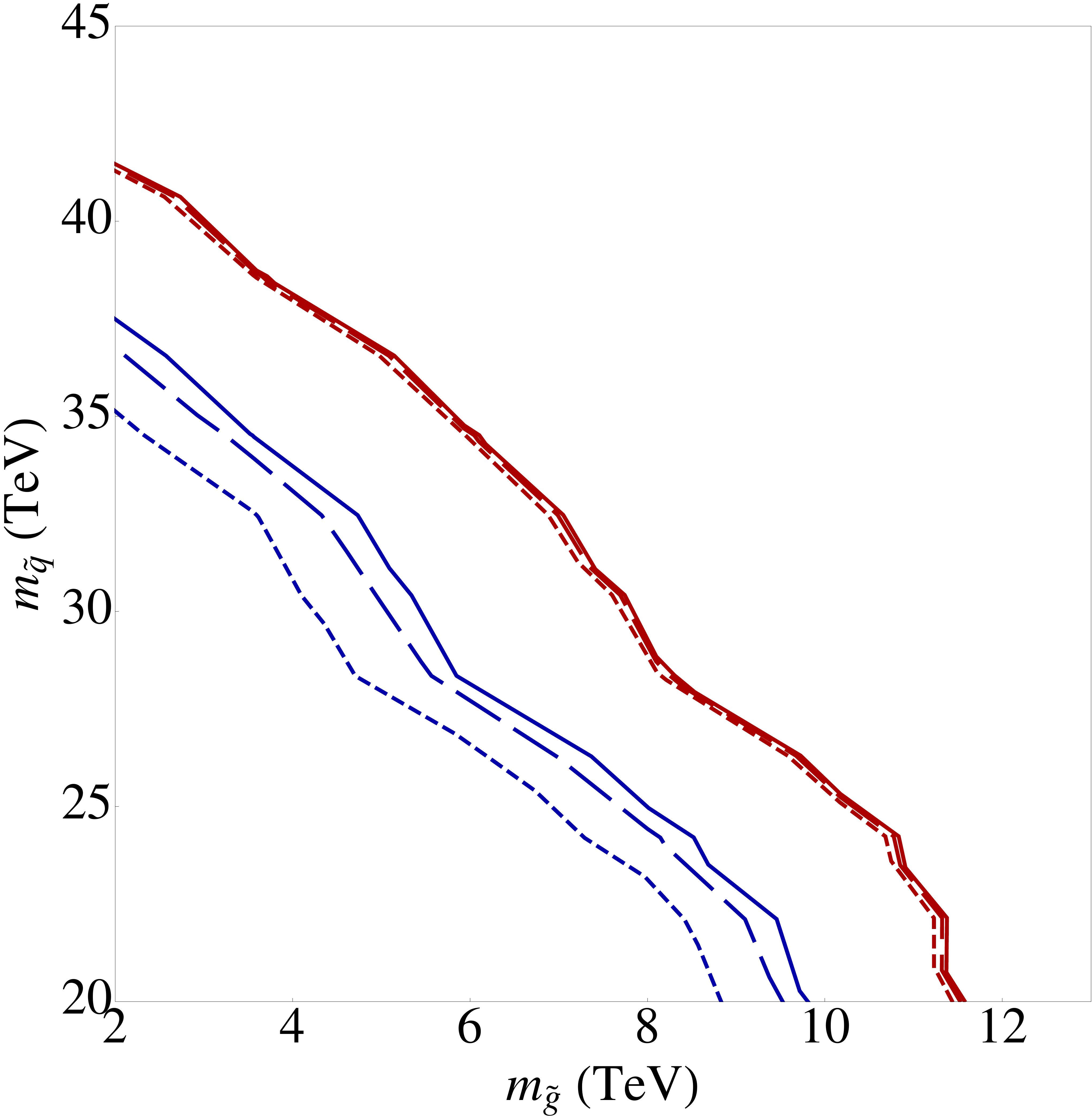}\label{Comp.FIG}}
  \caption{Experimental reach for squark-gluino associated production at a 100 TeV collider with 3 ab$^{-1}$ integrated luminosity.  Left panel: Experimental reach for spectra with a $\sim 100$ GeV LSP mass.  Right panel: Experimental reach for spectra with $m_{\gluino} - m_{\chioz} = 15$ GeV.  The solid, long dashed and short dashed lines are for and  $5,~10,~15$\% systematic uncertainty for the signal respectively. Blue lines indicate 5$\sigma$ discovery reach and red lines indicate 95\% exclusion limits. We assume 20\% systematic uncertainty in the background.}
\end{figure}
%%%%%%%%%%%%%%%%%%%%%%%%%%%%%%%%%%%%%%%%%

The solid, long-dashed and short-dashed lines correspond respectively to assuming systematic uncertainties of $5,~10$ and $15\%$ in the signal normalisation, while keeping the background systematic uncertainty fixed at $20\%$. The number of background events is quite low due to the hard leading jet $\pt$ and $\met$ cuts, so the projected reach is relatively insensitive to background systematic uncertainties.

The increased reach for the compressed spectra is due to the additional $\met$ resulting from the heavier LSP. We note that the entire neutralino-gluino coannihilation region (whose upper endpoint lies at $m_{\gluino} \approx m_{\chioz} \approx 8$ TeV \cite{Ellis:2015vaa}) can be excluded if the squark masses are $\lesssim 28$ TeV.

%There is some overlap between our results for the non-compressed spectra and the analysis of Cohen et. al \cite{Cohen:2013xda}, who considered both gluino pair production and associated production in similar spectra with squark masses $\lesssim 24$ TeV.
The results of the previous sections imply that gluino pair production is likely to be the discovery channel for coloured superpartners provided $m_{\gluino} \lesssim 14$ TeV.
However, for the compressed spectra, gluino pair production searches rapidly lose sensitivity.  As such, squark-gluino associated production could be a potential discovery channel
for spectra where the gluino and LSP are nearly degenerate.

\subsection{Squarks}
\label{sec:susy_squarks}

% not needed.
%\subsubsection{Squarks in simplified models.}

%% --------------------------------------------------------------------------------------
%% intro

While naturalness considerations motivate light stops, discussed in Section~\ref{sec:SUSY_stop}, and light
gluinos, discussed in Section~\ref{sec:susygluino}, supersymmetric partners of light-flavor quarks can have significantly larger masses at little extra fine-tuning cost.  However, models that include Dirac
gluinos~\cite{Kribs:2013oda} can accommodate light squark masses in a way that makes them a discovery mode
for BSM physics at hadron colliders.

This section summarizes the ``squark-neutralino'' simplified model discussed
in \cite{Cohen:2013xda, Cohen:2014hxa}, in which the first and second generation squarks
$\widetilde{q} = \widetilde{u}_L, \widetilde{u}_R$, $\widetilde{d}_L, \widetilde{d}_R,$ $\widetilde{c}_L,
\widetilde{c}_R,$ $\widetilde{s}_L, \widetilde{s}_R$ are the only kinematically accessible colored states.
All other SUSY particles are decoupled.  The relevant parameters of the model are the squark mass 
$m_{\squark}$, which is taken to be universal for the first two generations, and the neutralino mass
$m_{\chioz}$.  Squarks are pair-produced via strong interactions, and the only allowed decay is
to a light-flavor quark and the neutralino LSP.

Squark-neutralino simplified models have been probed at the LHC, operating at $\sqrt{s}=8$ TeV, by the
CMS~\cite{Khachatryan:2015vra} and ATLAS~\cite{Aad:2015iea} collaborations.  No significant excesses
have been observed, and limits on squark masses are approaching 1 TeV for neutralinos with masses up to
400 GeV.

The final state for the model under study is two high-\pt{} jets with significant \met.  Thus,
as with the gluino and associated gluino-squark searches in Section~\ref{sec:susygluino}, this model
is probed with a simple ``jets+\met'' search strategy, inspired by~\cite{ATL-PHYS-PUB-2013-002}.

Parton-level signal events were generated using \texttt{Madgraph5}  v1.5.10 \cite{Alwall:2011uj}.
All signals involve the pair production of SUSY particles and are matched using MLM matching up to 2 additional jets.
The $k_t$-ordered shower scheme with a matching scale of \texttt{qcut}=\texttt{xqcut}=$100\text{ GeV}$ was used.
Note that we do not account for any possible inadequacies inherent in the current Monte Carlo technology,
\emph{e.g.} electroweak gauge bosons are not included in the shower.

The gluinos and squarks were treated as stable at the parton level.  These events were subsequently decayed and
showered using \texttt{Pythia6} \cite{Sjostrand:2006za} and passed through the \texttt{Delphes} detector
simulation \cite{deFavereau:2013fsa} using the ``Snowmass" detector parameter card \cite{Anderson:2013kxz}.
Total production cross sections were computed at NLO using a modified version of
\texttt{Prospino} v2.1 \cite{Beenakker:1996ed, Beenakker:1996ch, Beenakker:1997ut}.

Background estimates are made using the ``Snowmass 2013'' background samples~\cite{Avetisyan:2013onh}.
Generated processes include $W/Z+$jets, $\ttbar$, single-top, diboson, $t+V$ and $\ttbar+V$, and Higgs.
QCD multijet backgrounds were not generated, thus the analysis makes stringent cuts on \met{} and related
quantities to ensure that QCD multijet backgrounds will be negligible.

The squark search is optimized in two different regions of the squark-neutralino mass plane.
The first search targets high-mass squarks with relatively-light LSPs using a straighforward jets+\met{} strategy.
The second search targets the ``compressed'' region where $m_{\squark}\approx m_{\chioz}$.

%% --------------------------------------------------------------------------------------

%% --------------------------------------------------------------------------------------
%% jets+met search

As with the jets+\met{} search for gluinos presented in section~\ref{sec:susygluino}, a standard event pre-selection is
defined by the following requirements:
\begin{itemize}
\item $\met/\sqrt{\Ht} > 15$ GeV$^{1/2}$
\item The leading jet \pt{} must satisfy $p_T^\text{leading} < 0.4\,\Ht$
\end{itemize}

After pre-selection, rectangular cuts on $\met$\ and $\Ht$ are simultaneously optimized to yield maximum signal significance.
The resulting requirements on \Ht{} and \met{} are typically a substantial fraction of the squark mass for low values of $m_{\chioz}$.
After optimization, the background is dominated by $W/Z + \text{jets}$, with smaller contributions from $\ttbar$ production.
All other backgrounds are negligible.  

The results of the squark search are shown in the solid lines in Fig.~\ref{fig:SQSQ_Comparison} for
four different collider scenarios.
The 14 TeV 300~\ifb{} limit with massless neutralinos is projected to be 1.5 TeV (corresponding
to 1022 events), while the 14 TeV 3000~\ifb{} limit is projected to be 1.7 TeV
(corresponding to 3482 events).  The 14 TeV LHC with 3000~\ifb{}
could discover a squark as heavy as 800 GeV if the neutralino is massless.
The 33 TeV 3000~\ifb{} limit with massless neutralinos is projected to be 3.4 TeV (corresponding to
3482 events), with discovery reach up to 1.4 TeV for massless neutralinos.

The 100 TeV 3000~\ifb{}
limit with massless neutralinos is projected to be 8.0 TeV (corresponding to 849 events), with
discovery reach up to 2.4 TeV if the neutralino is massless. Compared to the 14 and 33 TeV searches,
the squark reach degrades less rapidly as the neutralino mass is increased from the
massless limit.  The reduced cross section for light-squark production
and the lower jet multiplicity of the final state combine to reduce the mass reach for this model
relative to the stop or gluino searches.  

\begin{figure}[tbp]
  \centering
  \includegraphics[width=.48\columnwidth]{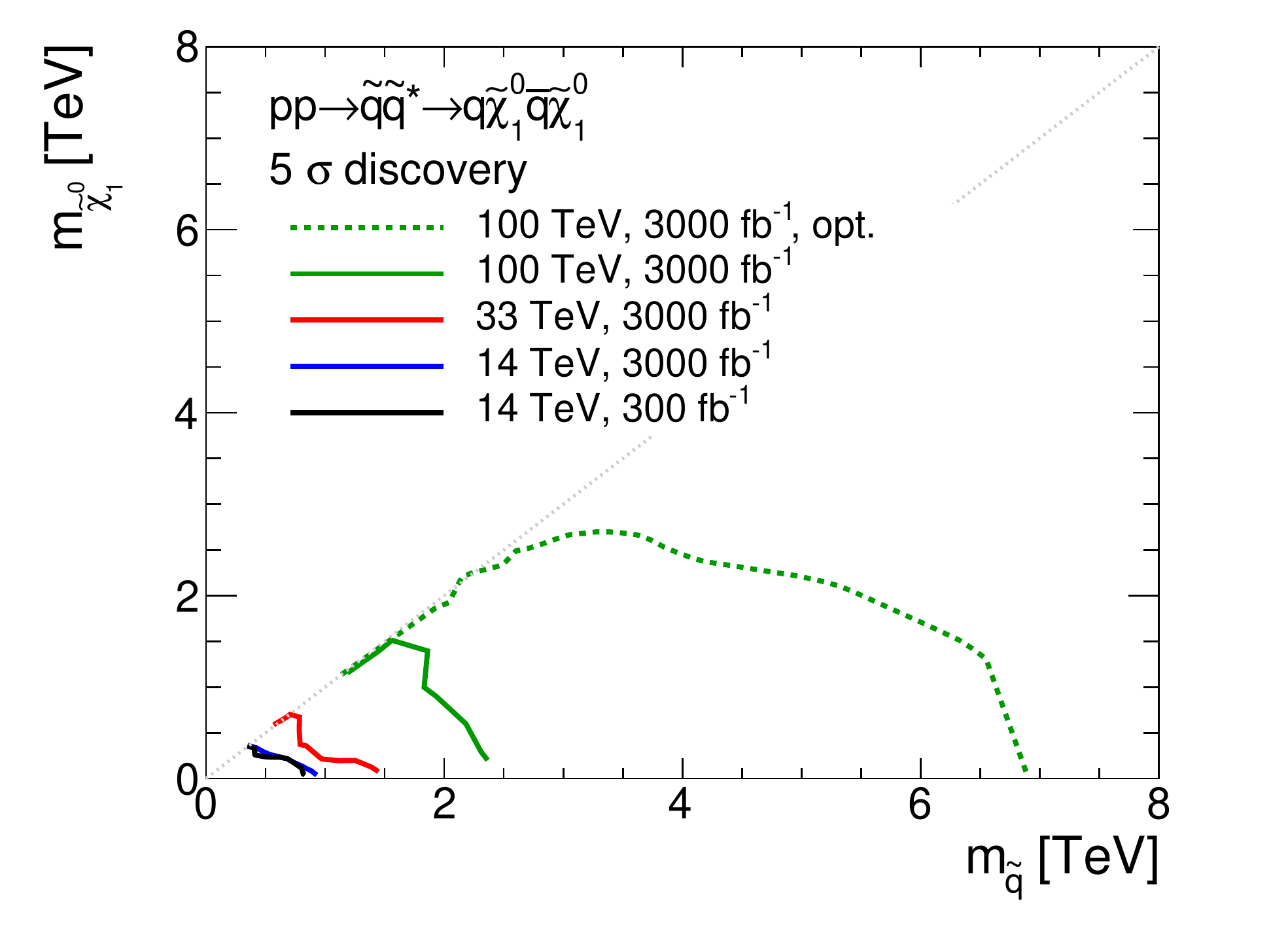}
  \includegraphics[width=.48\columnwidth]{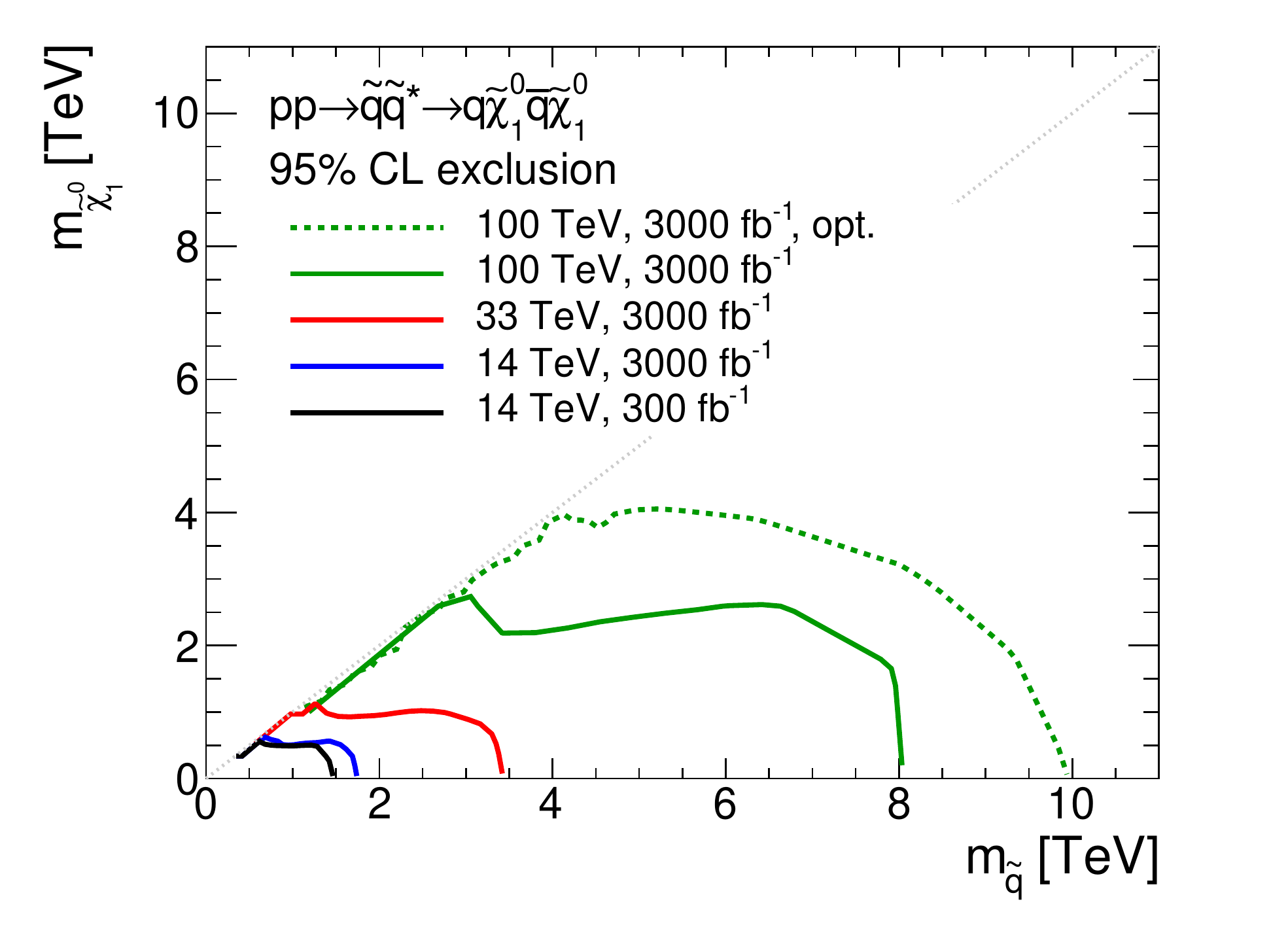}
  \caption{Results for the squark-neutralino model.  The left [right] panel shows the $5\,\sigma$ discovery reach [$95\%$ CL exclusion]
    for the four collider scenarios studied here.  A $20\%$ systematic uncertainty is assumed and pile-up is not included.  The dashed
  green line shows the results of a re-tuned search at $\sqrt{s}=$100 TeV.}
  \label{fig:SQSQ_Comparison}
\end{figure}

The poor performance of the search at 100 TeV motivated a re-analysis of this model for the 100 TeV scenario.
In the re-optimized analysis, the pre-selection requirements, which were optimized for the gluino-neutralino model described earlier,
are removed.  Events are required to have four jets with $\pt>500$ GeV, and must satisfy the following topological selection
requirements, motivated by the analysis in Ref.~\cite{Khachatryan:2015vra}:

\begin{itemize}
\item $(\vec{p}_{\mathrm{T}}^{\mathrm{\ miss}}-\sum_{\ell,j} \vec{p}_{\mathrm{T}})<100$ GeV
\item $\mathrm{min}\Delta\phi(p_{\mathrm{T}}^{\mathrm{miss}},$leading 4 jets$) < 0.6$
\end{itemize}

Requirements on \Ht{} and \met{} are then simultaneously optimized, as described earlier.  The results of the re-optimized
search are shown in the dashed green line of Fig.~\ref{fig:SQSQ_Comparison}.  The exclusion limits improve modestly, but the
discovery contours improve significantly, due partially to the increased signal acceptance of the new selection compared to
the previous jets+\met{} studies.  Further improvements are expected by implementing other analysis
techniques demonstrated in Ref.~\cite{Khachatryan:2015vra}, such as $\Ht$-binning and the use of variables like $M_{T,2}$.

%% --------------------------------------------------------------------------------------

%% --------------------------------------------------------------------------------------
%% Compressed search

The second analysis strategy targets the compressed region of the squark-neutralino plane, where:

\begin{equation}
  m_{\squark} - m_{\chioz} \equiv \Delta m \ll m_{\squark}.
\end{equation}

Due to the large LSP masses in this scenario, signal events often do not contain substantial \Ht, making a simple
jets+\met{} search ineffective.  In this case we rely on initial state radiation to boost the SUSY system, creating a
monojet+\met{} final state.

The compressed analysis is described in detail in Section~\ref{sec:susygluino}, and the
results of the search for four collider scenarios are shown in Fig.~\ref{fig:SQSQ_Compressed_Comparison}.  
For all four colliders, the \met-based strategy has the best performance and is used to quantify
the sensitivity.

At a 14 TeV collider, it is possible to exclude (discover) squarks
in the degenerate limit with mass less than $\sim$650 GeV (500 GeV) with 300~\ifb{} of data.
Increasing the integrated luminosity by a factor of 10 has a minimal impact on the
discovery reach for compressed squark models.
This search improves the exclusion (discovery)
reach near the degenerate limit by roughly 300 GeV (150 GeV) compared to the
jets+\met{}-based analysis described above; the jets+\met{} searches do not begin
to set stronger limits until $\Delta m\gtrsim$50 GeV.

For a 33 TeV collider, it is possible to exclude (discover)
squarks in the degenerate limit with mass less than $\sim$1.2 (0.7) TeV with 3000~\ifb{}
of data.  This does not substantially improves the discovery reach near the degenerate limit
compared to the jets+\met{} analysis, but does improve the exclusion reach by
roughly 200 GeV for $\Delta m\lsim 100$ GeV.

Finally, for the 100 TeV collider, it is possible to
exclude (discover) squarks in the degenerate limit with mass less than $\sim$4 TeV (3 TeV)
with 3000~\ifb{} of data. This improves the exclusion (discovery) reach near the degenerate
limit compared to the jets+\met{} analysis targeted at the non-compressed region
described above by roughly 1.5 TeV (1.8 TeV) for $\Delta m\lsim$200 GeV.

\begin{figure}[tbp]
  \centering
  \includegraphics[width=.48\columnwidth]{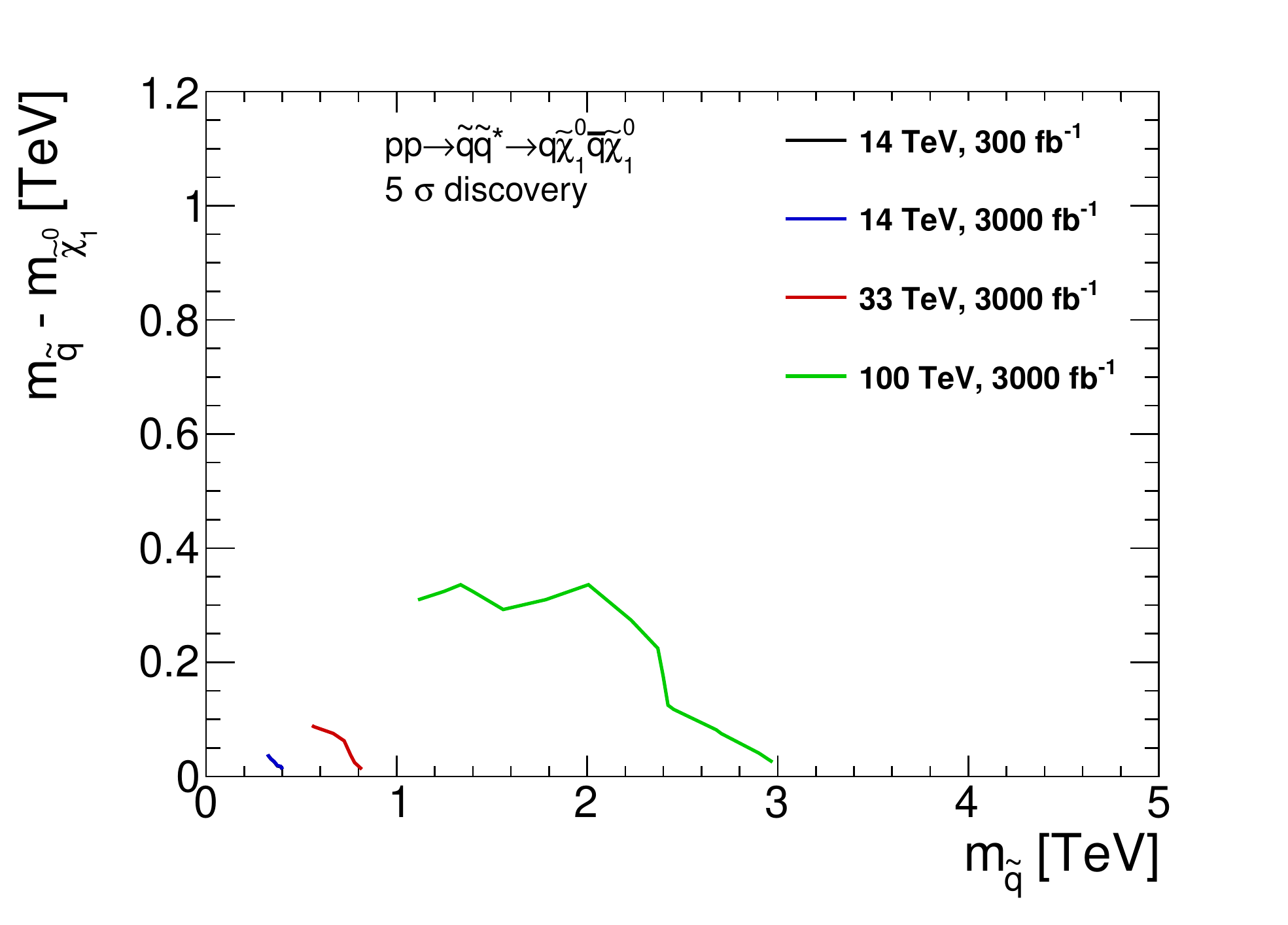}
  \includegraphics[width=.48\columnwidth]{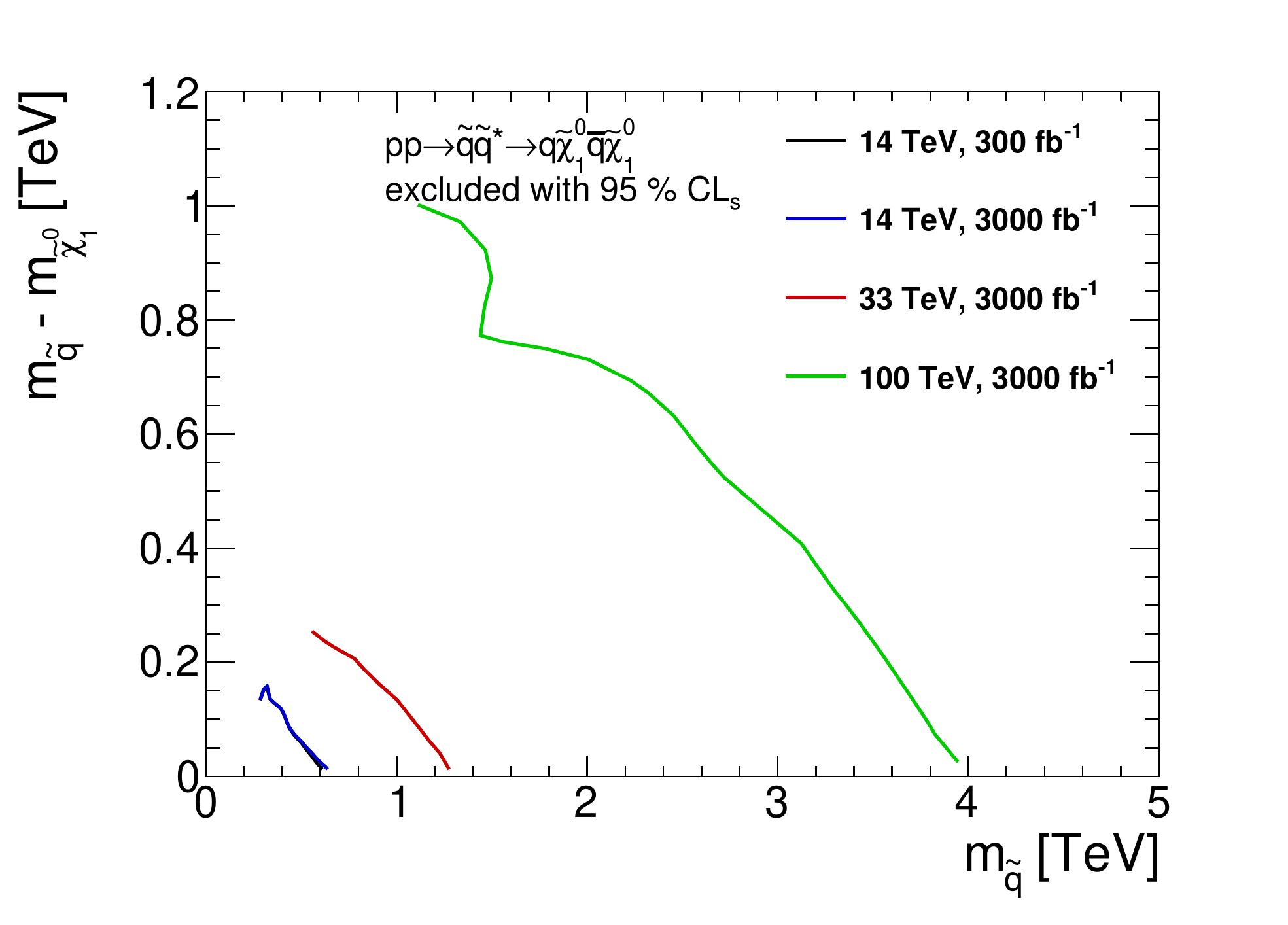}
  \caption{Results for the squark-neutralino model with light flavor decays for the analyses that target the compressed region of parameter space.
    The left [right] panel shows the $5\,\sigma$ discovery reach [$95\%$ CL exclusion] for the four collider scenarios studied here.
    A $20\%$ systematic uncertainty is assumed and pile-up is not included.}
  \label{fig:SQSQ_Compressed_Comparison}
\end{figure}

%% --------------------------------------------------------------------------------------

%{\flushleft Stefania Gori, Sunghoon Jung, Lian-Tao Wang, James D. Wells}

\subsection{Electroweakinos}% at a 100 TeV $pp$ collider}
\label{sec:susy_ewino}

This section describes the discovery prospects for electroweakinos -- Wino, Bino and Higgsino -- at a 100 TeV $pp$ collider.  For studies focussing on the dark matter aspect of electroweakinos, see Sec.~\ref{sec:DM}.
We focus on supersymmetric scenarios where electroweakinos have a mass at around the electroweak-TeV scale,
and all other superparticles are much heavier and beyond the collider reach.
We specifically consider scenarios in which the mass parameters of the electroweakinos are not too close to each other.
In this case, electroweakinos generally do not mix significantly with each other, leaving neutral and charged components
of Winos and Higgsinos almost degenerate and different kinds of electroweakinos well-separated in mass.
We call these nearly degenerate sets of states collectively the Lightest Supersymmetric Particle (LSP) ($\chi_1^{0,\pm}$)
or Next-to-Lightest Supersymmetric Particle NLSP ($\chi_2^{0, \pm}$)\footnote{In some work they are called the co-NLSP.}.

Here we very briefly summarize the work in Ref.~\cite{Gori:2014oua}\footnote{See also \cite{Acharya:2014pua} for a related work in the framework of a future 100 TeV collider.}
and focus on the direct production of NLSP pairs -- neutralino pair, chargino pair and neutralino-chargino pair --
and their subsequent decays to the LSP and a boson, either the Higgs boson $h$ or $W,Z$ gauge bosons,
producing a multiple lepton and missing transverse energy signature.

\begin{equation}
  \chi_2^{0, \pm}  \to \chi_1^{0, \pm} Z/h, \quad \chi_2^{0, \pm} \to \chi_1^{\pm, 0} W, \quad 
\end{equation}
\begin{equation}
  W \to \ell \nu, \quad Z \to \ell^+ \ell^-, \quad h \to ZZ^*, WW^* \to 4\ell, 2\ell 2\nu
\end{equation} 
Although final states involving hadronic jets are possible, multilepton signals typically provide the strongest discovery channels.
We divide multilepton signals into two opposite-sign leptons of any flavor (OSDL), two same-sign leptons of any flavor (SSDL),
three leptons ($3\ell$), and four leptons ($4\ell$), where leptons can be either electrons or muons.

For each simulated benchmark, we optimize the cuts on the following variables to maximize the statistical significance with an assumed luminosity of 3 ab$^{-1}$:
\begin{itemize}
\item \met
\item $p_T(\ell_2)/ p_T(\ell_1)$
\item $H_T$(jets)/$M_{\rm eff}$
\item $M_{\rm eff}^\prime = M_{\rm eff} - p_T(\ell_1)$
\item $M_T(E_T^{\rm{miss}},\ell\ell)$, the transverse masses between missing energy and various combinations of leptons
\item $E_T^{\rm{miss}}/M_{\rm eff}$
\end{itemize}
where $H_T$(jets) is the scalar sum of all jet $\pt$ (we do not veto any jets if present) and $M_{\rm eff}$ is the scalar $\pt$ sum of all jets,
leptons and missing energy.  We refer the reader to Ref.~\cite{Gori:2014oua} for more detailed discussions of the variables, cut optimization,
and other selection criteria that were considered.

We present results for the following cases: 
\begin{itemize}
\item Higgsino NLSP and Bino LSP (Higgsino-Bino) : $M_2 \gg \mu > M_1$.
\item Higgsino NLSP and Wino LSP (Higgsino-Wino) : $M_1 \gg \mu > M_2$.
\item Wino NLSP and Higgsino LSP (Wino-Higgsino) : $M_1 \gg M_2 > \mu$.
\item Wino NLSP and Bino LSP (Wino-Bino) : $\mu \gg M_2 > M_1$.
\end{itemize}
The mass of the heaviest electroweakino is fixed to be 5 TeV. Instead of following the simplified model approach, we take into account all predicted branching ratios of the NLSP to gauge bosons and the Higgs with various $\tan \beta$ and signs of electroweakino masses. Notably, for the first three cases with Higgsino as either the NLSP or the LSP, the branching ratios  do not depend sensitively on those parameters; and the branching ratios to the $Z$ and the Higgs boson are always the same~\cite{Jung:2014bda}. This is because the Higgsino system consists of two nearly degenerate neutralinos indistinguishable at colliders and summing their individual decays (only the sum is observable) leads to such a simple branching ratio relation. This can be derived from the Goldstone equivalence theorem, that holds generically in these scenarios as their mass separations are much larger than the electroweak scale, and from the Higgs alignment limit that we know from Higgs precision data. For the case of Wino-Bino, instead, the branching ratio of the NLSP depends sensitively on $\tan \beta$ and on the signs of mass parameters. 

We collect the $2\sigma$ exclusion bounds for the first three cases, with Higgsinos either LSP or NLSP, in Fig.~\ref{fig:discovery-1}. We do not specify the value of $\tan \beta$ and signs of mass parameters since the results almost do not depend on them. The $3\ell$ search (in red) provides the best overall sensitivity, but the SSDL (in yellow) can provide complementary sensitivity for the region with small mass-splitting. Maximum discovery reaches on the NLSP mass are between 1.5 and 2.3 TeV for massless LSP. The Wino-Higgsino case shows the best reach among the three cases because the Wino NLSP production rate is twice bigger than that of the Higgsino NLSP (see the right panel of the figure).

\begin{figure}
  \centerline{
    \includegraphics[width=0.33\columnwidth]{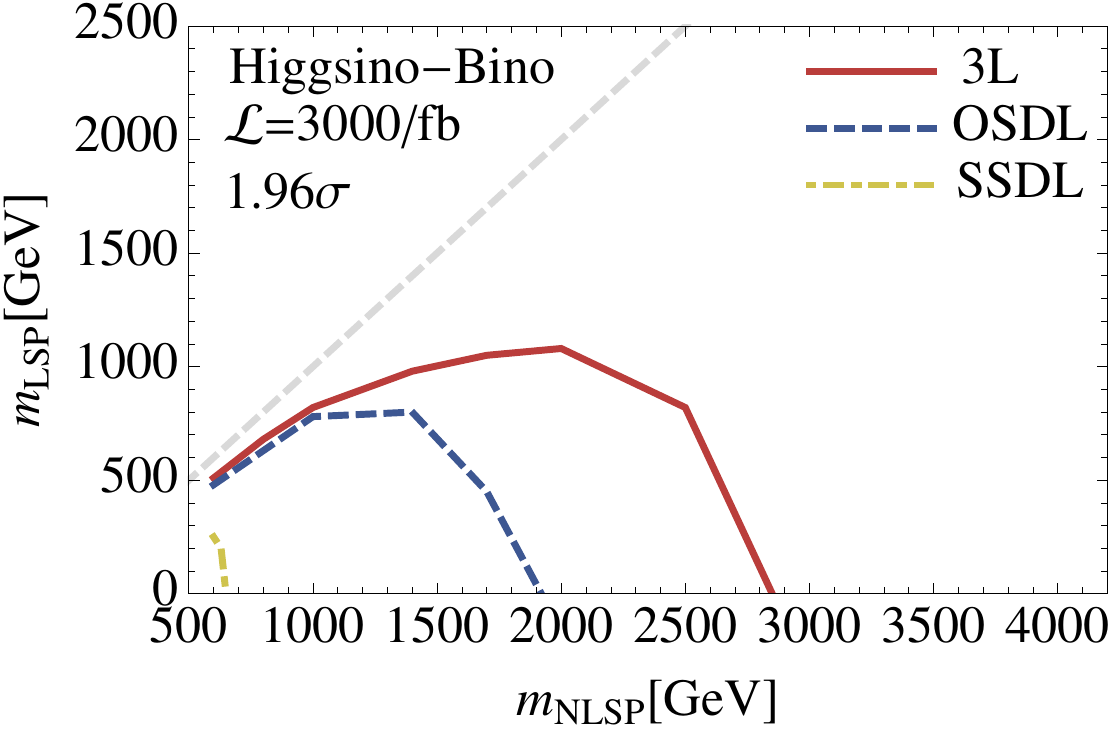} 
    \includegraphics[width=0.33\columnwidth]{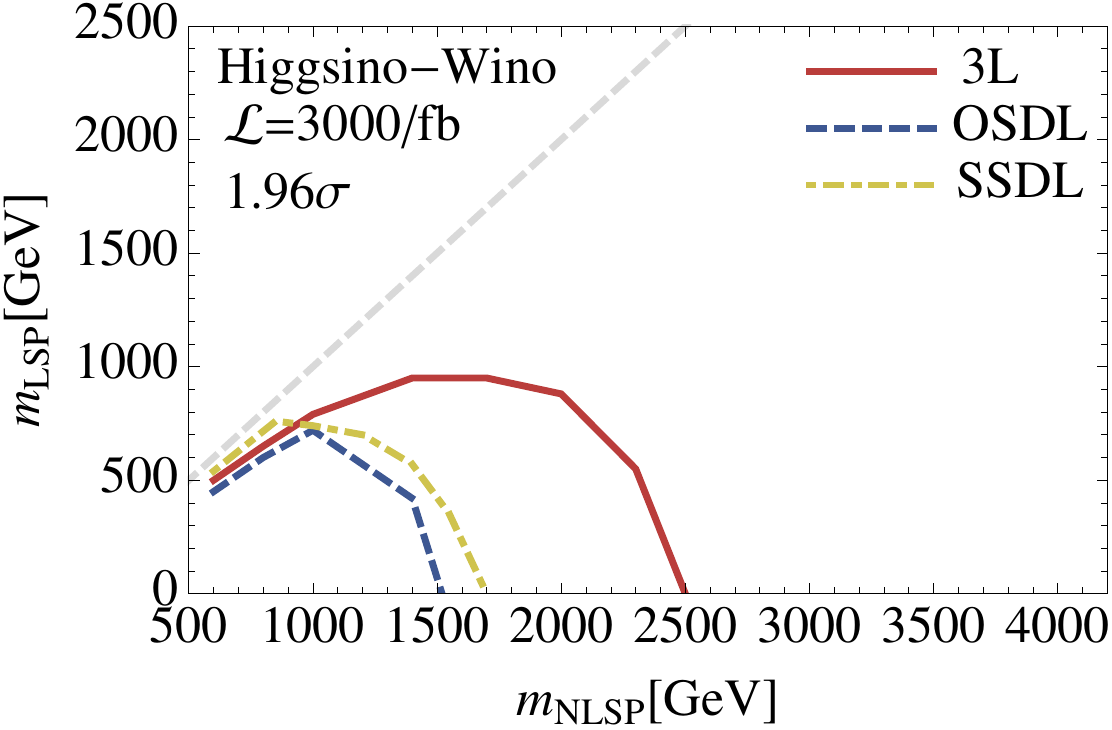} 
    \includegraphics[width=0.33\columnwidth]{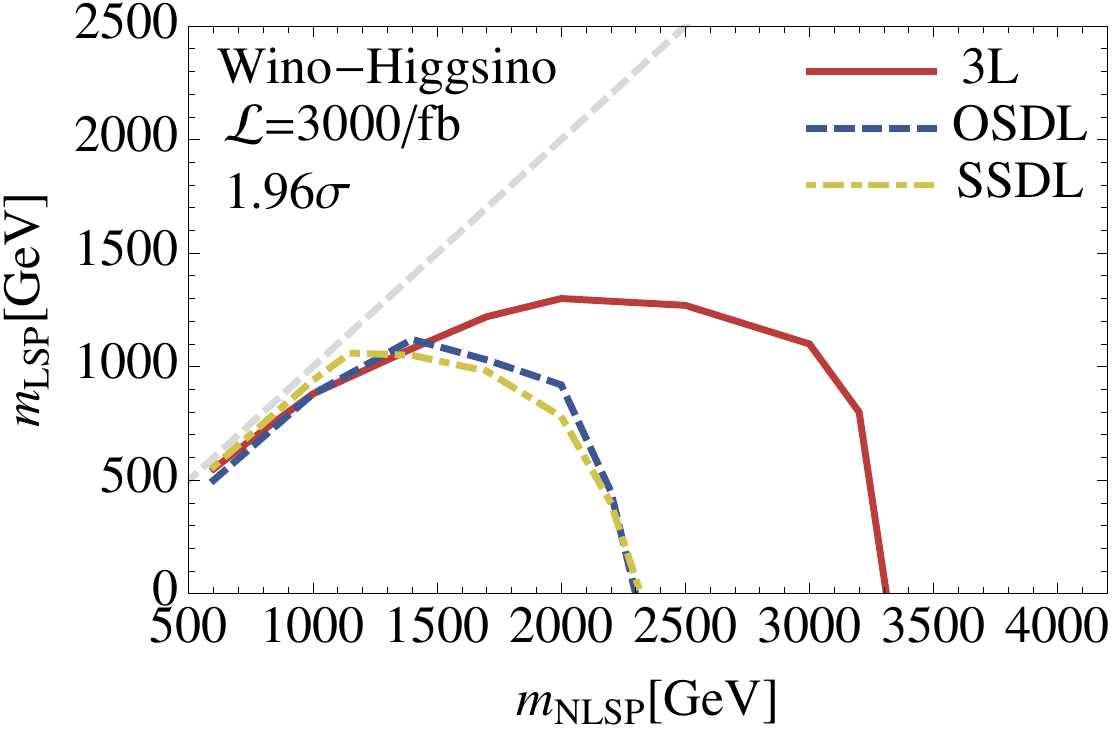} 
  }
  \caption{$2\sigma$ exclusion bounds of NLSP electroweakinos via $3\ell$ (red-solid), OSDL (blue-dashed) and SSDL(yellow-dotdashed) searches at a 100 TeV $pp$ collider with 3000 fb$^{-1}$. Three figures are for different NLSP-LSP combinations: Higgsino-NLSP and Bino-LSP (left), Higgsino-NLSP and Wino-LSP (middle), and Wino-NLSP and Higgsino-LSP (right). For the $5\sigma$ reach, see Ref.~\cite{Gori:2014oua}.}
  \label{fig:discovery-1}
\end{figure}

The results can also be interpreted to address whether thermal Dark Matter (DM) candidates of 1 TeV Higgsino or 3 TeV Wino~\cite{Hisano:2006nn,Cohen:2013ama,Fan:2013faa} can be discovered or excluded via electroweakino searches at a 100 TeV collider with $3\text{ ab}^{-1}$ of integrated luminosity.  The right panel demonstrates that an LSP Higgsino at 1 TeV can be excluded if the Wino has a mass lighter than $\sim 3$ TeV and not too close to 1 TeV. Wino DM, instead, cannot be probed with 3 ab$^{-1}$ luminosity (see the middle panel of the figure). Unfortunately, the discovery of the 1 TeV Higgsino (and 3 TeV Wino) DM with 3 ab$^{-1}$ data will be challenging (see the corresponding plots in~\cite{Gori:2014oua}).

The discovery and exclusion reach for the last case of Wino-Bino are collected in Fig.~\ref{fig:discovery-2}.
Four representative choices of additional parameters -- $\tan \beta$ and signs of mass parameters -- are considered. 
The four representative results differ significantly in the reach of the NLSP mass, in the shape of the reach curve,
and in the relative importance of $Z$ and $h$ boson contributions, primarily due to variations in the NLSP branching
ratios as the additional parameters change.

\begin{figure}
  \centering
  \includegraphics[width=0.48\columnwidth]{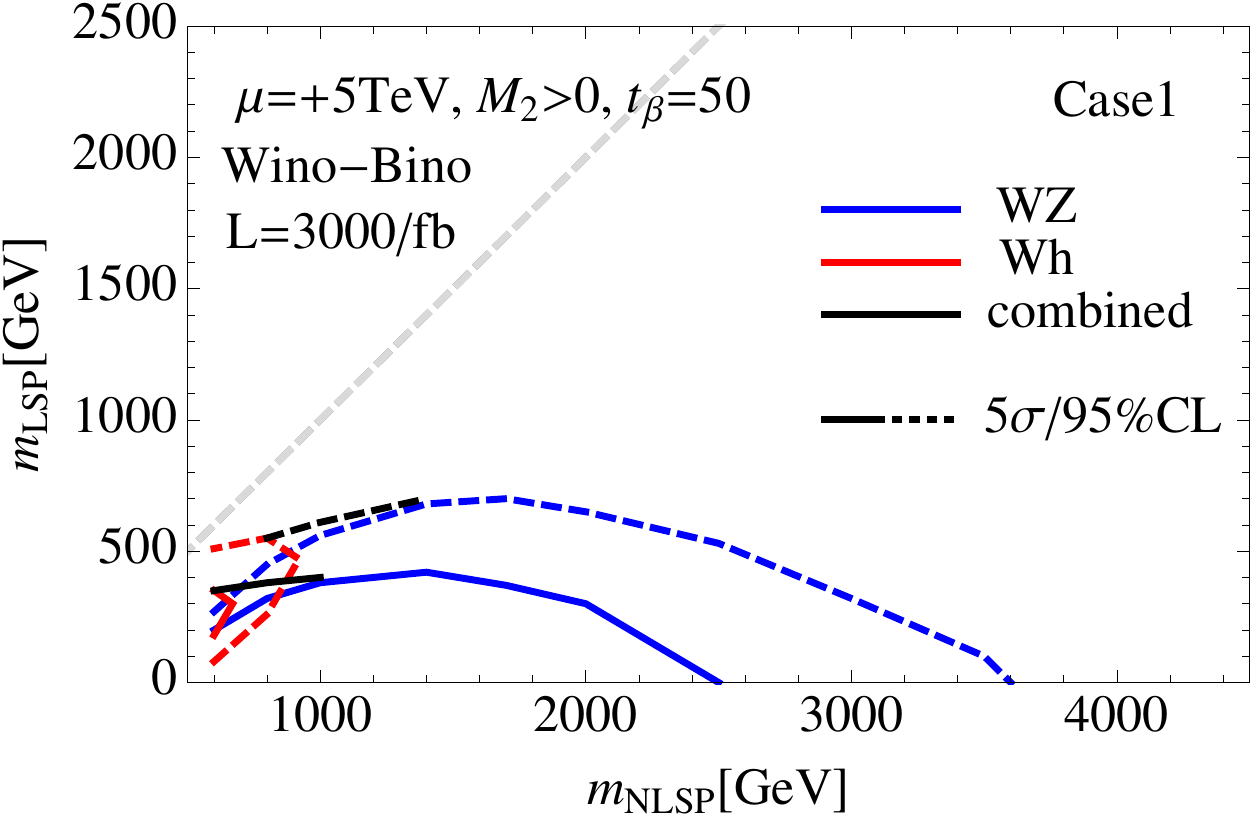} 
  \includegraphics[width=0.48\columnwidth]{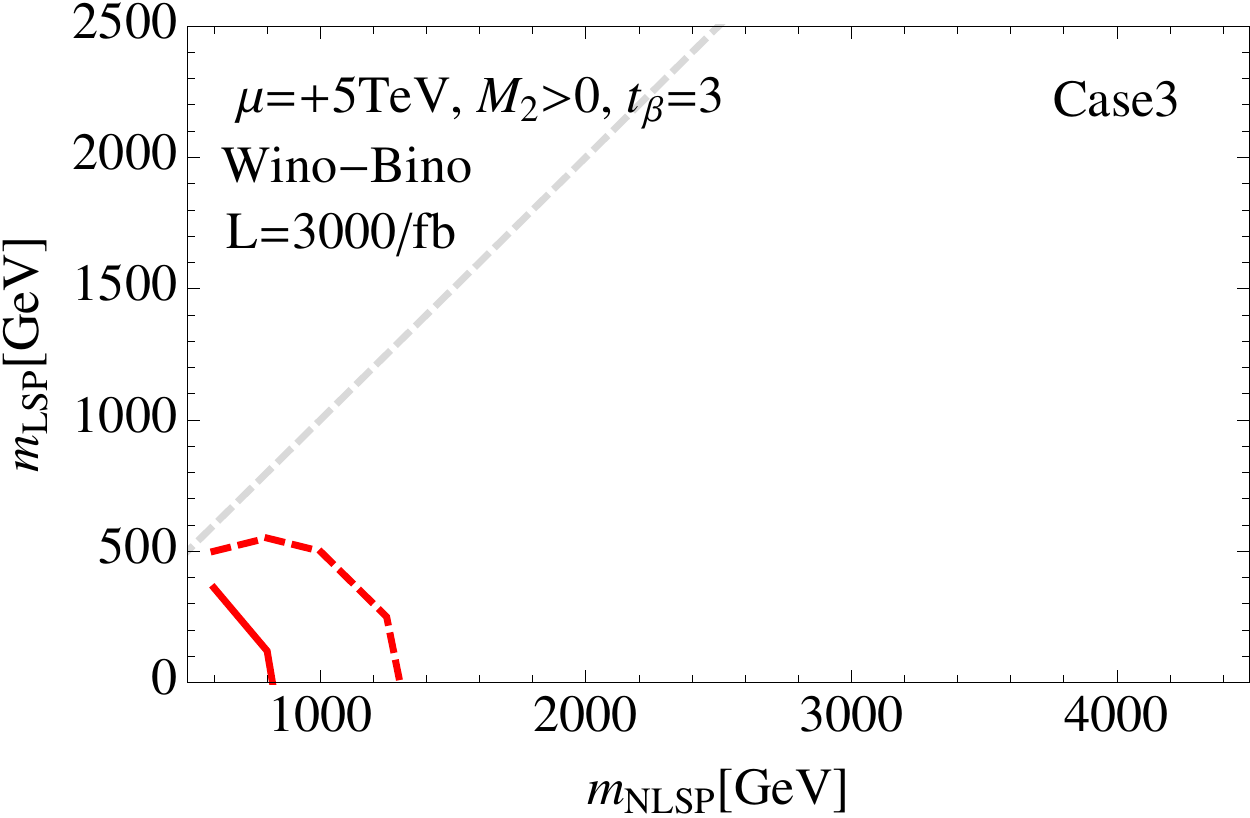}\\ 
  \includegraphics[width=0.48\columnwidth]{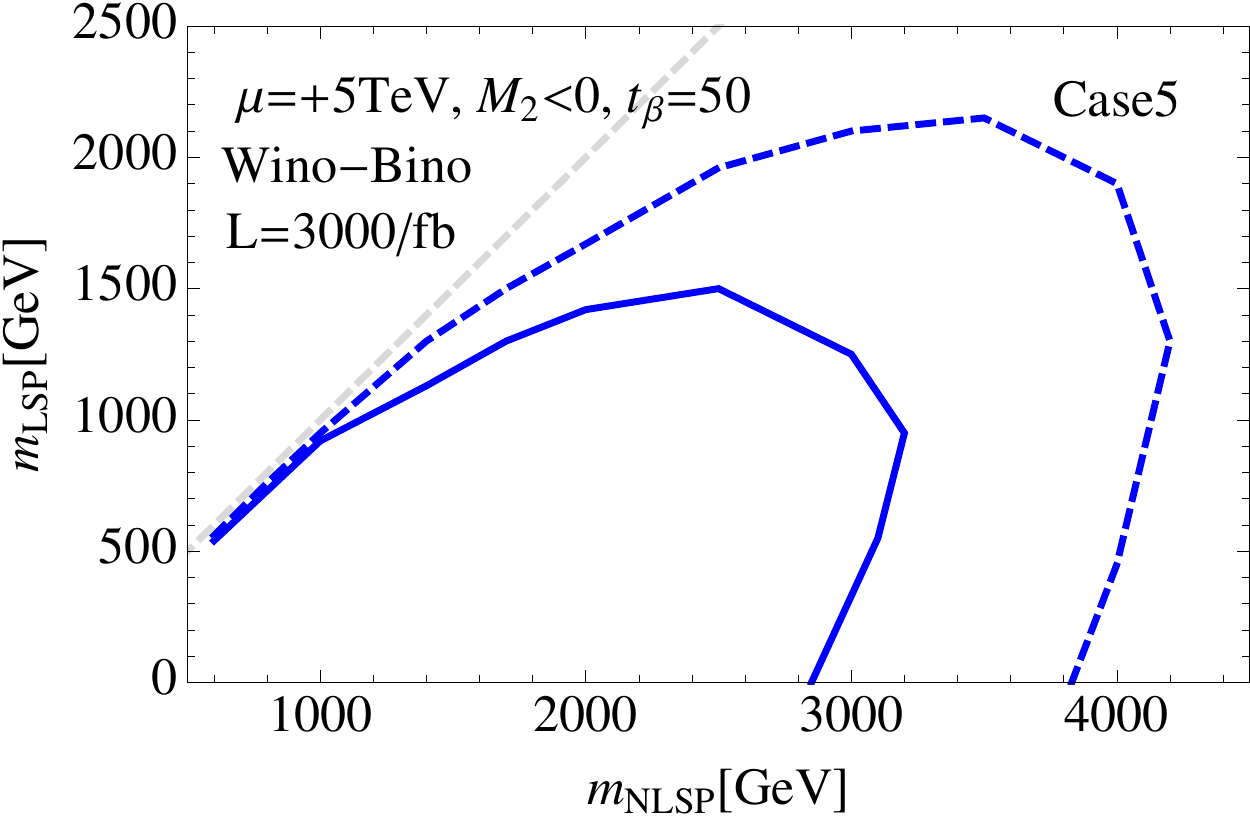}
  \includegraphics[width=0.48\columnwidth]{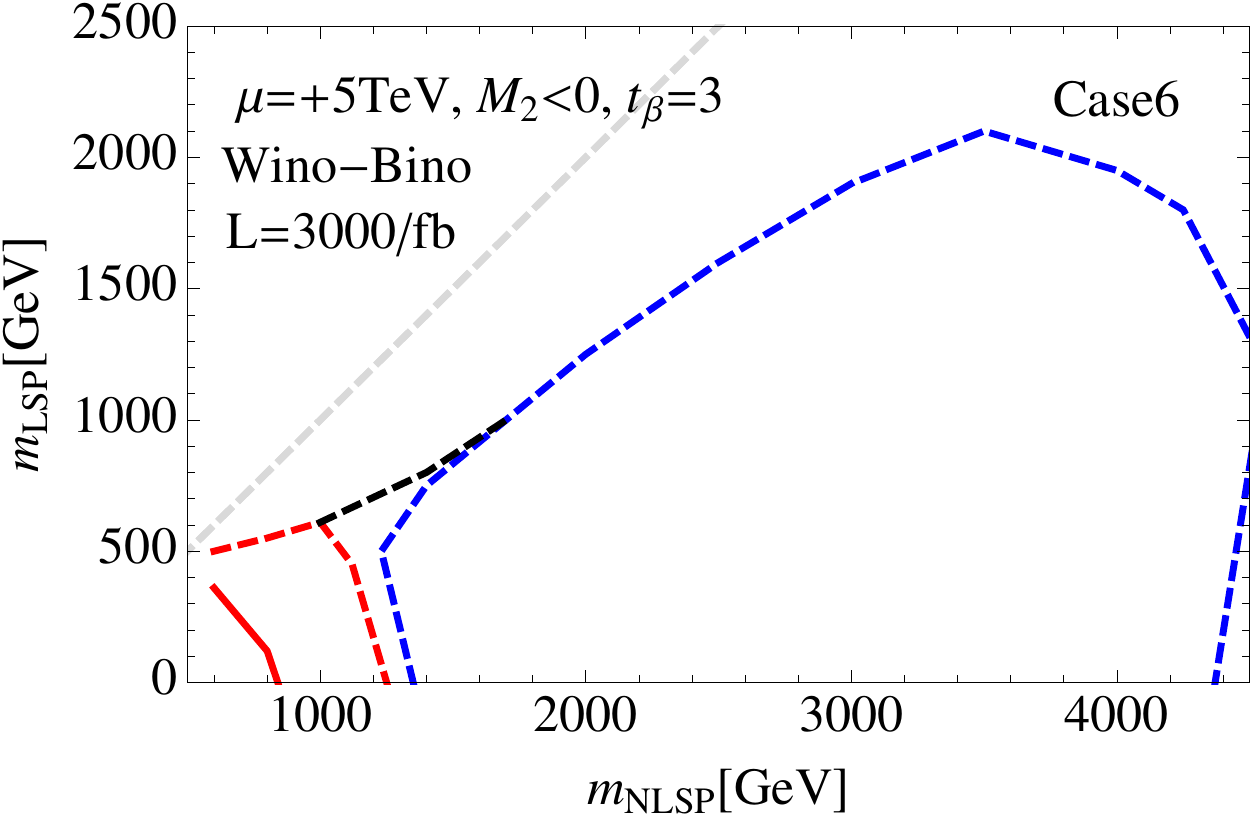} 
  \caption{$5\sigma$ discovery reach (solid) and 95\%CL exclusion (dashed)  for the case of Wino-NLSP and Bino-LSP with 3 ab$^{-1}$ at a 100 TeV $pp$ collider. Four representative choices of $\tan \beta$ and signs of mass parameters are shown. All multilepton channels are combined, but the $3\ell$ search contributes most. The contributions from intermediate $Z$(blue) or $h$(red) are separately shown to see the effects of NLSP branching ratios. For more results with different choices of the parameters, see Ref.~\cite{Gori:2014oua}.}
  \label{fig:discovery-2}
\end{figure}

The upper-right panel of Fig.~\ref{fig:discovery-2} demonstrates the importance of the Higgs boson contribution for small $\tan \beta$ and $\mu M_2 > 0$;
for other choices, there can be a (partial) cancellation between $\mu \sin 2\beta$ and $M_2$ terms for the Higgs partial width.
In other words, if the Higgsino is much heavier than the Wino,
such cancellation does not occur, making the decay to the Higgs boson always dominate, and the result becomes similar to the upper-right panel.
This effect is studied more extensively in Ref.~\cite{Baer:2012ts,Howe:2012xe,Arbey:2012fa}. When the branching ratio to the Higgs boson dominates,
the reach is relatively low because multi-lepton signals via the Higgs boson are suppressed by the small Higgs$\to WW/ZZ \to$multileptons branching ratios.

Other features of the curves in Fig.~\ref{fig:discovery-2} are driven by the branching ratio to $Z$ bosons, which depends on mass and other model parameters.  A detailed discussion
of the reach is provided in Ref.~\cite{Gori:2014oua}. In the optimal case, with almost 100\% branching ratio to the $Z$ boson, as in the lower-left panel,
multilepton signals can enable the discovery of NLSPs  with mass up to about 3 TeV for massless LSP with 3 ab$^{-1}$. 

Multilepton events with small angular separation between the leptons is a common feature of multi-TeV electroweakino production.
Such events are outside of the acceptance for isolated-lepton searches, but relaxing the requirements on lepton separation in $R$
can significantly improve the acceptance for high-mass signals.
For example, the luminosity needed to probe a 3.5 TeV Wino is reduced by a factor of two for $\Delta R(\ell,\ell)>0.05$ compared to $\Delta R(\ell,\ell)>0.1$.
Searches for an NLSP heavier than 3 TeV, which often produces collimated leptons, are also significantly improved by retaining events with near-by leptons.
This should be an important consideration for the design of the detectors at future $pp$ colliders.

In summary, a 100 TeV $pp$ collider, even with just  3 ab$^{-1}$ of integrated luminosity, can significantly improve the reach for electroweakinos compared to the LHC.
This provides an important probe of SUSY even in the difficult scenario in which the colored superpartners are heavy.
Of course, even if SUSY is discovered in other search channels, the discovery and studies of electroweakinos are crucial in understanding the nature of SUSY breaking.
Finally, even though the study presented here is in the context of SUSY, the general lesson is applicable to a broader range of possible new physics particles with only electroweak quantum numbers.

% chargino/neutralino (Acharya et al) http://inspirehep.net/record/1320763
% pMSSM and mono-jet http://inspirehep.net/record/1389857
%  Higgs and SUSY  http://inspirehep.net/record/1370688

\subsection{Long-lived Charged Particles}
\label{sec:SUSY_exotic}
Due to low backgrounds and increased production cross sections at 100 TeV, exotic processes may be a promising avenue for discovering new physics.  These exotic processes could include displaced vertices and long-lived charged or coloured objects.  In this section we focus on a particular example motivated in supersymmetry: long-lived charged sleptons.

We study the prospects for long-lived charged particle (LLCP) searches
at a 100\,TeV $pp$ collider, compared to the 14\,TeV LHC,
using time-of-flight measurements.
We use Drell--Yan pair-produced long-lived sleptons as an example.
A novel feature of 100\,TeV collisions is the significant energy loss
of energetic muons in detectors, which we utilize to discriminate
against fake LLCPs.
We find that the 14\,TeV LHC with an integrated luminosity
of 3\,ab$^{-1}$ is sensitive to LLCP sleptons with $m\lesssim 1.2\,\textrm{TeV}$,
and a 100\,TeV $pp$ collider with 3\,ab$^{-1}$ is sensitive up to $\sim4\,\textrm{TeV}$,
probing interesting dark matter scenarios, including in particular
slepton--neutralino  co-annihilating WIMP dark matter,
and superWIMP dark matter.

Long-lived charged particles (LLCPs), which are stable on collider--detector
timescales, require specific methods for triggering, reconstruction, and detection.
Thus they provide interesting, model-independent benchmarks for future collider experiments.
Furthermore, their discovery will have profound implications for particle physics
as well as for cosmology, where their long lifetime may affect the thermal history of the Universe.

Many extensions of the Standard Model predict LLCPs.
Supersymmetry contains LLCPs in large portions
of its parameter space.
In the slepton--neutralino co-annihilation scenario~\cite{Griest:1990kh,Ellis:1998kh,Citron:2012fg,Desai:2014uha},
dark matter (DM) consists of the lightest supersymmetric particle (LSP),
which is the neutralino  $\chioz$.
A charged slepton  $\slepton$ is the next-to-lightest SUSY 
particle (NLSP), and is almost degenerate with the LSP.
The slepton NLSP is then long-lived because its decay to the LSP 
is phase-space suppressed.
In the early Universe, the slepton remains in thermal equilibrium almost 
until the DM freezes-out, and the DM relic abundance 
is diluted through co-annihilations with the NLSP slepton. 
The correct relic abundance is obtained for slepton masses 
$m_{\slepton}\lesssim 600\,\textrm{GeV}$.

Another scenario of interest is superweakly-interacting massive (superWIMP)
DM~\cite{Feng:2003xh,Feng:2003uy}.  This scenario is naturally realized in
gauge-mediated SUSY breaking, where the gravitino $\tilde G$ is the LSP \cite{Ellis:1984eq,Moroi:1993mb,Ellis:2003dn} and 
a charged slepton is often the NLSP.  The NLSP slepton has a mass
$m_{\slepton}\sim1~\TeV$, and decays to the LSP with a lifetime
\begin{equation}
  \tau(\slepton \to l \tilde{G}) = 0.59\,\textrm{sec}
  \left(\frac{\textrm{TeV}}{m_{\slepton}}\right)^{5}
  \left(\frac{m_{\tilde{G}}}{\textrm{GeV}}\right)^2
  \left(1 -\frac{m_{\tilde{G}}^2}{m_{\slepton}^2} \right)^{-4}.
\end{equation}

The NLSP freezes out in the early Universe with a relic density larger 
than the observed value $\Omega h^2\simeq0.12$, and later decays to the LSP. 
The relic density is then diluted by the mass ratio $m_{\tilde G}/m_{\slepton}$.
Assuming that the NLSPs are right-handed sleptons ($\slepton_{\rm R}$),
and that $N_{\rm gen;LL}$ of them are co-NLSP ($1\le N_{\rm gen;LL}\le3$),
i.e. degenerate and long-lived, the gravitino relic density is numerically
given by~\cite{Feng:2015wqa}
\begin{equation}
  \Omega_{\tilde G} h^2 = N_{\rm gen;LL} \cdot 0.12 \, \frac{ m_{\slepton_{\rm R}} m_{\tilde{G}} } {M^2} \ ,
\end{equation}
where $M$ varies from 650~\GeV{} to 1.0~\TeV{} as the Bino mass  
varies from $m_{\tilde{B}} = \infty$ to $m_{\slepton_{\rm R}}$.
Figure~\ref{fig:superwimpparameters} shows the relic abundance and NLSP lifetime
in this scenario in the slepton--LSP mass plane.
The NLSP slepton with $m_{\slepton_{\rm R}}\gtrsim650\,\textrm{GeV}$ is cosmologically viable,
and they are observed as LLCPs at collider if $m_{\slepton_{\rm R}}\lesssim40\,\textrm{TeV}$.
We also show in this figure the expected reach of the 14\,TeV LHC and of a 100\,TeV collider,
which are the main results of Ref.~\cite{Feng:2015wqa}.

%%%%%%%%%%%%%%%%%%%%%%%%%%%%%%%%%%%%%%%%%%%%%%%%%%%%%%%%%%%%
\begin{figure}[tbp]\centering
  \includegraphics[width=0.4\textwidth]{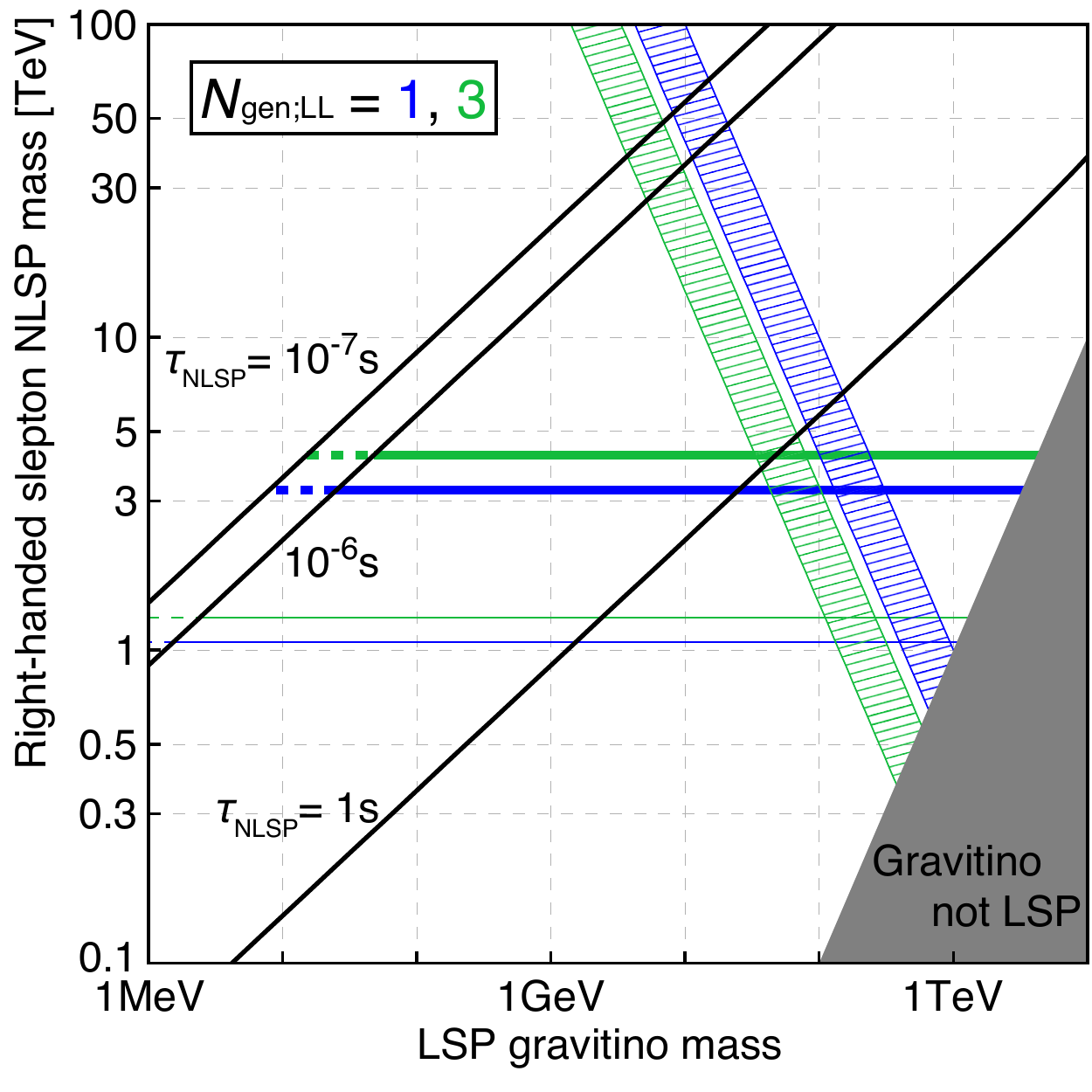}
  \caption{An overview of the parameter space in superWIMP scenarios with $\slepton_{\rm R}$-NLSP.
    In the blue (green) hatched region, gravitinos may saturate, depending on $m_{\tilde B}$, the DM relic density if
    $N_{\rm gen;LL}=1$ (3), i.e., one (three) among $\tilde e_{\rm R}$, $\tilde \mu_{\rm R}$, and $\tilde\tau_{\rm R}$ is long-lived.
    The black lines illustrate the lifetime of the NLSP slepton,
    and the horizontal lines are the expected exclusion limits
    of right-handed long-lived slepton searches: the thinner (thicker) lines are
    the 14\,TeV LHC (a 100\,TeV collider) with an integrated luminosity of 3\,ab$^{-1}$.}
  \label{fig:superwimpparameters}
\end{figure}
%%%%%%%%%%%%%%%%%%%%%%%%%%%%%%%%%%%%%%%%%%%%%%%%%%%%%%%%%

%\subsubsubsection{Collider experiments}\label{sec:susy_exotic_collider}
%LLCPs have been searched for by many collider experiments.

For collider experiments, muons constitute the main background to LLCP searches.
The only difference between a hypothetical LLCP and a muon is the (assumed) large mass of the former. 
Because of their large mass, LLCPs will typically be produced with a smaller 
value of relativistic $\beta$.  This velocity can be measured using the time-of-flight (ToF) to the outer detectors, 
or through the ionization energy loss, ${\rm d}E/{\rm d}x$, typically in the innermost layers of a silicon tracking detector.
Searches in 100\,TeV collisions can exploit a new handle for discriminating LLCPs from muons~\cite{Feng:2015wqa}.
Energetic muons with $p\gtrsim100\,\textrm{GeV}$ lose energy through radiative processes,
i.e., bremsstrahlung, electron pair-production, and photo-nuclear interactions~\cite{Groom:2001kq},
in addition to the ionization process, while LLCP's, with lower values of $\beta$, will radiate significantly less.
Therefore, by measuring the energy loss $E_{\rm loss}$ along the track of LLCP candidates, 
we can reduce the number of muon fakes.

We consider a $\slepton_{\rm R}$ LLCP with $N_{\rm gen;LL}=1$ 
as a benchmark model, and assume a worst-case scenario in which the 
only production process available is the Drell--Yan direct pair-production 
$pp\to(\gamma,Z)\to\slepton_{\rm R}\slepton_{\rm R}^*$.
For this scenario, the CMS (ATLAS) collaboration obtained a 
lower bound $m>346\,(286)\,\textrm{GeV}$ on the LLCP 
mass~\cite{Chatrchyan:2013oca,ATLAS:2014fka},
where the LLCP sleptons are identified by ToF and ${\rm d}E/{\rm d}x$ measurements.
Note that this is the most pessimistic limit; if heavier SUSY particles
%which we assume beyond the collider reach, 
can be produced, LLCPs coming from their cascade decays will also contribute 
to the signal, leading to more stringent limits.

The capabilities of future LHC runs and a 100\,TeV $pp$ collider are studied in 
Ref.~\cite{Feng:2015wqa}, utilizing the detector design and background samples from the 
Snowmass 2013 Community Summer Study~\cite{Anderson:2013kxz,Avetisyan:2013onh,Avetisyan:2013dta}.
Fig.~\ref{fig:LumiPlots} summarizes the results.
The 14\,TeV LHC is expected to probe the slepton--neutralino co-annihilation 
scenario with an integrated luminosity $\int\mathcal L=0.3\,{\rm ab}^{-1}$, 
and a high-luminosity run will discover or exclude 1\,TeV slepton LLCPs.
Meanwhile, a 100\,TeV $pp$ collider will access 3\,TeV with $\int\mathcal L=3\,\rm ab^{-1}$.

%%%%%%%%%%%%%%%%%%%%%%%%%%%%%%%%%%%%%%%%%%%%%%%%%%%%%%%%%%%
\begin{figure}[tbp]\centering
  \includegraphics[width=0.48\textwidth]{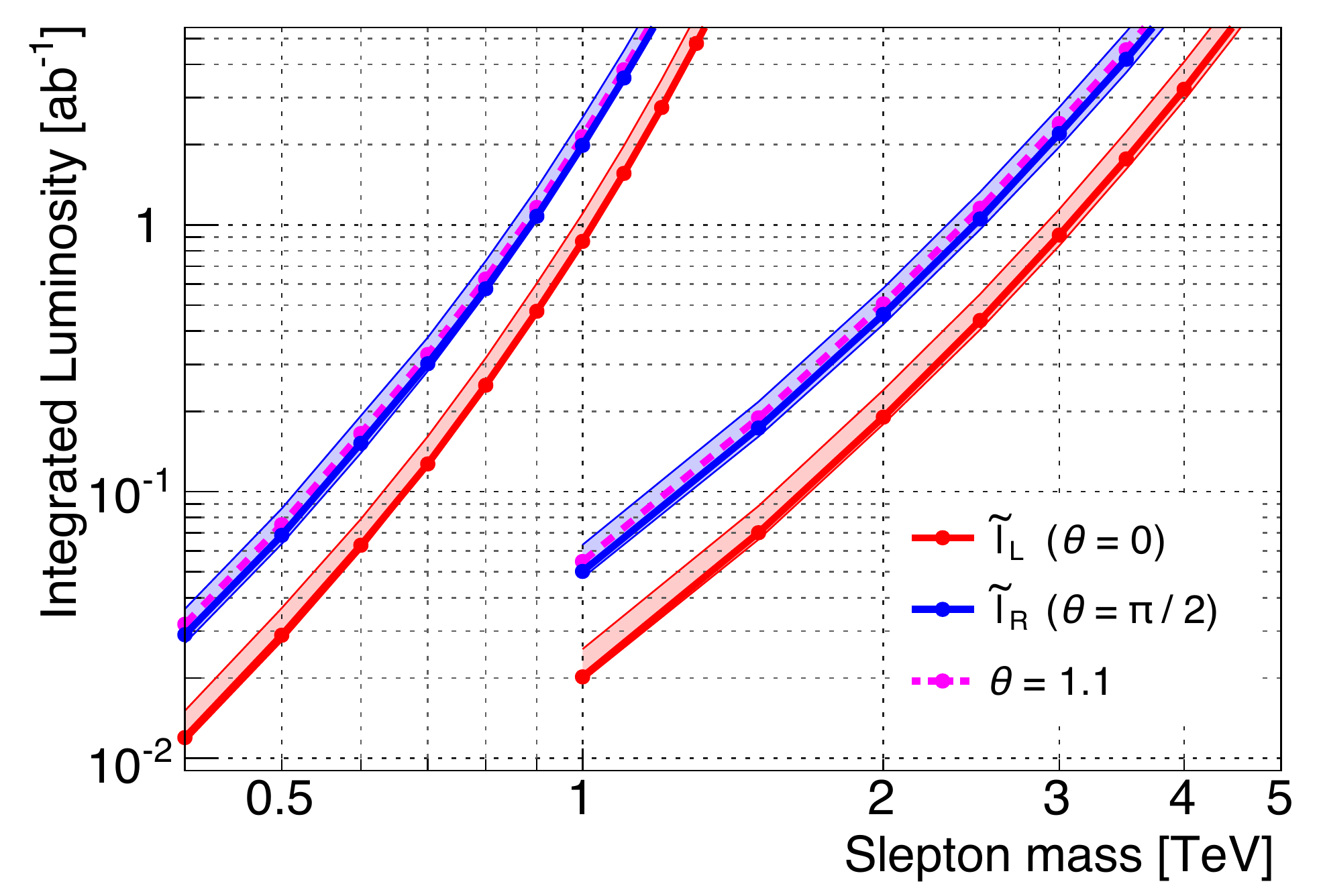}
  \hfill
  \includegraphics[width=0.48\textwidth]{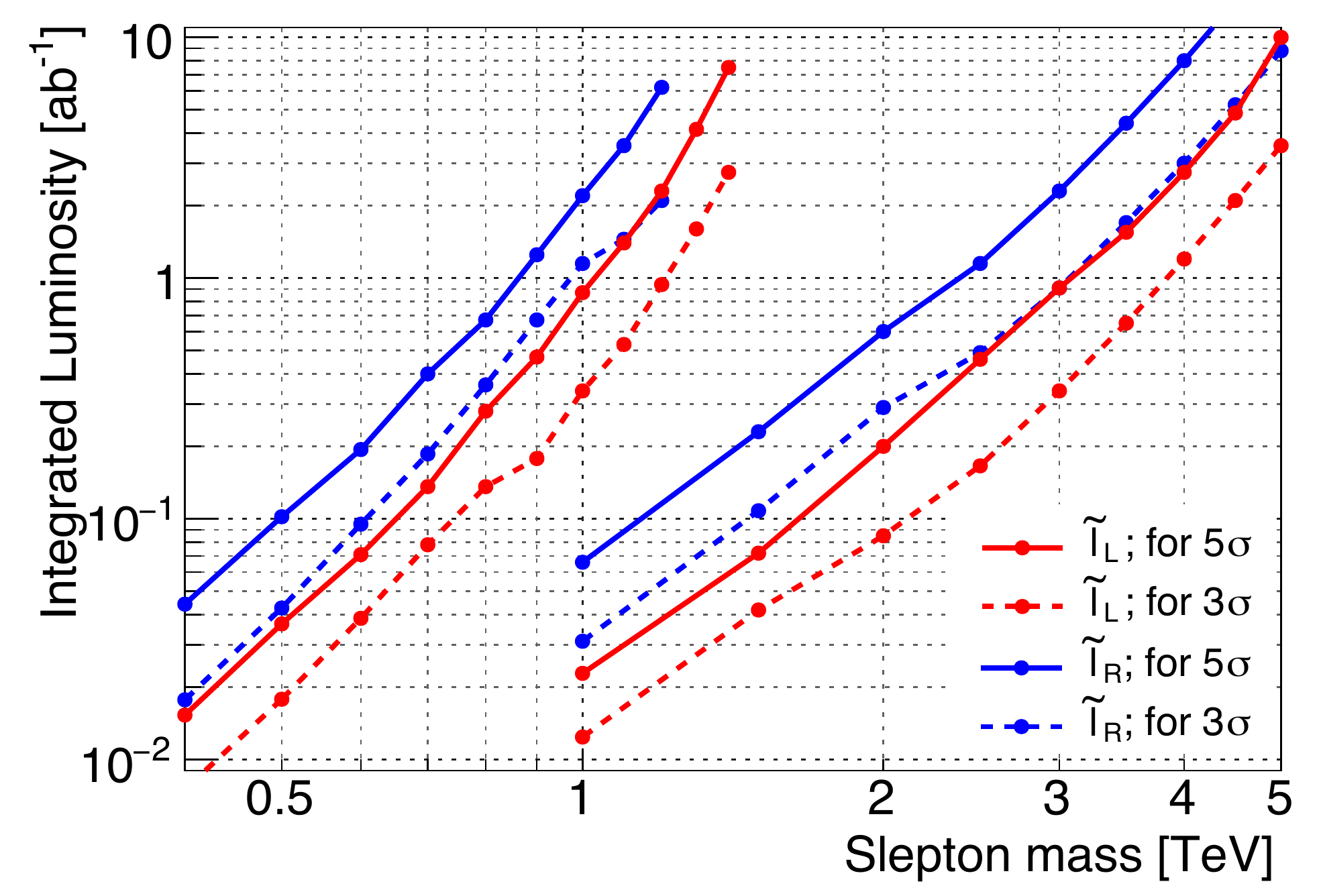}
  \caption{%
    (left)
    Expected capabilities of the 14\,TeV LHC and a 100\,TeV $pp$ collider on searches for long-lived sleptons with $N_{\rm gen;LL}=1$,
    plotted in terms of the integrated luminosity required for $95\%$ CL exclusion.
    Only the Drell--Yan pair-production of the LLCPs is assumed, and three values of the slepton left--right mixing angle $\theta$ are used.
    Note that, for a fixed $m_{\slepton}$, the Drell--Yan production cross section is maximized (minimized) at $\theta\simeq 0$ ($\theta\simeq 1.1$).
    The lines for $\slepton_{\rm L}$ ($\theta=0$) and $\slepton_{\rm R}$ ($\pi/2$) are
    drawn as bands, which show the $68\%$ statistical uncertainty of the expectation.
    (right)
    The integrated luminosity required for discovery.
    Solid (dashed) lines are for $5\sigma$- ($3\sigma$-) discoveries.%
  }
  \label{fig:LumiPlots}
\end{figure}
%%%%%%%%%%%%%%%%%%%%%%%%%%%%%%%%%%%%%%%%%%%%%%%%%%%%%%%%

In the 100\,TeV analysis, a qualitatively new event selection is introduced 
based on the energy deposit of a candidate LLCP in the calorimeter, $E_{\rm loss}$.
The energy loss of muons in matter, simulated with \texttt{Geant\,4.10}~\cite{GEANT4}, is shown 
in Fig.~\ref{fig:MuonEnergyLoss}.  Based on this, the signal LLCPs, which are
initially selected by $0.4<\beta<0.95$ and $P_{\rm T}>500\,\textrm{GeV}$,
are further required to have  $E_{\rm loss} \leq 30\,\textrm{GeV}$. 
The calorimeter is approximated as iron of 3\,m thickness.
The $E_{\rm loss}$ requirement removes 18\% of fake LLCPs.
%Because 
If each event is required to have two LLCPs, the muon-fake background is reduced by 34\%.
We note that pile-up may degrade the calorimeter resolution, and encourage a careful
study of this issue when designing future detectors.
Note that the $E_{\rm loss}$ cut is not introduced in the 14\,TeV LHC analysis, which has
a looser requirement of $\pt>100~\GeV$.

\begin{figure}[tbp]\centering
  \includegraphics[width=0.48\textwidth]{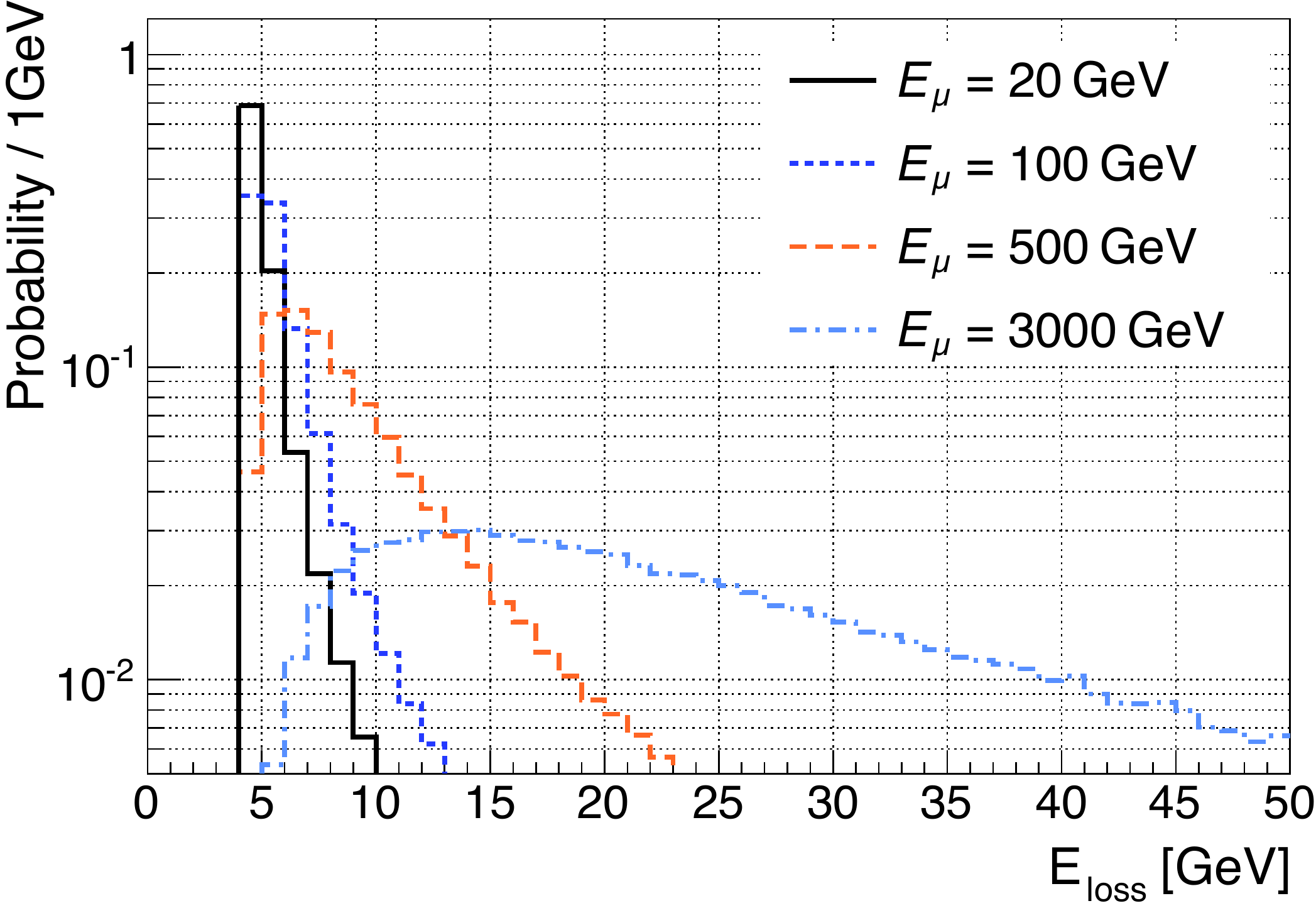}
  \hfill
  \includegraphics[width=0.48\textwidth]{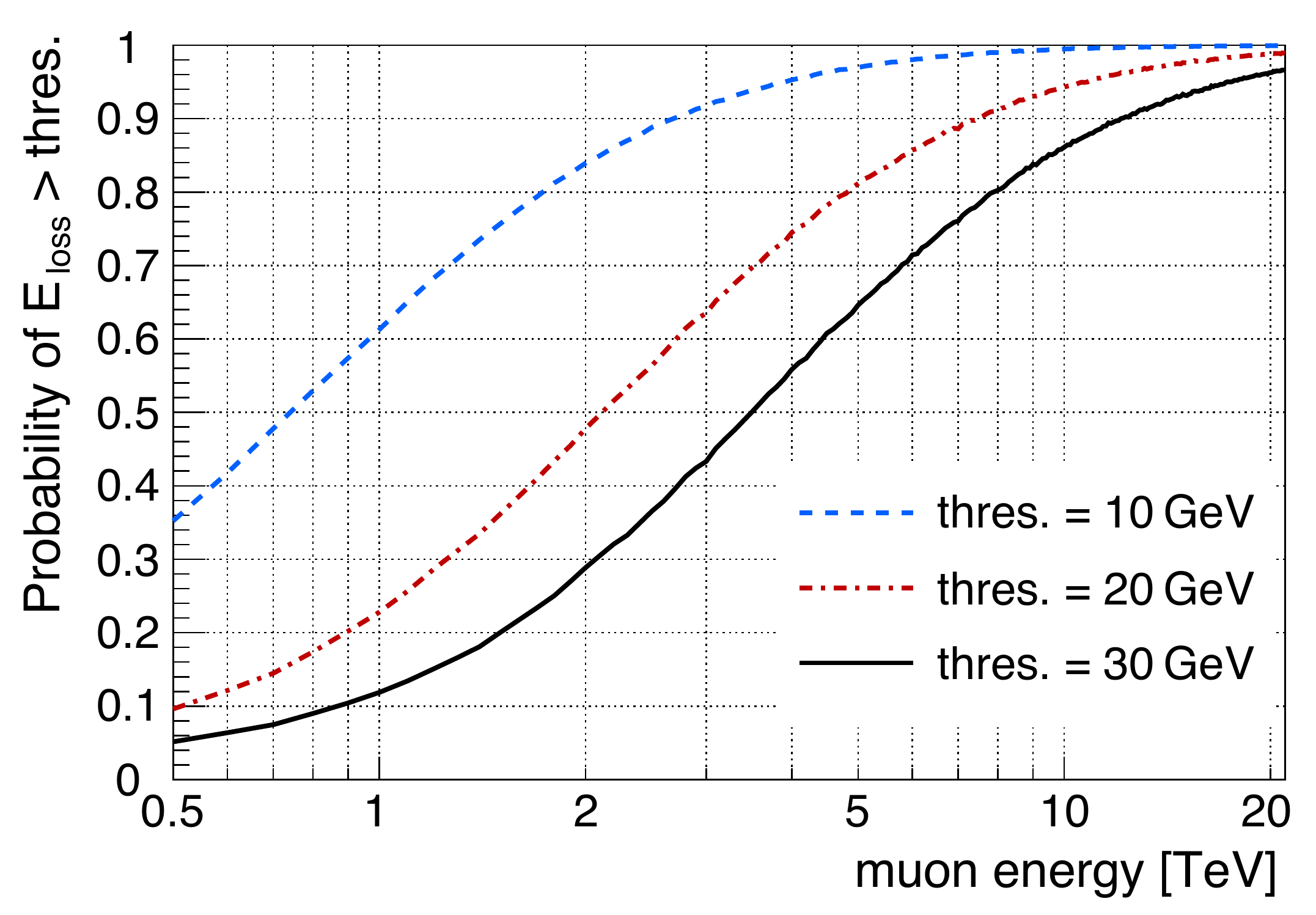}
  \caption{%
    (left)
    Energy deposit of a muon in the calorimeter, which is modeled as 3.0\,m iron.
    Note that this includes the ionization energy loss, which is 4.8--6.0\,GeV for $E_\mu=20$--3000\,GeV.
    (right)
    The probability that the energy loss of a muon in the calorimeter exceeds certain thresholds as a function of the muon energy.}
  \label{fig:MuonEnergyLoss}
\end{figure}

The resolution of the LLCP mass measurement is also discussed 
in~\cite{Feng:2015wqa}.  Good momentum resolution is essential to measure the mass of LLCPs.
We parameterize the momentum resolution of high-$\pt$ tracks by:
\begin{equation}
  \Delta \pt = A\oplus B\cdot \pt \oplus C \cdot p_{T}^2 \approx C\cdot p_{\rm T}^2.
\end{equation}
The relevant parameter $C$ in the ATLAS experiment was measured to be $C=0.168(16)/\textrm{TeV}$
for the barrel region of the muon spectrometer in early 7 TeV data~\cite{Salvucci:2012np}.
Stronger magnetic fields in the tracker, as well as larger detector 
dimensions, will improve the momentum resolution,
so we assume a value of $C=0.1/\textrm{TeV}$ for the 100\,TeV analysis, 
which gives the result in Fig.~\ref{fig:mass100}.

\begin{figure}[tbp]\centering
  \includegraphics[width=0.48\textwidth]{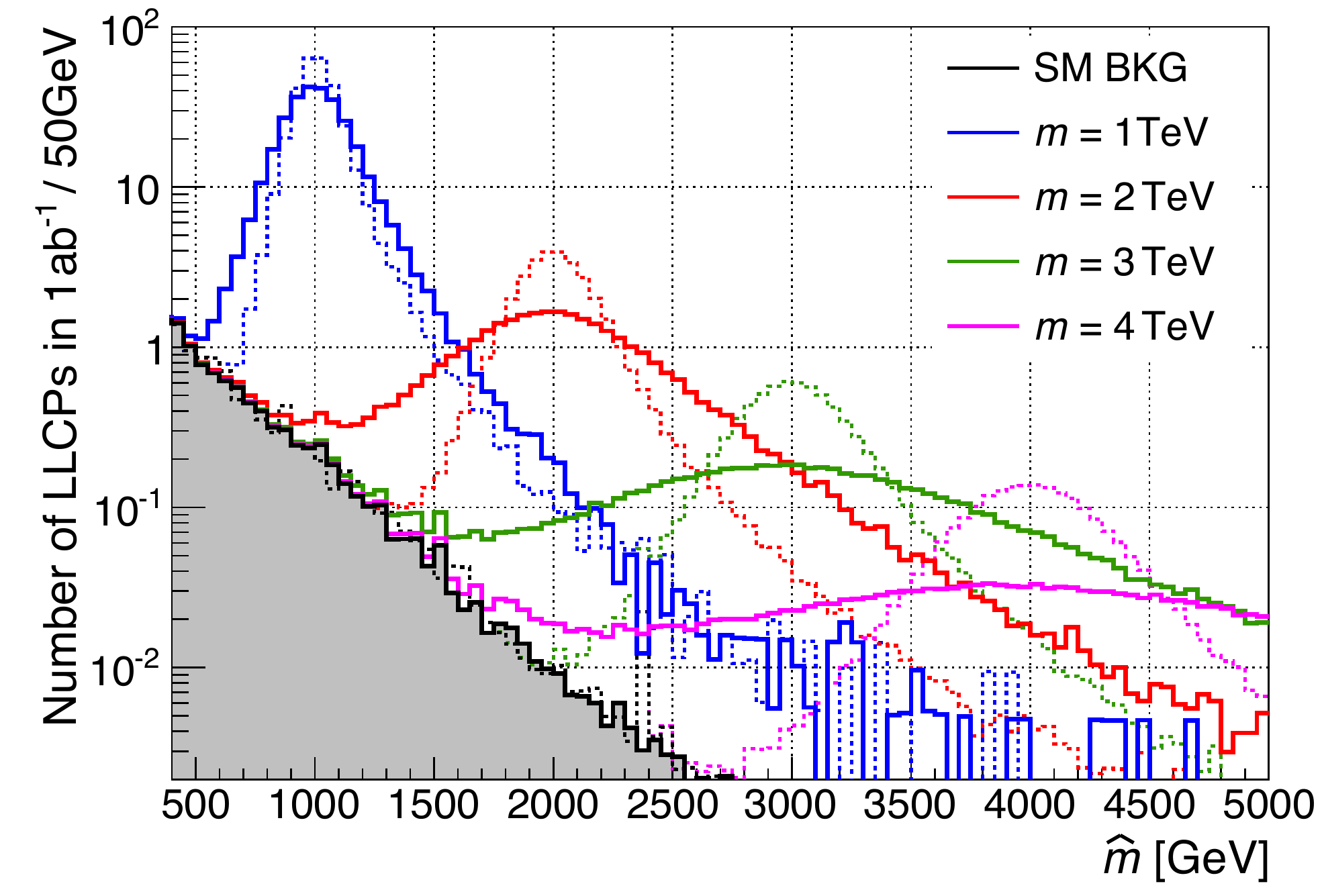}
  \caption{Distribution of the reconstructed LLCP mass in the 100\,TeV analysis.
    Solid lines are with $C=0.1/\textrm{TeV}$, while dotted lines with $C=0$ is drawn as a reference.}
  \label{fig:mass100}
\end{figure}

In summary, even for the most pessimistic scenario in which only the channel 
$pp\to(\gamma,Z)\to\slepton_{\rm R}\slepton_{\rm R}^*$ is available,
a 100\,TeV $pp$ collider with $\int\mathcal L=3\,{\rm ab}^{-1}$ has the 
capability to discover or exclude  LLCP sleptons with $m\lesssim 3\,\textrm{TeV}$.

\subsection{Indirect Probes}
\label{sec:SUSY_indirect}
At the 100 TeV collider it would also be possible to search for the indirect effects of supersymmetry on SM processes.  Two important factors for indirect probes of new physics are precision and energy.  Precision could be delivered at the 100 TeV collider due to the large number of SM events possible thanks to the large integrated luminosity.  For many processes this would render systematic uncertainties as the dominant limitation in indirect tests of new physics.  The high energy that can be achieved at the 100 TeV collider would also enhance the indirect constraints on new physics.  Indirect constraints on a multitude of new physics processes are possible, however we will focus here on a particular example in supersymmetry.

While light stop squarks would likely be directly observed at a 100 TeV collider, it may also be useful to search for their indirect effects.  In particular precision Higgs coupling constraints, particularly on the Higgs-glue-glue coupling, would provide a powerful probe of light stop scenarios.  In addition, as a 100 TeV collider would also be able to observe Higgs boson pair production, it is interesting to consider searching for the indirect effects of light stop squarks on Higgs pair production, as will be considered in this section.

As with single Higgs production, the dominant production mode for Higgs pairs is gluon fusion. The stop loop contributes to deviations from the SM di-Higgs rate that can be a powerful indirect signal of SUSY. The much larger di-Higgs cross section at a 100 TeV collider improves the sensitivity of such searches compared to the LHC.

The di-Higgs rate also constrains the Higgs potential, so considerable effort has gone into projecting collider sensitivity. Current LHC searches have used several final states \cite{Aad:2014yja,Aad:2015xja,CMS-PAS-HIG-13-032,CMS-PAS-HIG-14-013,Aad:2015uka,CMS-PAS-HIG-13-025}  and high luminosity projections for several channels have also been completed \cite{ATL-PHYS-PUB-2014-019,CMS:2015nat}. Phenomenological studies considering a 100 TeV machine have been done for the $bb\gamma\gamma$ \cite{Barr:2014sga,Azatov:2015oxa}, $4W$ \cite{Li:2015yia}, and $bb\,+$ leptons \cite{Papaefstathiou:2015iba} channels.

Of course, as detailed in Sec.~\ref{sec:SUSY_stop}, there also exist very powerful direct searches for superpartners, stops featuring prominently. However, these searches depend on particular assumptions about $R$-parity \cite{Brust:2012uf,Evans:2012bf,Bai:2013xla} or where the stop lies in the particle spectrum. Work on these limits include the cases of stops nearly degenerate with the top \cite{Fan:2011yu,Fan:2012jf,Csaki:2012fh,Han:2012fw,Kilic:2012kw,Czakon:2014fka}, part of a compressed spectrum \cite{LeCompte:2011fh,LeCompte:2011cn,Dreiner:2012gx,Bhattacherjee:2012mz,Drees:2012dd,Belanger:2012mk,Alves:2012ft,Krizka:2012ah}, or which decay into light superpartners like staus \cite{Carena:2012gp,Carena:2013iba}. Clearly, the insensitivity of the di-Higgs rate to these types of assumptions make it a valuable complementary search.

This is not to say that the di-Higgs approach is unconstrained. A simple EFT analysis \cite{Batell:2015koa} demonstrates that constraints on single Higgs production limit the deviation due to new colored particles in the di-Higgs rate (and that the two rates are anticorrelated). However, this analysis is inapplicable when one or both of the stops are light.  Even within the constraints imposed by single Higgs production, the shape of the di-Higgs distribution can reveal new physics. The SM di-Higgs rate experiences a cancellation at threshold, so new colored particles, like stops, can have large effects there. This motivates considering the differential cross section (in invariant mass or $p_T$ for example) close to threshold to see the greatest deviations.

\begin{figure}[t]
\centering
\includegraphics[height=2.6in,width=0.49\textwidth]{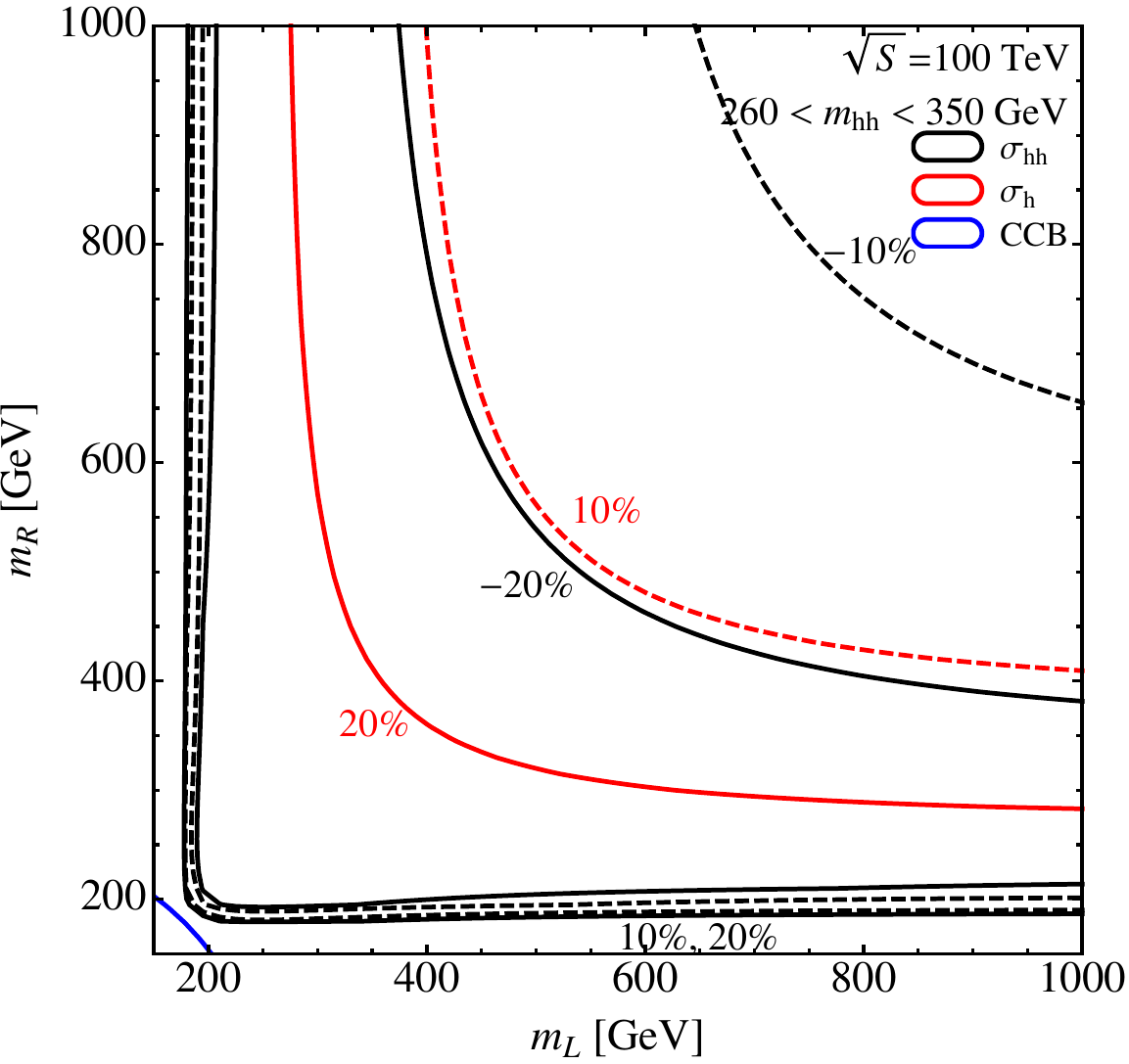} \includegraphics[height=2.6in,width=0.49\textwidth]{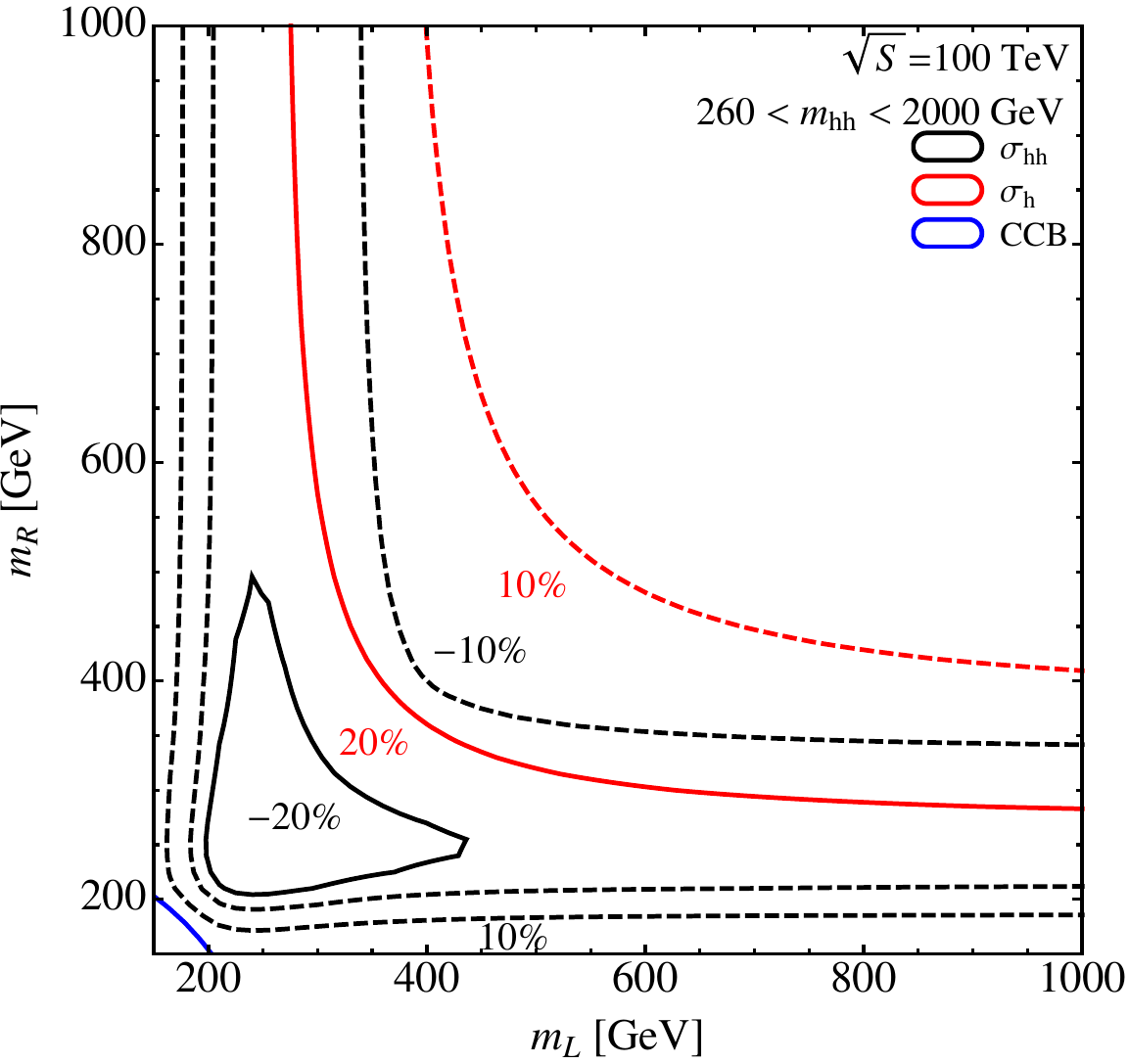}
\caption{Percentage corrections, relative to the SM, of the single Higgs (red) and di-Higgs (black) production cross sections at $\sqrt{s}=100$ TeV in a low energy bin with invariant masses $260<m_{hh}<350$ GeV (left) and $260<m_{hh}<2000$ GeV (right). Vanishing $A$-terms have been assumed and the physical left- and right-handed stop masses have been varied through the soft masses. The blue contour gives the approximate contour of color breaking vacuum constraint.}
\label{fig:m1m2100TeV}
\end{figure}

\begin{figure}[th!]
\centering
\includegraphics[height=2.6in,width=0.49\textwidth]{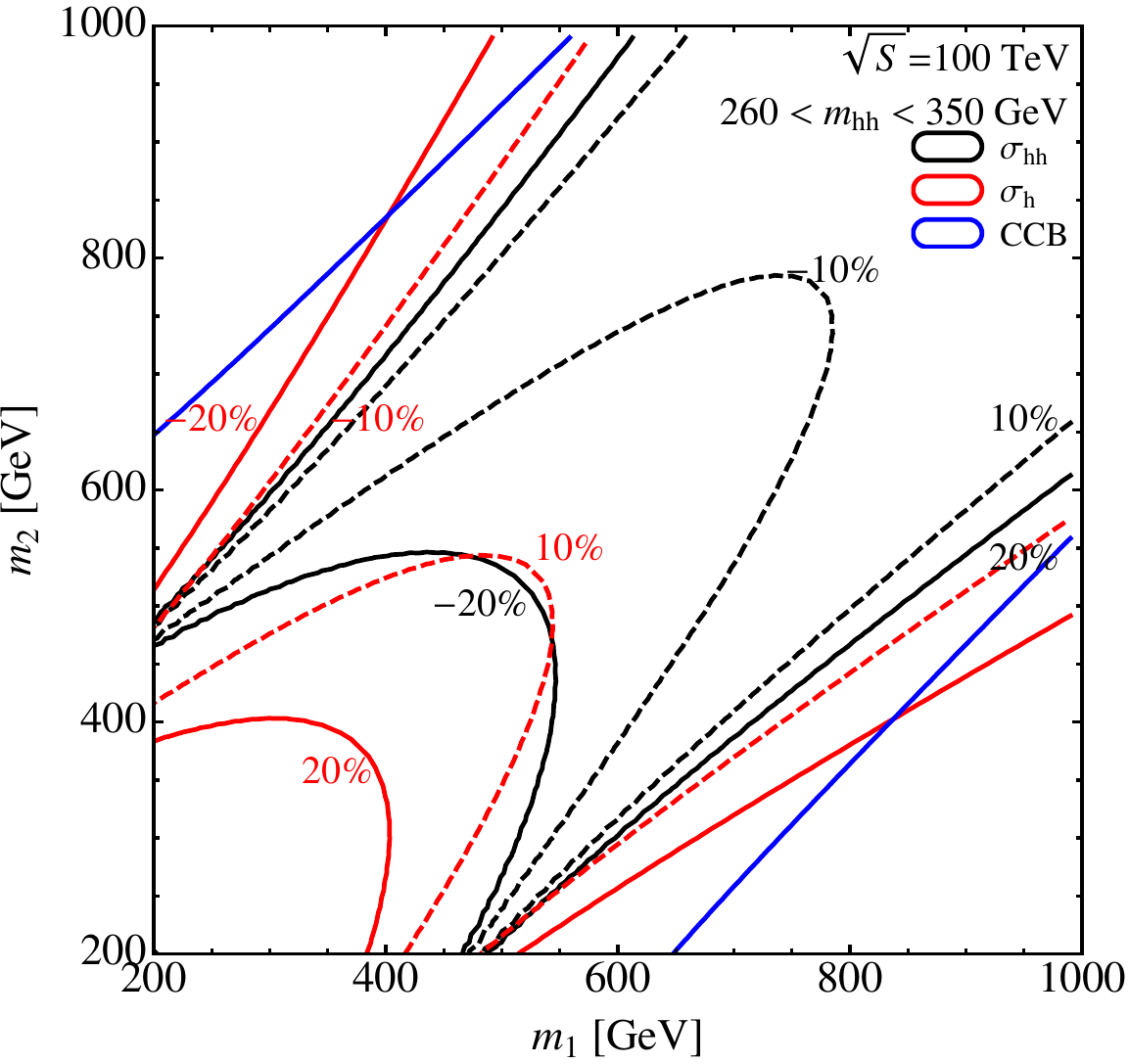} \includegraphics[height=2.6in,width=0.49\textwidth]{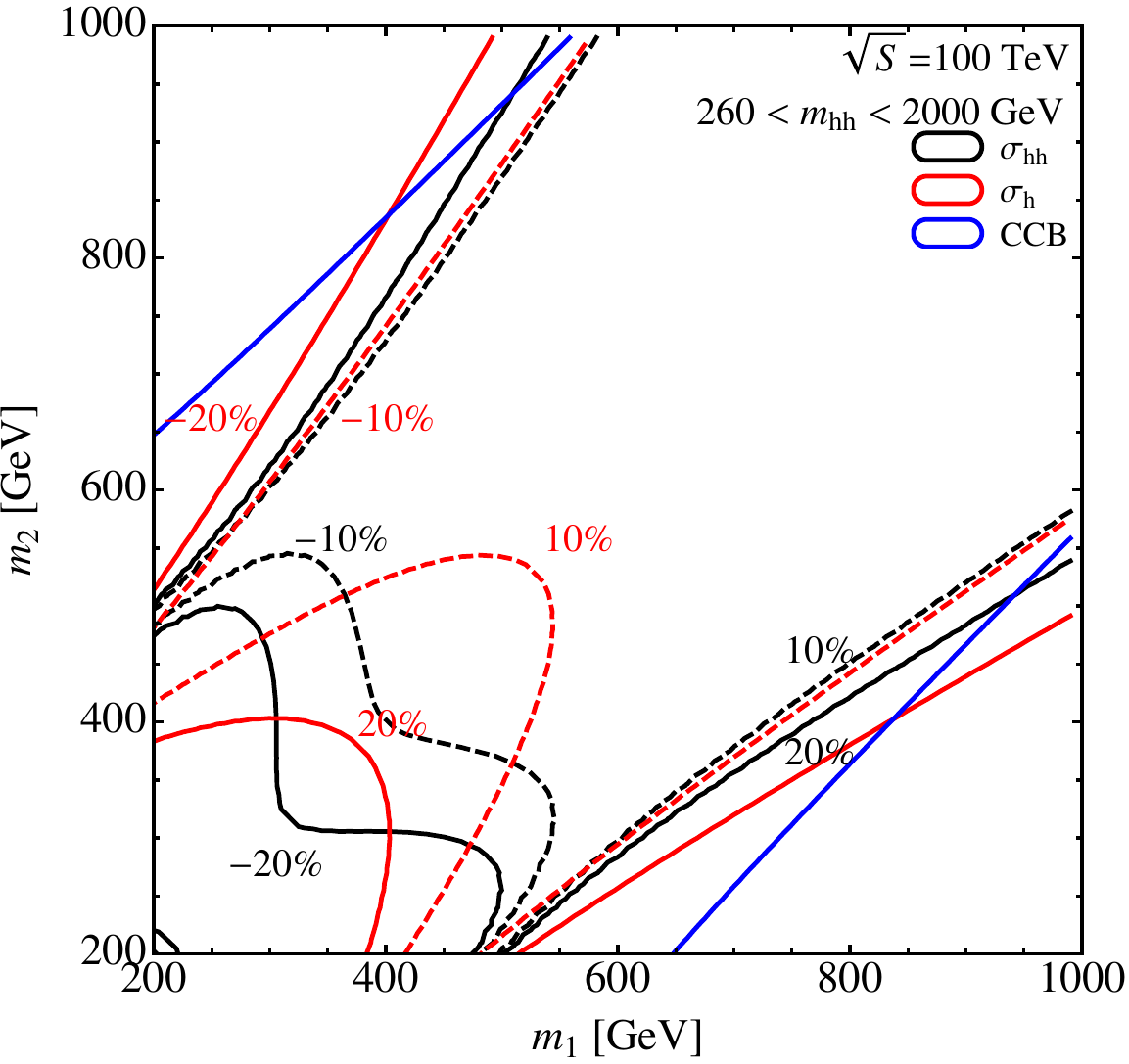}
\caption{Percentage corrections, relative to the SM, of the single Higgs (red) and di-Higgs (black) production cross sections at $\sqrt{s}=100$ TeV in a low energy bin with invariant masses $260<m_{hh}<350$ GeV (left) and $260<m_{hh}<2000$ GeV (right). Degenerate soft masses have been assumed, while the $A$-term is varied. The blue contour gives the approximate contour of color breaking vacuum constraint.}
\label{fig:mA100TeV}
\end{figure}

The calculated MSSM di-Higgs cross section \cite{Plehn:1996wb,Djouadi:1999rca,Belyaev:1999mx,BarrientosBendezu:2001di} has been used to explore this idea in \cite{Batell:2015koa}.  We consider contours of 10 and 20 percent deviation from the SM in the single Higgs and di-Higgs production at a 100 TeV collider. The deviations are considered in a small invariant mass bin close to threshold as well as the more usual large mass bin. The large bin analysis not only makes contact with previous studies, but highlights features in the spectrum at high invariant mass. These have greater effect on the total rate at a 100 TeV collider where the gluon luminosity at large masses are significantly increased.

To demonstrate the mass ranges in which observable effects in di-Higgs production could be observed at 100 TeV, we show contours of constant deviation in the pair production cross section at 100 TeV.  In Fig. \ref{fig:m1m2100TeV} the stop mixing is set to zero while the soft masses are varied. Fig. \ref{fig:mA100TeV} sets the left- and right-handed soft masses equal and then varied while the mixing parameter is also varied.  In regions where both stops are heavy and single Higgs production is SM-like the di-Higgs production is also SM-like. However, when the mass bin close to threshold is considered the fractional change in the di-Higgs rate increases dramatically, so that there are regions where the effects of the stops are seen in the di-Higgs rate close to threshold even while single Higgs deviations are small.

When one or both of the stops are light there are regions wherein the single Higgs production is SM-like, but the di-Higgs rate shows large deviations. This complementarity at low mass enables light stops to be detected even when direct searches fail and the single Higgs production appears SM-like do to a cancellation. The discriminating power of these studies depends on the high di-Higgs cross section at a 100 TeV collider. This can be easily seen by comparing these results to the 14 TeV results in \cite{Batell:2015koa}.

  \subsection{Model-Specific Interpretations}
  \label{sec:SUSY_model}
  Whereas the previous sections have focussed on a more model-independent approach, focussing on specific superpartners and in some cases simplified models, it is also worthwhile to consider 100 TeV measurements in the context of complete supersymmetric models.  The number of free parameters in the MSSM is too large to study in its entirety, thus we will consider two well-motivated scenarios that make specific predictions for the pattern of parameters, the `CMSSM' and `Mini-Split' supersymmetry.
  
  \subsubsection{Constrained Minimal Supersymmetric Standard Model}
The `CMSSM' \cite{Kane:1993td,Ellis:1996xu,Ellis:1997wva,Ellis:1998jk,Barger:1997kb,Ellis:2000we,Roszkowski:2001sb,Djouadi:2001yk,Chattopadhyay:2001va,Ellis:2002rp} assumes four input parameters, a universal scalar mass at the grand unification scale $m_0$, a universal gaugino mass $m_{1/2}$, a universal scalar trilinear soft term $A_0$, the usual Higgs mixing parameter $\tan \beta$, and also the sign of the $\mu$-term.  This framework is especially appealing from a phenomenological standpoint as it predicts the full set of soft parameters at the weak scale from these input parameters, after RG evolution, and this allows for the calculation and combination of diverse experimental constraints, from colliders to dark matter experiments.  The power of this approach is that successful phenomenological predictions in one area, for example dark matter, may be tested by constraining the predictions made from the same parameter set, in this case with colliders.  In particular, we will focus on testing regions of parameter space that predict the observed dark matter abundance by searching for the coloured particles whose masses are predicted once the parameters are set.

This material is based on \cite{Buchmueller:2015uqa}.  We analyze the
nature of the CMSSM parameter space for large values of $m_0$ and $m_{1/2}$,
considering the dark matter density prediction and the measurement
of the Higgs boson mass, which are the only constraints capable of imposing upper limits on $m_0$ and $m_{1/2}$.
Generally, bringing the relic dark matter density within the measured range
when the mass parameters are large requires some specific features in the
sparticle spectrum such as near-degeneracy between the LSP, the NLSP and perhaps
other sparticles, as this suppresses the relic dark matter density by introducing new
coannihilation channels during thermal freeze-out.  One such possibility is the narrow stop coannihilation
strip~\cite{Boehm:1999bj,Ellis:2001nx,Edsjo:2003us,DiazCruz:2007fc,Gogoladze:2011be,Ajaib:2011hs,eoz} where $\delta m = m_{\tilde t_1} - m_\chi$ is small.

\begin{figure}
\centerline{
\includegraphics[width=0.49\columnwidth]{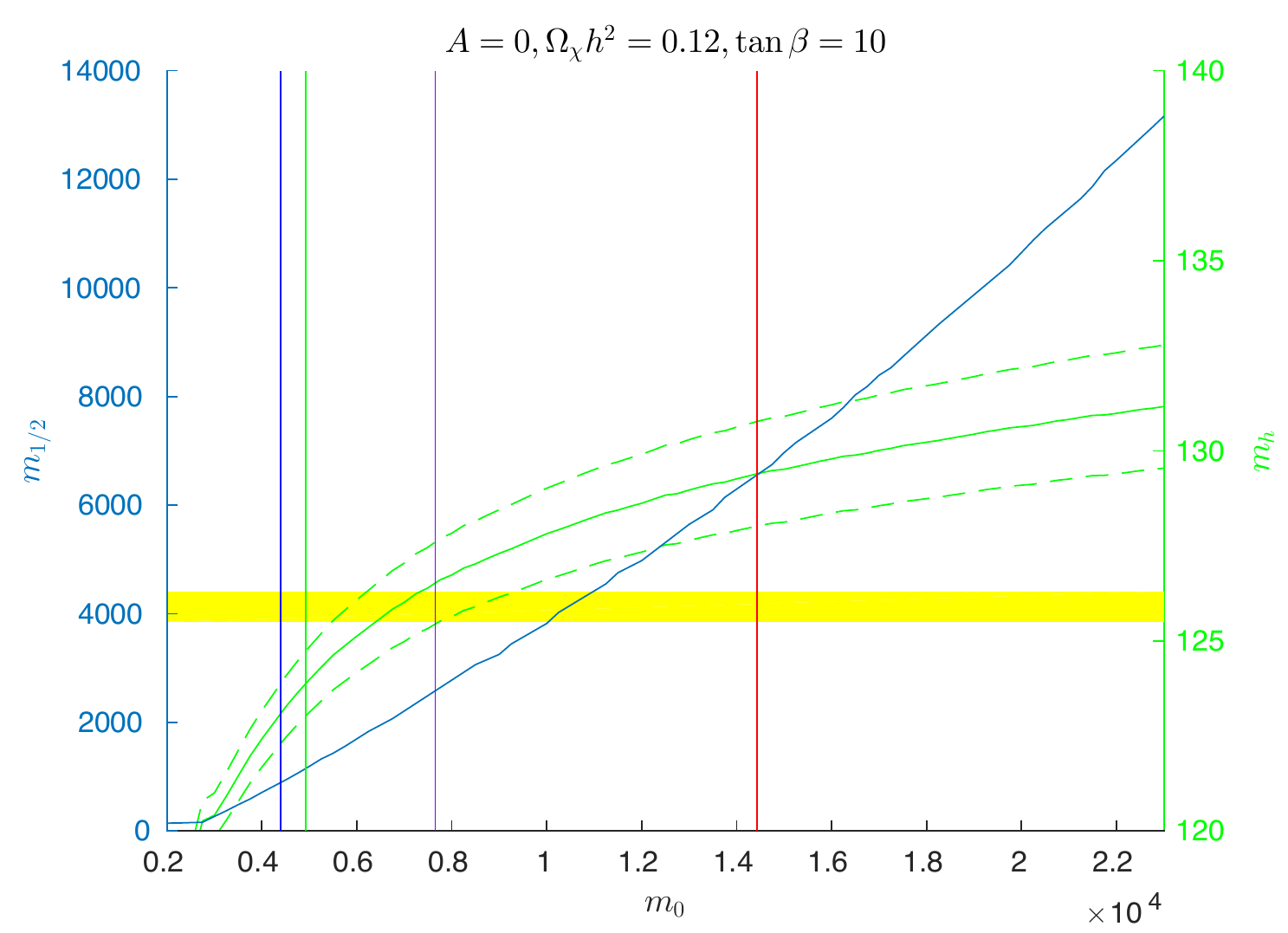} 
\includegraphics[width=0.49\columnwidth]{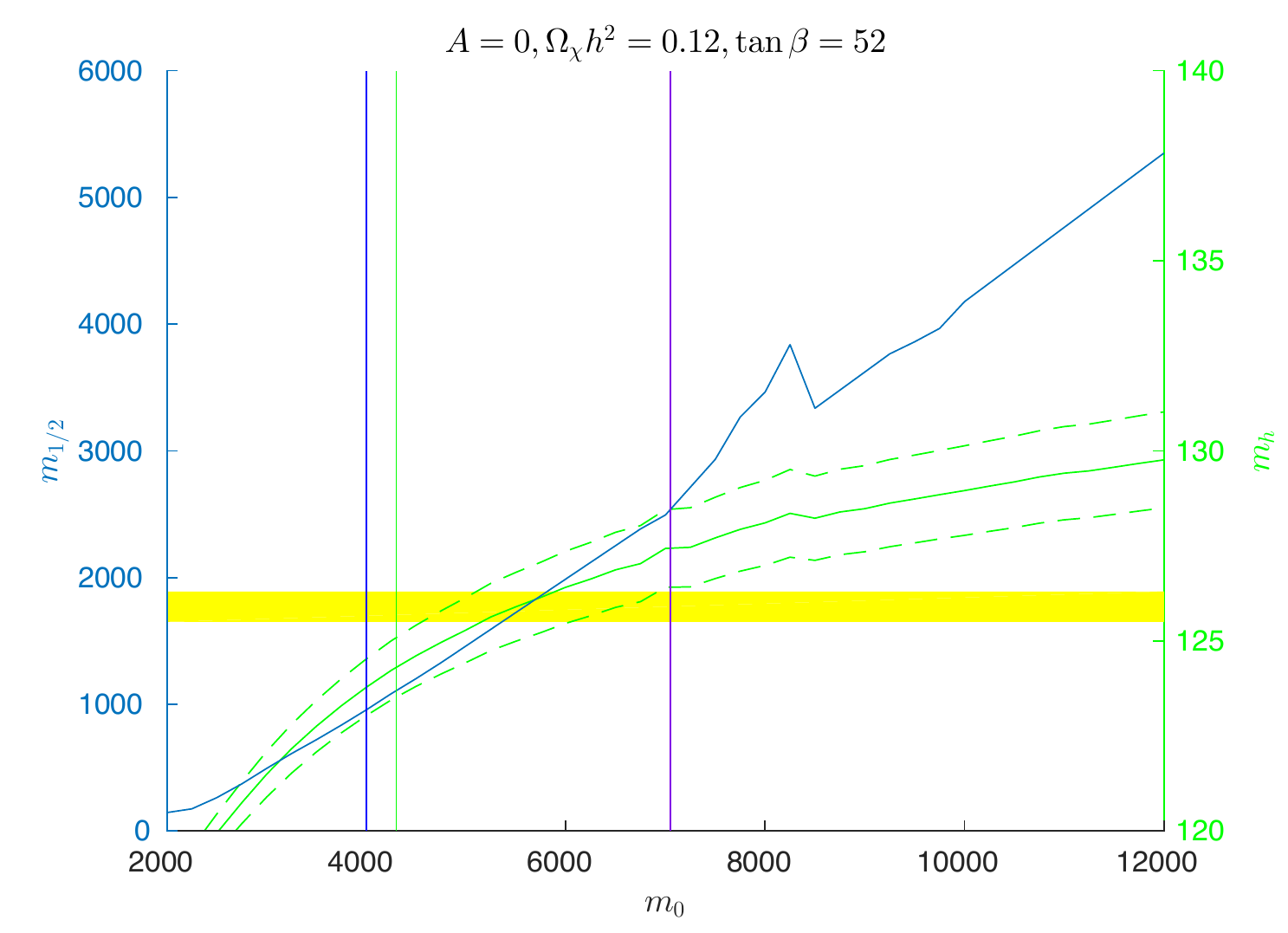}
}
\caption{The solid blue lines are the profiles in the $(m_0, m_{1/2})$ plane
of the focus-point strips for $A_0 = 0$ and $\tan \beta = 10$ (left panel),
and $A_0 = 0$ and $\tan \beta = 52$ (right panel).  The blue, green, purple and red lines are
particle exclusion reaches for particle searches at the LHC with
300 and 3000/fb at 14~TeV, 3000/fb with HE-LHC at 33~TeV
and 3000/fb at 100~TeV, respectively.
The solid lines are for generic $~/ \hspace{-.7em} E_T$ searches.  The solid (dashed) green lines are central values (probable ranges)
of $m_h$ calculated using {\tt FeynHiggs~2.10.0}, and the yellow band represents the experimental value of $m_h$.}
\label{fig:profiles}
\end{figure}
%
%%%%%%%%%%%%%% F I G U R E %%%%%%%%%%%%%%%%%%%%%%%%%%%%%%%%%%%%%%%%%%%%%%%%%%%

Another possibility is the focus-point strip of parameter space \cite{Feng:1999mn,Feng:1999zg,Feng:2000gh,Baer:2005ky,Feng:2011aa,Draper:2013cka}, appearing at larger values of
$m_0/m_{1/2}$, beside the boundary of the region where radiative electroweak symmetry breaking is consistent. Along the focus-point strip, the Higgsino component of
the neutralino LSP is enhanced, and its annihilations and coannihilations
with heavier neutralinos and charginos are enhanced. Various studies have
shown that the focus-point strip may extend to very large values of $m_0$
and $m_{1/2}$, with $m_0/m_{1/2} \sim 3$ and $A_0 \lsim m_0$.\footnote{However, it does not extend to arbitrarily large values of $m_0$ as the Higgs mass measurement constrains $m_0$.}

\begin{figure}
\centerline{
\includegraphics[width=0.49\columnwidth]{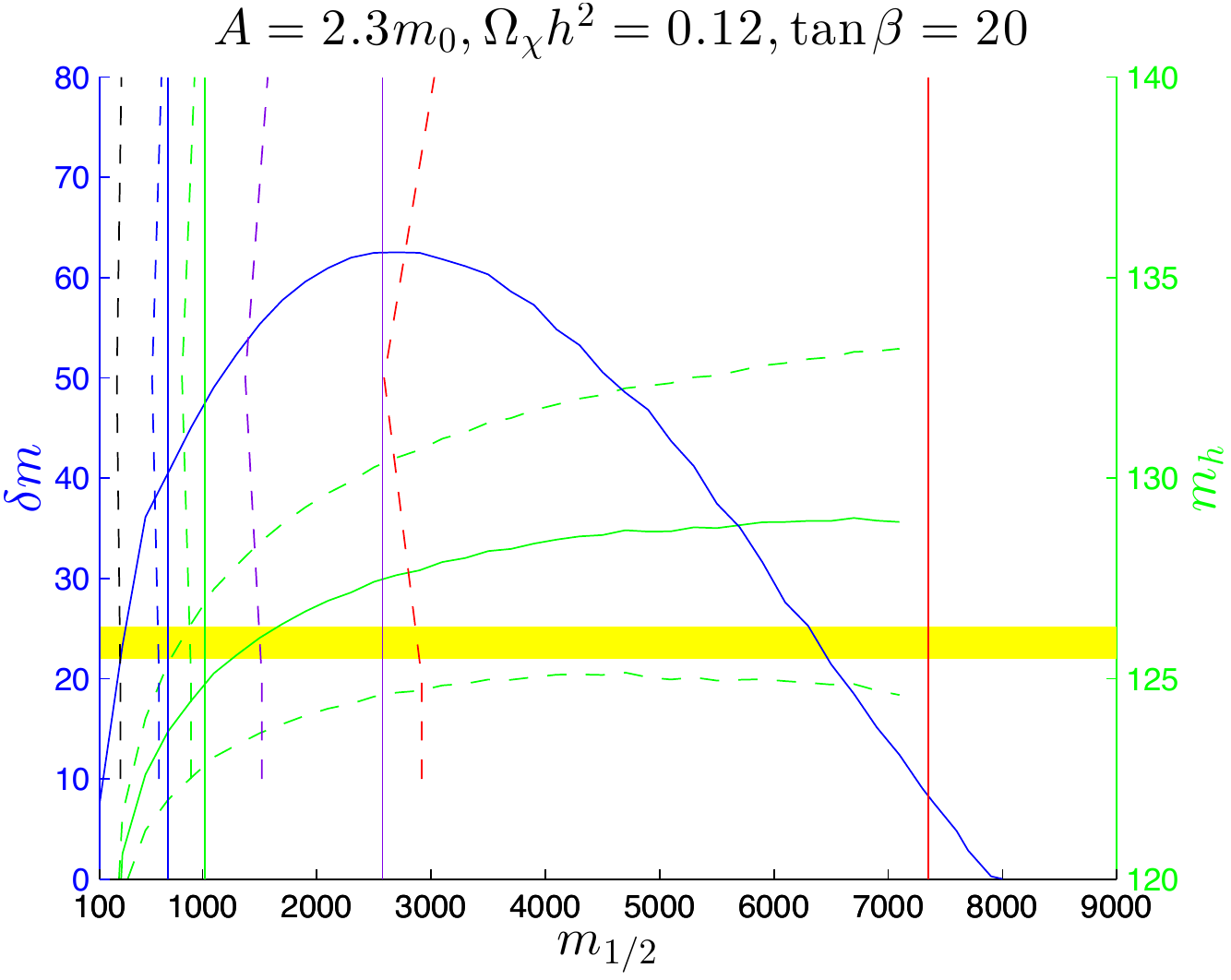}
\includegraphics[width=0.49\columnwidth]{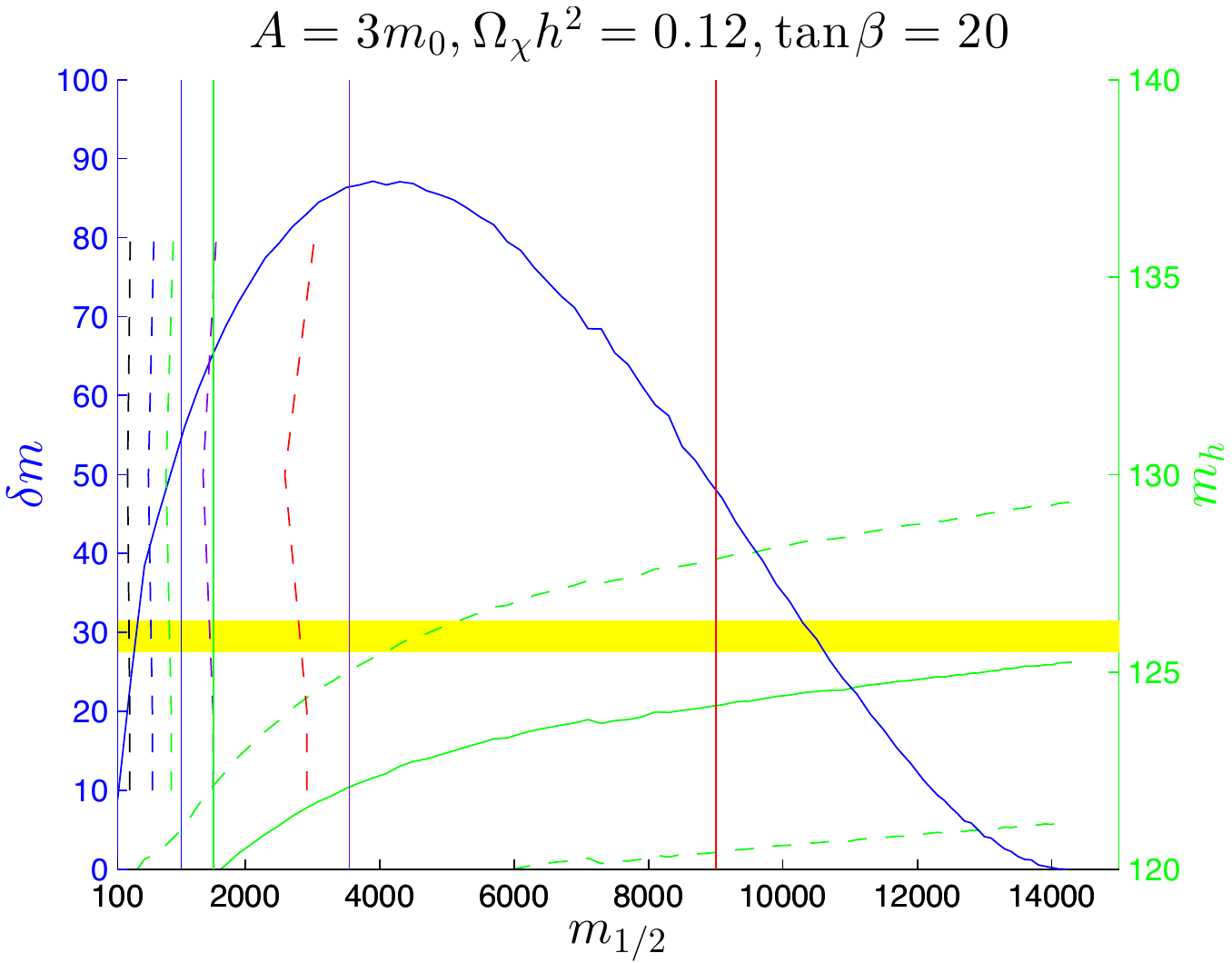}
}
\caption{The solid blue lines are the profiles in the $(m_{1/2}, \delta m \equiv m_{\tilde t_1} - m_\chi)$ plane
of the stop coannihilation strips for $A_0/m_0 = 2.3$ and $\tan \beta = 20$ (left panel),
and $A_0/m_0 = 3.0$ and $\tan \beta = 20$
(right panel).  The solid lines are for generic $~/ \hspace{-.7em} E_T$ searches and the dashed lines are for dedicated
stop searches, using the same colours as in Fig.~\ref{fig:profiles} (the dashed black line is for the LHC at 8~TeV).}
\label{fig:profiles2}
\end{figure}
%
%%%%%%%%%%%%%% F I G U R E %%%%%%%%%%%%%%%%%%%%%%%%%%%%%%%%%%%%%%%%%%%%%%%%%%%

Figs.~\ref{fig:profiles} and \ref{fig:profiles2} display the profiles of the focus-point strip and of the stop coannihilation strip, along their full lengths. 
Both pairs of plots show the Higgs mass values calculated using {\tt SSARD}\cite{Buchmueller:2015uqa} as inputs to {\tt FeynHiggs~2.10.0} (solid green lines).  Uncertainty estimates of $\pm 3$~GeV are also shown. Only portions of the
focus-point strips are compatible with the measured Higgs mass (yellow bands)
within these uncertainties, whereas for the stop
coannihilation strips there are significant additional uncertainties from
RGE running, and all of the strips are compatible with the measured Higgs mass.
In the cases of the stop coannihilation strips in the lower panels of Fig.~\ref{fig:profiles}, we also display as blue lines the
mass difference $\delta m \equiv m_{\tilde t_1} - m_\chi$ along the strips. In the examples shown,
this mass difference is generally $< m_W + m_b$, so that the branching ratio for two-body
${\tilde t_1} \to \chi + c$ decay usually dominates over that for four-body ${\tilde t_1} \to \chi + W + b + \nu$
decay. However, this is not always the case, as illustrated by examples in~\cite{eoz}.
The branching ratio for ${\tilde t_1} \to \chi + W + b + \nu$ decay may dominate when
$m_{\tilde t_1} - m_\chi > m_W + m_b$, as seen in the right panel of
Fig.~\ref{fig:profiles2}. Thus, a complete search for supersymmetry
at 100 TeV should include searches for both the ${\tilde t_1} \to \chi + c$ and 
${\tilde t_1} \to \chi + W + b + \nu$ decay signatures.

The (near-)vertical lines in Figs.~\ref{fig:profiles} and \ref{fig:profiles2} 
mark estimates from Ref.~\cite{Buchmueller:2015uqa} of the sensitivities of the LHC (black - 8~TeV, blue - 300/fb at 14~TeV,
green - 3000/fb at 14~TeV), 3000/fb at HE-LHC (purple) and 3000/fb at
100 TeV (red) along the stop coannihilation strips. The solid lines represent
the extrapolated reaches of the generic jets + $/ \hspace{-.7em} E_T$ searches, and the dashed lines in the lower panels represent
the extrapolated reaches of dedicated searches for ${\tilde t_1} \to c + \chi$ decays, which
lose some sensitivity as $\delta m$ increases because of the increase in the ${\tilde t_1} \to \chi + W + b + \nu$ decay
branching ratio. We see that the 100 TeV collider would be sensitive to the
full extents of the focus-point strips and of the stop coannihilation strip for $A_0 = 2.3 \, m_0$,
but not all the stop coannihilation strip for $A_0 = 3.0 \, m_0$: this is true in general for
$A_0/m_0 \gsim 2.5$. We note also that, as discussed in\cite{Buchmueller:2015uqa}, 
high-precision measurements of electroweak and Higgs observables could constrain the location 
along the dark matter strip, providing a potential consistency test of the supersymmetric model.
  
\subsubsection{Mini-Split Supersymmetry}
The spectrum of split supersymmetry (SUSY) contains lighter gauginos (bino, winos, gluinos) and heavier scalars (sfermions and Higgs bosons)~\cite{Wells:2003tf,ArkaniHamed:2004fb,Giudice:2004tc,Wells:2004di}. Among the gauginos, gluino production can be a useful way to search for split SUSY  at hadron colliders. Pure wino production is smaller by electroweak couplings. Pure bino production is very small since it has no direct couplings to gauge bosons. Unless gluinos are much heavier than other gauginos, gluino production can be the dominant production mode of split SUSY particles. In this section, we study gluino search propects at a 100 TeV $pp$ collider in split SUSY models~\cite{Jung:2013zya,Cohen:2013xda}.

Gluinos are pair produced at $pp$ colliders. Once produced, each gluino subsequently decays to the lightest gauginos (LSP) -- winos or binos -- via off-shell squarks as
\begin{equation}
\widetilde{g} \to \chi_1^0 j j.
\label{eq:decmode} \end{equation}
Pair production of gluinos then yields $\widetilde{g} \widetilde{g} \to \chi_1^0 \chi_1^0 jjjj$. This channel can be searched using an effective mass variable, $M_{\rm eff}$. The effective mass is defined as a scalar sum
\begin{equation}
M_{\rm eff} =  \sum_i p_T(i) + E_T^{\rm miss},
\end{equation}
where the sum runs over all jets with $p_T>50$ GeV and $\eta <5.0$. Cuts on the data that aid selecting signal over background include (see also~\cite{Hinchliffe:2001bz}):
\begin{itemize}
\item At least two jets with $p_T > 0.1\, M_{\rm eff}$.
\item Lepton veto.
\item $E_T^{\textrm{miss}} > 0.2 M_{\rm eff}$ and $p_T(j_1) < 0.35\, M_{\rm eff}$.
\item $\Delta \phi(j_1,E_T^{\textrm{miss}}) < \pi -0.2$ and $\Delta \phi(j_1,j_2)<2\pi/3$
\item $M_{\rm eff}> 1.5 M_{\widetilde{g}}$.
\end{itemize}
The $M_{\rm eff}$ spectrum depends only on the gluino mass and not on the LSP mass, as long as the gluino is more than 3 times heavier than the LSP~\cite{Jung:2013zya}. The discovery prospect in terms of the gluino mass is shown in Fig.~\ref{fig:reach-glupair}; the integrated luminosity needed for $5\sigma$ statistical significance is shown. 

\begin{figure}
\centerline{
\includegraphics[width=0.49\columnwidth]{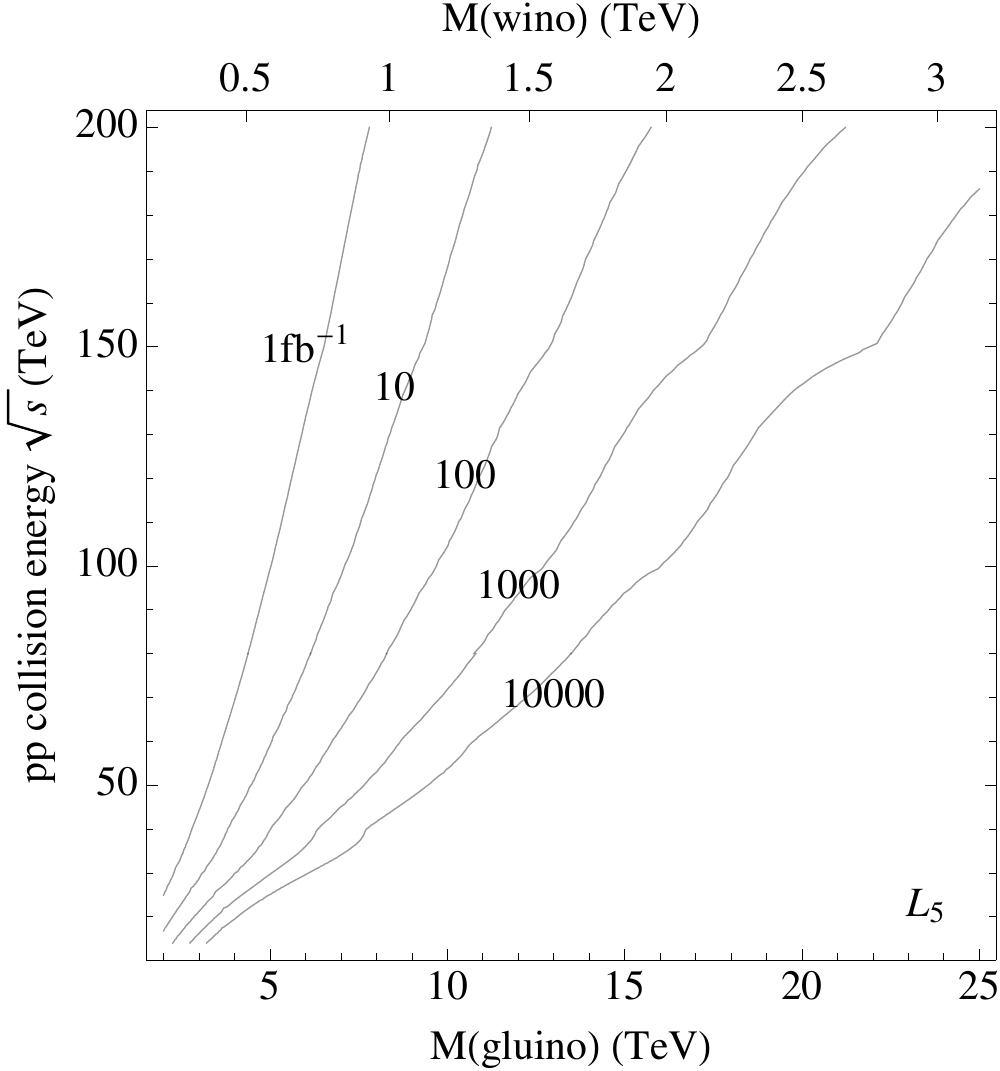} 
\includegraphics[width=0.49\columnwidth]{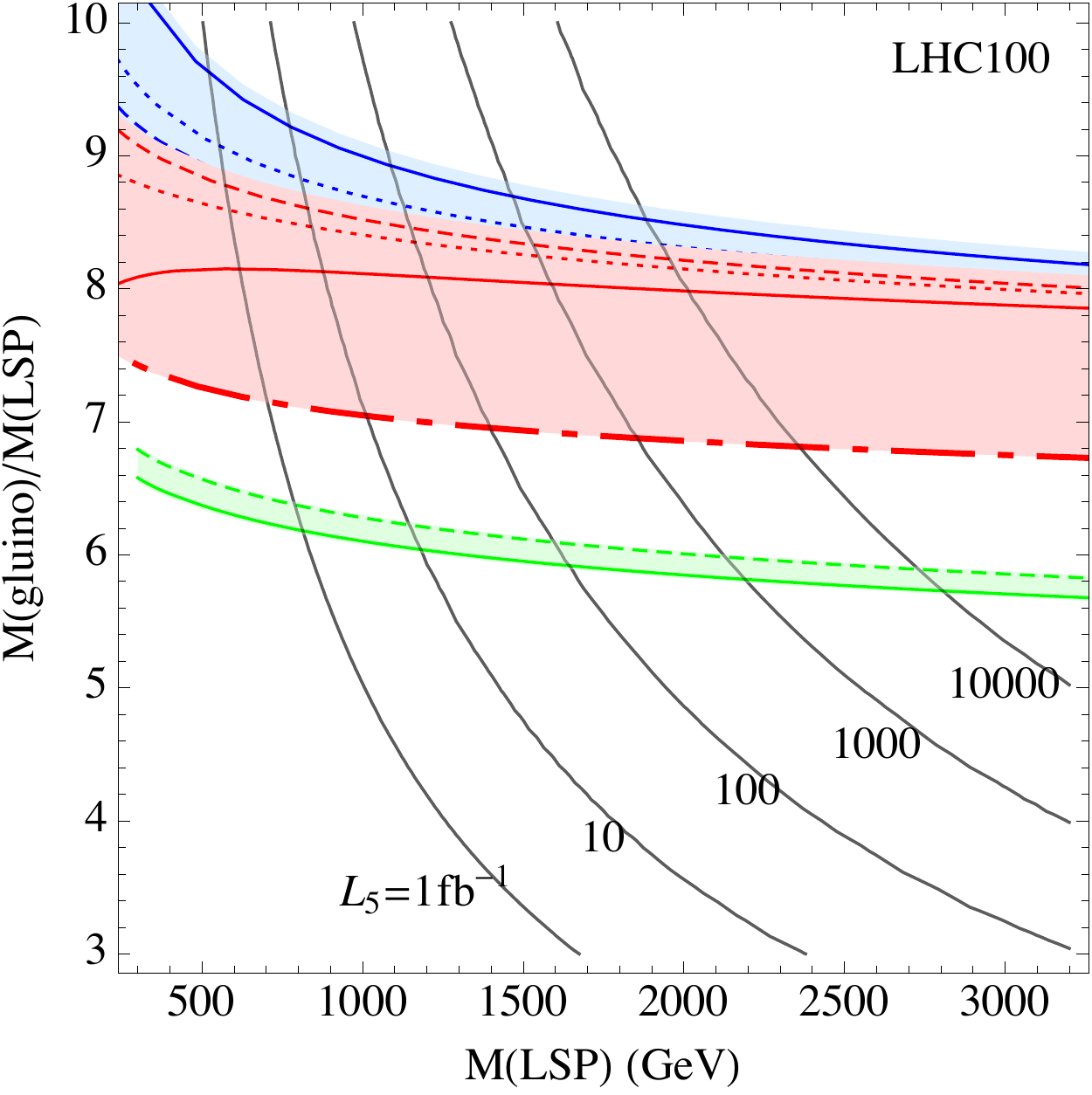} 
}
\caption{Left panel: The integrated luminosity (fb$^{-1}$) needed to discover gluino pairs in split SUSY models with $5\sigma$ statistical significance is contour plotted~\cite{Jung:2013zya}. It is assumed that the gluino is more than 3 times heavier than the LSP so that the $M_{\rm eff}$ analysis is valid. In the upper horizontal axis, the wino mass with the minimal AMSB relation is also shown. Right panel: The result at a 100 TeV $pp$ collider is shown in terms of the gluino-to-LSP mass ratio in split SUSY~\cite{Jung:2013zya}. Blue and red regions are predictions of AMSB models with squark masses and $\tan \beta$ varied. See text and Ref.~\cite{Jung:2013zya} for more details.}
\label{fig:reach-glupair}
\end{figure}

The results can be interpreted in three ways. First, at a 100 TeV $pp$ collider, up to about 7 (13) TeV gluinos can be discovered with 10 (1000) fb$^{-1}$ of data. This result applies to any split SUSY models as long as the gluino is 3 times heavier than the LSP and the decay mode in eq.(\ref{eq:decmode}) is dominant; for a smaller mass difference between the gluino and the LSP, the $M_{\rm eff}$ becomes a less useful observable for discovery and other strategies may be needed.

Second, if the minimal anomaly-mediated SUSY breaking (AMSB) model~\cite{Randall:1998uk,Giudice:1998xp} is considered as a particular example of split SUSY models, the reach on the gluino mass can be interpreted simultaneously as a reach on the mass of the wino dark matter candidate. In the minimal AMSB model, the wino is the LSP; squarks are one-loop factor heavier than the gluino and out of reach for the collider. The gluino-to-wino mass ratio, which is key to this interpretation of an indirect wino bound, is almost fixed, but varies from 8 to 9.5 for the wino mass between 3 TeV and 200 GeV, respectively~\cite{Jung:2013zya}. The variation is almost entirely due to the running of gauge couplings, and the Higgsino contribution to the quantum correction can be ignored. The wino mass corresponding to the gluino mass in this model is also shown in the upper horizontal axis in left panel of Fig.~\ref{fig:reach-glupair}. A 100 TeV $pp$ collider can probe wino mass up to 900 GeV (1.4 TeV)  with 10 (1000) fb$^{-1}$ of integrated luminosity. It is known that the $\sim 3$ TeV pure wino can be a thermal dark matter contributing to full dark matter density~\cite{Hisano:2006nn}, but astrophysical constraints may rule out this possibility~\cite{Cohen:2013ama,Fan:2013faa}. In any event, this very large wino mass is too heavy to be probed at a 100 TeV collider, yet a smaller wino mass, perhaps as a non-thermal dark matter source, can be indirectly probed to mass scales above a TeV as seen by this analysis.

Finally, the result can be interpreted more generally in terms of the reach on the gluino-to-LSP mass ratio. This is done in the right panel of Fig.~\ref{fig:reach-glupair}. The aforementioned minimal AMSB relation of the gluino and wino masses can be modified in general split SUSY models. The variations of sfermion masses between $1 \leq m_{\widetilde{f}} / m_{\widetilde{g}} \leq 4\pi \alpha_S$ and $3 \leq \tan \beta \leq 50$ with $|\mu|=4$ TeV lead to the blue and red regions; heavy squarks and Higgsinos modify the gluino mass and the wino mass at one-loop order, respectively. The dependence on the renormalization scale is much smaller. The region of $M_{\rm LSP}$ and $M_{\tilde g}/M_{\rm LSP}$ probed is under the integrated lumonsity contours in the figure. For more details of each curve, we refer to Ref.~\cite{Jung:2013zya}. This interpretation is most useful when a certain split SUSY model predicts a certain gaugino mass ratio.

\ifthenelse{\boolean{includetikzforreece}}{
  \subsection{Supersymmetry Post-Discovery at 100 TeV}
\label{sec:SUSY_post}
\def\Smiley{\tikz[baseline=-0.75ex,black]{
    \draw circle (2mm);
    \node[fill,circle,inner sep=0.5pt] (left eye) at (135:0.8mm) {};
    \node[fill,circle,inner sep=0.5pt] (right eye) at (45:0.8mm) {};
    \draw (-155:1.25mm) arc (-130:-50:1.8mm);
    }
}

\def\Sadey{\tikz[baseline=-0.75ex,black]{
    \draw circle (2mm);
    \node[fill,circle,inner sep=0.5pt] (left eye) at (135:0.8mm) {};
    \node[fill,circle,inner sep=0.5pt] (right eye) at (45:0.8mm) {};
    \draw (-155:1.25mm) arc (130:50:1.8mm);
    }
}
After any discovery of new particles at the LHC or at 100 TeV, we will want to undertake a detailed study of their properties. As one example, suppose that a gluino is discovered with a mass at the TeV scale, which decays through cascades to lighter electroweakinos. A 100 TeV collider would be an effective gluino factory: 3 ab$^{-1}$ of data would lead to $2 \times 10^7$ gluino pair production events if $m_{\tilde g} = 2$ TeV and $10^5$ events if $m_{\tilde g} = 5$ TeV \cite{Borschensky:2014cia}. Hence, a gluino discovery would be followed up by an extensive program of measuring gluino branching ratios and couplings. In this section, we will outline one particular aspect of such studies: a test of the MSSM prediction of the Higgs mass in split SUSY scenarios where the scalars are significantly heavier than the gluino.

%%%%%%%%%%%%%%%%%
\begin{figure}[tbp]\begin{center}
\includegraphics[width=0.45\textwidth]{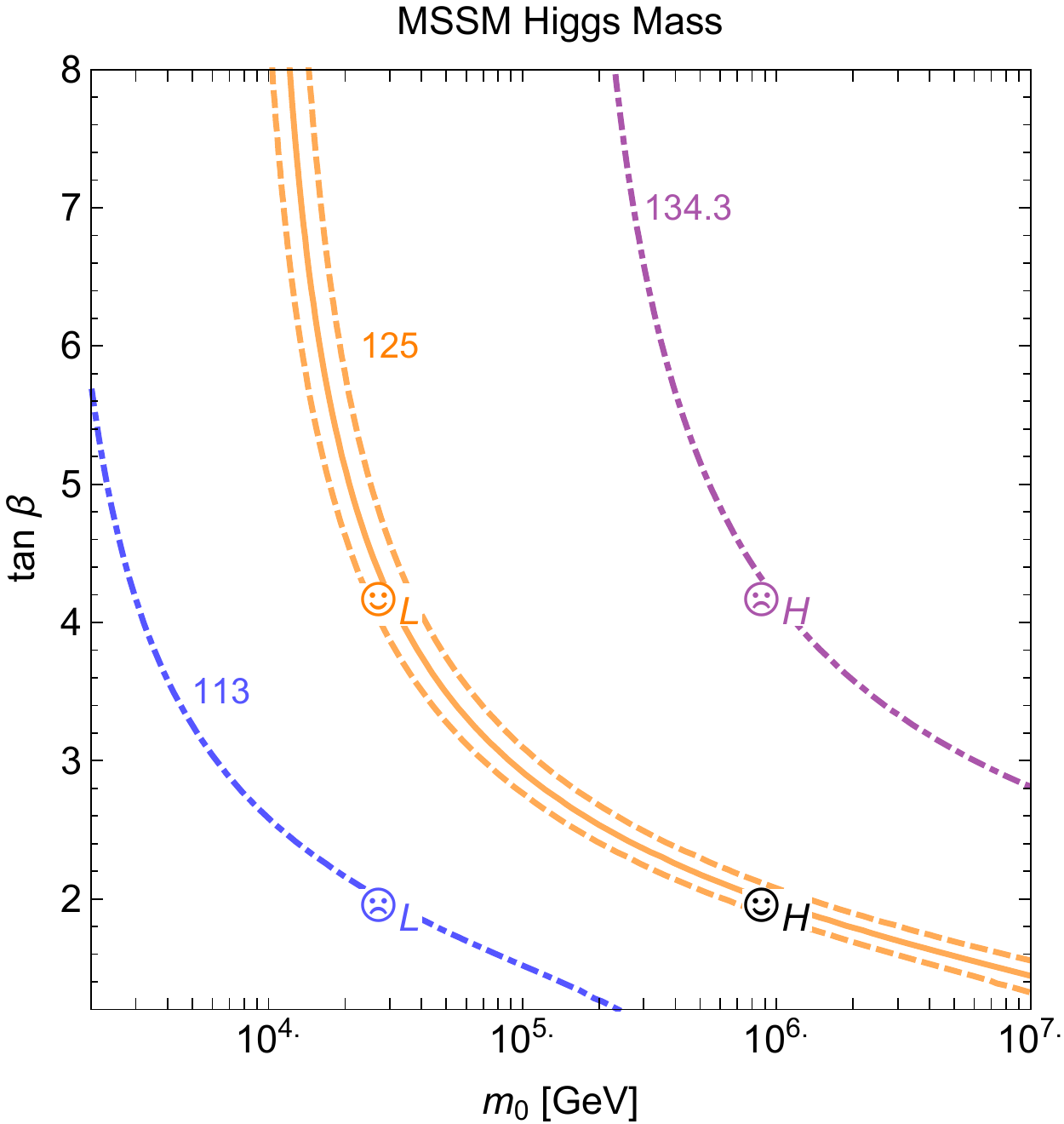}
\end{center}
\caption{The MSSM Higgs boson mass as computed with SusyHD
  \cite{Vega:2015fna}, taking $A_t = 0$ and fermion masses $m_{1/2} =
  1$ TeV. (Dependence on the details of fermion masses is mild.) The
  orange solid curve and the dashed orange curves show where the
  central value and the $\pm 1\sigma$ variations around it give a 125
  GeV Higgs mass. The dot-dashed purple and blue lines show where the
  Higgs mass is 134.3 and 113 GeV, respectively. We have marked four
  points, two giving the correct Higgs mass (smiling faces) and two
  giving an incorrect one (frowning faces), for a closer study.}
\label{fig:higgsmass}
\end{figure}%
%%%%%%%%%%%%%%%%%

The MSSM at tree level predicts that the Higgs mass is less than the
$Z$ mass, but loop effects can lift it
\cite{Barbieri:1990ja,Haber:1990aw,Casas:1994us,Carena:1995bx}. This
has led to extensive effort toward high-precision theoretical
calculations of the Higgs mass in the MSSM, recently reviewed in
\cite{Draper:2016pys}. The result primarily depends on the stop mass
matrix and $\tan \beta$. For relatively low stop masses, large values
of $\tan \beta$ and a sizable mixing parameter $A_t$ can achieve a 125
GeV Higgs mass, and the stops could be directly probed at 100
TeV. Here we will focus on the case of heavy scalars with small mixing
($A_t \approx 0$), which achieve a 125 GeV Higgs mass at relatively
low values of $\tan \beta$. Taking all scalars to have mass $m_0$ and
all gauginos to have mass 1 TeV, contours of Higgs boson masses in the
$(m_0, \tan \beta)$ plane are shown in Figure~\ref{fig:higgsmass}. The
orange curves show the region consistent with the measured Higgs
mass. We have singled out two points on this curve, one with 30 TeV
scalars and $\tan \beta \approx 4$ marked $\Smiley_L$ (the $L$ for
``low mass'' scalars, though they are still quite heavy!) and one with
1000 TeV scalars and $\tan \beta \approx 2$ marked $\Smiley_H$ (the $H$ for ``high mass''), for special attention. For comparison, we have also selected two other points at the vertices of a rectangle, $\Sadey_L$ and $\Sadey_H$, which predict a Higgs mass differing from the true value by about 10 GeV.

A future study should thoroughly explore the whole parameter space, but for now we highlight the low scalar mass at 30 TeV and high scalar mass at 1000 TeV as benchmarks, for both collider physics and model-building reasons. From the collider physics viewpoint, scalars that are modestly heavier than the gluino can be searched for directly in ${\tilde q}{\tilde g}$ associated production events \cite{Fan:2011jc}. Preliminary estimates for 100 TeV suggest that direct searches for this signal will probe first-generation squarks up to about 30 TeV \cite{Ellis:2015xba}. On the other hand, scalars that are {\em much} heavier than the gluino would imply measurably long gluino lifetimes \cite{ArkaniHamed:2004fb,Arvanitaki:2012ps,ArkaniHamed:2012gw}. For a 2 TeV gluino, the threshold at which lifetimes are measurable is roughly $m_0 \approx 1000$ TeV, corresponding to a 100 micron lifetime. Improved detector technology might push to lower lifetimes, but the lifetime goes as the fourth power of the scalar mass, so dramatic improvements in scalar mass reach are unlikely. Thus, the region of scalar masses $30~{\rm TeV} \lsim m_0 \lsim 1000~{\rm TeV}$ must be probed in a different way. The gluino branching ratio to gluon plus higgsino has been discussed as a key probe in this region \cite{Sato:2012xf} due to its logarithmic sensitivity to scalar masses \cite{Toharia:2005gm,Gambino:2005eh}. The parameter space and the possible probes are summarized in Figure~\ref{fig:gluinoscalarplane}.

%%%%%%%%%%%%%%%%%
\begin{figure}[tbp]\begin{center}
\includegraphics[width=0.65\textwidth]{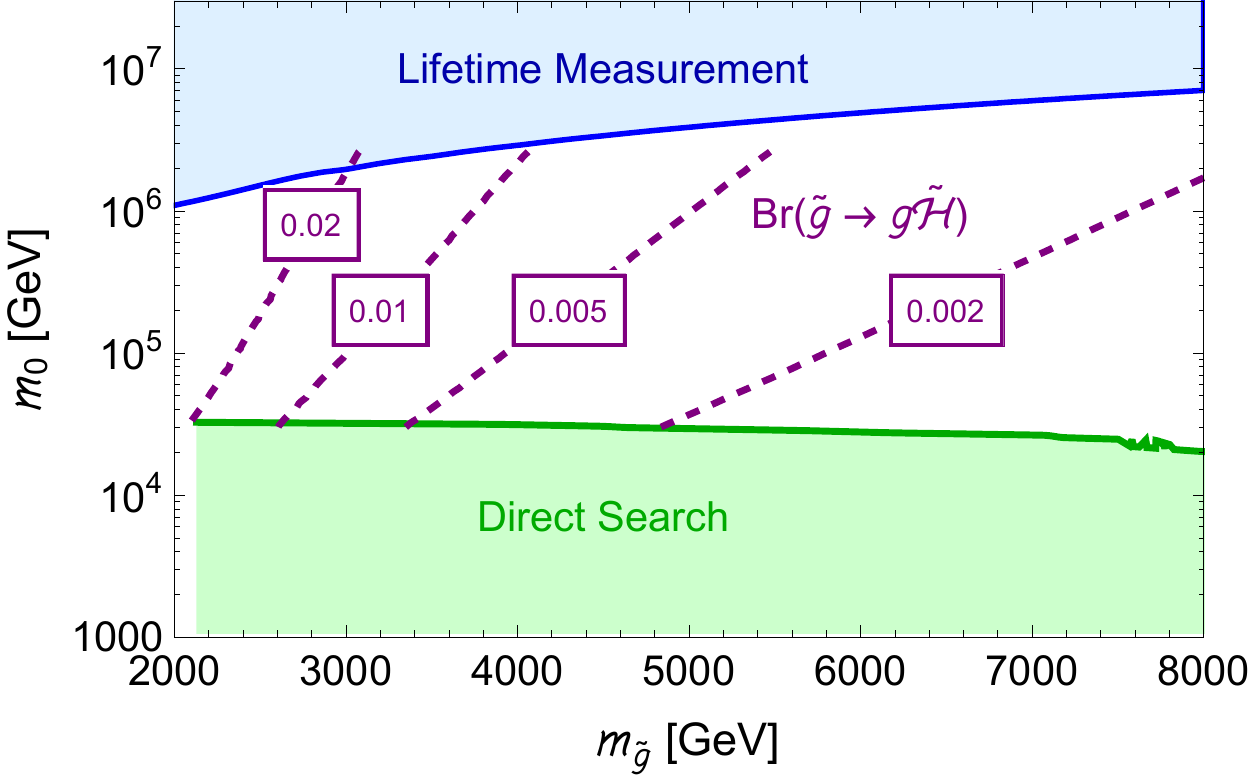}
\end{center}
\caption{Strategies for probing the SUSY scalar mass scale. The green region can be probed through associated squark-gluino production \cite{Ellis:2015xba}. The blue region leads to gluino lifetimes above 100 microns \cite{ArkaniHamed:2004fb,Arvanitaki:2012ps,ArkaniHamed:2012gw}. In the intermediate region, the scalar mass can be indirectly probed through gluino branching ratios (which are shown here for the choice of $\tan \beta$ that achieves a 125 GeV Higgs mass for the given $m_0$) \cite{Sato:2012xf}.}
\label{fig:gluinoscalarplane}
\end{figure}%
%%%%%%%%%%%%%%%%%

There is also a theoretical case for why the 30 TeV and 1000 TeV scalar mass scales are of particular interest. Many theories predict that scalars are roughly a loop factor heavier than gauginos, among them anomaly mediation without sequestered scalars \cite{Randall:1998uk,Giudice:1998xp,Wells:2003tf} and some moduli mediation scenarios \cite{Choi:2005ge,Choi:2005hd,Acharya:2007rc,Acharya:2008zi}. For weak-scale gauginos, these models predict scalars not far from the 30 TeV scale. A SUSY breaking scale near 30 TeV also allows gravitino or moduli decays just before BBN \cite{Coughlan:1983ci,Ellis:1986zt,deCarlos:1993wie,Moroi:1995fs,Kane:2015jia}. The higher 1000 TeV scale could be appealing from the point of view of flavor physics \cite{Wells:2004di,Altmannshofer:2013lfa,Baumgart:2014jya}. It is predicted in certain sequestered scenarios \cite{Blumenhagen:2009gk,Aparicio:2014wxa,Reece:2015qbf} that rely on approximate no-scale structure \cite{Cremmer:1983bf,Ellis:1983sf,Balasubramanian:2005zx,Conlon:2005ki}. These theories provide a strong motivation for distinguishing between the low and high scalar mass benchmarks when the scalars are neither light enough to directly produce nor heavy enough to cause the gluino lifetime to be measurable. 

\subsubsubsection{Gluino observables sensitive to scalar masses and $\tan \beta$}

Figure \ref{fig:higgsmass} makes it clear that we must probe both the scalar mass scale (especially the stop mass scale) and $\tan \beta$ in order to compare experimental results with the MSSM Higgs mass prediction. An earlier study has discussed the determination of the scalar mass in some detail, albeit at the LHC rather than 100 TeV \cite{Sato:2012xf}; we are not aware of a similarly detailed study on the determination of $\tan \beta$. We will assume that $M_3 > M_1, M_2, \mu$, so that the gluino can decay to any of the neutralinos and charginos. With a reasonable theoretical prior, this is at least an order-one fraction of the interesting parameter space. Any of the neutralinos and charginos will cascade promptly to the LSP. Gluino decays arise only from dimension-six operators generated by integrating out squarks. 

\subsubsubsection{Scalar mass measurement} 
\label{sec:scalarmassmeasure}

In the absence of a lifetime measurement, ratios of tree-level gluino decays probe the ratios of different squark masses but not the overall mass scale. However, the one-loop decay ${\tilde g} \to g {\tilde H}^0_{1,2}$ of a gluino to a gluon and a neutral higgsino has an additional {\em logarithmic} sensitivity to the scalar mass scale. This results from a loop diagram that begins with the four-fermion operator responsible for decaying a gluino to a higgsino, a top quark, and an anti-top quark; closes up the top loop; and adds a radiated gluon. As a result, the ratio of two- to three-body decays is a clean probe of the scalar mass scale, with the approximate dependence \cite{Toharia:2005gm}:
\begin{equation}
\frac{\Gamma({\tilde g} \to g {\tilde H}^0)}{\Gamma({\tilde g} \to t{\bar t} {\tilde H}^0)} \propto \frac{m_t^2}{m_{\tilde g}^2} \log^2 \frac{m_{\tilde t}^2}{m_t^2}. \label{eq:gluinoloopratio}
\end{equation} 
We will assume that the decay widths to the two neutral higgsinos are summed over, because they can be difficult to distinguish from one another experimentally. Resummation flattens out the scalar mass dependence at large $m_{\tilde t}$, but over the range we are interested in this is a relatively small effect \cite{Gambino:2005eh}. Furthermore, because both the numerator and the denominator depend in the same way on the stop mass and the top Yukawa coupling, this ratio is relatively insensitive to flavor-dependent physics (e.g. the stop mass compared to the first- and second-generation squarks) or to the value of $\tan \beta$. 

\subsubsubsection{Measurement of $\tan \beta$}

To measure $\tan \beta$ we can exploit Yukawa couplings $Y_u \propto 1/\sin \beta$ and $Y_d \propto 1/\cos \beta$ appearing in higgsino couplings. One probe is the rate of a gluino decay to higgsino relative to the rate to gauginos:
\begin{equation}
\frac{\Gamma({\tilde g} \to t{\bar t} {\tilde H}^0)}{\Gamma({\tilde g} \to t{\bar t} {\tilde B}^0)}, \frac{\Gamma({\tilde g} \to t{\bar t} {\tilde H}^0)}{\Gamma({\tilde g} \to t{\bar t} {\tilde W}^0)} \propto \frac{1}{\sin^2 \beta}. \label{eq:gluinotreeratio} 
\end{equation}
If the left- and right-handed stop masses are very different, measuring decays to both binos and winos can help to resolve the underlying physics. A disadvantage of this observable is that the dependence on $\tan \beta$ is mild over the range we are most interested in: systematic uncertainties in efficiencies at colliders of order 10\% could prevent us from drawing conclusions.

An observable with a steeper $\tan \beta$ dependence is the decay rate of the gluino to bottom quarks and a higgsino. In particular, if we can measure the ratio between two decays to higgsinos, we can obtain
\begin{equation}
\frac{\Gamma({\tilde g} \to b{\bar b} {\tilde H}^0)}{\Gamma({\tilde g} \to t{\bar t} {\tilde H}^0)} \propto \tan^2 \beta. \label{eq:gluinobbbarratio}
\end{equation}
The disadvantage is that the decay rate in the numerator is very small for the $\tan \beta$ values we are interested in. We could, alternatively, measure the ratio $\Gamma({\tilde g} \to b {\bar b} {\tilde H}^0)/\Gamma({\tilde g} \to g {\tilde H}^0)$. This has the same $\tan \beta$ dependence, is a larger ratio, and the events being compared may be more similar kinematically. The denominator is sensitive to the scalar mass scale, as noted above, but if this dependence has already been measured we can separate out the $\tan \beta$ dependence.

The observables $\Gamma(t{\bar t}{\tilde H}^0)/\Gamma(g {\tilde H}^0)$, $\Gamma(t{\bar t}{\tilde H}^0)/\Gamma(b {\bar b} {\tilde H}^0)$, and $\Gamma(t{\bar t}{\tilde H}^0)/\Gamma(t{\bar t}{\tilde B}^0)$, including resummation effects, are shown in Figure \ref{fig:gluinobranchingplots}. From the plot we see that, as expected, the first observable (in blue) is sensitive to the scalar mass scale but independent of $\tan \beta$, while the latter two observables (in green and red) are sensitive to $\tan \beta$ but only weakly depend on the scalar mass scale. (This very mild dependence is due to renormalization group mixing among the different dimension-six operators.) Notice that the $b{\bar b}{\tilde H}^0$ width has a much stronger sensitivity to $\tan \beta$ but is small---the curve has been rescaled by a factor of 10 to fit in the plot. 

%%%%%%%%%%%%%%%%%
\begin{figure}[tbp]\begin{center}
\includegraphics[width=0.47\textwidth]{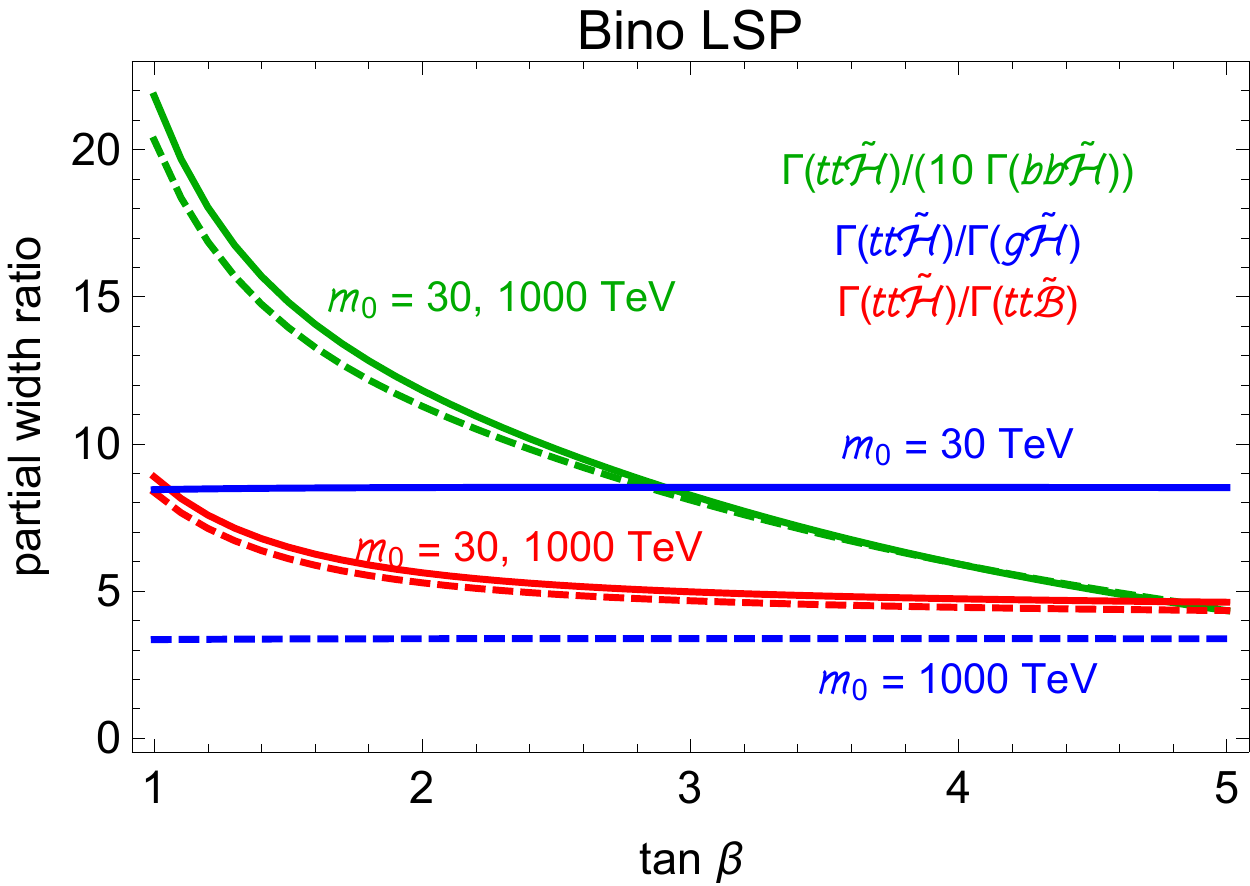}~~\includegraphics[width=0.49\textwidth]{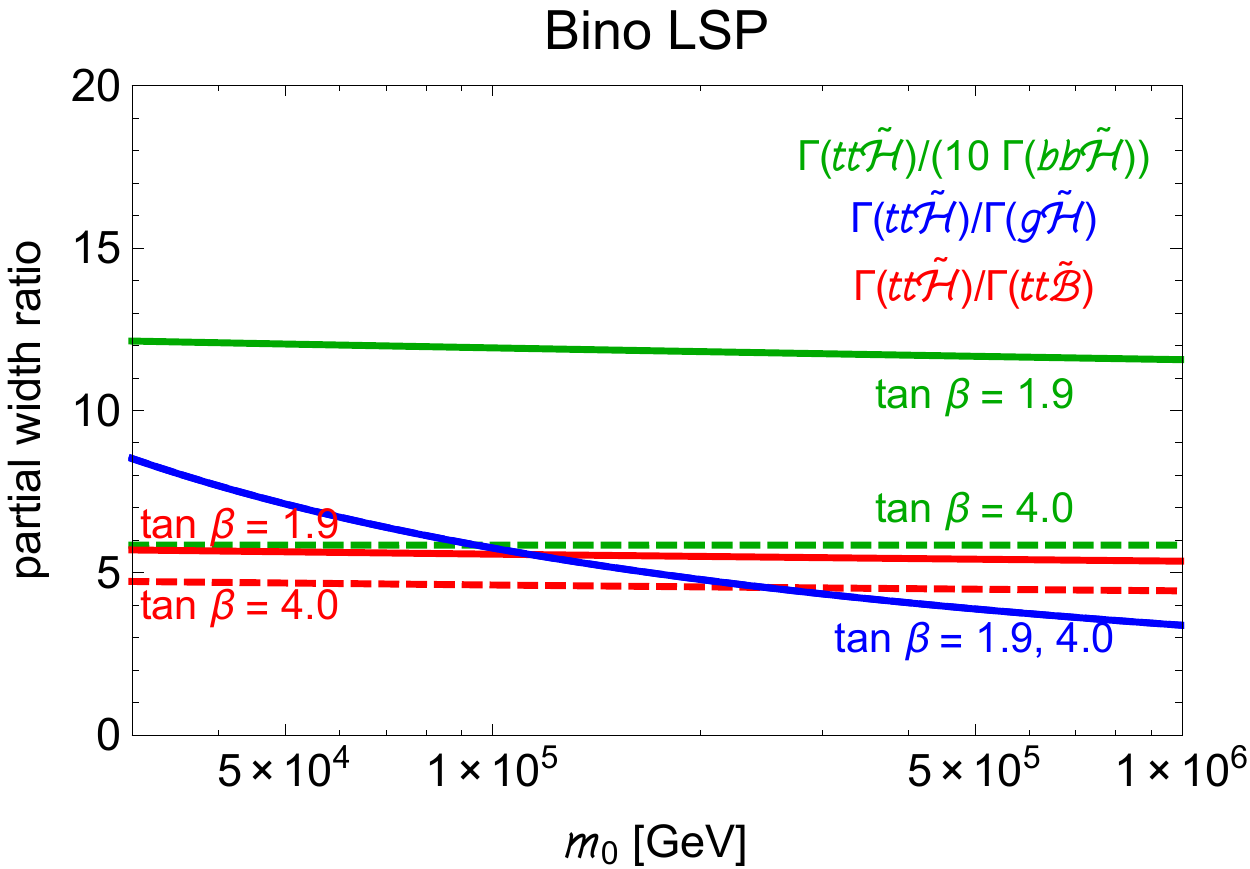}\\
\includegraphics[width=0.47\textwidth]{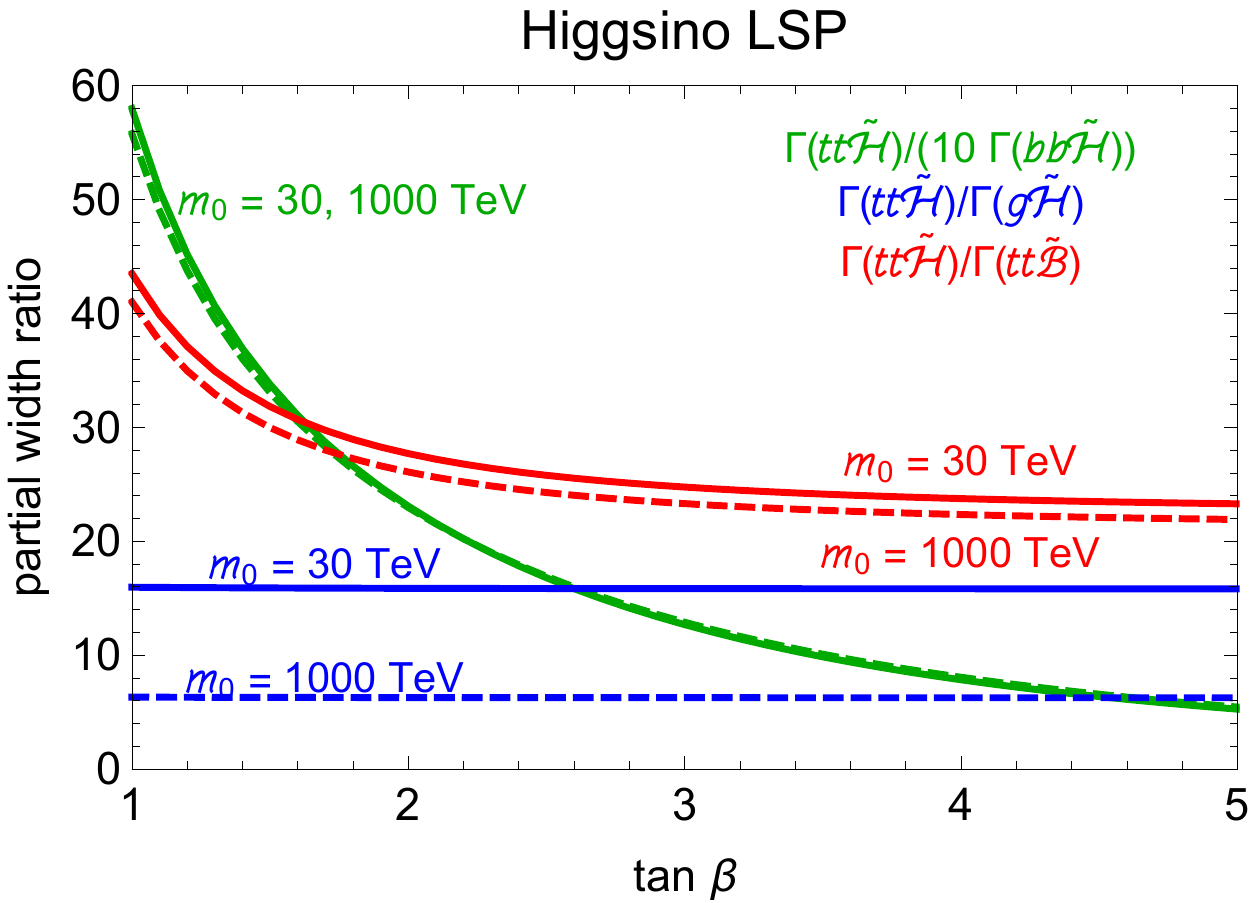}~~\includegraphics[width=0.49\textwidth]{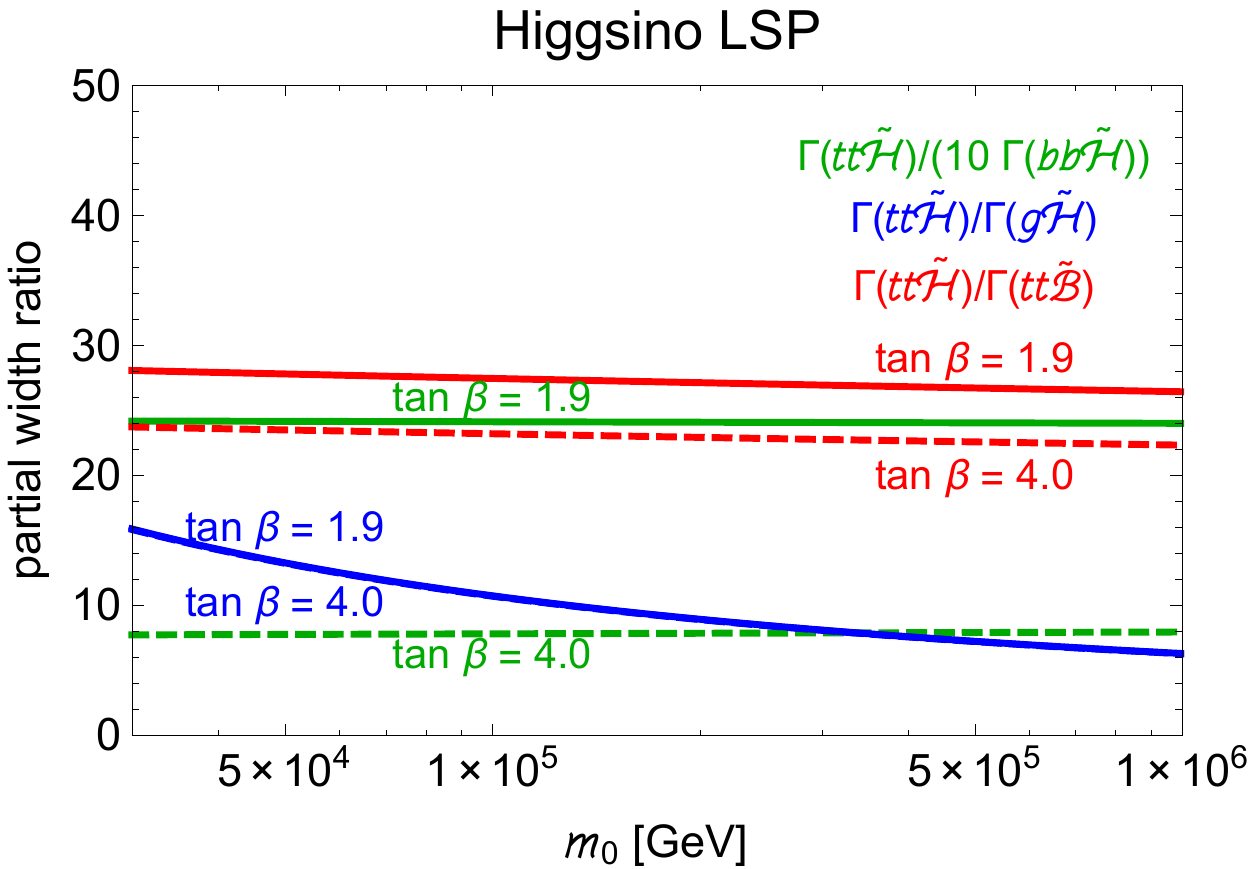}
\end{center}
\caption{Gluino branching ratios as probes of the SUSY scalar mass scale and $\tan \beta$. Top row: $M_1 = 200$ GeV, $M_2 = 400$ GeV, $\mu = 800$ GeV, and $M_3 = 2$ TeV. Bottom row: $M_1 = 700$ GeV, $M_2 = 1$ TeV, $\mu = 200$ GeV, and $M_3 = 2$ TeV. Notice that the $b{\bar b}$ width has been rescaled by a factor of 10 so that the green curve fits in the plot.}
\label{fig:gluinobranchingplots}
\end{figure}%
%%%%%%%%%%%%%%%%%

\subsubsubsection{Electroweak observables sensitive to $\tan \beta$}

We can also measure $\tan \beta$ through purely electroweak physics. The obvious place to look is decays purely in the electroweakino sector. See \cite{Han:2013kza} for a recent detailed discussion of branching ratios in this sector. Depending on the relative ordering of masses, some decays may be effectively absent or inaccessible, so different strategies are necessary. A less obvious probe of $\tan \beta$ arises from precision measurements of the Higgs boson.

\subsubsubsection{Higgsino LSPs}

If higgsinos are at the bottom of the spectrum, binos and winos will both promptly decay to higgsinos directly through the supersymmetric gauge interaction. In this case, the branching ratios carry very little information on $\tan \beta$. However, if events can be found (for instance in wino or gluino pair production) in which the decay ${\tilde H}^0_2 \to \ell^+ \ell^- {\tilde H}^0_1$ occurs (through an off-shell $Z$-boson), the dilepton invariant mass spectrum is sensitive to the higgsino mass difference, which is dependent on $\tan \beta$. However, this is a small effect: the leading mass splitting is $\sim m_Z^2/M_{1,2}$ and independent of $\tan \beta$, with a small subleading term of order $\mu m_Z^2/M_{1,2}^2 \sin(2\beta)$ (see e.g.~\cite{Martin:1997ns}). The fraction of events in which this $Z^*$ decay is observed is also weakly sensitive to $\tan \beta$.

\subsubsubsection{Higgsinos heavier than gauginos}
\label{sec:tanbetahiggsinosheavy}

Given a spectrum with $\mu > M_{1,2}$, decays of winos to binos (or vice versa if $M_1 > M_2$) may be observable, either in cascades from gluinos or higgsinos or from direct wino production when $M_2 > M_1$. The branching ratios in these decays depend on $\tan \beta$.  To understand this, integrate out the higgsino:
\begin{equation}
{\cal L}_{\rm eff} \supset \frac{g g'}{\mu} {\tilde B} {\tilde W}^i H_u \cdot T^i H_d + \frac{gg'}{2\mu^2} {\tilde B}{\bar \sigma}^\mu {\tilde W}^{i\dagger} \left(H_d^\dagger i \overset{\leftrightarrow}{D}_\mu \sigma^i H_d - H_u^\dagger i\overset{\leftrightarrow}{D}_\mu T^i H_u\right) + {\rm h.c.}
\end{equation}
With $M_2 > M_1$ for concreteness, the first term leads only to ${\tilde W}^0 \to h {\tilde B}$ while the second leads to ${\tilde W}^0 \to Z {\tilde B}$. Because the first term involves $H_u H_d$ and the second involves $\left|H_d\right|^2, \left|H_u\right|^2$, the $\tan \beta$ dependence of these widths will be different. In the limit $M_2/\mu \to 0, (m_h,m_Z)/M_2 \to 0$, and $M_1/M_2$ fixed, the ratio of decay widths scales as 
\begin{equation}
\frac{\Gamma({\tilde W}^0 \to h {\tilde B}^0)}{\Gamma({\tilde W}^0 \to Z {\tilde B}^0)} \approx \frac{4 \tan^2(2\beta) \mu^2}{M_2^2} \left(\frac{1+M_1/M_2}{1-M_1/M_2}\right)^2. \label{eq:winobinoratio} 
\end{equation}
This is potentially an interesting probe of $\tan \beta$.

\subsubsubsection{Higgsinos in the middle}

Now consider the spectrum $M_2 > \mu > M_1$. (The case $M_1 > \mu > M_2$ has similar physics, but binos are not directly produced, so the physics would have to be probed in gluino cascades.) Because winos decay to binos only through mixing with the higgsino, the overwhelming majority of decays will involve a two-step cascade ${\tilde W} \to {\tilde H} \to {\tilde B}$. The summed decay rates $\Gamma({\tilde W}^0 \to Z {\tilde H}^0_1) + \Gamma({\tilde W}^0 \to Z {\tilde H}^0_2)$ are independent of $\tan \beta$ at tree-level, but the individual amplitudes for these two processes go approximately as $\sin \beta \mp \cos \beta$. As a result, when $\tan \beta = 1$, some decays are entirely shut off: we find that ${\tilde W}^0 \to Z {\tilde H}^0_2,~h{\tilde H}^0_1$ occur and ${\tilde W}^0 \to Z {\tilde H}^0_1,~h{\tilde H}^0_2$ do not. A similar statement is true for the neutral higgsino decays to bino. Hence, for small $\tan \beta$, two-step decays involving two Higgses or two $Z$ bosons occur much more often than mixed decays with one $Z$ and one $h$. That is:
\begin{equation}
\frac{\Gamma({\tilde W}^0 \to Z h {\tilde B}^0)}{\Gamma({\tilde W}^0 \to Z Z {\tilde B}^0)+\Gamma({\tilde W}^0 \to h h {\tilde B}^0)} \propto \left(\frac{\sin \beta - \cos \beta}{\sin \beta + \cos \beta}\right)^2
\end{equation}
is a probe of how much $\tan \beta$ deviates from 1.

\subsubsubsection{Higgs boson branching ratios}

%%%%%%%%%%%%%%%%%
\begin{figure}[tbp]\begin{center}
\includegraphics[width=0.35\textwidth]{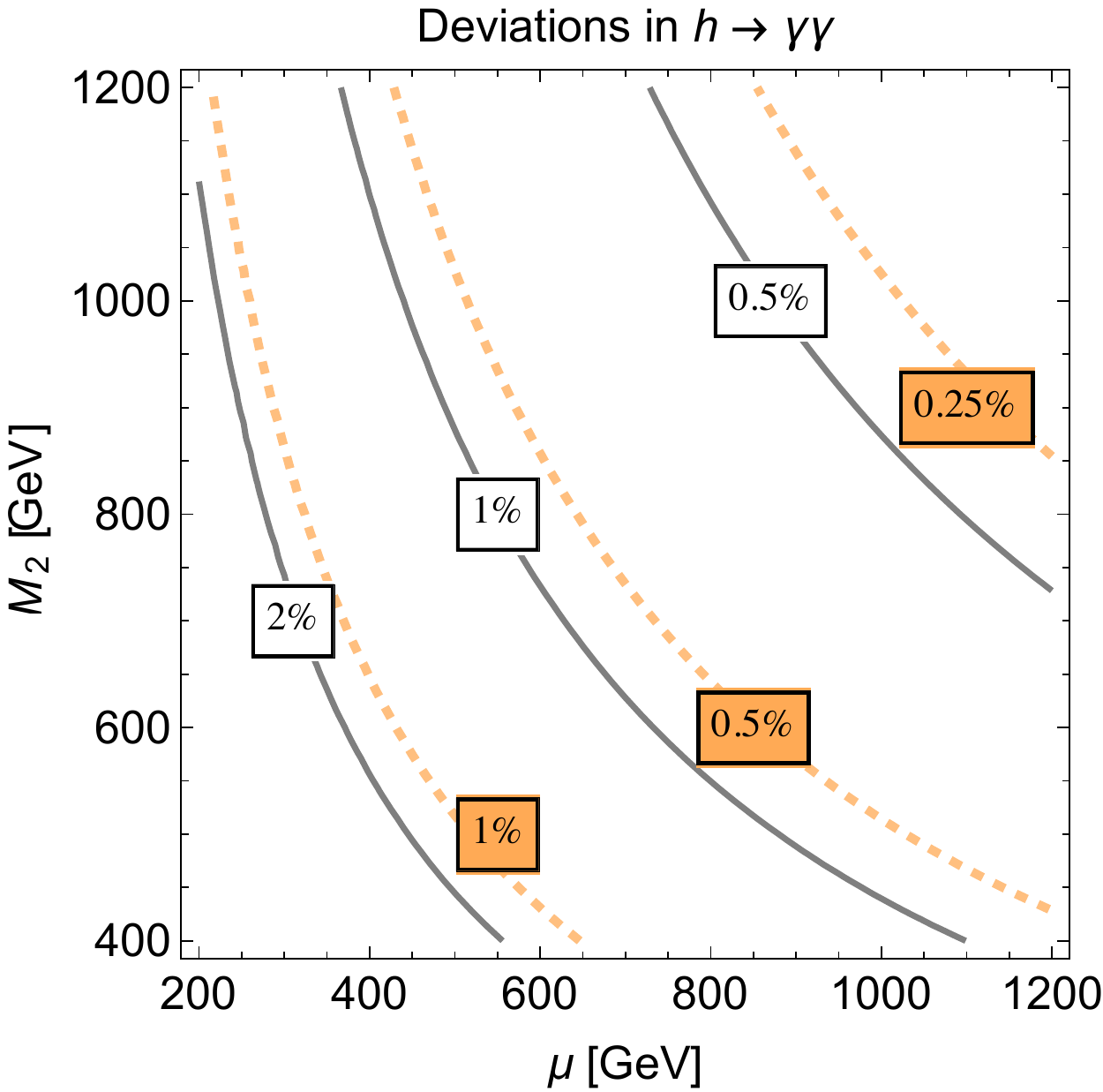}
\end{center}
\caption{Deviation in $h \to \gamma\gamma$ branching ratio from loops of charginos. Solid curves: $\tan \beta = 2$. Dashed orange curves: $\tan \beta = 4$.}
\label{fig:hgammagammaplot}
\end{figure}%
%%%%%%%%%%%%%%%%%

In split SUSY scenarios there is one light Standard Model-like Higgs boson. Its decays are affected only by loops of electroweakinos, which primarily modify the partial width to two photons (since this is already a small loop effect in the Standard Model). The deviation from the Standard Model width is given by
\begin{equation}
\frac{\Gamma(h \to \gamma\gamma)}{\Gamma(h \to \gamma \gamma)_{\rm SM}} \approx 1 + \frac{0.84 m_W^2 \sin(2\beta)}{\mu M_2 - m_W^2 \sin(2\beta)}.
\end{equation}
This is a small effect: only about a 2\% increase in the branching ratio when $\mu \approx M_2 \approx 500~{\rm GeV}$ and $\tan \beta \approx 2$, as illustrated in Figure \ref{fig:hgammagammaplot}. Neither the HL-LHC nor FCC-ee will measure the Higgs coupling to photons accurately enough to make use of this probe. However, there is a chance that 100 TeV can make a very precise measurement of the {\em ratio} $\Gamma(h \to \gamma\gamma)/\Gamma(h \to Z Z^*)$. Many systematics (e.g.~involving luminosity and production cross section) cancel in this ratio, so that hadron colliders can cleanly measure it \cite{ATLAS:2013hta,Peskin:2013xra}. If a sub-percent-level measurement of this ratio can be made, this could be an interesting alternative probe of $\tan \beta$ provided that the two charginos are not too heavy.

\subsubsubsection{Collider physics: measuring the observables} 

We will present some preliminary collider studies here, focusing mainly on SUSY backgrounds (i.e.~confusion among different decay modes). Of course, Standard Model backgrounds must be assessed, but cuts on missing $p_T$ and $H_T$ can help to reduce them, and in any case distinguishing different SUSY processes is a necessary step in measuring relative decay widths. A more extended study is in progress.% \cite{comesoon}.
 In simulating events we use Pythia \cite{Sjostrand:2014zea} supplied with a decay table computed by SUSY-HIT \cite{Djouadi:2006bz} and modified to include gluino decays as computed in \cite{Gambino:2005eh}. Jets are clustered using FastJet \cite{Cacciari:2005hq,Cacciari:2011ma}. For these preliminary studies we forego detector simulation.

\subsubsubsection{Measuring the scalar mass scale: an example with higgsino LSPs}

First we focus on a spectrum with $M_3 = 2~{\rm TeV}$, $M_2 = 1~{\rm TeV}$, $M_1 = 700~{\rm GeV}$, and $\mu = 200~{\rm GeV}$. For this spectrum, due to phase space factors, gluinos decay dominantly to third-generation quarks and higgsinos: depending on our choices of $m_0$ and $\tan \beta$, we have roughly $33-36\%$ $t{\bar t}{\tilde H}^0$ decays and $34-38\%$ $t{\bar b}{\tilde H}^-$ or ${\bar t}b{\tilde H}^+$ decays. Our first task is to measure $m_0$ via two-body decays as discussed in section \ref{sec:scalarmassmeasure}: ${\rm Br}({\tilde g} \to g{\tilde H}^0) \approx 5\%~(\Smiley_H, \Sadey_H)$ or $2\%~(\Smiley_L, \Sadey_L)$.

We attempt to identify events with a two-body decay on one side and a three-body decay on the other. Our final states, then, are
\begin{equation}
{\tilde g}{\tilde g} \to g + (t {\bar t}~{\rm or}~t{\bar b}~{\rm or}~{\bar t} b) + p_T^{\rm missing} + {\rm soft} + {\rm ISR/FSR}.
\end{equation}
We take advantage of the fact that, neglecting events with hard ISR, if we remove the gluon from the event we expect all other visible objects to have an invariant mass less than $M_{\tilde g}$. We find that the gluon is typically one of the two hardest jets in the event, so we select events in which removing one of the leading jets leaves a system with relatively low invariant mass. Then we attempt to test if the leading jet we removed is actually a gluon by requiring it to have little hard substructure as measured by $N$-subjettiness ratio variables $\tau_{N}/\tau_{N-1}$ \cite{Thaler:2010tr, Thaler:2011gf} (computed with the winner-take-all axis and $\beta = 1$ \cite{Larkoski:2014uqa}).

We have found a set of cuts with efficiency $\epsilon_{2\, {\rm body}} \approx 1.5 \times 10^{-3}$ on events containing a 2-body gluino decay but only $\epsilon_{3\, {\rm body}} \approx 1.3 \times 10^{-4}$ on other events:
\begin{align}
H_T > 2~{\rm TeV},  \quad\quad p_T^{\rm missing} > 1.6~{\rm TeV}, \quad\quad p_T(j_1) > 1~{\rm TeV},\\
N_{\rm jet} < 4, \quad\quad \left|\Delta \phi(j_2, p_T^{\rm missing})\right| > 1.8,\quad \quad M(j_{\rm gluon}~{\rm removed}) < 1.2~{\rm TeV},\\
\tau_2/\tau_1(j_{\rm gluon}) > 0.65, \quad \quad \tau_3/\tau_2 (j_{\rm gluon}) > 0.65, \quad \quad {\rm muons~near~}j_{\rm gluon}~{\rm vetoed}.
\end{align}
Here $j_{\rm gluon}$ is either the first or second highest $p_T$ jet in the event, chosen so that removing this jet and computing the mass of the others---denoted $M(j_{\rm gluon}~{\rm removed})$---gives the smallest result. We define $N_{\rm jet}$ as the number of jets with $\left|\eta\right| < 2.5$ and $p_T > 100~{\rm GeV}$ and $H_T$ as the scalar sum of the $p_T$ of those jets. We veto events with any muon of $p_T > 25~{\rm GeV}$ near our gluon candidate, which helps reduce the number of $b$- or $t$-jets faking our leading gluon. After these cuts, we can obtain samples that contain a significant fraction of 2-body decays: roughly 30\% for the points $\Smiley_L$ and $\Sadey_L$ and 60\% for the points $\Smiley_H$ and $\Sadey_H$. In 3 ab$^{-1}$ of data, these samples will contain somewhere around 3000 to 6000 events (depending on the two-body decay rate), so despite the relatively low efficiency, statistical uncertainties can be small.

%%%%%%%%%%%%%%%%%
\begin{figure}[tbp]\begin{center}
\includegraphics[width=0.5\textwidth]{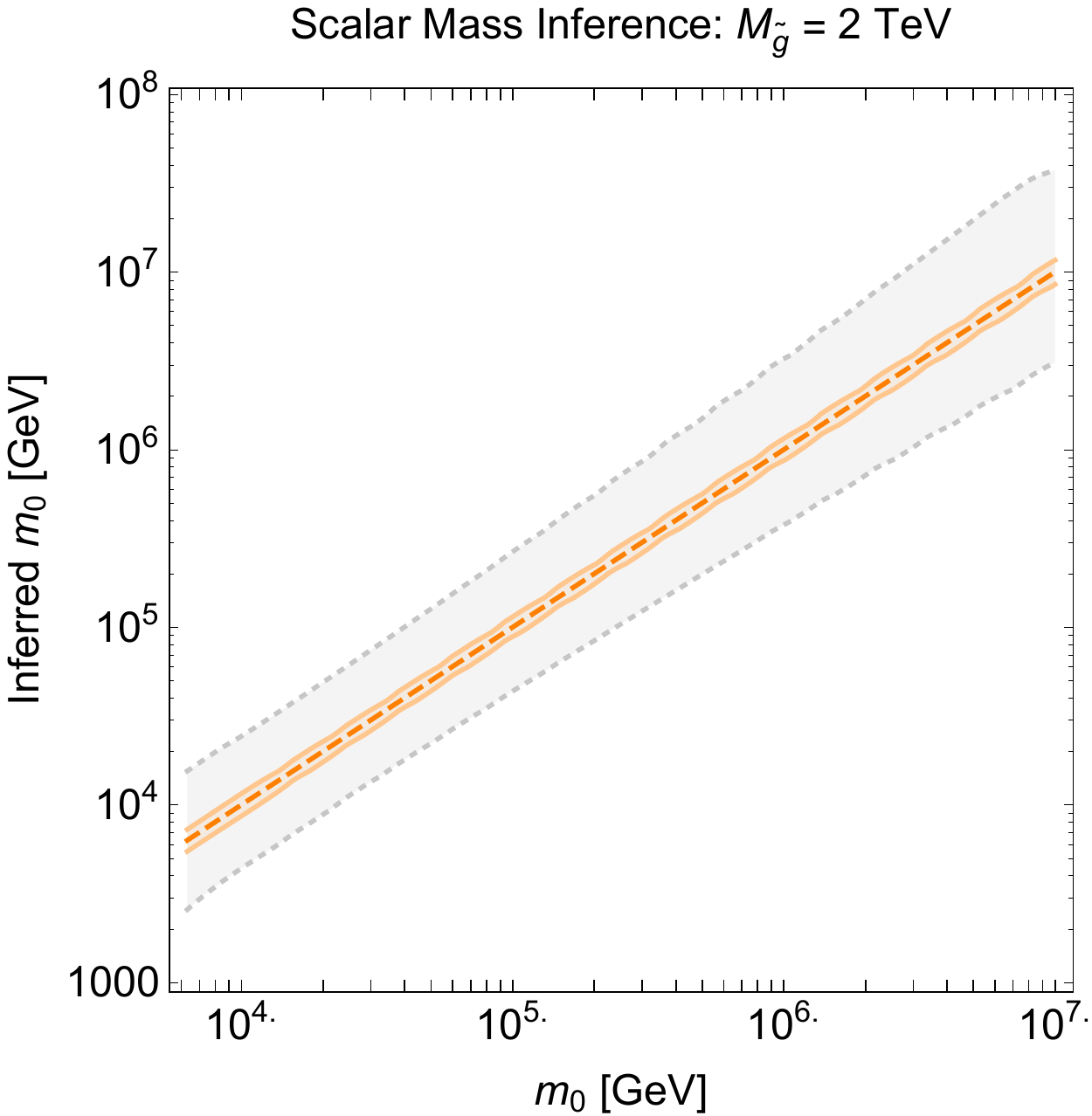}
\end{center}
\caption{Inference of the scalar mass scale $m_0$ from the measurement of the rate of 2-body decays ${\tilde g} \to g {\tilde H}^0$. The parameters are $M_3 = 2$ TeV, $\mu = 200$ GeV, $M_1 = 700$ GeV, and $M_2 = 1000$ GeV. The orange band represents $1\sigma$ statistical uncertainty with 3 ab$^{-1}$ of data, while the grey band corresponds to a 10\% systematic uncertainty on cut efficiencies times cross section times luminosity.}
\label{fig:M0EstimationPoint1}
\end{figure}%
%%%%%%%%%%%%%%%%%

The estimated performance of a simple cut-and-count analysis is presented in Figure \ref {fig:M0EstimationPoint1}. The orange band shows that statistical uncertainty alone can be quite small. The gray band represents a 10\% systematic uncertainty in the event rate. This corresponds to about a factor of 2 to 3 uncertainty in the scalar mass scale, bracketing $\Smiley_L$ to the range $13-74$ TeV and $\Smiley_H$ to $400-3200$ TeV. As we have emphasized above, measuring ratios (rather than simply counting events as we have done here) can help to cancel the sizable uncertainties in luminosity and production cross section. However, it may not be possible to eliminate other uncertainties, for instance in the efficiencies of some cuts. We emphasize that this is a preliminary analysis; the cuts can be further optimized, and a more sophisticated multivariate analysis would likely be effective. On the other hand, we have been somewhat optimistic in choosing a 2 TeV gluino mass, as both the production rate and the two-body branching fraction decrease for heavier gluinos. Further study will be required to optimize cuts in different kinematic regions and map out the expected statistical precision. 

\subsubsubsection{Measuring $\tan \beta$: an example with heavy higgsinos}

Now we turn to a spectrum with $M_3 = 2~{\rm TeV}$ but $M_1 = 200~{\rm GeV}$, $M_2 = 400~{\rm GeV}$, and $\mu = 800~{\rm GeV}$. In this case, as discussed in section \ref{sec:tanbetahiggsinosheavy}, the relative decay rate of neutral winos to $Z$ bosons and Higgs bosons can probe $\tan \beta$. We examine the case of electroweakino production: in 3 ab$^{-1}$ we expect about $7 \times 10^6$ wino pair production events (including both charged/charged and neutral/charged).

We search for ${\tilde W}^0 \to Z {\tilde B}^0$ in events with $Z \to \ell^+ \ell^-$, by requiring two opposite-sign same-flavor isolated leptons with $80~{\rm GeV} < m_{\ell\ell} < 100~{\rm GeV}$, two jets with $p_T > 30$ GeV, and $p_T^{\rm missing} > 200$ GeV. Opposite-sign opposite-flavor pairs in the $Z$ mass window are subtracted to eliminate contributions from $W$ bosons on both sides of the event. The efficiencies of the cuts are $\epsilon_{\tilde{W} \to Z \tilde {B}} \approx 8.7 \times 10^{-3}$ on events containing a neutral wino decaying to a $Z$ boson and bino but only $\epsilon_{\rm other} \approx 1.6 \times 10^{-4}$ on other events. After these cuts, we can obtain samples that contain a significant fraction of wino decaying to a $Z$ boson for $\tan \beta \approx 4$: roughly 55\% for the points $\Smiley_L$ and $\Sadey_H$. In 3 ab$^{-1}$ of data, these samples will contain somewhere around 1500 to 3000 events after the cuts, so again despite the relatively low efficiency, statistical uncertainties can be small.

%%%%%%%%%%%%%%%%%
\begin{figure}[tbp]\begin{center}
\includegraphics[width=0.5\textwidth]{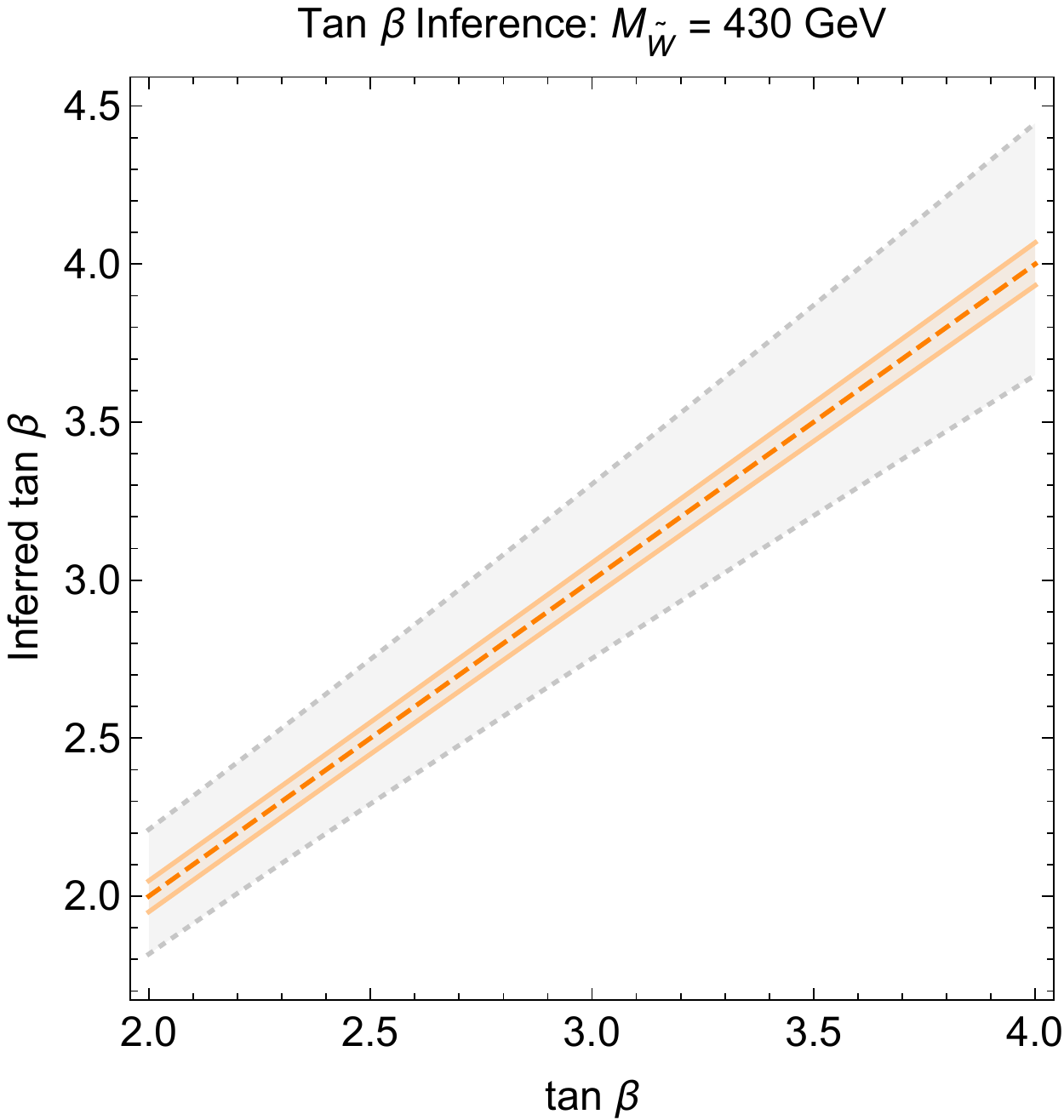}
\end{center}
\caption{Inference of $\tan \beta$ from the measurement of the rate of ${\tilde W}^0 \to Z {\tilde B}^0$ decays. The parameters are $M_3 = 2$ TeV, $M_1 = 200$ GeV, $M_2 = 400$ GeV, and $\mu = 800$ GeV. The orange band represents $1\sigma$ statistical uncertainty with 3 ab$^{-1}$ of data, while the grey band corresponds to a 10\% systematic uncertainty on cut efficiencies times cross section times luminosity.}
\label{fig:tanbetainferredpoint2}
\end{figure}%
%%%%%%%%%%%%%%%%%

The estimated performance of a simple cut-and-count analysis is presented in Figure \ref {fig:tanbetainferredpoint2}. Analogous to Figure \ref {fig:M0EstimationPoint1}, the orange band represents the small statistical uncertainty while the gray band represents a 10\% systematic uncertainty in the event rate. This corresponds to about a factor of 10\% variation in determining the value of $\tan \beta$.  In our analysis, we only considered the SUSY background and didn't take into account of the SM background. The result should be taken as a rough estimate to motivate a further refined analysis. 

%%%%%%%%%%%%%%%%%%%%%%%%%%%%%%%%%%
\subsubsubsection{Conclusions}
\label{sec:conclusion}
%%%%%%%%%%%%%%%%%%%%%%%%%%%%%%%%%%

A 100 TeV hadron collider has tremendous potential to address many deep fundamental questions in particle physics, such as the underlying mechanism that generates the observed Higgs mass. In this section, we have discussed how to use precision measurements of gluino and neutralino decays at a 100 TeV collider to test the Higgs mass explanation in the MSSM with gauginos at around a TeV. In the MSSM, the Higgs mass is raised from the tree-level predication by the loop contribution of heavy stops to the observed value. Direct searches and lifetime measurements, which have been discussed already in the literature, still leave untouched a large and interesting region of parameter space with scalar mass in the range (30--1000) TeV and $\tan \beta$ between 2 and 4. Among all observables in gaugino decays, the two-body decays of gluinos are loop-induced and logarithmically sensitive to the scalar mass scale. We have demonstrated that in a scenario with a higgsino LSP, the scalar mass could be inferred from measuring the gluino two-body decays with a factor of 2 to 3 uncertainty at a 100 TeV collider with 3 ab$^{-1}$ data. There are several different approaches to measure $\tan\beta$ from either gluino decay or wino or higgsino decays. We use winos decaying to a $Z$ boson in a scenario with a bino LSP and wino lighter than higgsino as an example to demonstrate the potential of determining $\tan\beta$ with a 10\% uncertainty. All the studies are still preliminary, yet they demonstrate the great potential of a 100 TeV collider in precision measurements of SUSY particles at around TeV scale or below and unraveling the mystery of the Higgs mass. A more thorough and refined analysis will be implemented and presented soon~\cite{comesoon}.

  % chargino/neutralino (Acharya et al) http://inspirehep.net/record/1320763
  % pMSSM and mono-jet http://inspirehep.net/record/1389857
  %  Higgs and SUSY  http://inspirehep.net/record/1370688

}

\subsection{Summary of Phenomenological Studies}
\label{sec:susy_summary}
The supersymmetric aspect of an experimental program at 100 TeV is very rich.  While comprising just the tip of the iceberg, the phenomenological studies of Secs.~\ref{sec:susy_xs} to \ref{sec:SUSY_post} demonstrate a varied frontier of exciting signatures.  In Fig.~\ref{fig:SUSY_summary} we summarise the results of these studies in the context of simplified models assuming $30 \text{ ab}^{-1}$ of integrated luminosity.

Let us revisit the broadly motivated mass ranges outlined in Sec.~\ref{sec:susy_intro}.  While supersymmetry may exist as a new spacetime symmetry at any energy, various considerations converge towards particular mass ranges for superpartners.  Dark matter considerations point towards gauginos and/or Higgsinos in the $\lesssim \mathcal{O}(10 \text{ TeV})$ range.  Gauge coupling unification is similarly suggestive for the gauginos, although the upper bound on gaugino masses is less robust than for thermal dark matter.  The Higgs mass points towards a range of scalar masses, however for $\tan \beta \gtrsim 4$, squarks, in particular stop squarks, should be expected below $\lesssim \mathcal{O}(\text{10's TeV})$.  If naturalness is desired, all coloured sparticles should be within $\lesssim \mathcal{O}(\text{few TeV})$, and the stops and gluinos as light as granted by current bounds.

\begin{figure}[tbp]
  \centering
  \includegraphics[width=0.8\columnwidth]{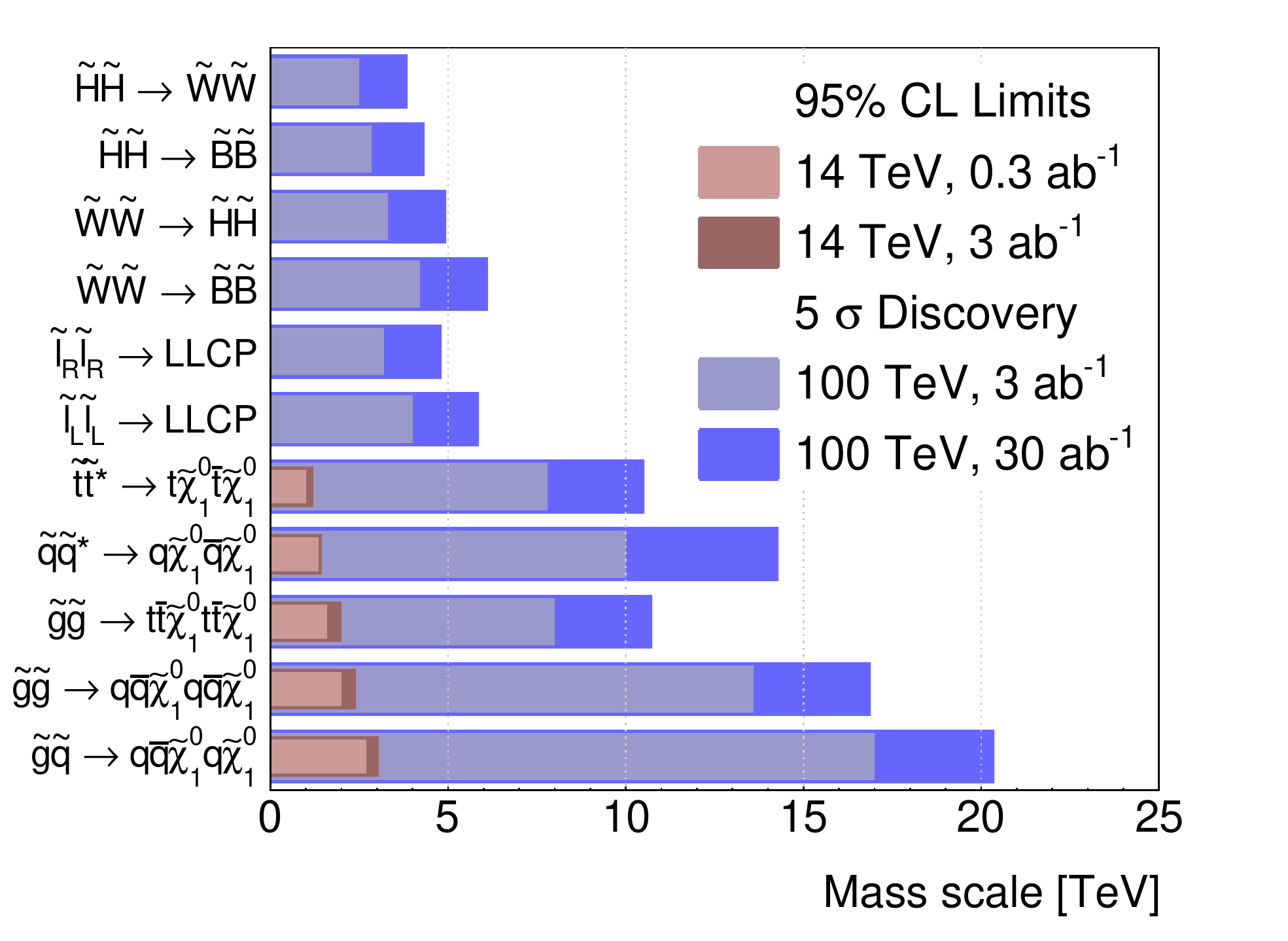}
  \caption{Sensitivity for simplified models considered in this section for the LHC, HL-LHC, and a $pp$ collider at $\sqrt{s}=100$ TeV with data samples of 3~\iab{} and 30~\iab.  The reach for strong-production at 14~\TeV{} is quantified by 95\% confidence level upper mass limits on the mass of squarks or gluinos (or both) when the LSP is massless, and is taken from ATLAS and CMS projections~\cite{ATL-PHYS-PUB-2013-002,ATL-PHYS-PUB-2013-011,ATL-PHYS-PUB-2014-010,CMS-PAS-SUS-14-012,CMS-PAS-FTR-13-014}, or from this document in the case of the $\gluino\gluino\rightarrow t\bar{t}\chioz t\bar{t}\chioz$ model.  Sensitivity for $\sqrt{s}=100$~\TeV and 3~\iab{} is quantified by the 5$\sigma$ discovery reach presented in this document.  The 30~\iab{} reach is from this document when available, otherwise it is projected from the 3~\iab{} reach using the Collider Reach web tool~\cite{ColliderReach}.}
  \label{fig:SUSY_summary}
\end{figure}

Comparing these expectations with the results in Fig.~\ref{fig:SUSY_summary} we see that much of the supersymmetric parameter space relevant to core puzzles in high energy physics, such as dark matter, grand unification, the Higgs mass, and naturalness, can be covered.  Let us now consider the context of such measurements in light of our current picture of fundamental physics at the LHC, particularly with respect to naturalness of the weak scale.  We will then look towards the future potential impact these measurements could have on our understanding of fundamental physics.

The primary goal of the LHC is the exploration of physics at the weak scale. Hence, testing the naturalness principle in the Higgs sector is a central issue.  At the time this document is written, the LHC verdict on naturalness is not yet final. Admittedly, data at $\sqrt{s}=8$ TeV and preliminary results at 13 TeV strongly disfavour the most straightforward implementations of natural low-energy supersymmetry, which favoured new weakly-interacting particles in the 100 GeV domain and strongly-interacting particles well below the TeV scale.

If the LHC reveals new phenomena, these discoveries will redesign the priorities of future high-energy physics. Although today it is impossible to say what those priorities will be, it is hard to imagine a new-physics scenario, supersymmetric or not, which will not motivate us to continue explorations towards energies higher those those of the LHC. As illustrated in this document, the physics program at 100 TeV is rich enough to provide an excellent tool to carry out such explorations at high energies.

If no discoveries are made at the LHC, the simplest versions of low-energy supersymmetry would be ruled out. This would be a momentous result, as supersymmetry has played a central role in the conceptual development of our field for decades. In this sense, the era of natural supersymmetry would come to an end.  However, in such an instance it would be incorrect to conclude that the naturalness principle is misguided. Excluding new dynamics at the weak scale would mean ruling out our favoured solutions to the naturalness problem, but not the problem itself, and knowing how nature deals with Higgs naturalness will remain a standing issue.  This reframing of the naturalness question would imply the loss of the logical connection between Higgs naturalness and new phenomena at the TeV scale. If this connection is lost, what would be so special about the energy scale explored by a 100 TeV collider and why should we expect new phenomena in that range?

In spite of its virtues at a more fundamental level, supersymmetry may not be the answer to Higgs naturalness. Speculations have been made about logical schemes that deal with Higgs naturalness without dynamics at the weak scale, such as the anthropic principle or cosmological relaxation. Intriguingly, even within these very different schemes, motivations for supersymmetry emerge, although at a scale different than the weak scale and also for different reasons.  In the context of unnatural setups, considerations discussed in Sec.~\ref{sec:susy_intro} about dark matter, gauge coupling unification, or the Higgs mass, or the limited cutoff that can be achieved in cosmological relaxation scenarios call for supersymmetry with a certain preference for the $\mathcal{O}(10\text{'s})$TeV range.  Fig.~\ref{fig:SUSY_summary} demonstrates that this energy range is prime territory for a 100 TeV collider.

To summarise, speculations about the role of supersymmetry in `unnatural' theories suggest that a future physics program should not be regarded as an extension of LHC searches, but rather as conceptually different. If the LHC is the machine of the {\it naturalness era}, future colliders would become the machine of the {\it post-naturalness era}. An era in which we are forced to change the focus of our basic questions about particle physics, in which we contemplate partly unnatural theories or theories where naturalness is realised in unconventional ways, and in which supersymmetry may enter in a new guise.  

\newpage
\section{Dark Matter} %\footnote{Editors: Schwaller, Harris}} 
\label{sec:DM}

%\section{Dark Matter searches and studies\footnote{Editors:
%  P.~Harris and P.~Schwaller; contributors: ....}}
%\label{ref:dm}

%\subsection{Introduction}
%
%
\subsection{Introduction}
Today there is an overwhelming amount of evidence from astrophysical observations that a large fraction of the observed matter density in the universe is invisible to us. This so called Dark Matter (DM) makes up 26\% of the total energy density in the universe, and more than 80\% of the total matter~\cite{Ade:2015xua}. Despite numerous observations of the astrophysical properties of DM, not much is known about its particle nature. The main constraints on a particle DM candidate $\chi$ are that it (see e.g.~\cite{Gelmini:2015zpa} for a more detailed discussion)
\begin{itemize}
	\item should gravitate like ordinary matter
	\item should not carry color or electromagnetic charge
	\item is massive and non-relativistic at the time the CMB forms
	\item are long lived enough to be present in the universe today ($\tau \gg \tau_{\rm universe}$) 
	\item does not have too strong self-interactions ($\sigma/M_{\rm DM} \lesssim 100~{\rm GeV}^{-3}$). 
\end{itemize}
While no SM particles satisfy these criteria, they do not pose very strong constraints on the properties of new particles to play the role of DM. In particular the allowed range of masses spans almost 80 orders of magnitude. Particles with mass below $10^{-22}\UeV$ would have a wave length so large that they wipe out structures on the kPc (kilo-Parsec) scale and larger~\cite{Hu:2000ke}, disagreeing with observations, while on the other end of the scale micro-lensing and MACHO (Massive Astrophysical Compact Halo Objects) searches put an upper bound of $2\times 10^{-9}$ solar masses or $10^{48}\UGeV$ on the mass of the dominant DM component~\cite{Griest:2013aaa,Alcock:1998fx,Yoo:2003fr}. 

Clearly we can not hope that any future collider will probe the full mass range allowed by astrophysical observations. However there is a very broad class of models for which theory motivates the \UGeV~-\UTeV~mass scale, and which therefore could be in range of a future hadron collider operating at a centre-of-mass energy around 100\UTeV. If at any point in the history of the early universe the DM is in thermal equilibrium with the SM particles, then we can estimate its relic density today by studying how it decouples from the SM, the so called freeze-out. For particles which are held in equilibrium by pair creation and annihilation processes, $(\chi \chi \leftrightarrow {\rm SM})$ one finds the simple relation that~\cite{Kolb:1990vq}
\begin{align}
	\Omega_{\rm DM}h^2 \sim \frac{10^{9}\UGeV^{-1}}{ M_{\rm pl}} \frac{1}{\langle \sigma v \rangle} \,, \label{eqn:relicD1}
\end{align}
where $\langle \sigma v \rangle$ is the velocity averaged annihilation cross section of the DM candidate $\chi$ into SM particles, $\Omega_{\rm DM}h^2 \approx 0.12$ is the observed relic abundance of DM~\cite{Ade:2015xua}, $M_{\rm pl}$ is the Planck scale and order one factors have been neglected. 

For a particle annihilating through processes which do not involve any larger mass scales, the annihilation cross section scales as $\langle \sigma v \rangle \sim g_{\rm eff}^4/M_{\rm DM}^2$, where $g_{\rm eff}$ is the effective coupling strength which parameterises the process. It follows that
\begin{align}
	\Omega_{\rm DM}h^2 \sim 0.12 \times \left(\frac{M_{\rm DM}}{2\UTeV}\right)^2 \left(\frac{0.3}{g_{\rm eff}}\right)^4 ,
\end{align}
i.e. that a DM candidate with a mass at or below the \UTeV~scale and which couples to the SM with a strength similar to the weak interactions naturally has a relic density in agreement with observations. There are several variations of this simple approximation which modify the preferred mass range, e.g. when the annihilation processes involve heavier states, when it is velocity suppressed, assisted by co-annihilation or increased through a resonance~\cite{Griest:1990kh}. Including these effects, one finds that weakly interacting massive particles (WIMPs) can reproduce the observed relic abundance when their mass is in the 10s of \UGeV~to few \UTeV~range. On one side this is the main reason why we hope to find evidence for DM at the LHC, but on the other hand it already tells us that a higher energy collider will be necessary to efficiently probe the WIMP paradigm for DM. 

As the mass of DM increases, Eqn.~\ref{eqn:relicD1} tells us that to maintain the observed relic abundance the annihilation cross section also has to increase. This becomes inconsistent with unitarity of the annihilation amplitudes at $M_{\rm DM} \lesssim 110\UTeV$, the so called unitarity bound on the mass of DM~\cite{Griest:1989wd,Blum:2014dca}. Most well motivated models of WIMP DM do not saturate this bound, but rather have upper limits on the DM mass in the few \UTeV~range. One main aspect of this document is to determine how well these models can be probed currently and with a future collider experiment. 

For DM masses at the lower end of the WIMP spectrum, one typically expects that annihilation proceeds through a mediator with $M_{\rm med} > 2 M_{\rm DM}$. Then the annihilation cross section is suppressed by $(M_{\rm DM}^4/M_{\rm med}^4)$. Assuming that no mediator particle exists with a mass below the Higgs mass, then this puts a lower bound of a few \UGeV~on the mass of the WIMP DM candidate, while an even wider range of DM masses becomes accessible if the mediator is lighter but very weakly coupled to the SM.

In the first part of this report, we will focus on WIMP scenarios where the relic density of DM is set by non-relativistic annihilation (freeze-out) to SM particles, either through a SM portal or through new mediators. Obviously if new mediators are present, searching for or constraining them directly might be possible before the DM itself becomes discoverable, and we will discuss how this interplay evolves when going to higher energies. 

\

There are DM models which do not fall into the WIMP category as defined above, but which are still relevant for DM searches at hadron colliders. Mainly these are models where the DM is in thermal equilibrium with the SM at some point, but the relic abundance is not determined by the usual freeze out mechanism. The best known examples are models of asymmetric DM (ADM), where the relic abundance is determined by an asymmetry in DM versus anti-DM in the early universe~\cite{Petraki:2013wwa,Zurek:2013wia}, possibly related to the baryon asymmetry of the universe, and models where the DM annihilates to an additional (lighter) state in the dark sector first, which later decay to SM particles: 
\begin{align}
	\chi \chi \to a a \quad \text{followed by}\quad a \to {\rm SM}\,.
\end{align}
Necessary ingredients in both cases are first an interaction to bring the dark sector into thermal equilibrium with the SM at early times, and furthermore a way to transfer entropy from the dark sector back to the visible sector after the relic abundance is set. For the ADM scenario this means that the symmetric abundance has to annihilate efficiently, either to SM particles as in the WIMP scenario (but with a cross section somewhat larger than in the WIMP case), or an annihilation into lighter, unstable particles of the dark sector. 

The entropy transfer must happen before the onset of big bang nucleosynthesis (BBN) at temperatures around $10$\UMeV. This puts an upper bound on the lifetime of the unstable dark sector particle of $\tau_a \ll 1$~s. From the collider perspective this means that the dark sector particles can either decay promptly, with a displaced vertex, or could be collider stable, and all three regimes need to be probed to say something conclusive about these non-WIMP scenarios. 

\

In the following introductory chapters, we will review the current bounds on DM from direct and indirect searches, from cosmology and from collider experiments, as well as how the sensitivity is expected to evolve in the next 20 to 30 years. Then we will discuss the prospects of a 100~TeV collider to probe the thermal WIMP scenario, starting with minimal and simplified models and moving on to some examples of UV complete models. Following that we discuss examples of non-WIMP DM scenarios which can be probed at hadron colliders and the possible benefit of a 100~TeV machine, before presenting our conclusions.

%\subsection{Status and Prospects of DM Searches}
\newcommand{\MET}{$\slashed{E}_T$~}
\newcommand{\Tev}{{\rm{TeV}}}
\subsection{Experimental searches for DM}
Searches for DM can be split into three separate classes: Direct detection, indirect detection, and collider based searches. Direct detection consists of the search for DM using a nuclear recoil. Indirect  detection consists of the class of searches looking for annihilation of DM in galactic collisions. This class of experiments consists of the set of satellite and ground based telescope experiments which search for excesses of either photons or antiparticles in space. Finally, there are collider searches, where production of DM is searched for via its missing energy signature. 
The two non-collider set of experiments can be compared to the collection of collider searches. For the case of a 100~TeV proton collider, the ultimate bounds of these searches are considered so as to put into context the comparative reach of the colliders. 

\subsubsection{Direct Detection}
Conventional direct detection probes the rate of DM nucleon interactions in earth based experiments. This is done through low rate, high sensitivity searches of low energy DM nucleon recoils. The searches yield a bound on the matter DM cross section $\sigma$. This cross section relies on two fundamental ingredients, the type of interaction, and the relative density of DM in the solar system. The searches compute the rate of DM nucleon collisions for a given recoil energy $E$, denoted $dR/dE$. This we can write as, 
\begin{align}
\frac{dR}{dE} & = \frac{\rho_{\rm DM}}{m_N m_{\rm DM}} \int_{v>v_{min}} v f(v) \frac{d \sigma(v,E)}{dE} dv\,
%\frac{dN}{dE} & = & \frac{\rho_{DM}}{m_{DM}} 
%\frac{\sigma_{0}}{2\mu^{2}_{DM}} \int_{v>v_{min}} d^{3}v \frac{f(\vec{v},t)}{v},\\
%\mu_{DM}       & = & m_{DM} m_{nucleon}/(m_{DM} + m_{nucleon})
\end{align}
where $\rho_{DM}$ is the local DM density, $f(v)$ is the local DM velocity profile for DM velocity $\vec{v}$, $m_{N}$ is the recoiling nucleus mass, $m_{DM}$ is the DM mass and $\frac{d \sigma}{dE}$ is the differential DM nucleus interaction cross section.  The rate measurement  can be translated to a cross section bound for a given DM mass through the fact that all parameters are known with the exception of the DM mass, $m_{DM}$, and the DM cross section. The other parameters, in particular the DM density, $\rho_{DM}$ and velocity profile, $f(v)$, are inferred from local galactic measurements combined with galactic simulations.  The current measurements for the DM density and velocity profile have a level of variability that is expected to improve over the coming years. However the variability itself motivates the use of a collider search to allow for precise determination of the DM properties. 

Direct detection searches can be split into two classes of DM interactions, spin-independent and spin-dependent interactions. Spin independent interactions consist of DM nucleon interactions that do not have any dependence on the spin structure of the mediator nucleon interaction. This includes interactions involving a scalar mediator or a vector mediator without an axial coupling. Spin-dependent DM occurs when the interaction model is sensitive to the spin structure of the nucleus. Direct detection has a much larger sensitivity to spin independent interactions due to the coherent enhancement of the cross section proportional to the square of the nucleus mass. 

An ultimate bound for direct detection comes from neutrino interactions in the detectors. This  background cannot be distinguished from DM interactions and thus is is irreducible~\cite{Cushman:2013zza}. This bound has served as a benchmark for DM searches and represents an ultimate goal for the next generation of direct detection experiments. This bound exists for both spin-dependent and spin-independent interactions, as shown in figure~\ref{fig:DDBounds}. Recently, the directional DM detection has demonstrated the capability to extend beyond the neutrino wall~\cite{Grothaus:2014hja}. However, there is currently no plan to build an experiment large enough to reach this boundary.
\begin{figure}[h]
\begin{center}
\includegraphics[width=0.49\textwidth]{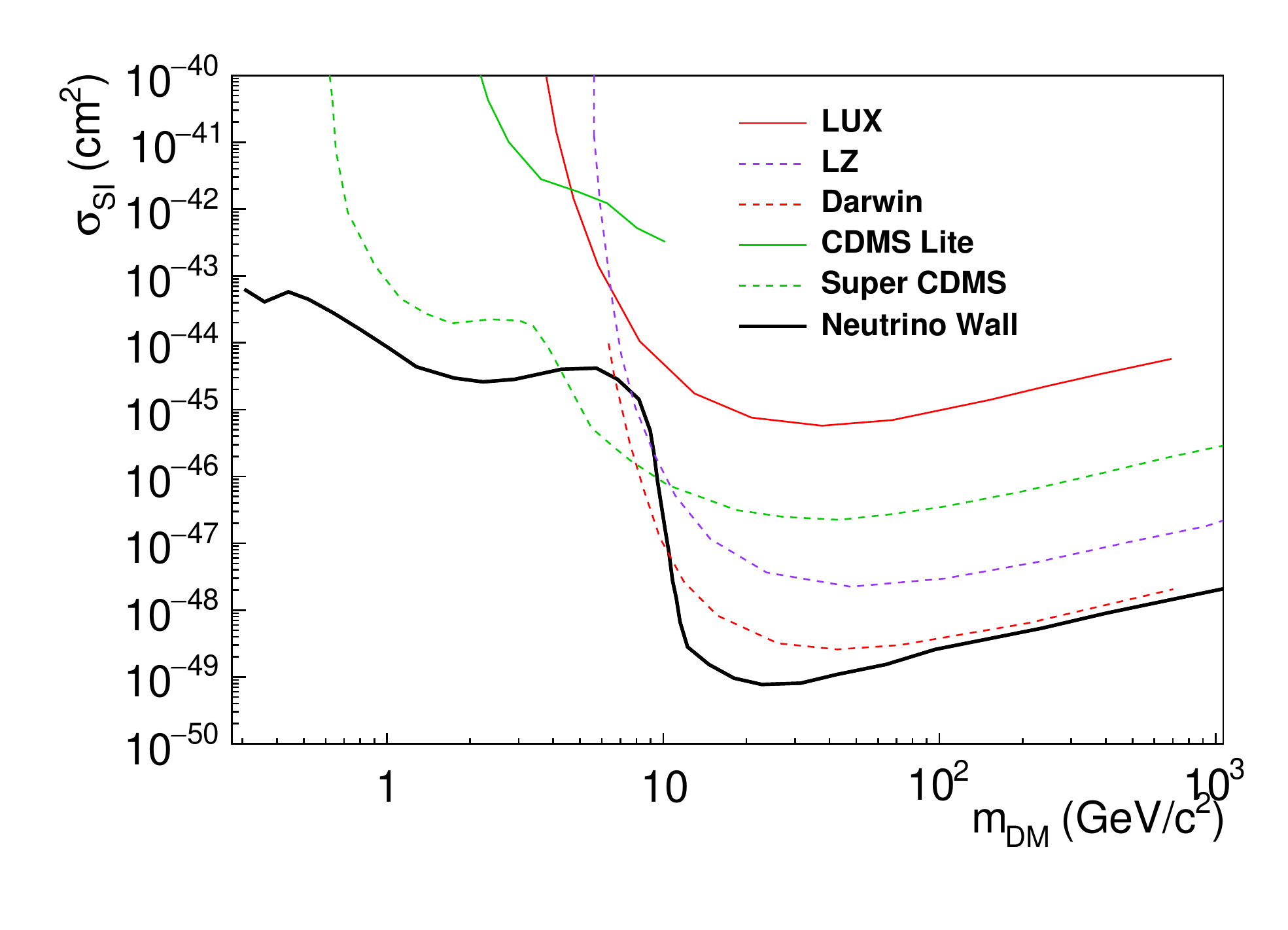} 
\includegraphics[width=0.49\textwidth]{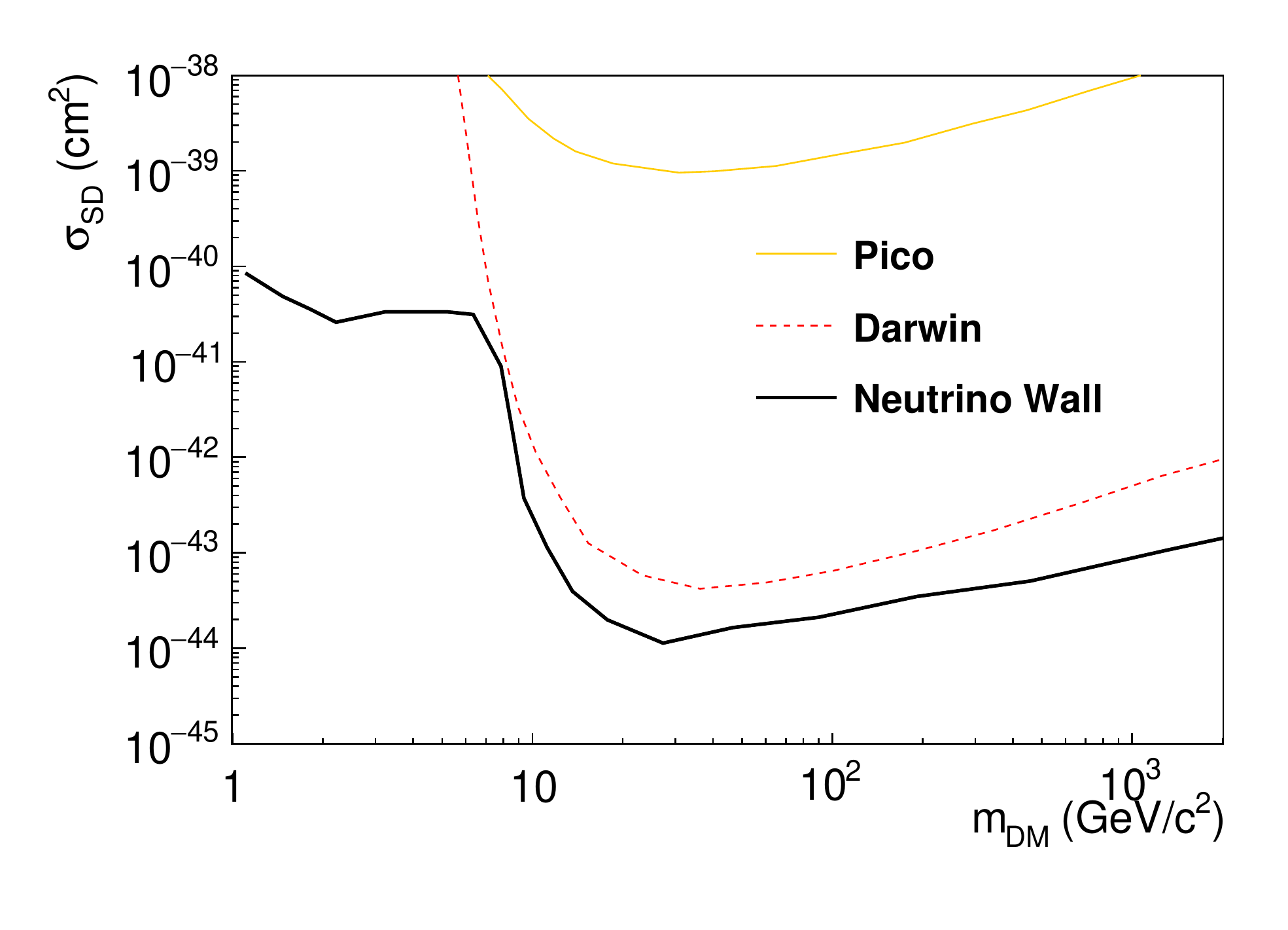} 
\caption{ Comparison of the current best bound (solid line) with upcoming experiments (dashed line), and the neutrino wall for spin independent(left) and spin dependent detection (right). For spin dependent interactions additional lines potentially exist for the LZ however they are currently not publicly available. 
}
\label{fig:DDBounds}
\end{center}
\end{figure}

The searches for direct detection have greatly improved over the past few years. This has been largely from the  development of two technologies, low energy cryogenic detectors, and large scale liquified noble gas detectors. Both these technologies are going through major upgrades in the detector size; further allowing for enhanced sensitivity. In particular, the CDMS detector is expected to extend the sensitivity to low mass DM to the sensitivity threshold near the neutrino wall. For high mass DM, the extension of the LUX detector (LZ and Darwin)\cite{Schumann:2015cpa,Akerib:2015cja}, and future extensions of liquid argon detectors  can potentially cross the neutrino wall~\cite{Aalseth:2015mba}, see Fig.~\ref{fig:DDBounds}. The same liquid noble gas detectors are sensitive to spin dependent DM~\cite{Aprile:2013doa,Akerib:2016lao}. Thus, allowing for the the extension of DM searches to the neutrino wall for all DM masses with both spin independent and spin dependent DM. 

\subsubsection{Indirect Detection}
Indirect detectors consist of space- and ground-based telescopes, which look for the products of DM induced interactions in and beyond the galaxy. Essentially, these searches consist in looking for the presence, on top of the ordinary cosmic rays, of possible anomalous fluxes of high energy photons, positrons, anti-protons and neutrinos which could be attributed to DM annihilations or decays. Further discrimination can be done by directional searches; explicitly searching for particles from dwarf galaxies or the center of the galaxy. Currently, the experiments can be divided into two sets of experiments:  particle based detectors, such as AMS, and photon based detectors, such as the FermiLAT satellite. In both cases, the quoted bound is on the interaction rate of particles at a given energy ($dN/dE$). For the case of photons, this can be written as 
\begin{eqnarray}
\frac{dN}{dE} & = & \frac{1}{4\pi}\frac{\langle\sigma v\rangle}{2m^{2}_{DM}}F(q^{2}) \rho_{DM}^{\prime} \,,\\
\rho^{\prime}_{DM}  & = & \int_{\Delta\Omega} d\Omega \int_{los} \rho_{DM}^{2}(r(l,\phi)) dl(r,\phi)\,.
\end{eqnarray}
Here $\rho^{\prime}_{DM}$ is the integral over the line of sight ($los$) of the square of the DM density $\rho_{DM}$, $F(q^{2})$ is the resulting fragmented particle distribution considering the initial particle produced, and $m_{DM}$ is the DM mass.  As with direct detection, the DM mass profile $\rho_{DM}$ is a necessary input into the calculation. These measurements also suffer from large uncertainties since the rates depend quadratically on $\rho_{\rm DM}$ and the integral often runs over regions where the density is poorly constrained. 

%However, for these measurements, the DM density variability is potentially reduced when compared with the local dark matter density. This is a result of the %larger region of space that is probed and the use of larger DM structures, such as dwarf galaxies. \\ 
%Fermi/HESS
Photon bounds coming from annihilating DM interactions consist of two classes of searches: continuum photon excess searches, and direct photon line searches. Continuum photon searches consist of searches of a broad excesses of photons over the predicted photon background. These searches have relatively large uncertainties since they require a precise knowledge of the photon background. When DM annihilates to a final state that is direct photons, photon line searches can be performed. These searches can exclude much smaller production cross sections since they consist of a classic bump hunt on top of the photon continuum background ~\cite{Fan:2013faa,Cohen:2013ama}. In both cases, the current results are driven by two experiments: FermiLAT, a low energy gamma ray satellite, and HESS, a high energy gamma ray telescope. The FermiLAT satellite dominates bounds for photon energies up to 1 TeV and the HESS telescope array is the dominant bound for energies above 1\UTeV. 

Figure~\ref{fig:IDBounds} shows the current bounds for the experiments. Both of the search regions are expected to improve with the upcoming launch of DAMPE~\cite{ChangJin:550} and Gamma-400\cite{Dokuchaev:2015ghx}. Further improvements at high energy will come with the Cherenkov Telescope Array (CTA)~\cite{cta}. These further extensions are shown in figure~\ref{fig:IDBounds}. The goal is to cover the model independent calculation of the relic density shown in figure~\ref{fig:IDBounds}. This allows for an exclusion/discovery benchmark of a large class of models. The relic density line can be avoided through models with p-wave annihilation or co-annihilation. In some models, direct photon line searches are more sensitive than broad spectrum searches. In figure~\ref{fig:IDBounds}, we also show the direct photon bounds searches and the extrapolated improvements~\cite{Abramowski:2013ax,Ackermann:2015lka} given a consistent level of improvement as that projected with the future continuum searches.  \\
%AMS anti proton 
The AMS anti-proton results are also shown~\ref{fig:IDBounds}, along with the band of variations coming from different astrophysical hypotheses. The AMS anti-proton results are already comparable to the existing FermiLAT bounds, and are expected to improve further. It is expected that AMS will continue to run for another 10 years. There are currently no projected upgrades of the AMS detector. 

\begin{center}
\begin{figure}[h]
\includegraphics[width=0.49\textwidth]{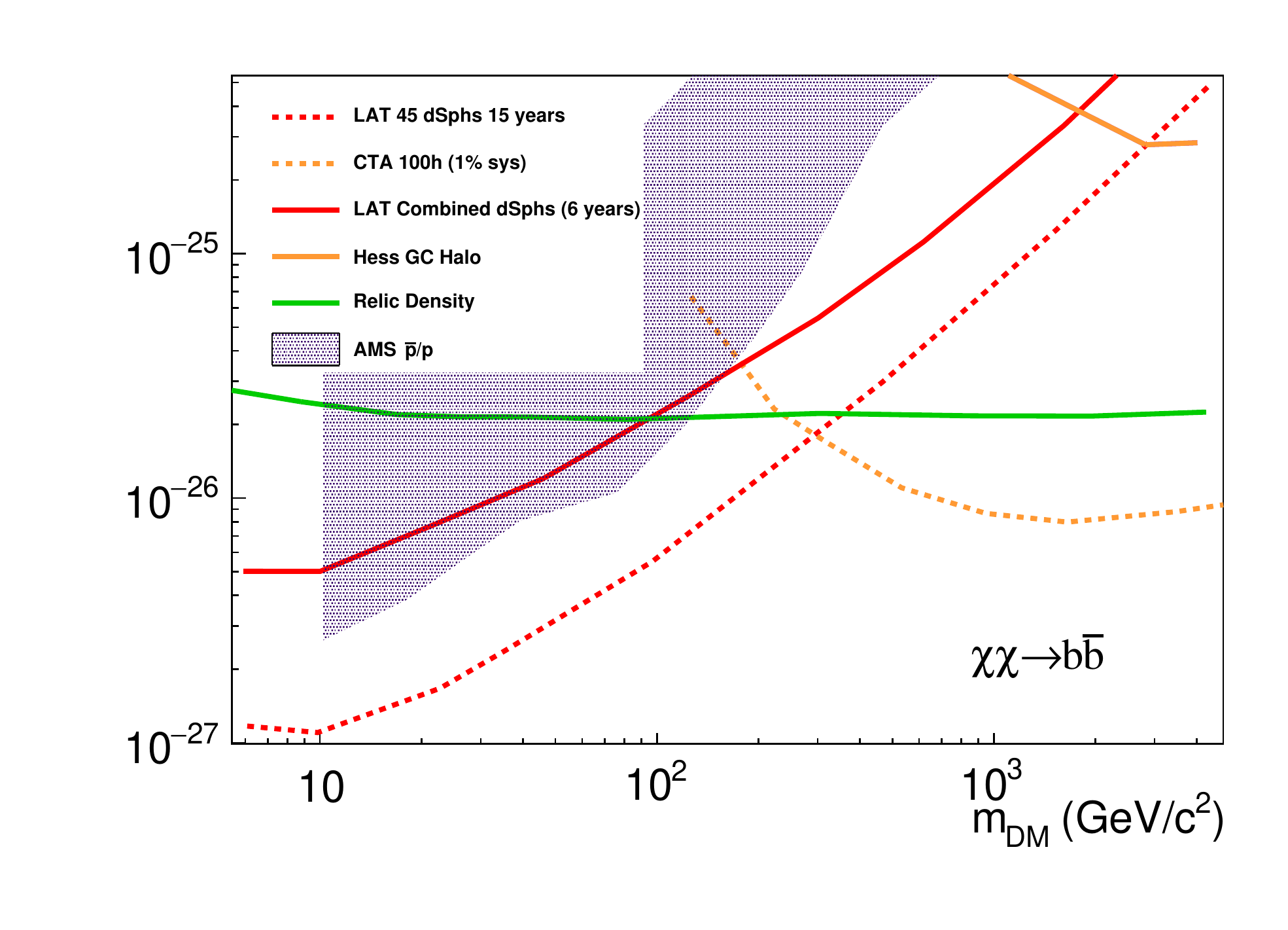} 
\includegraphics[width=0.49\textwidth]{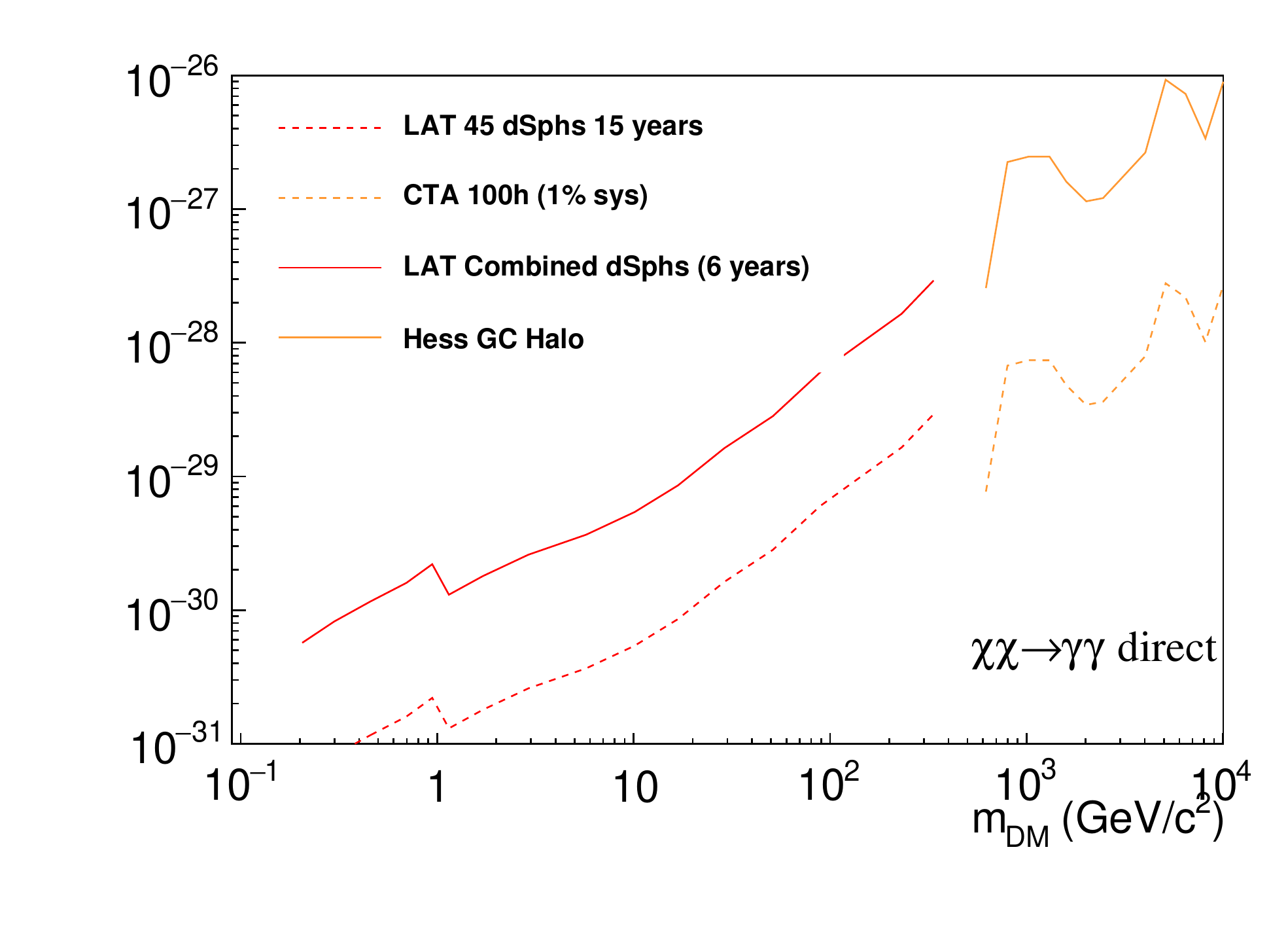} 
\caption{Indirect bounds coming from current AMS~\cite{Giesen:2015ufa,AMSTalk}, FermiLAT~\cite{Drlica-Wagner:2015xua}, and HESS~\cite{HESS:2015cda} results. Additionally, projected bounds based on 15 life years of FermiLAT and the CTA\cite{cta} experiment. Bounds are shown for $b\bar{b}$ final state for the continuum search(left) and for the direct photon search(right). 
}
\label{fig:IDBounds}
\end{figure}
\end{center}

\subsubsection{Relic Density}
The current measured relic density from cosmic microwave background(CMB) is  $\Omega_{DM}h^{2}=0.1198\pm0.0026$~\cite{Ade:2015xua}. This sets a benchmark for which models and constraints for DM can be compared. %Additionally, the distribution of the moments on the CMB itself can be used to a put a generic bound on DM. 
The annihilation cross section corresponding to the observed relic density is show in figure~\ref{fig:IDBounds}. For light DM the required values can be excluded by indirect detection for a number of relevant annihilation channels. 

For a large class of models, relic density constraints have the ability bound the allowed space of dark matter models. To illustrate the impact of the relic density, we consider the relic density bound for 4 types of mediators, a scalar, a pseudoscalar, a vector mediator and an axial mediator. These mediators are further discussed in the section on simplified models~\ref{sec:sms}. For scalar and pseudo-scalar mediators the couplings to quarks are taken to be proportional to the corresponding Higgs Yukawa couplings, $y_q$ as in models with minimal flavour violation \cite{D'Ambrosio:2002ex,Abdallah:2015ter}, and vector/axial mediators we take flavor universal couplings to all quarks. For simplicity's sake, we take the couplings to the quarks $g_{q}$ and the DM particles $g_{DM}$ to be unity ($g_{q}=g_{DM}=1$). This assumption is a bit naive given that for vector and axial mediators, the couplings are on the threshold of being physical. However, the large coupling also opens the allowed region of phase space of the DM models. In the sense that larger coupling means larger DM annihilation cross section, which means smaller DM density at the time of freeze out. This allows us to quote the resulting upper bounds on the DM production as conservative upper bounds. 

Figure~\ref{fig:relicdmbounds}, shows the DM bounds for the four sets of mediators computed using the MadDM~\cite{Backovic:2013dpa} program. The reach of the vector mediator is roughly 80~TeV, axial mediator is 8~TeV, scalar mediator is 6~TeV and pseudoscalar mediator is 40~TeV. The bounds all have a similar feature in that the reach in mediator is strongest for DM masses which are close to half the mediator mass (the resonant regime). These bounds can further be modified by the presence of additional particles that couple to the mediator. 

\begin{figure}
\begin{center}
\includegraphics[width=0.49\textwidth]{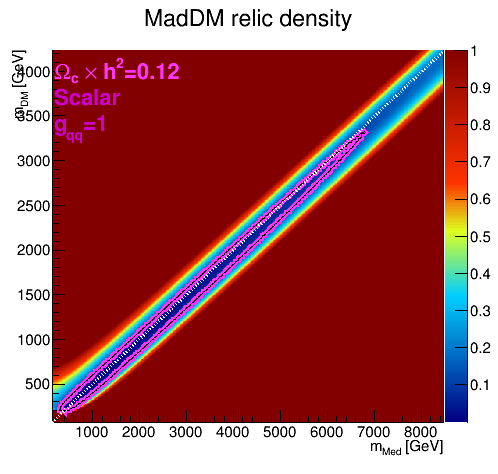} 
\includegraphics[width=0.49\textwidth]{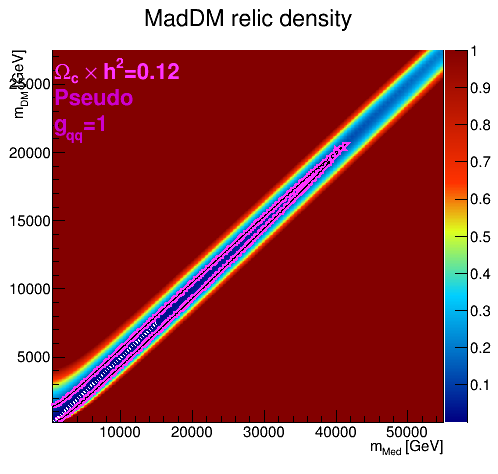} \\
\includegraphics[width=0.49\textwidth]{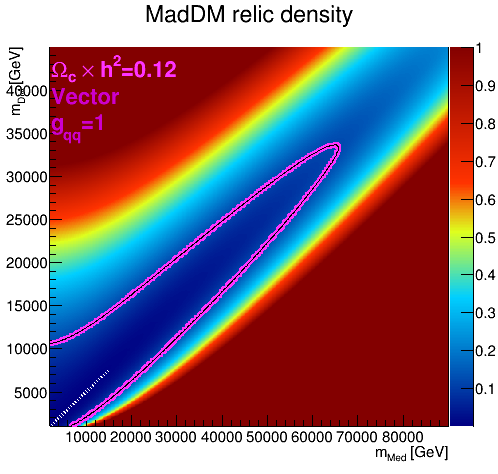} 
\includegraphics[width=0.49\textwidth]{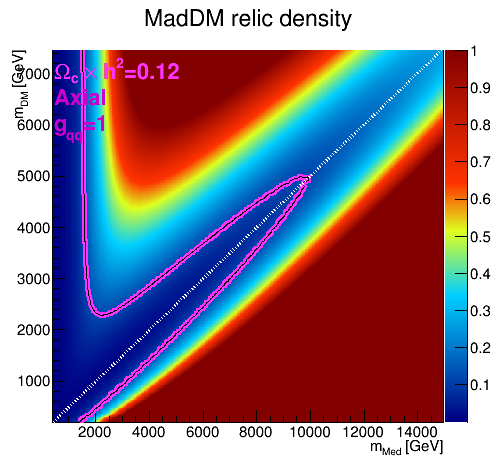} 
\caption{ 
  Relic density bound for the simplified models using a Scalar mediator (top-left), Pseudosalar mediator (top-right), Vector mediator (bottom-left), and Axial mediator (bottom-right).   
}
\label{fig:relicdmbounds}
\end{center}
\end{figure}

\subsubsection{Collider Production} 
In many models DM production at colliders proceeds via an additional particle, the so called mediator, which couples to both SM states and DM. 
This can be an s-channel mediator such as a new scalar particle or a vector boson, or a t-channel mediator such as a squark. 
Thus, excluding a Z or Higgs mediator, the search for DM at colliders consists of looking for the DM itself and at least one additional particle. 

% Direct Dark matter searches
When DM is produced directly, it will not interact with the detector and thus leaves a missing energy signature. The class of direct searches consist of missing energy signature along with an additional signature. In the case of proton-proton collisions, missing energy results as missing transverse energy due to the lack of conservation of the momentum in the collision axis. The most generic of these searches is the monojet DM search. The search consists of the selection of one or more jets and missing transverse energy recoiling against the jets. Additional missing energy searches consist of replacing the jet with another signature, such as as vector boson, photon or the Higgs boson (so called MET+X final state). 

For every MET+X collider search, the dominant background comes from $Z$+X, where the $Z$ boson decays to neutrinos $Z\rightarrow\nu\nu$. The current modelling uncertainty of the $Z$+X production is theoretically limited by the order of calculation precision for most regions of phase space. However, this can be overcome by modelling the Z$\rightarrow\nu\nu$ production through a combination of control regions  where no DM is present. The most advanced approach involves a simultaneous fit of $Z\rightarrow\ell\ell$, $\gamma+X$ control regions, with the theoretical predictions for the $Z+X/\gamma+X$ production ratio as an additional constraint. The full fit has been shown to model the distribution of $Z$+X differential production down to the percent level. This sets a benchmark for the level of precision considered in the rest of the document. It is likely that further development of these approaches in the ensuing years will allow for the preservation of such high precision out to high energies. This ensures that the DM searches will remain statistically limited.     %Note this is a plug for PH's work 

%Additional tagging of Direct searches (displaced etc....)
MET+X searches can be greatly enhanced by additional signatures that may occur in specific scenarios. For highly degenerate particles (in the co-annihilation regime), one can have long lived charged particles that decay into DM. This gives the classic MET+X signature with an additional signature: such as a short track resulting from the charged particle before it decayed. These additional tags have the ability to greatly reduce the background and further enhance the sensitivity of collider based searches.  

% Indirect DM searches
Since DM at the LHC involves additional non-Standard Model particles, one can indirectly search for DM by observing these additional new particles. For example, vector mediators will decay to quarks. Thus, one can search for vector mediators directly by looking for resonant di-jet production~\cite{Chala:2015ama}. While indirect searches implicitly require that all final states be probed at the LHC, a few final states stand out as particularly complementary. These include the di-jet resonant search and resonant diphoton searches.   

%Why people should stop complaining about triggers
Given that there is roughly 20-30 years of development before the 100 TeV collisions, it is likely that some of the current detector complications will be resolved. In particular, the triggering of events is expected to improve with time. To illustrate the expected level of improvement, consider the current MET triggers at the LHC. They are currently able to trigger MET with a threshold above 200~\UGeV. It has been predicted that at high luminosity running, MET triggers will become ineffective due the large amount additional collisions (pileup) that degrade the overall event resolution. However, this prediction will very likely become invalid due to new developments in track triggering~\cite{CERN-LHCC-2015-020,Butler:2020886} and advances in understanding the MET in dense environments~\cite{Bertolini:2014bba}. In this respect, the current LHC benchmarks for future sensitivity are  likely to be over conservative with respect to future developments.

\subsubsection{Current DM Related Excesses}
At the moment there are a few hints for new physics from astrophysics and collider experiments. First, the Fermi collaboration has confirmed an excess in gamma rays from the galactic center~\cite{TheFermi-LAT:2015kwa} which, as shown in several previous studies~\cite{Hooper:2010mq,Gordon:2013vta,Daylan:2014rsa,Calore:2014xka}, can be consistent with DM annihilating in the galactic center. In particular WIMP DM annihilating to massive SM particles in the mass range between 35\UGeV~and 310\UGeV~can successfully fit the excess~\cite{Agrawal:2014oha}. Since indirect observations of DM often remain inconclusive, a verification of the DM origin of this signal at a collider is desirable. The models discussed in~\cite{Agrawal:2014oha} are similar to the benchmark models that are considered in the following sections, and the preferred mass range should be testable at a collider. Currently, strong evidence exists already that the pseudoscalar interpretation is ruled out by the LHC~\cite{CMS-EXO-12-055}. Further interpretations are likely to be tested with a 100\UTeV~collider, if not already at the CERN-LHC. 

Furthermore the LHC experiments have recently reported an excess of events in the diphoton channel near 750\UGeV, consistent with a new resonance~\cite{ATLAS-CONF-2015-081,CMS-PAS-EXO-15-004}. The possibility that this resonance provides a window to DM was discussed for example in~\cite{Backovic:2015fnp,Knapen:2015dap,Han:2015cty,Bi:2015uqd,Bhattacharya:2016lyg,D'Eramo:2016mgv,Mambrini:2015wyu,Franceschini:2015kwy}. Within half a year from now the LHC should confirm or rule out the presence of this resonance, however it will be important to find out whether it will be sufficient to determine a connection to the dark sector, or whether a 100\UTeV~collider is necessary. 

Other long standing, potentially DM related excesses are the cosmic positron excess~\cite{Adriani:2008zr,Accardo:2014lma} and the annual modulation signal observed by the DAMA collaboration~\cite{Bernabei:2010mq}. The former suffers from the usual problem that an astrophysical origin of the signal is difficult to exclude, while the latter has not been confirmed by any other DM direct detection experiment so far, and is more and more in tension with constraints from these direct searches.

%\subsection{WIMP Dark Matter, Standard Model Mediators} 
\subsection{WIMP Dark Matter, Standard Model Mediators} 

DM that interacts with known particles through Standard Model mediators is the simplest and most minimal implementation of the WIMP scenario. Since the DM candidate has to be neutral and uncoloured, the most compact models introduce a single multiplet of the electroweak $SU(2)\times U(1)$ which should contain at least one neutral state. The smallest nontrivial, viable $SU(2)$ representations are a doublet with hypercharge $1/2$, a triplet and a fiveplet~\cite{Cirelli:2005uq}. These models introduce only one new parameter, the mass of the multiplet, such that their parameter space can be fully explored, as discussed in Secs.~\ref{sec:wgb1} - \ref{sec:wgb3}. Generically, for the correct abundance of thermal DM the mass of these WIMPs should be at the TeV scale~\cite{Cirelli:2005uq}. 

In principle models with more than one multiplet are also conceivable and motivated as low energy limits of more complicated BSM scenarios like the MSSM. The simplest such model consists of a $SU(2)$ singlet and a doublet~\cite{Mahbubani:2005pt,Cohen:2011ec,Joglekar:2012vc,Cheung:2013dua,Calibbi:2015nha,Cynolter:2015sua}, with other combinations also possible~\cite{Tait:2016qbg}. In Sec.~\ref{sec:wgb4} a model with a singlet, doublet and triplet is considered instead since this particle content is motivated by the chargino/neutralino sector of the MSSM. 

Maybe the simplest model in terms of particle content is that of a scalar $SU(2)$ singlet with a $Z_2$ parity symmetry, which couples to the SM only through a renormalizable coupling with the Higgs boson. The prospects for probing this model at a 100~TeV collider are discussed in Sec.~\ref{sec:DM_higgsportal}.

%\subsubsection{Weak Gauge Bosons}

%\documentclass[12pt]{article}
%\usepackage{amsmath,amssymb,float,fullpage,graphicx,hyperref,slashed}

%%%%%%%%%%%%%%%%%%%%%%%%%%%%%%%%%%%%%%%%%%%%%%%%%%%%%%%%%%%%%%%%%%%%%%%%
%%%%%%%%%%%%%%%%%%%%%%%%%%%%%%%%%%%%%%%%%%%%%%%%%%%%%%%%%%%%%%%%%%%%%%%%
%%\begin{document}
%
%\begin{center}
%  {{\LARGE\bfseries Dark Matter at the FCC-hh}\\\bigskip}
%  {{\large Matthew Low and Lian-Tao Wang}\\\bigskip}
%\end{center}

% *** GUIDELINES ***
%
%  1. What are cases where a 100 TeV collider makes a qualitative difference (rather than just improve limits by some factor)
%  2. Is there complementarity between a 100 TeV collider and other machines (ILC/TLEP/CEPC) or other experiments (direct/indirect detection)
%  3. What are possible benchmark scenarios/points which should be studied more carefully
%  4. Is the reach of 14 TeV LHC known, so that the gain at 100 TeV can be put in perpective

%%%%%%%%%%%%%%%%%%%%%%%%%%%%%%%%%%%%%%%%%%%%%%%%%%%%%%%%%%%%%%%%%%%%%%%%
\subsubsection{Weak Gauge Bosons 1: Wino, Higgsino DM }
\label{sec:wgb1}
%
% Note by Pedro: Merged into introduction of section
%
%Dark matter that interacts with known particles through standard model mediators is the simplest and most minimal implementation of the WIMP scenario.  In this case, the only new particles added are SU(2)$_L$ multiplets which interact with the standard model through their couplings to $W$'s and $Z$'s.  Here we only consider fermions, but scalars are possible too and favor different regions of parameter space~\cite{Cirelli:2005uq}.  Generically, for the correct abundance of thermal DM the mass of these WIMPs should be at the TeV scale~\cite{Cirelli:2005uq}.

The smallest multiplet is an SU(2)$_L$ doublet (also called the higgsino in the context of supersymmetry).  To have an electrically neutral state requires two doublets with hypercharges $Y=\pm 1/2$, thus we have two neutral Majorana states $\chi^0_1$, $\chi^0_2$, and one charged Dirac state $\chi^\pm$.\footnote{We assume the presence of additional operators to slightly split the neutral masses otherwise they would combine into a Dirac fermion which would be ruled out by direct detection.}  The various states are nearly mass degenerate with a small splitting arising from electroweak symmetry breaking effects.  In the high mass limit the charged fermions are heavier by $\Delta m \simeq$ 355 MeV~\cite{Thomas:1998wy}.

The neutral and charged states interact in pairs with the Standard Model via $W$'s and $Z$'s resulting in the interactions $\chi^0_1 \chi^0_2 Z$, $\chi^0_{1,2} \chi^\pm W^\mp$, $\chi^\pm \chi^\mp Z$, and $\chi^\pm \chi^\mp \gamma$.  At a hadron collider they are pair produced via Drell-Yan resulting in final states of pairs of invisible particles.  Even when a charged $\chi^\pm$ is produced the signal still looks like two invisible particles because the charged $\chi^\pm$ decays via $\chi^\pm \to \chi^0_1 + \pi^\pm$ and the $\pi^\pm$ has momentum $\sim \Delta m$ and is thus often undetectable.
%
%%%%%%%%%%%%%%
\begin{figure} [h]
\begin{center}
\includegraphics[width=0.42\textwidth]{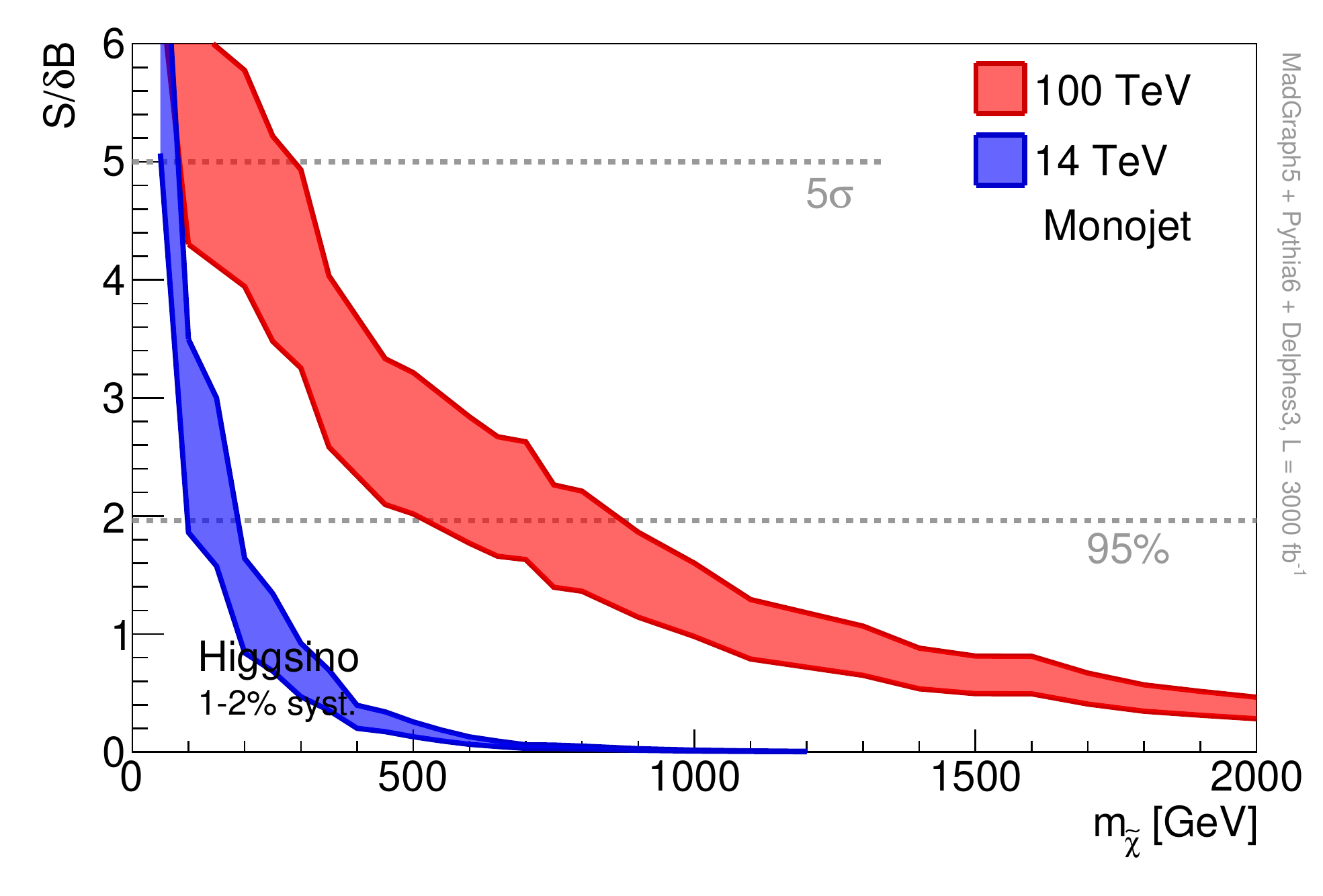} \quad\quad
\includegraphics[width=0.45\textwidth]{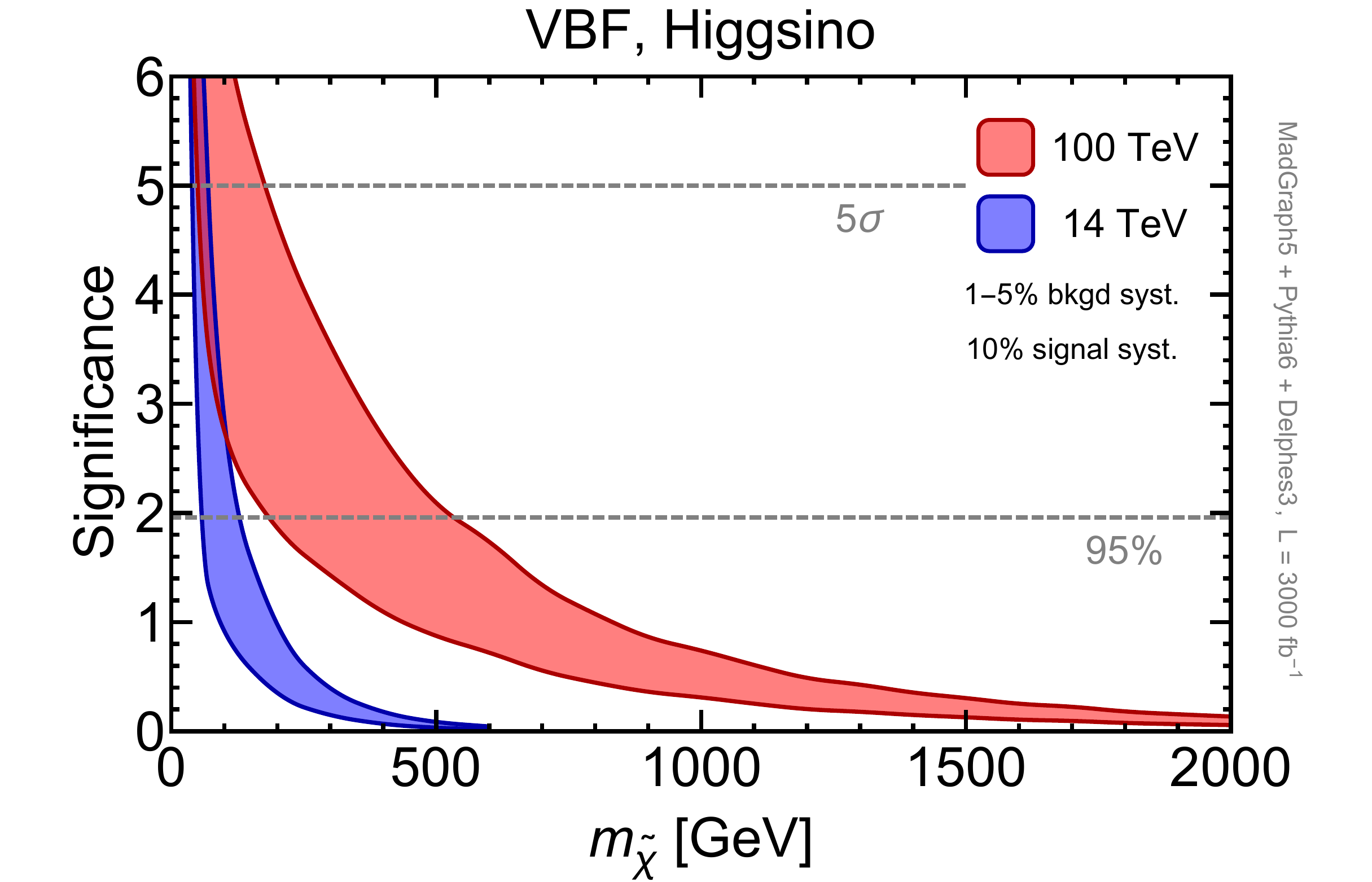}
\caption{Reach for higgsinos (SU(2)$_L$ doublets) in the monojet channel (left) and in the vector boson fusion channel (right).}
\label{fig:higgsino}
\end{center}
\end{figure}
%%%%%%%%%%%%%%

At a hadron collider, one needs additional objects in the event other than missing energy.  There are several possibilities: the initial state radiation of a jet (or a gauge boson), production in vector boson fusion, or tagging on the soft Standard Model objects in the final state.  Requiring the initial state radiation of a hard jet is called the monojet channel and looks for a high $p_T$ jet and large missing energy.  This scenario was studied in~\cite{Low:2014cba} and found to have a mass reach from 550 GeV to 850 GeV depending on the level of systematic uncertainty assumed, as shown in Fig.~\ref{fig:higgsino} (left).  Recasts of 8 TeV monojet searches have been performed and show that the mass reach at 8 TeV is less than 100 GeV~\cite{Schwaller:2013baa}.  In the vector boson fusion channel, one looks for two forward jets and missing energy.  This process typically has a lower rate than the monojet channel but one may have smaller backgrounds so it is not obvious apriori how the reach will compare to monojet.  This was studied in~\cite{Berlin:2015aba} and was found to have a mass reach of 150 GeV to 500 GeV, also shown in Fig.~\ref{fig:higgsino} (right).

The next case is an SU(2)$_L$ triplet with $Y=0$ (also called the wino in the context of supersymmetry).  Now there is one neutral state $\chi^0$ and one charged state $\chi^\pm$ with a mass splitting of $\Delta m \simeq$ 166 MeV~\cite{Ibe:2012sx}.  Both the monojet search and vector boson fusion searches can be performed and the mass reach is 0.9 TeV to 1.4 TeV, shown in Fig.~\ref{fig:wino}.  Again, the monojet channel is more sensitive than vector boson fusion. 

%%%%%%%%%%%%%%
\begin{figure} [h]
\begin{center}
\includegraphics[width=0.42\textwidth]{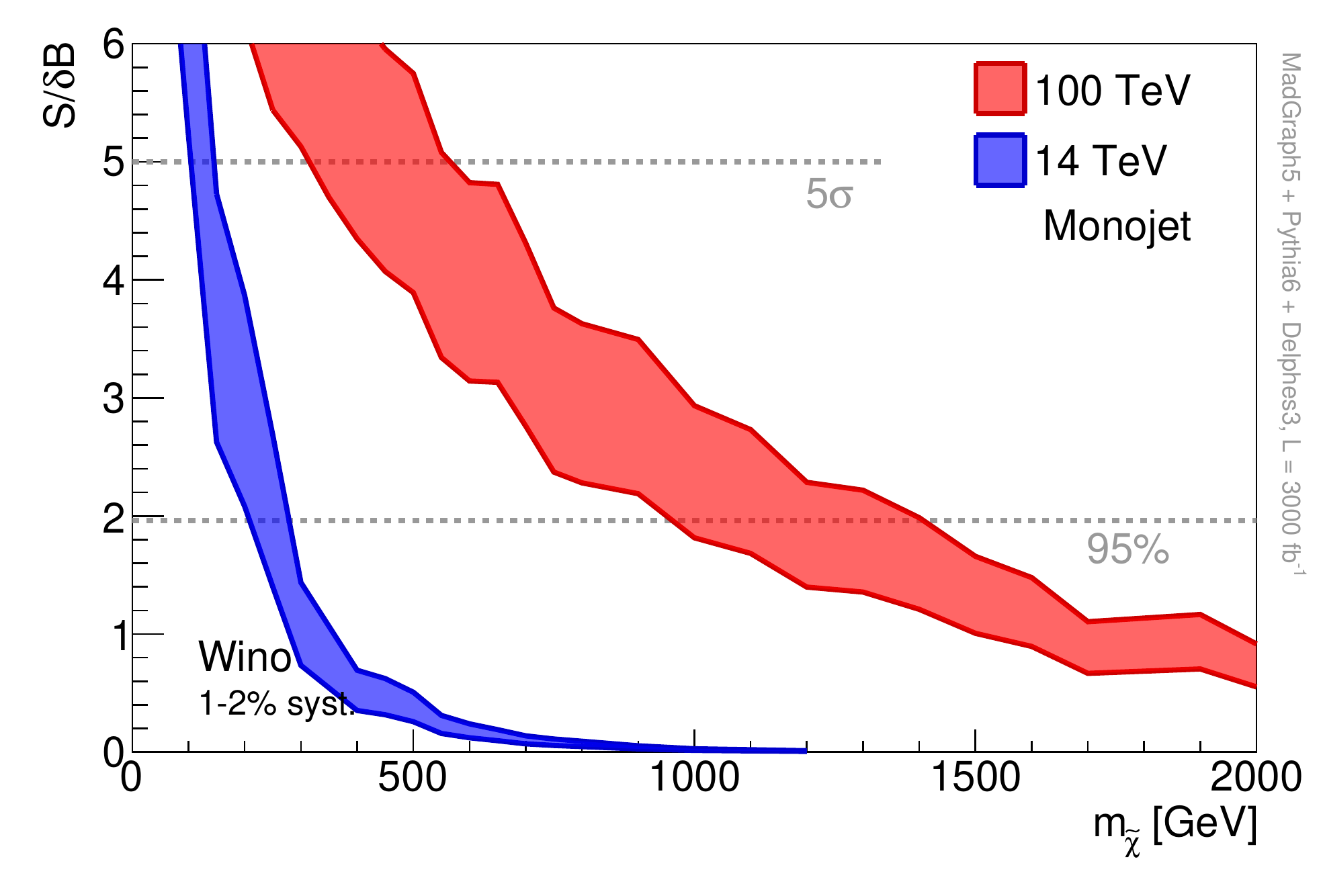} \quad\quad
\includegraphics[width=0.45\textwidth]{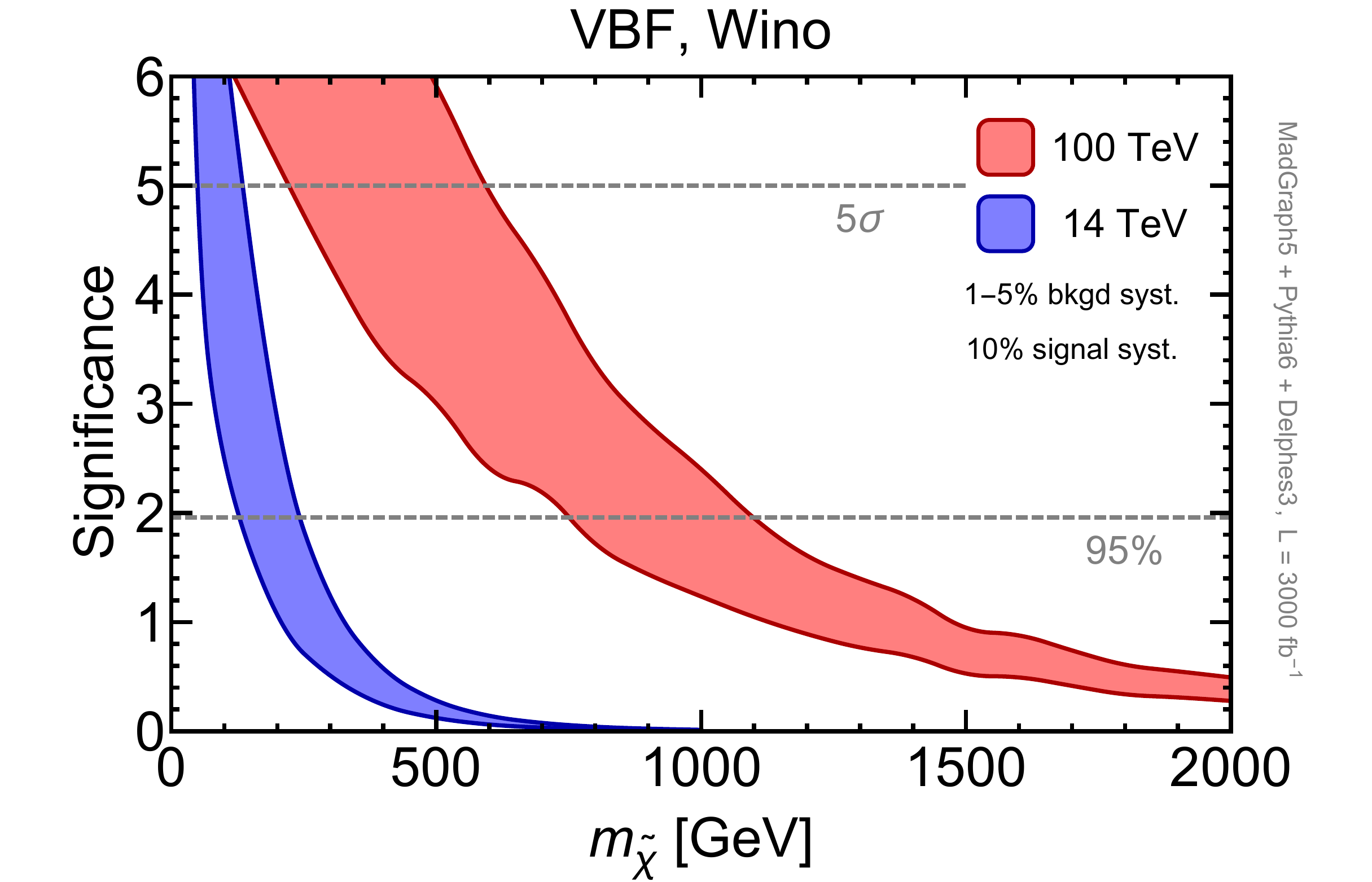}
\caption{Reach for winos (SU(2)$_L$ doublets) in the monojet channel (left) and in the vector boson fusion channel (right).}
\label{fig:wino}
\end{center}
\end{figure}
%%%%%%%%%%%%%%

An additional search that can be effectively utilized for the triplet case is the disappearing tracks search where one looks for a track from the charged state that suddenly disappears when it decays into the neutral state and a soft pion.  The triplet mass splitting of 166 MeV results in a lifetime of the $\chi^\pm$ of $c\tau \sim$ 6 cm which is long enough that some of the $\chi^\pm$'s will decay in the region where the detector is likely to have a tracker.  There are no physics backgrounds to this search, but there are a number of backgrounds arising from detector effects.  At the LHC, this is the most sensitive search for the pure wino and already sets limits of 270 GeV with 20 fb$^{-1}$~\cite{Aad:2013yna,CMS:2014gxa}.  One can extrapolate the LHC backgrounds to a 100 TeV collider, though it is important to keep in mind that the estimate is rough as the backgrounds could be much different at a future detector.  The extrapolation was performed in~\cite{Low:2014cba} and found to have a reach of 2.9 TeV.  This is shown in Fig.~\ref{fig:tracks} (left) with bands varying the background normalization up and down by a factor of 5.  See also~\cite{Cirelli:2014dsa} for similar studies on triplets.  Due to the shorter lifetime, this is typically not a useful channel for the doublet unless there is UV physics that decreases the splitting between the charged and neutral states, as shown in Fig.~\ref{fig:tracks} (right).

%%%%%%%%%%%%%%
\begin{figure} [h]
\begin{center}
\includegraphics[width=0.42\textwidth]{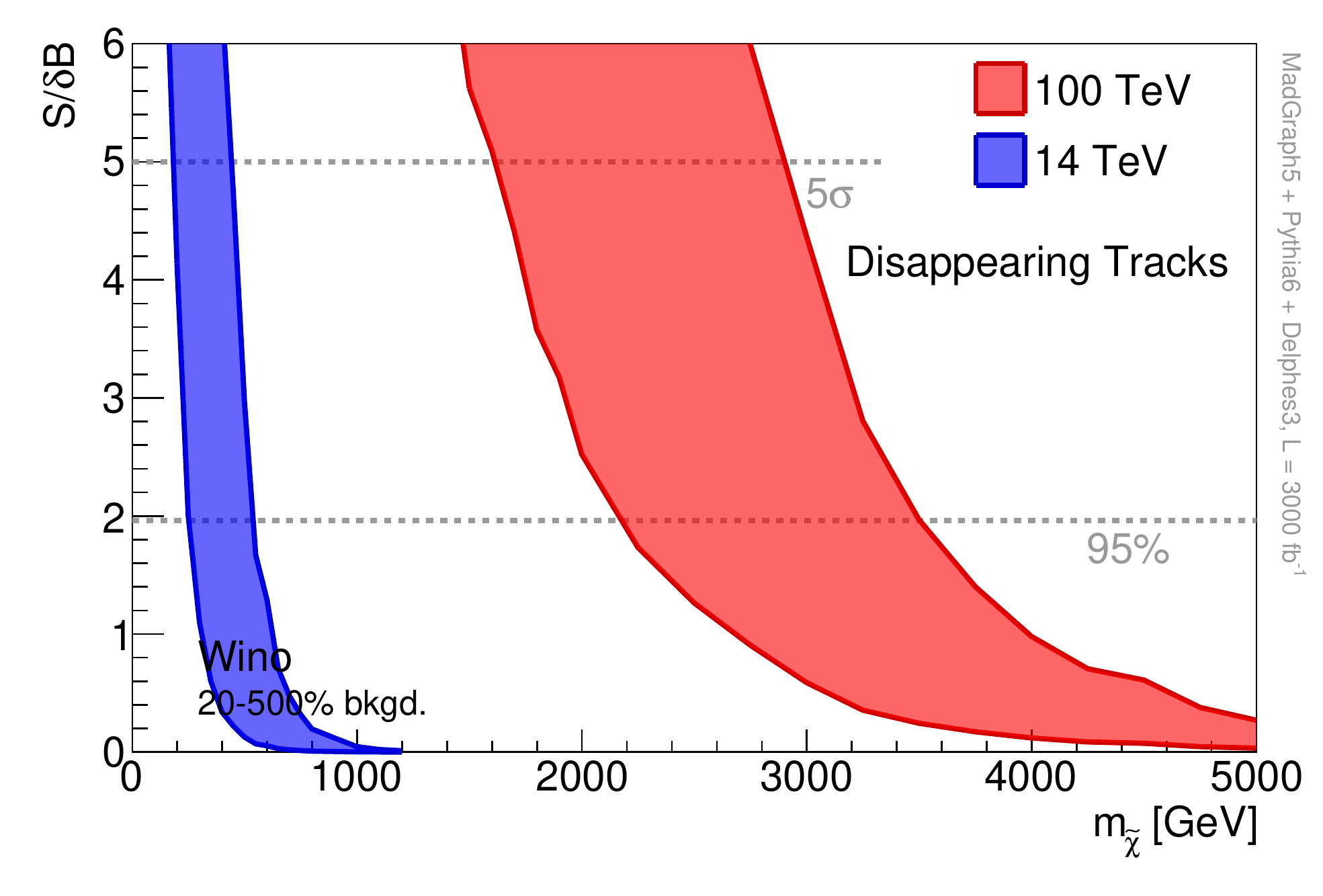} \quad\quad
\includegraphics[width=0.42\textwidth]{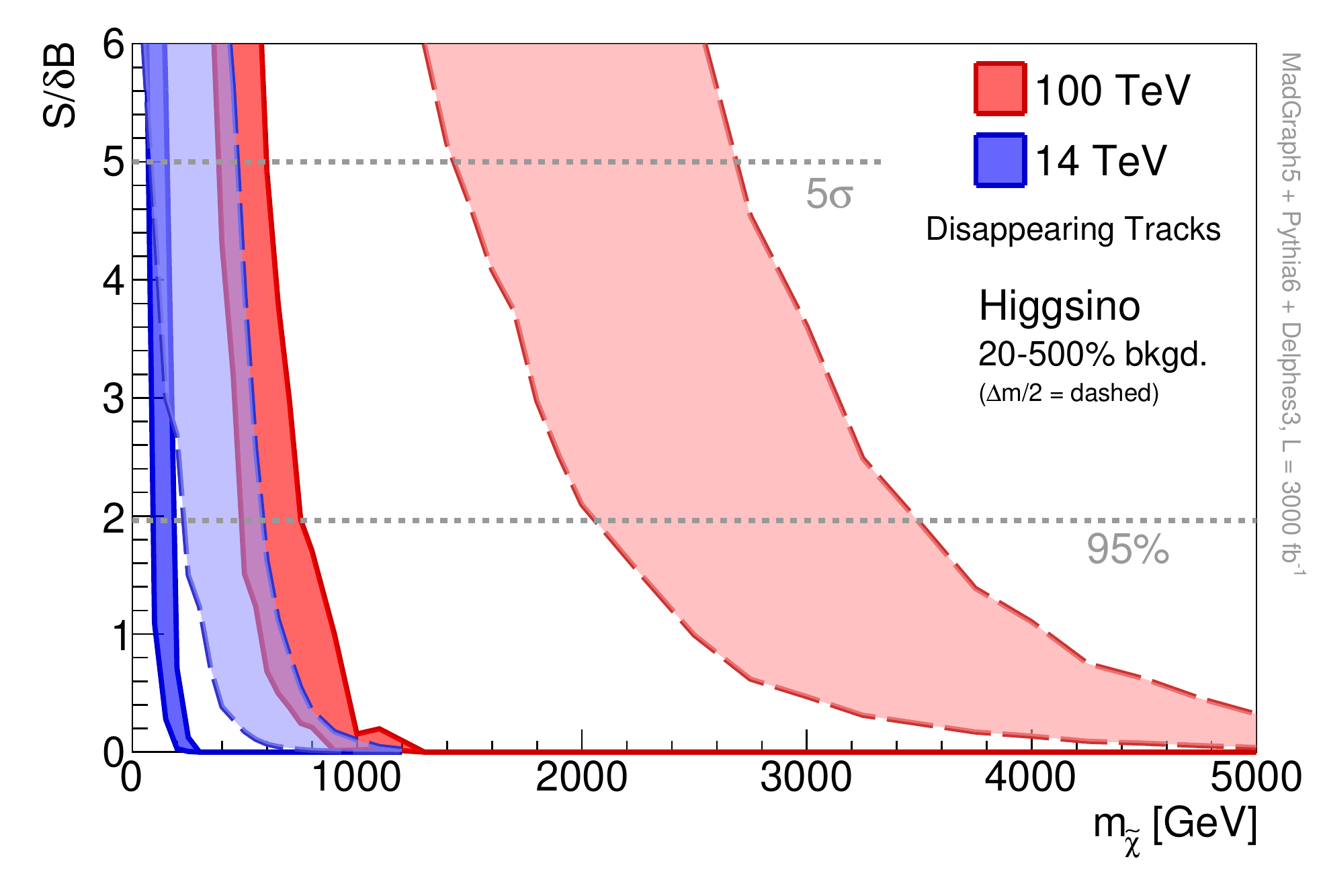}
\caption{Reach for winos (SU(2)$_L$ triplets) in the disappearing tracks channel (left) and for higgsinos in the disappearing tracks channel (right).}
\label{fig:tracks}
\end{center}
\end{figure}
%%%%%%%%%%%%%%

For both the monojet and vector boson fusion searches the projections were also computed for the 14 TeV LHC and it was found that a 100 TeV collider improves by about a factor of 5 (using an integrated luminosity of 3 ab$^{-1}$ at both 14 TeV and 100 TeV).  Performing these searches at 100 TeV would not be qualitatively different than at the LHC.  One search that offers the chance for a qualitative improvement is the disappearing tracks search, because this is very dependent on the yet-to-be-designed detector properties.  Given the high sensitivity of this channel one can envisage designing the detector to optimize this channel.  Concretely, current searches require a hard recoiling jet in disappearing track events to trigger on, but if one could trigger on the disappearing track itself, the rates would be increased.  Additionally, maintaining a high efficiency to select disappearing tracks is crucial.

Note that the study has been performed for $\sqrt{s}=$ 100 TeV.  To properly evaluate the mass reach at other collision energies would require dedicated studies, but simple estimates can be made.  By comparing parton luminosities one can see that the mass reach is linear (in center of mass energy) when the luminosity increases quadratically.  For a fixed luminosity, the mass reach increase is more mild and in some cases seen to be closer to 40\% for a factor of 2 in energy~\cite{Kotwal:2015rba}.

For the pure states described here there is an interesting complementarity with both direct detection and indirect detection.  Let us first consider the higgsino.  In order to have the correct thermal relic abundance, the higgsino mass should be 1 TeV.  In direct detection the rate vanishes at tree level because the coupling of neutralinos to Higgses arise from the mixing between higgsinos and binos or winos.  At one loop the rate is still suppressed due to an accidental cancellation~\cite{Hill:2013hoa}.  As shown above unless colliders can achieve systematic uncertainties below 1\% one is unable to exclude (not even discover) higgsinos.  
For winos, 3 TeV is the mass to satisfy the thermal relic abundance.  In direct detection the rates are again very small, but lie just above the neutrino coherent scattering rate, at $\sim$ 1.3 $\times$ $10^{-47}$ cm$^2$~\cite{Hill:2013hoa}, as discussed in more detail in the next section. 

%The complementarity of direct and indirect detection and collider searches for the wino scenario is discussed in more detail in the next section, while the case of DM which is a mixture of singlets, doublets and triplets is analysed in Sec.~\ref{sec:wgb4}, 
%

One can also consider DM state that are mixtures of binos, higgsinos, and winos.  The collider signatures depend strongly on the mass difference between the lightest neutralino and the other states.  For mass splittings of $\sim 20-50$ GeV, the reach is studied in~\cite{Low:2014cba} while for mass splittings of $\gtrsim$ 100 GeV the reach is studied in~\cite{Gori:2014oua}. This scenario is studied in more detail in Sec.~\ref{sec:wgb4}. A summary of the constraints on the pure, mixed and co-annihilating scenarios (c.f. Sec.~\ref{sec:coann}) is given in Fig.~\ref{fig:DM_wgb1_summary}.

These scenarios represent the worst possible cases in the sense that there are very few handles in the events.  Future directions that deserve more careful study are considering other particles in the spectrum that could increase the electroweakino rate or yield jets or leptons in their decays providing increased discrimination power.

%
%Refs.~\cite{Bramante:2014tba,Bramante:2015una} cover the complementarity between direct detection, indirect detection, and collider searches for thermal DM in much greater detail.  They argue that in the parameter space where the mass parameters of the singlet, doublet, and triplet are all below 4 TeV one can exclude the full parameter space by considering collider searches, direct detection, and indirect detection all together.  

%%%%%%%%%%%%%%
\begin{figure} [h]
\begin{center}
\includegraphics[width=0.65\textwidth]{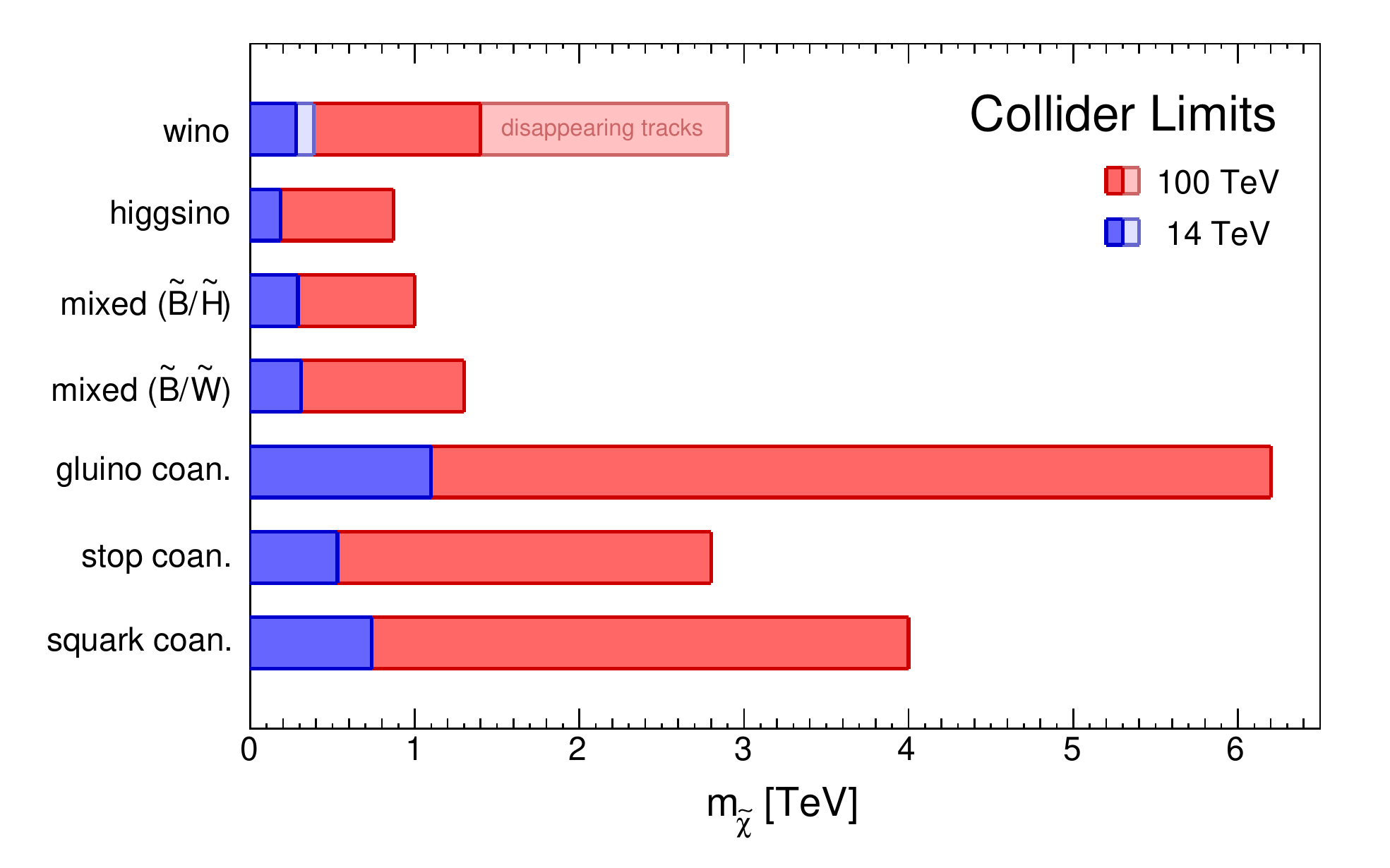}
\caption{Summary of reach for DM with SM mediators and through co-annihilation at 100 TeV.}
\label{fig:DM_wgb1_summary}
\end{center}
\end{figure}
\subsubsection{Weak Gauge Bosons 2: Wino DM }
\label{sec:wgb2}

As discussed above, an electroweak triplet with zero hypercharge is one of the most minimal DM models one can imagine~ \cite{Cirelli:2005uq,Cirelli:2009uv}, and is further motivated in models of high-scale supersymmetry~\cite{Wells:2003tf,ArkaniHamed:2004fb,Giudice:2004tc,Arvanitaki:2012ps,Hall:2011jd,Hall:2012zp,Hall:2013eko,Hall:2014vga} and other new physics scenarios~\cite{Chao:2012mx,Giudice:2004tc, Frigerio:2009wf,Farina:2013mla}.

We now summarise the status and prospects for the searches of an extra stable fermion triplet, focusing of course on the 100 TeV proton collider, but making explicit the comparison with other future colliders, as well as with direct and indirect DM detection experiments. Our discussion is based on ref~\cite{Cirelli:2014dsa} for the collider reaches, and it is updated with more recent results for DD \cite{Hisano:2015rsa}, as well as with preliminary ones for ID \cite{Cirelli:2016boh}.
\textbf{The model.}
The Lagrangian for the minimal Wino DM model reads
\begin{equation}
\mathcal{L} = \mathcal{L}_{\rm SM} + \frac{1}{2} \bar{\chi}(i \slashed{D} - M_\chi)\chi,
\label{Lagrangian}
\end{equation}
so that the only new parameter of this model is the $\chi$ mass $M_\chi$. If one demands $\chi$ to constitute 100\% of the DM via thermal freeze out, then also $M_\chi$ is fixed, to roughly 3 TeV\cite{Hryczuk:2014hpa}. We will also consider different values of $M_\chi$, to allow for different production mechanisms and for the possibility that $\chi$ does not constitute 100\% of the observed DM.

While at tree level the neutral and charged components of the triplet have the same mass, higher order corrections split the neutral Majorana fermion $\chi_0$ from the charged $\chi^\pm$. This mass splitting has been computed at the two-loop level in the SM \cite{Ibe:2012sx}, yielding to $M_{\chi^\pm} - M_{\chi_0} \simeq165$ MeV (stable to the level of 1 MeV for $M_\chi \gtrsim 1$ TeV)\footnote{Possible heavy New Physics contributions to $M_{\chi^\pm} - M_{\chi_0}$ are very suppressed, since the first effective operator contributing to a splitting arises at dimension~7.}.

%
%
%
%
%
%\paragraph{Direct production at current and future colliders.}
The direct pair production of DM particles receive contribution, in this model, not only from production of $\chi_0$, but also from that of $\chi^\pm$. In fact, the small mass splitting causes $\chi^\pm$ to decay into $\chi_0$ plus very soft pions, which are not reconstructed at the LHC, with a decay length at rest of $\sim 6$ cm. Since current detectors do not reconstruct tracks shorter than $O(30)$ cm, the bulk of the produced $\chi^\pm$ contributes to missing transverse energy in the same way of $\chi_0$. Still, a fraction of the $\chi^\pm$ can travel far enough to leave a track in the detector, and then decay to $\chi_0$ plus soft pions within it, thus yielding a \textit{disappearing track} signal that has no background within the SM \cite{Feng:1999fu}.

\begin{figure}[t]
\begin{center}
\includegraphics[width= 0.48 \textwidth]{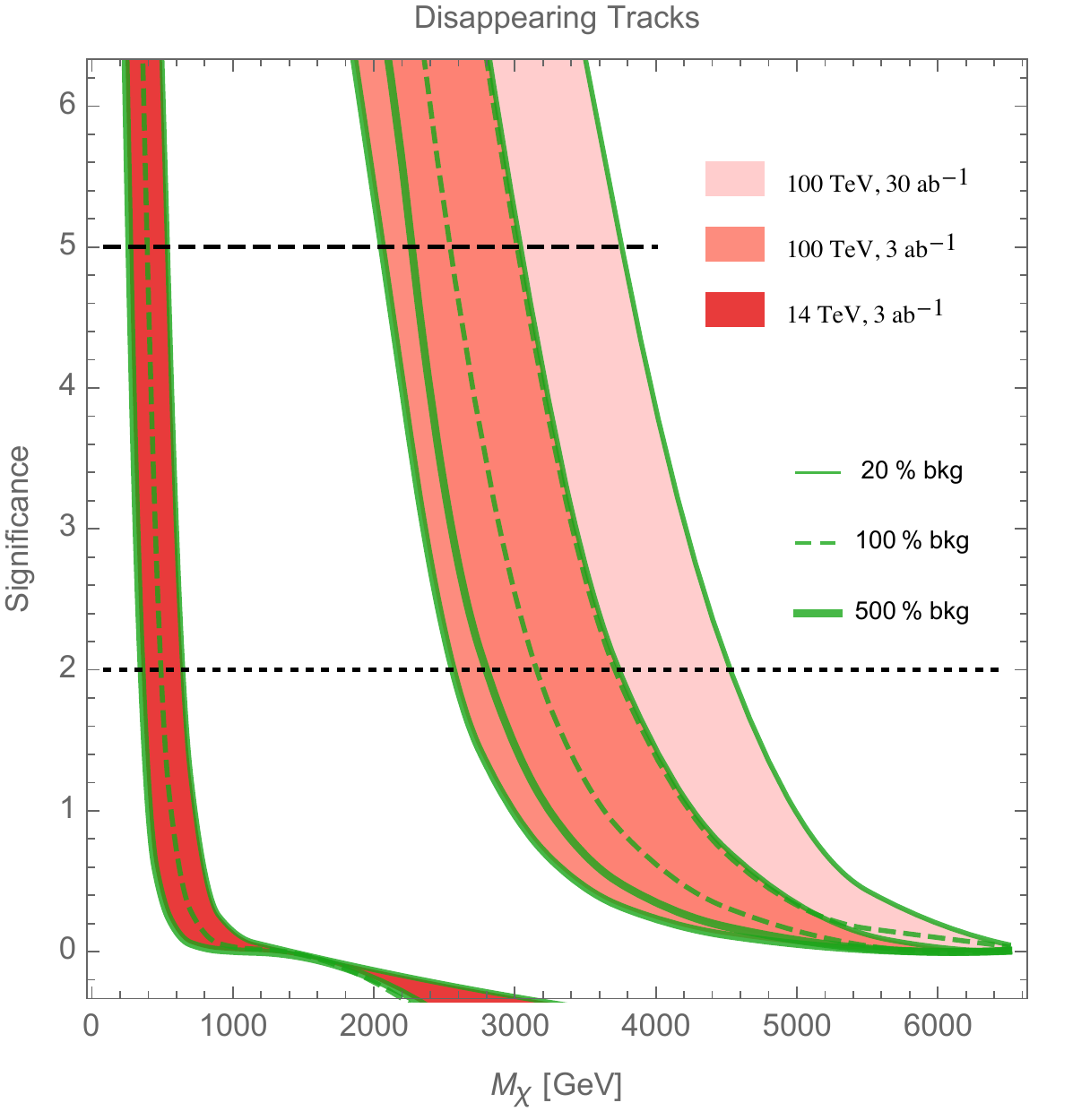} \quad
\includegraphics[width= 0.48 \textwidth]{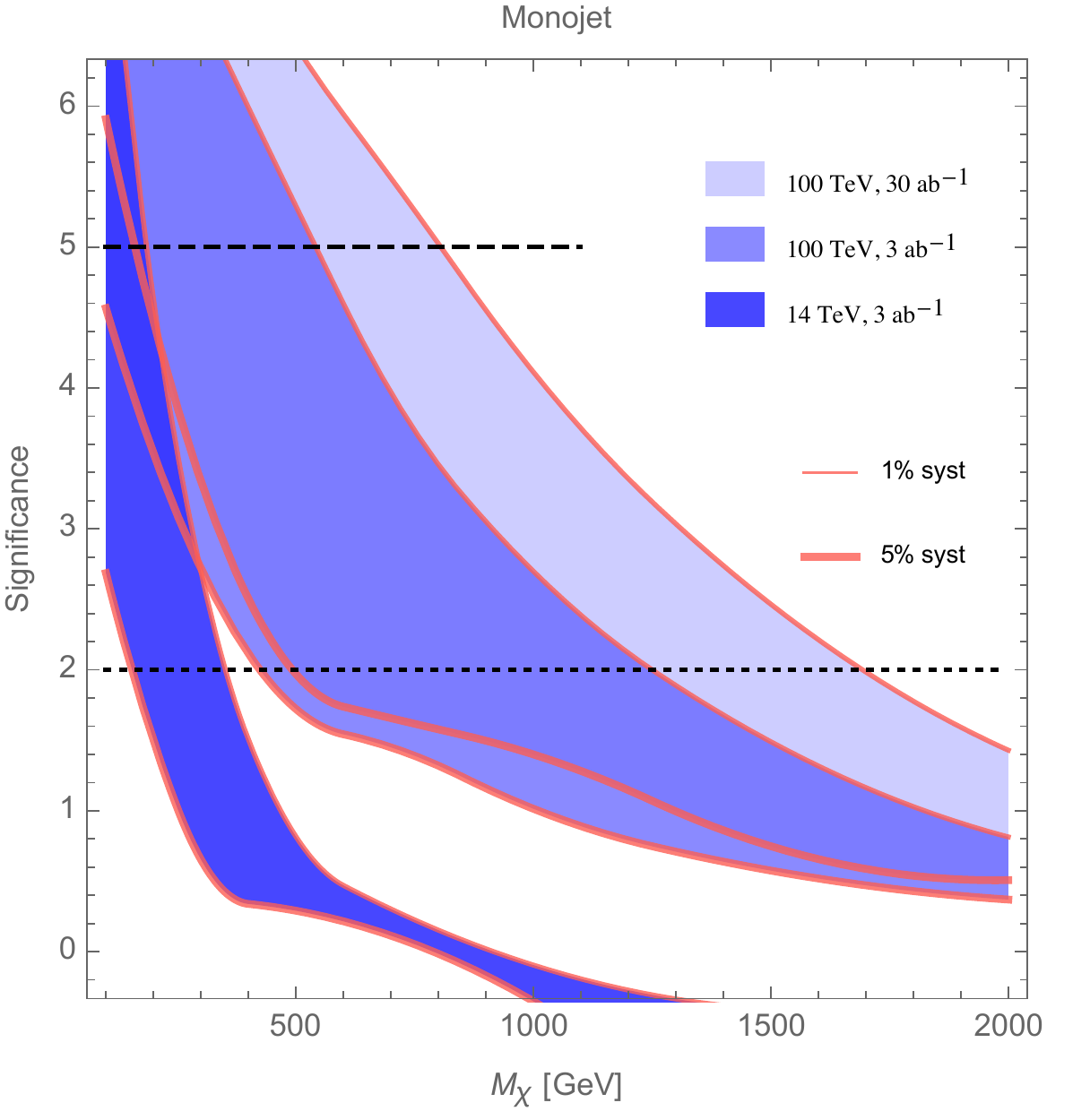}
\caption{\em \small \label{fig:ColliderReach} Reach of disappearing tracks (left) and monojet (right) searches~\cite{Cirelli:2014dsa}.}
\end{center}
\end{figure}

The current best probe of this model at colliders is indeed given by the ATLAS \cite{Aad:2013yna} and CMS\cite{CMS:2014gxa} searches for disappearing tracks, which obtained the bound
\begin{equation}
M_\chi > (260-270)\, \rm GeV.
\end{equation}

In ref. \cite{Cirelli:2014dsa},  the reach of the ATLAS search for disappearing tracks is extrapolated to the HL-LHC, as well as to the 100~TeV proton collider, for both 3 and 30 ab$^{-1}$ of integrated luminosity (see also ref. \cite{Low:2014cba}). The result of this procedure is shown in the left-hand plot of fig. \ref{fig:ColliderReach}. 
The background to this search comes from detector effects, and the red bands in the reach, for any given future benchmark, correspond to a conservative %-in our opinion-
quantification of the uncertainty coming from our extrapolation.
In the right-hand plot we show, for comparison, the expected reach in the ``standard'' monojet channel. Here the blue bands represent how the reach is expected to change according to the control that will be achieved over the systematics. The reach of other channels like vector boson fusion \cite{Cirelli:2014dsa,Berlin:2015aba} and monophoton \cite{Cirelli:2014dsa} is somehow weaker, but it will provide a useful complementarity.
Both for disappearing tracks and for the monojet searches we find a very good agreement with the results of ref. \cite{Low:2014cba}, and we refer the reader to ref. \cite{Cirelli:2014dsa} for more details.

While the region interesting for thermal WIMP DM is out of reach at any conceived future LHC stage, the 100~TeV collider has largely the potential to probe it, and say a final word over the existence of a pure-Wino (independently of DM). The only channel with the potential to discover thermal DM Winos is that of disappearing tracks, and it would benefit, at any future collider, from the capability of reconstructing tracks below the current length of $O(30)$~cm.

\textbf{Relation with future lepton colliders.}
Given that $\chi$ is a full EW multiplet, its contributions to EWPT are very suppressed, at the level of $W,Y \sim 10^{-7}$ \cite{Cirelli:2005uq}: this sensitivity target is not touched by LEP2 \cite{Barbieri:2004qk}, and looks out of reach at any proposed future lepton collider (see ref. \cite{Fan:2014vta} for the expected reaches of high energy positron collider and CPEC).
%A complete basis to parameterise EWPT is given by $S,T,W,Y$, with $Y$ the ``derivative'' of $S$, $W$ the second derivative of $T$
%LEP2 constraints on $W$ and $Y$ are in the ballpark of a few per-mille level% like the $\hat{S}$ and $\hat{T}$ ones
%Prospects for $\hat{S}$ and $\hat{T}$ at the FCC-ee and CPEC have been provided \textit{e.g.} in ref. \cite{Fan:2014vta}, see fig. 1.  Constraints on $\hat{S}$ and $\hat{T}$ are expected to improve by at most one order of magnitude with respect to the LEP2 ones. We did not find prospects for $W$ and $Y$ at FCC-ee or CPEC, but we would guess that it is hard to think an improvement by the $3--4$ orders of magnitudes needed to probe MDM.

%
%
%
%
%

\textbf{Direct and Indirect DM detection.}
The most recent cross section computation for the spin independent scattering of an EW fermion triplet with a nucleon gives \cite{Hisano:2015rsa} $\sigma_{\rm SI} = (2.3 \pm 0.5) \times 10^{-47}~{\rm cm}^2$. This value is out of reach at any current and planned experiment \cite{Cushman:2013zza}, for masses larger than 500 GeV. %Direct detection prospects for this model are therefore dim.

\begin{figure}[t]
\centering
\includegraphics[width=\textwidth]{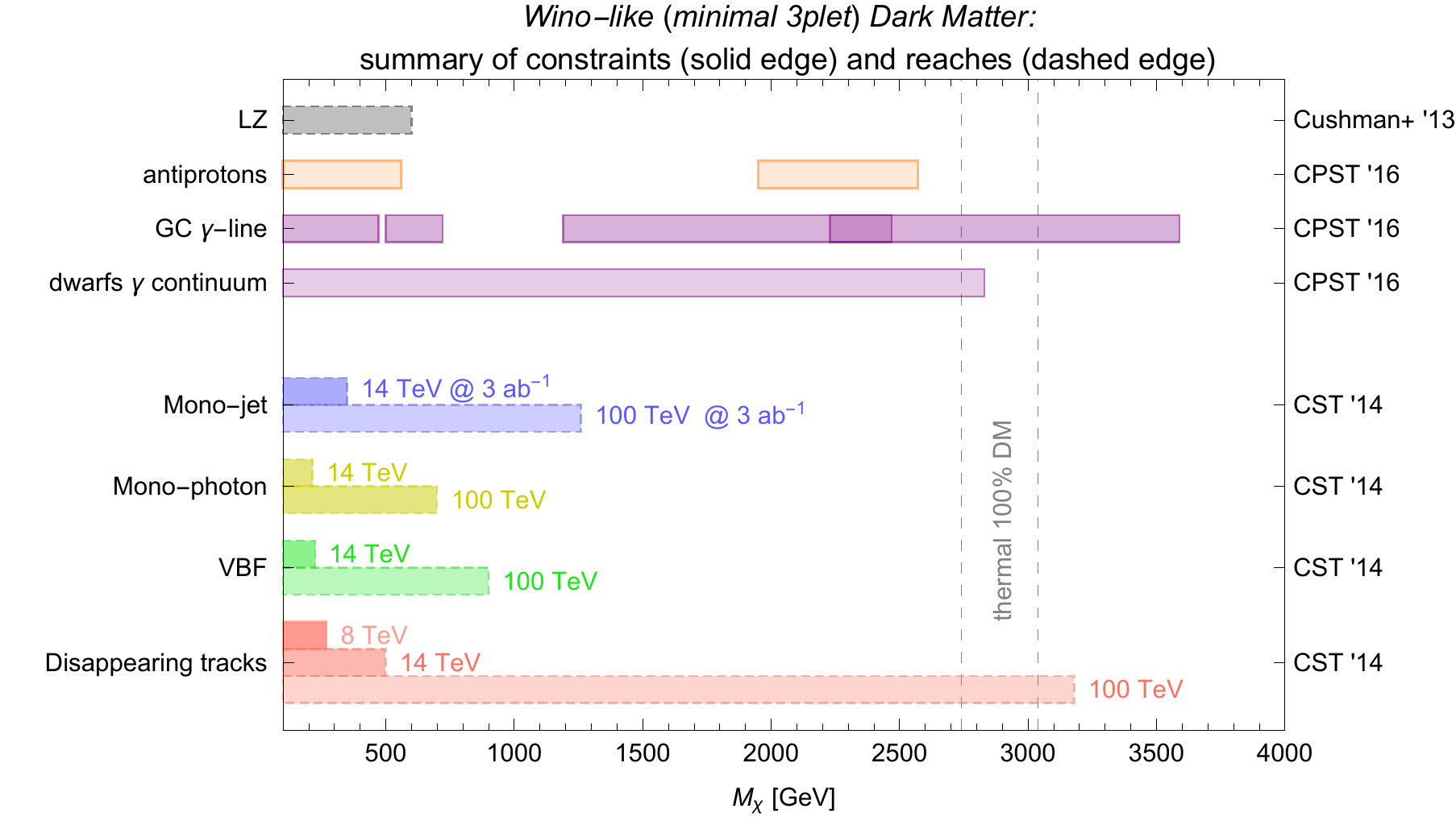}

\caption{\label{fig:summary1} %Current exclusions (continuous) and future reaches (dashed) of different probes of Wino-like DM.
For the GC gamma ray line search of HESS \cite{Abramowski:2013ax}, the lighter shading is the exclusion assuming a NFW profile, the darker one assuming a Burkert profile. The FERMI bound from a gamma continuum from dwarf spheroidal galaxies \cite{Ackermann:2015zua} also suffer from some astrophysical uncertainties \cite{Bonnivard:2015xpq}, so the lighter shading.}
\end{figure}

Concerning indirect detection,
gamma rays from the Galactic Center (as first recognized in \cite{Cohen:2013ama,Fan:2013faa} for lines) and dwarf spheroidal galaxies are, at present, the most promising probes.
We show the reach of two most relevant searches of this kind in fig. \ref{fig:summary1}, also to compare them with the previously discussed collider and DD reaches. We show there also the weaker reach of antiprotons from AMS-02 \cite{Cirelli:2016boh}, for comparison. As far as we know today, a very promising (gamma-ray) telescope to probe this model in the future appears to be CTA, expected to start taking data in 2018 \cite{Consortium:2010bc}. Whether it will exclude or not a pure-Wino, as 100\% of the DM, depends mostly on the control on the astrophysical uncertainties that will be achieved by then.

\subsubsection{Weak Gauge Bosons 3: Fiveplet DM }
\label{sec:wgb3}

While the doublet and triplet DM models discussed so far can decay to the SM through dimension 5 operators, a fiveplet of $SU(2)$ can only decay through a dimension 6 operator, thus guaranteeing a sufficiently long lifetime of the DM even if the global $Z_2$ symmetry which makes it stable is broken at the Planck scale.

We define the fiveplet, $\chi$ as 
$
\chi = \left( \chi^{++}, \chi^{+}, \chi^0, \chi^{-}, \chi^{--} \right).
$
At the renormalizable level, the size of the representation restricts the Lagrangian to
\begin{equation}
\mathcal{L} = \mathcal{L}_{\text{SM}} + c \bar{\chi} \left( \imath \slashed{D} - M \right) \chi,
\label{eqn_Lagrangian}
\end{equation}
where $M$ is the mass of the fiveplet and the constant, $c$, is $1/2$ or 1 depending on whether $\chi^0$ is Majorana or Dirac, respectively. 
%
%The fiveplet is charged under $SU(2)_L$, so the covariant derivative contains
%\begin{eqnarray}
%\left.t^a W^a_{\mu}\right|_{\text{fiveplet}} 
%= \begin{pmatrix} 2W^3_{\mu} & \sqrt{2}W^+_{\mu} & 0 & 0 & 0 \\ \sqrt{2} W^-_{\mu} & W^3_{\mu} &\sqrt{3} W^+_{\mu} & 0 & 0 \\ 
%0 &\sqrt{3} W^-_{\mu} & 0 & \sqrt{3} W^+_{\mu} & 0 \\
%0 & 0 & \sqrt{3} W^-_{\mu} &- W^3_{\mu} & \sqrt{2} W^+_{\mu} \\
%0 & 0 & 0& \sqrt{2} W^-_{\mu} & -2W^3_{\mu}\end{pmatrix} \nonumber.
%\end{eqnarray}
The mass degeneracy of the multiplet is broken at one loop by the gauge bosons. For masses of the multiplet $M \gg m_W$ the singly charged component lies $\sim 166$ MeV above the neutral component; the doubly charged state is heavier than the neutral state by $\sim 664$ MeV. These small mass splittings between the states of the multiplet leave little phase space for decays down to the neutral component. This implies that the charged states can have fairly long lifetimes and travel macroscopic distances at collider experiments. This will be the basis of our search strategy.

As providing a DM candidate is the motivation for this model, we want the neutral component to make up a significant portion of the observed abundance. The large quantum number of the fiveplet, along with the large number of states, allows for very efficient annihilations. This implies that in order to quench the observed relic abundance of DM through thermal freeze-out, the mass of the fiveplet must be heavy; nearly 9.6~TeV~if the Sommerfeld enhancement is included \cite{Hisano:2006nn, Cirelli:2009uv}. 
%There is little chance of discovering such a heavy, electroweakly-produced particle, even at a 100 TeV collider. In comparison, the projected exclusion reach for the the stop (which is strongly produced) is to a mass of $\sim 8$ TeV  \cite{Cohen:2014hxa}.
%
If the mass of the fiveplet is less than 9.6 TeV, the amount of DM left after freeze-out is less than the observed abundance, which leaves room for other sources of DM. There are also other mechanism that would allow for a lighter fiveplet to fulfill the relic abundance~\cite{Salati:2002md,Boucenna:2015haa,Fabbrichesi:2015bta,Aoki:2015nza}, such that we can treat the mass of the fiveplet as a free parameter with an upper bound of $9.6$\UTeV. 
%
%For example, there could be a modified expansion history during freeze-out \cite{Salati:2002md}. In Ref.~\cite{Boucenna:2015haa}, an asymmetric minimal DM model was considered. Using the asymmetry, rather than freeze-out, allows for a fiveplet with a mass between 2.5 and 10 TeV to yield the observed relic abundance. Alternatively, Ref.~\cite{Fabbrichesi:2015bta} studied a supersymmetric version of the fiveplet. In this case, the supersymmetric partner also is stable and adds to the relic abundance, allowing for the multiplet to be around 7 TeV in mass. As a last example of a light fiveplet quenching the relic abundance, we mention the model of \cite{Aoki:2015nza} which adds three heavy right handed neutrinos to the minimal DM set up. With this addition, they are able to generate masses for the light neutrinos via a type 1 see-saw. The right handed neutrinos can be produced by freeze-in through the Yukawa coupling. Decays of the heavy neutrinos down to the fiveplet can the boost the fiveplet abundance enough to achieve the correct abundance. Depending on the specifics of the model (such as the heavy neutrino mass) a fiveplet as light as 700 GeV could provide the right amount of DM.

Direct detection and indirect detection searches are able to constrain the model. A recent reevaluation of the nuclear matrix element was found to be lower than originally thought, leading to a spin independent cross section of $1.0\times10^{-46} \text{ cm}^2$ \cite{Junnarkar:2013ac,Farina:2013mla}. In this case, Xenon1T and LZ are projected to have a reach to $\sim350\UGeV$ and $\sim4000\UGeV$, respectively \cite{Cushman:2013zza}. Signals (or lack there of) of DM annihilations place the strongest bounds on the model. The Sommerfeld enhancement has a very large effect for the fiveplet and increases the cross section for annihilations into vector bosons \cite{Cirelli:2007xd,Cirelli:2009uv,DiLuzio:2015oha,Cirelli:2015bda,Garcia-Cely:2015dda, Aoki:2015nza}. With the non-observation of sharp gamma ray spectral features by H.E.S.S., a fermionic fiveplet can only make up all of the DM abundance for a small range of masses around 2.5~TeV~and 10~TeV, or be completely excluded, depending on the DM profile \cite{Abramowski:2013ax}. CTA will be able to exclude almost the entire mass range, even for an isothermal profile \cite{Garcia-Cely:2015dda,Cirelli:2015bda,Consortium:2010bc}.

The projected direct detection results depend on the DM abundance while indirect detection signals depends on both the abundance and the profile. However, collider bounds do not depend on astrophysical results. To this end, it is important study how the MDM fiveplet can be bounded by collider experiments. In Ref.~\cite{Ostdiek:2015aga}, the disappearing track searches done by ATLAS and CMS \cite{Aad:2013yna,CMS:2014gxa} are used to show the LHC has excluded a fiveplet below a mass of 267 (293) GeV~depending on whether it is Majorana (Dirac). Additionally, it was determined a Majorana (Dirac) fiveplet could be excluded at the 14 TeV~LHC up to a mass of 410-670 GeV~(465-745 GeV). In the following we review the disappearing track search strategy and extend the method to a 100 TeV~collider for the MDM fiveplet. 
%
%The disappearing track search is a modification of a mono jet strategy. The idea is to recoil a $\chi\bar{\chi}$ system off a hard jet. As the $\chi^0$ is DM, and not obsesrved in the detector, the experiment sees a jet recoiling off of `nothing' or missing transverse energy. However, it is not just the neutral components that are produced in the $\chi\bar{\chi}$ system. The components with charge eventually decay down to to neutral component via charged pions which are too soft to be detected. The decay length of the doubly charged component is around a millimeter while the singly charged component has decay length of a few centimeters. It is then possible for the charged particles to travel far enough in the detector to leave tracks and then decay, leading to missing energy. Thus, there appears to be a charged track which suddenly disappears before the edge of the tracker.

There have been a few other studies which extrapolate the ATLAS search \cite{Aad:2013yna} to future colliders \cite{Low:2014cba, Cirelli:2014dsa, Ostdiek:2015aga, Bramante:2015una}. The optimised cuts presented in~\cite{Cirelli:2014dsa} are used, which look for 
\begin{equation}
\begin{aligned}
p_{T,j_1} > 1 \UTeV, \;\;\;\slashed{E}_T > 1.4 \UTeV, \;\;\;  \Delta \phi_{j,\slashed{E}_T}^{\text{min}} > 1.5,\\
p_{T,\text{track}} > 2.1\UTeV, \;\;\;0.1 < \left| \eta_{\text{track}} \right| < 1.9, \text{ and }\\30\text{ cm} < \text{transverse track length} < 80 \text{ cm}.
\end{aligned}
\label{eqn_cuts}
\end{equation}
The variable $\Delta \phi_{j,\slashed{E}_T}^{\text{min}}$ is the azimulthal angle between the any jet with $p_T > 500\UGeV$ and the missing energy. The requirements on the transverse track length and $\eta_{\text{track}}$ come from the ATLAS search and are used in the future collider extensions as a method to estimate the background. There is no obvious Standard Model process which mimics this signal; the background comes from $p_T$-mis-measured tracks and hadrons with large momentum transfer interactions with pieces of the detector. The measured signal and background thus depend heavily on the specifics of the detector. In their search, ATLAS gives the observed shape of the $p_T$-mis-measured tracks as $d\sigma/d p_T^{\text{track}} = (p_T^{\text{track}})^{-a}$ where $a=1.78\pm0.05$. This shape is normalized to the background at $\sqrt{s}=8\UTeV$ reported in \cite{Aad:2013yna} and then scaled to the ratio of the $Z(\nu \bar{\nu}) + \text{jets}$ cross section passing initial cuts on $p_{T,j_1}$, $\slashed{E}_T$, and $\Delta \phi_{j,\slashed{E}_T}^{\text{min}}$ at $\sqrt{s}=8$ and 100 TeV. There is much inherent uncertainty in this method of extrapolating the background. Measuring the spectrum of the $p_T$-mis-measured tracks at the current run of the LHC could help in this regard. To be conservative, we allow for a range of the background cross section, larger or smaller by a factor of 5.

The model is implemented using \textsc{FeynRules2.0} \cite{Alloul:2013bka}, and generated events using \textsc{MadGraph5$\_$aMC@NLO} \cite{Alwall:2014hca} for $\chi\bar{\chi}$ production with up to two extra partons. The events were matched with the MLM scheme, hadronized, and showered using \textsc{Pythia 6.4} \cite{Sjostrand:2006za}, and fast detector simulation was done with \textsc{Delphes} \cite{deFavereau:2013fsa}  using \textsc{FastJet} \cite{Cacciari:2005hq,Cacciari:2011ma} to cluster the jets with the anti-$k_T$ algorithm \cite{Cacciari:2008gp}. The default ATLAS card was used for \textsc{Delphes}, modified so the neutral component of $\chi$ would add to the missing energy. See Refs.~\cite{Ostdiek:2015aga} and \cite{Bramante:2015una} for more details about the analysis.

Our results are summarized in Fig.~\ref{fig_results} assuming 15 ab$^{-1}$ of integrated luminosity. The left panel shows the expected number of events passing the cuts as a function of the transverse distance travelled by the track. This shows that the heavier mass points are harder to find not just because the production cross section decreases, but also because the tracks do not travel as far. For a given jet momentum that the $\chi\bar{\chi}$ system recoils off, the heavier DM points do not get as much of a boost. In order to travel between 30 and 80 cm (with a decay length of a few cm) the system needs to be quite boosted.

\begin{figure}[t]
\begin{center}
\includegraphics[width=3.25in]{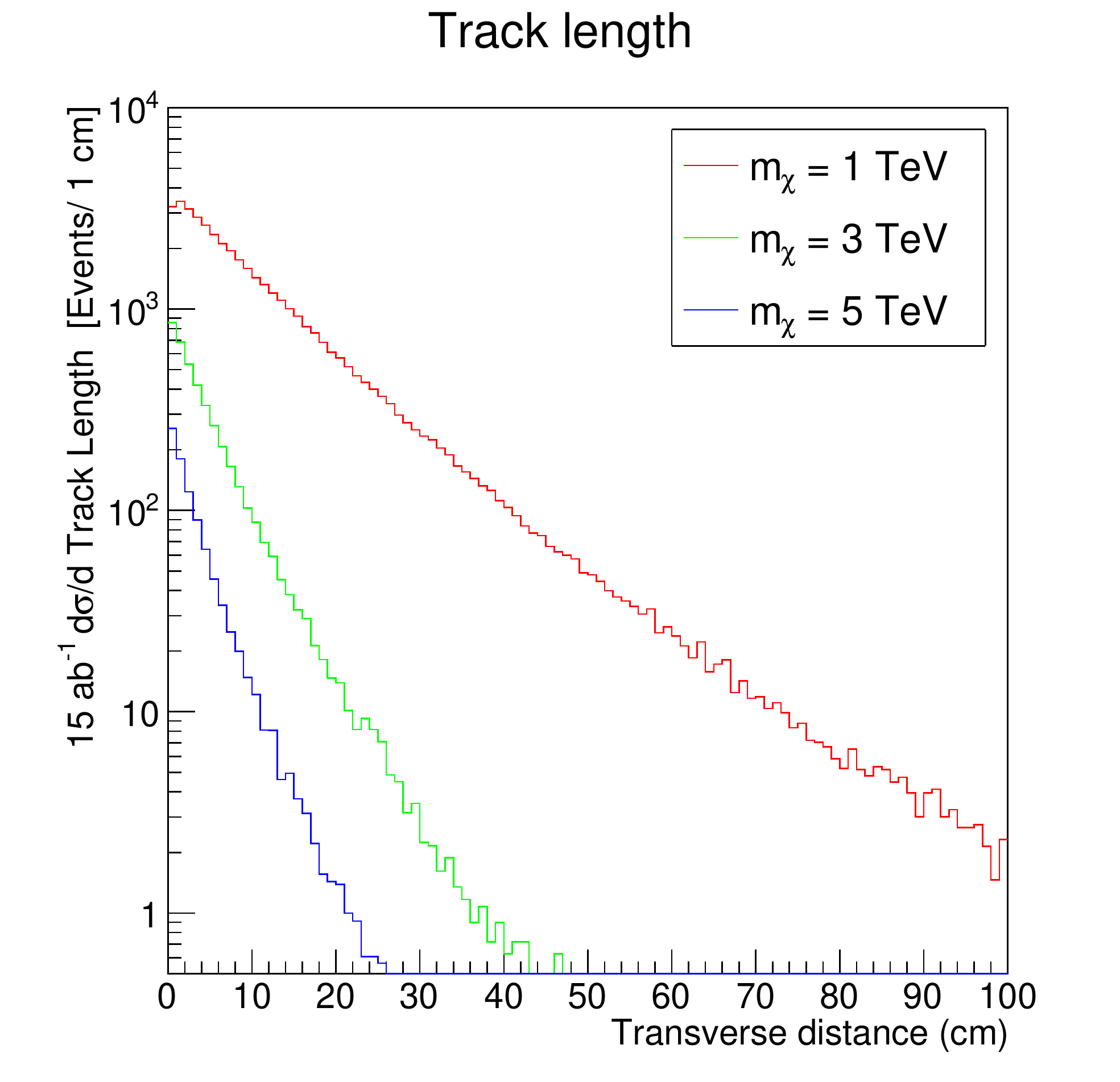}
\includegraphics[width=2.9in]{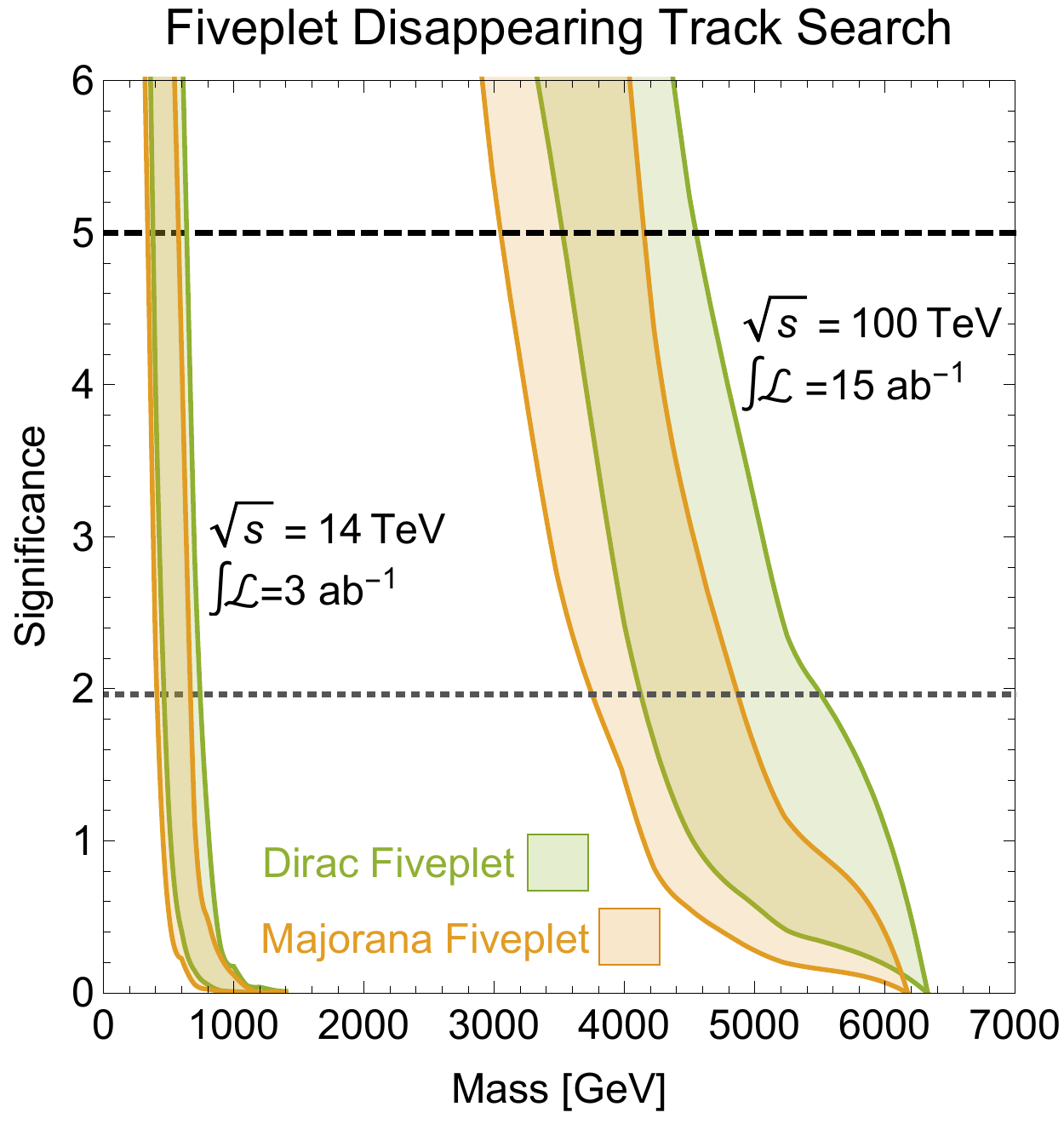}
\caption{Left: The number of events passing the cuts in Eq.~\eqref{eqn_cuts} as a function of transverse track length. The different colors are for different masses of a Majorana Fiveplet. The larger masses have lower production cross section and do not receive as large of a boost from the jet, so do not travel as far. The expected background for track lengths between 30 and 80 cm is around 2 events. Right: Reach for the $\sqrt{s}=14\UTeV$ LHC and a future $\sqrt{s}=100\UTeV$ collider. The bands are generated by varying the background between $20\%$ and $500\%$ of the extrapolated value.}
\label{fig_results}
\end{center}
\end{figure}

In the right panel, we plot the significance as a function of the fiveplet mass. This is computed using 
\begin{equation}
\text{Significance}=\frac{S}{\sqrt{B+\alpha^2 B^2 + \beta^2 S^2}}
\end{equation}
where S and B are the number of signal and background events. The background and signal systematics are incorporated into $\alpha$ and $\beta$ and are conservatively given values of $\alpha=20\%$ and $\beta = 10\%$ \cite{Low:2014cba, Cirelli:2014dsa}. The bands in the plot are generated by varying the number of background events between $20\%$ and $500\%$ of the $\sim2$ events expected from the extrapolation. %The results are summarized in Table \ref{tab:results}. 

A 100 TeV~collider can greatly extend the search for minimal DM. The discovery reach of the Majorana fiveplet is between 3.1--4.2 TeV~while the exclusion reach is 3.8--4.9 TeV. For the case of the Dirac fiveplet, the discovery and exclusion reaches are 3.5--4.5 TeV and 4.1--5.5 TeV. These results are about a factor of 7 higher than the estimated reach at the LHC. These mass reaches are important in terms of complementarity with the other DM experiments. Depending on the DM profile, there is a possible gap in coverage in the indirect detection experiments for a fiveplet mass of $\sim2.5\UTeV$, which will be covered by a 100 TeV collider. In addition, the projected LZ results could reach a fiveplet with a mass up to 4 TeV. With a possible higher mass reach than this, the 100 TeV collider can exceed direct detection results without the question of the current relic abundance.

\subsubsection{Weak Gauge Bosons 4: Thermal Relic Neutralino DM }
\label{sec:wgb4}
%%%%%%%%%%%%%%%%%%%%Section Thermal Neutralinos%%%%%%%%%%%%%%%%%%%%%%%%%

Weakly interacting DM are one of the few physics scenarios
which can feature an \textit{upper} limit on the particle masses. This
argument is based mostly on a combination of the experimentally
observed relic density constraint and on the fact that the DM
states interact weakly. In that spirit, the neutralino/chargino sector
of the MSSM is a way to interpolate between singlet, doublet, and
triplet $SU(2)$ representations of the DM state. The question
is to what degree a 100~TeV proton-proton collider, together 
with future direct and indirect detection experiments, can cover the
relic neutralino surface with all scalar superpartners, \textsl{i.e.}
squarks, sleptons, and the heavy Higgs boson decoupled to masses above
8~TeV. The main challenge to collider searches are nearly mass
degenerate ($\lesssim 5 \%$ mass difference) states in the
neutralino/chargino sector which lead us to a dedicated analysis with
very soft leptons and photons combined with extremely hard initial
state radiation jets at a 100~TeV hadron collider.

In Refs.~\cite{Bramante:2014tba,Bramante:2015una} it is for example
demonstrated how nearly pure bino DM, which freezes out to
the observed relic abundance through co-annihilation, can be uncovered
by a 100~TeV hadron collider.  Figure~\ref{fig:relicneutcompl} shows
the main collider-related result: for almost pure wino DM as
well as for bino-like co-annihilating DM with small couplings
to the Standard Model, a future 100~TeV hadron collider allows for
full coverage of the relic neutralino surface.\medskip

%-----------------------------------------------------------
\begin{figure}
\centering
\includegraphics[scale=.74]{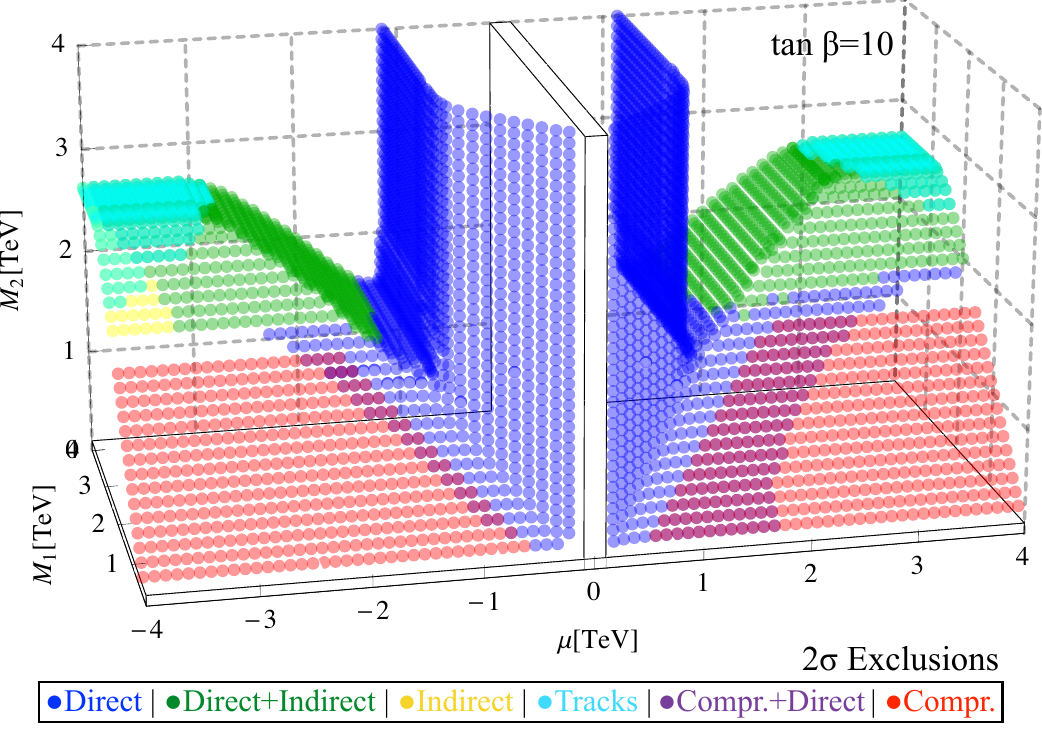}
\caption{Relic neutralino surface defined by a thermal primordial freeze-out to abundance
  $\Omega h^2 \simeq 0.12$, with all non-neutralino superpartners
  decoupled, and including Sommerfeld-enhancements to freeze-out
  annihilation cross-sections. Regions inside the central boxes are
  excluded by LEP~II~\cite{lepii}. We also show
  future direct detection and indirect detection prospects.  The
  compressed and charged track collider studies referred to as
  ``Compr." and ``Tracks" are described in the
  text~\cite{Bramante:2014tba,Bramante:2015una}.}
\label{fig:relicneutcompl}
\end{figure}
%-----------------------------------------------------------

To define the relic neutralino surface the thermal relic abundances
were calculated using \texttt{DarkSE}~\cite{Hryczuk:2011vi}, which
incorporates Sommerfeld-enhancements to the relic abundance code of
\texttt{DarkSUSY}~\cite{Bechtle:2012zk}. In addition, the
annihilation cross-section to nearly-pure wino freeze-out were checked
with \texttt{MicrOMEGAs}~\cite{Belanger:2013oya}, modified to
include Sommerfeld enhancement. MSSM mass spectra were generated with
\texttt{SuSpect}~\cite{Djouadi:2002ze}, where loop corrections from
decoupled SUSY particles were turned off, but electroweakino
charged-neutral electroweak custodial mass splittings were set before
matrix diagonalization, as described in
Ref.~\cite{Bramante:2015una}.\medskip

%-----------------------------------------------------------
\begin{figure}
\includegraphics[scale=.74]{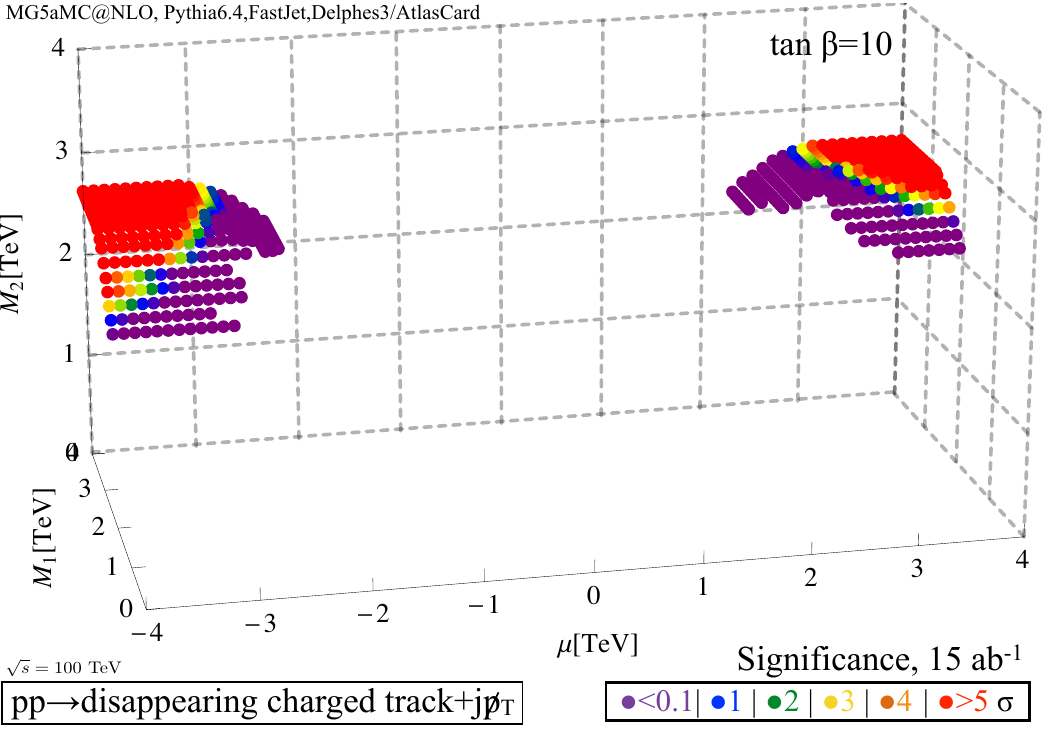} 
\includegraphics[scale=.74]{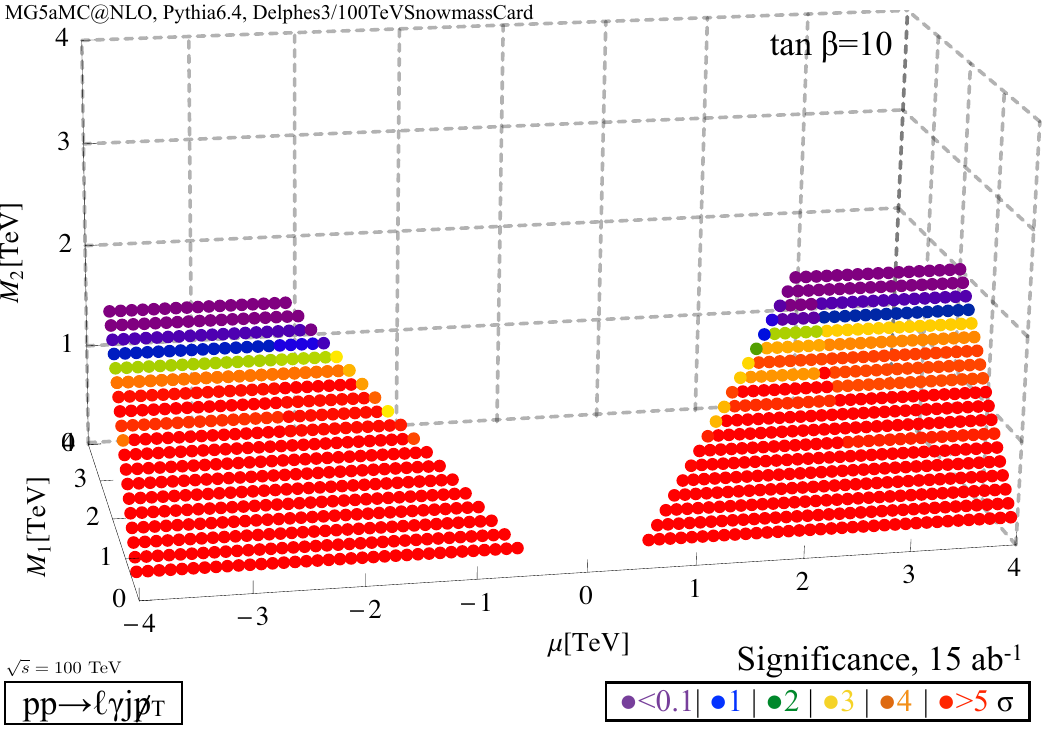}
\caption{The significance reach over thermal relic neutralino parameter space is shown, for both charged tracks and compressed ($\gamma + \ell + j + {\rm MET}$) searches at a 100 TeV collider, after 15 ${\rm ab^{-1}}$ of data. In the case of compressed searches, a larger parameter space can be probed than what is shown above (see \cite{Bramante:2014tba}).}
\label{figure:colliderstudies}
\end{figure}
%-----------------------------------------------------------

In Figure \ref{fig:relicneutcompl} we show the
parameter space probed at a 100~TeV hadron collider, alongside the
$2\sigma$ relic neutralino reach of future direct and indirect
detection experiments. On the collider side, it first includes a study
of sensitivity to disappearing charged tracks for thermal relic
neutralinos. In SUSY parameter space with a predominantly wino LSP,
the mass splitting between the LSP and lightest chargino becomes as
small as $ 160$ MeV. With such a small inter-state mass splitting,
decays of the chargino to LSP through an off-shell $W$ boson are
suppressed and can leave $\mathcal{O}(10~{\rm mm})$ long charged
tracks in the detector that vanish to missing transverse energy
(MET). To estimate the background to a disappearing charged track
search, Ref.~\cite{Bramante:2015una} matched the data-driven
background of an ATLAS charged tracks study~\cite{Aad:2013yna}, by
simulating $pp \rightarrow Z(\nu \bar{\nu})+{\rm jets}$ events at
center-of-mass energies 8~TeV and 100~TeV. The ratio of these events
passing kinematic cuts on missing transverse momentum and jet momentum
were used along with the number of background events found by ATLAS,
to project the number of background events at a 100~TeV
collider. Signal events featuring at least one chargino paired with a
partner electroweakino and jets, were simulated using
\texttt{MG5aMC@NLO}~\cite{Alwall:2014hca} with MLM
matching~\cite{Mangano:2006rw}, combined with
\texttt{Pythia6.4}~\cite{Sjostrand:2006za} and
\texttt{DELPHES3}~\cite{deFavereau:2013fsa}. A dedicated set of cuts
makes use of the strengths of a 100~TeV hadron collider with excellent
detector performance~\cite{Bramante:2015una}:
\begin{enumerate}
\item at least two jets with $|\eta| < 2.5$ and $p_T$ greater than
  1~TeV and 0.5~TeV, respectively;
\item at least one disappearing  track with $p_T > 2.1$~TeV,
  $0.1 < |\eta| <1.9$, and length 30-80~cm;
\item MET in excess of 1.4~TeV.
\end{enumerate}
The significance shown in Figure~\ref{figure:colliderstudies} after 15
ab$^{-1}$ is based on the estimate $S/\sqrt{B + \alpha^2 B^2 + \beta^2
  S^2}$ with $\alpha =2$ and $\beta=0.1$ setting the signal and
background uncertainty. We see that the charged track search indeed
covers the wino LSP region of the relic neutralino surface.\medskip

A second parameter region which needs to be targeted by a 100~TeV
collider is nearly pure gauge singlet DM, which freezes out
to the observed relic abundance by co-annihilating with a heavier
partner. Such a nearly-pure bino LSP arises in relic neutralino
parameter space where $M_2 < 2 ~{\rm TeV}$ and $\mu > 1 ~{\rm
  TeV}$~\cite{Bramante:2014tba,Bramante:2015una}. Compressed
electroweakino searches target the production of electroweakinos
$5-50$~GeV heavier than lighter electroweakino states. As the heavier
electroweakinos decay to the LSP, they emit soft state leptons and
photons with $p_T \sim 5-50$~GeV. Such soft leptons and photons as
part of the collider signature allow for a smaller MET cut than
traditional ``jet $+$ MET" DM searches, boosting the mass reach of
compressed DM
searches~\cite{Schwaller:2013baa,Han:2014kaa,Bramante:2014dza,Han:2014xoa,Low:2014cba,Baer:2014kya,Bramante:2014tba,Han:2015lma,Han:2015lha,Bhattacharyya:2009cc,Giudice:2010wb,Calibbi:2013poa,Gori:2014oua,Chakraborti:2015mra,Izaguirre:2015zva}.
In Ref.~\cite{Bramante:2015una}, signal events of the type $pp
\rightarrow \chi^\pm \chi^0 + {\rm jets}$ were simulated along with
the dominant background $pp \rightarrow \gamma W^\pm +{\rm jets}$. The
cuts applied in this study, which employed \texttt{MG5aMC@NLO},
\texttt{Pythia6.4}, and \texttt{DELPHES3}, with the default-valued
Snowmass future detector card~\cite{Anderson:2013kxz} were:
\begin{enumerate}
\item exactly one photon and exactly one lepton with $p_T = [10-60]~{\rm GeV}$ and $|\eta| < 2.5$, separated by $\Delta R > 0.5$, and a photon-lepton MT2~\cite{Bramante:2015una} of $M_{T2}^{(\gamma,\ell)} < 10$~GeV;
\item at least one jet, with $|\eta| < 2.5$ and $p_T > 0.8$~TeV, but
  no more than 2 jets with $p_T > 0.3$~TeV;
\item MET in excess of 1.2~TeV.
\end{enumerate}
A 10-50 GeV size mass splitting between the LSP and NLSP guarantees
that leptons and photons emitted in NLSP to LSP decays will be
soft. This should be contrasted with the expected transverse momentum
of leptons and photons coming from SM $pp \rightarrow \gamma W^\pm j$
events, which is the dominant background at a 100~TeV collider. The
requirement that the $W$ boson produce $\sim$ TeV of MET suppresses
low-$p_T$ phase space for the accompanying lepton. For similar
reasons, because background events are boosted, emitted photons will
also tend to have high $p_T$. Thus the increased energy of the 100~TeV
environment makes this search particularly incisive.

The significance found in this study is shown in
Figure~\ref{figure:colliderstudies}, where the background and signal
systematic uncertainties were taken to be $\alpha=\beta=0.05$.  Its
mass reach extends to nearly $m_\chi \sim 2$ TeV, as shown in
Figure~\ref{figure:colliderstudies}, exactly complementing indirect
and direct searches. This complementarity is a consequence of the MSSM
thermal relic DM LSPs flipping from being nearly pure bino to
nearly pure wino around $M_2 \approx 2$ TeV. For $M_2 \lesssim 2$ TeV,
the NLSP wino is inaccessible by direct and indirect searches, but
would be produced at a 100~TeV collider.\medskip

To illustrate the complementarity of 100~TeV collider searches and
direct as well as indirect detection experiments we illustrate the
parameter space excludable by future liquid xenon direct detection
searches, and by the Cerenkov Telescope Array's search for gamma ray
lines emitted from the central kiloparsecs of the Milky Way
galaxy~\cite{Carr:2015hta} in Figure~\ref{figure:directandindirect}.
We also show present bounds from LUX~\cite{Akerib:2013tjd},
XENON100~\cite{Aprile:2013doa}, and HESS~\cite{Abramowski:2013ax}.
Future direct detection constraints were set using
\texttt{MicrOMEGAs} output along with the Xenon direct detection
reach projected for the next
decade~\cite{Cushman:2013zza}. Constraints from indirect detection of
DM annihilation in the galactic center were determined with
\texttt{MicrOMEGAs} and the one-loop, Sommerfeld-enhanced
annihilation rates of Ref.~\cite{Hryczuk:2011vi}. The projected
indirect reach is calculated using the sensitivity forecasted
in~\cite{Carr:2015hta}, assuming a standard Einasto DM density profile
in the Milky Way.

The picture in all three fields of DM searches turns out to
be similar: current experiments, including the LHC, are able to
significantly cut into the relic neutralino surface. However, a full
coverage of the surface, along with a comprehensive test of weakly
interacting DM, is only guaranteed by the next generation of
experiments: a 100~TeV hadron collider combined with $n$-ton xenon
detectors and CTA.  This complementarity, which requires two dedicated
collider search strategies, is the central message of
Figure~\ref{fig:relicneutcompl}.

%-----------------------------------------------------------
\begin{figure}
\includegraphics[scale=.74]{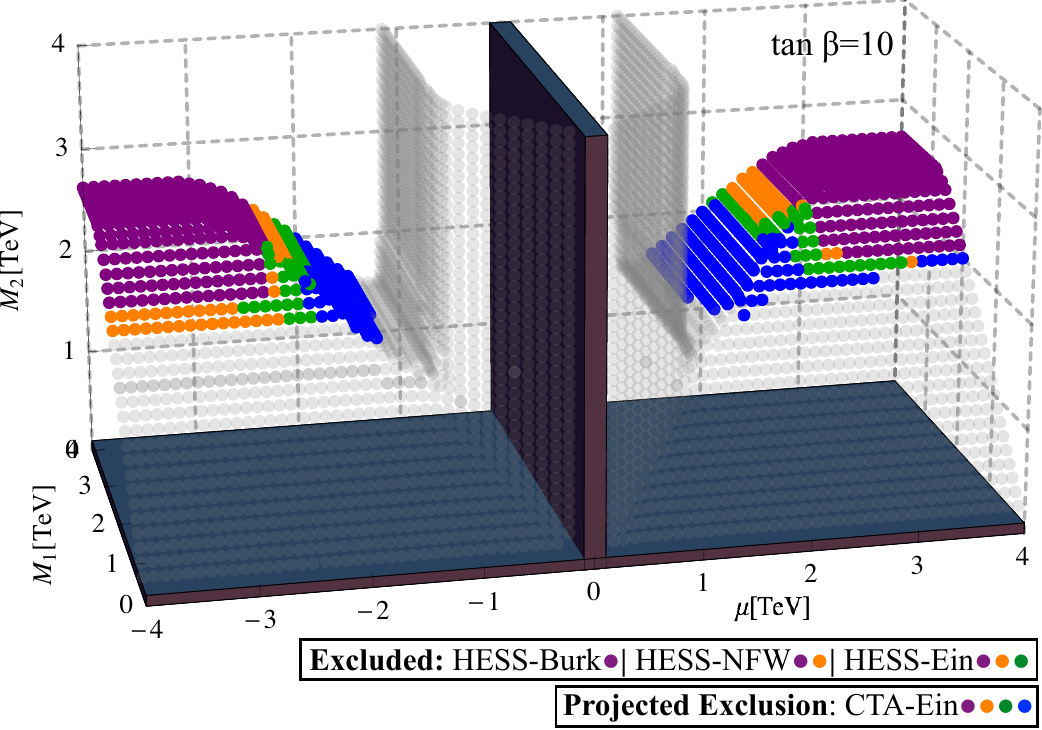} 
\includegraphics[scale=.74]{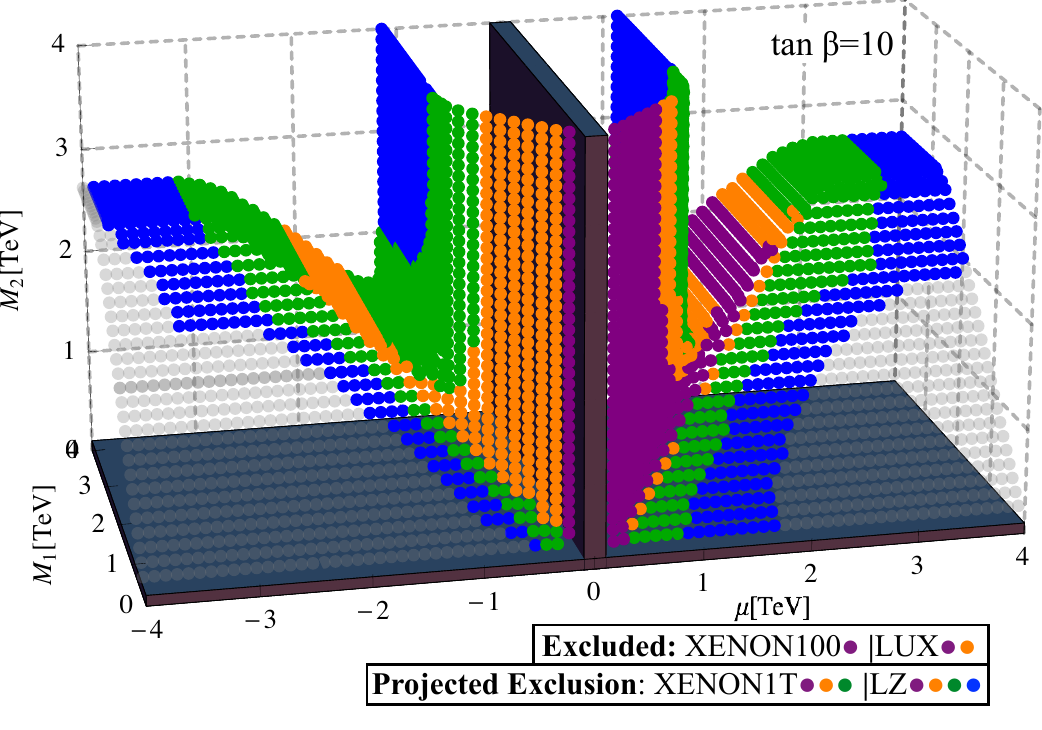}
\caption{Present indirect and direct detection limits on thermal relic neutralinos are shown, along with future constraints. Codes and calculations employed in the production of this plot are described in the caption of Figure \ref{fig:relicneutcompl}.}
\label{figure:directandindirect}
\end{figure}
\subsubsection{Higgs Portal}
\label{sec:DM_higgsportal}

The Higgs boson provides a unique low-dimension portal between the Standard Model and a dark sector via interactions of the form $|H|^2 \mathcal{O}$, where $\mathcal{O}$ is a gauge-invariant operator with $\Delta_{\mathcal{O}} \lesssim 2$. The classic example is $\mathcal{O} = \phi^2$ were $\phi$ is neutral under the SM but enjoys a $Z_2$ symmetry \cite{Silveira:1985rk, McDonald:1993ex, Burgess:2000yq, Patt:2006fw, Barger:2007im}. Here we will consider such a scalar Higgs Portal with interactions
\begin{equation} \label{eq:portal}
\mathcal{L} = \mathcal{L}_{SM} - \frac{1}{2} \partial_\mu \phi \partial^\mu \phi - \frac{1}{2} M^2 \phi^2 - c_\phi |H|^2 \phi^2
\end{equation}
where $H$ is the SM-like Higgs doublet and $\phi$ is a scalar neutral under the Standard Model. After electroweak symmetry breaking the theory consists of
\begin{equation} \label{eq:brokenportal}
\mathcal{L} = \mathcal{L}_{SM} - \frac{1}{2} \partial_\mu \phi \partial^\mu \phi - \frac{1}{2} m_\phi^2 \phi^2 - c_\phi v h \phi^2 - \frac{1}{2} c_\phi h^2 \phi^2
\end{equation}
where $m_\phi^2 = M^2 + c_\phi v^2$ in units where $v = 246$ GeV. 

This portal interaction respects an unbroken $Z_2$ symmetry $\phi \to - \phi$. If the $Z_2$ symmetry is exact the Higgs Portal furnishes a DM candidate \cite{Silveira:1985rk,McDonald:1993ex,Burgess:2000yq,Davoudiasl:2004be,Patt:2006fw}. Higgs Portal DM is highly predictive in the sense that the coupling $c_\phi$ is determined as a function of $m_\phi$ if $\phi$ is required to provide the entirety of the observed DM abundance, as illustrated in Fig.~\ref{fig:thermDM}. While thermal abundance corresponds to small values of $c_\phi$, larger values are allowed if $\phi$ only accounts for some fraction of the DM or is produced non-thermally in the early Universe.
 
% ---------------------------------------------
\begin{figure}[tbp] 
   \centering
 \includegraphics[width=4.0in]{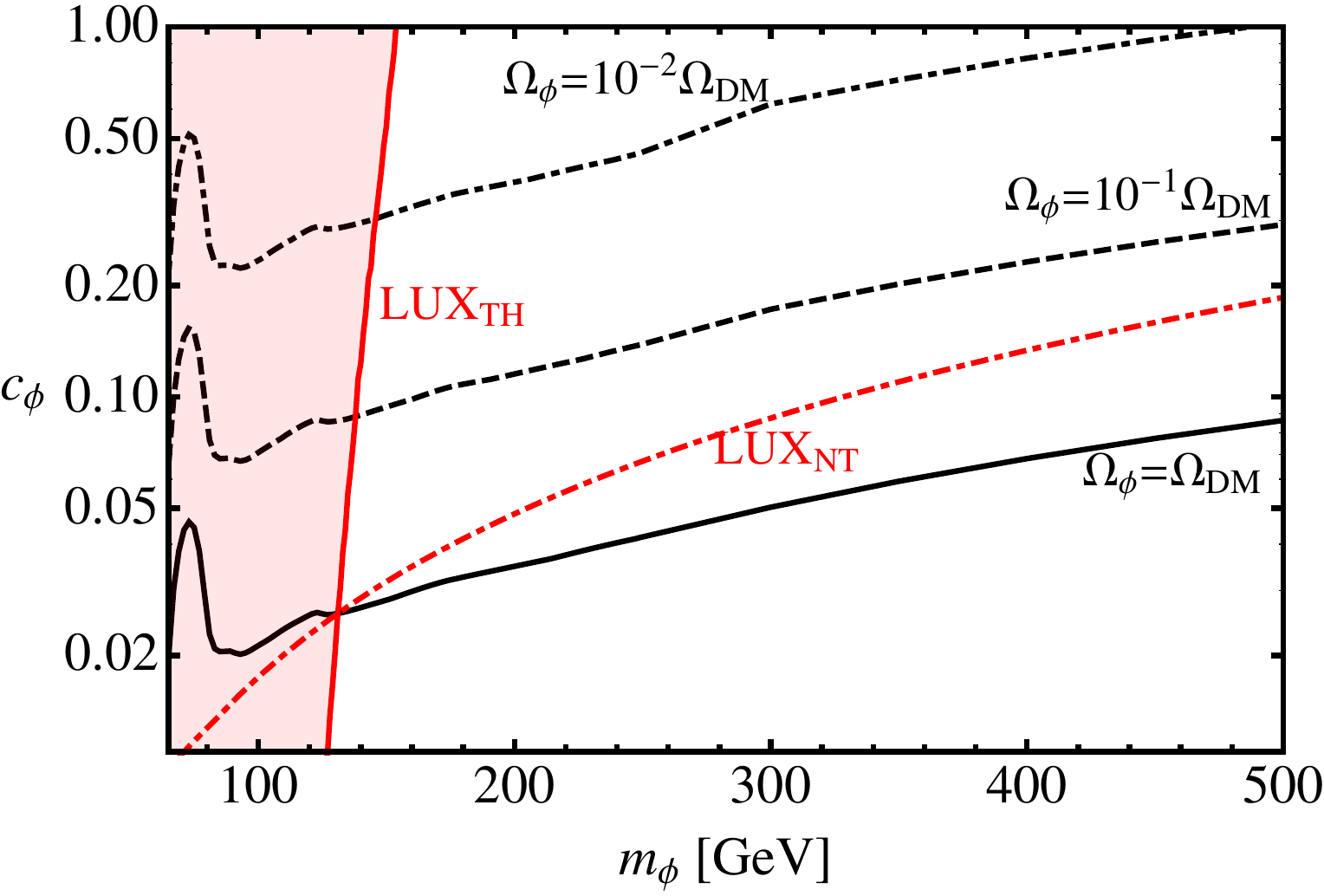}
  { \caption{Contours of relic DM density from freeze-out through the Higgs Portal.  Constraints on the parameter space from the LUX direct detection experiment \cite{Akerib:2013tjd} are shown in dotdashed red ($\text{LUX}_{\text{NT}}$) where we assume non-thermal processes give the observed DM relic abundance in regions where thermal freeze-out over- or under-produces DM.  The solid red line ($\text{LUX}_{\text{TH}}$) and shaded region show the parameter space excluded given a standard thermal history, where $\phi$ may comprise only a fraction of DM. }
   \label{fig:thermDM}}
\end{figure}
% ---------------------------------------------

Although it only communicates with the SM via the Higgs sector, current direct detection experiments are already sensitive to Higgs Portal DM. Current bounds on $c_\phi$ from the LUX experiment \cite{Akerib:2013tjd} are shown in Fig.~\ref{fig:thermDM}, both in the case that $\phi$ comprises the entirety of the observed DM abundance (assuming late-time dilution or production in regions of parameter space where thermal freeze-out over- or under-produces DM) and in the case where $\phi$ simply has a thermal abundance (in which case it typically comprises only a fraction of the observed DM abundance). Significantly, when Higgs Portal DM has a thermal history, predicted direct detection rates are almost independent of the Higgs Portal coupling, and the predicted rate largely becomes a function of the mass alone. This raises the prospect of strong complementarity between direct detection and collider probes of Higgs Portal DM, where rates for $\phi$ production scale with positive powers of $c_\phi$. 

When $m_\phi < m_h/2$ this scenario may be very efficiently probed at colliders via the Higgs invisible width \cite{Eboli:2000ze,Low:2011kp,Djouadi:2011aa,Englert:2011us, Djouadi:2012zc, Englert:2013gz, Aad:2014iia,Chatrchyan:2014tja,Bernaciak:2014pna}, since the Higgs can decay on-shell into $\phi$ pairs and the smallness of the SM Higgs width ensures the rate for $pp \to h + X \to \phi \phi + X$ is large for a wide range of $c_\phi$. When $m_\phi > m_h/2$, however, the Higgs cannot decay on-shell to $\phi \phi$, and $\phi$ pair production instead proceeds through an off-shell Higgs, $pp \to h^* + X \to \phi \phi + X$. The cross section for this process is then suppressed by an additional factor of $|c_\phi|^2$ as well as two-body phase space. In this regime, $pp$ colliders such as the LHC and a 100~TeV proton collider can provide the best means of probing Higgs Portal DM.
 
With this in mind we assess the reach of $pp$ colliders at $\sqrt{s} = $ 14 \& 100 TeV, with an eye towards constraining the region $m_\phi > m_h/2$ where hadron machines provides sensitivity complementary to electron-positron colliders and direct detection experiments. We implement the Higgs Portal in \texttt{FeynRules} with $m_h = 125$ GeV, generating signal and background events at leading order using \texttt{MadGraph5} \texttt{v1.5.8}~\cite{Alwall:2011uj}, showering with \texttt{Pythia\,8.186}~\cite{Sjostrand:2007gs} tune 4C, and simulating detector effects in \texttt{Delphes\,v3.1.2} with the default CMS detector card (for 14 TeV) and the Snowmass detector card ~\cite{Anderson:2013kxz} (for 100 TeV). We consider various channels, including vector boson fusion, gluon fusion with an associated jet, and $t \bar t$ associated production. In the case of gluon fusion with an associated jet, events generated with \texttt{MadGraph} are re-weighted to more accurately reflect the $p_T$ spectrum of the associated jet. 

The pre-selection and analysis cuts used in these channels are detailed in \cite{Craig:2014lda}.  A simple cut-and-count analysis is performed, determining the exclusion significance of a search in terms of signal events $S$ and background events $B$ passing cuts via $S / \sqrt{S+B}$, and the discovery significance via $S /\sqrt{S}$. For $\sqrt{s} = $ 14 TeV an integrated luminosity of 3 ab$^{-1}$ is assumed, while for $\sqrt{s} =$ 100 TeV scenarios with 3 ab$^{-1}$ and 30 ab$^{-1}$ are considered, respectively. Systematic uncertainties in the signal and background estimates are neglected; systematic uncertainties in background determination could have a substantial impact at $\sqrt{s} = 100$ TeV since $S/B$ is quite small, but one expects data-driven determination of $Z$+jets and other backgrounds to substantially lower systematic uncertainties by the 100 TeV era.

% ---------------------------------------------
\begin{figure}[htbp] 
   \centering
 \includegraphics[width=2.8in]{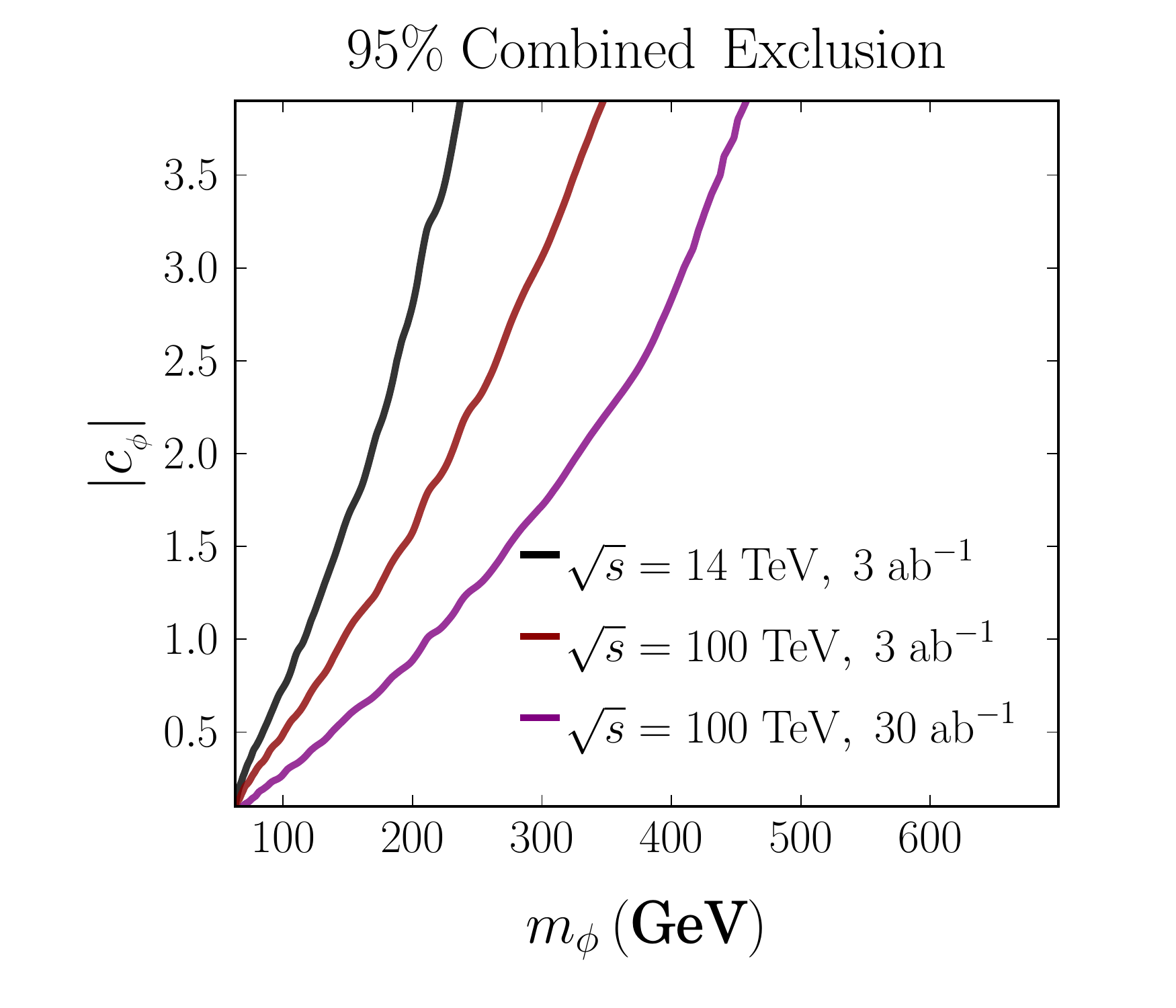} 
  \includegraphics[width=2.8in]{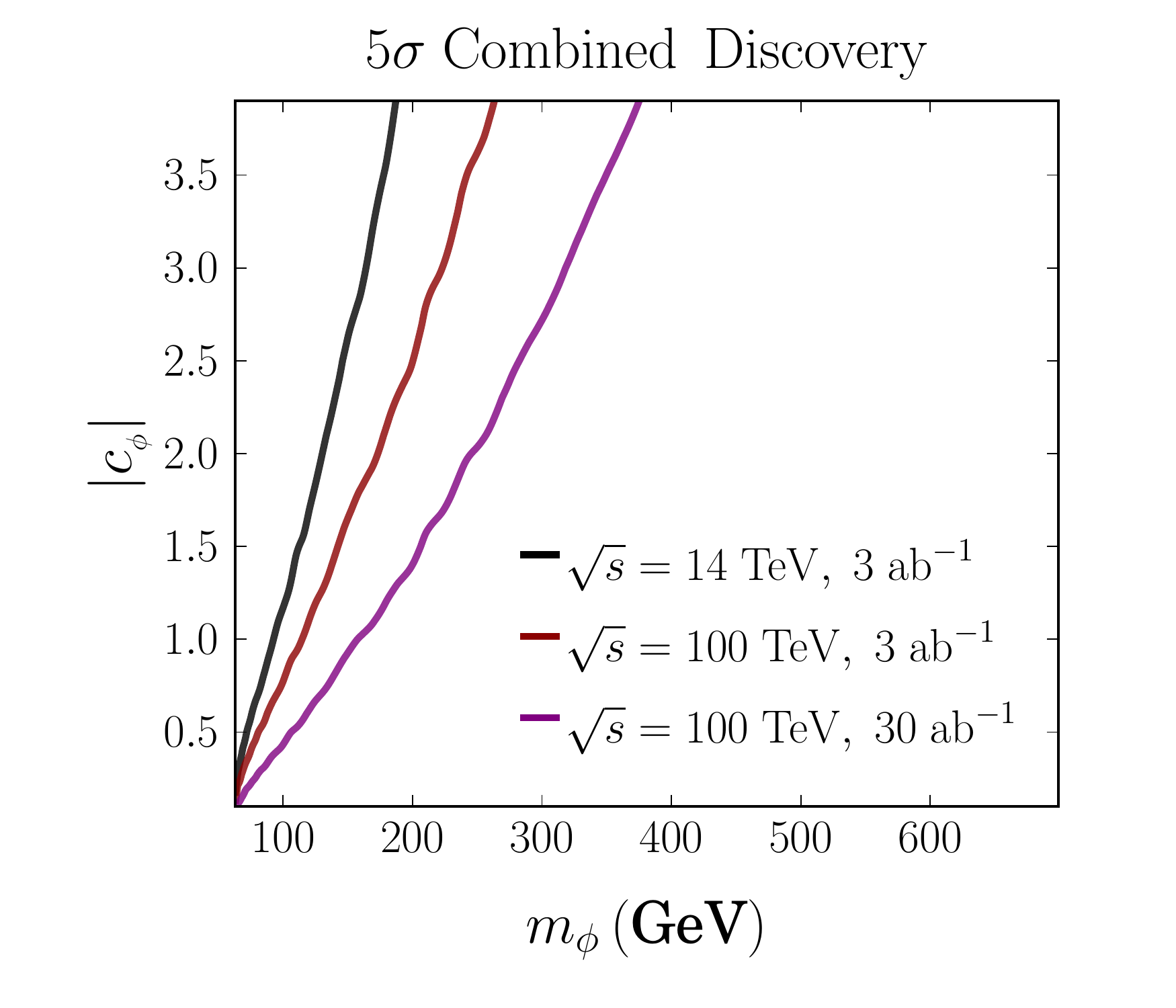} 
{   \caption{Left: Approximate 95$\%$ exclusion reach from the combination of VBF, $ggH$ and $t\bar{t}H$ channels with 3 ab$^{-1}$ at $\sqrt{s} = 14$ and 3, 30 ab$^{-1}$ at $\sqrt{s} = 100$ TeV determined from $S/\sqrt{B} = 1.96$, neglecting systematic errors and correlations between channels. Right: Approximate 5$\sigma$ discovery reach from the same combination at $\sqrt{s} = 14, 100$ TeV.}
   \label{fig:combined}}
\end{figure}
% ---------------------------------------------

To estimate the reach of a concerted Higgs Portal search program, we present the approximate combined reach of VBF, monojet, and $t \bar t$ searches at $\sqrt{s} = 14$ and 100 TeV  in  Fig.~\ref{fig:combined}. We obtain the combination by adding the significance of the VBF,  monojet, and $t \bar t$ channels in quadrature, neglecting possible correlations between the two channels. As the cross section is suppressed at high center-of-mass energies by the off-shell Higgs propagator, the improvement in limits between $\sqrt{s} =14$ TeV and 100 TeV at comparable integrated luminosity is due in part to improved separation of signal from Standard Model backgrounds. 

\begin{figure}[htbp] 
   \centering
 \includegraphics[width=2.8in]{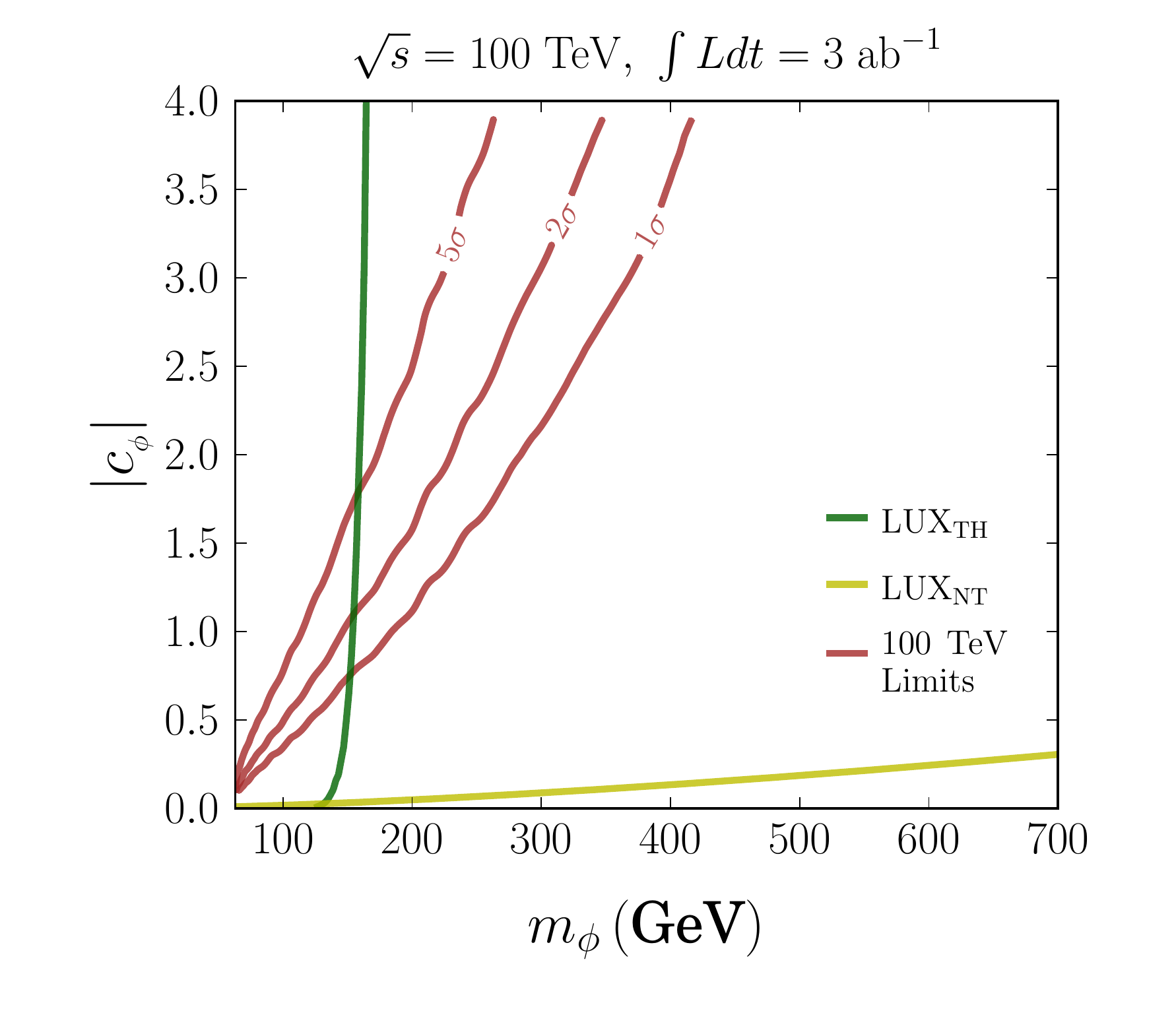}
  \includegraphics[width=2.8in]{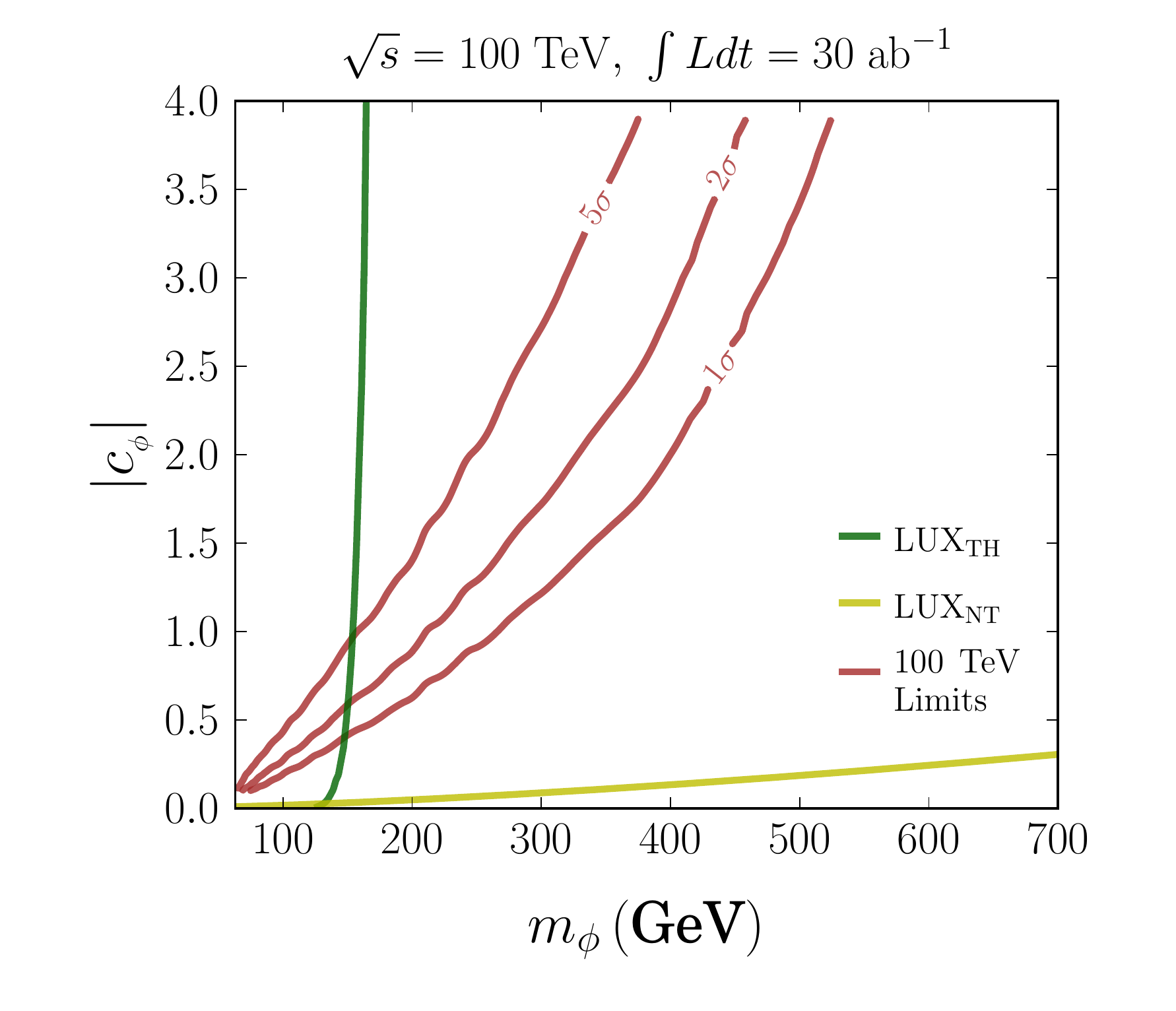}  
{   \caption{Combined reach of direct searches in VBF, $ggH$ and $t\bar{t}H$ channels at $\sqrt{s} = 100$ TeV for 3 ab$^{-1}$ (left) and 30 ab$^{-1}$ (right) compared to DM direct detection. In each plot the red lines denotes the $1\sigma$ exclusion, $2\sigma$ exclusion, and $5\sigma$ discovery reach from direct searches at $\sqrt{s} = 100$ TeV. The region to the left of the green (yellow) line denotes the LUX exclusion for Higgs Portal DM with thermal (non-thermal) abundance given by $c_\phi, m_\phi$. }
   \label{fig:summary}}
\end{figure}

The complementarity between collider searches at 100 TeV and direct detection experiments is illustrated in Fig.~\ref{fig:summary}. Collider searches are not competitive with DM direct detection for small couplings, but at $c_\phi \gtrsim 1$ can exceed the exclusion and discovery reach of the LUX direct detection experiment when the Higgs portal state possesses its natural thermal abundance. In the event of a signal in future direct detection experiments, this also suggests that direct evidence for Higgs Portal states may be obtained through searches at colliders.   In summary, \begin{itemize}
\item  If the $Z_2$ symmetry is exact and the Higgs Portal DM saturates the observed DM density (which may require a non-thermal history), then direct detection probes are likely to be most sensitive.
\item  If the $Z_2$ symmetry is exact and a standard thermal history is assumed then in regions where $\Omega_\phi \leq \Omega_{DM}$ colliders and direct detection experiments provide complementary probes, sensitive to different parameter regions due to a different scaling behavior with the portal coupling $c_\phi$.
\item  If the $Z_2$ symmetry is approximate and only stabilizes $\phi$ on the timescale $\tau \gtrsim 10^{-8}$s but is allowed to decay in the early Universe, or if the $Z_2$ symmetry is exact but $\phi$ has hidden sector decays to other neutral states then colliders are the only probes of the Higgs Portal coupling, with electron-positron colliders constraining $m_\phi > m_h/2$ and proton-proton colliders constraining $m_\phi \geq m_h/2$
\end{itemize}
%
%\bibliography{portalbib}
%\bibliographystyle{jhep}
%
%\end{document}  

\subsection{WIMP Dark Matter, BSM Mediators (Simplified Models)}

%\subsubsection{Simplified Models}
%\subsubsection{Simplified Models Introduction} 
For a large class of models, the search for DM can be simplified to a search for a generic mediator that couples to Standard Model particles\cite{Malik:2014ggr,Abercrombie:2015wmb,Buchmueller:2013dya,Buchmueller:2014yoa,DiFranzo:2013vra,Harris:2014hga,Buckley:2014fba,Haisch:2015ioa,Chala:2015ama,Khoze:2015sra}. The choice of mediator can be used to span the class of different experiements that search for DM.  The simplest split of the mediator class is between spin 0 and spin 1 mediators. For spin 0 mediators, the couplings of the mediator to Standard Model particles are assumed to be yukawa. The prodcution modes thus resemble the Higgs production. This complements the searches for heavy scalar mediators. Additionally, in the case where electroweak symmetry breaking is present, the search directly parallels the heavy higgs searches. For spin 1 mediators, the couplings are flavor universal with equal coupling strength to each of the quarks. Production modes of these resemble Standard Model $Z$ boson production and searches with this model parallel $Z^{\prime}$ production.

The mediators can further be split by the type of coupling structure. In the case of spin 0 mediators, this can be split into scalar and pseudoscalar couplings. While the production cross sections, and sensitivity do not change by much at the collider. Pseudoscalar mediators are velocity suppressed with direct detection, and enhanced with indirect detection. This significantly changes the sensitivity of these experiements. Furthermore, the bounds from relic density can also change significantly.  For spin 1 type mediators, the split in coupling structure yields vector and axial-vector mediators. Again, the sensitivity for collider experiments does not change by a large amount; however, the sensitivity with direct detection changes drastically in direct detection. Axial-vector mediators can only be probed with spin-dependent direct detection , whereas vector mediators can be probed with spin-independent direct detection. Mixed coupling structures are not considered, since they can often be determined from reinterpreted bounds of the purely coupled mediator searches; section~\ref{sec:smsmjdj} explicitly considers interesting combinations of vector and axial-vector mediators. The full lagrangians for these simplified models are described in~\ref{sec:sms}.

For all mediator types, no mixing with Standard Model particles is assumed. Mixing of additional particles, such as a heavy scalar with the Higgs boson, can lead to additional costraints coming from precision measurements of the Higgs couplings. Mixing parameters typically require a completed model and are thus ignored so as to be generic. 
\subsubsection{Simplified Model Collider Bounds }
\label{sec:sms}
In DM searches at hadron colliders, the putative dark particles are pair-produced in collisions of the visible sector particles -- 
the Standard Model quarks and gluons. In the set-up studied here~\cite{Harris:2015kda}, there are no direct interactions between 
the SM sector and the DM particles. Instead these interactions are mediated by an intermediate degree of freedom --
the mediator field. In general, one can expect four types of mediators, 
scalar $S$, pseudo-scalar $P$, vector $Z'$ or axial-vector $Z''$. The corresponding four classes of simplified models 
describing elementary interactions of these four mediators with the SM quarks and with the dark sector fermions $\chi$ 
are 
\begin{align}
\label{eq:LS} 
\mathcal{L}_{\mathrm{scalar}}&\supset\, -\,\frac{1}{2}m_{\rm MED}^2 S^2 - g_{\rm DM}  S \, \bar{\chi}\chi
 - \sum_q g_{SM}^q S \, \bar{q}q  - m_{\rm DM} \bar{\chi}\chi \,,
 \\
 \label{eq:LP} 
\mathcal{L}_{\rm{pseudo-scalar}}&\supset\, -\,\frac{1}{2}m_{\rm MED}^2 P^2 - i g_{\rm DM}  P \, \bar{\chi} \gamma^5\chi
 -\sum_q  i g_{SM}^q  P \, \bar{q}  \gamma^5q  - m_{\rm DM} \bar{\chi}\chi\,,
 \\
 \label{eq:LV} 
\mathcal{L}_{\mathrm{vector}}&\supset \, \frac{1}{2}m_{\rm MED}^2 Z'_{\mu} Z'^{\mu} - g_{\rm DM}Z'_{\mu} \bar{\chi}\gamma^{\mu}\chi -\sum_q g_{SM}^q Z'_{\mu} \bar{q}\gamma^{\mu}q - m_{\rm DM} \bar{\chi}\chi\,,
 \\
 \label{eq:LA} 
\mathcal{L}_{\rm{axial}}&\supset\,  \frac{1}{2}m_{\rm MED}^2 Z''_{\mu} Z''^{\mu} - g_{\rm DM} Z''_{\mu} \bar{\chi}\gamma^{\mu}\gamma^5\chi -\sum_q g_{SM}^q Z''_{\mu} \bar{q}\gamma^{\mu}\gamma^5q - m_{\rm DM} \bar{\chi}\chi\,.
\end{align}
The coupling constant $g_{\rm DM}$ characterizes the interactions of the messengers with the dark sector particles,
which for simplicity we take to be Dirac fermions $\chi$, $\bar{\chi}$, the case of scalar DM particles is a straightforward extension 
of these results. 

The coupling constants linking the messengers to the SM quarks are collectively described by $g^q_{\rm SM}$,
\begin{eqnarray}
{\rm scalar \,\,\& \,\, pseudo-scalar\, messengers:} && \quad g_{\rm SM}^q \equiv\,  g_q\, y_q\,=\, g_q\, \frac{m_q}{v}\,, 
\label{eq:gdef}\\
{\rm vector \,\,\,\& \,\, axial-vector\, \, messengers:} && \quad g_{\rm SM}^q \, =\, g_{\rm SM}\,.
\label{eq:gdef2}
\end{eqnarray}
For scalar and pseudo-scalar messengers the couplings to quarks are taken to be proportional to the corresponding Higgs Yukawa couplings, $y_q$ as in models with minimal flavour violation \cite{D'Ambrosio:2002ex}, and we keep the scaling 
$g_q$ flavour-universal for all quarks.
For axial and vector mediators  $g_{\rm SM}$ is a gauge coupling in the dark sector which we also
take to be flavour universal.
The coupling parameters which we can vary are thus $g_{\rm DM}$ plus either $g_q$ or $g_{\rm SM}$, the latter choice depending on the messengers.\footnote{In Ref.~\cite{Harris:2014hga}, $g_{\rm DM}$ is parameterised for (pseudo-)scalar
messengers as $g_{\rm DM}\,=\, g_\chi\, m_{\rm DM}/v$ to look symmetric w.r.t. \eqref{eq:gdef}, and  
$g_\chi$ is treated as a free parameter. Here we do not impose this requirement and leave $g_{\rm DM}$ as the free parameter.}

In general, the simplified model description of the dark sector is characterised by five parameters:
the mediator mass $m_{\rm MED}$, the mediator width 
$\Gamma_{\rm MED}$, the dark particle mass $m_{\rm DM}$, and the mediator-SM and the mediator-Dark sector couplings, 
$g_{\rm SM}$, $g_{\rm DM}$. Out of these, the mediator width $\Gamma_{\rm MED}$,
does not appear explicitly in the simplified model Lagrangians \eqref{eq:LS}-\eqref{eq:LA} and should be specified separately.
$\Gamma_{\rm MED}$ accounts for the allowed decay modes of a given mediator particle into other particles from the visible 
and the dark sector.
In a complete theory, $\Gamma_{\rm MED}$ can be computed from its Lagrangian, but in a simplified model we can instead determine only the
so-called minimal width $\Gamma_{\rm MED, min}$, i.e. the mediator width computed using the mediator interactions with the SM quarks 
and the $\bar{\chi}$, $\chi$ DM particles defined in Eqs.~\eqref{eq:LS}-\eqref{eq:LA}. 
Importantly  $\Gamma_{\rm MED, min}$ does not take into account the possibility of the mediator to decay into
e.g. other particles of the dark sector, beyond $\bar{\chi}$, $\chi$, which would increase the value of $\Gamma_{\rm DM}$.
In Ref.~\cite{Harris:2014hga} the role of $\Gamma_{\rm MED}$ is investigated as an independent parameter in the simplified models
characterisation of dark sectors by using a simple grid for $\Gamma_{\rm DM}=\{1,2,5,10\}\times \Gamma_{\rm MED, min}$, it is known that this can reduce the sensitivity substantially. We instead adopt a reduced simplified description where the width is set to its minimal computed
 value $\Gamma_{\rm MED, min}$ which amounts to larger signal cross-sections 
 (we will also check that $ \Gamma_{\rm MED, min} < m_{\rm MED}/2$). For our simplified models we have
\begin{equation}
\label{eq:GVA}
\Gamma_{\rm MED, min}\,=\, \Gamma_{\chi\overline{\chi}} \,+ \,\sum_{i=1}^{N_f} N_c\, \Gamma_{q_i\overline{q}_i}  
\end{equation}
where $\Gamma_{\chi\overline{\chi}}$ is the mediator decay rate into two DM fermions, and the sum  is over the SM quark flavours.
Depending on the mediator mass, decays to top quarks may or may not be open i.e. $m_{\rm MED}$ should
 be $> 2m_t$ for an open decay. The partial decay widths of vector, Axial-vector, scalar and pseudo-scalar mediators into fermions are
 given by,
\begin{eqnarray}
\label{eq:GV}
\Gamma^{V}_{f\overline{f}} &=& \frac{g^2_{f}(m_{\rm MED}^2+2m_f^2)}{12 \pi m_{\rm MED}}\sqrt{1-\frac{4m_f^2}{m_{\rm MED}^2}} \quad ,
\quad
\Gamma^{A}_{f\overline{f}} \,=\, \frac{g^2_{f}(m_{\rm MED}^2-4m_f^2)}{12 \pi m_{\rm MED}}\sqrt{1-\frac{4m_f^2}{m_{\rm MED}^2}} \\
\label{eq:GS}
\Gamma^{S}_{f\overline{f}} &=& \frac{g^2_{f}}{8\pi }\,m_{\rm MED}\,\left(1-\frac{4m_f^2}{m_{\rm MED}^2}\right)^\frac{3}{2}
\quad ,\quad
\Gamma^{P}_{f\overline{f}} \,=\, \frac{g^2_{f}}{8\pi}\,m_{\rm MED}\,\left(1-\frac{4m_f^2}{m_{\rm MED}^2}\right)^\frac{1}{2}
\end{eqnarray}
where $m_f$ denotes masses of either SM quarks $q$ or DM fermions $\chi$ and the coupling constant $g_{f}$ denotes
either $g_{\rm SM}$ or $g_{\rm DM}$.

\begin{center}
\begin{figure}[ht]
\includegraphics[width=0.3\textwidth]{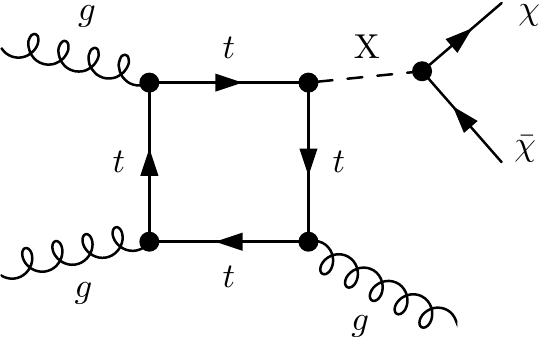} \hspace{0.5cm} %\hspace{1.5cm}
\includegraphics[width=0.28\textwidth]{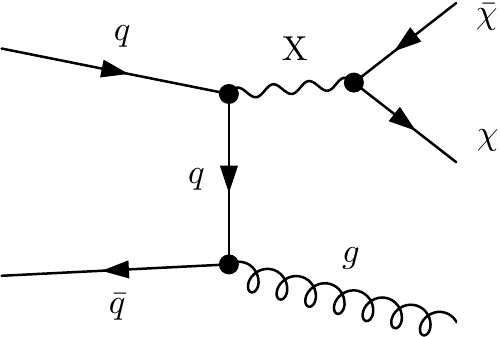}
\caption{Representative Feynman diagrams for gluon and quark induced mono-jet plus MET processes. The mediator X can be a scalar, pseudo-scalar, vector or axial-vector particle. The gluon fusion process involves the heavy quark loop which we compute in the microscopic theory, while the quark-anti-quark annihilation is a tree-level process at leading order.
}
\label{fig:feyn}
\end{figure}
\end{center}

For the simplified DM searches, the most universal DM search can be done by performing the jets+MET search (so-called monojet search)\cite{Abazov:2003gp,Aaltonen:2012jb, Chatrchyan:2012me,ATLAS-CONF-2012-084,Khachatryan:2014rra,Diehl:2014dda,Feng:2005gj,Cao:2009uw,Beltran:2010ww,Goodman:2010yf,Goodman:2010ku,Fox:2011pm,Haisch:2012kf}.Depending on the choice for the mediator field different production mechanisms will contribute. 
For vectors and axial-vectors the dominant mechanism is the
quark-antiquark annihilation at tree-level.  For scalars and pseudo-scalars on the other hand, 
the loop-level gluon fusion processes are more relevant. The representative Feynman diagrams for both channels are
shown in Fig.~\ref{fig:feyn}. In comparing DM collider searches with direct and indirect detection experiments it is important to keep in mind that 
our collider processes and limits continue to be applicable for discovery of any dark sector particles escaping the detector.
Hence dark particles produced at colliders do not have to be the cosmologically stable DM.

Regarding the possible origin and the UV consistency of the simplified models \eqref{eq:LS}-\eqref{eq:LA},
the scalar and pseudo-scalar messenger fields in our simplified models \eqref{eq:LS}-\eqref{eq:LP}
are singlets under the Standard Model. The simplified models \eqref{eq:LS}-\eqref{eq:LP} can arise from two types of the 
more fundamental theories. The simplest theories of the first type are the two-Higgs-doublet models \cite{Branco:2011iw}.
In this case the mediators would originate from the second Higgs doublet. The other type of models
giving rise to our simplified models are even simpler in the sense that scalar mediators (and the dark sector particles 
they are coupled to) can be genuinely neutral under the SM but mix with the neutral component of the Higgs\cite{Silveira:1985rk, Schabinger:2005ei, Patt:2006fw, Englert:2011yb}.These models provide a direct connection of the dark sector with Higgs physics and can link 
the origin of the electroweak and the DM scales~\cite{Englert:2013gz,Hambye:2013sna,Carone:2013wla,Khoze:2014xha}.
The simplified dark sector models with vector and axial-vector mediators in Eqs.~\eqref{eq:LV}-\eqref{eq:LA}
can also be derived from appropriate first-principles 
theories.  Since the mediators are spin-one particles, these UV models would necessarily require the mediators 
to be gauge fields and the DM to be charged under these gauge transformations.
A classification of anomaly-free 
extensions of the Standard Model  Abelian $U(1)'$ factor was given in \cite{Carena:2004xs}
and can be used for constructing an example of a consistent gauge-invariant vector and axial theories of the type \eqref{eq:LA}.

\paragraph{Dark matter projections}
\label{sec:detmc}
Difficulty exists in correctly modeling the production of the backgrounds at 100~TeV. In particular, the knowledge of the gluon pdfs, the influence of higher order QCD effects, and corrections coming from the electorweak Sudakovs. At 100~TeV collider energies, emission 
of additional radiation will result in copious jet-production around the Electroweak scale. This will require delicate handling 
with respect to matching and merging of parton shower and matrix element emissions. Given the likely 
timescale of construction, and the rapid improvement in theoretical tools, none of the above issues should be regarded as significantly 
likely to negatively affect the physics program at a 100~TeV collider. For this study, we probe the sensitivity of the monojet search at the 100~TeV collider. The dominant backgrounds for events in either the LHC or the future collider will come from $Z\rightarrow\nu\bar{\nu}$,$W\rightarrow\ell\nu$, and $t\bar{t}$ production.  To simulate a hypothetical study, all samples are done using aMC@NLO~\cite{Alwall:2014hca} with 0,1,2 jets merged with the excepton $W$+jets, where the second jet was not produced. 

For the signal we use MadGraph for the Vector/Axial simplified models and a combination of
MCFM~\cite{Fox:2012ru,MCFMweb} and VBFNLO~\cite{Arnold:2008rz,Arnold:2011wj,Baglio:2014uba} for the production of Scalar/Pseudoscalar mediators in association with one
and two-jets. The output LHE events are then merged using the CKKW-L
interface of Pythia 8~\cite{Sjostrand:2007gs}. NNPDF3.0~\cite{Ball:2014uwa} PDF's are
used for the generation of all Monte-Carlo samples. This scheme of generation allows for the full use of the second jet in the discrimination of signal and background.

The signal extraction is performed with a full shape analysis of the MET distribution following a selection of the monojet final state.  The dominant backgrounds combine from $Z\rightarrow\nu\bar{\nu}$ production. The second largest background is comes from the $W$ boson production where a lepton is either fails the lepton identification or is out of the acceptance of the detector. The third largest background comes from $t\bar{t}$ production where again a lepton from one of the W boson decays is outside of the detector volume.  For the $Z\rightarrow \nu\bar{\nu}$ background, the $Z\rightarrow\mu^+\mu^-$
control region is used to model the background. For the $W\rightarrow \ell \nu$,
top and diboson backgrounds, we use the single lepton control region. For each of these control regions the full statistical
uncertainty on the shape is propagated per bin on each of the
backgrounds with an additional one percent uncertainty uncorrelated
per bin to account for additional modelling uncertainties. For all
but the tail bins of the shape uncertainties on the \MET spectrum are
roughly 1\% with the dominant uncertainty resulting from the
additional one percent modelling uncertainty. The signal is
profiled using the standard limit extraction ($\mathrm{CL}_{s}$) \cite{Read:2002hq,Cowan:2010js}. Additional
nuisances are placed on the background normalization for lepton
efficiencies and luminosity. The overall uncertainy setup is extremely conservative since more advanced approaches are in use at the LHC. Also, it is likley that advances in the understanding of higher order electorweak  and QCD corrections will be able to further constrain these backgrounds to sub-percentage precision. 

Detector effects for a pseudo future high energy detector, and LHC detector are simulated requiring the same jet and MET resolutions as the CMS detector with the one exception that the detector has an added lepton acceptance extended up to $|\eta| < 4.0$ and $|\eta| < 5.5$ for the 14~TeV and 100~TeV detectors respectively~\cite{Butler:2020886,ATLAS:1502664,Khachatryan:2014gga}.  Effects from pileup are taken to account to match the expected conditions for high luminosity running at the LHC. 
\begin{center}
\begin{figure}[ht]
\includegraphics[width=0.5\textwidth]{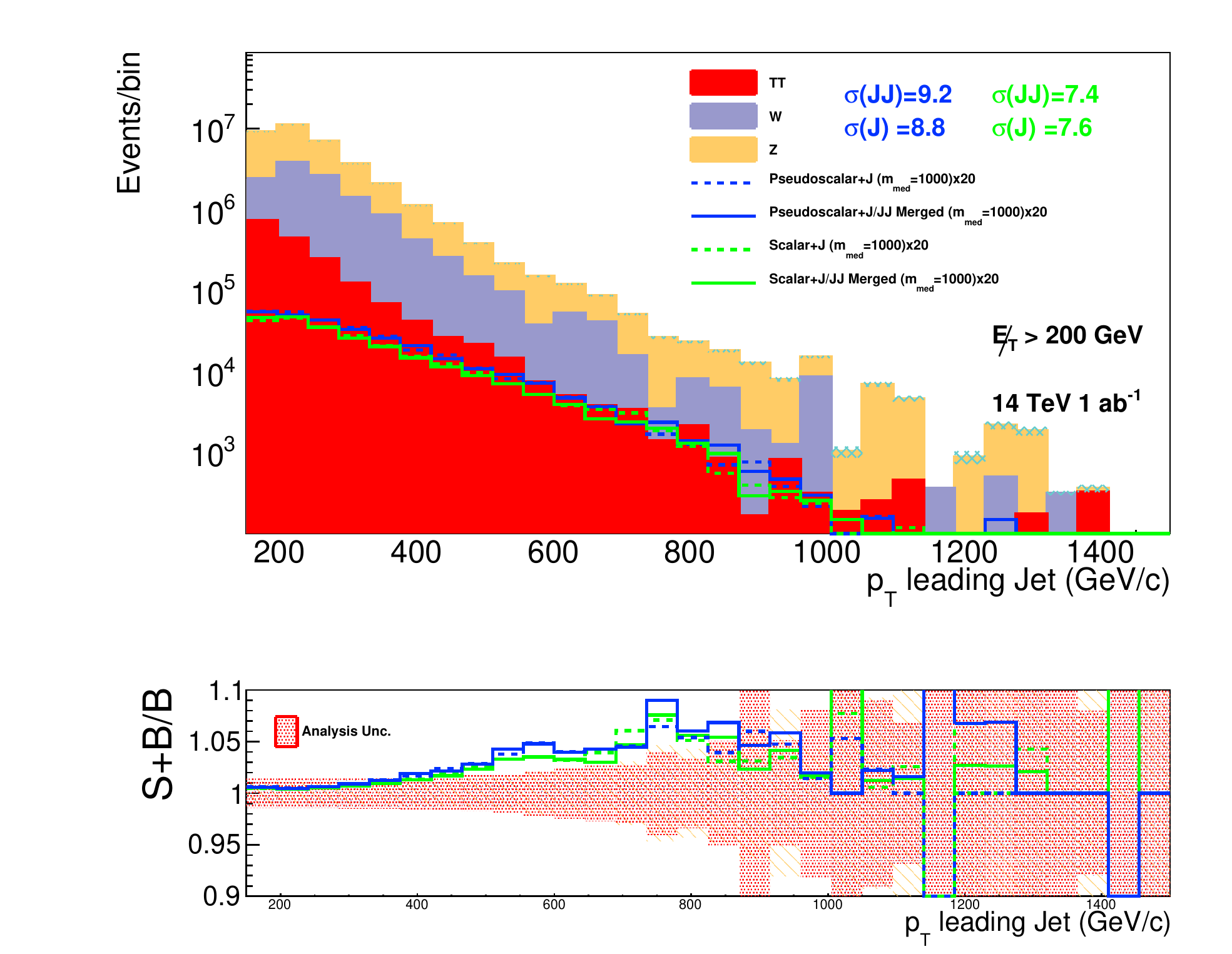} \hskip-0.5cm
\includegraphics[width=0.5\textwidth]{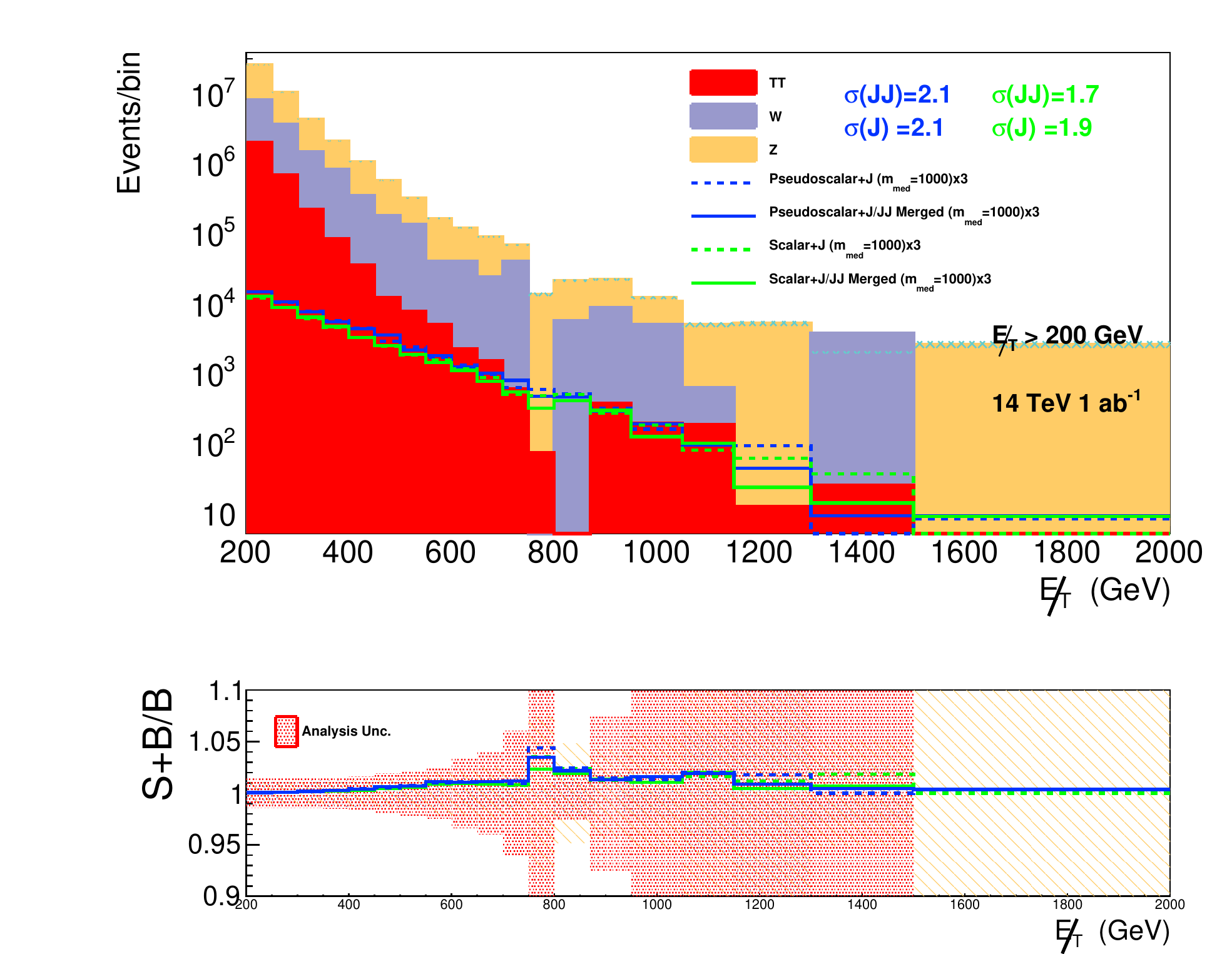}
\includegraphics[width=0.5\textwidth]{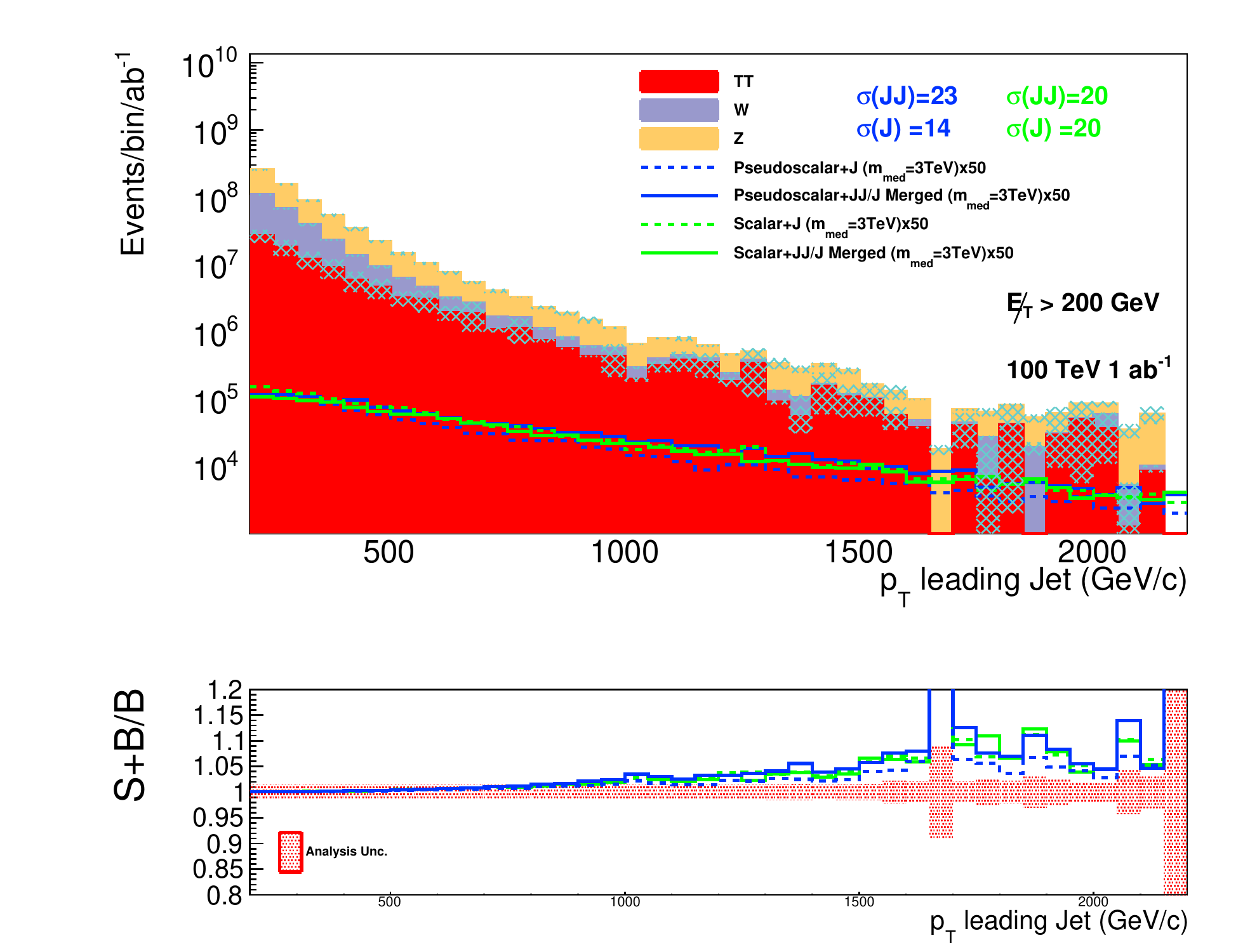} \hskip-0.5cm
\includegraphics[width=0.5\textwidth]{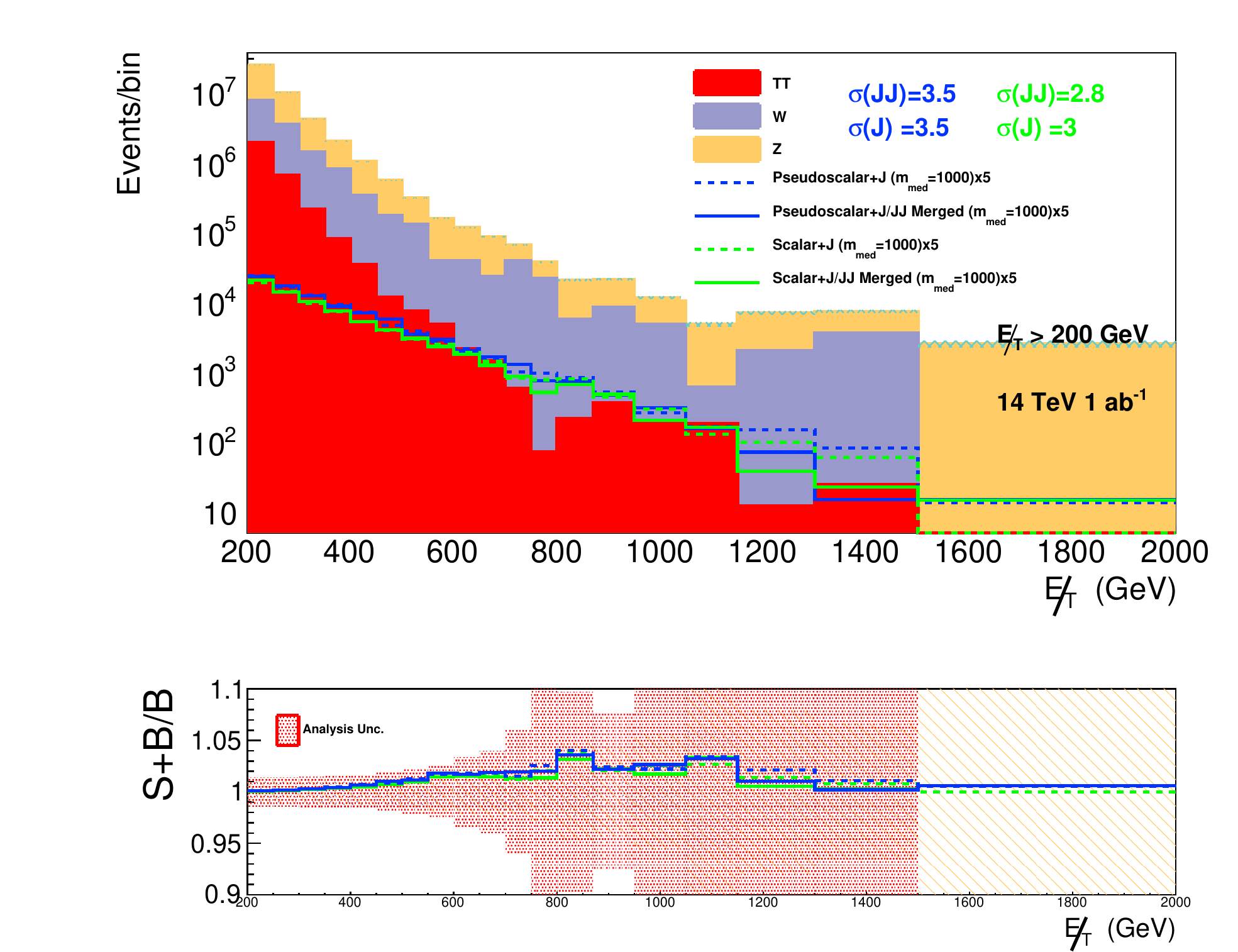}
\caption{Kinematic distributions for signal scalar and mediator models and the SM backgrounds at $14$\UTeV (top) and $100$\UTeV (bottom)
assuming 1~ab$^{-1}$ of integrated luminosity. We show four kinematic 
variables: $p_T$ of the leading jet (left)and the missing energy \MET (right). Ratios of $(S+B)/B$ are shown for each observable. The red bands indicate the uncertainties on the background distributions. The accordingly color-coded numbers for sig(JJ) and sig(J) give the statistical significance to disfavour the presence of the signal using the $\mathrm{CL}_s$ method. }
\label{fig:kin14TeVscal}
\end{figure}
\end{center}

Kinematic distributions for our simplified models of dark sectors alongside the main SM backgrounds 
are shown in Fig.~\ref{fig:kin14TeVscal}. The distributions are shown as functions of two kinematic variables, $p_T$ of the leading jet, and missing energy \MET. The  event selection cuts imposed for the distributions in Fig.~\ref{fig:kin14TeVscal} are $\slashed{E}_T \geq 200$ GeV and $\min ( \Delta\phi_{\slashed{E}_T,j_i}) \geq 0.5$, where $i$ runs over all jets in each event. 

\paragraph{Direct and Indirect Detection Limits}
\label{sec:limits}
Comparisons for direct and indirect detection cross sections be determined from the Lagrangians Eq.~(\ref{eq:LS})-(\ref{eq:LA}) giving,
\begin{equation}
\sigma_{\mathrm{\chi p}}^V = \frac{9}{\pi} \frac{g_{\rm DM}^2 g_{\rm SM}^2 \rho^2}{m^4_{\mathrm{MED}}}
\label{eq:16}
\end{equation}
and 
\begin{equation}
\sigma_{\mathrm{\chi p}}^A = \frac{3}{\pi}\frac{g_{\rm DM}^2 g_{\rm SM}^2 a^2 \rho^2}{m^4_{\mathrm{MED}}},
\label{eq:18}
\end{equation}
with $a \simeq 0.43$ \cite{Cheng:2012qr, Buchmueller:2013dya} and the reduced mass $\rho =m_{{\rm DM}} m_{p}/(m_{\rm DM} + m_{p})$, for the cross section of a DM particle scattering spin-independently (vector mediator) or spin-dependently (axial-vector mediator) from a proton.

The cross section for a DM particle scattering from a nuclei via a scalar mediator of Eq.~(\ref{eq:LS}) is given by 
\cite{Kurylov:2003ra,Hisano:2010ct, Cheung:2013pfa}
\begin{equation}
\sigma_{\mathrm{\chi p}}^S =  \frac{\rho^2}{\pi}  \left | \sum_{q=u,d,s} f^{p}_q \,\frac{m_p}{m_q}\left( \frac{g_{DM} g_q y_q}
{m^2_{\mathrm{MED}}} \right )    
+ \frac{2}{27} f_{\mathrm{TG}} \,\sum_{q=c,b,t} \frac{m_p}{m_q}\left( \frac{g_{DM} g_q y_q}{m^2_{\mathrm{MED}}} \right ) \right |^2,
\label{eq:17}
\end{equation}
where $f^{p}_u= 0.019$, $f^{p}_d=  0.045$, $f^{p}_s= 0.043$ and 
$f_\mathrm{TG}\simeq 1-\sum_{q=u,d,s}f^{n}_{q} $ \cite{Hoferichter:2015dsa,Crivellin:2013ipa}
and $m_p$ is the proton mass.

When comparing the expected sensitivity for the LHC and a 100~TeV collider for DM searches to those of Direct Detection it is interesting to compare the 
expected impact of the neutrino wall~\cite{Cushman:2013zza,Buchmueller:2014yoa}.  We take their interaction cross section to be indicative for the ultimate reach of DD experiments~\cite{Cushman:2013zza,Buchmueller:2014yoa}.  
For a pseudo-scalar mediator, taking existing limits into account \cite{Ackermann:2011wa,Abdo:2010ex}, indirect detection experiments can result in stronger limits than direct detection experiments \cite{Zheng:2010js, Boehm:2014hva}. For the simplified model of Eq.~(\ref{eq:LP}), we use the velocity-averaged DM annihilation cross section into $\bar{b}b$,
\begin{equation}
\left < \sigma v \right >_{\bar{b}b}^P = \frac{N_C}{2 \pi} \frac{(y_b g_{b})^2 g_{DM}^2  \, m_{\rm DM}^2 } {(m_{\rm MED}^2 - 4 m_{\rm DM}^2)^2 + m_{\rm MED}^2 \Gamma_{\rm MED}^2} \sqrt{1 - \frac{m_b^2}{m^2_{DM}}},
\label{eq:19}
\end{equation}
which allows us to derive a limit on the parameters in the $\bar{b}b$ channel \cite{Ackermann:2011wa}. 

\paragraph{Results} 
\label{sec:results}
Results are obtained scanning over a spectrum of signal models at 14 TeV and 100 TeV. A predicted luminosity of 1~ab$^{-1}$ is used for both
analyses, so the sensitivity can be compared directly.  We note that this  amount of integrated luminosity is a rather modest amount compared to what is likely 
to be collected at a future collider, the LHC bounds on the other hand represent a qualitative upper bound given the run plans over the next 10 years. 

Figure~\ref{fig:exclCrossS} presents the total cross section which the analysis excludes for each of the four mediator types defined in Eqs.~\eqref{eq:LS}-\eqref{eq:LA}.We define our cross sections by setting $g_{\mathrm{DM}} = g_{\mathrm{SM}} = 1$ and select the mediator mass as indicated in the legend of each figure respectively. 
As an illustrative example we have chosen a relatively small characteristic value of 100 GeV, although the results 
obtained for other kinematically accessible values of DM mass were found to be similar.  
The kinematics of the process are then completely specified once the couplings $g_{\rm DM}$ and $g_{\rm SM}$  are set, since this fixes the minimal width of the mediator \cite{Harris:2014hga}. The excluded cross section is then related to the predicted cross section as follows, 
\begin{equation}
\label{eq:mu}
\sigma = \mu ~ \sigma(g_{\rm DM} = 1, g_{\rm SM}=1, m_{\rm MED}),
\end{equation} 
With the kinematics of the model fixed we set a limit on $\mu$ defined above using the $\mathrm{CL}_s$-method, again  assuming $1~\mathrm{ab}^{-1}$ of data. Values with $\mu<1$  indicate the excluded couplings and  width are smaller than the tested model, and the point is then excluded. In Fig.~\ref{fig:exclCrossS} we also distinguish between the mono-jet (shown in green) and the multi-jet-based analyses (shown in yellow). It can be seen that the new multi-jet-based analysis is more powerful and provides a considerable improvement at 14 and at 100 TeV.

\begin{center}
\begin{figure}[ht]
\includegraphics[width=0.51\textwidth]{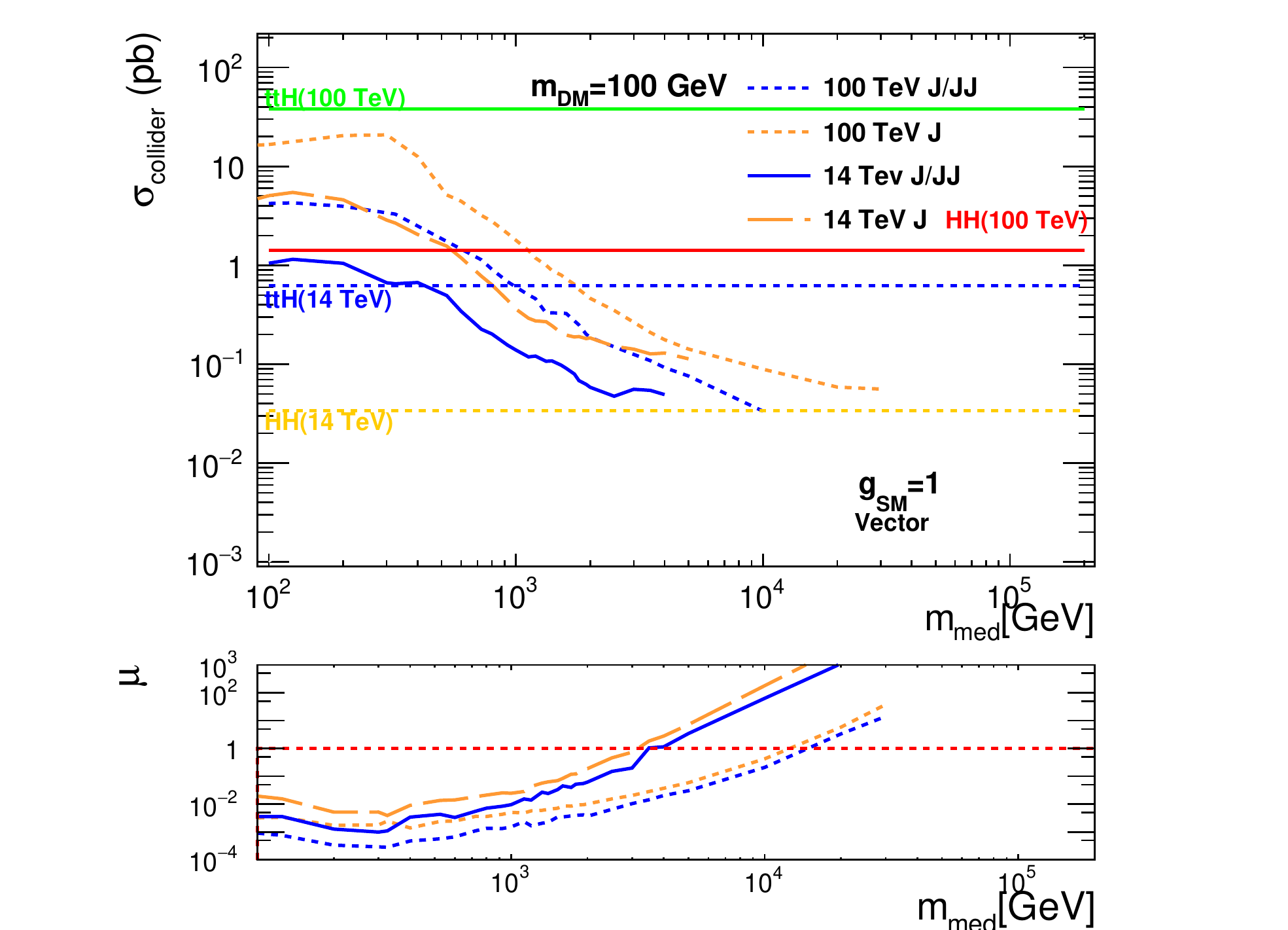} \hskip-0.9cm
\includegraphics[width=0.51\textwidth]{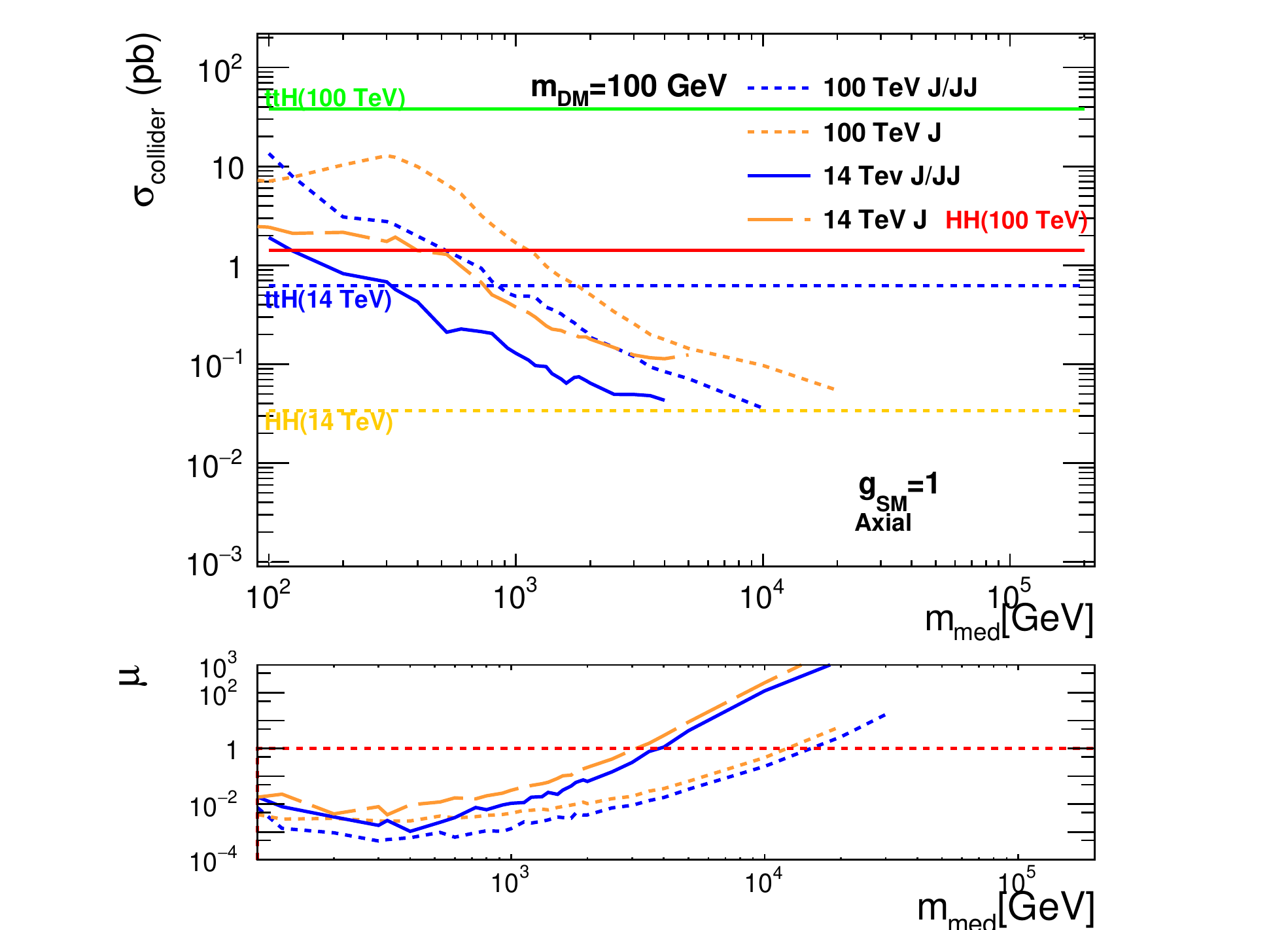}\\
\includegraphics[width=0.51\textwidth]{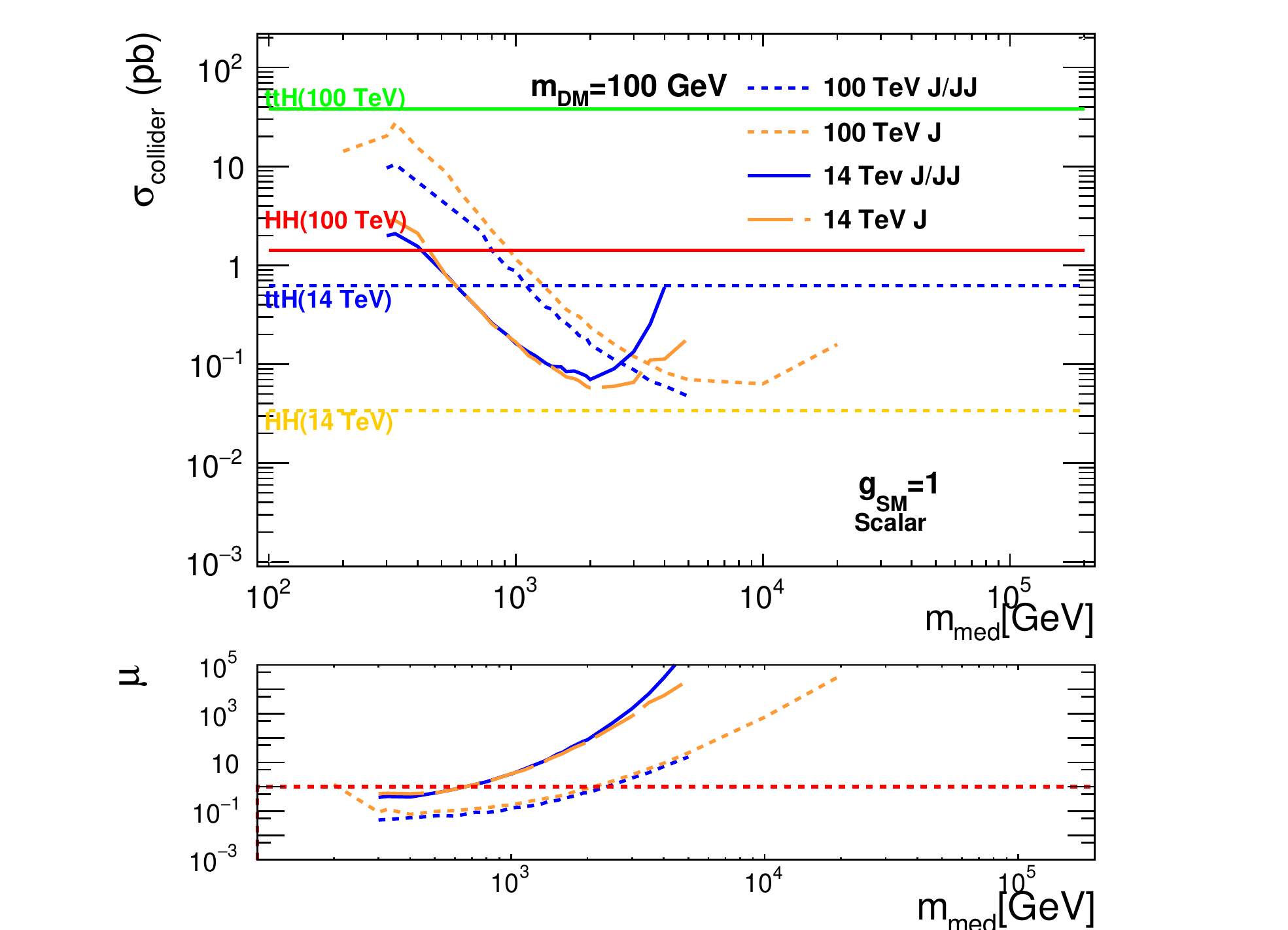} \hskip-0.9cm
\includegraphics[width=0.51\textwidth]{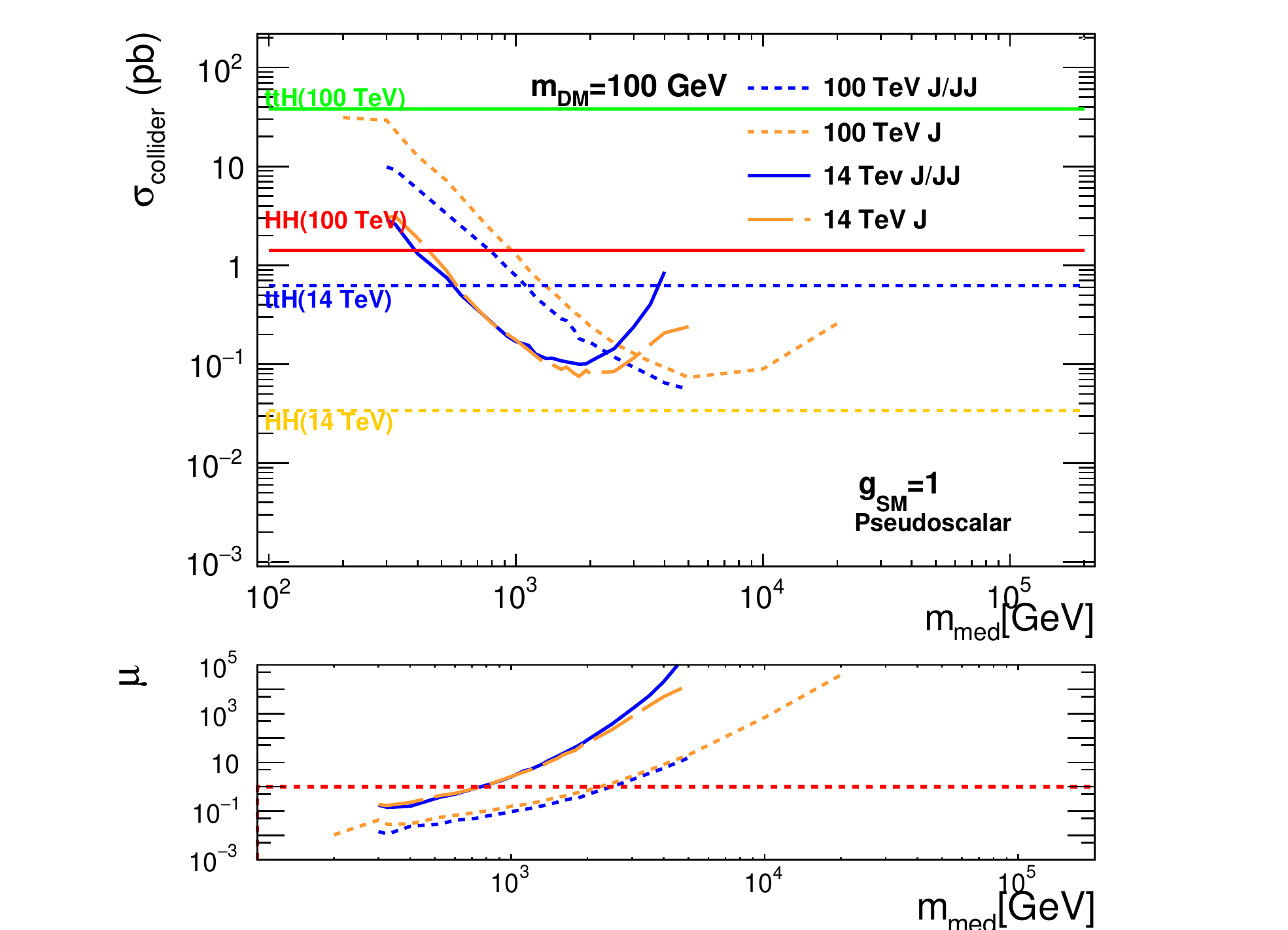}
\caption{Cross section exclusion limits as a function of mediator mass for a fixed DM mass   at a given coupling. We show results for vector (upper left panel), axial-vector (upper right panel), scalar (lower left panel) and pseudoscalar (lower right panel)}
\label{fig:exclCrossS}
\end{figure}
\end{center}
For the case of scalar and pseudo-scalar mediators at 14 TeV there is a cross-over for mediators heavier
than $\simeq 1$ TeV, which is absent at 100 TeV. This corresponds to exactly the regions of phase space in which the off-shell 
effects dominate. The one-jet sample has access to the significant cross section which arises from the tails of the Breit-Wigner distribution, whereas the 
multileg sample does not. This region therefore has large theory errors using the multi-leg sample. However, we note that the region of phase space 
for which the multi-leg sample breaks down is far from the values of $\mu =1$, so this region of phase space is of limited importance in regards to setting 
limits on model parameters. Finally we note that Fig.~\ref{fig:exclCrossS} also includes cross sections for interesting SM predictions which the 100~TeV collider and Run II of the LHC will investigate. We present the cross sections for $t\overline{t}H$ and $HH$ and show their relative size compared to our DM predictions. 

\begin{center}
\begin{figure}[ht]
\includegraphics[width=0.5\textwidth]{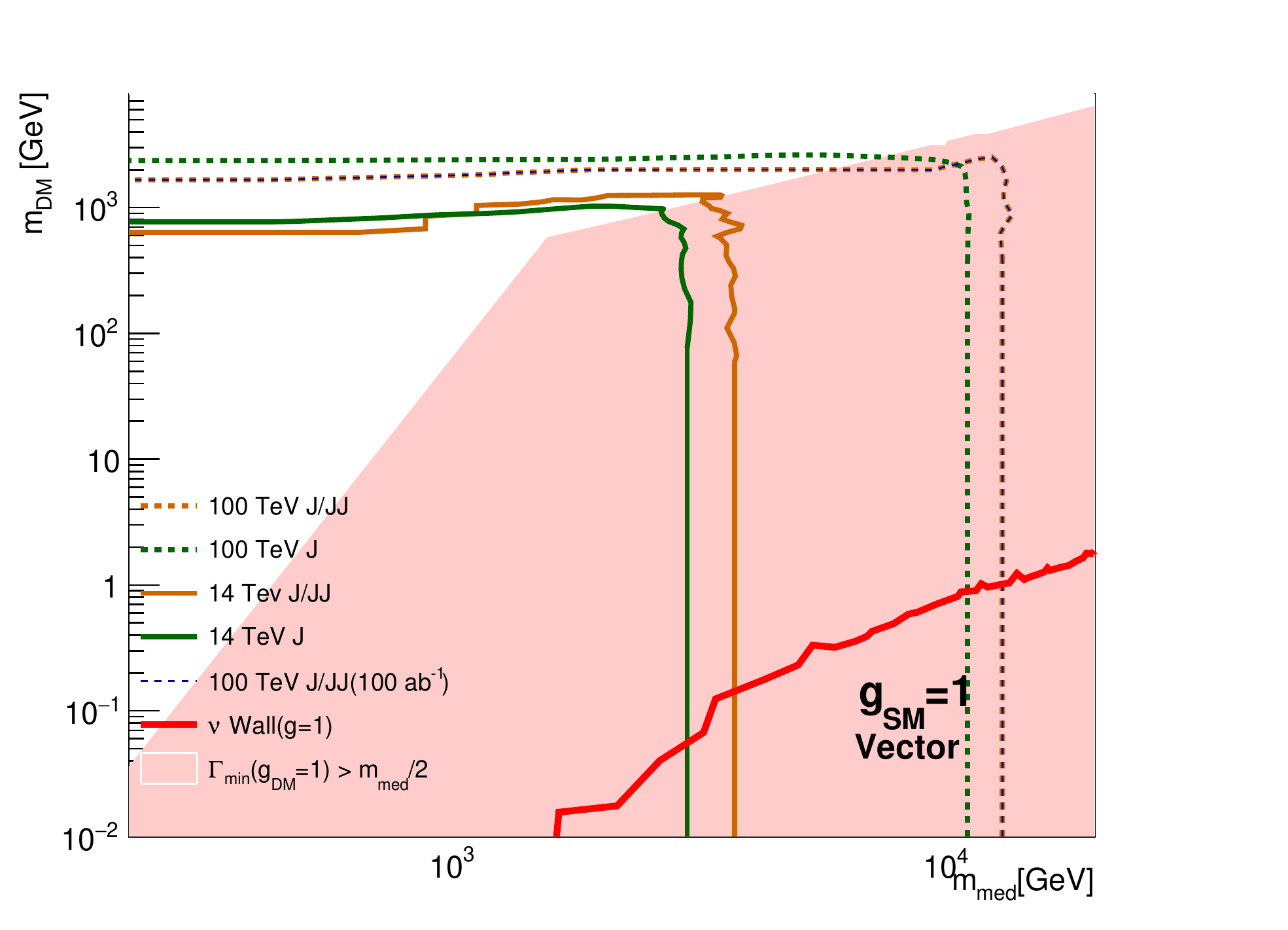} \hskip-0.6cm
\includegraphics[width=0.5\textwidth]{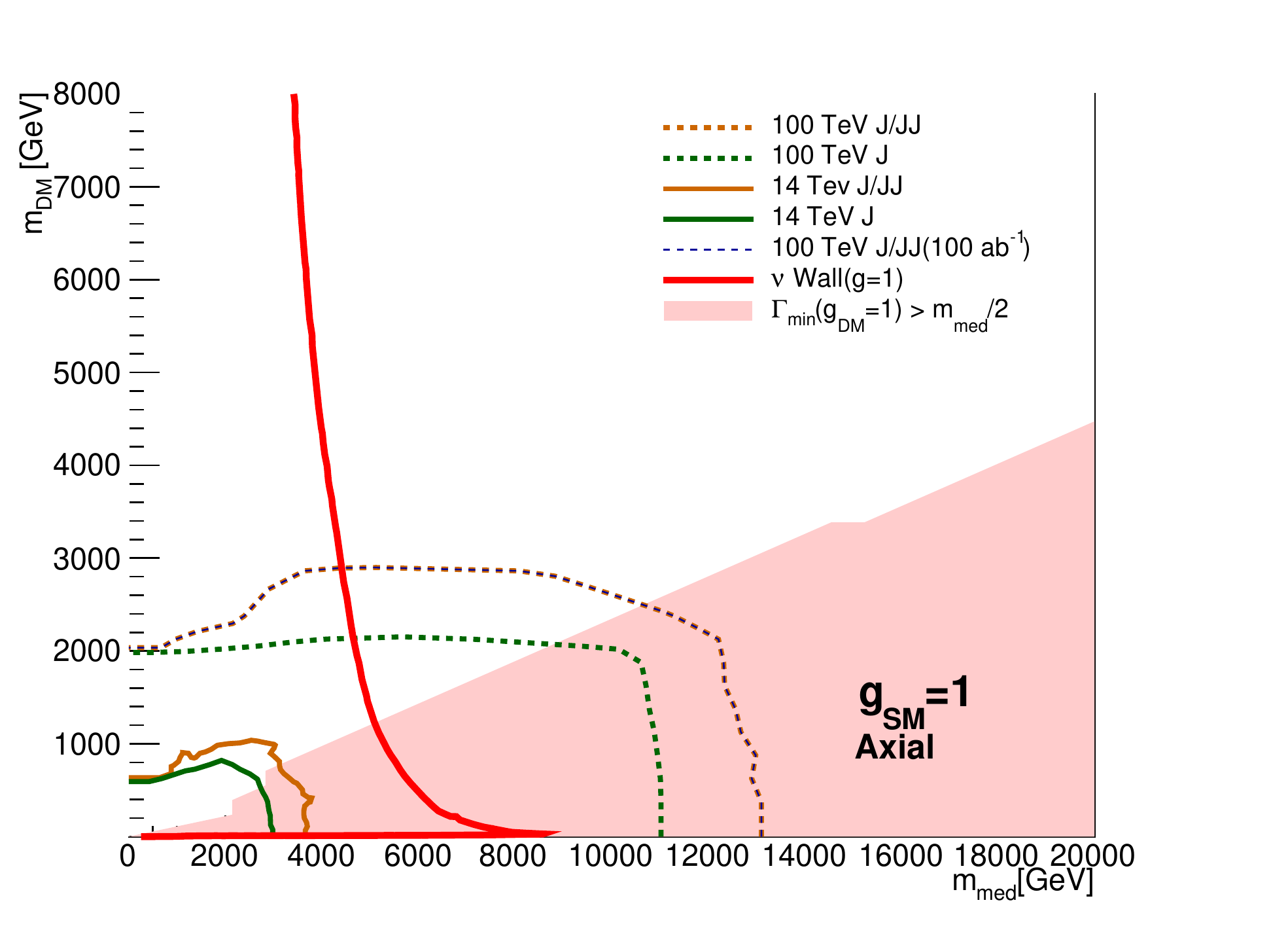}
\caption{Mass limits for vector mediator models (left panel) and axial-vector models (right pannel) at $14$ and $100$ TeV colliders 
using the multi-leg and a single-leg analysis. We also show the neutrino wall limit of the direct detection. }
\label{fig:exclMass:9}
\end{figure}
\end{center}

\begin{center}
\begin{figure}[ht]
\includegraphics[width=0.5\textwidth]{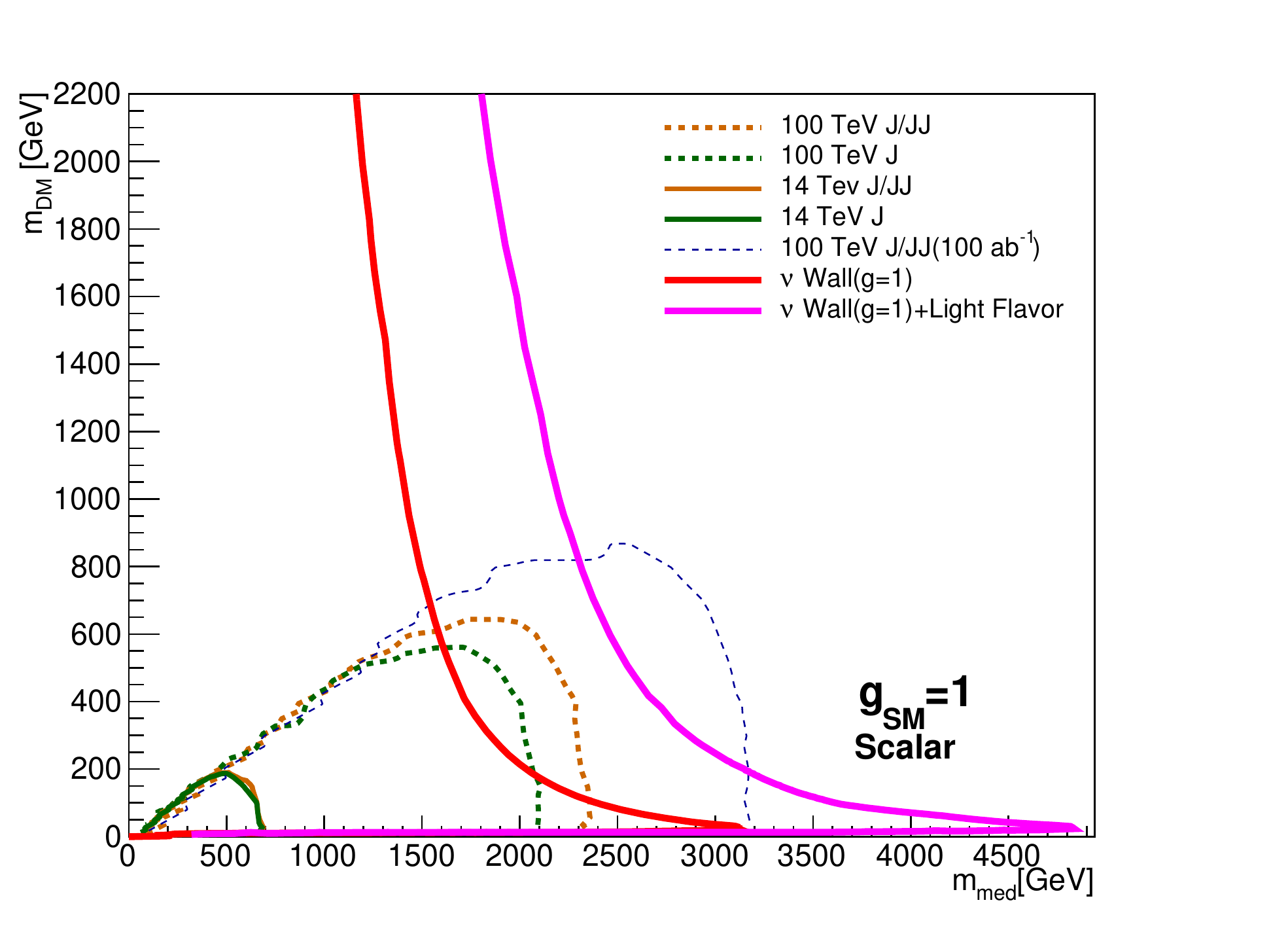} \hskip-0.6cm
\includegraphics[width=0.5\textwidth]{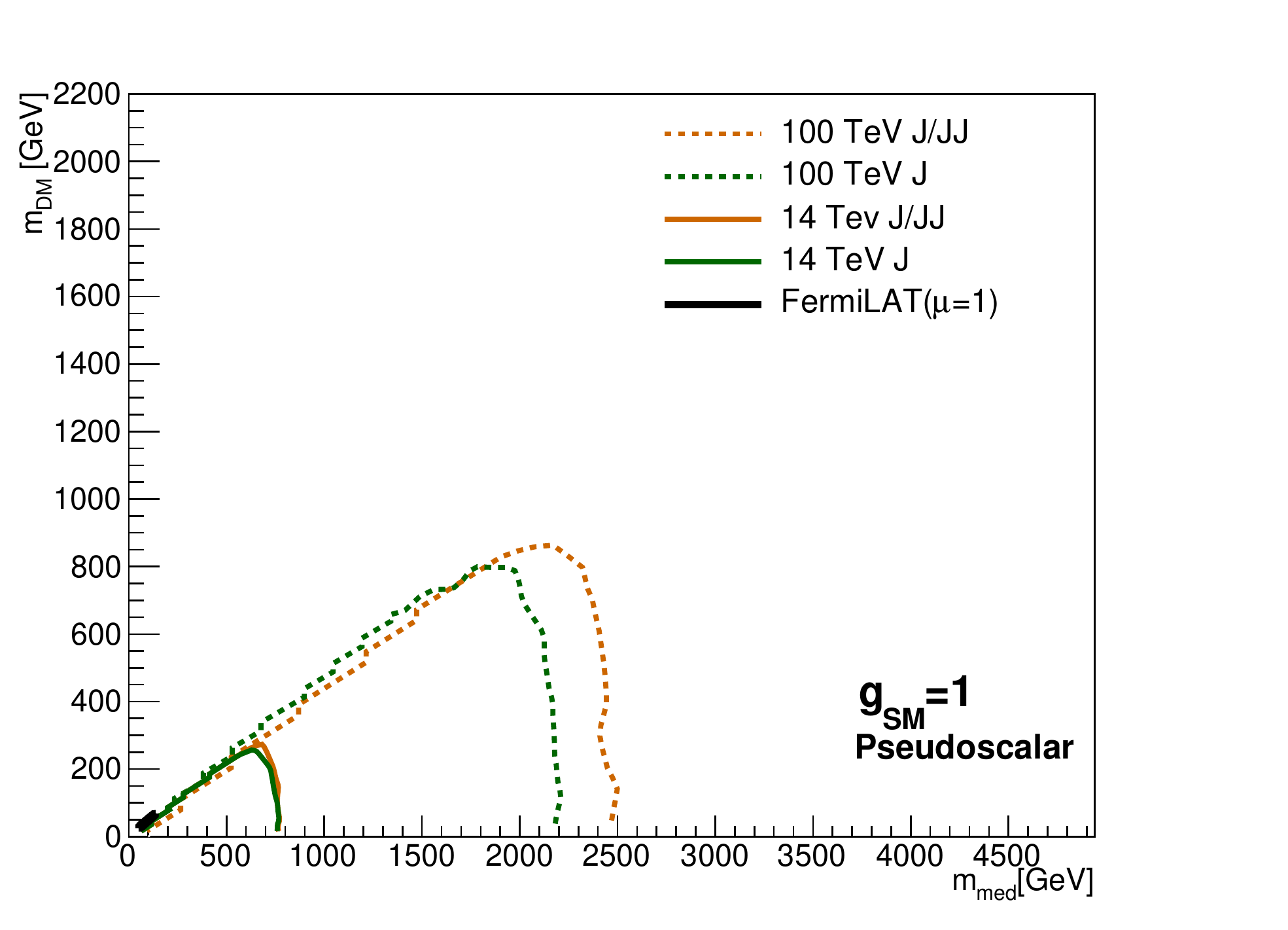}
\caption{Mass limits for scalar mediator models (left panel) and pseudo-scalar models (right pannel) at $14$ and $100$ TeV colliders 
using the multi-leg and a single-leg analysis. The neutrino wall affecting the direct detection experiments is shown in the left plot
and the indirect detection limit for pseudo-scalars using FERMI-LAT data~\cite{Ackermann:2011wa} is shown as a tiny speck in the lower left of the plot on the right.}
\label{fig:exclMass:10}
\end{figure}
\end{center}

In Figures~\ref{fig:exclMass:9} and \ref{fig:exclMass:10} we show these exclusion contours for the simplified model analysis using a
 fixed value of the mediator couplings, $g_{\rm DM}=g_{\rm SM}=1$ for all 4 mediator models of Eqs.~\eqref{eq:LS}-\eqref{eq:LA}. To enable the direct comparison between different experiments/techniques, these figures show all five exclusion contours -- 
 the 14 TeV and the 100 TeV limits, using both the one-jet and the multi-jet 
 analysis, together with the DD/ID non-collider limits/projections. 
 
It is interesting to note the dependence of the DD limits in the scalar mediator case on the number of quark degrees of freedom it couples to.
Unlike the production mechanism at collider searches which is sensitive only to the heavy top quark, the DD limits are sensitive also to 
light degrees of freedom thanks to the cancellation of the quark mass in the $y_q/m_q$ factor in Eq.~\eqref{eq:17}. Thus, the DD limits are quite sensitive to choice of flavors that mediator couples to in the simplified model.
 The magenta contour in in Figure~\ref{fig:exclMass:9}  represents the inclusion of interactions with all quark flavors (as in the simplified model in Eq.~\eqref{eq:LS}). For a different choice of the simplified model, for example with only the top quark couplings to the mediator, the DD contour
 is shown in red. The difference between the red and magenta contours in the scalar mediator case in Fig.~\ref{fig:exclMass:9} 
 shows the sensitivity of the DD limits to a range of simplified models; at the same time the collider searches are are primarily sensitive to
 the scalar-to-top couplings\footnote{We note that in the previous figures the $\nu$-wall curve corresponds to the magenta curve.}. For this parameter choice we note that the collider constraints lie below the neutrino wall for 1 ab$^{-1}$, as the 100~TeV collider collects more data the wall can be breached. As an example we plot the expected limit given 100 ab$^{-1}$ of 100~TeV data for the scalar mediator. 
  
\subsubsection{Comparison with Relic Density}
Finally, in the context of simplified models, we can compare the sensitivity of the four mediator types with the relic density bounds. The relic density bound serves as a qualitative upper bound for the simplified models~\cite{Pree:2016hwc}. If full coverage can be obtained over the range of the allowed space given the relic constraints, the simplified model can probe all allowed space consistent with the relic density. Such models can be modified to circumvent the relic density constraint. However, most modifications of the simplified model which embed them in more realstic models lead to tighter constraints on the relic density. 

The bounds from a 100 TeV collider,  the neutrino wall, and the projected bounds from indirect detection are shown in figure~\ref{fig:smsbounds}. From these bounds, we observe that the allowed mediator masses that preserve the relic density are exlucded by direct detection for vector mediators. The axial mediators are nearly excluded by the collider bounds, and with additional data will be excluded. The allowed scalar region is excluded up to roughly 3~\UTeV, and the pseudoscalar is excluded up to 3.5~\UTeV. The allowed regions for both the scalar are not completely covered. The pseudoscalar, in particular, poses the largest challenge to be constrained by either indirect detection or collider constraints. It should be noted that both the direct photon line and indirect Fermi and HESS projections are shown for the indirect bounds in figure~\ref{fig:smsbounds}. 

\begin{figure}[ht]
\begin{center}
\includegraphics[width=0.45\textwidth]{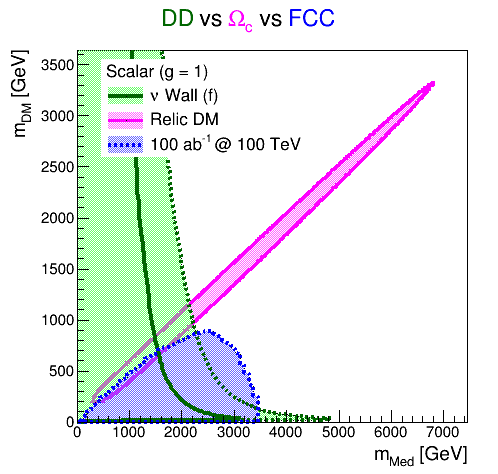} \hskip-0.6cm
\includegraphics[width=0.45\textwidth]{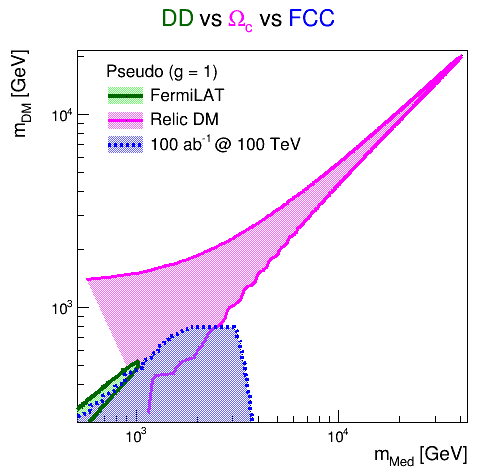}\\
\includegraphics[width=0.45\textwidth]{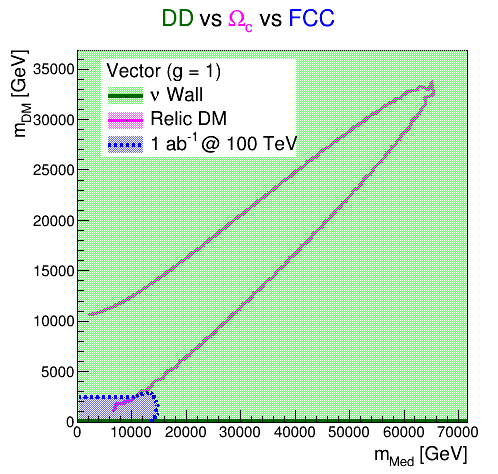}
\includegraphics[width=0.45\textwidth]{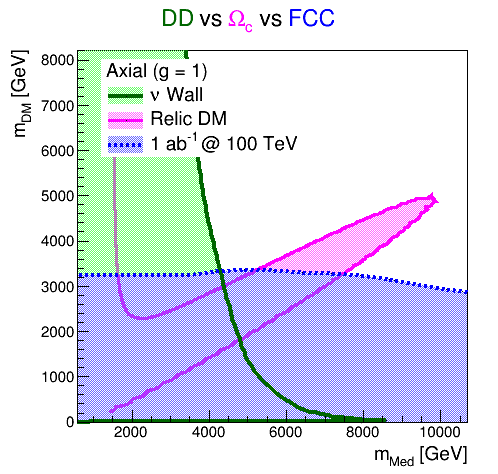}
\caption{Mass limits for scalar mediator models (top left panel), pseudo-scalar models (top right panel), vector models (bottom left panel), and axial models (bottom right panels) at $100$ TeV colliders. The neutrino wall affecting the direct detection experiments is green for all plots expluding the pseudo-scalar mediator, where the projected indirect detection limit using FERMI-LAT and HESS projections data~\cite{HESS:2015cda} is shown. The relic density is additionally computed all allowed mediator and DM masses are contained within the relic density lines. }
\label{fig:smsbounds}
\end{center}
\end{figure}

%\subsubsection{Mediator vs. Direct Search}
\subsubsection{Probing Thermal DM with Monojets and Dijets}
 \label{sec:smsmjdj}

Simplified models offer a useful framework to focus on the interactions of the DM particles, while at the same time being flexible enough to allow for a rich phenomenology~\cite{Abdallah:2015ter}. As pointed out in Refs.~\cite{Frandsen:2012rk,Fairbairn:2014aqa,Chala:2015ama}, one of the central implications of assuming the presence of a new mediator is that one can probe the model not only with collider searches based on missing energy in association with SM particles, but also with dedicated searches for the mediator particles themselves, which make use of the fact that any mediator produced from SM particles in the initial state can also decay back into SM states. Combining both kinds of searches it is possible to constrain the visible and invisible decay modes of the assumed mediator and hence probe a wide range of mediator masses.

In the present study we demonstrate this complementarity for a 100 TeV circular proton collider and compare the resulting constraints to the parameter space compatible with WIMP freeze-out. For concreteness, we consider the case of a spin-1 mediator, which could e.g.\ be the massive gauge boson of an additional broken $U(1)'$ gauge symmetry. As discussed in~\cite{Kahlhoefer:2015bea}, it is important that the couplings of the mediator are chosen in a way that preserves gauge invariance and that perturbative unitarity is not violated in the parameter regions under consideration. Following~\cite{Kahlhoefer:2015bea}, we therefore assume that the WIMP is a Majorana fermion and that the mediator has only vectorial couplings to SM quarks:
\begin{equation}
\mathcal{L} \supset  - g_q \sum_q Z'^\mu \, \bar{q} \gamma_\mu q - \frac{g_\text{DM}}{2} Z'^\mu \, \bar{\chi} \gamma_\mu \gamma^5 \chi \; .
\end{equation}
This choice suppresses constraints from electroweak precision observables, searches for dilepton resonances and DM direct detection experiments, which would otherwise rule out most of the parameter space compatible with thermal freeze-out. In other words, we %focus on the case where colliders will have the largest impact.
focus on a typical case that the 100~TeV collider will have to tackle if no DM detection arises in the next decade. 

\textbf{Constraining simplified models.}
To calculate the expected sensitivity of the 100~TeV collider for simplified models we assume an integrated luminosity of $\mathcal{L} = 10\:\text{ab}^{-1}$ at a centre-of-mass energy of $100\:\text{TeV}$. For comparison, we also show the reach of the Large Hadron Collider with $\mathcal{L} = 300\:\text{fb}^{-1}$ at $14\:\text{TeV}$ (LHC14).

\noindent {\bf Monojets: } We implement the analysis
strategy suggested in~\cite{Xiang:2015lfa}, which essentially
corresponds to a scaled-up version of the most recent CMS
analysis~\cite{Khachatryan:2014rra}. Most importantly, the analysis
cuts require missing transverse energy ($\slashed{E}_T$) in excess of
$2.6\:\text{TeV}$. We simulate the expected signal using
\texttt{MadGraph~v5}~\cite{Alwall:2014hca}
and \texttt{Pythia~v6}~\cite{Sjostrand:2006za}. In existing
monojet searches, detector effects play a rather small role, leading
to a modest reduction of the monojet cross section by about
20\%~\cite{Aad:2015zva}, and we assume that these effects are of
similar size in future colliders.

We also simulate the dominant SM background, which arises from invisibly decaying $Z$-bosons, $pp \rightarrow j+Z(\rightarrow \nu\bar{\nu})$. In addition to statistical uncertainties, we include 1\% systematic uncertainties, implying that statistical and systematic uncertainties are of comparable magnitude. Denoting the expected number of background events by $B$, we can then potentially exclude a given set of parameters at $95\%$ CL if the predicted number of signal events $S$ violates the inequality $S^2/(S + B + (0.01\,B)^2) < 3.84$~\cite{Fox:2011pm}. We find that the 100~TeV collider can probe any physics contributions in excess of $\sigma_\text{crit} = 0.15\:\text{fb}$.

For the LHC we implement the analysis strategy proposed in~\cite{Monojets}, which requires $\slashed{E}_T > 800\:\text{GeV}$. For the cuts that we employ, and assuming 2\% systematic uncertainties, we find that the LHC will be able to probe monojet cross section larger than $\sigma_\text{crit} = 0.6\:\text{fb}$.

\noindent {\bf Dijets: } The search for dijet events coming from the $pp\to Z^{'} \to jj$ process probes the large-$m_{\text{DM}}$ parameter region~\cite{Chala:2015ama}. We simulate this signal by means of \texttt{MadGraph~v5}, \texttt{Pythia~v6} and \texttt{Delphes~v3}~\cite{deFavereau:2013fsa}. The background expectations after imposing the cuts adopted in the CMS dijet analysis~\cite{CMS:kxa} are extracted from~\cite{Yu:2013wta}. We apply these cuts to the signal using MadAnalysis~\cite{Conte:2014zja}. The reach of the 100~TeV collider to this signal is then estimated by applying the CL$_s$ method to the distribution of the dijet invariant mass $m_{jj}$, neglecting systematic uncertainties. This approach allows us to probe even broad resonances. We proceed in the same way for LHC14.

\begin{figure}[tb]
\centering
\includegraphics[width=0.42\textwidth]{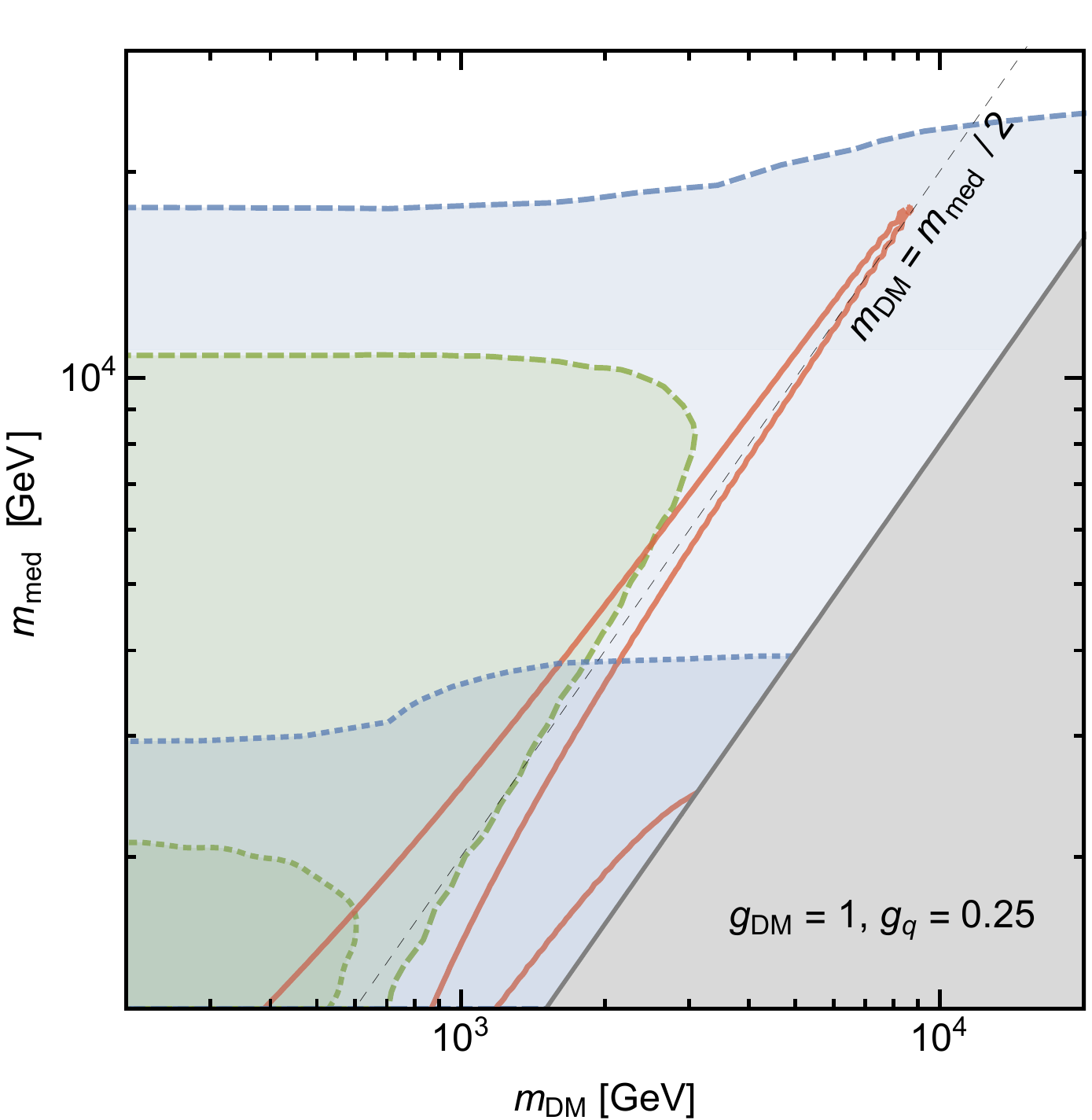}\qquad
\includegraphics[width=0.42\textwidth]{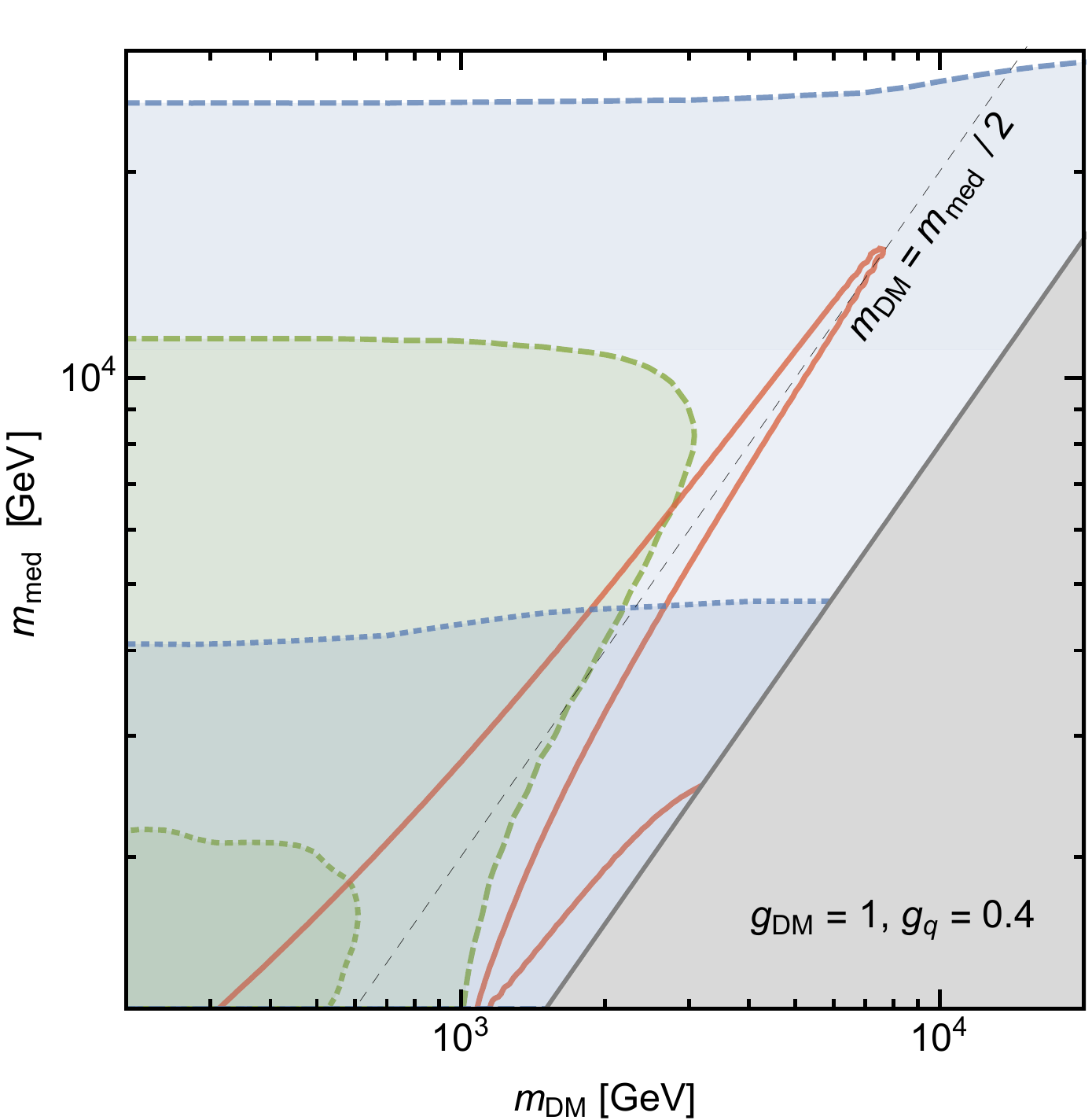}
\caption{Expected sensitivity of monojet (green) and dijet resonance (blue) searches at the 100~TeV collider (dashed lines) compared to the expected sensitivity of the LHC at 14 TeV (dotted lines) and the parameter values that reproduce the observed relic abundance (red, solid). The grey regions are excluded by perturbative unitarity (cf.~\cite{Kahlhoefer:2015bea}).  }
\label{fig:m-m}
\end{figure}

\noindent {\bf Combination: }In Fig.~\ref{fig:m-m} we show several examples for the expected
sensitivity of the 100~TeV collider compared to the projection for LHC14
TeV and the parameters that reproduce the observed DM relic abundance
($\Omega_\text{DM} / h^2 = 0.119$) as calculated with \texttt{micrOMEGAs
  v4}~\cite{Belanger:2014vza}. Note that we do not display direct
detection bounds, which are strongly suppressed for the coupling
combinations that we consider and therefore turn out not to be
competitive to the bounds from colliders. Theoretical constraints from perturbative unitarity (cf.~\cite{Kahlhoefer:2015bea})
are also shown.

\begin{figure}[tb]
\centering
\includegraphics[width=0.46\textwidth]{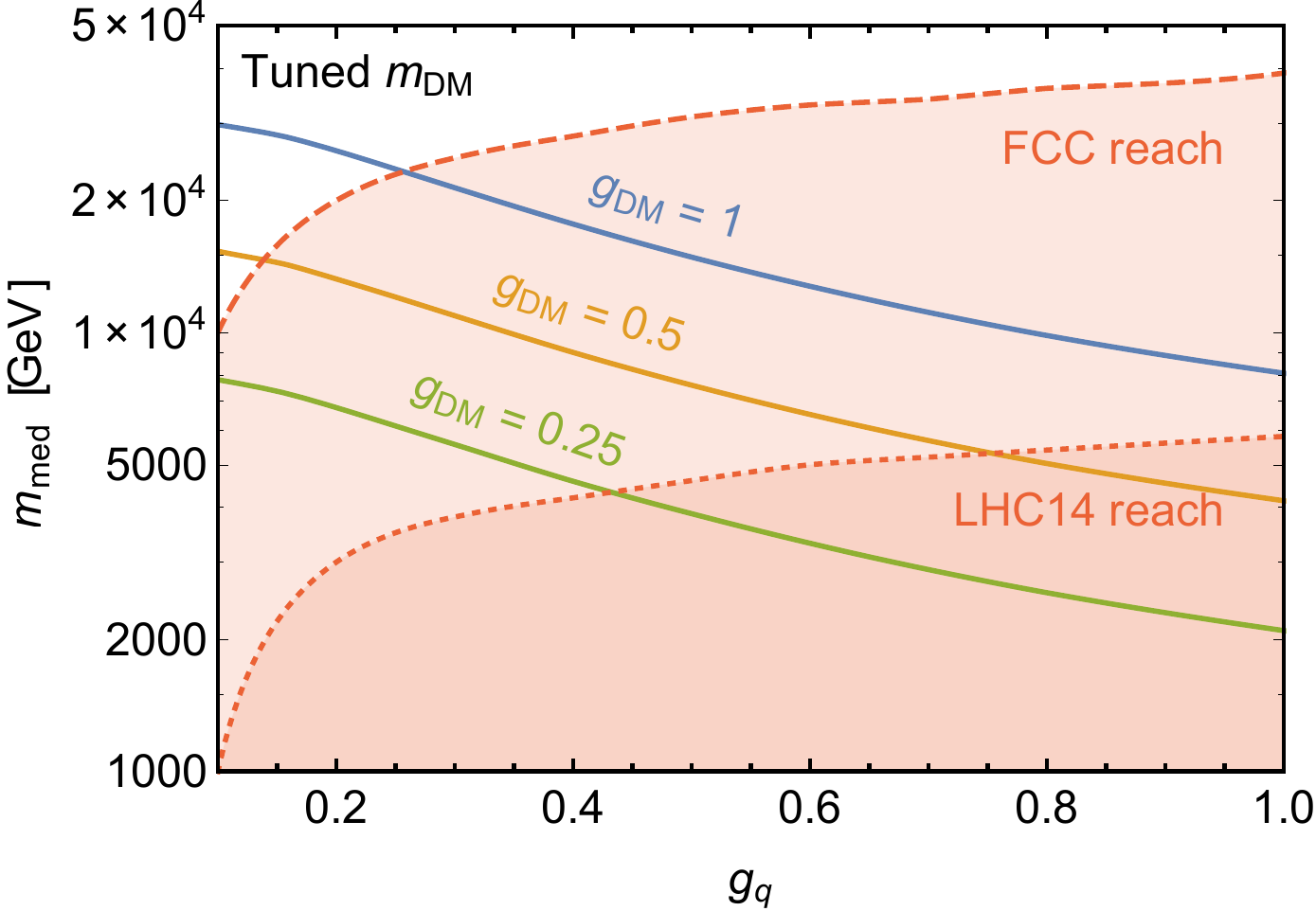}\quad
\includegraphics[width=0.46\textwidth]{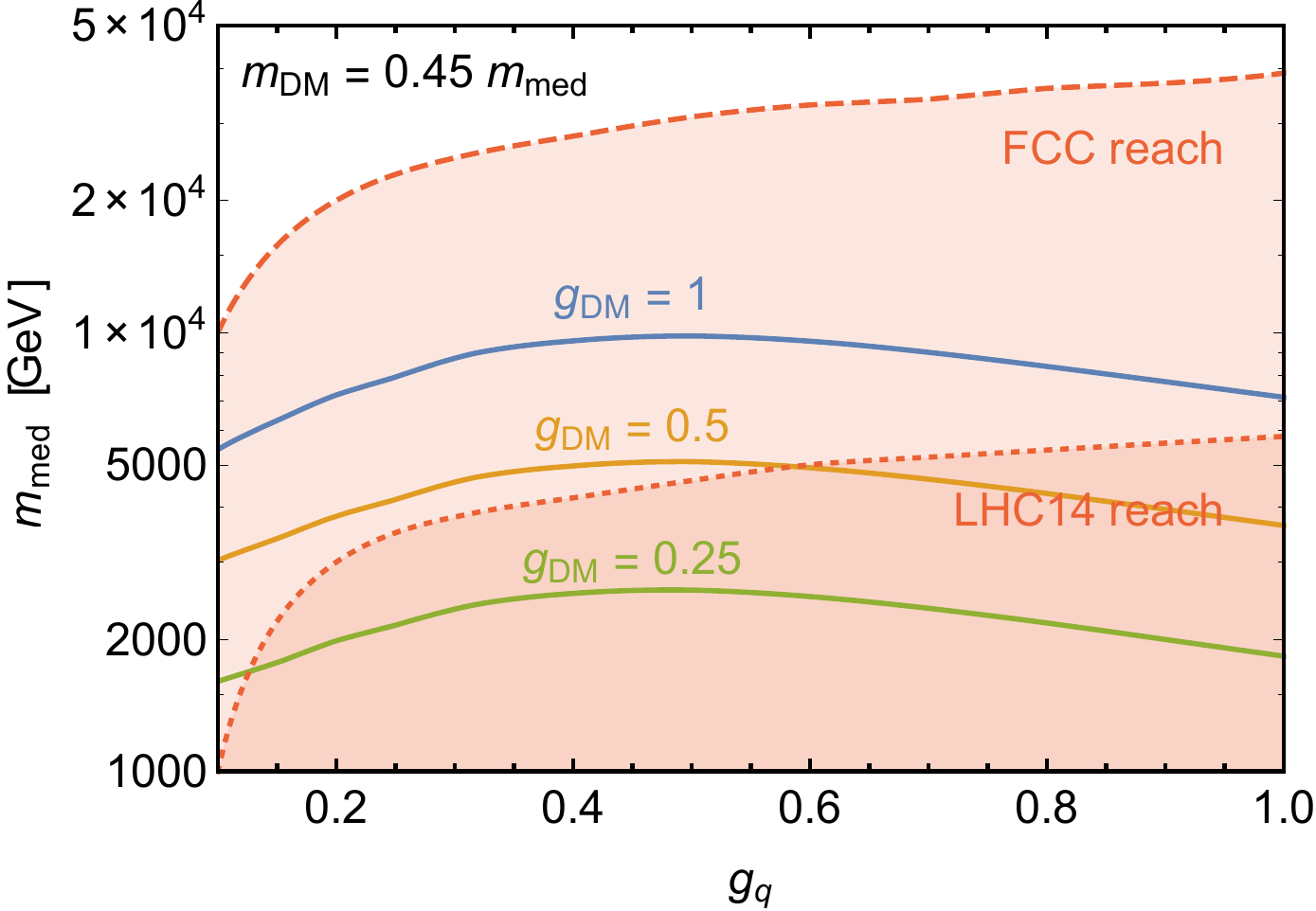}
\caption{Largest value of $m_\text{med}$ compatible with the observed
  relic abundance (for several values of $g_\text{DM}$; solid lines)
  in comparison with the largest mediator mass that can be probed by
  the 100~TeV collider (dashed lines) and LHC14 (dotted lines) as a function of the quark coupling $g_q$. In
  the left panel we tune $m_\text{DM}$ in such a way that the resonant
  enhancement is maximised, whereas in the right panel we fix
  $m_\text{DM} = 0.45 \, m_\text{med}$.}
\label{fig:m-g}
\end{figure}

\textbf{Probing the resonance region.}
For the cases considered in Fig.~\ref{fig:m-m}, the 100~TeV collider is sensitive to all parameter points compatible with the observed DM relic
abundance. However, for small couplings and $m_\text{DM} \approx m_\text{med} / 2$ there
is a strong enhancement of the DM annihilation cross section due to
the mediator going on-shell in the process $\chi \chi \rightarrow q
\bar{q}$. As a result, very large mediator masses can still be
compatible with the observed DM relic abundance. This is quantified in
Fig.~\ref{fig:m-g}, which displays the maximal
$m_\text{med}$ allowed by $\Omega_\text{DM} / h^2 \le 0.119$ as a
function of $g_q$ for several values of $g_\text{DM}$ (solid lines).
In the left panel we tune $m_\text{DM}$ for each value of $g_q$ and
$g_\text{DM}$ to maximise the resonant
enhancement during thermal freeze-out, which typically requires $m_\text{DM}$
slightly below $m_\text{med}/2$ due to the kinetic energy of the
annihilating DM particles. For small couplings, the resonance is very
narrow and the resonant enhancement is possible only at the expense
of a large fine-tuning on $m_\text{DM}$.  In the right panel we
therefore show an example without excessive tuning (setting
$m_\text{DM} = 0.45\, m_\text{med}$) such that the effect of the
resonance is reduced rather than enhanced for very narrow resonances
and smaller mediator masses are required to reproduce the
observed relic abundance.

The crucial observation is that dijet resonance searches at the 100~TeV collider 
potentially possess the sensitivity to probe even the resonance
region. Fig.~\ref{fig:m-g} also shows the maximum mediator mass that
can be probed by the 100~TeV collider (dashed curve) and LHC14 (dotted curve). We
conclude that for most of the values of $g_\text{DM}$ and $g_q$ that we
consider, the 100~TeV collider can probe \emph{all} mediator masses and DM masses
compatible with thermal freeze-out. This conclusion is made explicit in Fig.~\ref{fig:g-g}, where
we show the potential reach of the 100~TeV collider and LHC14 as a function of the two
couplings $g_q$ and $g_\text{DM}$. We observe that only highly-tuned
corners of parameter space (with $m_\text{DM} \approx m_\text{med} /
2$ and $g_q \ll g_\text{DM}$) can potentially evade detection at a 100 TeV collider. In the largest fraction of the allowed parameter space, on the
other hand, the 100 TeV collider will be able to probe the assumption of thermal
freeze-out for the simplified model that we consider. Notably, this conclusion applies even to the scenario where the analyzed DM particle constitutes only a fraction of the total relic density, which requires even smaller mediator masses. 

\begin{figure}[t]
\centering
\includegraphics[width=0.46\textwidth]{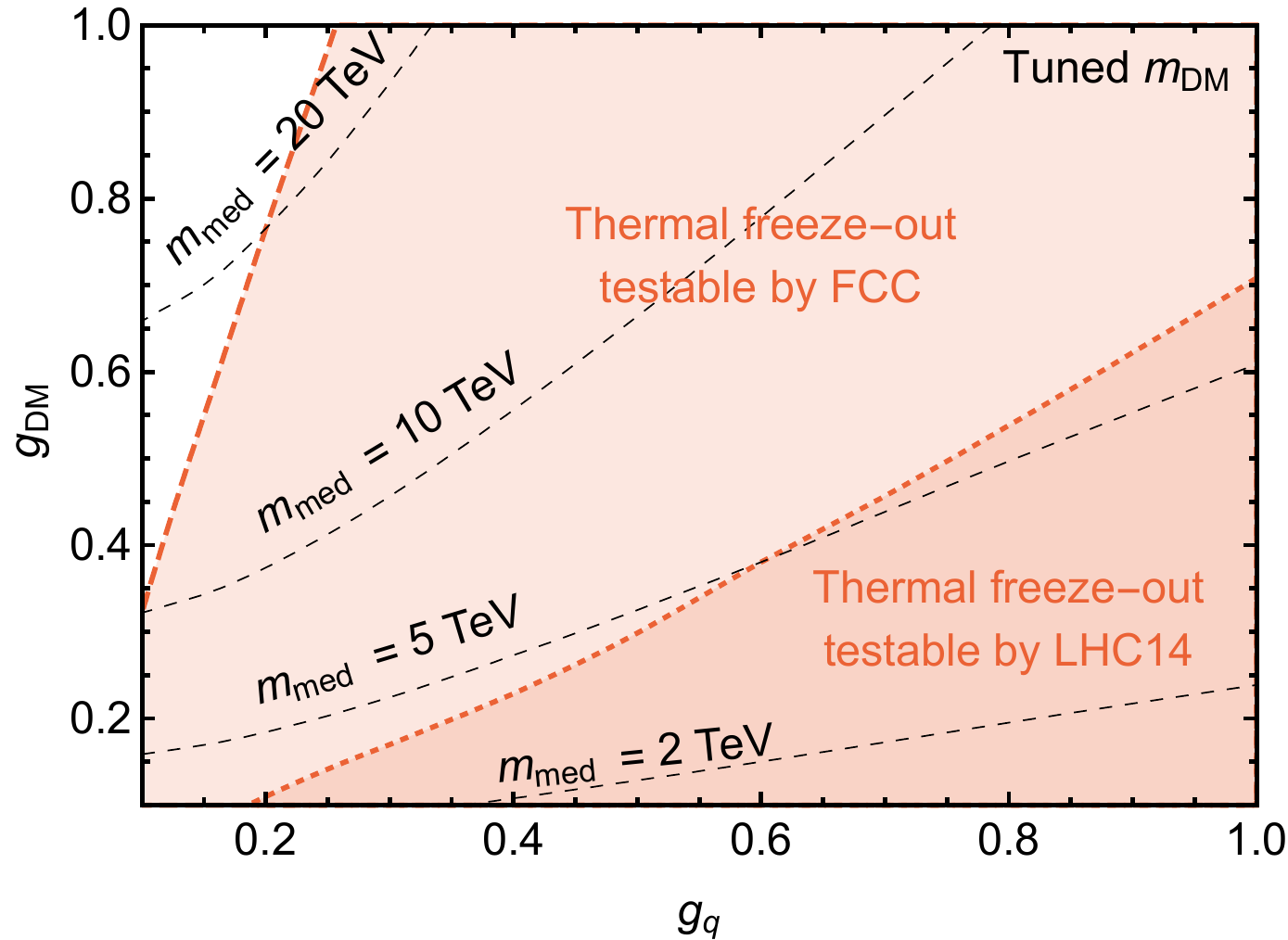}\quad
\includegraphics[width=0.46\textwidth]{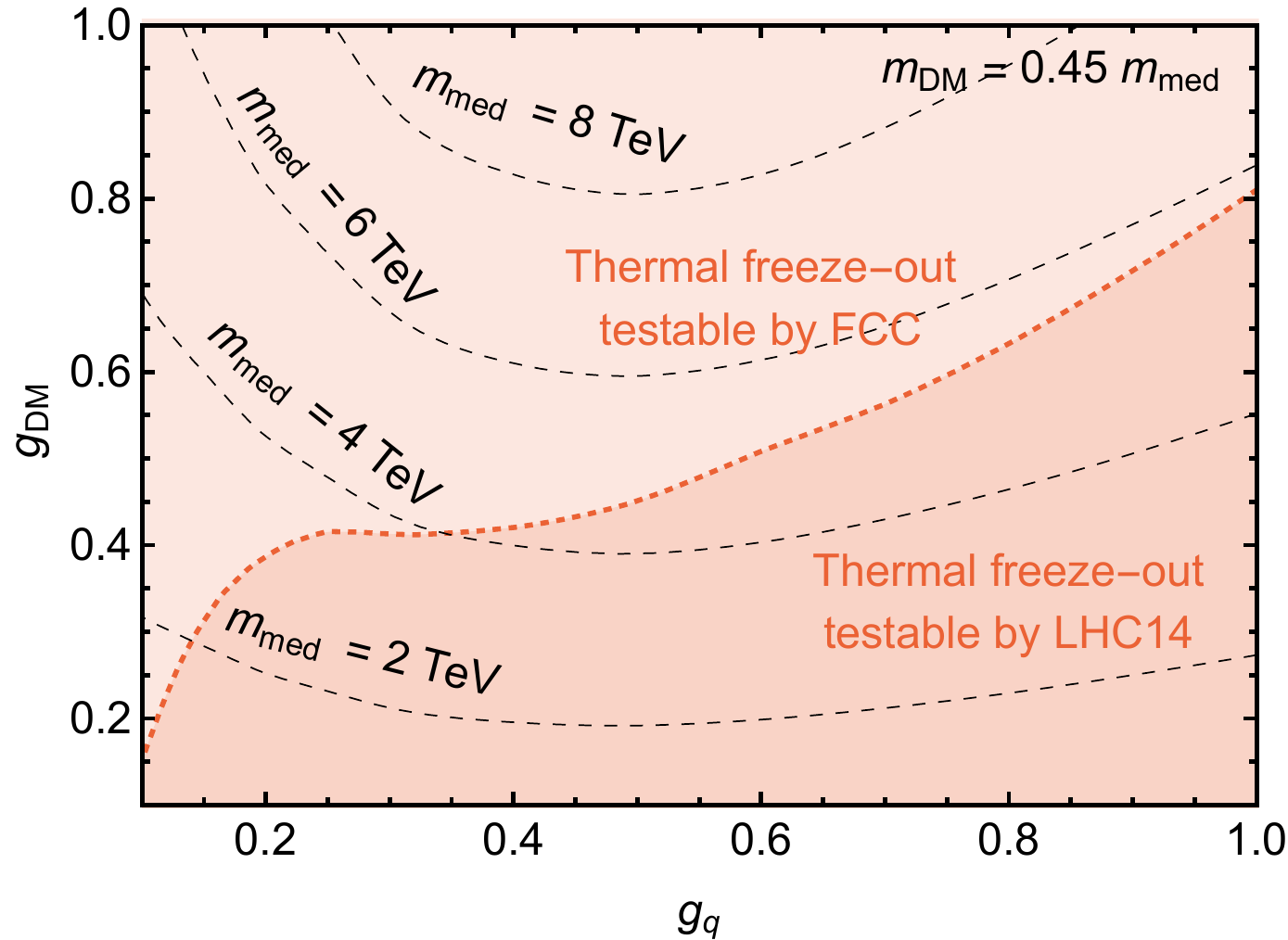}
\caption{Expected sensitivity at a 100 TeV collider (dashed) and LHC14 (dotted) for thermal WIMPs in the $g_q$-$g_\text{DM}$ parameter plane, when setting $m_\text{med}$ to the largest value compatible with the observed relic abundance (as indicated by the black dashed lines). In the left panel we tune $m_\text{DM}$ in such a way that the resonant enhancement is maximised, in the right panel, we fix $m_\text{DM} = 0.45 \, m_\text{med}$.}
\label{fig:g-g}
\end{figure}

\def\eref#1{{Eq.~(\ref{eq:#1})}}
\def\fref#1{{Fig.~\ref{fig:#1}}}
\def\sref#1{{Sec.~\ref{sec:#1}}}
\def\ifb{{\ \rm fb}^{-1}}
\def\iab{{\ \rm ab}^{-1}}

\subsubsection{Light Mediators: Dark Photons at a 100~TeV~collider}
\label{sec:dm_darkp}

A 100 TeV~collider extends the mass reach of direct searches for new physics over the LHC and other lower-energy colliders.  
There are also precision studies that can only be performed at such a high energy machine, such as high-invariant-mass measurements of the Drell-Yan (DY) spectrum. These allow for model-independent searches for new TeV-scale electroweakly charged states through their effect on the RG evolution of SM gauge couplings~\cite{Alves:2014cda}. 
A somewhat less appreciated possibility is that the 100 TeV~collider can discover new physics that produces only low-energy objects,  %discovery avenue for new physics are searches for 
%is the possible discovery of 
%very rare low-energy processes, 
such as certain exotic Higgs decays~\cite{Curtin:2013fra} or low-mass hidden sectors. 
One might think that lepton colliders are better suited to discover such new physics, since reconstruction of low-$p_T$ events with multiple soft objects is greatly aided by the much cleaner final state compared to hadron colliders.  However, if the final state is conspicuous enough, e.g.~by producing multiple leptons and/or photons, then the enormous production rate and luminosity at a 100 TeV~collider (and to a lesser extent the HL-LHC) can easily outweigh this advantage, and allow access to Beyond the Standard Model (BSM) couplings that are several orders of magnitude smaller than what is possible at the statistically limited lepton collider experiments. 

The 100 TeV~collider can therefore act as an \emph{intensity frontier experiment} for the study of, for example, light hidden sectors. One of the best examples to demonstrate this capability are BSM theories with dark photons~\cite{Holdom:1985ag,Galison:1983pa,Dienes:1996zr,Curtin:2014cca}. This will also demonstrate the complementarity of future lepton and hadron colliders in offering different experimental probes of the same BSM scenario. Here we very briefly summarize the work in~\cite{Curtin:2014cca}, which studied the experimental reach of current and future lepton and hadron colliders to study BSM sectors with dark photons.

The minimal benchmark model for dark photons features a dark $U(1)_D$ gauge symmetry that mixes with hypercharge via a small kinetic mixing $\epsilon$:
 \begin{equation}\label{eq:KM}
\mathcal{L} \supset -\frac{1}{4} \,\hat B_{\mu\nu}\, \hat B^{\mu\nu} - \frac{1}{4} \,\hat Z_{D\mu\nu}\, \hat Z_D^{\mu\nu}  + \frac{1}{2}\,\frac{\epsilon}{\cos\theta} \,\hat Z_ {D\mu\nu}\,\hat B^{\mu\nu} + \frac{1}{2}\, m_{D,0}^2\, \hat Z_D^\mu \, \hat Z_{D\mu}\, .
\end{equation}
Here $\hat B, \hat Z_D$ are the hypercharge and $U(1)_D$ gauge bosons with non-canonical kinetic terms. The kinetic mixing can be eliminated via a field redefinition to make the kinetic terms canonical. This results in fields charged under SM hypercharge effectively acquiring a ``milli-charge'' under the $U(1)_D$ gauge interaction. This gives rise to an effective mass mixing between the dark photon $Z_D$ and the SM $Z$-boson through the acquired milli-charge of the SM Higgs. Diagonalizing the gauge boson mass matrix yields a Lagrangian containing $Z_Df \bar f$ and $h Z_D Z$ couplings of $\mathcal{O}(\epsilon)$. In the absence of other hidden-sector states lighter than the dark photon, $Z_D$ decays dominantly to SM fermion pairs with gauge-like  branching ratios. This results in sizable branching fractions to SM leptons, and should be contrasted with e.g.~new scalar states that couple to SM fermions with Yukawa-like interactions and decay dominantly to third generation fermions.\footnote{See \cite{Curtin:2014cca} for a MadGraph \cite{Alwall:2011uj} implementation, as well as tables of branching ratios and decay widths.}

The dark photon mass term $m_{D,0}^2$ in \eref{KM} arises most simply by introducing a ``dark Higgs'' $S$ that breaks $U(1)_D$ via a vacuum expectation value in analogy to the SM Higgs mechanism. This expands the scalar sector of the SM as follows:
\begin{equation}
\label{eq:HM}
V_0 (H,S) =   -\mu^2|H|^2 +  \lambda |H|^4 -\mu_S^2 |S|^2 + \lambda_S |S|^4 +
   \kappa  |S|^2|H|^2\, ,
\end{equation}
where the renormalizable coupling $\kappa$ induces mixing between the SM Higgs and $S$, and is assumed to be small so as not to greatly modify the properties of the observed SM-like Higgs boson. Apart from Higgs coupling modifications, this mixing also induces $h Z_D Z_D$ terms of $\mathcal{O}(\kappa)$. (This term is also generated by kinetic mixing, but only at $\mathcal{O}(\epsilon^2)$.)

Historically, most of the effort in searching for dark photons has been devoted to masses in the range $\UMeV \lesssim m_{Z_D} \lesssim 10 \UGeV$, through techniques as diverse as  precision QED measurements, rare meson decays, supernova cooling, collider
experiments, and beam dumps~\cite{Bjorken:2009mm,Batell:2009yf,Essig:2009nc,Freytsis:2009bh,Essig:2010xa,Blumlein:2011mv,Andreas:2012mt,Pospelov:2008zw,Reece:2009un,Aubert:2009cp,Hook:2010tw,Bjorken:1988as,Riordan:1987aw,Bross:1989mp,Babusci:2012cr,Archilli:2011zc,Abrahamyan:2011gv,Merkel:2011ze,Dent:2012mx,Davoudiasl:2012ig,Davoudiasl:2012ag,
  Davoudiasl:2013aya,Endo:2012hp,Balewski:2013oza,Adlarson:2013eza,Agakishiev:2013fwl,Blumlein:2013cua,Andreas:2013lya,Battaglieri:2014hga,Merkel:2014avp,Lees:2014xha,Adare:2014ega,Kazanas:2014mca,Echenard:2014lma,Gorbunov:2014wqa,NA482}. 
%The focus on this mass range is, in part, motivated by certain data anomalies.  
%For example, various DM-related anomalies motivate(d) 
%the existence of DM-$Z_D$ interactions, see e.g.~\cite{ArkaniHamed:2008qn,Pospelov:2008jd,Finkbeiner:2007kk,Fayet:2004bw,Loeb:2010gj,Kaplinghat:2015aga}, and $\sim 10-100$~MeV dark photons can resolve the $\sim 3.6\sigma$ discrepancy between the observed and SM value of the muon anomalous magnetic moment~\cite{Pospelov:2008zw,Bennett:2006fi,Davier:2010nc} 
%(this explanation is now disfavored if the dark photon dominantly decays directly to SM particles).  
However there is no theory reason not to extend the searches to the entire experimentally accessible range~\cite{Curtin:2013fra,
  Gopalakrishna:2008dv,Davoudiasl:2012ag,Davoudiasl:2013aya,Chang:2013lfa,Falkowski:2014ffa,Cline:2014dwa,Hoenig:2014dsa,Curtin:2013fra}. Generation of dark photon masses in natural theories has been studied in~\cite{ArkaniHamed:2008qp,Cheung:2009fk,Baumgart:2009tn,Morrissey:2009ur,Essig:2009nc}, and in many cases can allow for natural dark photons above 10 GeV equally well. 

\begin{figure}[!t]
\begin{center}
\includegraphics[width=0.9\textwidth]{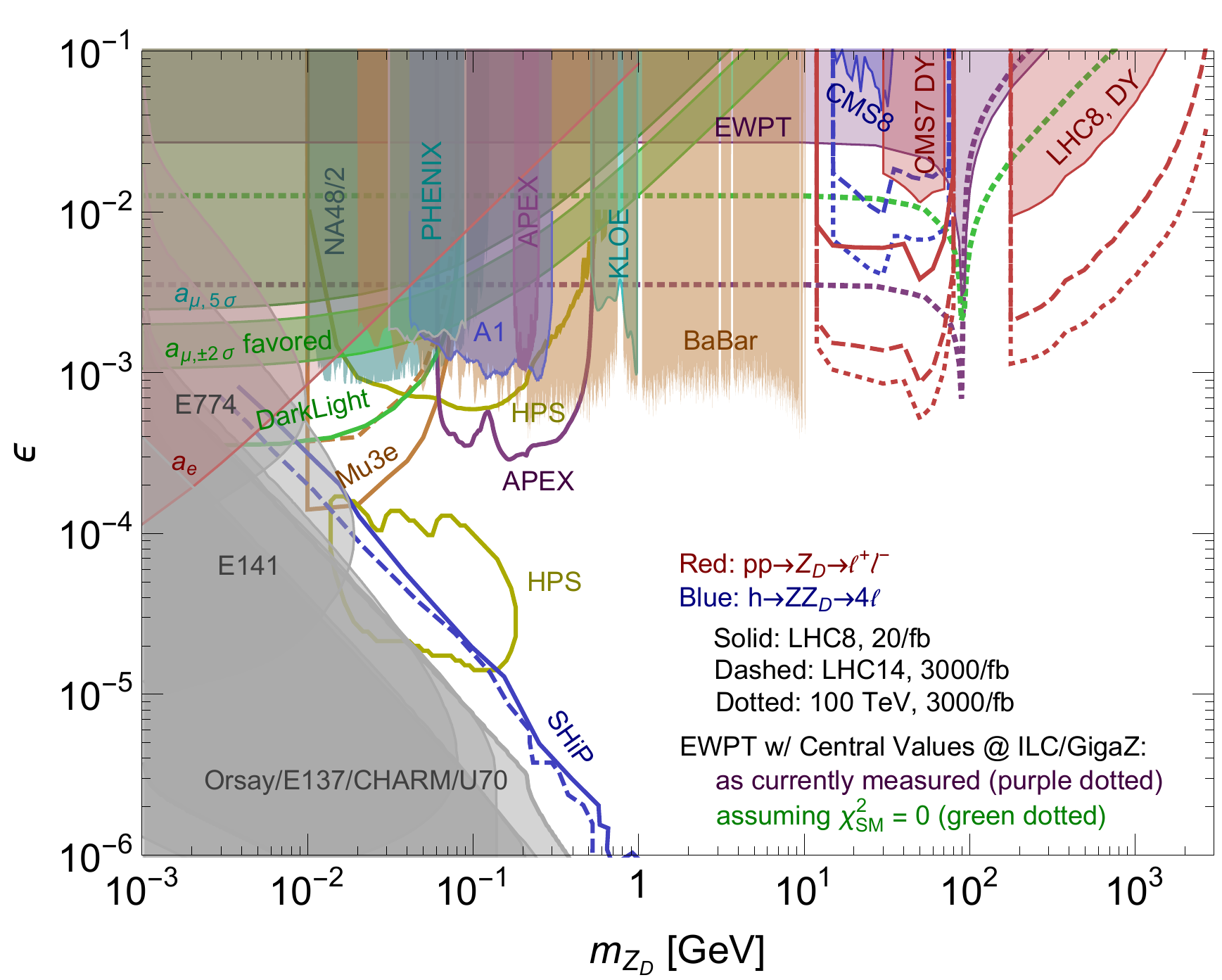} 
\end{center}
\caption{Summary of dark photon constraints and prospects (see \cite{Curtin:2014cca} for references).  
High-energy colliders (LHC14, 100~TeV, ILC/GigaZ) are uniquely sensitive to dark photons with $m_{Z_D}\gtrsim 10$~GeV, 
while precision QED observables and searches at $B$- and $\Phi$-factories, beam dump experiments, and 
fixed target-experiments probe lower masses.  
Dark photons can be detected at high-energy colliders in a significant part of open parameter space 
in the exotic decay of the 125 GeV~Higgs boson, $h\to Z Z_D\to 4\ell$, (blue curves) 
in Drell-Yan events, $pp\to Z_D\to\ell\ell$, (red curves) and through 
improved measurements of electroweak precision observables (green/purple dashed curves).  
Note that all constraints and prospects assume that the dark photon decays directly to SM particles, except 
for the precision measurements of the electron/muon anomalous magnetic moment and the electroweak observables.  
%If, in addition to kinetic mixing, the 125 GeV Higgs mixes with the dark Higgs that breaks the dark $U(1)$, then the decay 
%$h\to Z_D Z_D$ would set constraints on $\epsilon$ that are orders of magnitude more powerful than other searches down to 
%dark photon masses of $\sim 100$~MeV, see \fref{zdzdlimits2}. 
Figure taken from \cite{Curtin:2014cca}. Drell-Yan projections are rescaled from the LHC results of \cite{Cline:2014dwa, Hoenig:2014dsa}, and we anticipate some further improvement at high masses may be possible. 
}
\label{fig:summaryplot}
\end{figure}

In our study of dark photon signatures, it is useful to distinguish the \emph{hypercharge portal}, i.e. the kinetic mixing $\epsilon$ in \eref{KM}, from the \emph{Higgs portal}, i.e. the Higgs mixing $\kappa$ in \eref{HM}. Both refer to renormalizable operators connecting the SM to a hidden BSM sector, and both generate distinct leading signals that may be observed at future lepton and hadron colliders. Note that while some aspects of the Higgs portal coupling were studied already above, that study was restricted to the case where the Higgs is the only mediator to the DM sector, while here the mediator couples to the Higgs portal, thus allowing for mixing with the Higgs and a vast range of complementary signatures not studied above. 

The most promising signatures of the hypercharge portal that were studied in \cite{Curtin:2014cca} are 1. electroweak precision observables (EWPOs) sensitive to the hypercharge portal, 2. the process $p p \to Z_D \to \ell^+ \ell^-$ production via DY-like direct production and 3. exotic Higgs decays via the hypercharge portal, $h \to Z Z_D \to 4 \ell$. 
%
%the following:
%\begin{enumerate}
%
%\item Electroweak precision observables (EWPOs) are sensitive to the hypercharge portal and their measurements at future $e^+e^-$ colliders could probe kinetic mixings $\epsilon \sim \mathrm{few} \ \times \ 10^{-3}$ in a  model-independent fashion, independently of the dark photon decay modes.
%
%\item $p p \to Z_D \to \ell^+ \ell^-$ production via DY-like direct production through the hypercharge portal  can probe $\epsilon\gtrsim 4 \times 10^{-4}$ at a 100~TeV $pp$ collider with 3000 fb$^{-1}$ data. This is the best probe if Higgs mixing is small or not present.  
%  
%\item Exotic Higgs decays via the hypercharge portal, $h \to Z Z_D \to 4 \ell$, can also be probed at the 100 TeV collider. These searches are slightly less sensitive than measurements of DY-like production, but are crucial to distinguish a possible dilepton signal due to dark photons from a generic $Z'$ without kinetic mixing.
%
%\end{enumerate} 
Their reach is compared in the plane of dark photon mass vs.~kinetic mixing in \fref{summaryplot}. Ideally, detection in all three channels, at both future 100 TeV~and lepton colliders, would allow for a detailed diagnosis of the dark sector.

\begin{figure}[t]
\begin{center}
\hspace*{-10mm}
\begin{tabular}{cc}
\includegraphics[width=8cm]{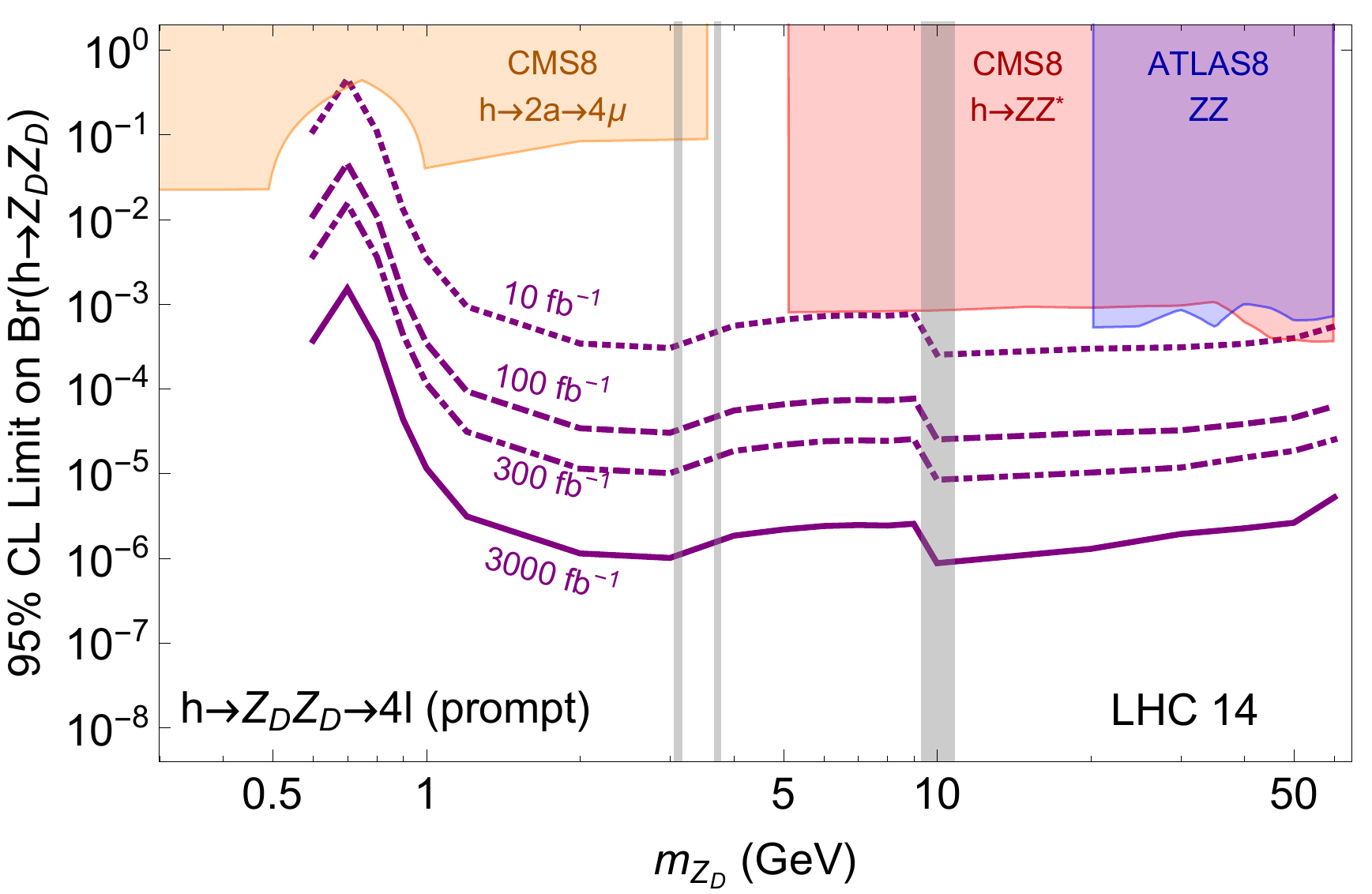}
&
\includegraphics[width=8cm]{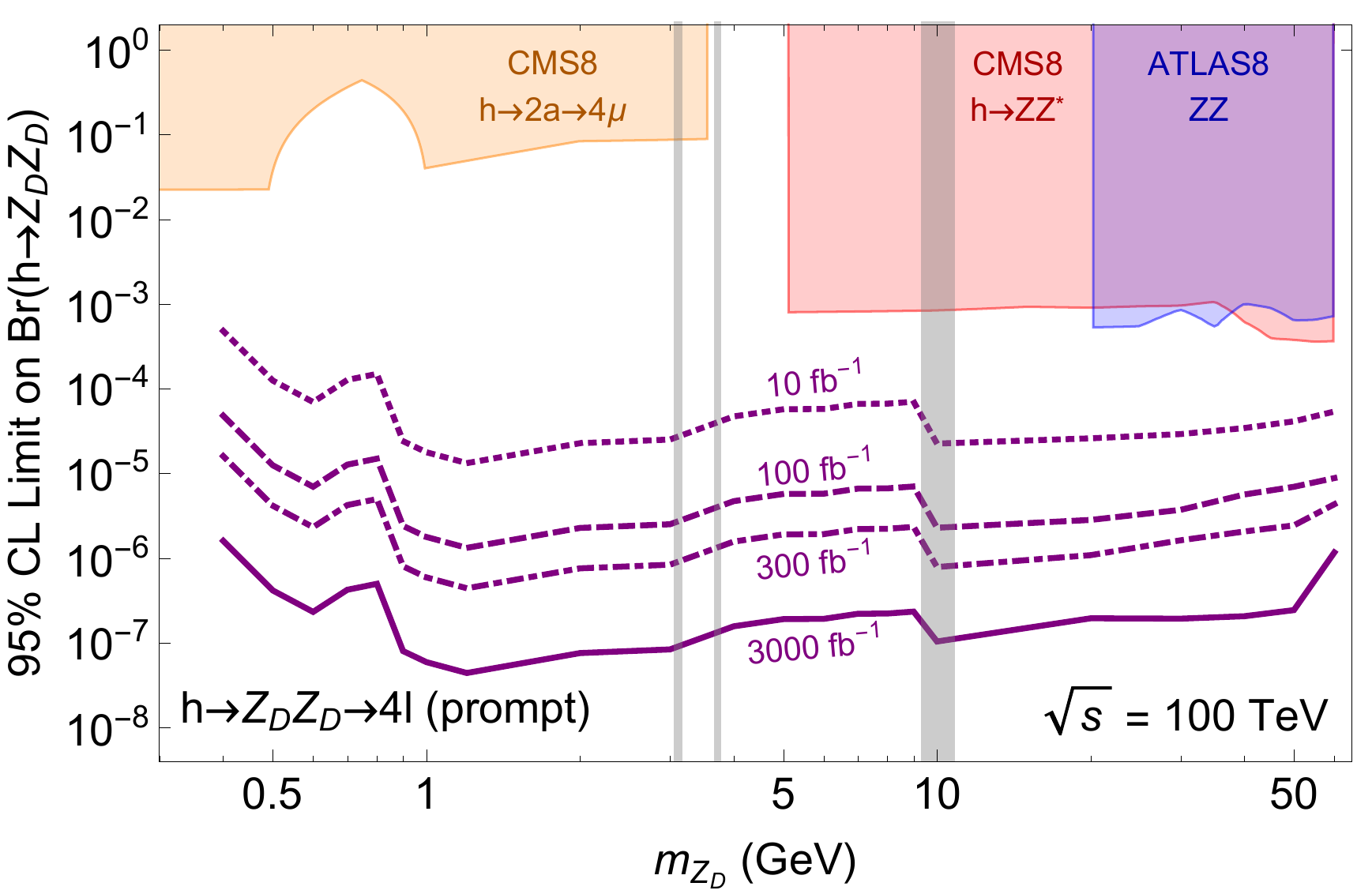}
\\
\includegraphics[width=8cm]
{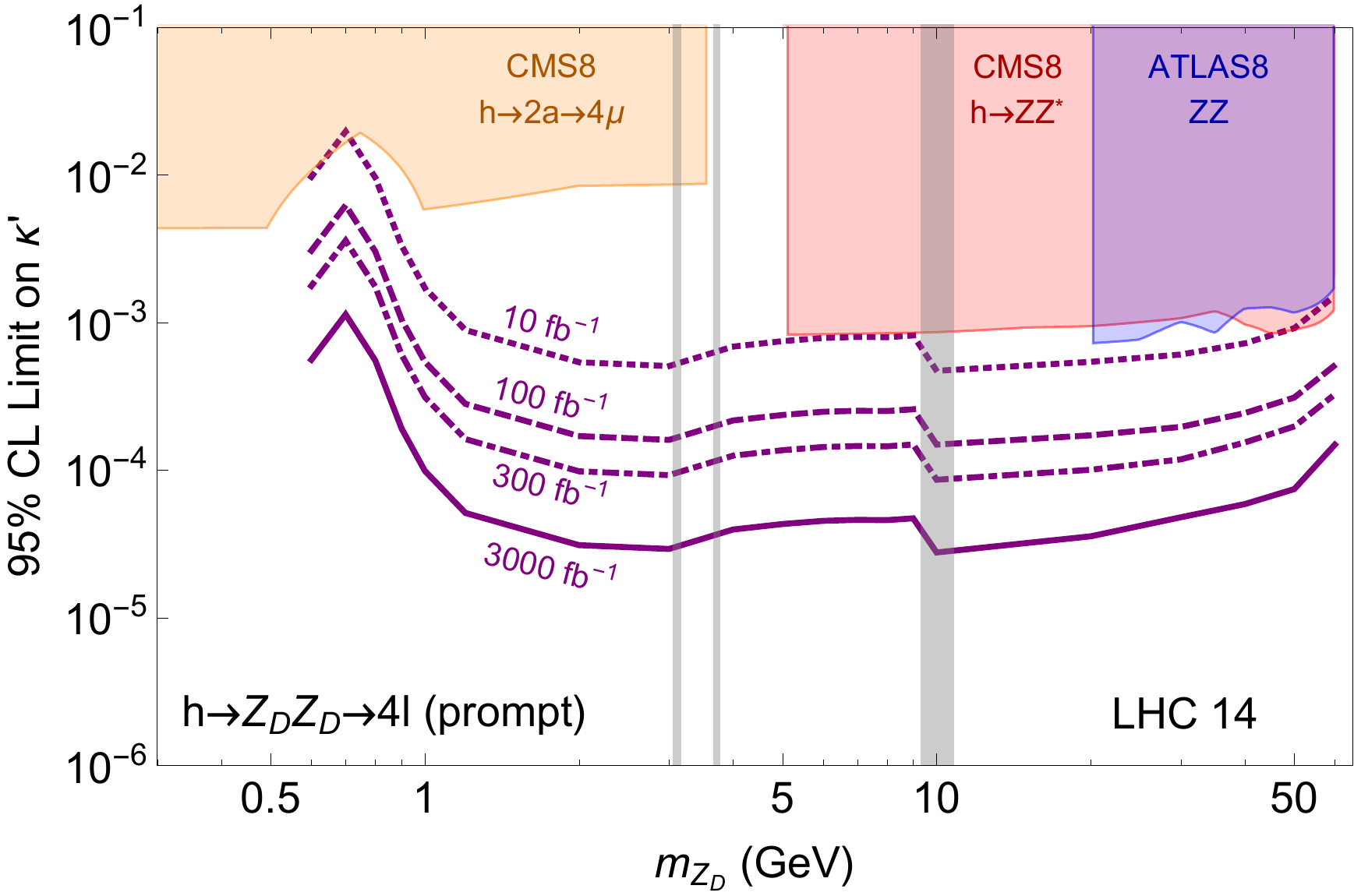}
&
\includegraphics[width=8cm]{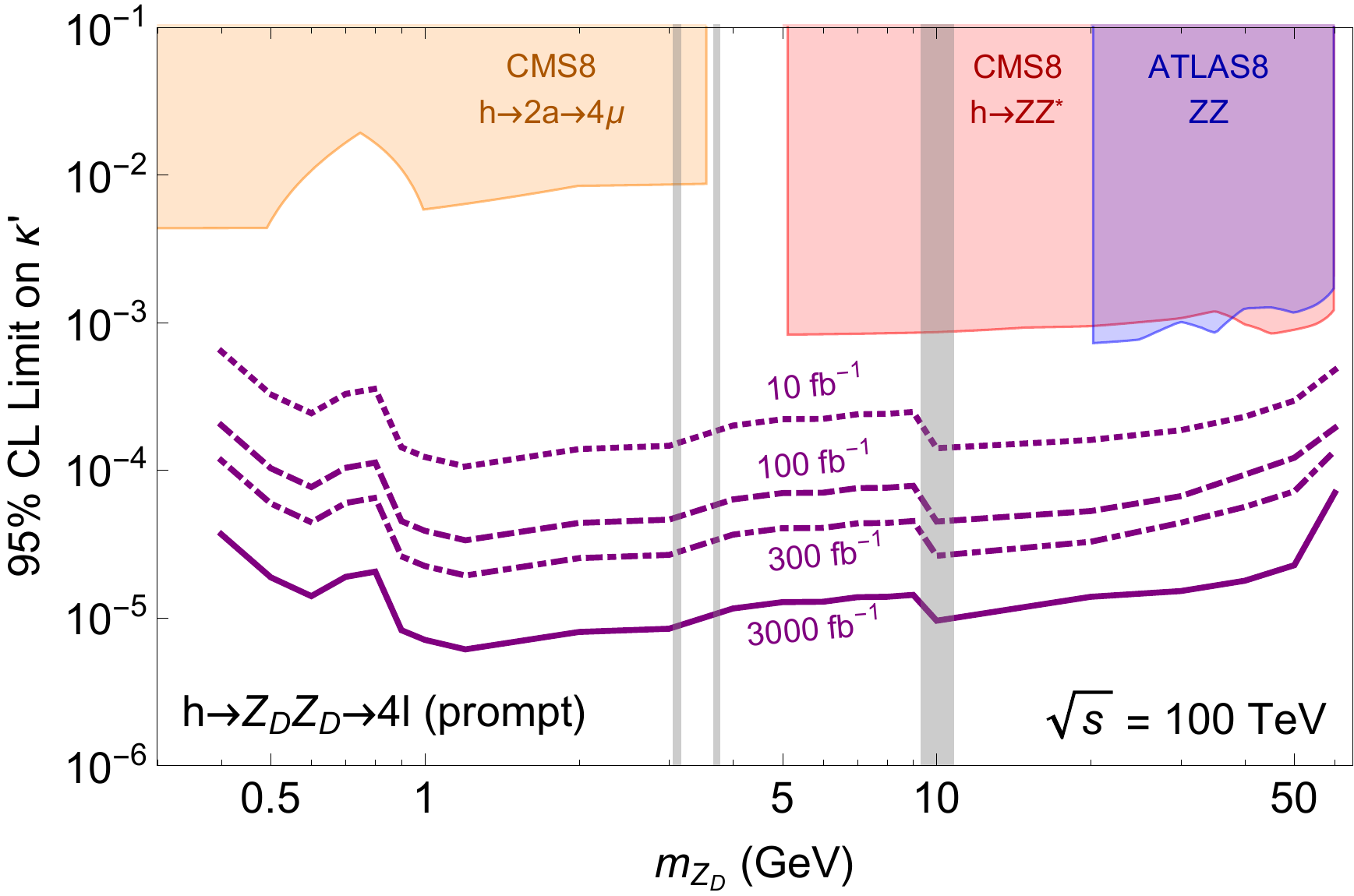}
\end{tabular}
\end{center}
\caption{{\small 
 Expected 95\% CLs limits on the total exotic Higgs decay branching ratio, $\mathrm{Br}(h\to Z_D Z_D)$ (top), and the effective Higgs mixing parameter $\kappa'$ (bottom) at the LHC (left) and a 100 TeV pp collider (right).  Gray bands correspond to regions where quarkonium background may invalidate these projections. The limits obtained in \cite{Curtin:2013fra} from a recast of LHC Run 1 results are shown in red ($h\to Z Z^*\to 4\ell$ search by CMS \cite{CMS:xwa}) and blue (ATLAS $ZZ$ cross section measurement \cite{ATLAS:2013gma}) shaded regions. The limit from the CMS 8 TeV~$h \to 2a \to 4\mu$ search \cite{CMS:2013lea} is shaded in orange, assuming the efficiencies for pseudoscalar and dark photon decay to muons are the same. Figure from \cite{Curtin:2014cca}.
      }}
\label{fig:BrZdZdlimit}
\end{figure}

If there is non-negligible mixing between the SM-like and the dark Higgs, additional measurements are possible via the Higgs portal. It is useful to define the effective mixing parameter
\begin{equation}\label{eq:kappaprime}
\kappa' = \kappa \  \frac{m_h^2}{|m_h^2-m_s^2|} \ .
\end{equation}
The most promising Higgs portal signature is the exotic Higgs decay $h \to Z_D Z_D$, with branching fraction that scales as $\mathcal{O}({\kappa'}^2)$. As long as the $Z_D$ are produced on-shell, they decay via the hypercharge portal without $\epsilon$ affecting the $Z_D Z_D$ production cross section. As \fref{BrZdZdlimit} makes clear, this is the scenario where the advantages of the 100 TeV~collider as an intensity frontier experiment are most apparent. Branching ratios as low as $\mathrm{Br}(h \to Z_D Z_D) \sim 10^{-7} (10^{-8})$ can be probed with 300 fb$^{-1}$ ($3\iab$) when $Z_D$ decays promptly, since the search for the $4 \ell$ final state with two $m_{Z_D}$ resonances and one $m_h$ resonance is practically background-free.

Prompt $Z_D$ decay implies $\epsilon \gtrsim 10^{-5}$. Detection of prompt $h \to Z_D Z_D \to 4\ell$ is therefore already sensitive to kinetic mixings much smaller than what can be probed directly. That sensitivity is greatly extended when expanding the search to include \emph{long-lived particles} produced in exotic Higgs decays. This is separately motivated in theories of  Neutral Naturalness \cite{Craig:2015pha, Curtin:2015fna} and more generally in Hidden Valleys \cite{Strassler:2006im, Strassler:2006ri, Strassler:2006qa, Han:2007ae}, of which the dark photon scenario is a particular example. 

\fref{zdzdlimits2} illustrates the sensitivity to kinetic mixing achievable if dark photon decays within a 1 or 10m detector volume could be reconstructed at the LHC or a 100 TeV~collider (assuming prompt lepton efficiencies and expected signal-to-background).  Different contours indicate different assumptions made for the exotic Higgs decay branching ratio $\mathrm{Br}(h \to Z_D Z_D)$, which can be relatively large even if kinetic mixing is tiny. The enormous rate of Higgs production at a 100 TeV collider compensates for the overwhelming fraction of dark photons that escape the detector for very small kinetic mixing, allowing $\epsilon$ as small as $\sim 10^{-10}$ to be probed. This opens a window onto a broad swath of otherwise inaccessible parameter space, and relies on having available a production mechanism for dark sector states that is separate to the coupling which controls their decay to SM particles. Searches with sensitivity to the displaced dilepton final state are already underway at the LHC \cite{CMS:2014hka, Aad:2014yea}.

\begin{figure}[t]
\begin{center}
\includegraphics[width=0.48\textwidth]{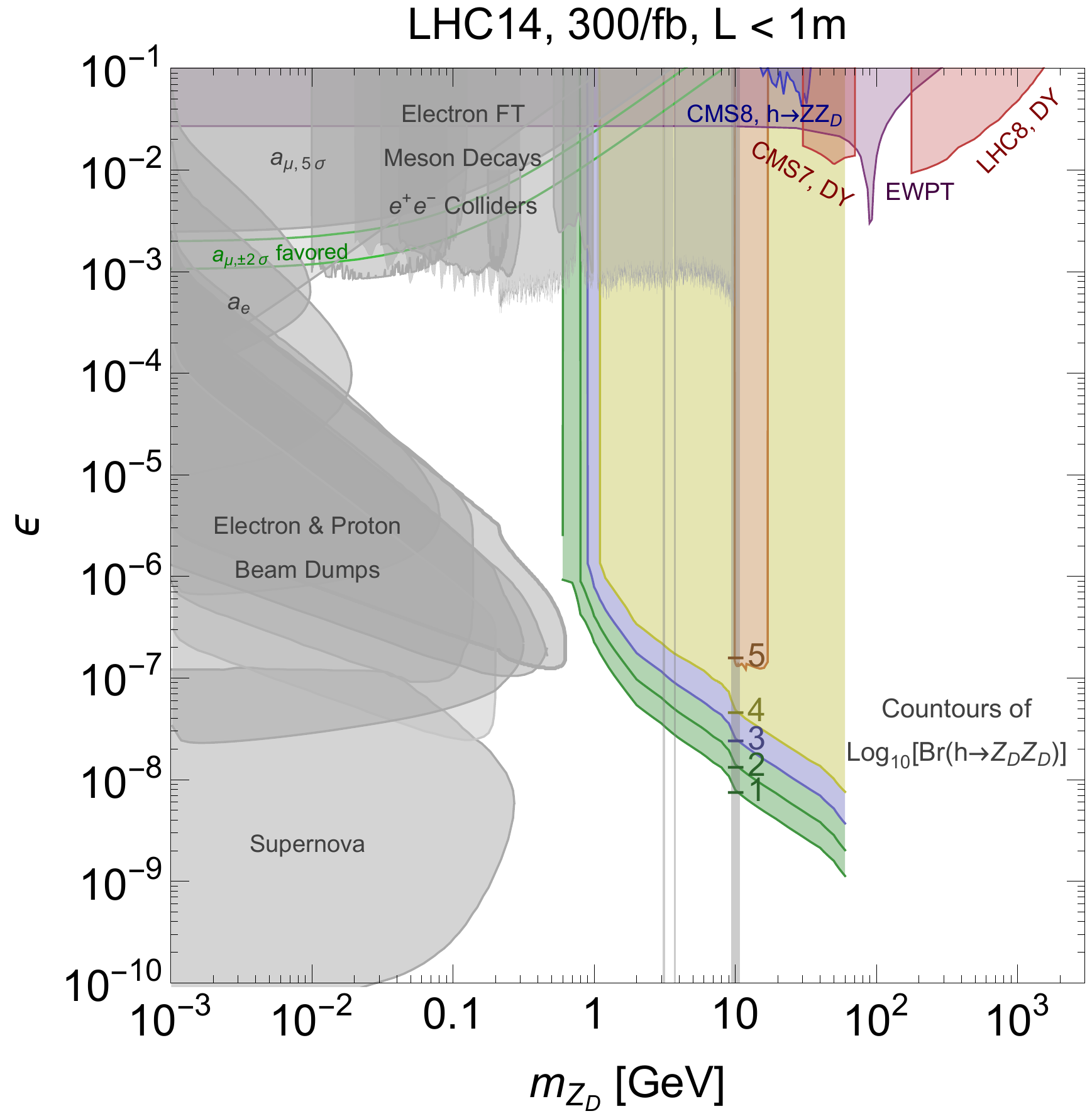} 
%~~\includegraphics[width=0.48\textwidth]{./DM_BSM_portal_3/figures/A-visible-Higgs-Zd-Zd-LHC14-3000ifb-v1.pdf} \\
%\vskip 4mm
\includegraphics[width=0.48\textwidth]{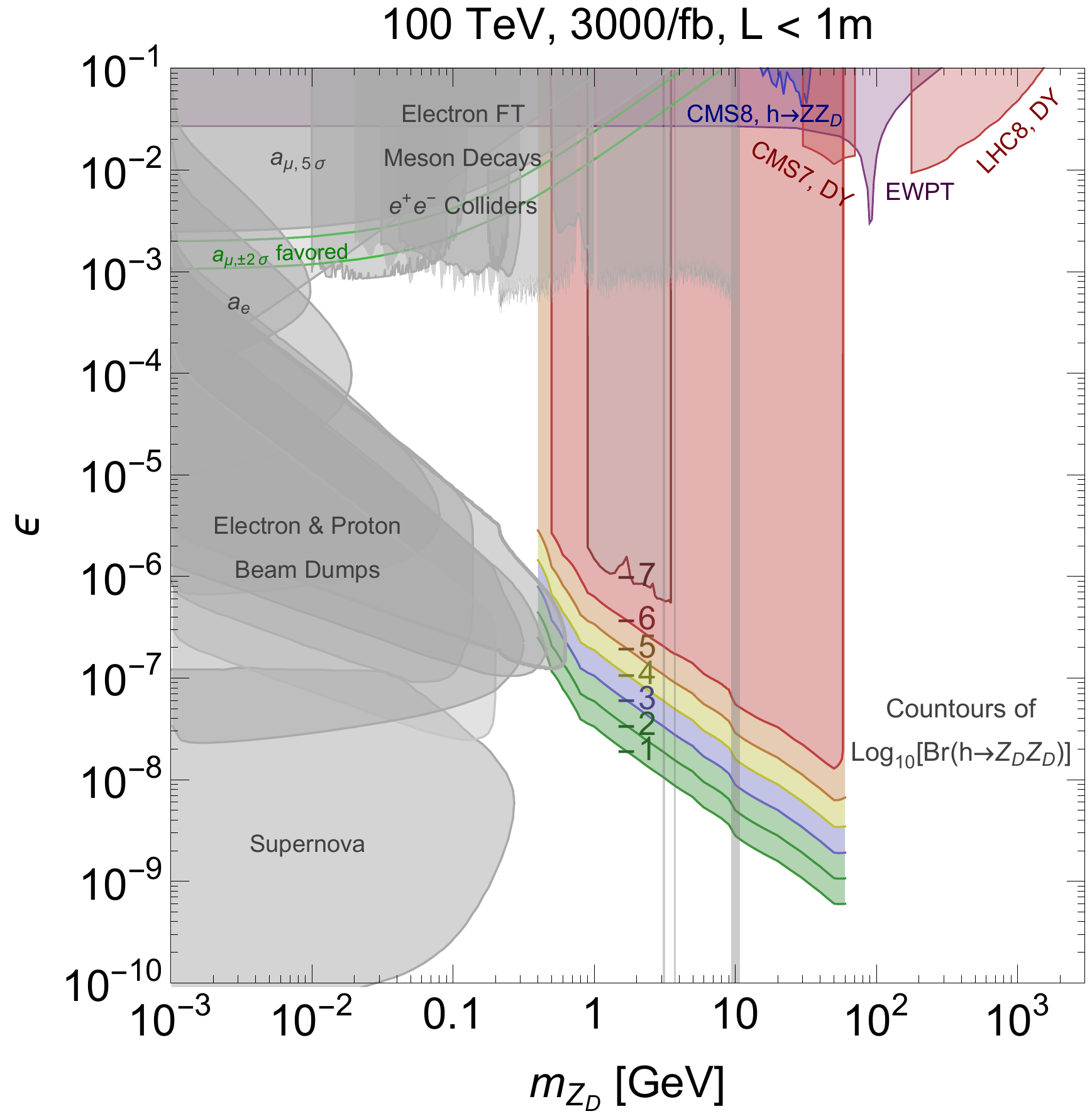} 
%~~\includegraphics[width=0.48\textwidth]{./DM_BSM_portal_3/figures/A-visible-Higgs-Zd-Zd-100TeV-3000ifb10m-v1.pdf} 
\end{center}
\caption{ 
Estimate of expected 95\% CLs limits on $\epsilon$ for
  different $\mathrm{Br}(h \to Z_D Z_D)$ at the LHC (\emph{left}) and a 100 TeV~collider (\emph{right}),
  assuming a displaced lepton jet search has the same sensitivity to
  decays within the given distance from the interaction point as a
  prompt $Z_D Z_D$ search (see \fref{BrZdZdlimit}).  A detector size
  $L$ of $1$~m is assumed for all plots. Gray shaded regions show current constraints (see \cite{Curtin:2014cca} for references). 
  }
\label{fig:zdzdlimits2}
\end{figure}

For the Higgs portal, important complementarities with the capabilities of  future lepton colliders can also be identified.
\begin{itemize}
\item A sizable $\mathrm{Br}(h \to Z_D Z_D)$ is generated through non-negligible mixing of the SM-like Higgs with the dark Higgs $S$. This leads to potentially detectable Higgs coupling deviations at lepton colliders.
\item Direct production of the SM-singlet dark Higgs $s$ is possible at both lepton and hadron colliders, either directly or through exotic decays of the SM-like Higgs boson. If $s$ can only decay via its mixing with the Higgs, the dominant final state will be third-generation fermion pairs $\bar b b$, $\tau^+ \tau^-$ and $\bar c c$. If $s$ is light enough to be produced at lepton colliders, the energy of its decay products may be so low that reconstruction of these final states without resonances of light leptons is difficult at a 100 TeV~hadron collider. In that case, the clean environment of a lepton collider could prove invaluable in searching for $s \to \bar b b, \tau^+ \tau^-$ signals.
\end{itemize}

Finally, the potentially stunning sensitivity of a 100 TeV~collider to dark photon signatures is only possible if future detector designs allow for the recording and reconstruction of relatively soft objects, with $p_T \sim \mathcal{O}(20 \UGeV)$, as well as low-mass $\sim \mathcal{O}(10 \UGeV)$ displaced decays.  If these requirements are satisfied, a 100~TeV~collider could easily discover hidden sector signatures that cannot be probed by any other means.

%
%
%\bibliography{zdark}
%\bibliographystyle{JHEP}
%
%
%
%\end{document}

\subsection{WIMP, Non-Minimal Models}

%\documentclass[12pt]{article}
%\usepackage{amsmath,amssymb,float,fullpage,graphicx,hyperref,slashed}

%%%%%%%%%%%%%%%%%%%%%%%%%%%%%%%%%%%%%%%%%%%%%%%%%%%%%%%%%%%%%%%%%%%%%%%%
%%%%%%%%%%%%%%%%%%%%%%%%%%%%%%%%%%%%%%%%%%%%%%%%%%%%%%%%%%%%%%%%%%%%%%%%
%%\begin{document}
%
%\begin{center}
%  {{\LARGE\bfseries Dark Matter at the FCC-hh}\\\bigskip}
%  {{\large Matthew Low and Lian-Tao Wang}\\\bigskip}
%\end{center}

% *** GUIDELINES ***
%
%  1. What are cases where a 100 TeV collider makes a qualitative difference (rather than just improve limits by some factor)
%  2. Is there complementarity between a 100 TeV collider and other machines (ILC/TLEP/CEPC) or other experiments (direct/indirect detection)
%  3. What are possible benchmark scenarios/points which should be studied more carefully
%  4. Is the reach of 14 TeV LHC known, so that the gain at 100 TeV can be put in perpective

%%%%%%%%%%%%%%%%%%%%%%%%%%%%%%%%%%%%%%%%%%%%%%%%%%%%%%%%%%%%%%%%%%%%%%%%
\subsubsection{Gluino, Stop Coannihilation}
\label{sec:coann}

A non-minimal scenario that can thermally produce the correct relic abundance and may be testable at a future collider is co-annihilation.  For concreteness we consider the DM to be a bino (electroweak singlet) and the co-annihilator to be a colored sparticle with a mass $m = 1.05 \; m_{\tilde{\chi}}$ where $m_{\tilde{\chi}}$ is the bino mass.  The near mass degeneracy allows the bino annihilation rate (with the co-annihilator) to increase which decreases the relic abundance.  The mass that gives the correct relic abundance depends on the splitting between the bino and co-annihilator.  The collider rate, on the other hand, is determined by the mass of the co-annihilator.

The first co-annihilator we consider is the gluino which is a color octet fermion.  At 100 TeV one finds a reach for the bino of 5.8 TeV to 6.2 TeV.\footnote{Actually the reach is 6.1 TeV to 6.5 TeV on the gluino and is not too sensitive to the bino mass.}  The reach is shown in Fig.~\ref{fig:coan} (left).  In this figure the upper $x$-axis shows the bino mass assuming $m_{\tilde{g}} = 1.05 \; m_{\tilde{\chi}}$ while the lower $x$-axis shows the $m_{\tilde{g}} - m_{\tilde{\chi}}$ value for which one finds the correct relic abundance~\cite{deSimone:2014pda}.  In this projection the gluinos are pair produced and assumed to decay via $\tilde{g} \to \tilde{\chi} + \text{undetected}$.  The most effective search is in the monojet channel.

%%%%%%%%%%%%%%
\begin{figure} [h]
\begin{center}
\includegraphics[width=0.42\textwidth]{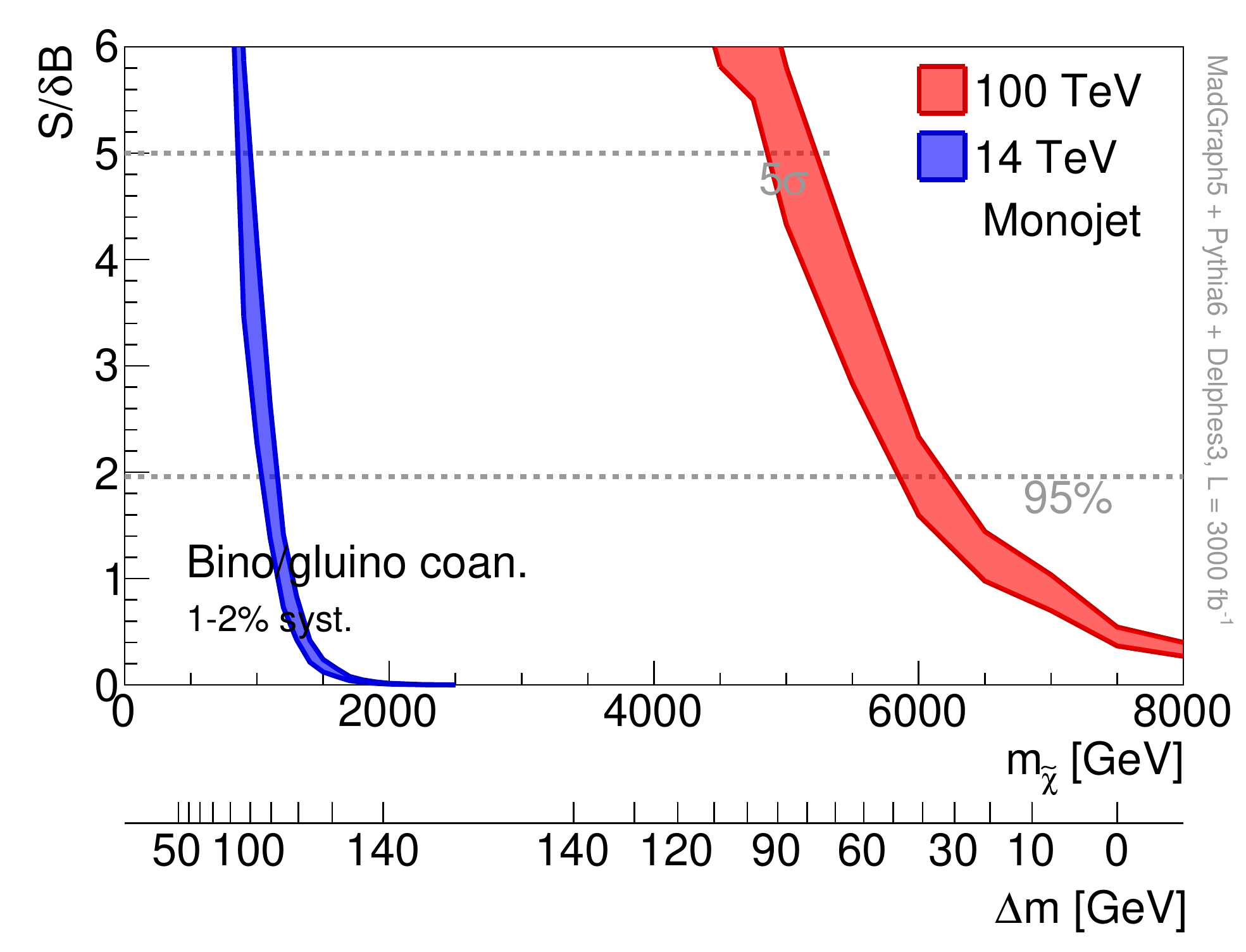} \quad\quad
\includegraphics[width=0.42\textwidth]{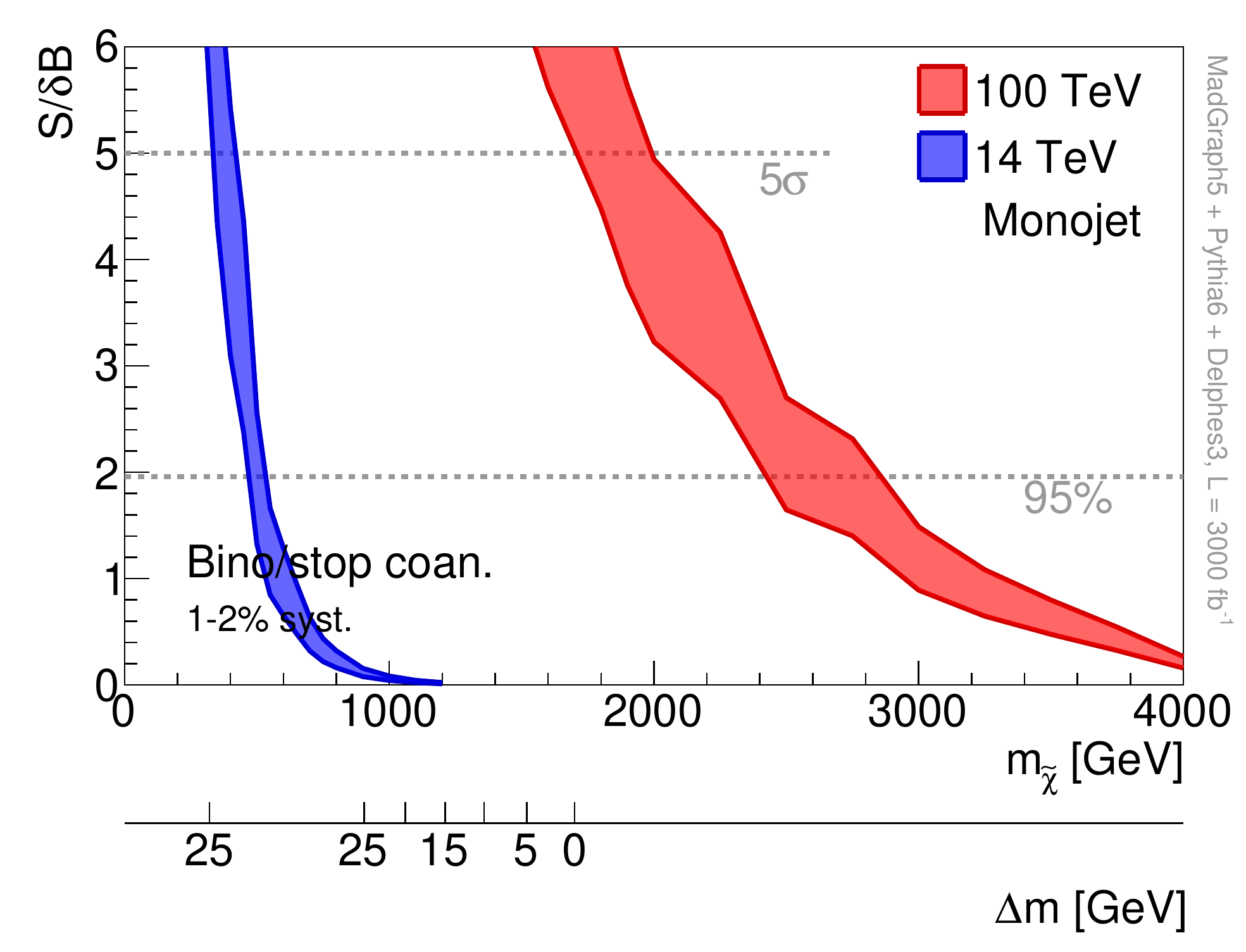}
\caption{Reach for binos that co-annihilate with gluinos (left) and and binos that co-annihilate with stops (right).}
\label{fig:coan}
\end{center}
\end{figure}
%%%%%%%%%%%%%%

The second co-annihilator we consider is the right handed stop which is a color triplet scalar.  The expected exclusion for the bino is 2.4 TeV to 2.8 TeV and in fact the discovery reach is 1.7 TeV to 2.1 TeV and is shown in Fig.~\ref{fig:coan} (right).  The $x$-axes, the same as in the gluino case, show the bino mass assuming $m_{\tilde{t}} = 1.05 \; m_{\tilde{\chi}}$ and the $m_{\tilde{t}} - m_{\tilde{\chi}}$ value for the relic abundance~\cite{Harigaya:2014dwa,deSimone:2014pda}.  The stops are pair produced and assumed to decay via $\tilde{t} \to \tilde{\chi} + \text{undetected}$ so the monojet channel is used.

Relative to 14 TeV, the increase is reach is about a factor of 5.  Importantly, however, the factor of 5 is enough to cover the thermal relic region for stops and come fairly close to covering the region for gluinos.

One obvious direction for future study is to consider also looking for the decay products of the gluino or stop.  This may be challenging because the most likely decays are $\tilde{g} \to \tilde{\chi} jj$ via an offshell squark and $\tilde{t} \to \tilde{\chi} b jj$ or $\tilde{t} \to \tilde{\chi} b \ell\nu$ via an offshell top, where the jets and leptons will have a momentum $\sim \Delta m$ which is $\mathcal{O}(10)$ GeV in the preferred parameter region.  It is not clear how feasible it will be to tag such low $p_T$ objects at 100 TeV.  On the other hand, in the monojet channel the $\tilde{g}\tilde{g}$ or $\tilde{t}\tilde{t}^*$ system recoils against a hard jet, so one can expect more energetic decay products when using certain selection criteria.

%\bibliographystyle{unsrt}
%\bibliography{fcc-neutralinos}
%\end{document}

\subsubsection{MSSM Dark Matter}

A crucial question in the development of a new collider program is whether, beyond increasing our sensitivity and mass reach to
new phenomena, the design energy and luminosity can provide us with definite answers to our most pressing questions. DM
motivated SUSY is a very compelling framework to ask ourselves this question for a high energy hadron collider project, such as the
100 TeV proton collider. SUSY is one of the best motivated theories of physics beyond the SM. The DM relic density provides us with well-defined
constraints on the mass and the nature of the WIMP candidate. In the MSSM with neutralino LSP, $\chi^0_1$ masses above
3.5-4~TeV are strongly disfavoured by the universe overclosure bounds, thus defining a mass scale well within the reach of the 100 TeV collider
energy. In addition, direct DM searches set constraints probing more and more in depth into the MSSM parameter space, in
particular for values of the $\mu$ parameter below 1-1.5~TeV. These constraints are complementary to those derived from direct searches
at the LHC and indirect sensitivity from the Higgs sector. 

For this report, the sensitivity of an 100~TeV $pp$ collider to DM-motivated
MSSM has been study by scans of the 19-parameters pMSSM where the SUSY particle masses have been independently varied up to 20~TeV.
The pMSSM has been extensively used to assess the current and projected coverage of the MSSM parameter space by the LHC
searches~\cite{Arbey:2013iza,Cahill-Rowley:2014twa,Aad:2015baa,CMS:2015ebf}.
Generated pMSSM points have been checked against low-energy and flavour physics constraints and the lightest Higgs boson mass has
been required to be in the range 119$< M_{h^0}<$129~GeV. In addition, we require the neutralino relic density not to exceed the
PLANCK CMB result, when accounting for systematic uncertainties. This allows for additional, non-SUSY contributions to the relic density
from CMB. For each accepted pMSSM point, sets of inclusive SUSY events have been generated and the physics objects computed after a
parametric detector simulation.
\begin{figure}[h!]
  \begin{center}
    \includegraphics[width=0.5\linewidth]{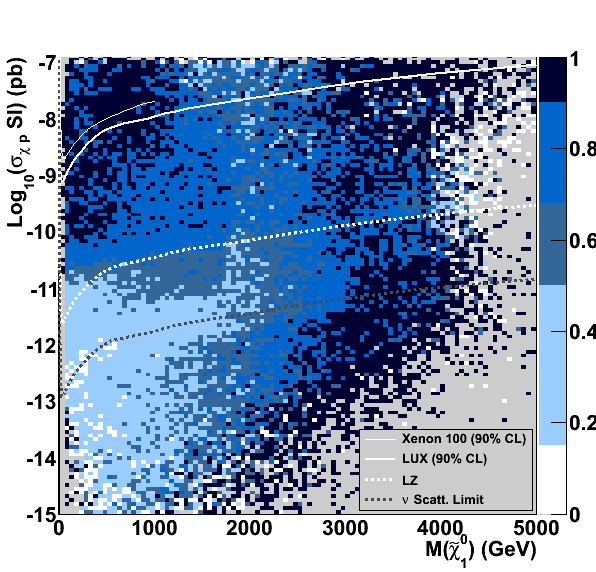}%
    \caption{Fraction of DM-motivated pMSSM points with neutralino LSP and SUSY masses up to 20~TeV excluded by searches at a 100~TeV $pp$ collider
      in the neutralino scattering cross section vs.\ neutralino mass plane. Current and projected limits from DM direct detection experiments are
      overlayed.}
  \label{fig:DDpSIMN1100TeV}
  \end{center}
\end{figure}

Searches in jets+MET, leptons+MET, monojets and monoW/Z+MET have been evaluated using analysis strategies 
derived from those currently performed on the LHC data, but re-optimising the kinematical cuts. Results have been obtained in terms of
the fraction of pMSSM points that could be excluded in case no excess of events is observed with a given integrated luminosity of 100~TeV
$pp$ data and when combining with current and future DM direct detection data. Results are summarised in Figure~\ref{fig:DDpSIMN1100TeV} and
Table~\ref{tab:results} showing how the combination of 100 TeV collider and future DM direct detection experiments can virtually saturate the
MSSM parameter space, if SUSY is responsible for (at least part of) DM.
\begin{table}
  \begin{center}
    \begin{tabular}{|c|c|c|c|c|c|}
      \hline
      $\sqrt{s}$ & $L$ & Collider & +LUX & +LX & +3rd Gen. \\
      (TeV)      & (ab$^{-1}$) & (MET) & DM & DM & DM      \\
      \hline
      100 & 1.0  & 0.63 & 0.65 & 0.73 & 0.90 \\
      100 & 3.0  & 0.67 & 0.69 & 0.75 & 0.91 \\
      100 & 5.0  & 0.69 & 0.72 & 0.76 & 0.92 \\
      \hline
    \end{tabular}
  \end{center}
  \caption{Fraction of DM-motivated pMSSM points with neutralino LSP and SUSY masses up to 20~TeV excluded by searches at a 100~TeV $pp$ collider
    and DM direct searches.}
  \label{tab:results}
\end{table}

In gravitino DM models with thermal leptogenesis, the interplay of gravitino relic density, re-heating temperature of the Universe after the
inflationary phase, $T_{RH}$ and the gluino mass, $M_{\tilde{g}}$, determine an upper bound on the gluino mass relevant to the HL-LHC and also
a high energy hadron collider~\cite{Arvey:2015nra}.
With a sensitivity to the gluino mass up to $\sim$10~TeV, the 100~TeV collider can fully probe these
models for $T_{RH} > 3 \times 10^8$~GeV (see Fig.~\ref{fig:MGLRHT20TeV}).  
\begin{figure}[h!]
  \begin{center}
    \includegraphics[width=0.5\linewidth]{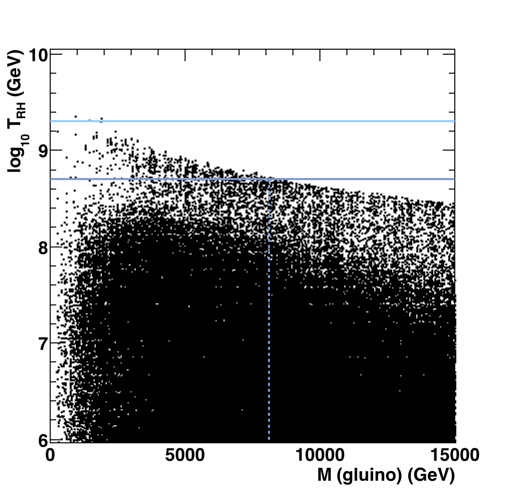}%
    \caption{Distribution of DM-motivated pMSSM points with gravitino LSP, neutralino NLSP and SUSY masses up to 20~TeV in the re-heating temperature
      vs.\ gluino mass plane. A constraint on the gluino mass can exclude gravitino LSP models with thermal leptogenesis requiring re-heating
      temperatures above $\sim 5 \times 10^8$~GeV.}
  \label{fig:MGLRHT20TeV}
  \end{center}
\end{figure}

%\end{document}

%
\subsection{Beyond WIMP DM}
\label{sec:DM_beyond_wimp}
As mentioned in the introduction, there is a variety of DM models where the observed relic density is obtained by a mechanism different from WIMP freeze-out, but which nevertheless are testable in collider experiments. 

Here we will discuss an example of an asymmetric DM (ADM) model, a scenario with a composite hidden sector, a model of super-WIMPs and a variation of supersymmetry where the abundance of the DM candidate, the gravitino, is set by decays of the next-to lightest supersymmetric particle. All these models lead to collider signatures that could be detectable at a hadron collider but which are different from the usual DM search channels. Their detectability should therefore also be taken into account when discussing future collider experiments. 

What this section is not, but what would be highly desirable for the future, is a full classification of beyond-WIMP DM signatures which are relevant for colliders. In particular the testability of the ADM paradigm (and not just particular models) at a future collider should be analyzed further. 

\subsubsection{Asymmetric DM through the Higgs Portal}
The Higgsogenesis scenario introduced in Ref.~\cite{Servant:2013uwa} is one of the most compact ADM models. The DM sector consists of a pair of a vector-like $SU(2)$ doublet of fermions $X_2$ and a neutral fermionic singlet $X_1$ (the DM candidate), and is thus similar to the singlet doublet models mentioned above and to the Bino/Higgsino scenario, which can be probed at a 100\UTeV~collider up to masses of order 1.2\UTeV, see Fig.~\ref{fig:DM_wgb1_summary}  and Ref.~\cite{Low:2014cba} for more details. 

The basic idea is to use the chemical potential of the Higgs to transfer an asymmetry between the SM and the DM sector. After an asymmetry is generated in the visible sector, but before electroweak symmetry breaking, the Higgs carries a nonzero charge asymmetry, which is transferred to the DM sector by an operator
\begin{align}
	{\cal L} \supset \frac{1}{\Lambda_2} (H^\dagger X_2)^2 +{\rm h.c.}\,,
\end{align}
which is possible for values of $\Lambda_2$ up to the GUT scale. A small Yukawa coupling $y_H \bar{X}_2 X_1 H$ allows $X_2 \to X_1 H$ decays, which transfer the asymmetry to $X_1$ once the temperature drops below $M_{X_2}$, and which should happen after the transfer operator freezes out. The inverse process, where an asymmetry generated in the dark sector is transferred to the SM, is also possible with this mechanism. 

In Ref.~\cite{Servant:2013uwa} it was shown that this mechanism can give the correct DM relic abundance for a range of DM masses from 10\UGeV~to 10\UTeV, which should at least partially be in range of a future hadron collider. A more detailed study would be welcome. 

Note that while the Higgsogenesis mechanism is minimal in the sense that no complicated transfer sector is possible, the problem of annihilating the symmetric component is not yet addressed. Indeed Ref.~\cite{Servant:2013uwa} introduces an auxiliary mediator $\phi$ such that the DM can annihilate through the process $X_1 \bar{X}_1 \to \phi \phi$, with $\phi$ later decaying to SM particles. In this case the constraints from light mediator searches (see Sec.~\ref{sec:sms}) become relevant, and additional studies for light scalar portals at 100\UTeV~are needed. 

Instead if one demands that the symmetric component of DM in ADM models annihilates directly into SM particles, then one can use a similar approach as for WIMP particles, namely by either classifying the annihilation channels using effective operators or, as done in the WIMP section above, using the particles that mediate the annihilation. The main difference is that now the annihilation has to be \textbf{stronger} than in the WIMP case, since the relic abundance should be dominated by the asymmetric contribution, otherwise it would just be another WIMP scenario. For ADM annihilating to SM quarks, this was studied in~\cite{Buckley:2011kk,MarchRussell:2012hi} using effective operators to parameterise the interactions. An update of this study using the simplified model approach discussed above, and including projections to 100\UTeV~hadron colliders, would be useful. 

Alternatively if the ADM candidate is part of a complex, possibly composite dark sector, then the fast annihilation of the symmetric component into unstable dark sector states can be a natural consequence of the model, and such an example is discussed in the next section.

\subsubsection{Dark QCD, Hidden Valley DM}
\label{sec:DM_darkQCD}
Models where the DM candidate is a stable composite state of a QCD like dark sector are well motivated, simply by comparison with the case of the proton in QCD. Stability of DM would be guaranteed by global DM number conservation, and the mass scale can be generated through dimensional transmutation from a small coupling at a high scale. 

Such models were originally considered in the context of parity symmetric "mirror world" scenarios~\cite{Lee:1956qn,Kobzarev:1966qya,Foot:1991bp,Berezhiani:1995am}, where the DM would be composed of mirror protons. However in those scenarios the only interactions of the visible world with the dark sector are gravitational, such that they are not relevant for collider phenomenology (and indeed are difficult to verify overall). 

In~\cite{Strassler:2006im,Han:2007ae} so called Hidden Valley were introduced where a QCD like confining hidden sector communicates with the SM through heavy mediators, and DM models based on this general construction were introduced e.g. in~\cite{Kribs:2009fy,Blennow:2010qp,Frandsen:2011kt,Bai:2013xga}. The main idea is that the DM candidate is a composite baryonic bound state made out of dark quarks of a "dark QCD" which, similar to the proton in the SM, is stable because it carries a conserved DM number, and with a mass of order of the \UGeV~scale set by the dark QCD confinement scale $\Lambda_D$. A heavy \UTeV~scale mediator $X$ is responsible for sharing the asymmetry between the SM and the dark sector. Fast annihilation of the symmetric DM component is now guaranteed by the equivalent of proton anti-proton annihilations into pions in QCD, i.e. the DM annihilates to dark pions. The heavy mediators itself can allow the dark pions to decay back to SM particles, therefore no additional light mediators have to be introduced by hand. 

Observability of such a dark sector is mainly determined by whether the mediator particle can be produced at a collider. If this is the case, the phenomenology can be quite spectacular, since the strong dynamics in the hidden sector can produce dark jets, events with many displaced vertices, final states with many heavy flavours, and more~\cite{Han:2007ae,Strassler:2008fv}. 

Let us first consider one characteristic signature dubbed "emerging jets" in~\cite{Schwaller:2015gea}, where the mediator $X$ is pair produced and decays to quarks and dark quarks. While each quark will undergo a regular shower and hadronization process and give rise to jets, the dark quarks will shower in the dark sector first and produce dark jets made out of mostly dark pions. These dark pions naturally have lifetimes of order milimeters to meters, and therefore decay back to SM particles throughout the detector, collimated within a "dark jet". The strategy proposed in~\cite{Schwaller:2015gea} is to reconstruct regular multi-jet events and then search for emerging jets, i.e. jets with few or no tracks pointing back to the interaction point, in the multi-jet sample. It was found that requiring two emerging jets in one event almost fully removes the QCD backgrounds, which mostly come from long lived neutral mesons decaying in the detector. At the 14\UTeV~LHC this class of models can be probed for mediator masses up to1.5\UTeV~which, by applying naive parton luminosity scaling\footnote{http://collider-reach.web.cern.ch}, implies a reach of up to 9\UTeV~at a 100\UTeV~machine with 3~${\rm ab}^{-1}$. 

A variation of the above scenario occurs when the dark pions in the shower have different lifetimes, ranging from prompt decays to particles stable enough to escape the detector. In this case one gets a jets plus missing energy signal, with the missing energy correlated with the jet directions. Ref.~\cite{Cohen:2015toa} proposes a search for such events coming from the decay of a $Z'$ mediator into hidden sector particles, which utilises the transverse mass computed from the jets and missing energy to distinguish from QCD backgrounds. At the LHC the projected reach is 3.5\UTeV, which should scale up to 20\UTeV~at a 100\UTeV~collider. 

Finally if the hidden sector communicates with the SM mainly through the Higgs portal, then exotic Higgs decays involving displaced vertices might be the leading tool to probe such dark sectors~\cite{Strassler:2006ri,Curtin:2013fra}. A 100\UTeV~machine is an ideal tool to search for exotic Higgs decays, simply due to the large amount of Higgs bosons produced there (this is discussed in the accompanying Higgs at 100\UTeV~document). 

The main point of this section is to introduce some examples of DM models which give rise to signatures which are different from the well known missing energy searches and also different from standard mediator searches. Detectability at a hadron collider in these cases is mainly due to particles associated with the DM sector, but not the DM itself\footnote{This is also true e.g. for the disappearing track search which is crucial for the Wino scenario discussed above.}, and this is also the reason why new search strategies are possible and necessary. 

A possible classification of signatures should be possible in two steps: First, one specifies how the dark sector communicates with the SM. This is very similar to the WIMP case, so the classification of SM and BSM mediators done above can be applied here as well. The second step is to classify the dynamics in the dark sector itself. This includes distinguishing perturbative and non-perturbative dark sectors, and a classification of additional symmetries which could for example give rise to a hierarchy of lifetimes or further constrain the interactions of the mediator.

\subsubsection{Radiating DM}

\label{sec:dm_rad}
%----------------------------------------------------------------------------
\textbf{Radiation in the Dark Sector}
%----------------------------------------------------------------------------
%
At the spectacular partonic center of mass energies afforded by a 100~TeV
collider, radiative processes reach unprecedented levels. This is true not only
for QCD and electroweak interactions, but also for any new physics sector that
contains light particles with appreciable couplings.  A particularly
well-motivated example is self-interacting DM~\cite{Spergel:1999mh,
Tulin:2013teo}, which could potentially resolve shortcomings of our present
understanding of cosmic structure formation on dwarf galaxy scales (see also
\cite{Vogelsberger:2014kha, Sawala:2014xka}).  Moreover, DM self-interactions
can also lead to Sommerfeld enhancement in DM annihilation \cite{Hisano:2003ec,
Cirelli:2008pk, ArkaniHamed:2008qn}, and possibly even to the formation of DM
bound states \cite{Shepherd:2009sa, Altmannshofer:2014cla}.

Perhaps the leading candidate for the mediator of DM self-interactions is a dark photon $A'$---the gauge boson of a corresponding to a new
local $U(1)'$ symmetry in the dark sector.  To be phenomenologically relevant,
$A'$ is typically light (MeV--GeV), has relatively strong couplings to DM
($\alpha' \sim 0.01$--0.1), and tiny couplings to the SM sector through
kinetic mixing with the photon.  The dark sector Lagrangian in such a scenario
reads
\begin{align}
  \mathcal{L}_\text{dark} \equiv
  \bar{\chi} (i\slashed{\partial} - m_\chi + i g_{A'} \slashed{A'}) \chi
    - \frac{1}{4} F'_{\mu\nu} F'^{\mu\nu}
    + \frac{1}{2} m_{A'}^2 A'_\mu A'^\mu
    - \frac{\epsilon}{2} F'_{\mu\nu} F^{\mu\nu} \,.
  \label{eq:L-radiating-DM}
\end{align}
Here, $\chi$ is the fermionic DM particle with mass $m_\chi$, $g_{A'} = \sqrt{4
\pi \alpha'}$ is the $U(1)'$ gauge coupling, $m_{A'}$ is the dark photon mass,
and $\epsilon$ is the kinetic mixing parameter, which is typically $< 10^{-3}$.
We remain agnostic about the origin of the dark photon mass---it could originate
from a dark sector Higgs mechanism or from the St\"uckelberg mechanism.

When DM is produced with a large boost at a collider, there is a high
probability that additional collinear $A'$ bosons are radiated (see
fig.~\ref{fig:dark-radiation-diagram}) \cite{Buschmann:2015awa}. This is
particularly true if the DM itself and the dark photon are rather light, for
instance on the order of GeV, where direct detection bounds are weak. The higher
the center of mass energy of the process, the more $A'$ bosons are radiated, as
illustrated in fig.~\ref{fig:n-A'}.  These $A'$ bosons eventually decay to
observable SM particles through the kinetic mixing term in
eq.~\eqref{eq:L-radiating-DM}, with the exact branching ratios depending
sensitively on $m_{A'}$ (see fig.~2 in ref.~\cite{Buschmann:2015awa}).
Depending on the value of $\epsilon$, the decays can be either prompt or
displaced. Phenomenologically, the final state of the process $p p \to \bar\chi
\chi + n A'$ thus consists of two ``jets'' of collimated $A'$ decay products,
plus missing energy.

\begin{figure}
  \begin{center}
    \includegraphics[width=0.4\textwidth]{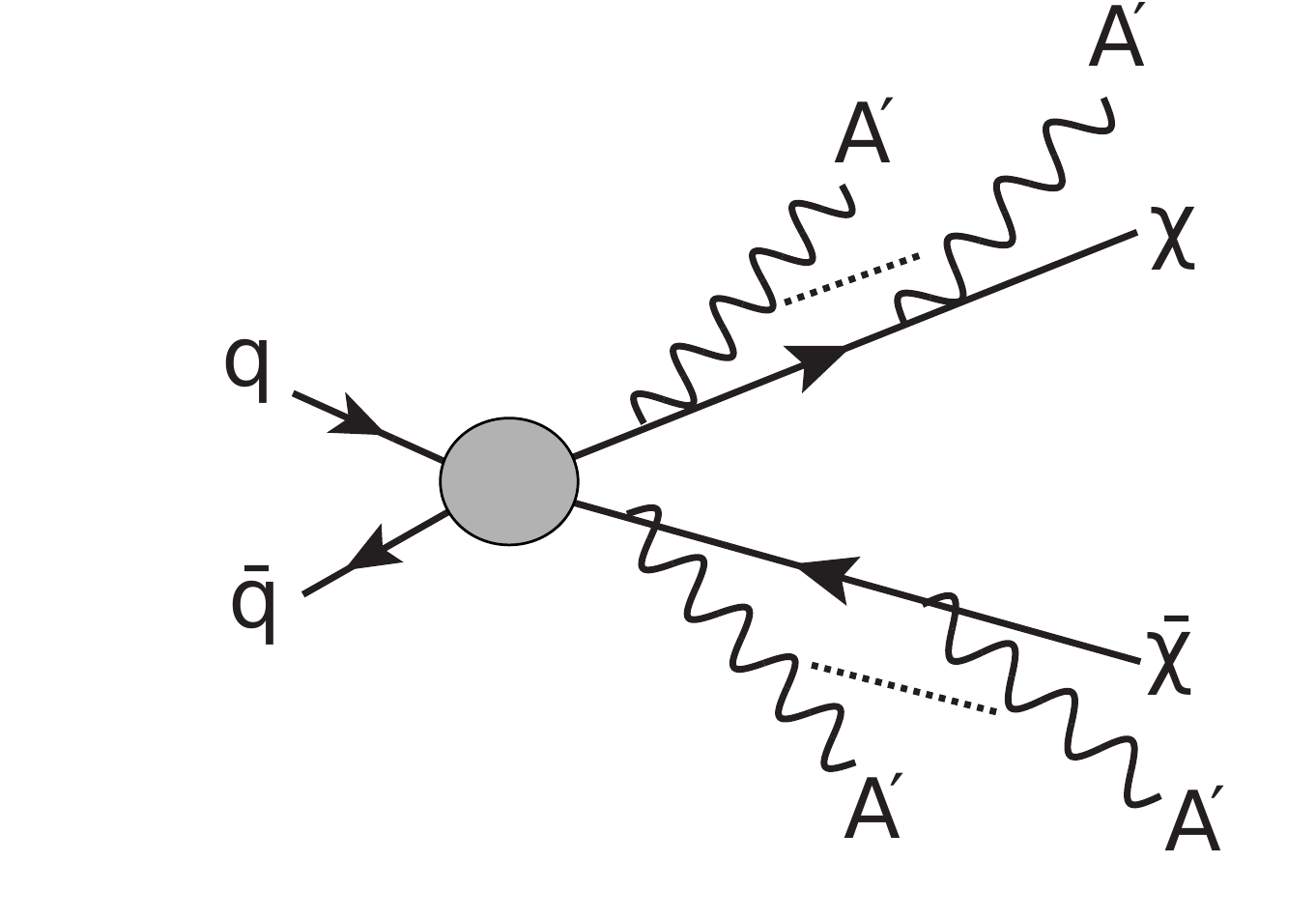}
  \end{center}
  \caption{The process that gives radiating DM its name: production of
    two DM particles $\chi$, followed by the emission of several soft or collinear
    dark photons $A'$~\cite{Buschmann:2015awa}.}
  \label{fig:dark-radiation-diagram}
\end{figure}

\begin{figure}
  \begin{center}
    \includegraphics[width=0.6\textwidth]{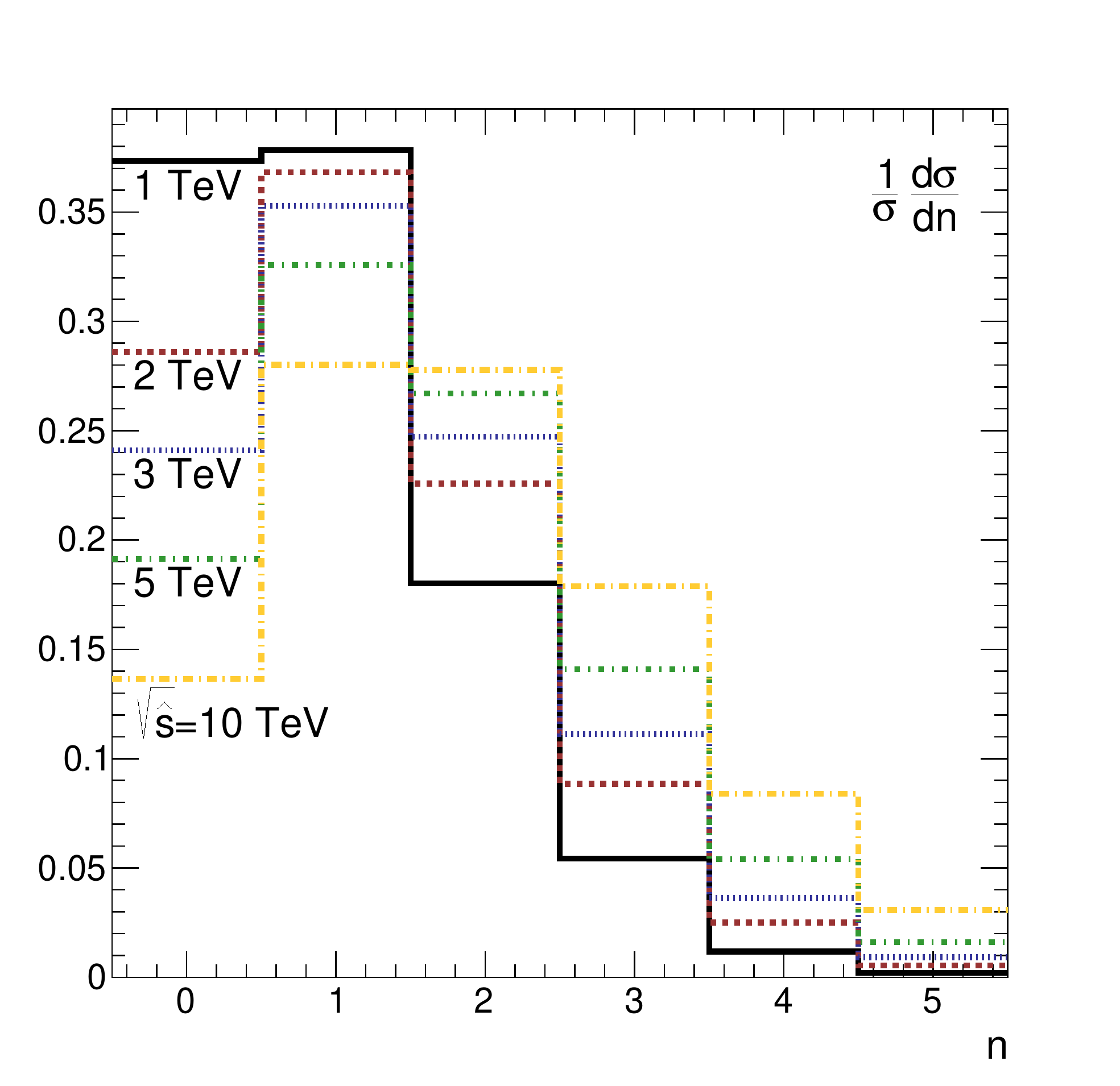}
  \end{center}
  \caption{The distribution of the number of dark photons $A'$ radiated
    in each DM pair production process $p p \to \bar\chi \chi$
    for several values of the partonic center of mass energy $\sqrt{\hat{s}}$.
    We have used a dark fine structure constant of $\alpha' = 0.05$, a DM mass
    of $m_\chi = 4$~GeV, and a dark photon mass $m_{A'} = 1.5$~GeV.
    The computation was carried out in Pythia~8~\cite{Carloni:2010tw,
    Carloni:2011kk, Sjostrand:2014zea}, using the hidden valley model implemented
    therein. See ref.~\cite{Buschmann:2015awa} for an analytic treatment
    of dark radiation.}
  \label{fig:n-A'}
\end{figure}

%----------------------------------------------------------------------------
\noindent\textbf{Phenomenology of Radiating Dark Matter}
%----------------------------------------------------------------------------
%
Depending on the decay modes, these $A'$ jets can be classified into one of the
following categories:
\begin{itemize}
  \item {\bf Lepton Jets.} If all $A'$ bosons radiated by a DM particle decay
    leptonically, a large number of collimated leptons is expected.
    Such lepton jets have been discussed previously for instance in
    refs.~\cite{ArkaniHamed:2008qp, Cheung:2009su, Katz:2009qq, Bai:2009it,
      Baumgart:2009tn, Chan:2011aa, Falkowski:2010gv, Curtin:2013fra, Gupta:2015lfa,
    Autran:2015mfa}.  Experimentally, lepton jets have been searched for
    in refs.~\cite{Aad:2012qua, Aad:2014yea}.  SM backgrounds to these searches
    are extremely low, making them a particularly sensitive probe of new physics.
    This is especially true for $m_{A'} \lesssim 2 m_\pi$, where all $A'$ bosons
    decay leptonically. At larger $m_{A'}$, the branching ratio for $A' \to \ell^+\ell^-$
    varies between 20\% and 70\%.

  \item {\bf Mixed Jets.} If some of the $A'$ in the dark photon jet decay
    leptonically and others decay hadronically, we expect a QCD-like jet with
    anomalously large lepton content. This signature can be distinguished from
    ordinary QCD jets by looking for an anomalously large energy deposit in the
    electromagnetic calorimeter and/or the muon system. Moreover, for displaced
    $A'$ decays, the occurrence of displaced vertices is a smoking gun signature.

  \item {\bf Hadronic Jets.} If all collimated $A'$ bosons decay to hadrons,
    they closely resemble a QCD jet.  Nevertheless, if the kinetic mixing parameter
    $\epsilon$ is so small that $A'$ decays are displaced, a separation from
    SM backgrounds is possible.
\end{itemize}

%----------------------------------------------------------------------------
\noindent\textbf{Why go to 100~TeV}
%----------------------------------------------------------------------------
%
It is clear from fig.~\ref{fig:n-A'} that a search for radiating DM
could greatly benefit from the increased center of mass energy afforded by a
100~TeV collider: at higher $\sqrt{\hat{s}}$, the probability that a DM
particle radiates at least one $A'$ is much higher than at the LHC, and in many
events, several $A'$ bosons will be emitted per DM particle, making the
signature even more spectacular.  Of course, the relation between the partonic
center of mass energy $\sqrt{\hat{s}}$ and the collider energy $\sqrt{s}$
depends on the details of the DM production process.  If DM is produced through
an $s$-channel mediator with mass at the electroweak scale, $\sqrt{\hat{s}}$ is
typically similar to the mediator mass if that mass is kinematically
accessible.  Naturally, at a 100~TeV collider, much heavier mediators can be
probed than at the LHC.  For off-shell production, e.g.\ through colored
$t$-channel mediators such as squarks, $\sqrt{\hat{s}}$ is determined by the
valence quark PDFs. Once again, a 100~TeV machine would have a significant edge
over the LHC. In fact, dark radiation cannot be probed in such $t$-channel
scenarios at the LHC at all. At 100~TeV, such restrictions are removed,
allowing a 100~TeV collider to probe radiating DM in all of the most important
electroweak-scale production channels.

%----------------------------------------------------------------------------
\noindent\textbf{Some Thoughts on Search Strategies and Detector Design}
%----------------------------------------------------------------------------
%
The sensitivity of a search for radiating DM hinges crucially on the analysis
cuts imposed. Questions to consider in designing a search for radiating
DM are
\begin{itemize}
  \item {\bf Is there a signal in the tracking detector?} Most
    prompt $A'$ decays will lead to such signals, but for displaced decays of
    longer lived $A'$ bosons, it will usually be absent. Moreover,
    some subdominant
    decay modes ($A' \to K^0 \bar{K}^0$ and $A' \to \pi^0 \gamma$) do not yield
    a tracker signal.

  \item {\bf Is there a signal in the calorimeters?} Once again, the presence
    of such a signal depends on the $A'$ decay mode: $A' \to \mu^+ \mu^-$ is not
    visible to the calorimeters, while all other decay modes are.

  \item {\bf What fraction of the decay energy is deposited in the electromagnetic
    calorimeter?} (as opposed to the hadronic calorimeter). For short-lived
    $A'$, this fraction allows to distinguish different decay modes. For displaced
    decays, however, even a decay like $A' \to e^+ e^-$ can deposit its energy
    mostly in the hadronic calorimeter.
\end{itemize}

It is clear from these considerations that the sensitivity of a search for
radiating DM depends sensitively on the detector design. In particular, the
values of the kinetic mixing parameter $\epsilon$ that can be probed efficiently
in the search for displaced decays change with the radial size of the detector.
This statement can be quantified by considering that the $A'$ decay rate to
leptons is given by
\begin{align}
  \Gamma(A' \to \ell^+\ell^-) =
    \frac{1}{3} \alpha \epsilon^2 m_{A'}
    = \frac{1}{8 \times 10^{-6}\,\text{cm}}
      \bigg( \frac{\epsilon}{10^{-3}} \bigg)^2
      \bigg( \frac{m_{A'}}{\text{GeV}} \bigg)
\end{align}
in the limit $m_\ell \ll m_{A'}$ (see ref.~\cite{Buschmann:2015awa} for a more
detailed discussion of $A'$ decays).  For this reason, it may be useful, for
instance, to have one rather compact detector and one fairly large one (like at
the LHC). Thinking even further, a dedicated search for radiating DM
in the small $\epsilon$ region might benefit from dedicated muon detectors
placed far away $\gtrsim 10$~m from the interaction point. Of course, with such
a system it would be difficult to achieve good angular coverage.

%============================================================================
%\bibliographystyle{JHEP}
%\bibliography{radiating-dm-100-TeV}
%============================================================================

%\end{document}

%  {\color{red} contribution by Kopp et al, coming in 2-3 days}

\subsubsection{SuperWIMPs and Gravitino DM}
\label{sec:DM_superW}
Many extensions of the SM introduce additional particles which are only very weakly coupled to the SM, and which are potential DM candidates. Prime examples are the axion as a solution of the strong CP problem, and the gravitino which arises in supersymmetric models involving gravity. In some cases the mechanism that sets the relic density of such particles is accessible at colliders, and we will discuss one such example now, following~\cite{Feng:2005gj} (see also~\cite{Hamaguchi:2004df,Feng:2004yi,Buchmuller:2007ui,Arcadi:2014tsa}). 

Consider a super-WIMP (SWIMP) with mass $m_{\rm SWIMP}$ which is stable on cosmological time scales, but which is not thermalized in the early universe, and furthermore a weakly interacting particle $L$ which freezes out with relic abundance $\Omega_{L}$. If $L$ decays to the SWIMP with a lifetime short enough to not upset nucleosynthesis, then the super-WIMP can be the DM candidate with abundance set by 
\begin{align}
	\Omega_{\rm SWIMP} h^2 = \frac{m_{\rm SWIMP}}{m_{L}} \Omega_{L} \,.
\end{align}
The difference from a WIMP scenario is that now $L$ can be charged, and this is the case for example in supersymmetric models where $L$ can be a charged slepton, or a KK-lepton in extra-dimensional models. Signatures of these scenarios now include heavy stable charged particles travelling through the collider and displaced decays of $L$ in the detector which can give rise to either displaced vertices or kinked tracks. Since $L$ is the natural end product of any new particle which is produced and which carries the DM symmetry, these signatures will also appear in combination with jets and leptons. Another exciting possibility is that $L$, if charged, can loose energy rapidly and therefore might get stuck in the detector or in the surrounding material~\cite{Feng:2004yi}, where it can be trapped and analyzed before it decays. 

The reach for these signatures should scale with the center of mass energy, since they are free from SM backgrounds. Therefore a 100\UTeV~collider can certainly probe $m_L$ in the multi-TeV range, maybe even reach 10s of TeV, provided that future detectors have at least the same capabilities as the LHC experiments. 

\subsection{DM Summary}

In this section, the reach of a 100~TeV collider for large classes of DM models was explored and compared to the reach of the 14~TeV LHC, indirect and direct DM detection experiments and other collider and laboratory experiments. As one would have expected, a 100~TeV machine vastly increases the mass range up to which DM models can be probed, and in several cases the upper mass limit indicated by the observed DM density is reached. In other words:
\begin{itemize}
	\item There are well defined DM models whose parameter space can be fully probed at a 100~TeV collider. 
\end{itemize}
Maybe the simplest example is the Wino (SU(2) triplet) scenario studied in Secs.~\ref{sec:wgb1}, \ref{sec:wgb2}, \ref{sec:wgb4}. Here one important aspect is that monojet searches alone can not cover the theoretically motivated DM mass range, however the combination with either indirect detection or with disappearing track searches is sufficient to fully probe the viable parameter space. 

Monojet and related missing energy searches are essential to establishing the presence (or absence) of DM at a collider, and in many scenarios they extend the reach beyond parameter space accessible in direct detection, as can be nicely seen for example in Figs.~\ref{fig:exclMass:9}, \ref{fig:exclMass:10}. However also here it is important to notice that full coverage of the viable parameter space is often only possible by combining these searches with other channels. A nice example is provided in Fig.~\ref{fig:m-g}, where the search for the mediator in the dijet channel is necessary to probe the tuned region of parameter space where resonant annihilation allows for very large DM and mediator masses. 

Fortunately a hadron collider allows the study of many signal channels in parallel, and standard multijet + MET searches can easily be combined with disappearing track or displaced object searches. The importance of those more exotic signatures is not only highlighted by their ability to close of the parameter space of some of the minimal models discussed in Secs.~\ref{sec:wgb1}-\ref{sec:wgb4}, but also because they give access to a broader range of DM models which might not easily show up in missing energy signatures, like some of the examples discussed in Sec.~\ref{sec:DM_beyond_wimp}. 

\begin{table*}[t]     
\begin{center}                                                                                                                                                                              
\begin{tabular}{c|c|c}                                                                                                                                                                  
\hline\hline                                                                                                                                                                                        
  Final State & Analysis & section  \\                                                                                                                                  
\hline 
jet+MET           & Wino, Higgsino DM & \ref{sec:wgb1} - \ref{sec:wgb4}  \\
jet+MET           & Higgs Portal             & \ref{sec:DM_higgsportal}   \\
jet+MET           & Simplified Vector/Axial & \ref{sec:sms} - \ref{sec:smsmjdj} \\
jet+MET           & Simplified Scalar/Pseudo & \ref{sec:sms} - \ref{sec:smsmjdj} \\
jet+MET           & Gluion/stop coannihalation & \ref{sec:coann}  \\
VBF jets +MET  & Wino, Higgsino DM & \ref{sec:wgb1} - \ref{sec:wgb2} \\ 
VBF jets +MET  & Higgs Portal            &  \ref{sec:DM_higgsportal} \\ 
photon+MET    & Wino                        & \ref{sec:wgb2} \\
Disappearing tracks  & Wino,Higgsino & \ref{sec:wgb1} - \ref{sec:wgb2} \\
Disappearing tracks  & Fiveplet DM     & \ref{sec:wgb3}  \\
Disappearing tracks  & Relic-Neutralino & \ref{sec:wgb4}  \\
lepton+$\gamma$+MET & Relic-Neutralino & \ref{sec:wgb4}  \\
$Z_{D}\rightarrow ll$+($Z_{D}\rightarrow ll$) & Dark Photons & \ref{sec:dm_darkp}, \ref{sec:dm_rad}  \\
%$Z_{D}\rightarrow ll$+($Z_{D}\rightarrow ll$) & Dark Photons & 3.7.3  \\
displaced jets & Dark QCD/Hidden Valley  & \ref{sec:DM_darkQCD}  \\
long lived charged particle & Super-WIMPS/Gravitino & \ref{sec:DM_superW}  \\
dijet                 & Simplified Vector/Axial & \ref{sec:sms} - \ref{sec:smsmjdj} \\
\hline
\end{tabular}                                                                                                                                                                                       
\caption{Overview of the final states and the associated model, with a link to the respective section. }                                                                               
\label{tab:darkmatterfinalstate}       
\end{center}                                                                                                                                                                        
\end{table*}                                                                                                                                                                                        
%Note this should have the right labels

A list of final states which are relevant for DM searches at a 100~TeV collider is given in Tab.~\ref{tab:darkmatterfinalstate}. Here information is provided which models are probed by which final states and, for models which are testable in several final states, which ones are the most sensitive. It should be emphasised that this list is not complete but instead based on the models and channels for which studies are available, and in particular scenarios for which only one channel is sensitive would benefit from further studies. 

While the overall prospects for observing DM at a 100~TeV hadron collider are promising, one should not forget that there are models which are notoriously difficult to probe. The worst offender is one of the simplest models, namely the Higgs portal with a singlet scalar DM candidate, discussed in Sec.~\ref{sec:DM_higgsportal}. The good news here is that this model also has a clear prediction for the direct detection rate, which is accessible in the next generation of experiments. Further studies of such a DM candidate might then require a linear collider, so this should be seen as additional motivation to study the complementarity of high energy lepton colliders with a 100~TeV hadron collider.

%Feel free to introduce subsections. Each subsection should be a
%separate file, to be included with an 
%\verb| \input{subsection}| statement.
%
%As labels for the subsections, try to use things like
%\verb|sec:DM_xxx|. Ditto for references to equations, figs:
%\verb|\label{eq:DM_xxx}|, etc

\newpage
\section{Other BSM Signatures} %\footnote{Editors: Torre, Golling}} 
\label{sec:beyond}
%\begin{itemize}
%\item M.~Mangano, A.~Thamm, R.~Torre, A.~Wulzer, {\it Resonances at 100 TeV: jets vs leptons} {\bf should be part of the introduction}
%\item T.~Han, M.~Mangano, Z.~Qian, {\it 1511.06495} {\bf to be edited}
%\end{itemize}
This section is devoted to the assessment of the potentials of a future circular hadron collider at 100 TeV (to which we will sometimes refer simply as FCC) in terms of BSM signature that are not ``typical'' of supersymmetric and dark matter models. Of course, most of the signatures discussed here can be relevant for certain such models, however they do not constitute their key ingredient nor their smoking gun signatures in a way clarified in the following discussion. In particular the goal of this section can be summarized in two points:
\begin{itemize}
\item Study morivated and generic signatures that allow to test new physics. With motivated and generic we mean signatures that are shared by large classes of new physics scenarios.
\item Define collider and detector benchmarks that allow the highest possible sensitivity to these motivated and generic signatures and that can help in a concrete assessment of the needed design of the future facilities.
\end{itemize}

The results will be presented as a list of more or less detailed studies of concrete BSM models, which predict a particular signature, shared by broad BSM scenarios. The cases presented in what follows generally fall into three broad categories:
\begin{itemize}
\item Single production of new particles;
\item Pair production of new particles;
\item Precisione measurement aimed at constraining NP indirectly.
\end{itemize}
However, since single and pair production can constitute, in some cases, like for instance fermionic top partners, inseparable signatures of motivated BSM scenarios, we prefer to separate the results into three sections devoted to new bosonic resonances, which are signatures typically interesting for single production and new fermionic resonances, which, depending on the scenario can be interesting both in the single production (in association with other SM particles) and pair production. Finally we will devote a last section to the non-resonant signatures, which are aimed at constraining new physics indirectly. We can already get a preliminary idea, which will be refined in the following sections, of the reach of a 100 TeV collider on high invariant mass objects, by considering the single production of a general heavy resonance $R$.
\begin{figure}[htbp]
\begin{center}
\includegraphics[scale=0.36]{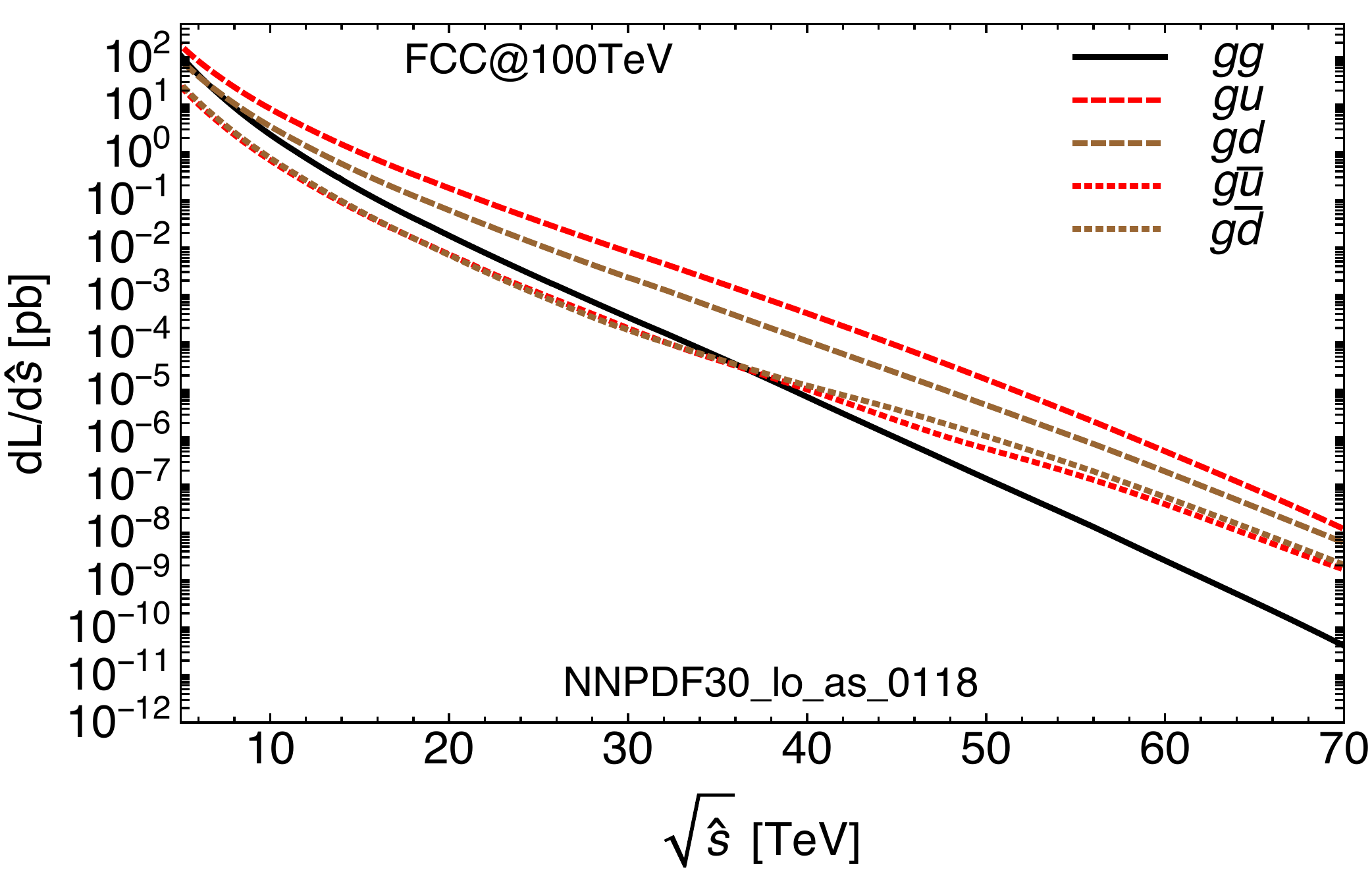}\hspace{0.3cm}
\includegraphics[scale=0.36]{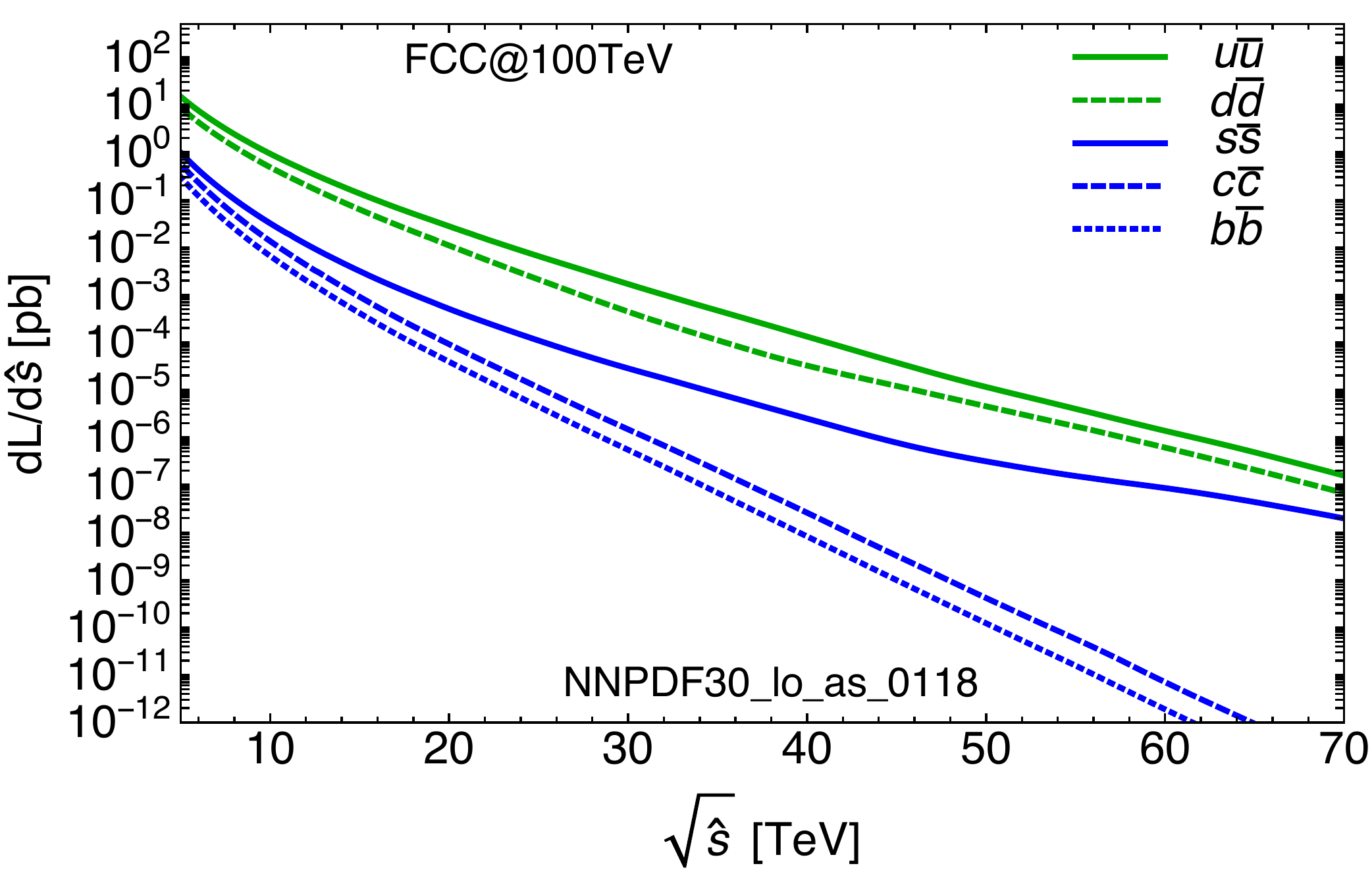}
\includegraphics[scale=0.36]{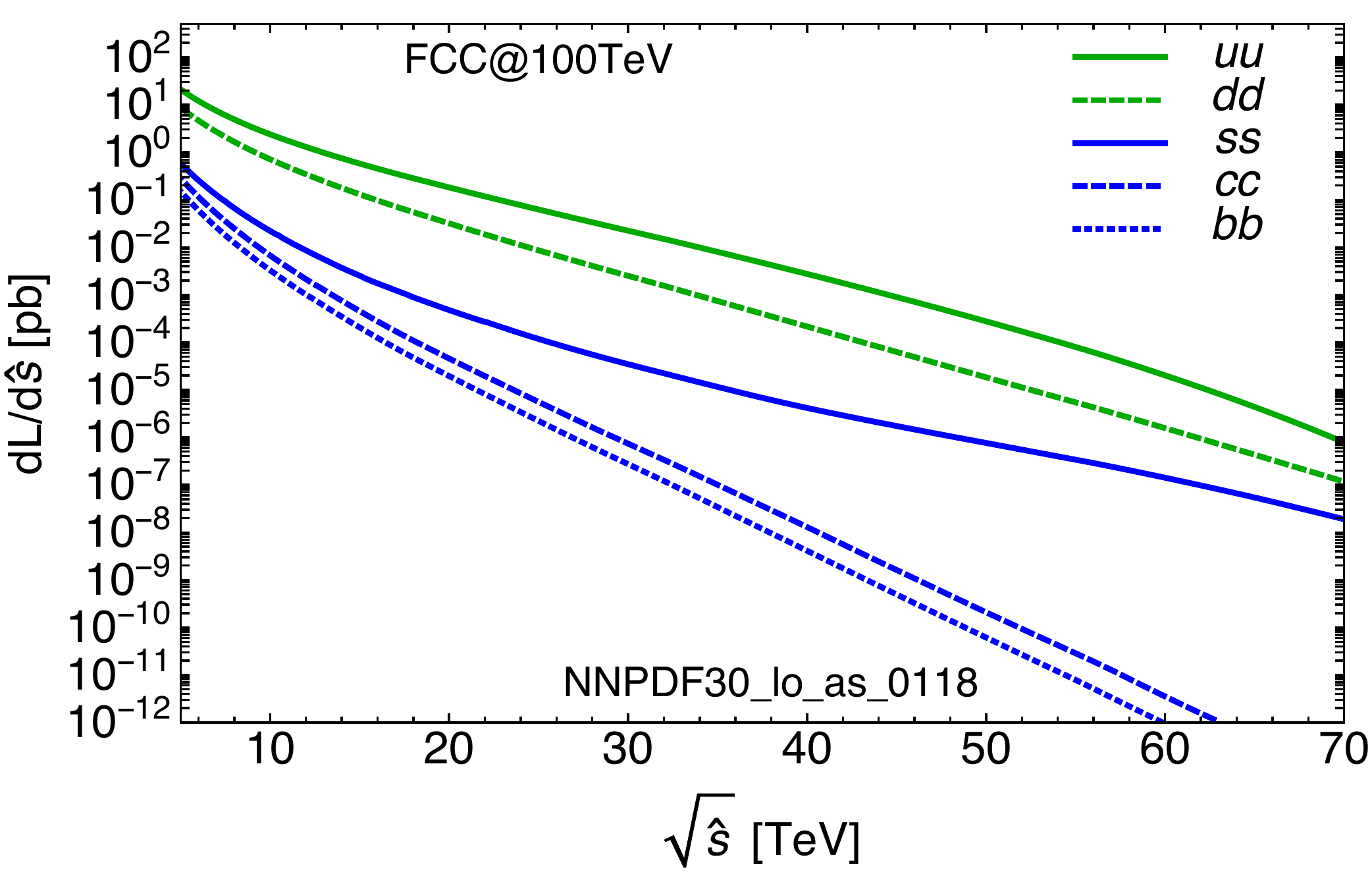}\hspace{0.2cm}
\end{center}
\caption{Parton luminosities $d\mathcal{L}/d\hat{s}$ as functions of the partonic center of mass energy $\sqrt{\hat{s}}$ for a 100 TeV proton-proton collider, computed using the NNPDF30\_LO\_as\_0118 PDF set \cite{Ball:2014uwa}. {\it Upper left:} $gg$, $qg$ and $\bar{q}g$ initial states; {\it Upper right:} $q\bar{q}$ initial states; {\it Lower:} $qq$ initial states.}
\label{partonlumis}
\end{figure}
For the production of a narrow resonance $R$, which can be described as a $2\to1$ processes, the inclusive tree level production cross-section can be written as
\begin{equation}
\sigma(p p \to R+X) = \sum_{i,j \,\in\, p}\frac{\Gamma_{R\,\to \,i j}}{M_{V}} \frac{16\pi^{2}(2J_{R}+1)}{N_{\text{pol}}}\frac{C_{R}}{C_{i}C_{j}}\frac{dL_{i j}}{d\hat{s}}\Bigg|_{\hat{s}=M_{R}^{2}}\,,
\label{BreitWignerCS}
\end{equation}
where $\Gamma_{R\,\to\, i j}$ represent the partial widths of the corresponding decay process $R\to ij$, \mbox{$i,j=\{g,q,\overline{q},W_{L,T},Z_{L,T},\gamma\}$} are the colliding partons in the two protons, and $dL_{i j}/d\hat{s}|_{\hat{s}=M_{R}^{2}}$ is the corresponding parton luminosity evaluated at the resonance mass. The factor $J_{R}$ is the spin of the resonance, $C_{R}$ the dimension of its color representation, $C_{i,j}$ are the dimensions of the color representations of the two partons and $N_{\text{pol}}$ is the number of polarization states of the incoming partons contributing to the production. This last quantity is equal or smaller than the sum over polatization $(2s_{i}+1)(2s_{j}+1)$, where $s_{i},s_{j}$ are the spins of the incoming partons. For instance, in the case of a scalar produced by gluon fusion, like the Higgs boson, $N_{\text{pol}}=2$, since only the $(+-)$ and $(-+)$ polarization configurations of the initial gluons contribute to the production of a $J=0$ state. If needed, the cross-section in eq.~\eqref{BreitWignerCS} can be corrected by a $k$-factor to take into account higher order radiative corrections. In Figure \ref{partonlumis} the parton luminosities $dL_{i j}/d\hat{s}$ as function of $\hat{s}$ are shown for quark and gluon partonic initial states, and in particular for $gg, gq, q\bar{q}$ and $qq$ configurations. There are two kind of corrections to the expression of the production cross section in eq.~\eqref{BreitWignerCS}, which come from width effects, suppressed by $\Gamma/M$ and from the effect of parton luminosities varying too fast, within a region corresponding to the resonance width, to be considered constant. In this latter case, approximating the integral over the parton luminosities with their value at the resonance mass fails, generating a large off-shell tail at low masses. This threshold effect usually corresponds to the region where the parton luminosities start to decrease faster than exponentially, which roughly corresponds, in Figure \ref{partonlumis}, to the point where the curves change their convexity.

%\subsubsection{New colored resonances}

A simple application of the formula in eq.~\eqref{BreitWignerCS} for the production of new resonances is given by the production of exotic colore resonances. In Figure \ref{fig:colored} we show the production cross sections\footnote{The widths relevant to compute these production cross sections using eq.~\eqref{BreitWignerCS} are set to the value corresponding to interactions fixed by a (model-dependent) dimensionless coupling that we have set to one for illustration \cite{Han:2010rf}.} for some different colored resonances corresponding to charged and neutral color-octet vectors, charged color-sextet vectors with fractional charge and excited quarks with spin $3/2$. From the figure it is clear that production cross sections of the order of fb or hundreds of ab are expected for colored states in the mass range 25-50 TeV. Considering the large integrated luminosity planned for a 100 TeV collider, of several inverse ab, some of these states should be accessible up to masses even above $50$ TeV, depending on their production mechanism. As it is clear from Figure \ref{partonlumis}, the color-octet vectors, being produced by $q\bar{q}$ have the lowest production cross section, followed by the excited quark, produced by $qg$ and the diquark sextets, produced by $qq$. In general, all these colored resonances are expected to decay back to di-jets. Assuming an integrated luminosity of several inverse ab a 100 TeV collider should be able to extend the reach on colored resonances of the LHC from a few TeV, to the 30-60 TeV region.

Another interesting possibility is the production of new gauge bosons, such as $Z'$ and $W'$ vectors. These are typically produced by Drell-Yan $q\bar{q}$ annihilation and decay to two leptons or lepton-neutrino depending on the charge, leading to final states that are effectively zero background in the multi-TeV region. The typical cross sections of different $Z'$ and $W'$ models \cite{Langacker:2008yv} are shown in Figure \ref{fig:WpZp}. Again the reach of the LHC, which is around 5 TeV for such states, can be significantly extended at a 100 TeV collider, up to 25-35 TeV. The case of new vector bosons arising as composite resonances is also very interesting. In Figure \ref{fig:vector_resonance_reach} we show the cross section of a $\rho$-like state arising in minimal composite Higgs models \cite{Agashe:2014kda} for typical values of the parameters. In this case decays to two electroweak gauge bosons are typically enhanced compared to di-lepton and lepton-neutrino final states, corresponding to a reach that extends up to about 20 TeV. Scenarios presenting new $Z'$ and $W'$ bosons are studied in more detail in Sections \ref{Rizzo}, \ref{Doglioni}, \ref{Auerbach}, respectively in the di-lepton and lepton neutrino, di-jet and $t\bar{t}$ final states. Resonances arising from a strong sector responsible for electroweak symmetry breaking (EWSB) are discussed in section \ref{Thamm}, where the reach from direct searches in DY production is compared with the indirect reach of leptonic colliders and in section \ref{Vignaroli}, where the VBF production is studied in the $tb$ final state.

%A high energy hadron collider is a QCD machine. Any new states with
%QCD interactions would be copiously produced via quark and gluon
%partons. Some such exotic states have been systematically classified in
%ref.~\cite{Han:2010rf}, and the LHC experiments have been actively
%searching for them \cite{Aad:2014aqa,Chatrchyan:2013qha}. The
%non-observation at the LHC sets  bounds on their mass, bounds that
%will extend well beyond a few TeV after the LHC energy increase to
%$13-14$ TeV. 
%This mass reach would be substantially extended by the 100 TeV collider. 

\begin{figure}[t!]
\centering
\includegraphics[scale=0.70]{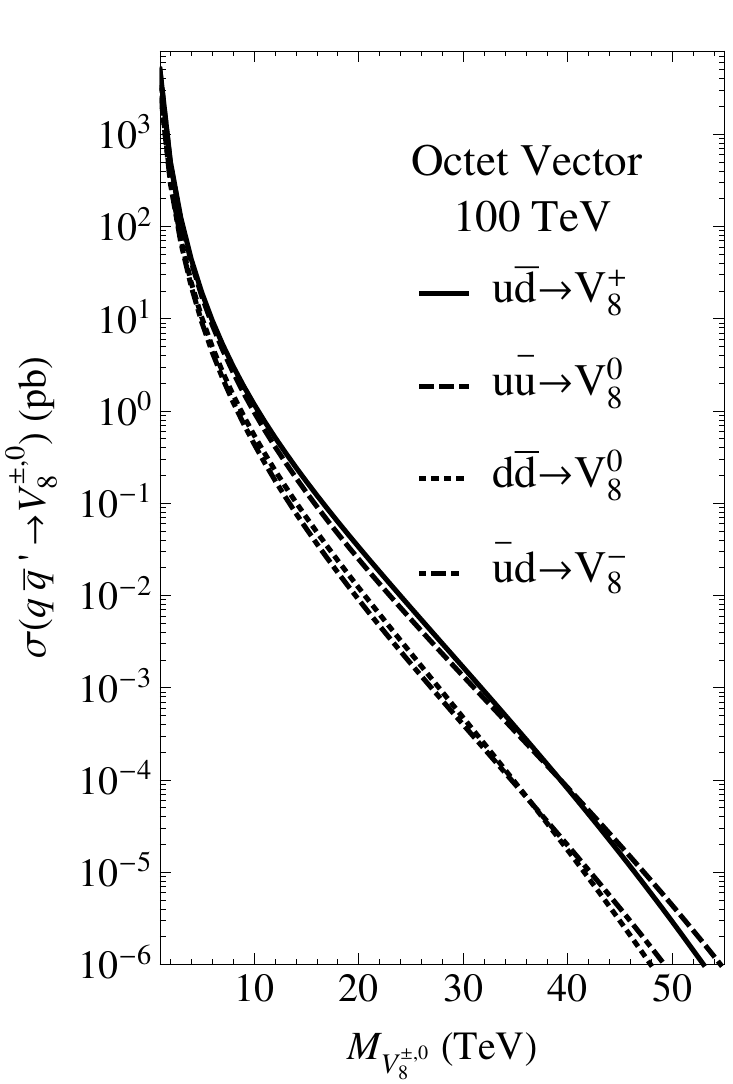}
\includegraphics[scale=0.70]{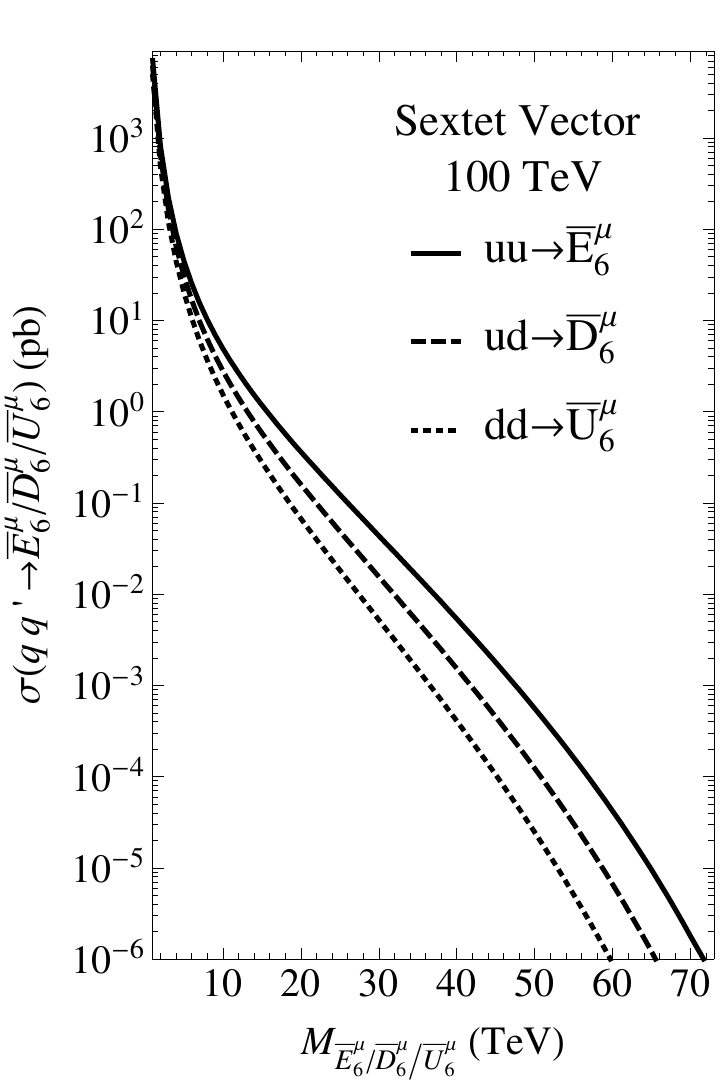}
\includegraphics[scale=0.70]{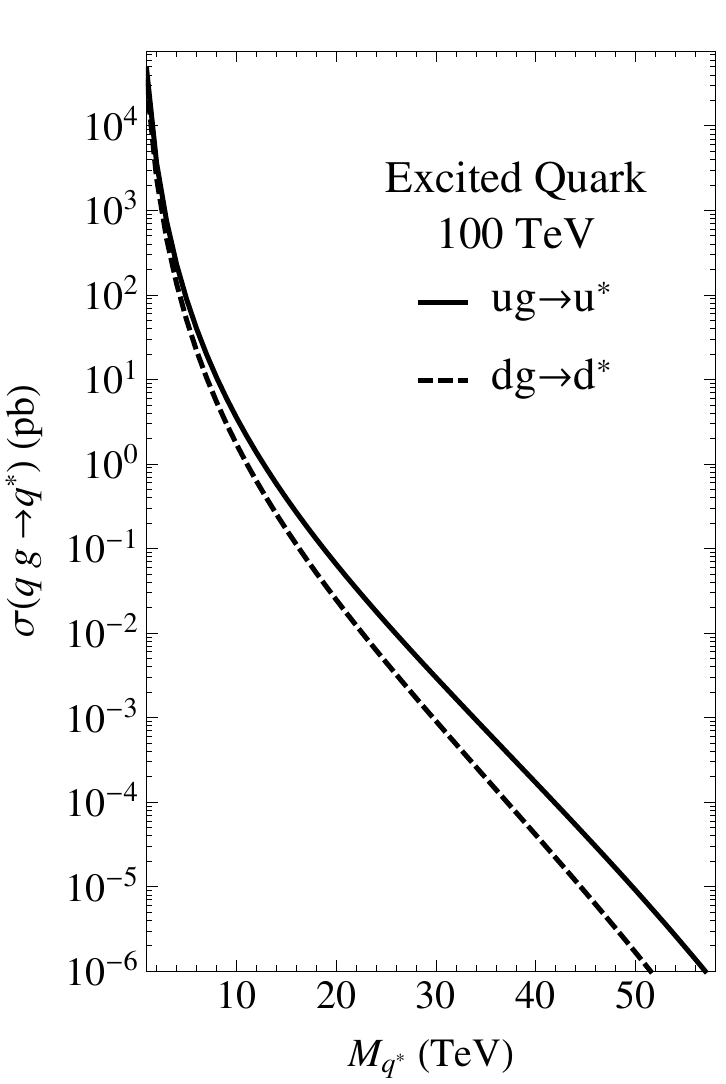}
\caption{Production cross sections for exotic colored resonances at 14 TeV and 
  100~TeV. {\it Left:} charged and neutral color-octet vector states; {\it Center:} fractionally charged color-sextet vector states (di-quark-like); {\it Right:} spin-3/2 excited quark states.} 
\label{fig:colored}
\end{figure}

\begin{figure}[h!]
\begin{center}
  \includegraphics[scale=0.4]{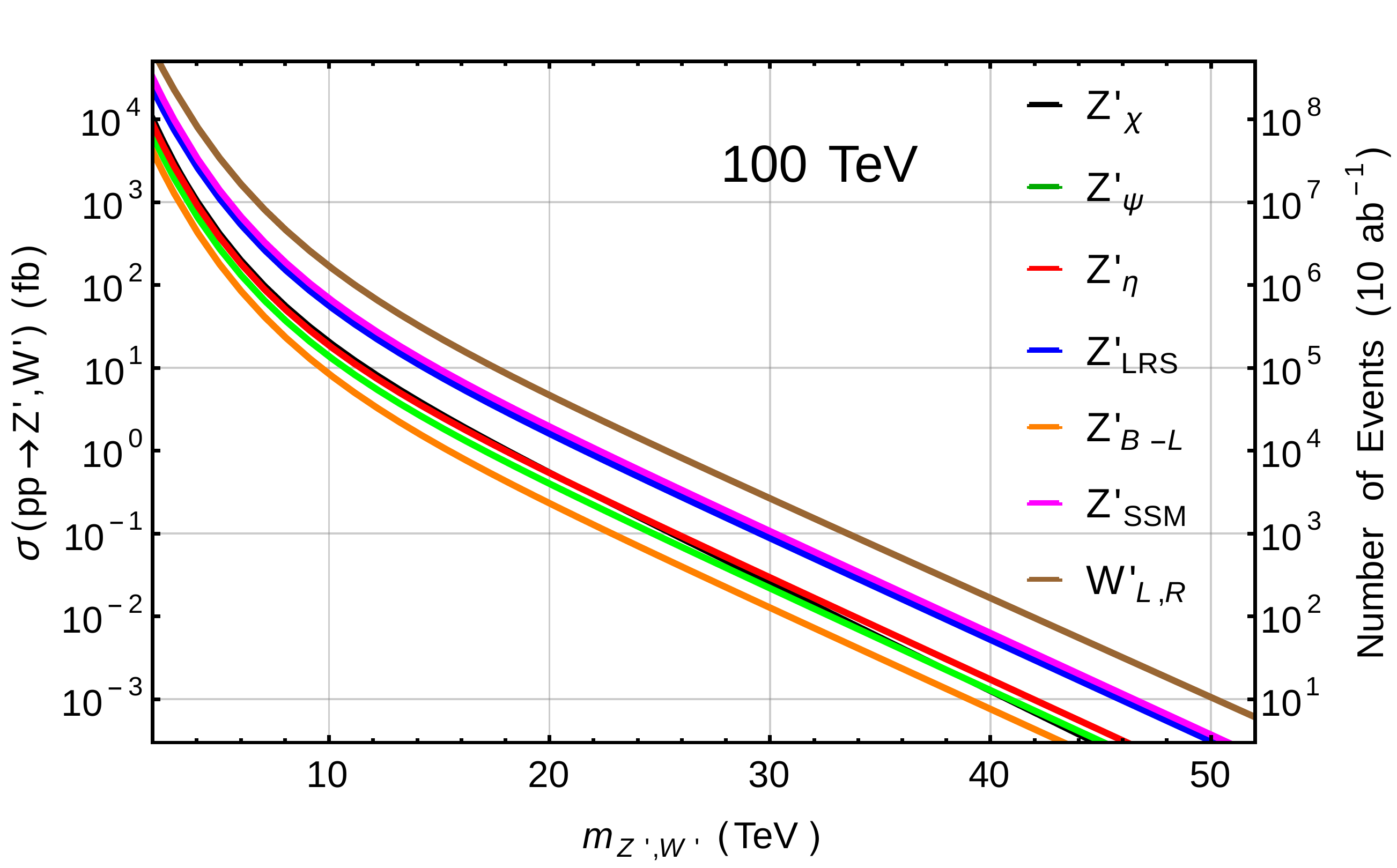}
\end{center}\caption{Production cross section of
$W'$ and $Z'$ gauge bosons in various models \cite{Langacker:2008yv}
  at 100~TeV.} 
\label{fig:WpZp}
\end{figure}

\begin{figure}[h!]
\centering
\includegraphics[scale=1.25]{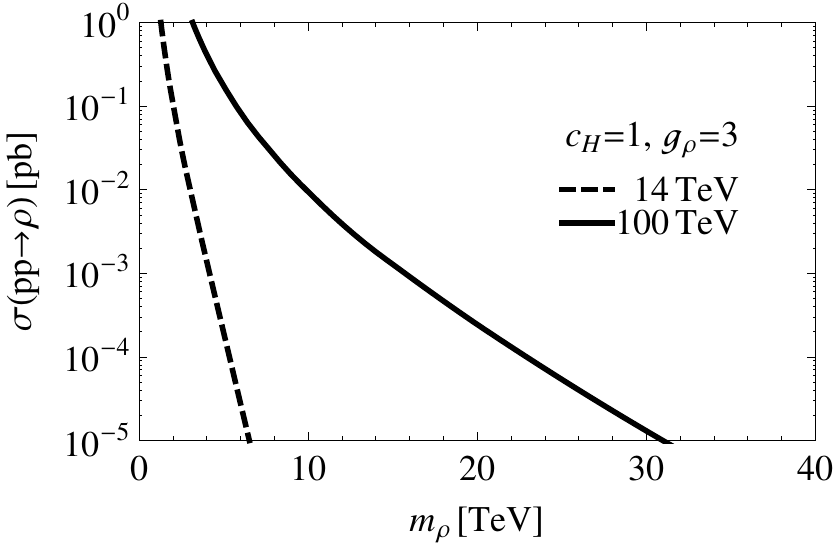}
\caption{Production cross section of a $\rho$-like vector resonance arising in minimal composite Higgs models
   \cite{Agashe:2014kda} at 100~TeV. Typical values of the relevant parameters have been chosen.}
\label{fig:vector_resonance_reach}
\end{figure}

\begin{figure}[h!]
\begin{center}
  \includegraphics[scale=0.32]{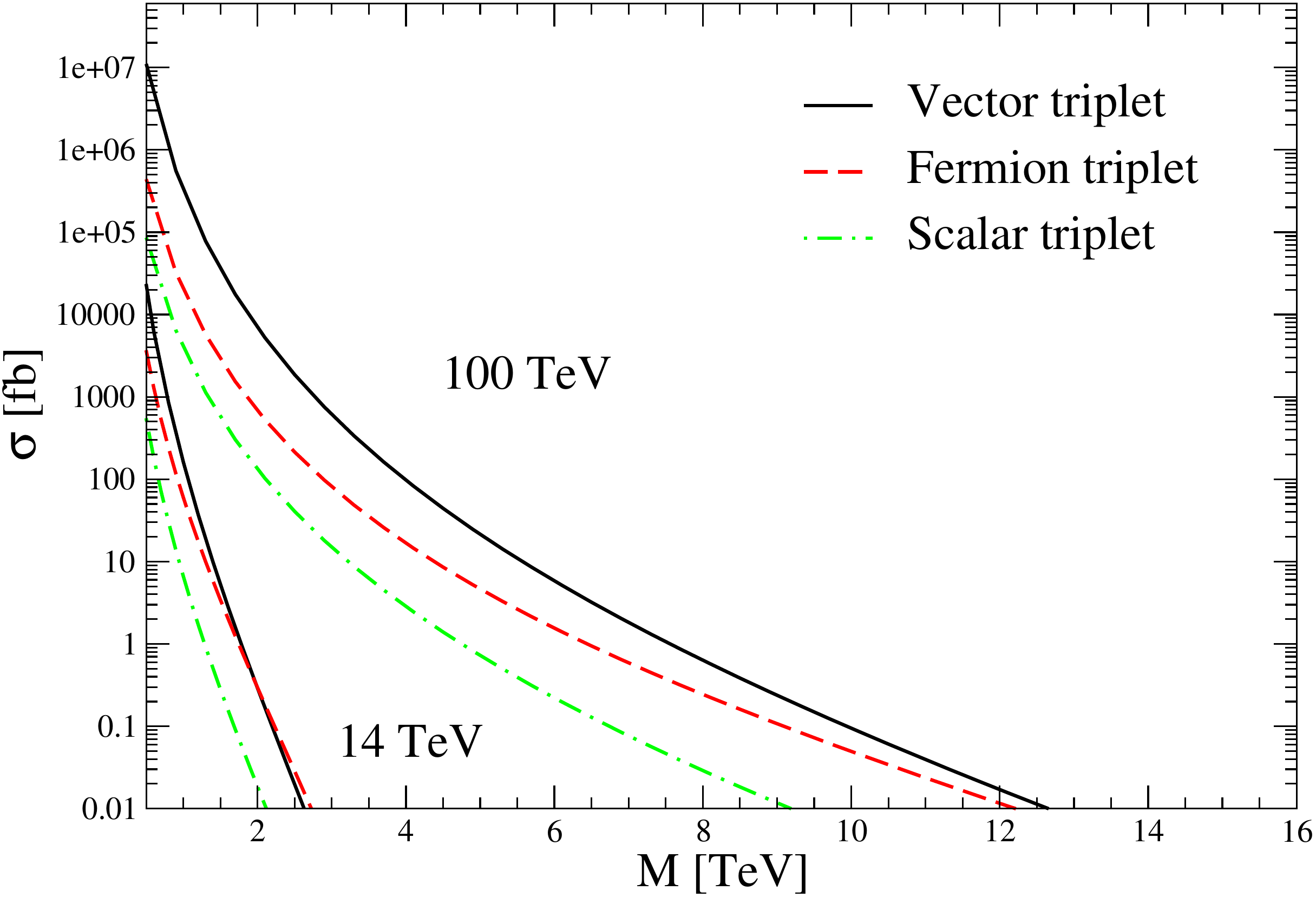}
    \includegraphics[scale=0.32]{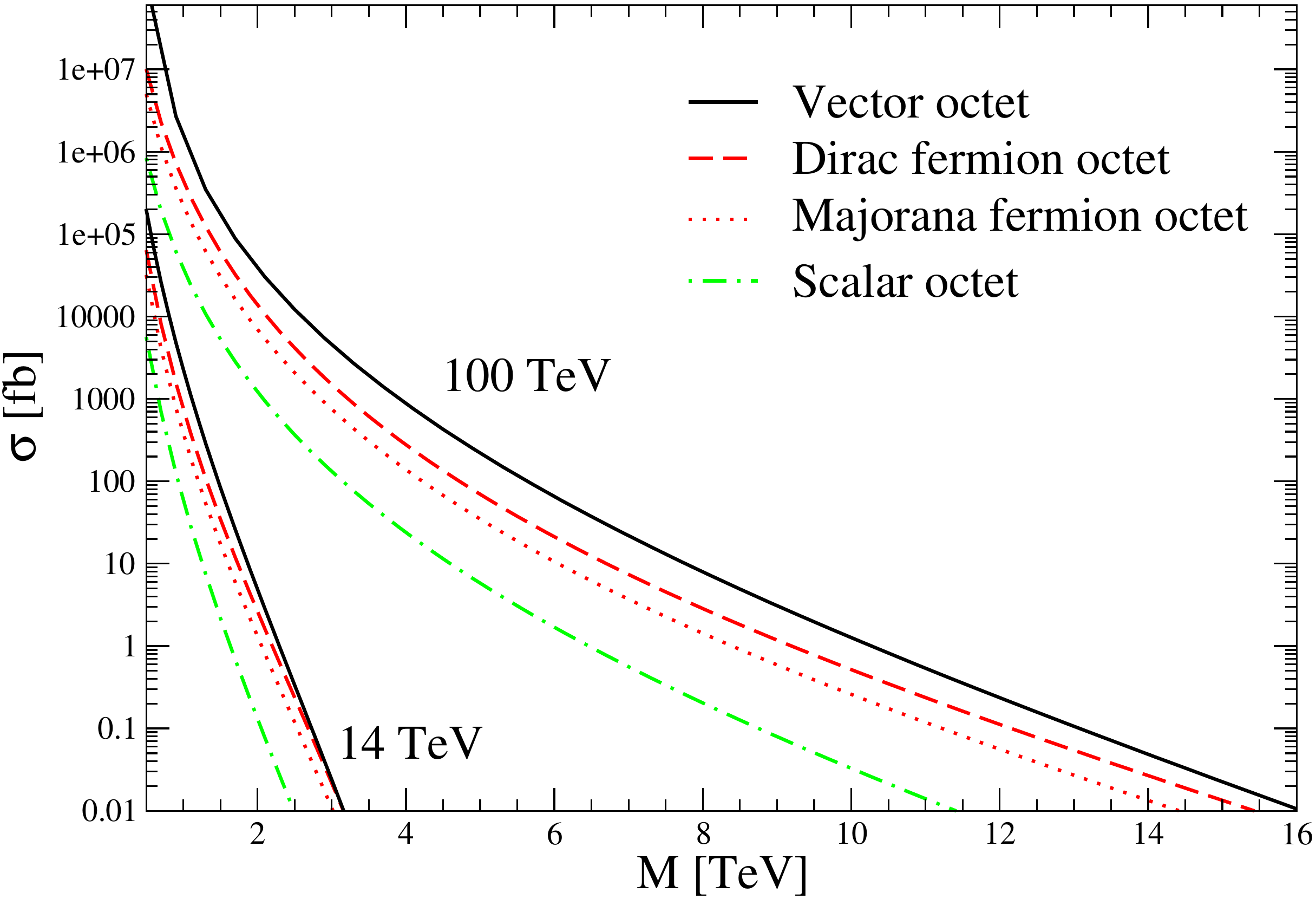}
\end{center}
\caption{Pair production cross sections of new colored states. {\it Left:} color-triplets; {\it Right:} color-octets. The different dashing represent spin-1 (solid), $1/2$ (dashed), and 0 (dot-dashed).}
\label{fig:Tp}
\end{figure}

%\subsubsection{Pair Production of Exotic Color States}
Up to now we have given some examples of the reach of a 100 TeV collider on single resonance production. Of course also pair productions are extremely favored by the large available center of mass energy. Some of the very well know candidates for pair production searches are the so-called top partners, which can be either scalar, fermions or even vectors \cite{Cai:2008ss}, depending on the model, whose role is compensating the large sensitivity of the Higgs mass parameter to the top loop contribution responsible for the hierarchy problem of the electroweak scale. On top of these particles, other states with different quantum numbers can be related to naturalness, like for instance color octet fermions, as the gluino in SUSY. All these states related to naturalness will be discussed extensively in this BSM part of the report, both in the SUSY section, focusing, for what concerns colored particles, on stops, sbottoms and gluinos, and in this section, focusing on fermionic partners of the top, usually referred to as just top partners. In order to give a preliminary idea of the reach of a 100 TeV collider on these particles, we show in Figure \ref{fig:Tp} the typical production cross sections for both color-triplets \cite{Chen:2012uw} and color-octets \cite{Chen:2014haa} scalars, fermions and vectors. In the case of color-triplet particles, the spin-0, $1/2$, and 1, refer to stop-like, $T'$-like and color-triplet vector top partners respectively. In the case of color-octet they correspond to states that are techni-meson-like, gluino-like (both Majorana or Dirac) and KK-gluon-like. Differently to what happens for the single production of resonances, which essentially depend on a free coupling of the new theory, the pair productions are completely fixed by QCD interactions and, once the quantum numbers under the color gauge group are fixed, their production rate can be determined model independently. Of course, the bigger the color charge and the spin, the bigger the pair production cross section. Provided that the decay channels can be efficiently discriminated from background, the large rates are expected to lead to mass reach that extend from about $5$ TeV for color-triplet scalars, to about 15 TeV for color-octet vectors. Concerning colored resonances, fermionic top-partners are discussed in more details in Section \ref{Wulzer}, where both single and pair production are considered and in Section \ref{Salvioni}, where the top-partners arising in Twin Higgs models are studied in signatures involving displaced decays with a prompt $t\bar{t}$ pair.

In order to compare the reach on colored resonances produced in pairs, with un-colored ones, we consider the pair production of new heavy leptons. These particles may be motivated in BSM scenarios related to the mechanism of neutrino mass generation. In Figure \ref{fig:Seesaw} we show the heavy leptons pair production cross section. We consider both a triplet \cite{Foot:1988aq} and a singlet of $SU(2)_{L}$, denoted by $T$ and $N$. In the case of the triplet, analogously to what happens for the colored particles, the pair production cross section is entirely fixed by electroweak interactions \cite{Arhrib:2009mz,Li:2009mw}. The states in the triplet are expected to be almost degenerate in mass and typically decay into the $T^\pm \to W^\pm\nu, Z\ell^\pm,
h\ell^\pm$ and $T^0 \to W^\pm\ell^\mp, Z\nu, h\nu$ final states, as discussed in Section~\ref{Ruiz}. From the Figure, we see that depending on the decay channel the reach on these particles, which at the LHC is limited to around the TeV, can be extended at a 100 TeV collider in the range of $5-8$ TeV. Concerning the SM singlet $N$, we consider its production in association with a SM lepton, which depends on the details of the mixing matrix between $N$ and the SM neutrinos. In the Figure we show the $N\ell^\pm$ production cross section normalized to a mixing matrix equal to the identity, which would correspond to the production cross section of a doublet of $SU(2)$. Some of these signatures involving heavy leptons are discussed in detail in Section \ref{Ruiz}.

\begin{figure}[h!]
\begin{center}
  \includegraphics[scale=0.4]{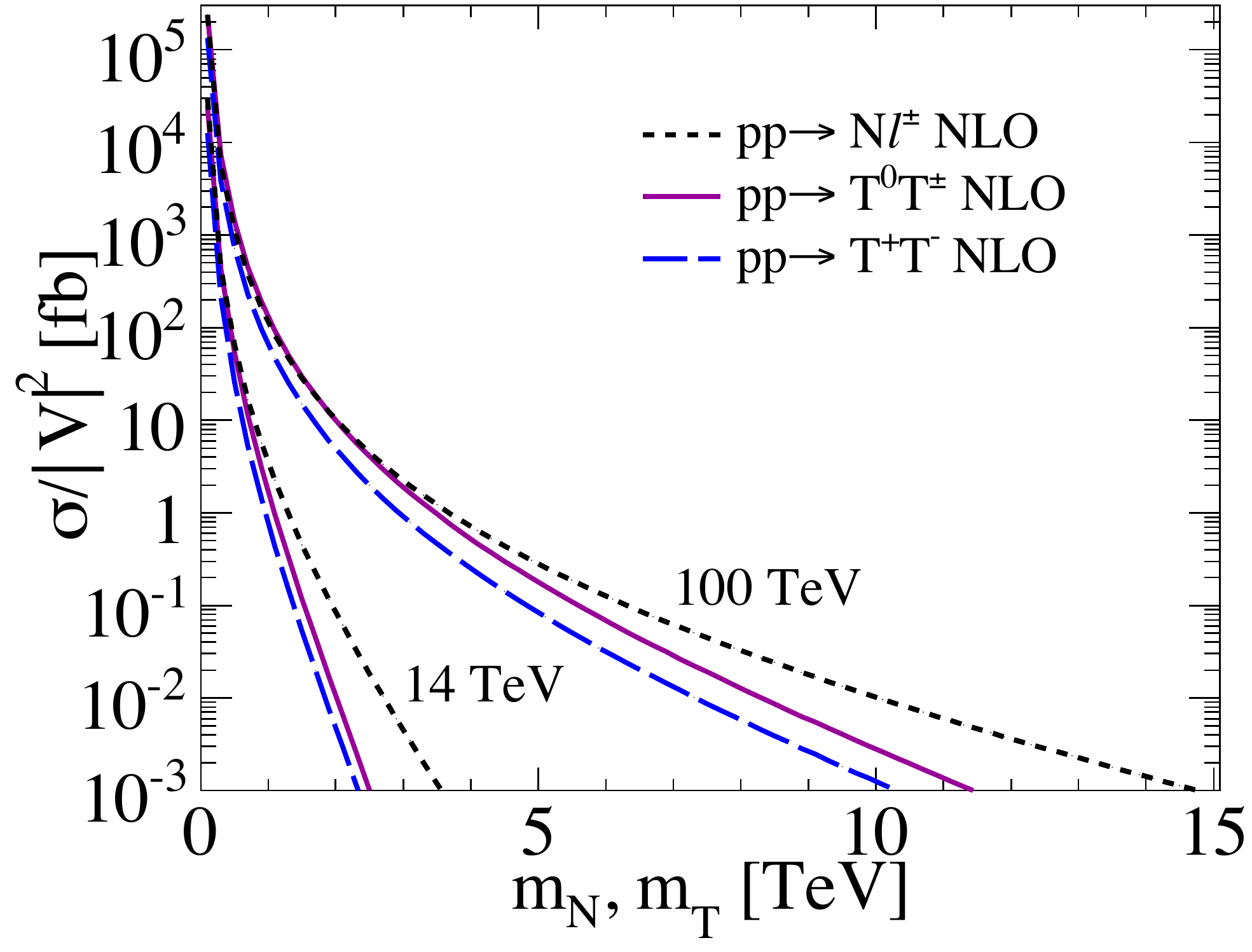}
\end{center}
\caption{Pair production of new heavy leptons at 14 and 100~TeV, for an SU(2) triplet ($T^{\pm,0}$)  and for a singlet state $N \ell^\pm$ via mixing. See section \ref{Ruiz} for details.}
\label{fig:Seesaw}
\end{figure}

\subsection{New Bosonic Resonances}

\subsubsection{New Gauge Bosons in Dilepton Final States}\label{Rizzo}
%\begin{itemize}
%\item T.~Rizzo, {\it Identifying New Gauge Bosons at a 100 TeV Collider} {\bf contribution received}
%\end{itemize}

If a new gauge boson is discovered the next step will be to identify its origins within some underlying UV theory. 
A necessary step along this road is the determination of the new gauge boson couplings to the various fields of the Standard Model.  
Here we discuss a set of measurements that have been proposed to access this information and their potential use 
at a 100 TeV hadron collider.

``Identification'' is the first step after a new discovery, \ie the determination of what it is that has been found. We have 
been experiencing an example of this procedure in action in the ongoing program to probe the detailed nature of the 
$125$ GeV Higgs boson. In going from the LHC to a 100 TeV collider, the window for the discovery of new gauge bosons is 
enormously increased as is summarized for the usual canonical scenarios in both Table~\ref{rates} and Fig.~\ref{fig1rizzo}.  
While discovery in the clean Drell-Yan leptonic channels will only require order 10's of events, the determination of the 
various couplings necessitates event statistics which are order $10-100$ times larger. This implies that only new gauge bosons 
with masses about $10$ TeV or more below their corresponding discovery reach will be amenable to coupling extractions for a 
given value of the integrated luminosity.  

\begin{table}
\centering
\begin{tabular}{|c|c|c|c|} \hline\hline
Model    &  1 ab$^{-1}$  & 10 ab$^{-1}$  & 100 ab$^{-1}$  \\
\hline
SSM      &  23.8 & 33.3  & 41.3  \\
LRM     &   22.6 & 31.5  & 39.5  \\
$\psi$   &  20.1 & 29.1  & 37.2  \\
$\chi$    & 22.7 & 30.6  & 38.2  \\
$\eta$   &  20.3 & 29.8  & 38.0  \\
I          & 22.4 & 29.2  & 36.2  \\
\hline\hline
\end{tabular}
\caption{Discovery reach in the leptonic decay mode for various $Z'$ models \cite{Langacker:2008yv} in TeV at $\sqrt s=$100 TeV for 
different integrated luminosities. Exclusion reach are roughly $\simeq 3.5$ TeV larger in all cases.}
\label{rates}
\end{table}

Frequently, to simplify the situation as much as possible in the case of a new $Z'$, it is assumed that the couplings 
are generation independent, that $Z-Z'$ mixing can be neglected (which is a very good assumption for large masses) and that the gauge charge to which the $Z'$ couples commutes with the corresponding SM generators. Under these assumptions the couplings of the $Z'$ to the SM fermions are given in terms of only 5 independent parameters corresponding to the SM fermion representations: $Q_L$, $L_L$, $u_R$, $d_R$ and $e_R$. Surrendering any of these simplifying assumptions enlarges the set of independent parameters that need to be determined. On the other hand, in the case of a $W'$, the most important quantity to determine is the helicity of its couplings to the SM fermions, which separates potential models into two broad categories.

\begin{figure}[htbp]
\begin{center}
\includegraphics[scale=0.36]{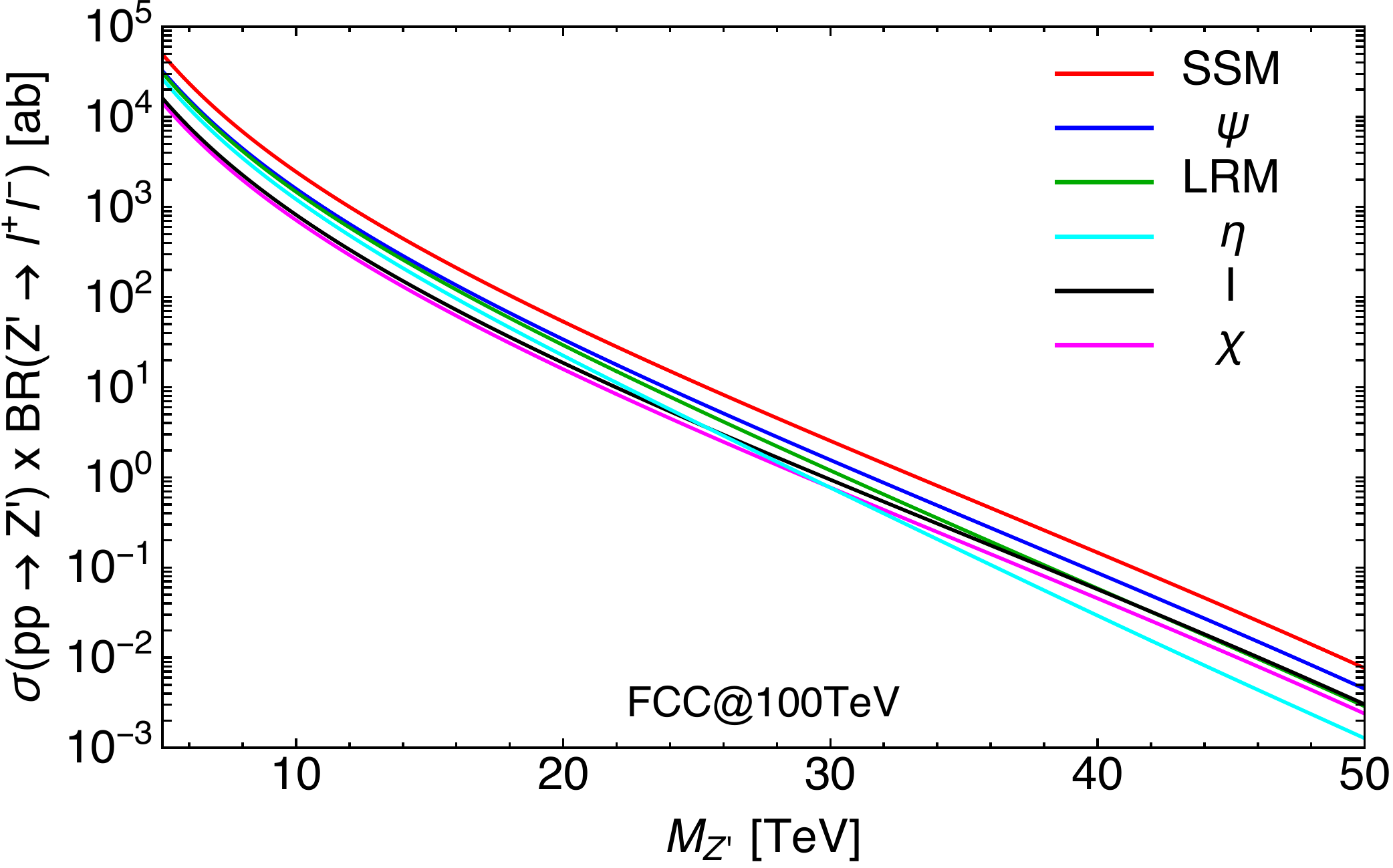}\hspace{0.3cm}
\includegraphics[scale=0.354]{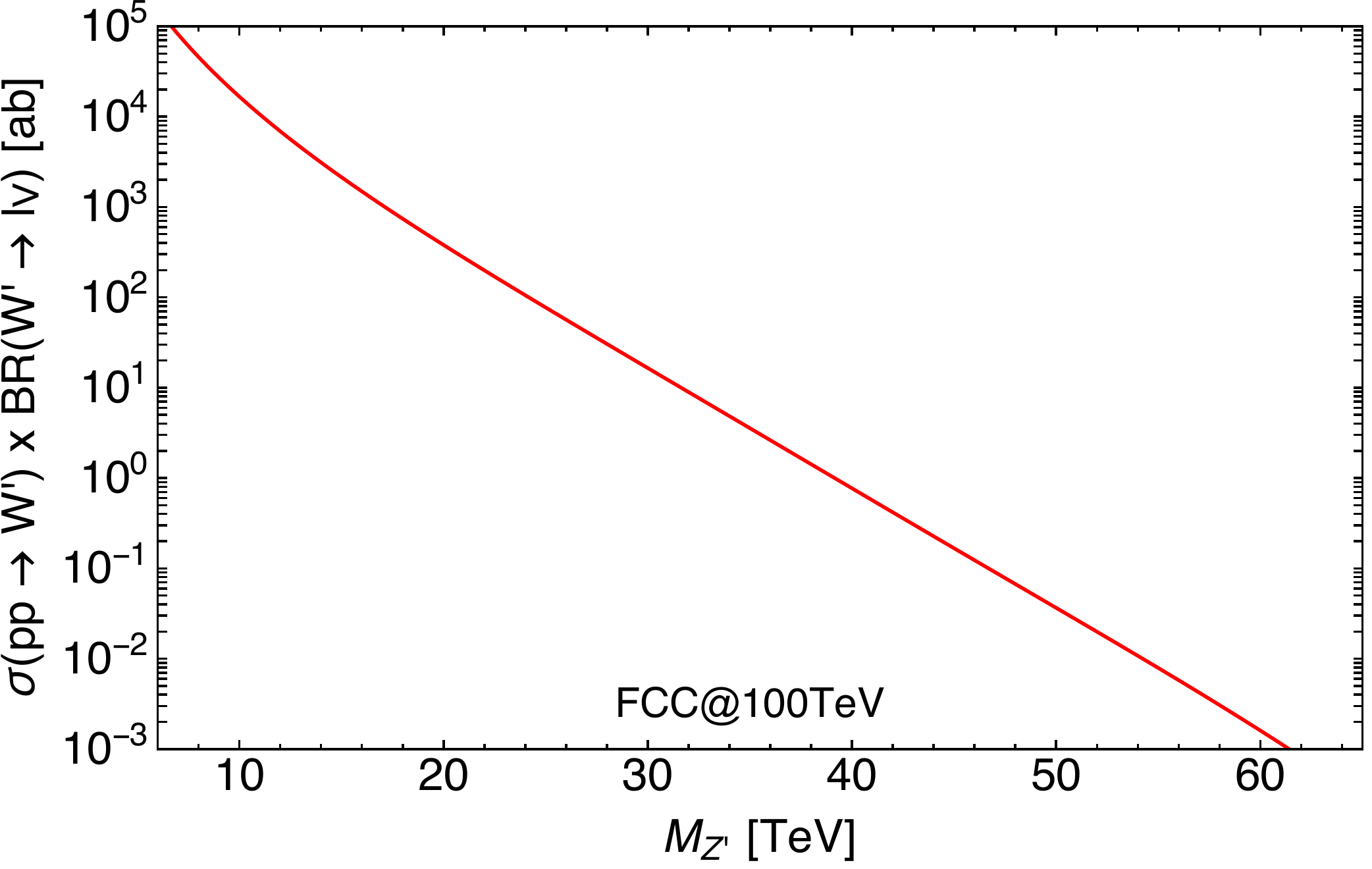}
\end{center}
\caption{{\it Left:} Cross section times leptonic branching fraction, $\sigma(pp\to Z')\times \text{BR}(Z'\to l^{+}l^{-})$, as functions of the mass for the $Z'$ models indicated in the legend at $\sqrt s=$100 TeV. The legend is ordered following the order of the curves at $M_{Z'}=20$ TeV; {\it Right:} same quantity for the SSM/LRM $W'$ (with $g_R=g_L$ and Dirac neutrinos).}
\label{fig1rizzo}
\end{figure}

In order to extract the values of the $Z'$ couplings in as model-independent of a way as possible we cannot assume that the new gauge bosons will only decay into the known SM particles. This implies that measurements which depend on the new gauge bosons width, such as the production cross section times leptonic branching fraction, $\sigma B_\ell$, cannot be used for this purpose. 
Several of these decay-independent observables (which mostly employ high $p_T$ lepton triggers) have been proposed and were discussed in detail in ref.~\cite{Rizzo:2014xma}, which we will summarize below. First we consider those observables that employ the dilepton discovery mode to extract coupling information.

\begin{itemize}

\item The most obvious way to by-pass the shortfall of $\sigma B_\ell$ as a useful observable is to rescale it 
by the $Z'$ width, \eg, $\sigma B_\ell \Gamma_{Z'}$ so that it only depends on the couplings above in the narrow width approximation (NWA). The typical values of $\Gamma_{Z'}/M_{Z'}$ in the models in Table~\ref{rates} are of the order of $1\%$ or so. One finds that this observable, $\sigma B_\ell \Gamma_{Z'}$, varies by a factor of about $20$ for just these six familiar sample models thus showing its strong coupling sensitivity. Since this quantity makes use of the discovery channel, only a few hundred events are necessary to obtain a reasonably reliable determination with small (statistical) errors.{\footnote {30 times the events needed for discovery corresponds to roughly a reduction of at least $10$ TeV in mass from the discovery reach for the same integrated luminosity.}} The main difficulty with employing this observable occurs when the $Z'$ width  is substantially smaller than the dilepton mass resolution so that the value of $\Gamma_{Z'}$ cannot be trivially determined.    

\item A familiar observable is the leptonic forward-backward asymmetry, $A_{FB}$, which can be obtained from the 
lepton's angular distribution. In the $Z'$ rest frame, with $z=\cos \theta^*$ being defined between the initial 
quark, $q$, and outgoing $l^-$ direction, this distribution is given by $\sim 1+z^2 +8 A_{FB} z/3$. Note that a non-zero 
value of $A_{FB}$ requires vector and axial-vector couplings of the $Z'$ to both the quarks and leptons. A problem arises 
in that the $q$ direction cannot always be identified with the boost direction of the $Z'$, which itself is not always 
cleanly defined (without applying a suitable cut). Similarly, the rapidity coverage for the  
leptons can be critical especially if it more restricted than the typical ATLAS/CMS value of $|\eta_\ell|<2.5$ as larger 
scattering angles, which show the greatest sensitivity, will have a reduced contribution. All this results in a 
dilution of $A_{FB}$ which can be partially compensated for using Monte Carlo. Again, only a few hundred events are needed 
to obtain a reasonable estimate of $A_{FB}$ but this increase in required statistics for a fixed luminosity implies 
a significant reduction in the reach for coupling information extraction. Since both $u\bar u$ and $d\bar d$ 
initial states will, in general, contribute to $Z'$ production the knowledge of the PDFs (and their evolution) will be  
important in the use of $A_{FB}$ to extract $Z'$ coupling information. Going beyond NWA it is possible that information on $A_{FB}$ 
can be obtained in the interference region below the $Z'$ peak but this will also require a significant increase in integrated 
luminosity.

\item The last observable that makes direct use of the $\ell=e,\mu$ discovery channel is the central rapidity ratio ($r_y$), 
\ie the fraction of events with lepton rapidities below some cut value compared to those above that same value. $r_y$ is 
sensitive to the $u\bar u$/$d\bar d$ admixture in the $Z'$ couplings and so is also quite sensitive to the $Z'$ mass due to 
the running of the PDFs. However, in comparison to the previously discussed observables, $r_y$ provides somewhat weaker 
information on the $Z'$ couplings. 

\item The $Z'\to \tau^+\tau^-$ mode is also potentially powerful, particularly in the case of generation-independent couplings,  
as the $\tau$ polarization, $P_\tau$, (as can be determined in single-prong decays) can be used to extract the coupling ratio 
$v_\ell/a_\ell$. Even in the 1-parameter $E_6$ models, $P_\tau$ can take on values over its entire allowed range 
$-1\leq P_\tau \leq 1$. However, this mode suffers from the obvious $\tau$ identification issues in this highly boosted regime and 
its possible effectiveness for coupling extraction at 100 TeV will require further study.   

\end{itemize}

To get a rough idea of the separation of the various models en route to coupling extraction provided by these observables, 
consider the top two panels in Fig.~\ref{fig2} which assumes $L=5$ ab$^{-1}$ and $M_{Z'}=15$ TeV. We see that these six 
models are all distinguishable from each other except for the two $E_6$ models, $\psi$ and $\eta$, using just these 
variables alone.

\begin{figure}[htbp]
\begin{center}
\includegraphics[scale=0.36]{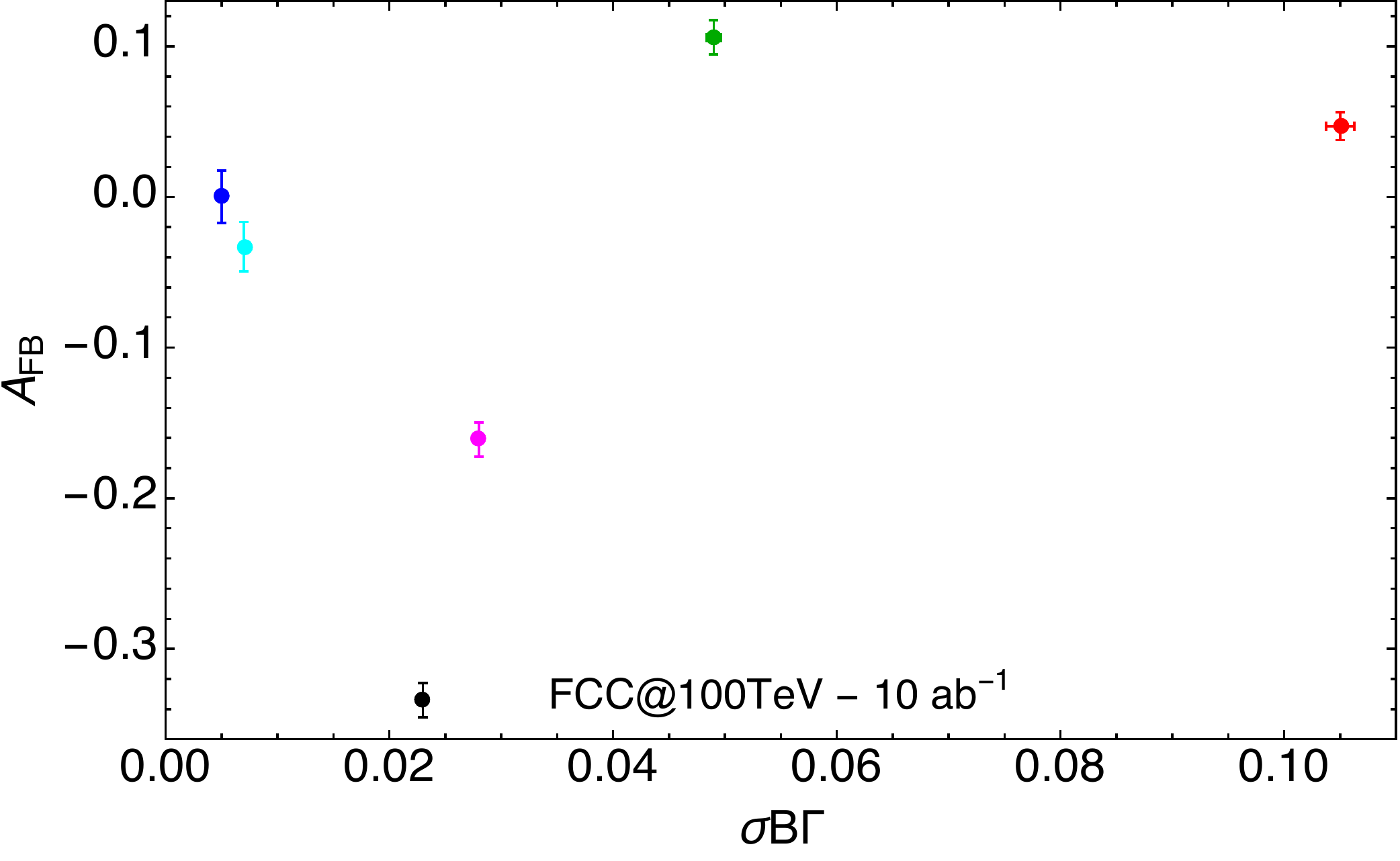}\hspace{0.3cm}
\includegraphics[scale=0.36]{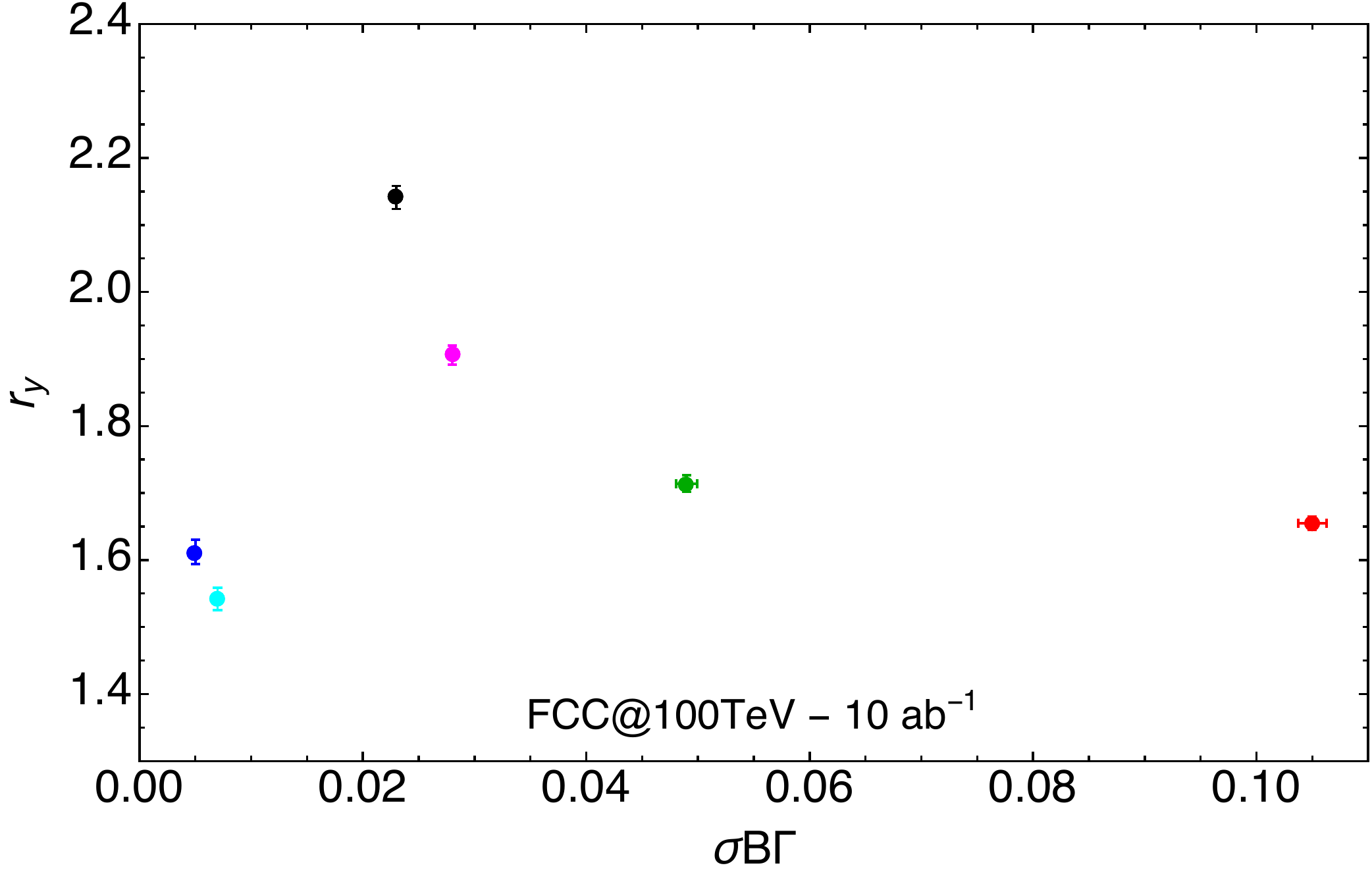}\\
\vspace{0.3cm}
\includegraphics[scale=0.36]{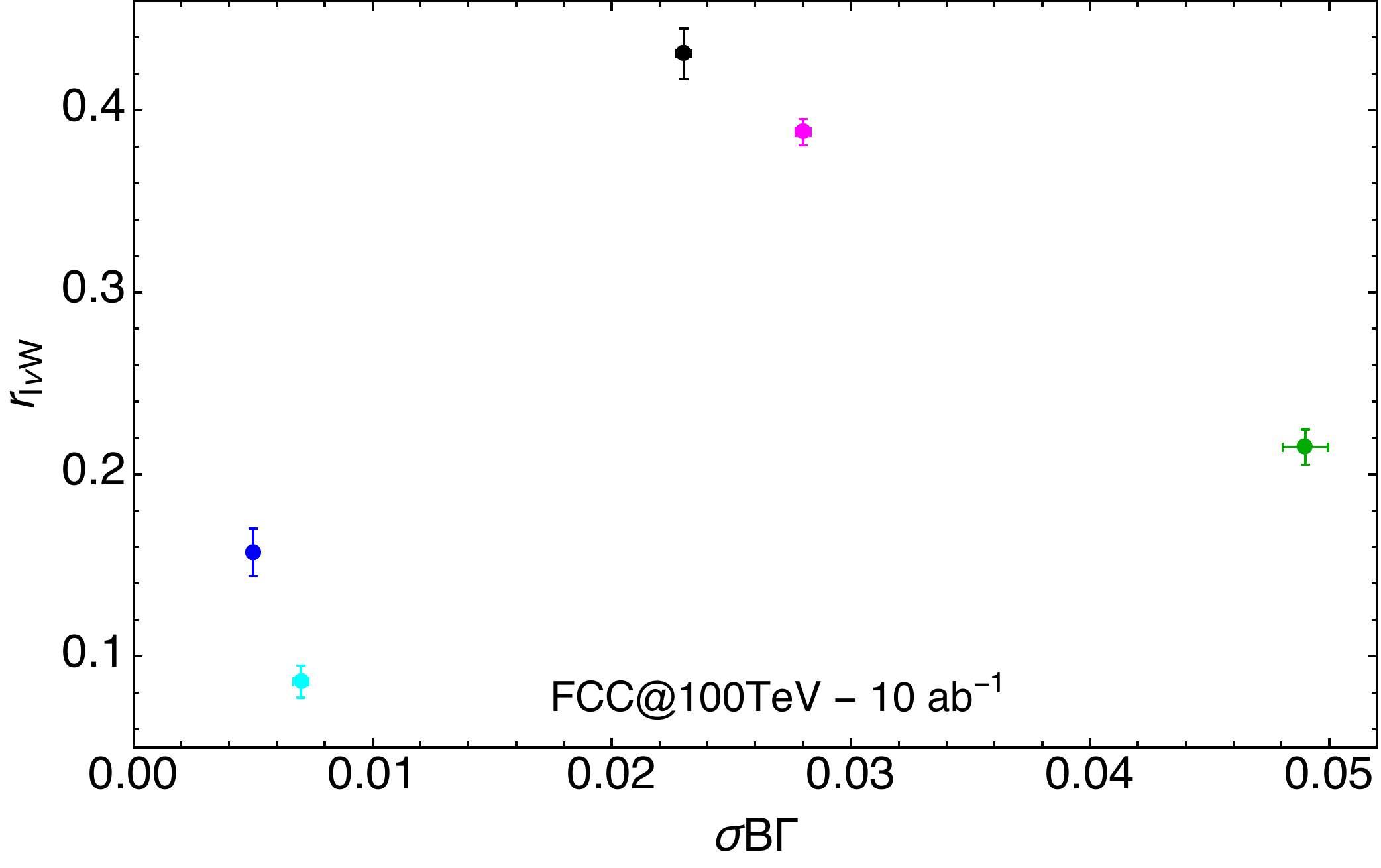}
\end{center}
\caption{Comparisons of the values of the observables $A_{FB}$, $r_{y}$ and $r_{l\nu W}$ for $L=10$ ab$^{-1}$ and $M_{Z'}=15$ TeV for the six models shown in Table~\ref{rates} and Fig.~\ref{fig1rizzo} above with the same color coding (from left to right: blue, cyan, black, magenta, green and red, see Fig.~\ref{fig1rizzo} for the corresponding models).}
\label{fig2}
\end{figure}

Once we go beyond the dilepton channel, numerous possibilities to obtain coupling information are available each with their 
own strengths and weaknesses.  

\begin{itemize}

\item Still restricting ourselves to 2-body decays to SM fermions, $Z'\to t\bar t$ (using boosted top techniques) may be 
useful if the top polarization can be measured as it is sensitive to the ratio of the LH- and RH-couplings at the 
$Z't\bar t$ vertex. The use of this variable at $100$ TeV requires further study. See section \ref{Auerbach} for a study of resonances decaying into the $t\bar{t}$ final state.

\item 3-body decays of the $Z'$ can be useful, \eg in the absence of $Z-Z'$ mixing the decay $Z'\to \ell \nu W$ occurs by 
$W$ emission off of a lepton leg. The $W$, being coupled to the leptons in a LH manner, projects out the LH 
$Z'$ coupling to leptons as well. Although this decay rate suffers from both 3-body phase space and coupling factors in 
comparison to $Z'\to \ell^+ \ell^-$, it is also $~\log ^2(M_{Z'}/M_W)$ enhanced due to the infrared and collinear 
singularities in the relevant diagrams. This enhancement can be quite important for mass ratios of order $200$ that we 
are considering here. For example, for a $Z'$ mass of 15 TeV in the LRM the cross section for the $l^\pm \nu W^\mp$ final 
state is about $50$ ab. One difficulty at these energies is the rather large boost of the final state $W,Z$ and its small 
opening angle with respect to the lepton from which it was emitted, \ie isolation issues. Further, if the $W,Z$ are found 
through their dijet decay modes (for statistical reasons) this will not easily allow for $W,Z$ separation and will   
likely appear as a fat single jet. However, this final state deserves further study since the ratio $r_{\ell \nu W}=
\Gamma(Z'\to \ell \nu  W)/\Gamma(Z'\to \ell^+ \ell^-)$ can provide reasonable coupling information as shown in Fig.~\ref{fig2}. 

\item The $Z'$ can be produced in association with another SM state, \eg $Z'W^\pm$, by ``initial state radiation''. This 
additional gauge boson can be used as a probe to again ``project out'' certain combinations of the leptonic $Z'$ couplings. 
These channels suffer from some of the same issues as in the case of $r_{\ell \nu W}$ as $W^\pm$ and $Z$ final 
states will be essentially impossible to distinguish in their dijet decay modes while employing their leptonic decays will 
require increased luminosity to compensate for the smaller branching fractions. More study is needed. 

\end{itemize}

\begin{figure}[t!]
\begin{center}
\includegraphics[scale=0.45]{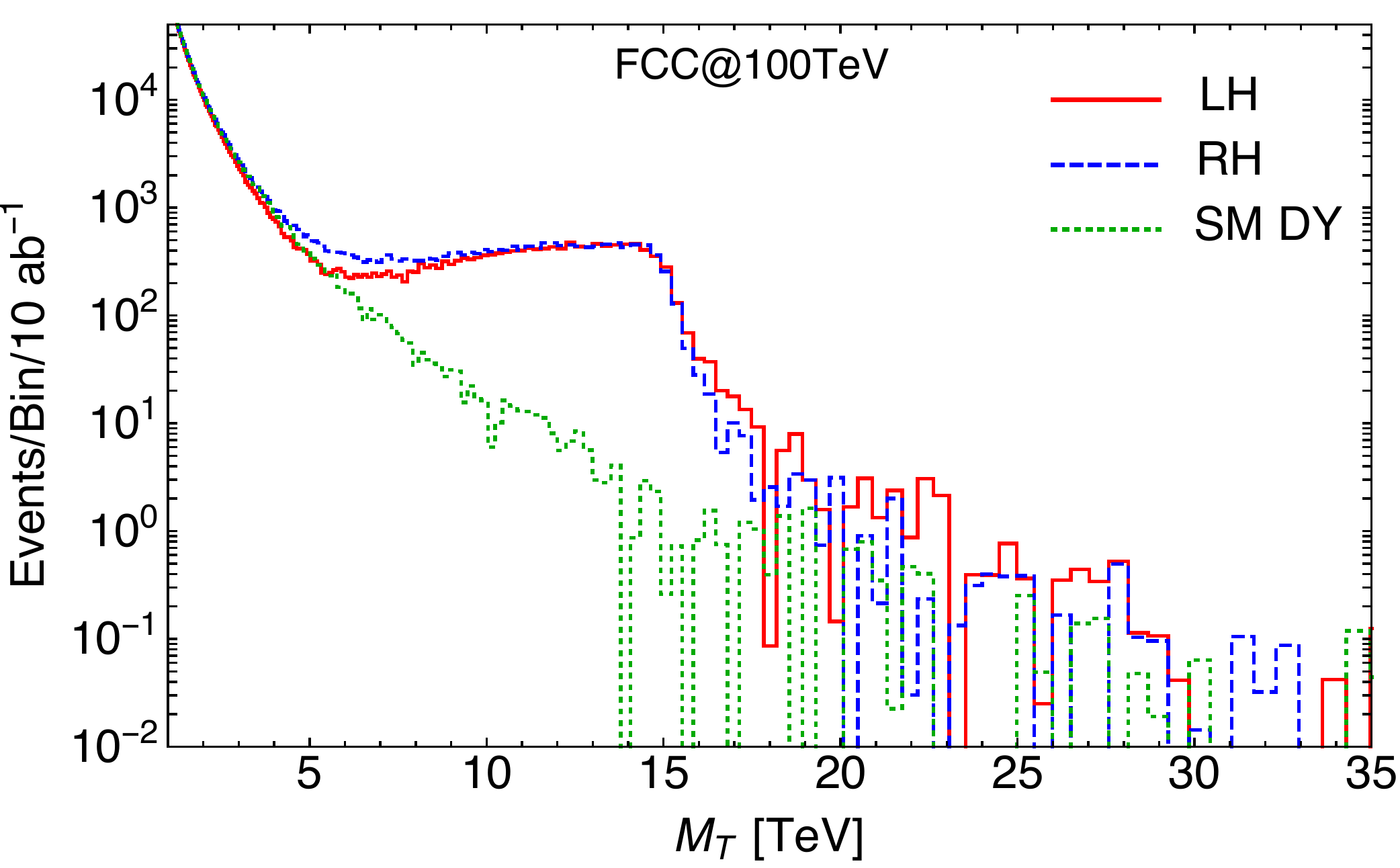}
\end{center}
\caption{Comparison of the $W'$ leptonic transverse mass distribution for LH(red,solid) or RH(blue,dashed) couplings assuming a mass of 
15 TeV and an integrated luminosity of 10 ab$^{-1}$. The Drell-Yan SM background appears is in green(dotted).}
\label{fig3}
\end{figure}

Now let us very briefly mention the case of a new $W'$. If the $W'$ couples LH and/or neutrinos are Dirac states then the 
$W'\to \ell \nu$ mode is the standard discovery channel (the case of $W'$ decays involving RH neutrinos will be discussed in Section \ref{BuphalDev}). For integrated luminosities of 1(10,100) ab$^{-1}$ the discovery 
reach is found to be 31.6(39.1, 46.7) TeV. As mentioned above, the main issue here is whether the $W'$ couples 
in a LH or RH manner. Employing {\it only} this mode in the NWA, however, a purely LH or RH $W'$ are indistinguishable.  
One possibility to get around this, similar to that discussed above, is to make use of the $W'\to t\bar b$ mode and then 
determine the polarization of the top. Another is to go off-resonance in the transverse mass ($M_T$) region below the $W'$ 
Jacobian peak and examine the $W-W'$ interference; this interference is absent(destructive) if the $W'$ is RH(LH).  This is 
particularly noticeable when $M_T \simeq 0.4 M_{W'}$, as can be seen in Fig.~\ref{fig3}, even for low integrated luminosities. 
Using this technique the $W'$ coupling helicity can be determined for masses up to roughly $10$ TeV below the discovery reach. 
Of course if the neutrinos are Majorana fields and the $W'$ couples in a RH manner as in the LRM then $W'\to \ell N$ is the 
discovery channel where the heavy $N$ itself decays to dijets and a charged lepton. This signature will be studies in details in Section \ref{BuphalDev}.

%\subsubsection{Di-tau final states ---- To be removed}

%\subsubsection{LFV resonances}

\subsubsection{Di-jet Resonances and Calorimeter Requirements}\label{Doglioni}
%\begin{itemize}
%\item C.~Doglioni, A.~Boveia et al., {\it Di-jet benchmarks (alro relation to DM group)} {\bf reminder sent to contributors (expected end of January)}
%\end{itemize}
%In the interest of advancing the frontier of high-energy physics beyond the scope of the Large Hadron Collider (LHC), preliminary investigations are being made into the design of a FCC-hh which would support proton collisions at energies up to 100 TeV. If the planned runs of the LHC in the coming decade yield evidence of new physics beyond the Standard Model, such a collider could be geared toward probing these phenomena in unprecedented detail. Even if no new discoveries precede its construction, the new machine would provide access to a new energy regime in which it is expected that events from new physics phenomena could be observed.

In this preliminary stage of the design of a future 100 TeV collider, we seek to estimate the necessary specifications of detection devices that will be able to accommodate the planned high energies and provide suitably precise measurements of any new phenomena that occur. In this subsection we discuss the kinematic properties of resonant processes involving jets, with particular attention to the performance of hadronic jets and to the calorimeter containment of very energetic particles. The benchmark model chosen for this study is a resonant new particle coupling to quarks and gluons, that would manifest as a local excess over the QCD background. 

The sensitivity of a new physics search for a resonant process is linked to an accurate measurement
of the jet energy, and therefore to the calorimeter resolution. Broader local excesses are more difficult
to distinguish over the QCD background. In this contribution we use simulations of new physics events 
(specifically, decays of excited quarks) and modify the calorimeter energy resolution in the detector simulation,
in order to observe the effect of the smearing on the width of the signal excess in the dijet invariant mass. 
Firstly we give an overview of the simulation and software used for this study, including an overview of
the benchmark model chosen. Then we include further detail on the calorimeter smearing and the event selection. 
The last part of this contribution contains the results of this study and conclusions towards future studies.  

%https://cds.cern.ch/record/682130/files/phys-92-010.pdf

%\section{Monte Carlo simulation and tools}
%\label{sec:MCTools}

The benchmark model used in this study is quark compositeness \cite{Baur:1987ga,Baur:1989kv}: excited up and down quarks, and relative antiparticles, 
are simulated from proton-proton collisions at 100 TeV.  Only gauge interactions are included in the benchmark model. 
The excited quark masses are assumed to be 10 and 40 TeV. Cross sections for this model at 100 TeV as a function 
of the $q^{*}$ mass are shown in Fig.~\ref{fig:xsecs} for different parton distribution functions \cite{Martin:2009iq,Kretzer:2003it,Pumplin:2002vw, Lai:2010vv}. 
The Pythia event generator~\cite{Sjostrand:2007gs} using the MRST2008LO~\cite{Martin:2009iq} and the default tune for parton shower, 
interfaced to the Sacrifice steering software \cite{Sacrifice}, are used for the event generation. 
Following the event generation, Delphes 3.1.2 \cite{deFavereau:2013fsa} is used to apply detector effects using the standard FCC detector card
to the signal and perform jet finding, using the anti-kt algorithm \cite{Cacciari:2005hq} with radius 0.5. The events were analysed in MadAnalysis~\cite{Conte:2012fm} 
and ROOT 5.34.18 \cite{Brun:1997pa}. 

\begin{figure}[t!]
	\begin{center}
		\includegraphics[scale=0.4]{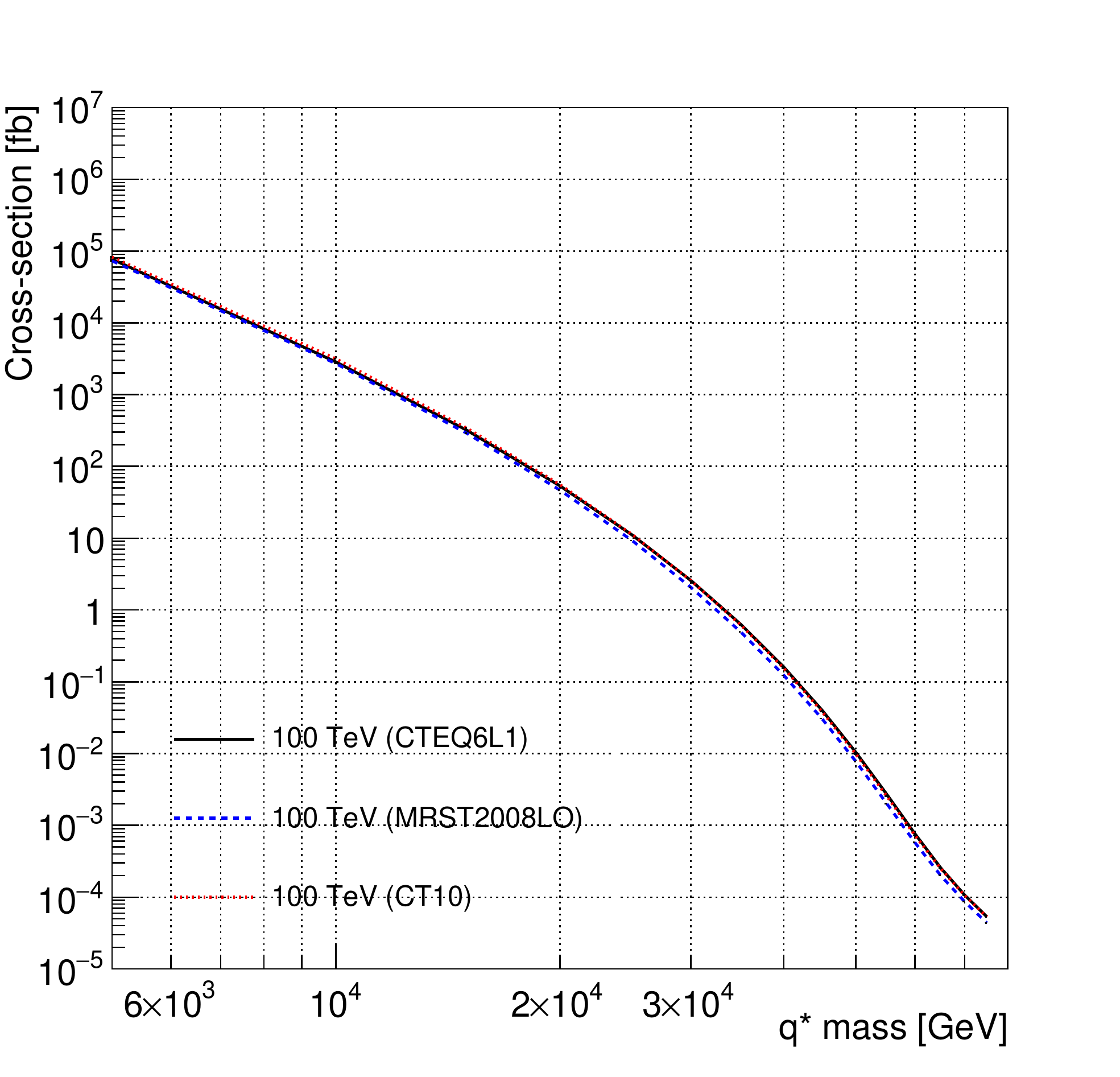}
		\caption{Excited quark production cross-sections as a function of mass.}
		\label{fig:xsecs}
	\end{center}
\end{figure}

%\section{Event smearing}
%\label{sec:EventSmearing}

\noindent The energy resolution for a calorimeter in this study is parameterized as:
\begin{equation}
\frac{[\sigma]}{E} = \frac{50}{E} \oplus c \%\,,
\end{equation}
where $c$ is the constant term that is varied in this study to model possible effects e.g. from calorimeter punch-through,
using the Delphes \texttt{SimpleCalorimeter} module.  
%\section{Event selection}
%\label{sec:EventSelection}
Events were selected according to the following criteria \cite{Aad2016}:
\begin{itemize} 
	\item Leading and subleading jets must have \pt{} $>$ 50 GeV and rapidity $|y| <$ 2.8;
	\item Half the rapidity separation ($y^{*}$) of leading and subleading jets must be below 0.6. 
\end{itemize} 
Analysis of data from simulated 10 TeV $q^{*}$ decays smeared with Delphes indicates that an increase in the constant term of the jet
energy resolution broadens the width of the dijet invariant mass signal. To quantify the broadening, 
the core of the dijet mass distribution after smearing and event selection is fitted using a Gaussian as shown in Figs.~\ref{fig:Gaussian10}. 
The width of the Gaussian is then divided by the dijet mass to obtain a relative resolution, and plotted 
as a function of the constant term broadening in Fig.~\ref{fig:Broadening}. 
The relative mass resolution ranges from $2.5\%$ without any smearing to approximately $7\%$ in the case of a $15\%$ constant term, 
for a $q^{*}$ with a mass of 40 TeV . 

\begin{figure}[t!]
	\begin{center}
		\includegraphics[scale=0.34]{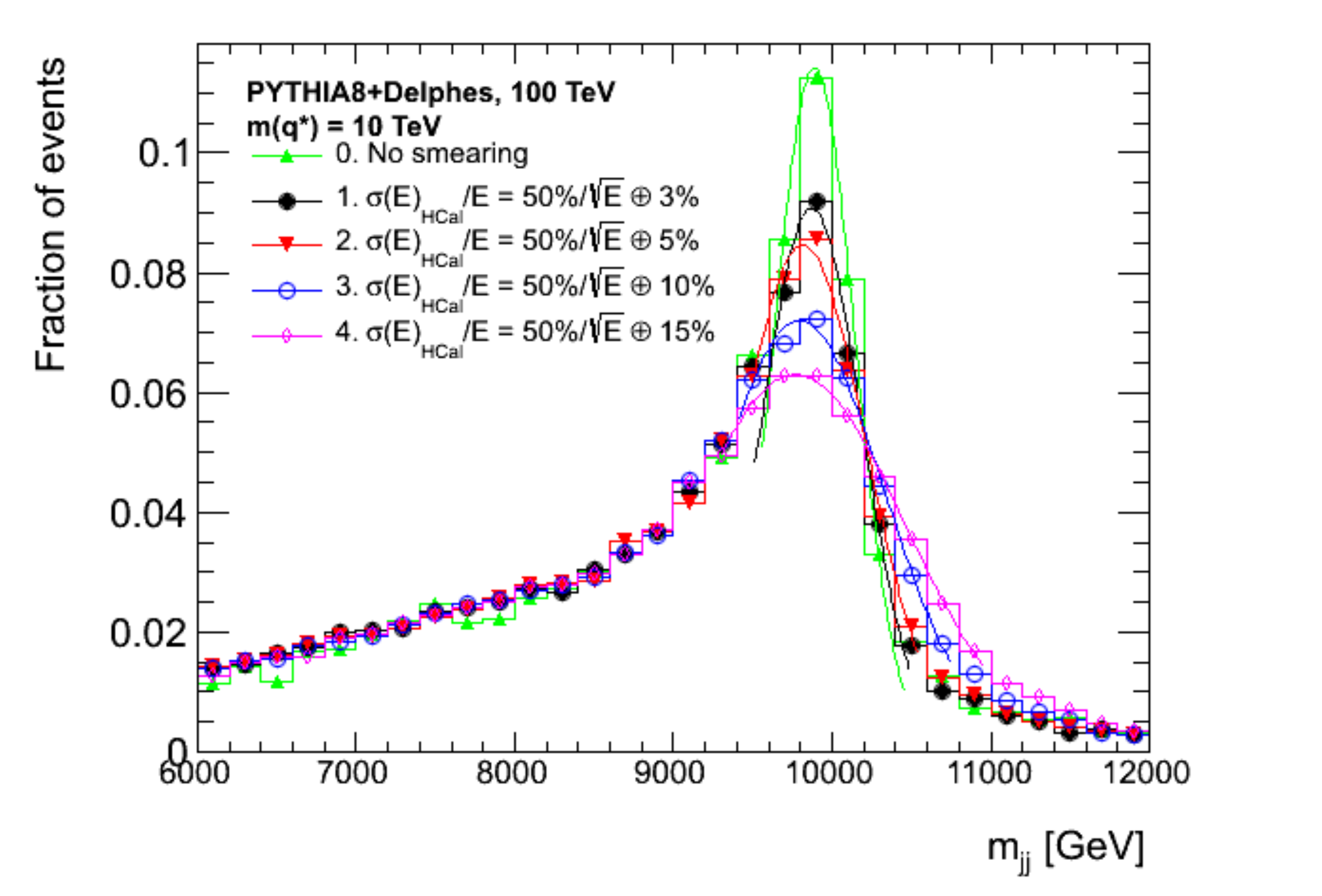}\hspace{-0.8 cm}
		\includegraphics[scale=0.34]{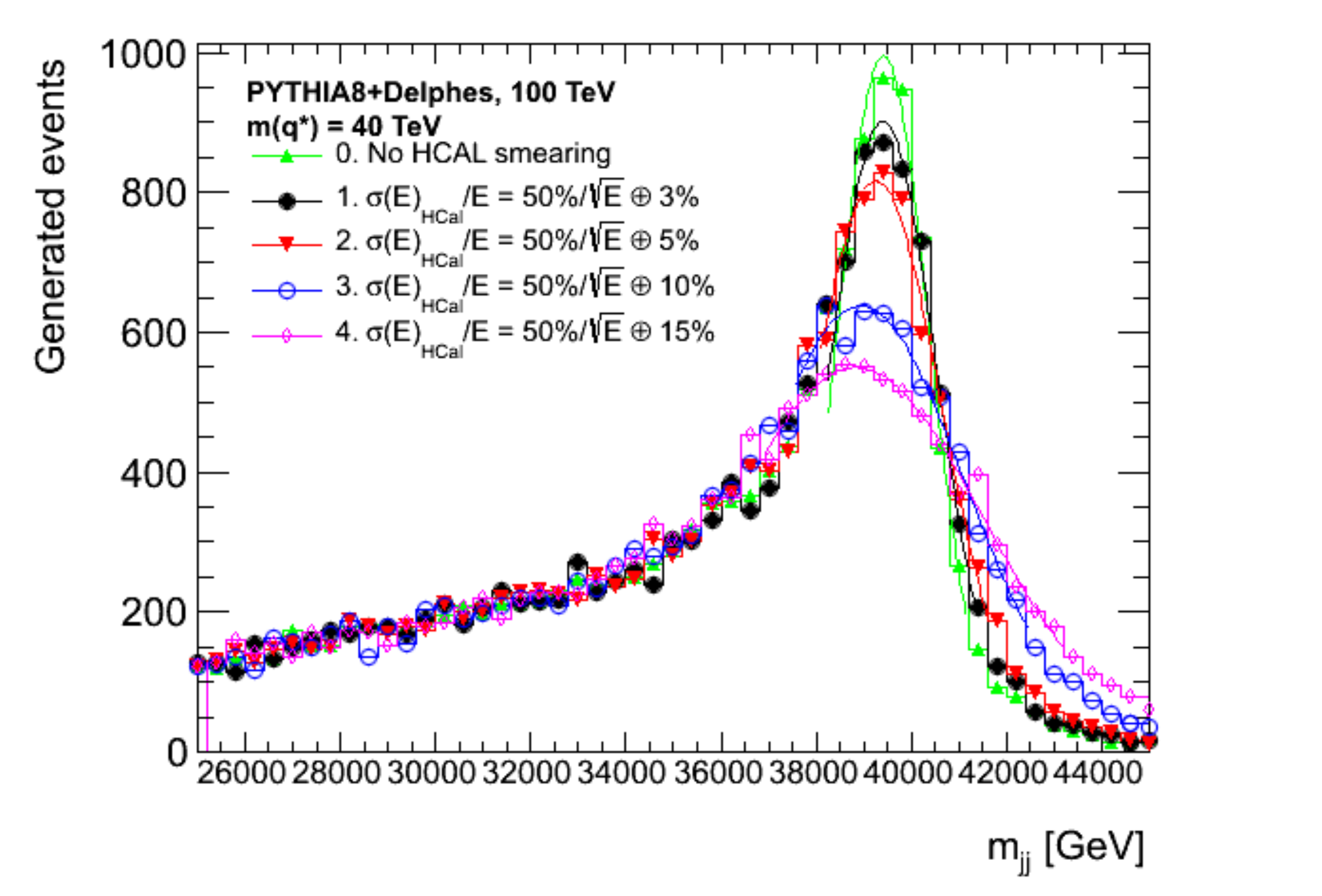}
		\caption{Dijet invariant mass peak for a $q^{*}$ after event selection, fitted using a Gaussian shape, after smearing with different values of the constant term of the jet energy resolution. {\it Left:} 10 TeV $q^{*}$; {\it Right:} 40 TeV $q^{*}$.}
		\label{fig:Gaussian10}
	\end{center}
\end{figure}

%\begin{figure}[htbp]
%	\begin{center}
%		
%		\caption{Dijet invariant mass peak for a 40 TeV q* after event selection, fitted using a Gaussian shape, after smearing with different values of the constant term of the jet energy resolution.}
%		\label{fig:Gaussian40}
%	\end{center}
%\end{figure}

\begin{figure}[htbp]
	\begin{center}
		\includegraphics[scale=0.4]{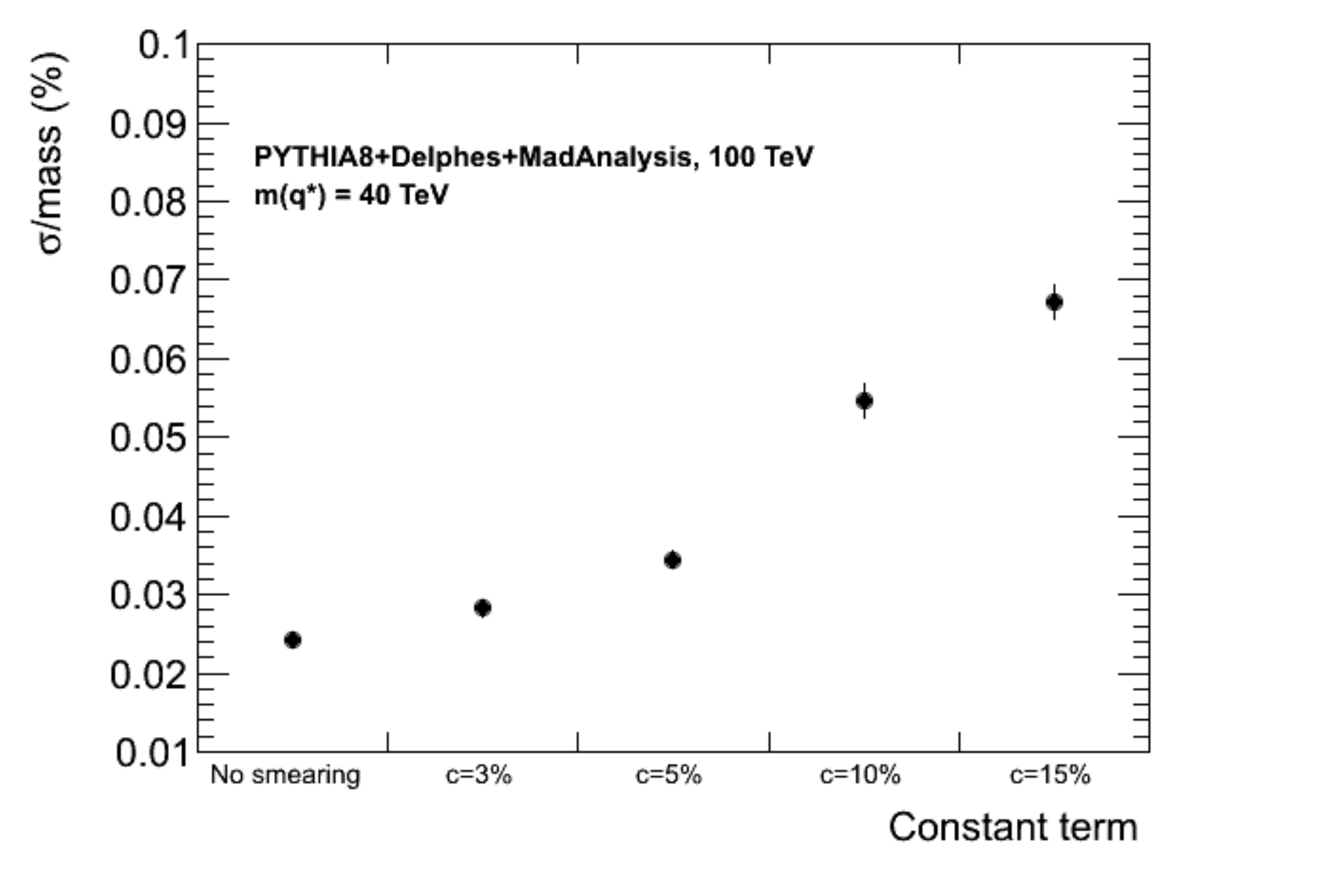}
		\caption{Relative mass resolution as a function of jet energy resolution constant term for a 40 TeV $q^{*}$.}
		\label{fig:Broadening}
	\end{center}
\end{figure}

%\section{Outlook}
%\label{sec:outlook}

Figure \ref{fig:Gaussian10} show that a Gaussian shape does not provide a good description of the signal peak. The broadening of the dijet mass peak is driven by both PDF effects, dominant in the higher mass samples, and by the choice of the jet algorithm. Although initially informative, this study should be extended to a more systematic analysis of different effects that broaden the dijet invariant mass peak in order to improve the peak parameterisation, and to the study of the impact of this broadening on the search sensitivity.

\subsubsection{Resonances in the $jj$ Final State}\label{Yu}
Estimates for the sensitivity of a 100~TeV $pp$ collider to color
singlet $Z^\prime_B$ and color octet $G'$ vector boson dijet
resonances have been performed in ref.~\cite{Yu:2013wta}, while
studies of other resonant and non-resonant scenarios are performed in
refs.~\cite{Apanasevich:2013cta, Kong:2013xta, Agashe:2013kxa}.  The
color singlet $Z^\prime_B$ particle is a dijet resonance predicted in
models with gauged baryon number~\cite{Dobrescu:2013cmh}, whose
phenomenology is encapsulated by a flavor-universal coupling to quarks
$g_B / 6$ and the $Z^\prime_B$ mass.  The coloron $G'$ arises in
extended $SU(3)_C$ color models as a heavy cousin of the SM gluon, and
also couples universally to quarks with a coupling $g_s \tan \theta$.
The two models also exhibit different dijet resonance peak structures
as a result of different final state radiation, and serve to
complement the discussion in the previous section regarding broadening
effects and peak sensivity via explicit simulation of color-singlet
and color-octet resonance models.

We simulate QCD continuum background and $Z^\prime_B$ and $G'$ signals
using MadGraph 5~\cite{Alwall:2011uj} and passed to
PYTHIA~\cite{Sjostrand:2006za} for parton showering and hadronization:
we use MLM~\cite{Mangano:2002ea} matching between QCD two-jet and
three-jet final states.  Events are clustered using \textsc{FastJet}
v.3.0.2~\cite{Cacciari:2011ma} by the anti-$k_T$
algorithm~\cite{Cacciari:2008gp} with distance parameter $R = 0.5$ and
basic detector simulation effects are used to smear the jet energies
and momentum reconstruction.  We do not include any interference
between signal and background for the resonance searches, and so the
background sample is identical for each of the BSM searches.  The
dijet invariant mass is constructed following the CMS 8~TeV
analysis~\cite{Khachatryan:2015sja}, where the two leading $p_T$ jets
are used as seed jets.  Then, subleading jets within $\Delta R = 1.1$
are added to the closest seed jet to form two wide jets.

We analyze the dijet search sensitivity at the 100~TeV $pp$ collider.
Using our samples for QCD background and signal, we conduct a
resonance search using a Crystal Ball fit on the signal distribution
to identify the peak structure of the
resonance~\cite{Dobrescu:2013cmh}.  To estimate the statistical
significance $\sigma = N_S / \sqrt{N_S + N_B}$ of this signal peak, we
compare the number of signal events within 3 standard deviations of
the Gaussian core of the Crystal Ball fit to the number of QCD events
in the same mass window: we do not include systematic uncertainties,
though these are certainly important when the resonance becomes very
weakly coupled.

The results for the $Z^\prime_B$ and coloron resonances are shown in
fig.~\ref{fig:dijet_sensitivity} for $5\sigma$ discovery sensitivity
using 3 ab$^{-1}$ and 10 ab$^{-1}$ integrated luminosity.  We have
reproduced the current exclusion limits from
Ref.~\cite{Dobrescu:2013cmh} in the gray region.  For the right panel,
the coloron has a total width larger than 15\% of its mass above the
curve marked ``Wide resonance,'' while for couplings below the line
labeled ``Non-minimal models,'' the ultraviolet completion of the
extended color sector requires additional particles, such as
vectorlike quarks or a second coloron, to retain perturbative gauge
couplings.

We see that a $Z^\prime_B$ boson can be discovered as heavy as 32~TeV,
depending on its coupling to quarks $g_B$, while the 100~TeV $pp$
collider also have discovery sensitivity to couplings as small as $g_B
\sim 0.2$ for lighter $Z^\prime_B$ resonances.  Colorons can be
discovered as heavy as 42~TeV, and couplings as small as $\tan \theta
\sim 0.02$ can also be seen.  The sensitivity prospects of the 100~TeV
$pp$ machine for low $\mathcal{O}(\text{TeV})$ resonance masses,
however, strongly depend on the dijet trigger threshold, which in turn
depends on improvements in trigger bandwidth and limits from detector
hardware.

\begin{figure*}[t]
\begin{center}
\includegraphics[width=0.47\textwidth, angle=0]{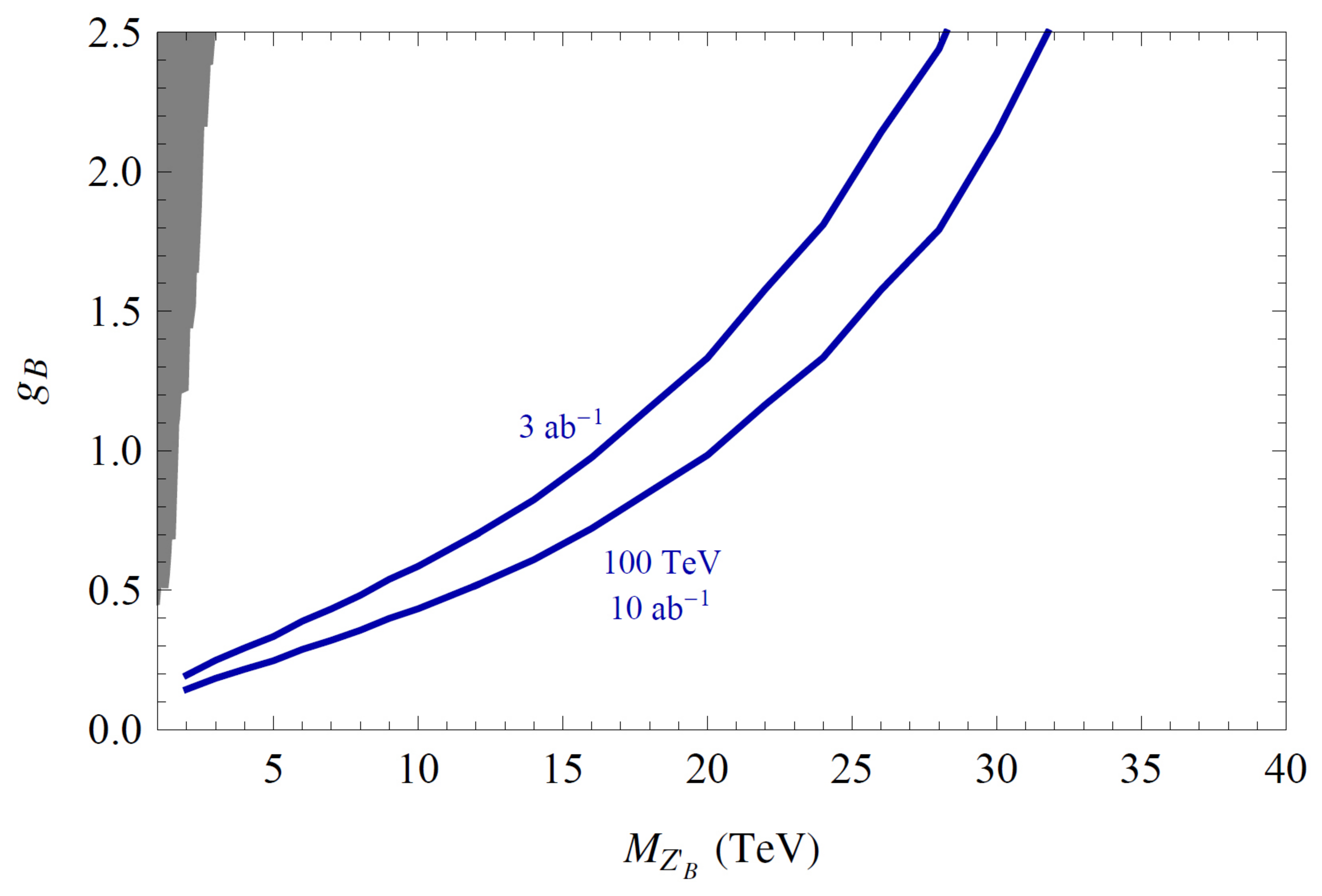}
\hspace{0.2cm}
\includegraphics[width=0.47\textwidth, angle=0]{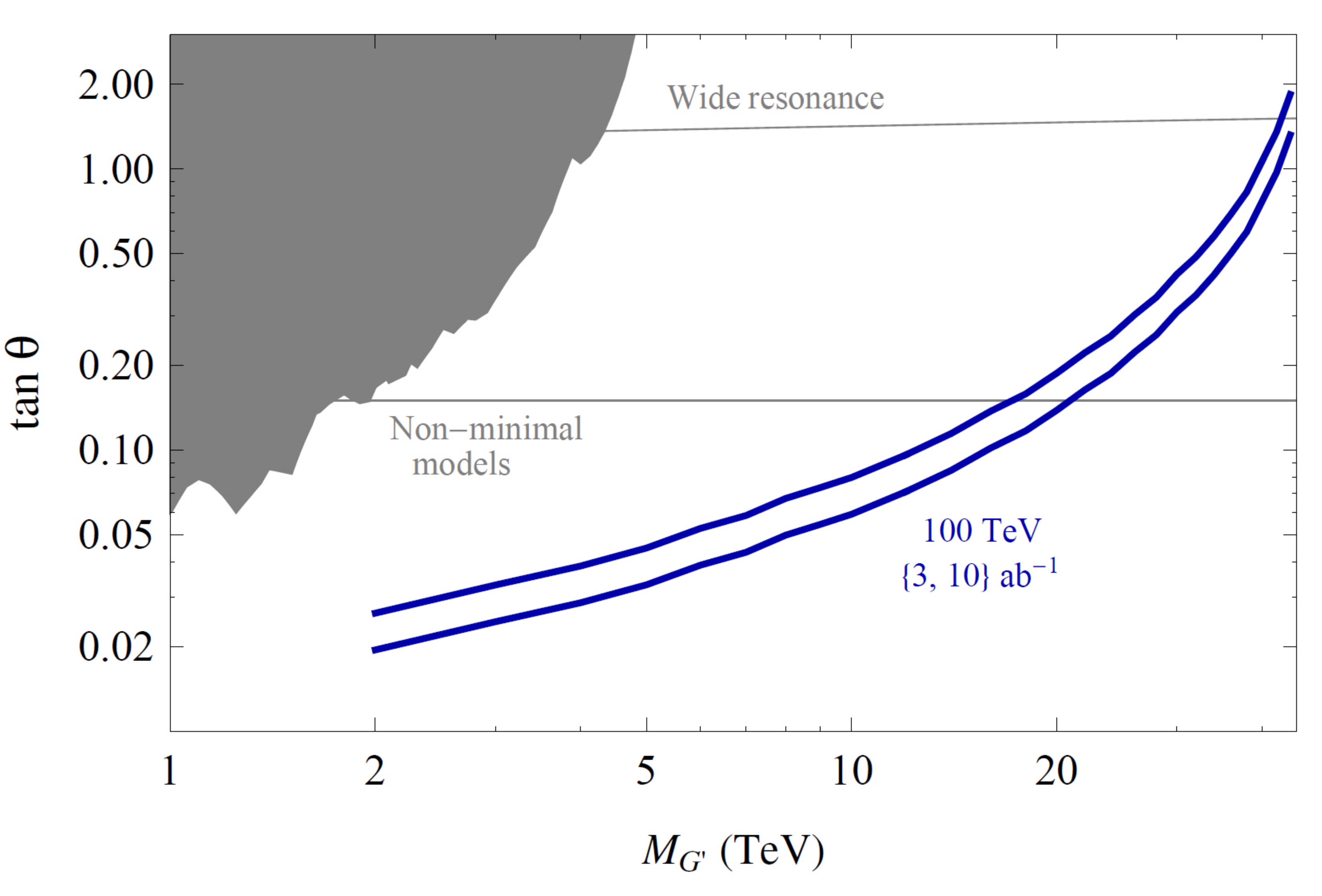}
\caption{Projected $5\sigma$ discovery sensitivity for (left) $Z^\prime_B$ and
  (right) coloron dijet resonances, where the current exclusion bounds
  are shaded gray.}
\label{fig:dijet_sensitivity}
\end{center}
\end{figure*}

\subsubsection{Resonances in $t\bar{t}$ Final State}\label{Auerbach}
%List of contributions:
%\begin{itemize}
%\item B.~Auerbach, S.~Chekanov, J.~Love, J.~Proudfoot, A.V.~Kotwal, {\it Sensitivity to new high-mass states decaying to  $t\bar{t}$ at a 100~TeV collider} {\bf contribution received}
%\end{itemize}
The sensitivity of a 100 TeV $pp$ collider to heavy particles decaying to top-antitop $(t\bar{t})$ final states
has been studied in ref.~\cite{Auerbach:2014xua}. The existence of such particles was discussed in the framework of a
generic Randall-Sundrum  model \cite{PhysRevLett.83.3370}. This model predicts  a number of heavy particles, such
as an extra $Z'$ gauge boson (see ref.~\cite{Langacker:2008yv} for a review) or Kaluza-Klein (KK) excitation of the gluon $g_{KK}$ \cite{Lillie:2007yh}.
The studies used a complete suite of leading-order and next-to-leading order  Monte Carlo samples from the HepSim repository \cite{Chekanov:2014fga} to understand the backgrounds expected for top decays for transverse energies above 3 TeV.
No detector simulation was used.

The studies used dijet invariant mass distributions to extract  $t\bar{t}$ resonance signals above 8 TeV. The dijet masses were reconstructed using the  anti-$k_T$ algorithm \cite{Cacciari:2008gp} with a distance parameter of 0.5. The studies of the sensitivity were performed in a fully boosted regime, i.e. by looking at the invariant mass of two jets arising from the $t\bar{t}$ system. This approach is challenging due to large collimation of decay products from top quarks, and large background expected from the SM jets.
It should also be noted that even leptonic top-quark decays is a challenge at such transverse momentum, since leptons from $W$ decays are often within the vicinity of boosted $b$-quark jets.

The analysis used several popular discriminating variables that reduce SM backgrounds, such as $N$-subjettiness characteristics~\cite{Ellis:2009su,Thaler:2010tr}, the jet $k_T$ splitting scales \cite{Butterworth:2002tt},  jet eccentricity \cite{Chekanov:2010vc},
the effective radius of jets and jet masses. In addition, $b$-tagging was used assuming a $70\%$  $b$-tagging efficiency.

Figure~\ref{h_jetjet5} shows the dijet masses after double $b$-tagging and jet shape
cuts optimized to increase the signal-over-background ratios. This figure was used
to estimate sensitivity, which is equivalent to ``2$\sigma$ evidence'' value of $\sigma \times\text{BR}$ for the signal  calculated
using the $CL_b$ method as implemented
in the {\sc MClimit} program \cite{Junk:1999kv}.
Figure~\ref{exclu_summary_5} shows the sensitivity limits for $\,\mathrm{Z^{\prime}}$ and $g_{KK}$ particles simulated using the PYTHIA8 model \cite{Sjostrand:2006za}.
It should be noted that PYTHIA8 generates the boosted $t\bar{t}$ topology similarly,
but the decay widths and the production rates of $Z'$ and $g_{KK}$ are different.
The width of the $Z'$ boson was set to $\Gamma/M=3\%$, while the width of $g_{KK}$ is substantially larger, $\Gamma/M=16\%$. The $g_{KK}$ production rate
is more than a factor of ten larger than that of $Z'$ boson.

\begin{figure}[t!]
\begin{center}
\includegraphics[scale=0.38, angle=0]{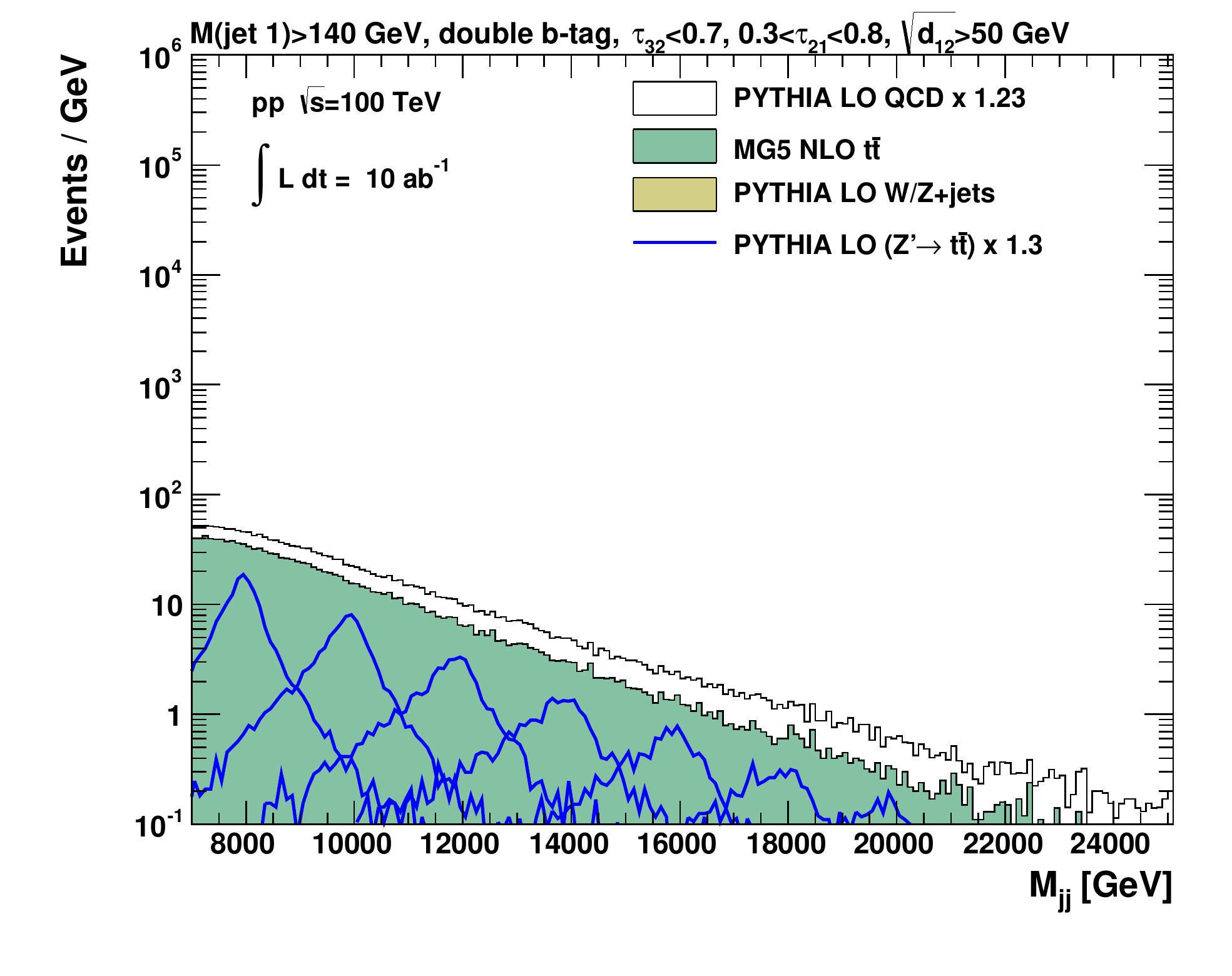}
\includegraphics[scale=0.38, angle=0]{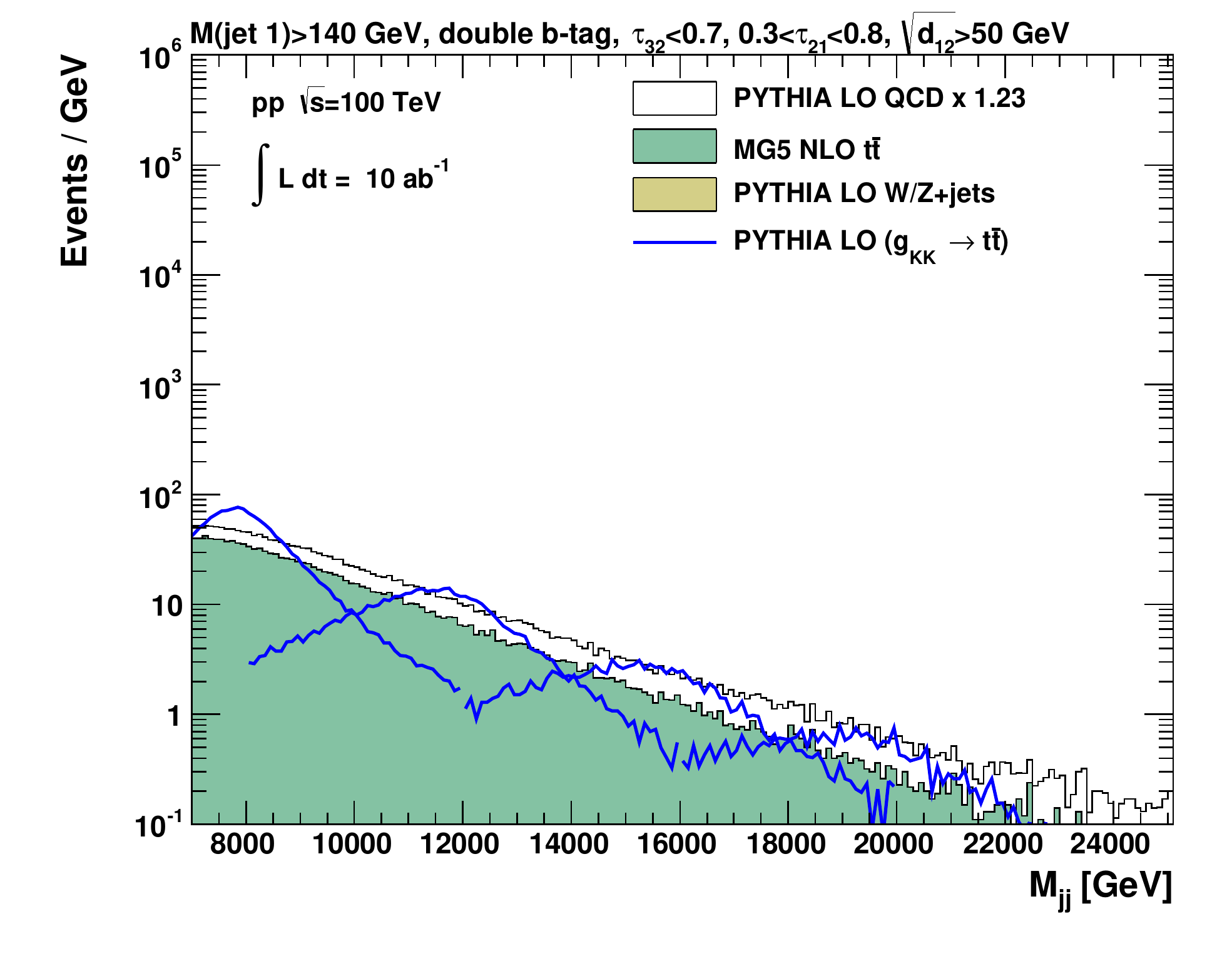}
\caption{Dijet mass distributions after jet shape and $b$-tagging requirements on both jets (see ref.~\cite{Auerbach:2014xua} for details). No subjet requirements were imposed on the second jet. The expectations for resonant processes are shown with the lines. The background histograms are stacked. {\it Left:} Dijet mass distribution with $Z' \to t\bar{t}$ signal; {\it Right:} Dijet mass distribution with $g_{KK} \to t\bar{t}$ signal.}
\label{h_jetjet5}
\end{center}
\end{figure}

\begin{figure}[t!]
\centering
\includegraphics[scale=0.45]{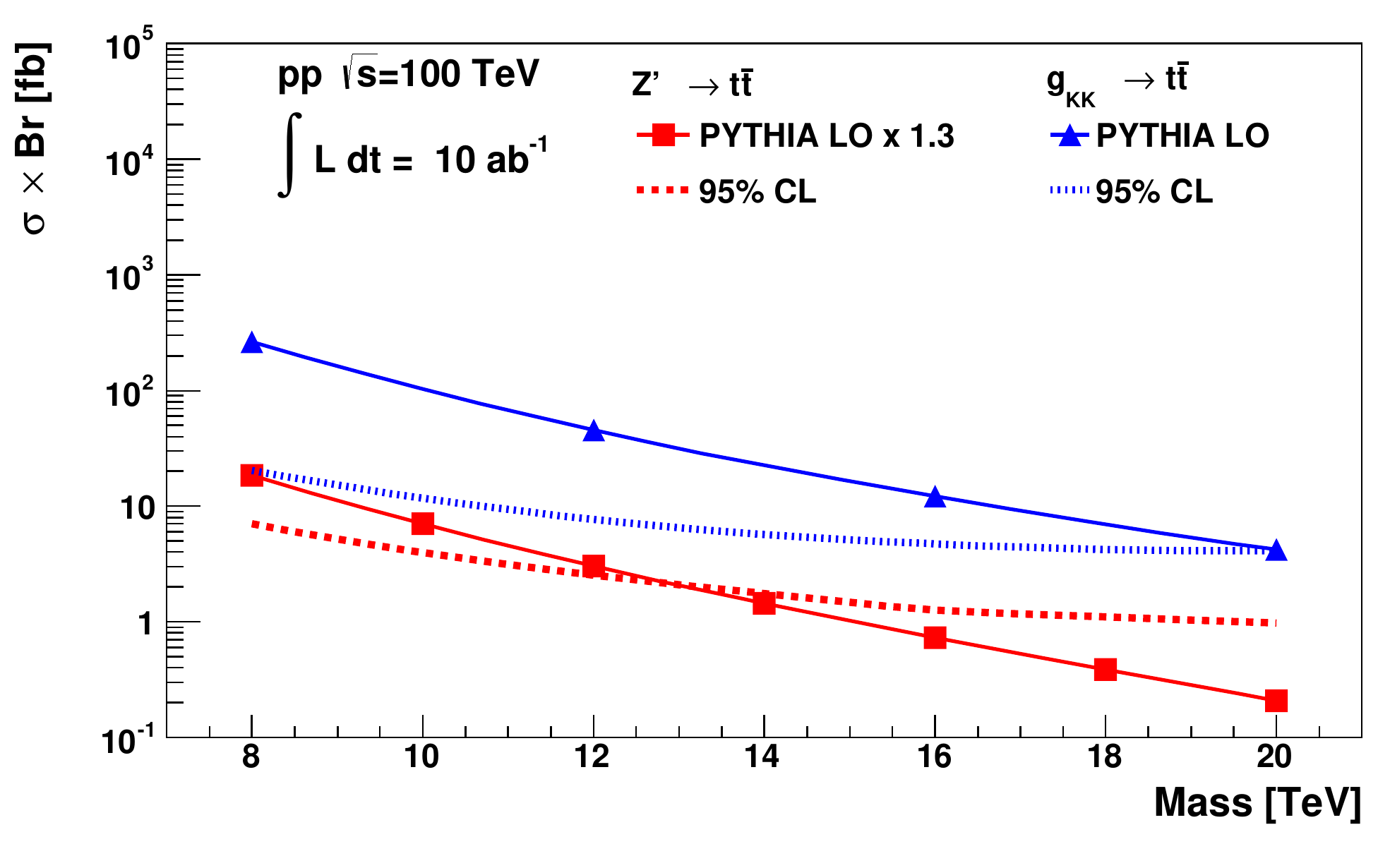}
\caption{
95\% CL sensitivity estimates  for a 100~TeV collider with an integrated luminosity of $10$ ab$^{-1}$
for $Z'$ and  $g_{KK}$ bosons decaying to $t\bar{t}$ 
using the ``fully-boosted'' regime without 
resolving separate decay products of top quarks. Details of the selection cuts are given in ref.~\cite{Auerbach:2014xua}. 
The sensitivities are given  after  applying jet substructure selections \protect{\cite{Auerbach:2014xua}} 
and double $b$-tagging. 
}
\label{exclu_summary_5}
\end{figure}

The discriminating variables based on jet substructure and $b$-tagging  can increase the signal-over background ratio by several orders of magnitude, as shown in Table~\ref{tab_cuts2}.
This increases the  sensitivity on the $\sigma \times \text{BR}$ of $Z'$  and $g_{KK}$ bosons by more than a  factor of ten.
A requirement for a high-momentum muon  inside boosted jets can improve the signal-over-background ratio as shown in Table~\ref{tab_cuts2},
but it  significantly reduces statistics, thus it does 
not lead to a competitive limit compared to the selection based on 
a combination of $b$-tagging and jet substructure variables. 
Figure~\ref{h_erdmann} illustrates the rejection factor for QCD background events as a function of the efficiency
of top-quark reconstruction \cite{Auerbach:2014xua}. 
Identifying top quarks with a cut on a high-momentum muon near or inside a jet is less performant compared to 
jet substructure techniques once branching ratios are taken into account.  
Even simple jet substructure techniques, such as a cut on $\tau_{32}$ 
(the ratio of the $N$-subjettiness variables $\tau_3/\tau_2$ \cite{Ellis:2009su,Thaler:2010tr}) 
and the splitting scale $\sqrt{d_{12}}$  \cite{Butterworth:2002tt},  can overperform the leptonic channel in terms of the background rejection and signal efficiency. 

It should be pointed out that the $95\%$ CL sensitivity estimates  for a 100~TeV collider with the integrated luminosity of $10$~$\mathrm{ab^{-1}}$
are rather general, as long as the widths of the $t\bar{t}$ resonances are similar to  those discussed in this analysis. 
Table~\ref{tbl:mubd} shows the values of $\sigma \times\text{BR}$ for theory and experimental sensitivity as a function of resonance masses
used in Fig.~\ref{exclu_summary_5} for different values of integrated luminosity.
It can be seen that a 100~TeV collider  with an integrated luminosity of 10~$\mathrm{ab^{-1}}$  can be sensitive
to a $g_{KK}$ resonance with a mass of 17~TeV, assuming the LO QCD cross section for the $g_{KK}$ production.
The study also shows that  the assumed integrated luminosity is sufficient to be sensitive to $Z'\to t\bar{t}$ decays with mass of $13$~TeV. 

\begin{table}[t]
\begin{center}
\begin{tabular}{|c|c|c|c|c|c|}
No cuts & JS2    &   $b$-tag &  $b$-tag+JS1 & $b$-tag+JS2 & $b$-tag+JS1+$\mu$ \\ \hline
0.0007  & 0.007 &   0.16     &  0.19                 & 0.21                 & 0.36 \\
\end{tabular}
\end{center}
\caption{The signal-over-background (S/B) ratio for a $Z'$ with mass 10~TeV for different combinations of the selection cuts \cite{Auerbach:2014xua}. The abbreviation "JS2" indicates the jet substructure cuts applied for both jets, while "JS1" indicates the jet substructure cuts for a single jet. The last column shows a combination of $b$-tagging, jet substructure selection and a reconstruction of a high-momentum muon that carries more than $35\%$ of jet transverse momentum. Although the S/B ratio is the largest for the last column, the statistics expected for $10$~ab$^{-1}$ is not sufficient to obtain a competitive $95\%$ CL sensitivity compared to other selections.
}
\label{tab_cuts2}
\end{table}

It is useful to estimate how the sensitivity would improve with integrated luminosity.
The results discussed above were extrapolated to higher values of luminosity using a similar technique. For an integrated luminosity of 30~ab$^{-1}$, the $Z'$ mass reach would increase to 16~TeV, while the mass reach for $g_{KK}$ would increase to 19.5~TeV. More details on this analysis can be found in ref.~\cite{Auerbach:2014xua}. 
  
\begin{figure}[t]
\begin{center}
\includegraphics[scale=0.4]{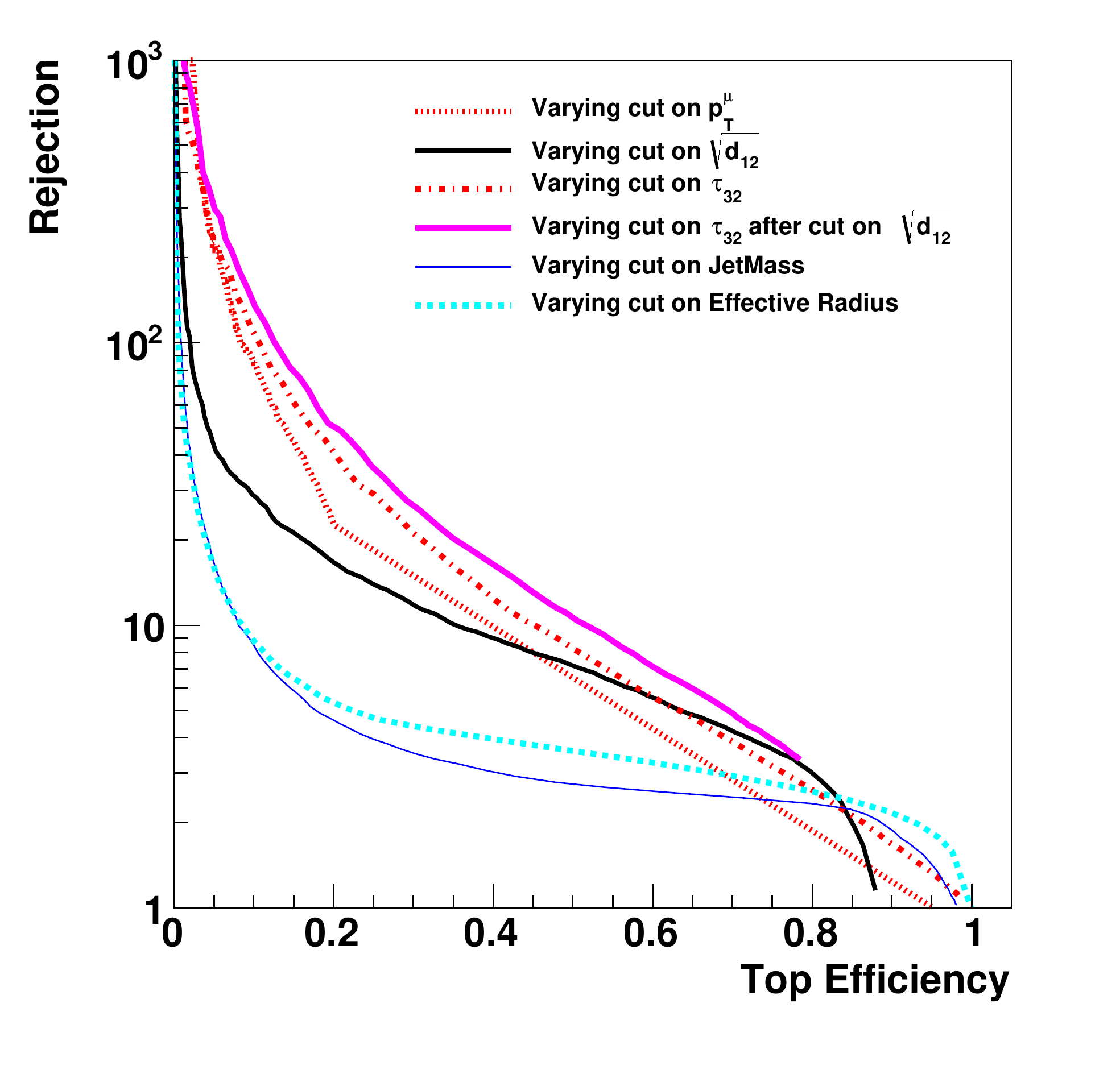}
\caption{
Rejection factor for QCD jets versus efficiency of reconstruction of top quarks for different variables used to select top jets.
}
\label{h_erdmann}
\end{center}
\end{figure}

\begin{table}[t]
%\begin{ruledtabular}
\begin{center}
\begin{tabular}{lcccc}
%\begin{tabularx}{\textwidth}{lcccc}
\hline
\hline
mass & \multicolumn{4}{c}{$\sigma \times\text{BR}$ (fb)} \\
(TeV)    & $Z'$ (th.)  & $Z'$ (exp.) & $g_{KK}$ (th.) & $g_{KK}$ (exp.)  \\
\hline
8  & 18.46 & 7.00  & 262.3 & 20.2 \\
10 & 7.03 & 3.97  & & \\
12 & 3.02 & 2.54 & 45.4 & 7.7 \\
14 & 1.44 & 1.75 & & \\
16 & 0.73 & 1.27 & 12.2 & 4.7 \\
18 & 0.39 & 1.10 & &  \\
20 & 0.21 & 0.98  & 4.2 & 4.1 \\
\hline
\hline
\end{tabular}
\end{center}
\caption{Values of $\sigma \times\text{BR}$ for theory and experimental sensitivity as a function of resonance mass shown in Fig.~\ref{exclu_summary_5}.}
\label{tbl:mubd}
\end{table}

\subsubsection{Composite Resonances: Direct vs Indirect Probes}\label{Thamm}
%\begin{itemize}
%\item A.~Thamm, R.~Torre, A.~Wulzer, {\it Composite Resonances: direct vs indirect probes} {\bf contribution received}
%\end{itemize}
In this subsection, we study the expected direct reach of a 100 TeV collider on heavy vector triplets \cite{Pappadopulo:2014qza} and compare it to the expected indirect reach of various proposed future lepton colliders in a minimal composite Higgs model \cite{Thamm:2015zwa}.

The comparison of the discovery and exclusion prospects of a 100 TeV collider and various proposed lepton colliders is crucial to gain a deeper understanding of the expected impact of these experiments. It is most conveniently done within a small parameter space of an explicit model. Here we choose a minimal composite Higgs model \cite{Agashe:2004ib,2014arXiv1401.2457B,Panico:2015jxa}. This is not only a theoretically well-motivated scenario but also predicts direct and indirect signs of new physics which can be studied at the different colliders. All new physics effects are comprehensively discussed in ref.~\cite{Thamm:2015zwa} and we refer the reader to this reference for further details. Here we report a brief summary of this study. 

The strongest indirect constraints on composite Higgs models come from electroweak precision tests. However, their impact depends heavily on the details of the model. In order to remain as much as possible agnostic on these details and therefore more model independent, we do not focus, when considering indirect effects, on electroweak precision tests, measured both at LEP (with some improvement from Tevatron and LHC) and possibly improved at future leptonic machines. Instead we concentrate on indirect effects originating from the modification of the Higgs couplings because they are largely model-independent. In fact, for all models based on the minimal coset $SO(5)/SO(4)$, the Higgs coupling to electroweak gauge bosons is universally predicted to deviate from the SM expectation by $k_V = \sqrt{1-\xi}$, where $\xi = v^2/f^2$, $f$ is the decay constant of the pseudo-Nambu-Goldstone boson Higgs and $v$ is the scale of EWSB. We will thus take the sensitivity of future lepton colliders to $k_V$ as a good model-independent measure of their reach on composite Higgs models.

Direct signatures stem from top partners and electroweak vector resonances. Top partners are generally more model-dependent: their mass controls the generation of the Higgs potential and thus the level of fine tuning required to achieve EWSB and a light Higgs boson \cite{DeSimone:2012ul}. The prospect for top partner searches at a 100 TeV collider are discussed in Section \ref{Wulzer}. In the analysis we perform here, aiming at being as much as possible model independent, we focus on vector resonances which are associated with the current operators of the SM gauge group. In particular, we study a colorless triplet under $SU(2)_L$ with zero hypercharge. The simplified model for the heavy vector triplet, studied in detail in model B of ref.~\cite{Pappadopulo:2014qza}, depends on only two parameters: the mass $m_\rho$ of the vector triplet and the new coupling $g_\rho$ which describes the self interactions of the heavy vector and parameterizes the couplings to SM particles. The two parameters are related to $\xi$ by
\begin{equation}
\xi = \frac{g_\rho^2 v^2}{m_\rho^2}\,.
\end{equation}
Thus the indirect reach on $\xi$ can be compared to the direct reach on $m_\rho$, and exclusion bounds are set in the $(m_{\rho},\xi)$ or, analogously, in the $(m_\rho,g_\rho)$ plane.

\begin{figure*}[t!]
\begin{center}
\includegraphics[scale=0.35]{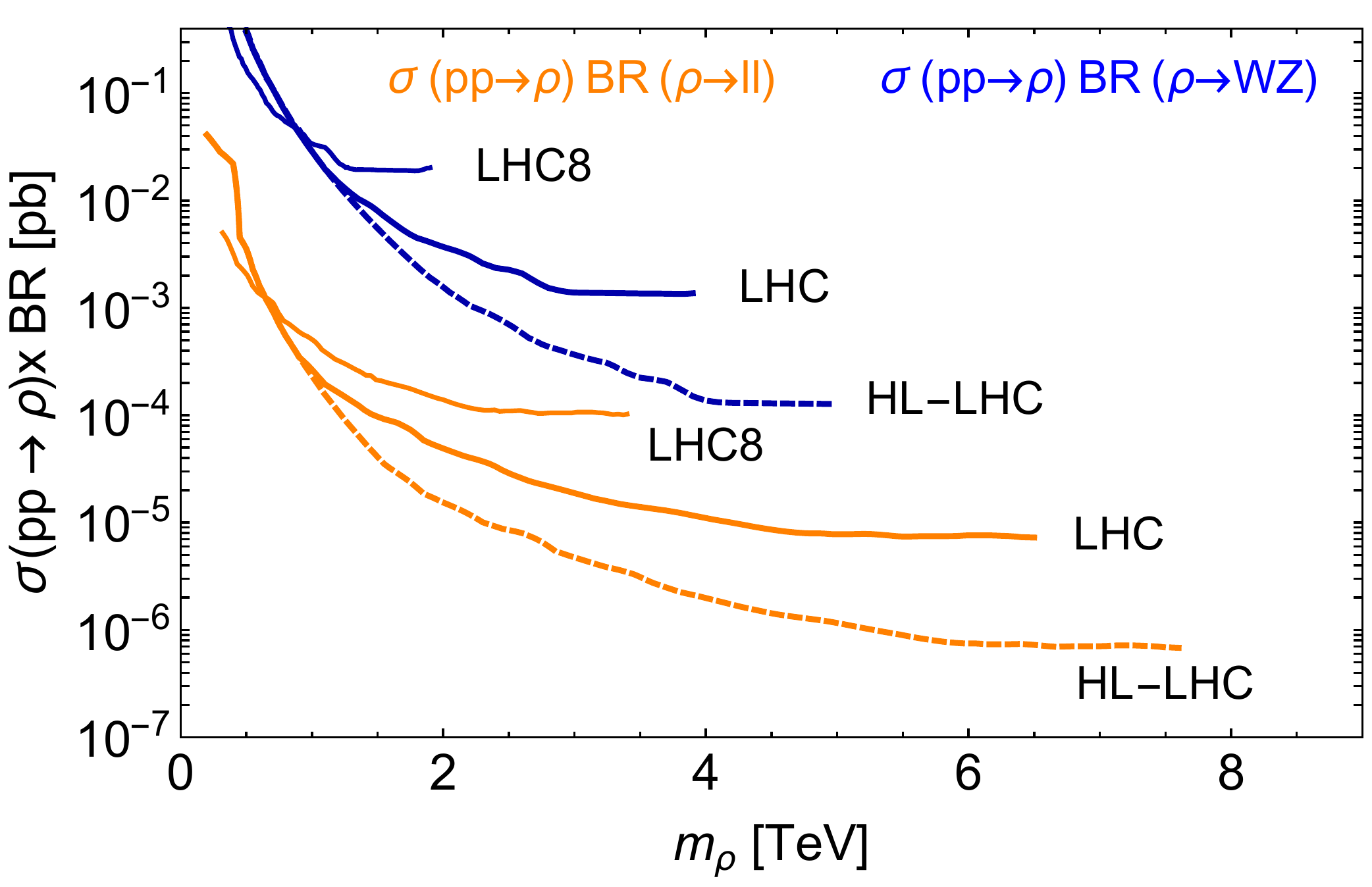}
\hspace{0.5cm}\includegraphics[scale=0.35]{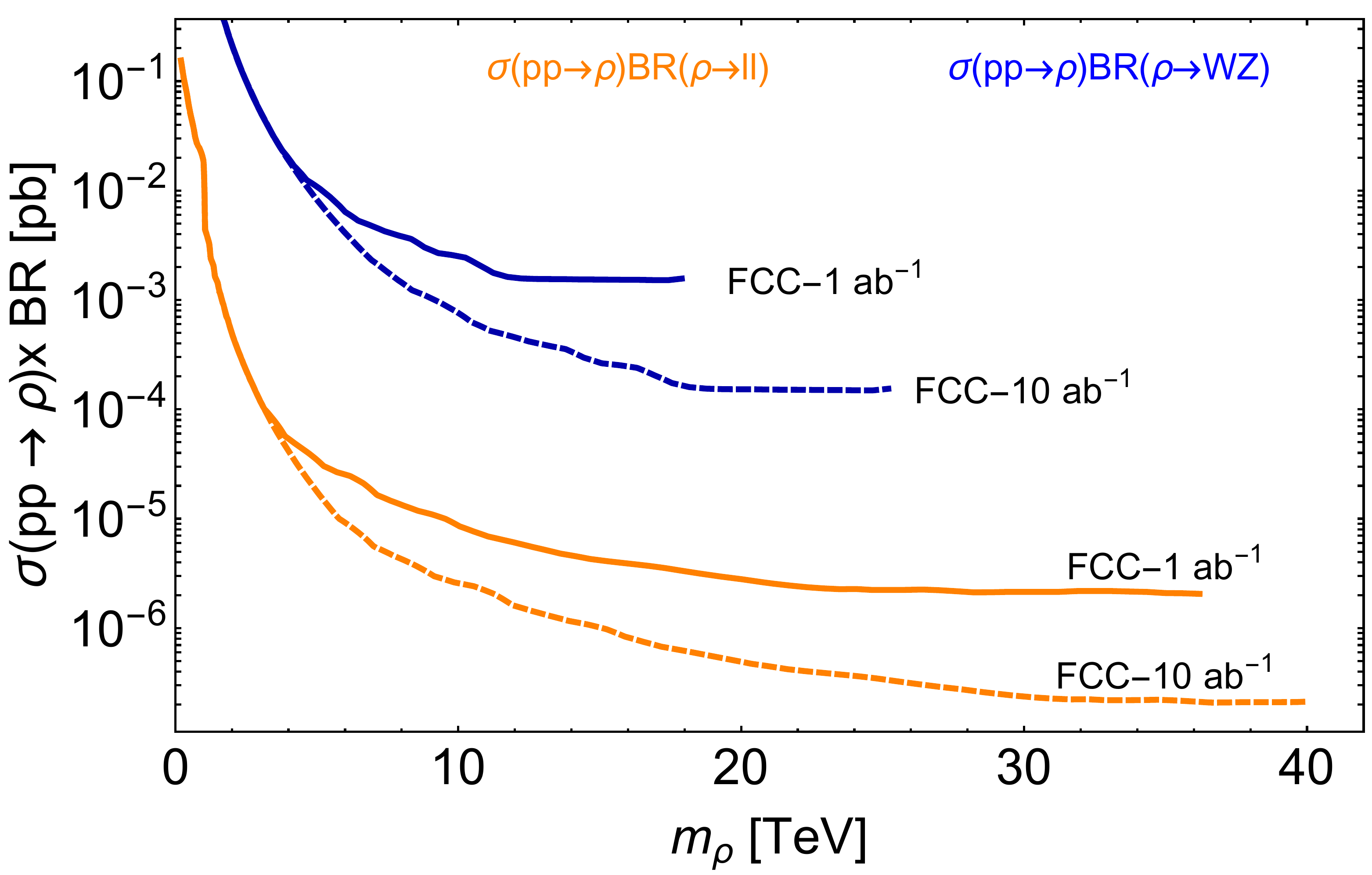}\vspace{-0.2cm}
\caption{{\it Left:} Bounds on $\sigma\times \text{BR}$ from LHC at $8\,$TeV (LHC8) with $20\,$fb$^{-1}$ (solid) and corresponding extrapolations to LHC at $14\,$TeV with $300\,$fb$^{-1}$ (solid) (LHC) and $3\,$ab$^{-1}$ (dashed) (HL-LHC); {\it Right:} extrapolation of LHC8 to a 100 TeV collider with $1\,$ab$^{-1}$ (solid) and $10\,$ab$^{-1}$ (dashed). The two analyses of refs.~\cite{CMS-PAS-EXO-12-061} (CMS di-leptons, orange, lower curves) and \cite{Khachatryan:2014xja} (CMS fully leptonic di-bosons, blue, upper curves) are considered.}\label{fig:extrapolatedbounds}\vspace{-0.4cm}
\end{center}
\end{figure*}
Figure \ref{fig:extrapolatedbounds} shows the LHC bounds on $\sigma \times$BR at $8\,$TeV with $20\,$fb$^{-1}$, expected limits at the $14\,$TeV LHC with $300\,$fb$^{-1}$ (LHC) and $3\,$ab$^{-1}$ (HL-LHC) and at a 100 TeV collider with $1\,$ab$^{-1}$ and $10\,$ab$^{-1}$. Blue curves represent CMS bounds on $WZ$ in a fully leptonic final state \cite{Khachatryan:2014xja}, while the more sensitive orange curves depict CMS limits from opposite sign di-lepton searches \cite{CMS-PAS-EXO-12-061}. We verified that the corresponding ATLAS searches in refs.~\cite{ATLAScollaboration:2014eb} and \cite{ATLAScollaboration:2014uc} yield similar results. The current bounds have been extrapolated to larger center-of-mass energies and different integrated luminosities with the procedure described in ref.~\cite{Thamm:2015zwa}. As it can be seen from the figures, the limits approach a constant value at high invariant masses. This is expected as the region corresponds to the zero background regime. An increase of the integrated luminosity by a factor of 10 at a constant center-of-mass energy improves the exclusion bound by a factor of 10 for large masses and and by a factor $\sqrt{10}\sim 3$ for intermediate masses where the background becomes sizeable, and the limit scales as the square root of the number of background events. At very low masses, the bounds become unreliable due to a subtlety in the extrapolation procedure which gives a conservative, but not strongest, bound (see ref.~\cite{Thamm:2015zwa} for details). Finally note in the low mass limit, that bounds from the $14\,$TeV LHC are weaker than at $8\,$TeV, and a similar feature can be observed when comparing a 100 TeV collider to the LHC at $14\,$TeV. This is due to the larger background expected at higher center-of-mass energies. The growing cross-section at higher energy colliders will compensate for this effect, however, and will eventually result in stronger limits on the model parameters in the entire mass range.

These bounds on $\sigma \times$BR can be translated into excluded regions in the $(m_\rho,g_\rho)$ and $(m_{\rho},\xi)$ parameter space shown in the left and right panels of Fig.~\ref{fig:mrho_grho_plane} respectively. Both plots show the relevant parameter space for a $100\,$TeV collider. The strong coupling $g_\rho$ is constrained to be larger than the SM couplings but still within the perturbative regime, $1 \leq g_\rho \leq 4\pi$, and $\xi \le 1$. Regions which violate these conditions are theoretically excluded and shaded in grey in the exclusion plots. Current direct limits for $8\,$TeV are shown in dark violet, while extrapolated bounds from the $14\,$TeV LHC and a 100 TeV collider are depicted in medium and light colours. The 100 TeV collider bound refers to an integrated luminosity of $10\,$ab$^{-1}$, while the dashed violet line shows the bound at $1\,$ab$^{-1}$. 

\begin{figure*}[t!]
\begin{center}
\includegraphics[scale=0.23]{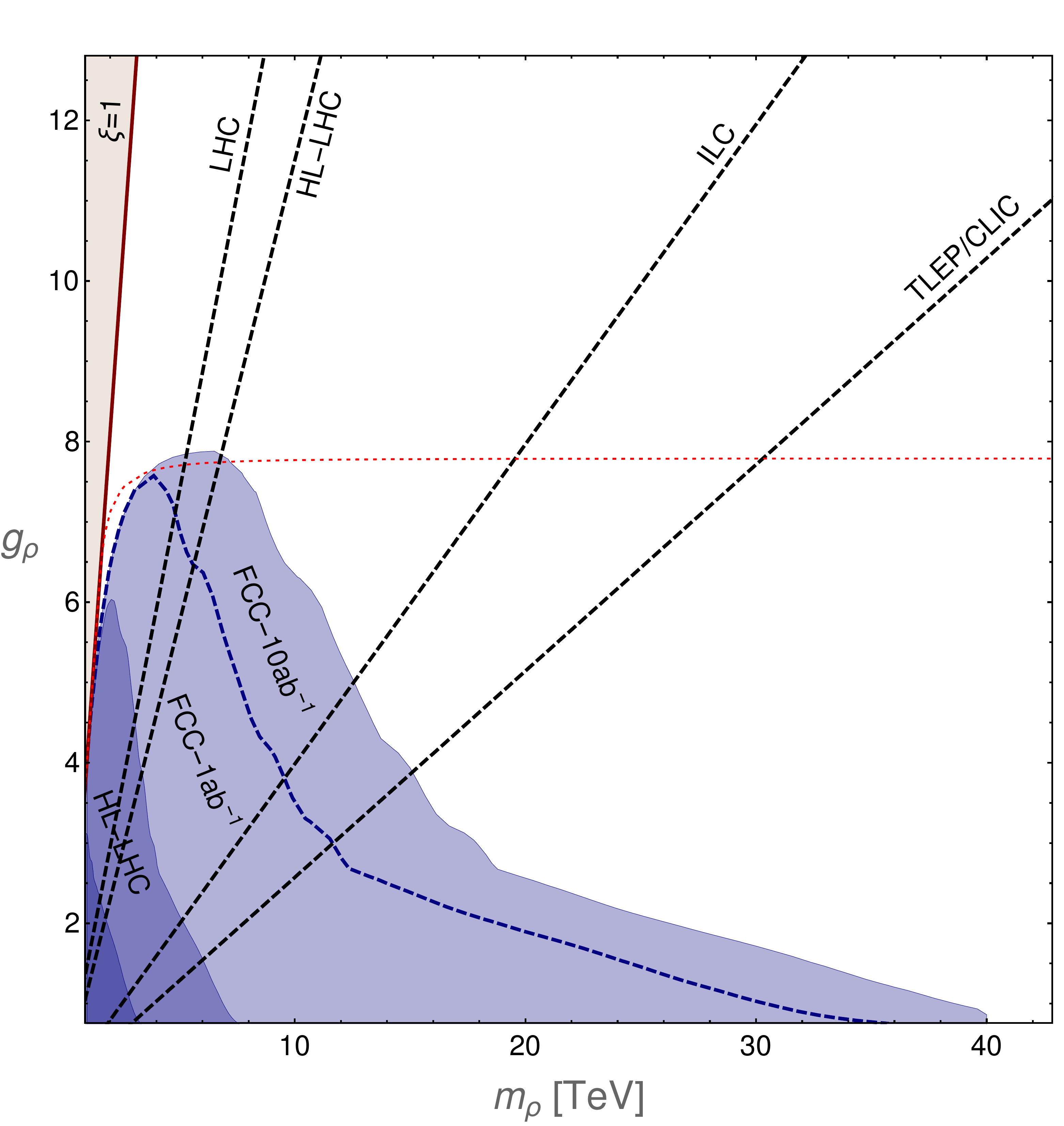}\hspace{4mm}
\hspace{0.5cm}\includegraphics[scale=0.23]{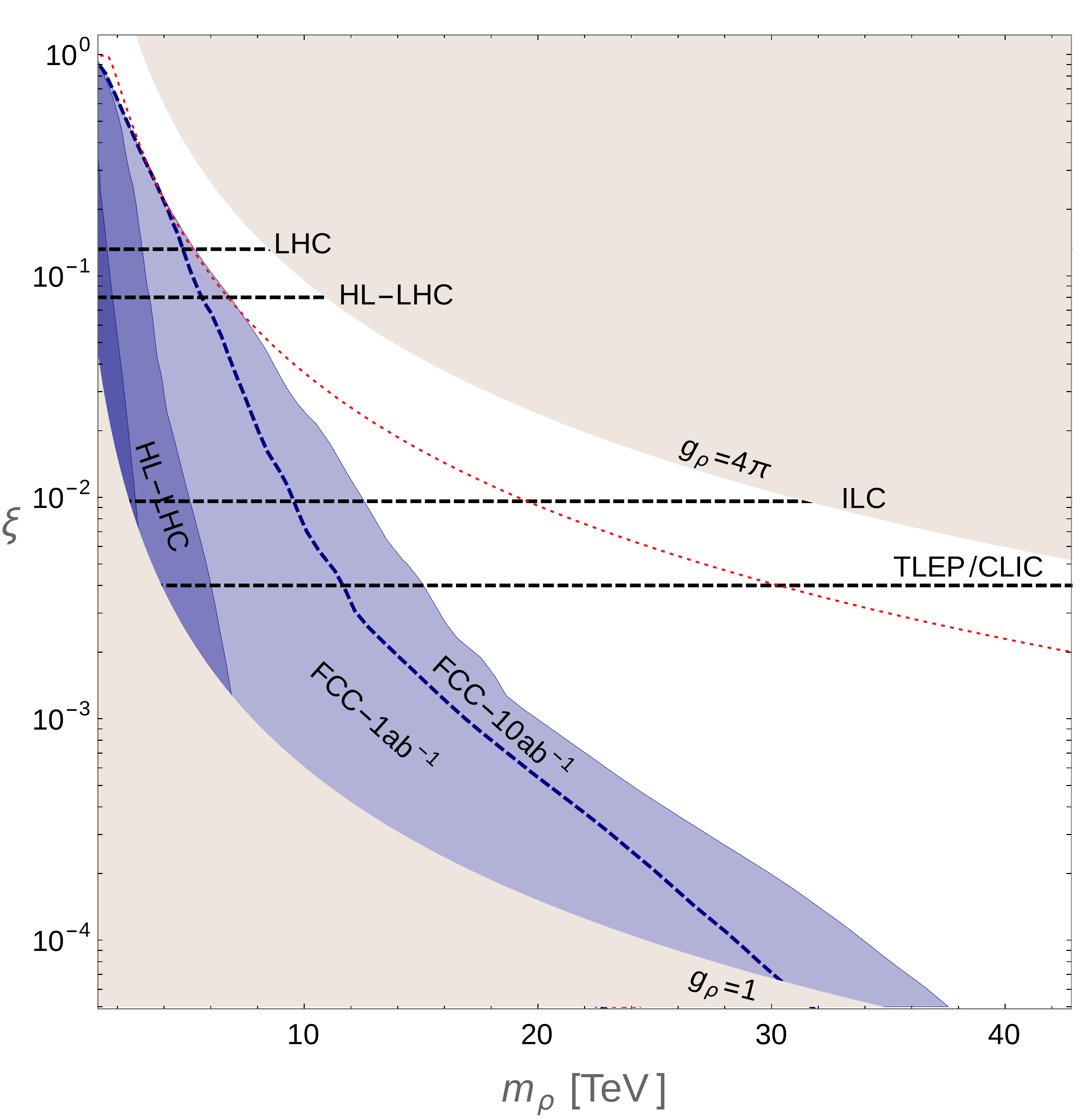}
\caption{Comparison of direct and indirect searches in the parameter space of the MCHM. {\it Left:} comparison in the $(m_{\rho},g_{\rho})$ plane; {\it Right:} comparison in the $(m_{\rho},\xi)$ plane. See the text for details. }\label{fig:mrho_grho_plane}\vspace{-0.5cm}
\end{center}
\end{figure*}

The shape of the direct bounds can be understood as follows. The coupling of the heavy triplet to SM fermions scales as $g^2/g_\rho$, where $g$ is the SM $SU(2)_L$ gauge coupling. Since the dominant production mode is Drell-Yan, we see that the vector becomes effectively weakly coupled for large values of $g_\rho$. This explains the weaker mass reach at large couplings. Furthermore, the scaling behaviour accounts for the far mass reach of the di-lepton channel for low couplings. The kink at intermediate masses is due to the transition between the regions where di-boson searches dominate the exclusion (low masses, large coupling) and where di-boson searches dominate the exclusion (high masses, low coupling). The triplet coupling to SM bosons goes as $g_\rho$ and hence di-boson channels are more sensitive for larger values of $g_\rho$.

From the plots we can infer, as expected, that an increase in the center-of-mass energy of the collider enhances the mass reach significantly. In fact, only a $100\,$TeV collider has the capability to access the multi-TeV region. An increase in luminosity improves the mass reach only slightly but is considerably more effective in the reach for larger $g_\rho$.

Note that resonances become broad for large $g_\rho$ because their coupling to longitudinal vector bosons and the Higgs grows which increases the intrinsic width as $g_\rho^2$. Broad resonances are harder to detect and since a narrow resonance has been assumed in our analysis we expect the actual limits to be even weaker than ours in the large coupling regime. To estimate the region where finite width effects should start to become relevant we included the fine red dotted curves which depict the boundary to the region where the widths exceeds $20\%$ of the mass. In the region above the red line the width is even larger and our bounds are no longer reliable (see ref.~\cite{Pappadopulo:2014qza} for details). 

Indirect constraints are depicted as black dashed lines and show the expected $2\,\sigma$ errors on $\xi$, corresponding to twice the error on $k_V\simeq 1-\xi/2$, obtained from single Higgs production. The values are taken from refs.~\cite{CMS-NOTE-2012/006,ATL-PHYS-PUB-2013-014,Dawson:2013bba}. In the $(m_{\rho},\xi)$ plane, the limits simply corresponds to horizontal lines and translate into straight lines with varying inclination in the $(m_{\rho},g_\rho)$ plane. In particular, the plots show the LHC reach with $300\,$fb$^{-1}$ and $3\,$ab$^{-1}$ corresponding to $\xi > 0.13$ and $\xi>0.08$ respectively, and the expected reach of the ILC and a leptonic FCC at $\sqrt{s} = 500\,$GeV and $\sqrt{s} = 350$\,GeV corresponding to $\xi>0.01$ and $\xi>0.004$. Note that CLIC with 2 ab$^{-1}$ is expected to have a sensitivity comparable to the leptonic FCC.

In conclusion, the plots demonstrate that direct and indirect searches are complementary and probe the parameter space of a composite Higgs model from different directions. While direct searches are more powerful in the low coupling regime, indirect searches win for large couplings.

\subsubsection{Hunting the Flavon}\label{Bauer}
%\begin{itemize}
%\item M.~Bauer, T.~Schell, T.~Plehn, {\it Hunting the Flavon } 
%\end{itemize}

In spite of a major effort in theoretical and experimental particle
physics over many decades, the hierarchies in the quark and lepton
masses and mixing angles is not explained by the Standard Model.  A
high-energy collider going significantly beyond LHC energies will, for
the first time, have the chance to systematically probe a dynamic
origin of this flavor structure. We know various theories, which
address the flavor structure for example through abelian flavor
symmetries~\cite{Froggatt:1978nt,Leurer:1992wg,Leurer:1993gy,Huitu:2016pwk},
loop-suppressed couplings to the Higgs~\cite{Georgi:1972hy}, partial
compositeness~\cite{Kaplan:1991dc}, or wave-function
localization~\cite{Gherghetta:2000qt,Grossman:1999ra,Blanke:2008zb,Casagrande:2008hr,Bauer:2009cf}. All
of these mechanisms introduce flavor-violating couplings and new,
heavy degrees of freedom, which are usually expected to be too heavy
to be produced at the LHC.  For instance, partial compositeness or
warped extra dimensions predict vector-like heavy quarks and colored
spin-one resonances with large cross sections, as discussed elsewhere
in this report. Unfortunately, these resonance features are often not
uniquely pointing to flavor models.

A 100~TeV machine will for the first time directly probe parameter
space not excluded by quark flavor experiments.  For low flavor
breaking scales, which are well motivated if the flavor sector is
related to electroweak symmetry
breaking~\cite{Bauer:2015fxa,Bauer:2015kzy} or dark
matter~\cite{Calibbi:2015sfa}, the FCC-hh has the potential to
discover the dynamical degree of freedom of flavor symmetry breaking.
In our discussion following Ref.~\cite{Bauer:2016rxs} we focus on
flavon couplings directly related to the flavor breaking mechanism
induced by a minimal Froggatt-Nielsen model.

%------------------------------------------------------------
\begin{figure}[t!]
  \centering
  \includegraphics[width=.4\textwidth]{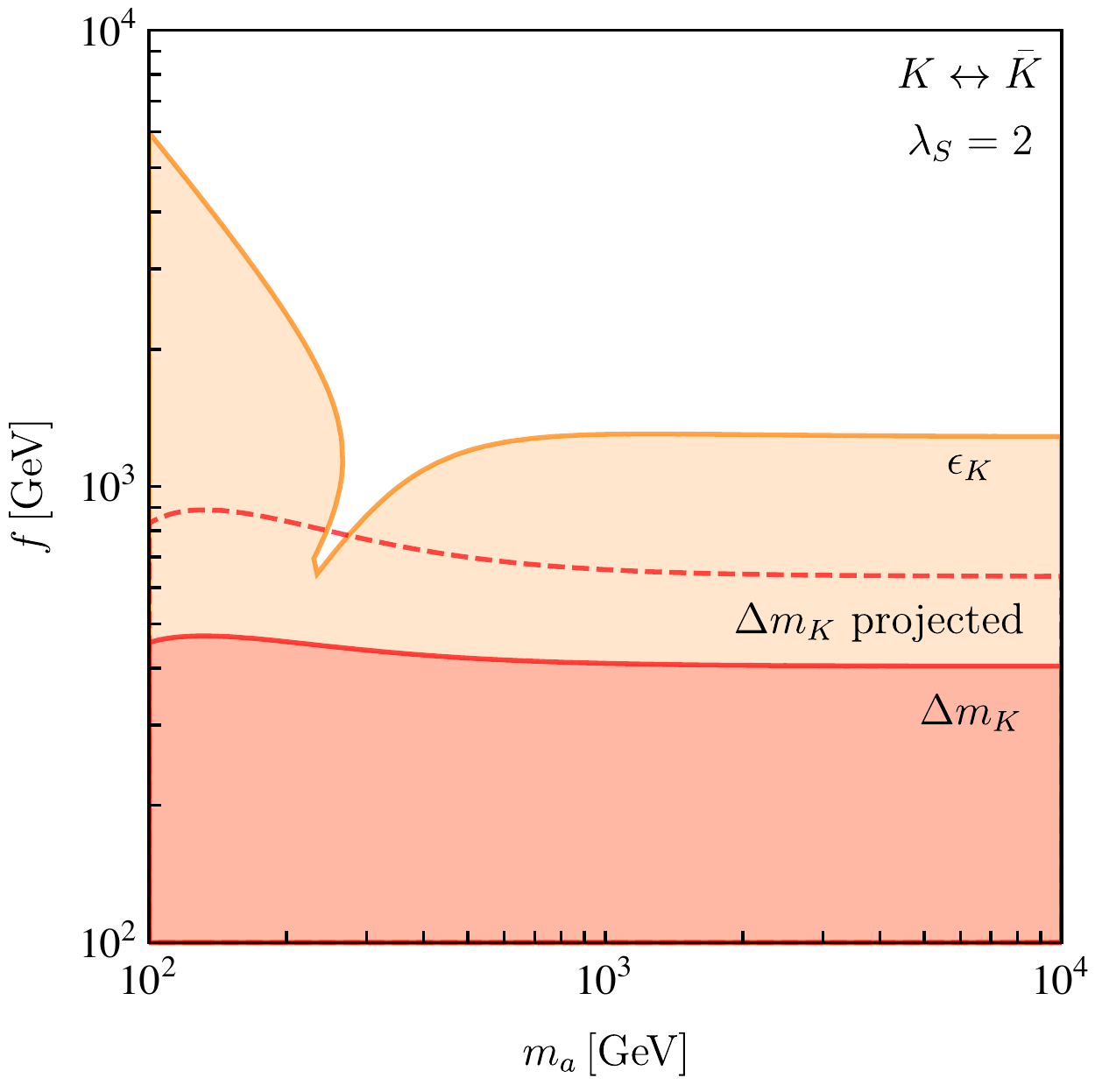}
  \hspace*{0.10\textwidth}
  \includegraphics[width=.4\textwidth]{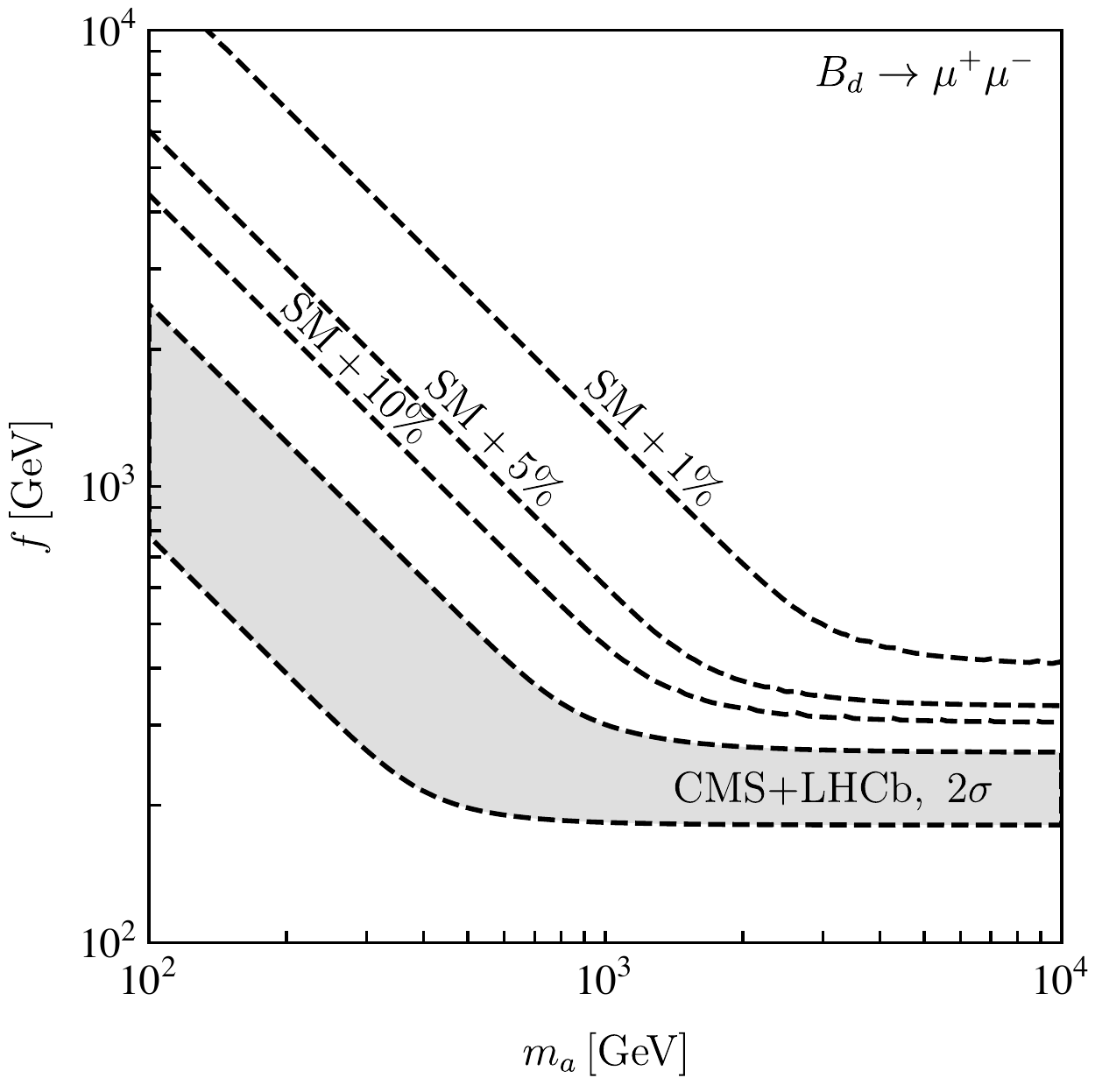}
  \caption{Left: regions in the $m_a-f$ plane excluded by flavon
    contributions to $\epsilon_K$ (orange) and $\Delta m_K$ (red) for
    our benchmark point and $\lambda_S=2$. The dashed red contour
    corresponds to the excluded region based on projected improvements
    in $\Delta m_K$. Right: parameter space where the branching ratio
    for $B_d\rightarrow \mu^+\mu^-$ stays within the $2\sigma$
    confidence interval (shaded gray), as well as contours of $1\%$,
    $5\%$ and $10\%$ enhancement with respect to the SM prediction.
    Figure from Ref.~\cite{Bauer:2016rxs}.}
\label{fig:KK} 
\end{figure}
%------------------------------------------------------------

Before we discuss potential FCC searches, we briefly review the
current and future indirect constraints in the quark and lepton
sectors. On the quark side, the flavon mass and couplings are
constrained by the non-observation of new physics in meson mixing and
semi-leptonic meson decays.  Future improvements in meson mixing
analyses are unliely to significantly change the typical current
constraints~\cite{Charles:2013aka}, and CP-violation in $K-\bar K$
mixing will remain the strongest bound. We show its impact on the
flavon parameter space in the left panel of Figure~\ref{fig:KK}. In
the semi-leptonic decay $B_d\rightarrow \mu^+\mu^-$ possible per-cent
deviations from the SM prediction could hint at a flavon, while
current limits from $B_d$ and $B_s$ decays from CMS and LHCb
measurements~\cite{Chatrchyan:2013bka,
  Aaij:2013aka,CMS:2014xfa,CMS:2014xfa} are weaker than bounds from
meson mixing. In the right panel of Figure~\ref{fig:KK}, we show the
current best fit point as well as contours of constant deviation from
flavon exchange in $B_d\rightarrow \mu^+\mu^-$ searches.

In contrast to the quark sector, the next generation of experiments
measuring lepton flavor violation will gain immense sensitivity over
the current experiments. The current bound from the radiative decay
$\mu\to e\gamma$ will be improved by an order of magnitude by the
MEG~II experiment~\cite{Baldini:2013ke}. In addition, the
DeeMe~\cite{Natori:2014yba}, COMET~\cite{Kuno:2013mha}, and
Mu2e~\cite{Abrams:2012er} experiment project an improvement of up to
four orders of magnitude in $\mu \rightarrow e$ conversion. Finally,
the bound on $\text{Br}(\mu \rightarrow 3e)$ should be improved by
five orders of magnitude by the Mu3e experiment~\cite{Kiehn:2015eta}.  To
illustrate this improvement we choose a benchmark point where the
flavor structures in the quark and lepton sectors are generated by the
same minimal parameter setup. On the one hand, this allows us to
directly assess the different experimental projections. Given that
this link is a strong assumption, we should aim at an independent
coverage of the flavon parameter space through leptonic and hadronic
observables.  In the left panel of Figure~\ref{fig:MuEConvbounds}, we
show the current constraint and impact of future limits from $\mu
\rightarrow e$ conversion for which we find the strongest future
limits. At least for relatively small flavon masses the experimental
test of the lepton sector will soon surpass the flavor physics reach.

%------------------------------------------------------------
\begin{figure}[t]
  \centering
  \includegraphics[width=.4\textwidth]{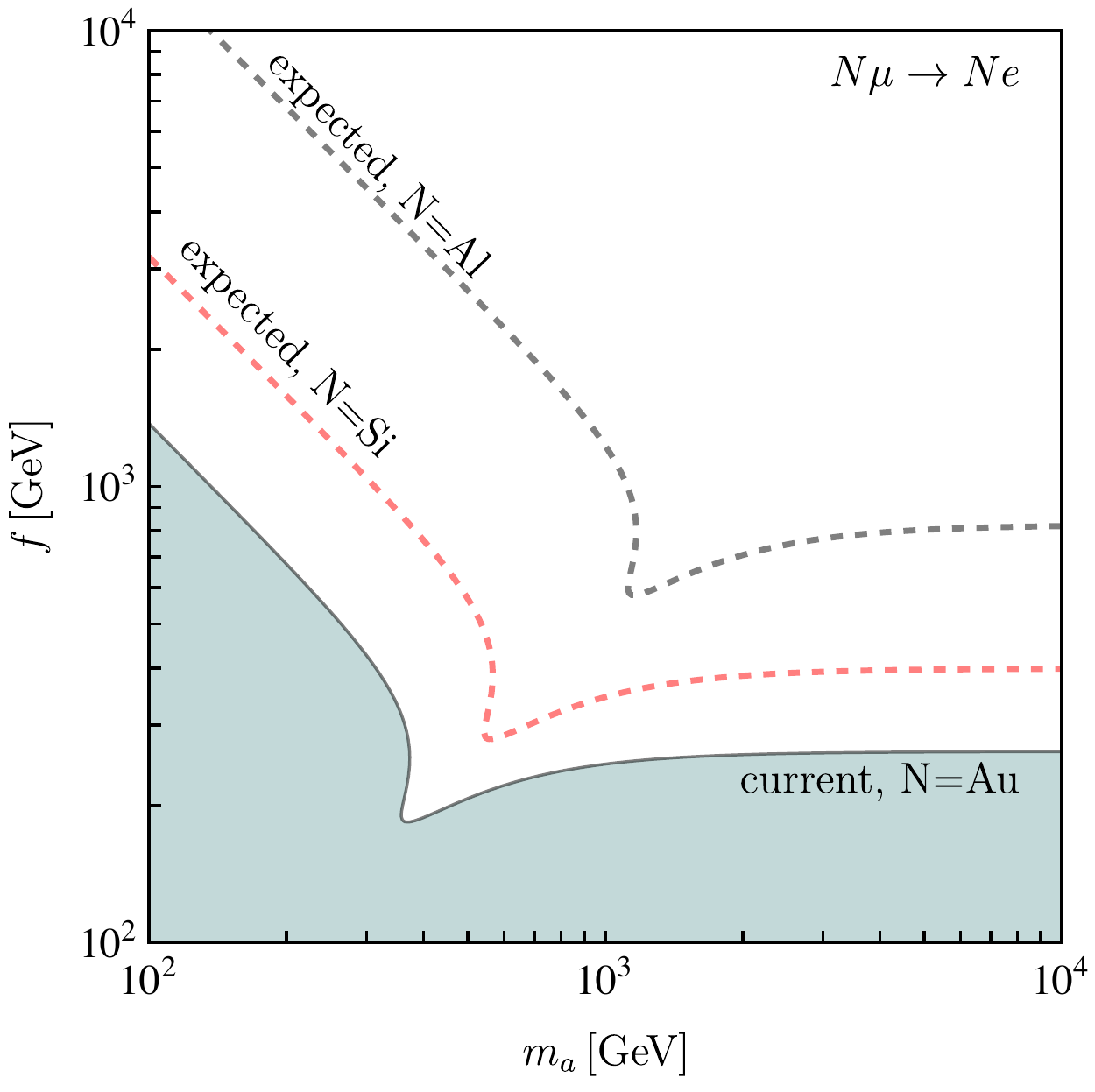}
  \hspace*{0.10\textwidth}
  \includegraphics[width=.4\textwidth]{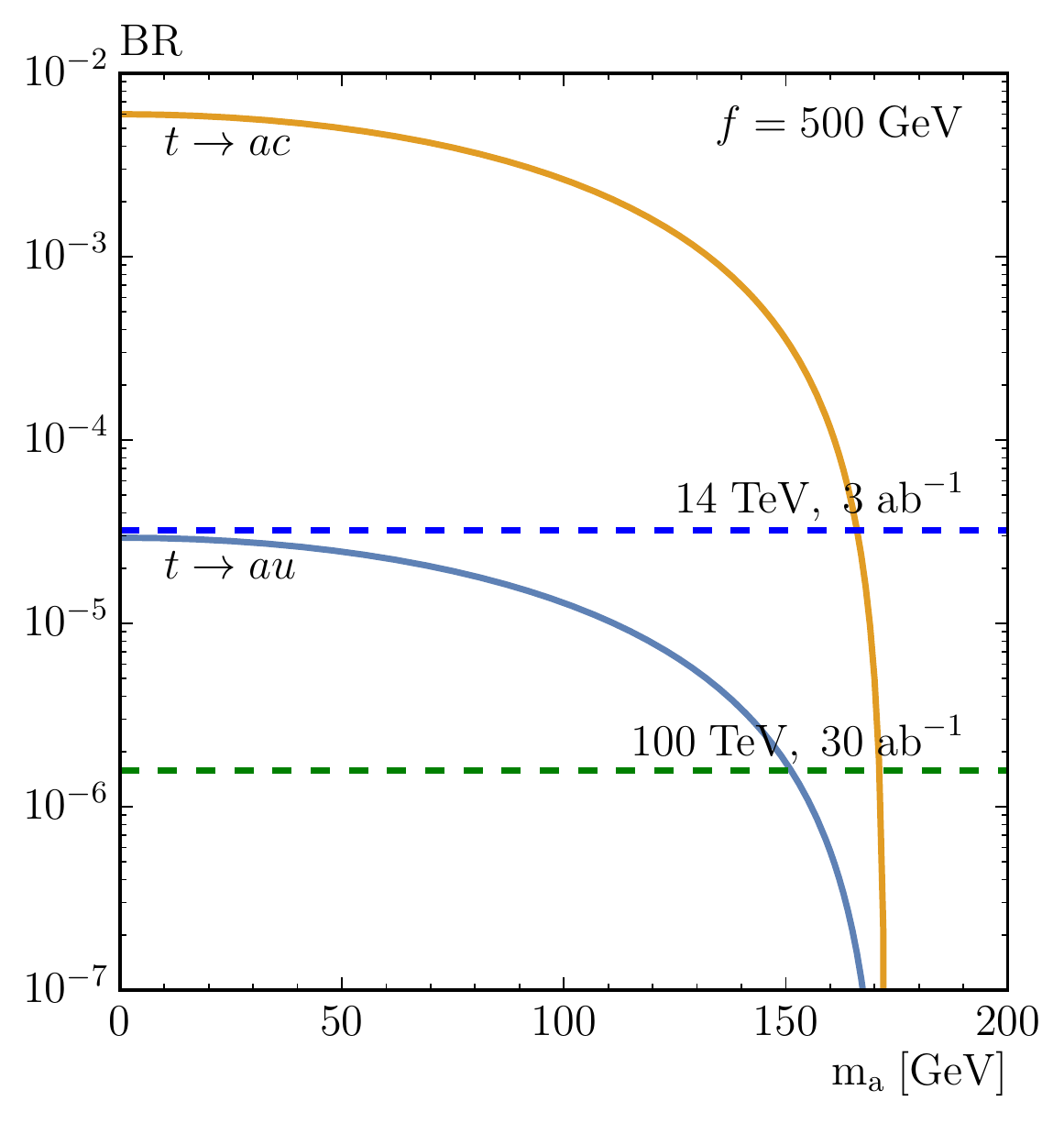}
  \caption{Left: Regions in the $m_a-f$ plane excluded by flavon
    contributions to the conversion $N \mu \rightarrow N e$. Right:
    top branching ratios into a flavon and a jet as a function of the
    flavon mass, assuming a fixed VEV of $f = 500$ GeV.  Figure 
    from Ref.~\cite{Bauer:2016rxs}.}
\label{fig:MuEConvbounds} 
\end{figure}
%------------------------------------------------------------  

Collider searches at the FCC will complement the quark flavor reach in
particular in the weak regime where the scalar flavon effects are
partly cancelled by the additional pseudo-scalar contributions. Such
searches are particularly challenging due to the absence of flavon
couplings to electroweak gauge bosons and top quarks (in our simple
setup).

For our collider signatures we rely on flavon-specific flavor
off-diagonal coupling to charm and top quarks. In the light-flavon
region, the LHC sets a bound from rare top decays, which will be
significantly improved at the FCC-hh, as discussed in the SM part of this report
 \cite{SM-report}. The corresponding branching ratios
and projected limits are shown in the right panel of
Figure~\ref{fig:MuEConvbounds}. The small production cross sections at
the LHC, shown in the left panel of Figure~\ref{fig:prod}, make it
impossible to probe flavons heavier than the top quark even at high
LHC luminosities.  Flavon production cross sections at the 100~TeV
collider are typically larger by two orders of magnitude. In
particular flavon production with heavy initial-state sea quarks
become gain relevance.  Still, we find that $s$-channel resonance
searches at 100~TeV are barely sensitive due to top and QCD
backgrounds~\cite{Bauer:2016rxs}.

We therefore propose to search for flavons in associated production
with a top quark. For the leading flavon decay $a \to t$+jet we arrive
at a same-sign top pair signature with an additional jet,
\begin{align}
  pp  \to t_\ell a \to t_\ell t_\ell \bar{c} \; ,
\end{align} 
with a partonic $gc$ initial state. It leads to two same-sign
leptons, two $b$-jets, and one additional jet. We simulate the hard
process with \textsc{Madgraph5}+\textsc{Pythia8}+\textsc{Delphes3}~
\cite{Alwall:2014hca,Sjostrand:2014zea,deFavereau:2013fsa,Cacciari:2005hq,Cacciari:2011ma}
and find a signal rate of $5.4\times
10^{-3}~\text{pb} \times (500~\text{GeV}/f)^2$ for $m_a
=500~\text{GeV}$. The irreducible SM background,
\begin{align}
pp \to b b W^+ W^+ j
\end{align}
has a leading order cross section of $5.7\cdot 10^{-7}~$pb and is
therefore negligible. Instead, we need to consider $t_{\ell}\bar{t}Z
j$ and $t_{\ell}\bar{t}W^+j$ production, with at least one leptonic
top decay and a leptonically decaying weak boson. These backgrounds
are significantly larger, $\sigma_{t_{\ell}\bar{t}W^+j} =
0.33~\text{pb}$ and $\sigma_{t_{\ell}\bar{t}Zj} = 0.48~\text{pb}$.  To
isolate the signal, we
\begin{itemize}
\item[-] require two isolated same-sign leptons with
\begin{align}
R_\text{iso} = 0.2 \,,\qquad 
I_\text{iso} =0.1 \,,\qquad 
p_{T,\ell} > 10~\text{GeV} \,,\qquad 
|\eta_\ell| < 2.5 \; ;
\end{align}
\item[-] veto a third lepton with any opposite-sign combination
  giving $|m_{\ell^+ \ell^-} - m_Z|< 15$~GeV;
\item[-] identify the hardest anti-$k_T$ jet~\cite{fastjet} ($p_T >
  40$~GeV, and $|\eta_j| < 2.5$) jet with $p_{T,j} > 100~$GeV as our
  $c$-candidate;
\item[-] require, among the non-$c$ jets, at least two $b$-tags with a parton-level $b$-quark within $R < 0.3$ and an assumed tagging efficiency 50~\%;
\item[-] require $\slashed{p}_T > 50$~GeV;
\item[-] require $m_t < m_{T2} < m_a$;
\item[-] onsider two $b$-jet charge tagging efficiencies, as described below.
\end{itemize}
Since the missing transverse momentum has to be distributed between
the flavon decay and the top decay, we define two branches by
assigning each $b$-quark to the leptons and minimizing $\Delta
R_{\ell_1b_i}+\Delta R_{\ell_2b_j}$.  We further assign the hard
$c$-jet to the top candidate with the smaller $\Delta y_{(\ell
  b),j}$. For most signal events we expect $m_t < m_{T2} < m_a$, which
allows us to search for an excess of events over the background that
provides side-bands at high value of $m_{T2}$ ~\cite{Lester:1999tx,
  Barr:2003rg}. We show the corresponding distribution in the left
panel of Figure~\ref{fig:sst_gc}.

%------------------------------------------------------------
\begin{figure}[t]
  \centering
  \includegraphics[width=0.4\textwidth]{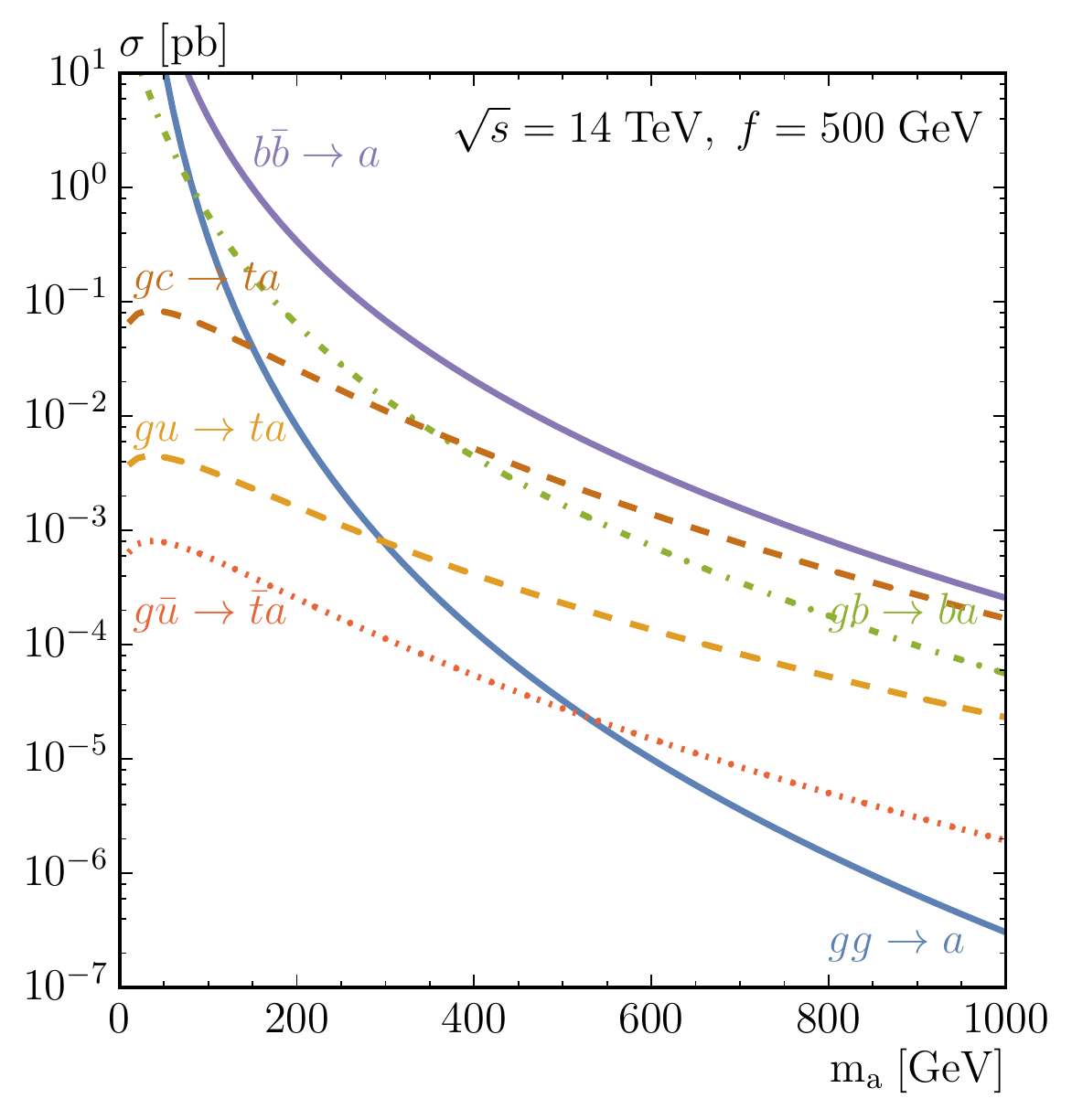}
  \hspace*{0.10\textwidth}
  \includegraphics[width=0.4\textwidth]{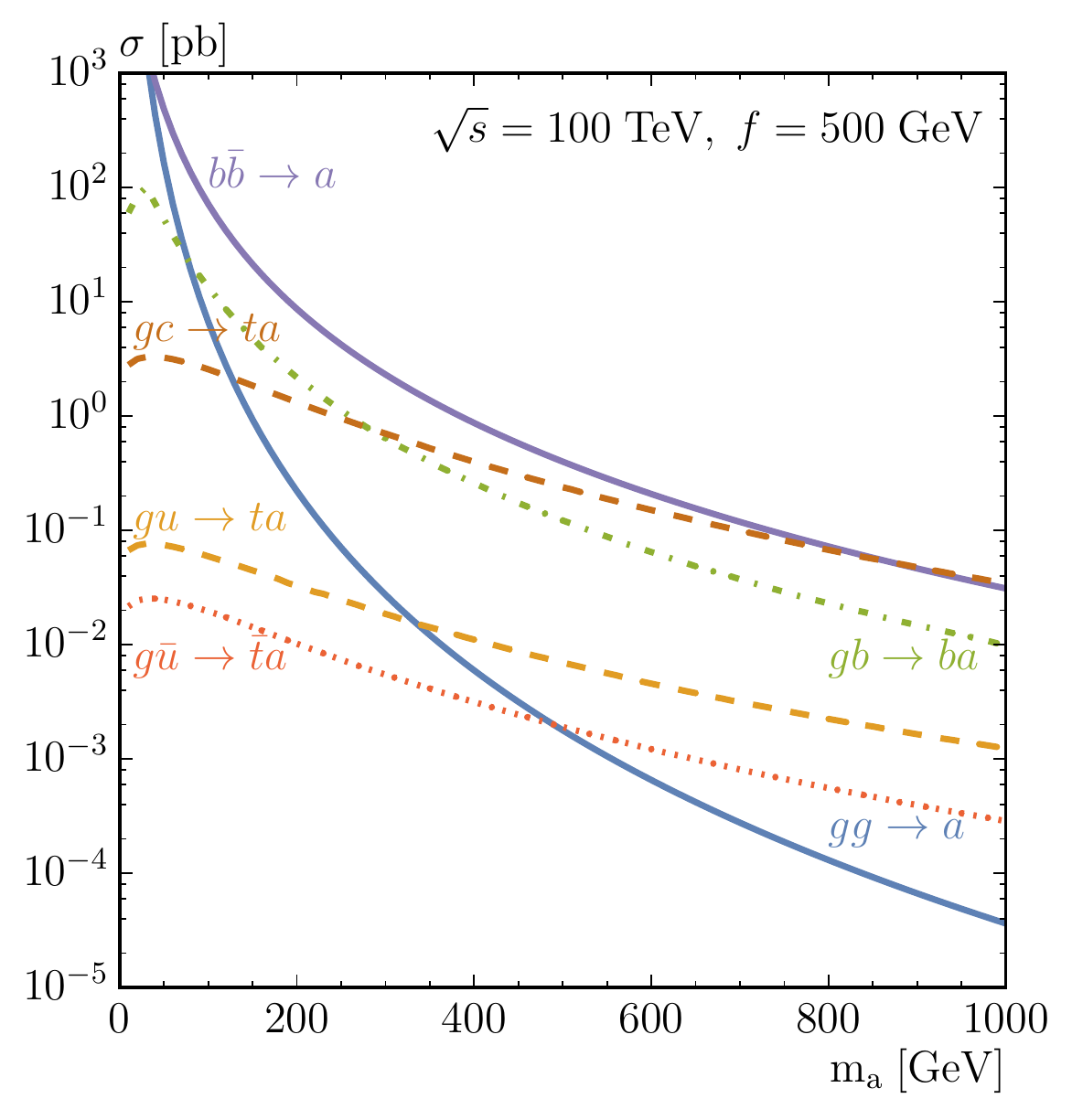}
\caption{Flavon production cross sections in the different channels
  for the $14~$TeV LHC and a $100~$TeV hadron collider using the
  \textsc{MSTW2008} PDF set~\cite{Martin:2009iq}. Couplings are
  evaluated at $\mu = m_a$ or $\mu = m_a+m_t$ with
  \textsc{CRunDec}~\cite{Schmidt:2012az}. Figure from
  Ref.~\cite{Bauer:2016rxs}.}
\label{fig:prod}
\end{figure}
%------------------------------------------------------------

A final, distinctive feature of the signal is that both leptons
originate from tops, so the two $b$-jets should be tagged with the
same charge~\cite{Alves:2006df}. Recent ATLAS
studies~\cite{ATL-PHYS-PUB-2015-040} show that a $b$-$\bar{b}$
distinction is possible with $\epsilon_S = 0.2$ and $\epsilon_B =
0.06$. For our analysis we assume two scenarios: a conservative
estimate based on these ATLAS efficiencies, and a more optimistic case
for which we assume an improved mis-tagging rate of $\epsilon_B =
0.01$ and an overall $b$-tagging efficiency of 70~\%.

%------------------------------------------------------------
\begin{figure}[t]
\centering
 \includegraphics[width=0.41\textwidth]{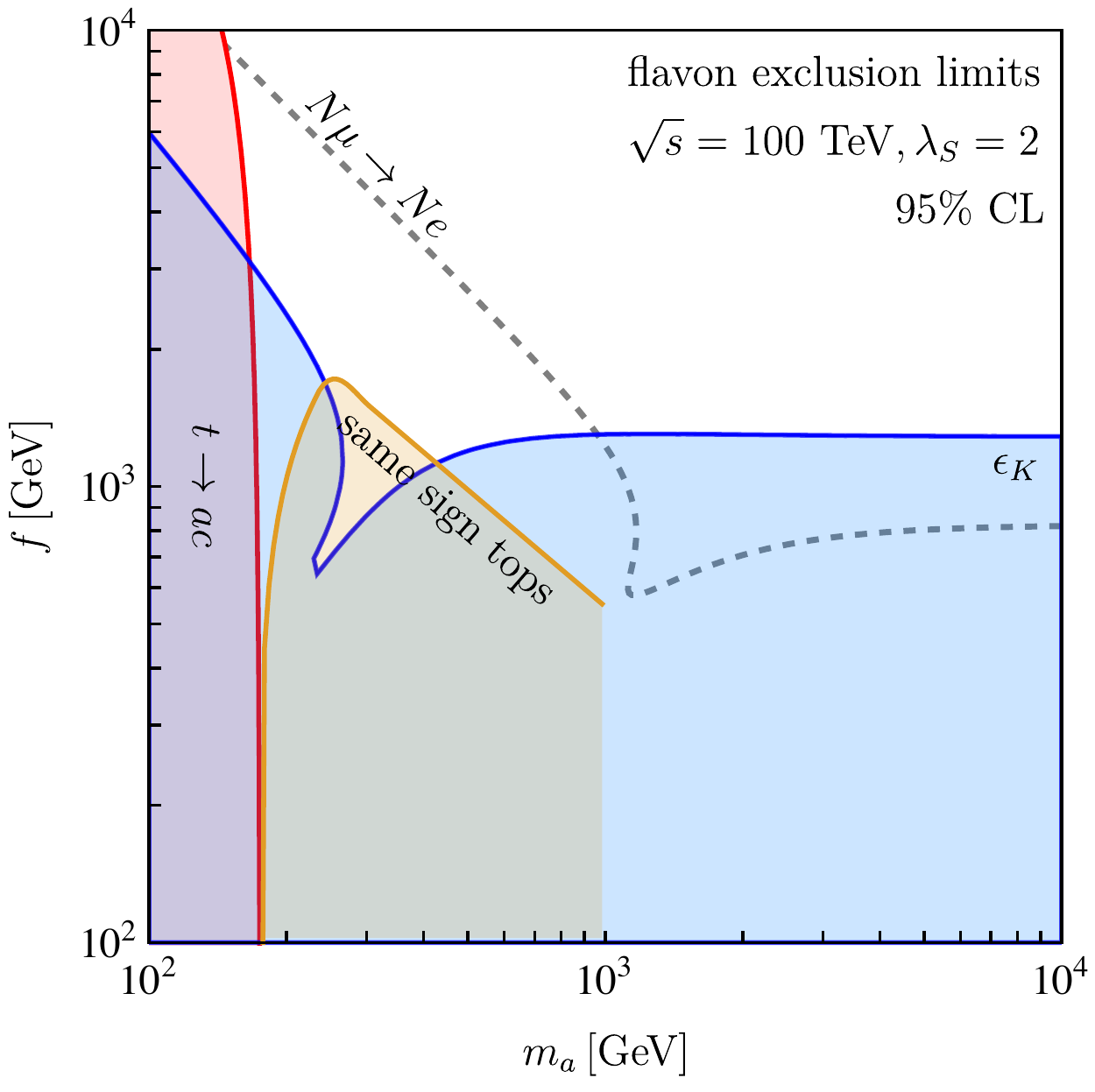}
\caption{Regions in the $m_a-f$ plane which can be probed by quark
  flavor physics ($\epsilon_K$), by lepton flavor physics ($\mu \to e$
  conversion), and by a 100~TeV hadron collider. For the latter we
  show the reach of anomalous top decays and same-sign top
  production. Figure from Ref.~\cite{Bauer:2016rxs}.}
\label{fig:sst_gc}
\end{figure}
%------------------------------------------------------------

To summarize our findings and illustrate the competitive reach of a
100~TeV collider, we show its projected 95\% CL reach in the
associated production channel in Figure~\ref{fig:sst_gc}. In addition,
we show the projected reach of indirect quark flavor and lepton
flavor experiments. We see that experiments sensitive to the quark and
lepton sectors nicely overlap in the parameter space of our
universally challenging benchmark model. The combination of direct and
indirect searches in the quark and lepton sectors will for the first
time give us the opportunity to test the dynamic nature of the flavor
structure in the Standard Model. Just like any collider search, the
FCC-hh will provide us with conclusive information about the nature of
the flavon up to TeV-scale flavon masses.

%%%%%%%%%%%%%%%%%%%%%%%%%%%%%%%%%%%%%%%%%%%%%%%%%%%%%%%%%%%%%%%%%
%%%%%%%%%%%%%%%%%%%%%%%%%%%%%%%%%%%%%%%%%%%%%%%%%%%%%%%%%%%%%%%%%

\subsubsection{$W'\to tb$ in Weak Boson Fusion}\label{Vignaroli}
%List of contributions:
%\begin{itemize}
%%\item A.~Ismail, I.~Low, M.~Low, L-T.~Wang, {\it WW Resonances in VBF} {\bf reminder sent to contributors, they never replied, so contribution will probably not appear}
%\item A.~Kotwal, {\it Study	of resonant double-Higgs production	in the vector boson fusion	process at a 100 TeV pp collider} {\bf in BSM Higgs section}
%\item K.~Mohan and N.~Vignaroli, {\it $W'\to tb$ in weak boson fusion at the FCC} {\bf contribution received}
%\end{itemize}
In this section we discuss the motivations, summarize the main results and suggest possible improvements of the study presented in ref.~\cite{Mohan:2015doa}, where a first estimate of the reach of a 100 TeV collider on a $W^{'}$ vector resonance produced via weak-boson-fusion and decaying dominantly into $tb$ was presented. As we pointed out in the previous subsection, vector resonances $V$ are a general prediction of many BSM scenarios and in particular of compelling models of Higgs compositeness \cite{Agashe:2004rs}, where they emerge from new strongly interacting dynamics which also generates the Higgs. Naturally, one thus expects a strong interaction of the vector resonances with the Higgs and the would-be Goldstone bosons, {\it i.e.} the longitudinal $W_L, Z_L$ bosons. The mass hierarchy  of the SM quarks can indeed be explained through variations in the size of the mixing of SM quarks with the strong sector. Consequently, the interaction of  vector resonances with SM light quarks is typically small. Heavier quarks, such as the top and bottom, have a sizable mixing with their composite partners in the strong sector and are partially composite particles \cite{Kaplan:1991dc}. The light generations have instead a negligible degree of compositeness. The light quark couplings to vector resonances is thus small and is inversely proportional to the $V$ coupling to the $W_L/Z_L$ bosons, which we denote as $g_V$ \cite{Contino:2006nn, Vignaroli:2014bpa, Pappadopulo:2014qza}. This implies that for larger $g_V$ couplings, corresponding to the regime of a more strongly coupled BSM dynamics, the vector resonance production via Drell-Yan, which is the main production mechanism at the LHC, is suppressed by $\sim g^{2}/g^2_V$. In the large $g_V$ regime, the alternative vector-boson-fusion (VBF) production mechanism, which is instead enhanced by $g^2_V$, becomes thus relevant (as shown in Fig.~\ref{fig:xsec}) and can allow to directly test a strongly-coupled (but yet perturbative) regime that could otherwise be difficult to test via the DY channel. The sensitivity of the VBF production, due to its t-channel nature, increases considerably with the center-of-mass energy of a $pp$ collider (Fig.~\ref{fig:xsec}). A future 100 TeV collider, as we estimated in ref.~\cite{Mohan:2015doa}, can give a unique opportunity to test a wide range of vector resonance masses for large $g_V$ coupling. As it is clear from the previous subsection, this particular choice of parameter space cannot be easily probed at the LHC \footnote{Also considering that $V$ resonances become typically broad in this regime.}, even with 3000 fb$^{-1}$ \cite{Thamm:2015zwa}. 

%Motivated by the partial compositeness scenario, we have considered a large interaction of the $W^{'}$ with $tb$. 
For our analysis of the channel in Fig.~\ref{fig:feyn2} at a 100 TeV collider, we have considered a two-site effective description of a Minimal Composite Higgs Model (MCHM) \cite{Agashe:2004rs} with partial compositeness (see ref.~\cite{Mohan:2015doa} for details on the model). The relevant terms of the Lagrangian read as follows:
%\begin{align}
%\begin{split}
\begin{equation}\label{eq:LV2}
\begin{array}{lll}
\mathcal{L}_V  &= & \displaystyle-g_2 M_{W} \cot\theta_2 W^{'+}_{\mu}W^{-\mu}h  \vspace{2mm}\\
&&\displaystyle+ i\frac{g_2}{c_W}\cot\theta_2 \frac{M^2_W}{M^2_{W'}} \left[ Z^{\mu} W^{+\nu } \left(\partial_\mu W^{' -}_{\nu} -\partial_\nu W^{'-}_{\mu} \right) \right.\vspace{2mm}\\
&&\displaystyle + Z^{\mu} W^{'  +\nu} \left(\partial_\mu W^{ -}_{\nu} -\partial_\nu W^{ -}_{\mu} \right) + W^{' + \mu} W^{ - \nu } \left(\partial_\mu Z_{\nu} -\partial_\nu Z_{\mu} \right) \left. \right. \big ] \vspace{2mm}\\
&&\displaystyle- \frac{g_2}{\sqrt{2}} \tan\theta_2 W^{' +}_{\mu}\left( \bar{q}^u_L \gamma^{\mu} q^{d}_L + \bar{\nu}_{lL} \gamma^{\mu} l^{-}_L \right)\vspace{2mm}\\
&&\displaystyle+ \frac{g_2}{\sqrt{2}}  W^{' +}_{\mu}\left( \bar{t}_L \gamma^{\mu} b_L\right)\left( s^2_L \cot\theta_2-c^2_L \tan\theta_2\right) + \text{H. c.}
\end{array}
\end{equation}
%\end{split}
%\end{align}
where $g_2= e/\sin\theta_W$, $s_W(c_W) \equiv \sin\theta_W (\cos\theta_W)$ and $q=(q^u, q^d)$ represents a doublet of the first or second generation of quarks. The parameter $s_L$ ($c_L=\sqrt{1-s^2_L}$) represents the degree of compositeness of the 3rd generation $(t_L, b_L)$ doublet. Motivated by the partial compositeness scenario, we have considered a relatively large value $s_L=0.7$. For such a value the $W^{'}\to tb$ BR is about 0.6 in the regime $g_V \gtrsim 3$, relevant to this analysis. The $\theta_2$ parameter in the Lagrangian determines the rotation which diagonalizes the mixing between a composite $W^{*}$ resonance from the strong sector and an elementary $W$ boson, which leads to the $W^{'}$ and to the SM $W$ mass-eigenstates.
$\theta_2$ controls the interactions of the vector resonances. In particular, the $V$ coupling to $W_L/Z_L$ bosons is given by $g_V = g_2 \cot\theta_2$.

\begin{figure}
\centering
\includegraphics[scale=0.4]{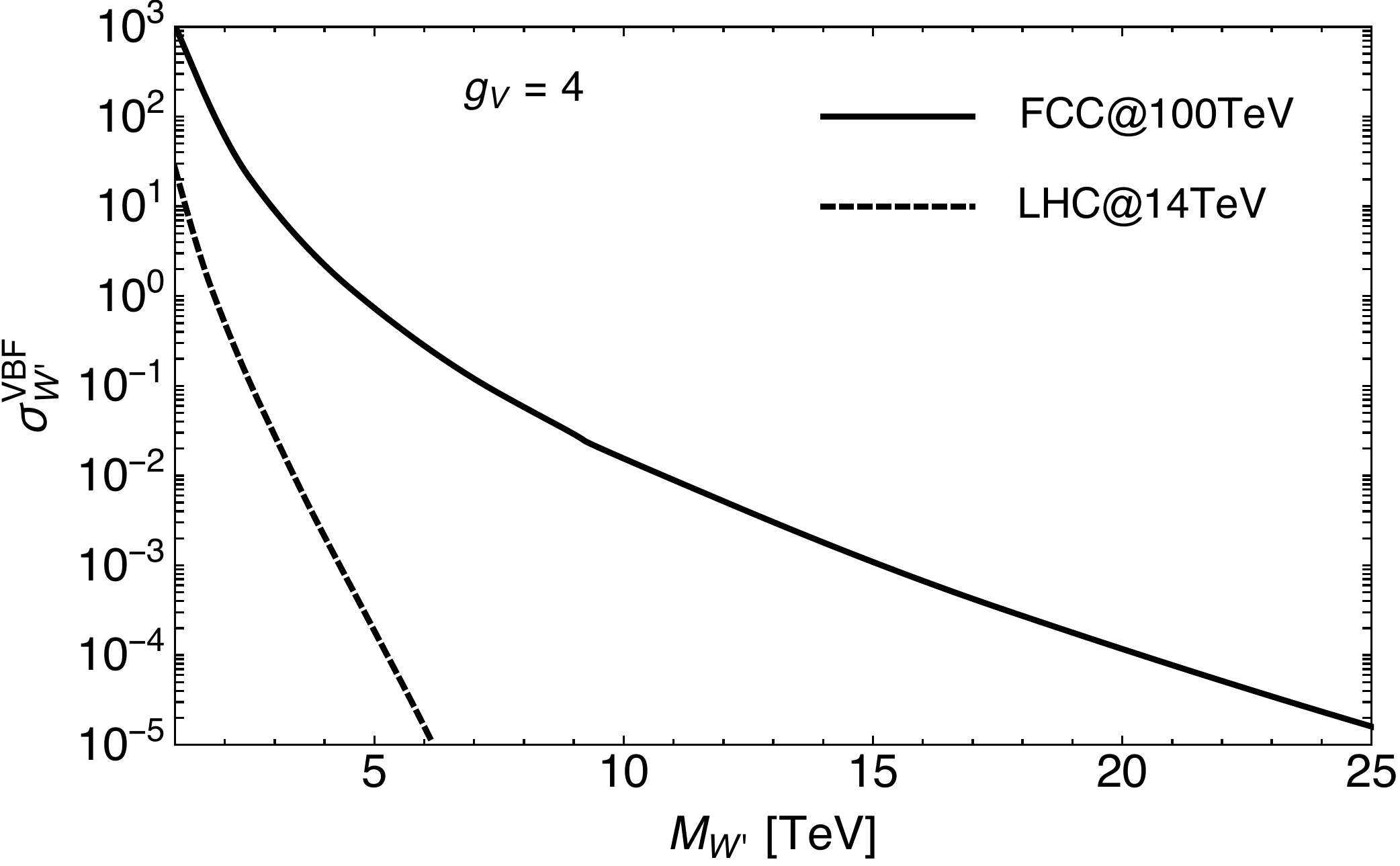} \vspace{3mm}\\
\includegraphics[scale=0.36]{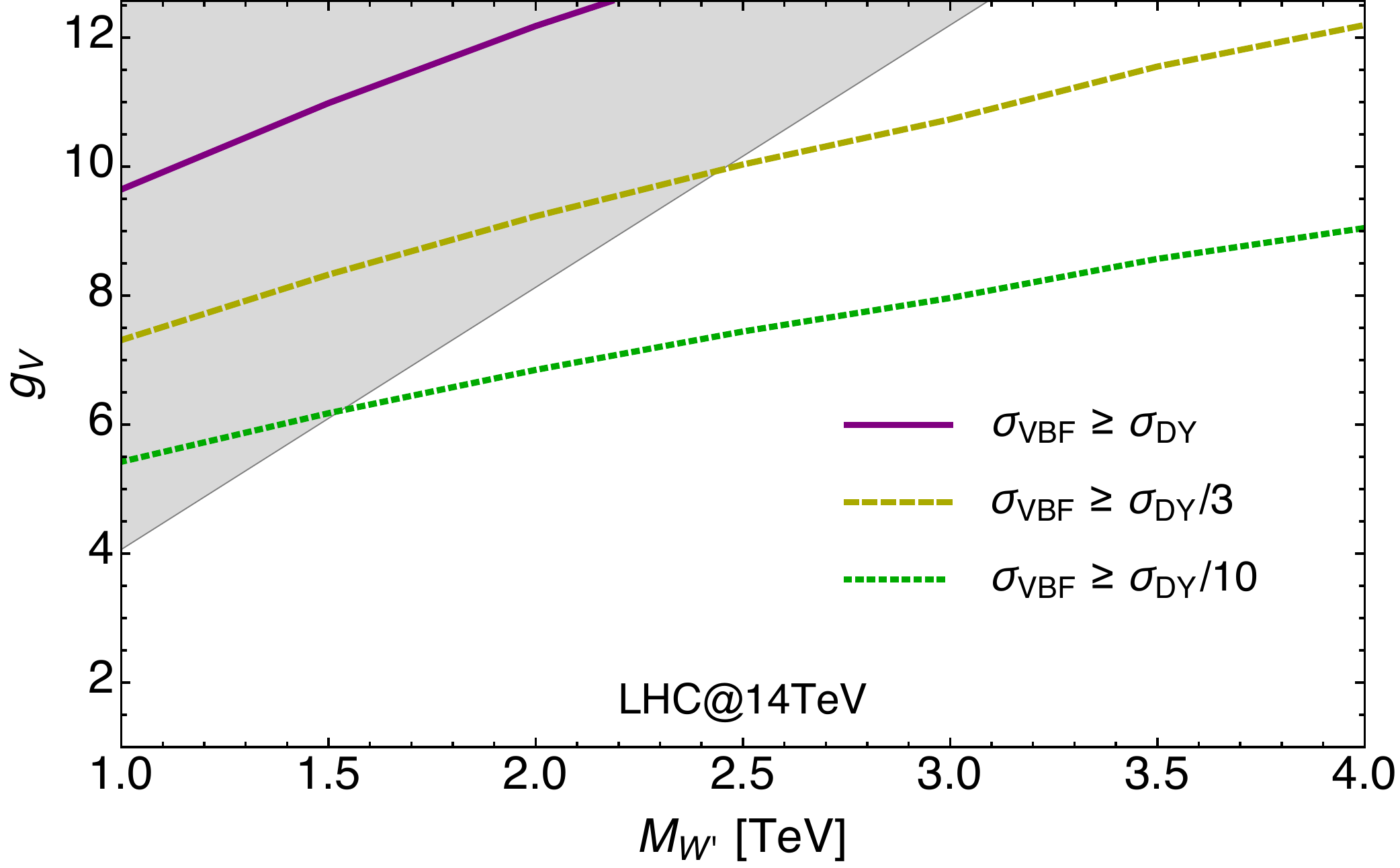} \hspace{5mm}
\includegraphics[scale=0.36]{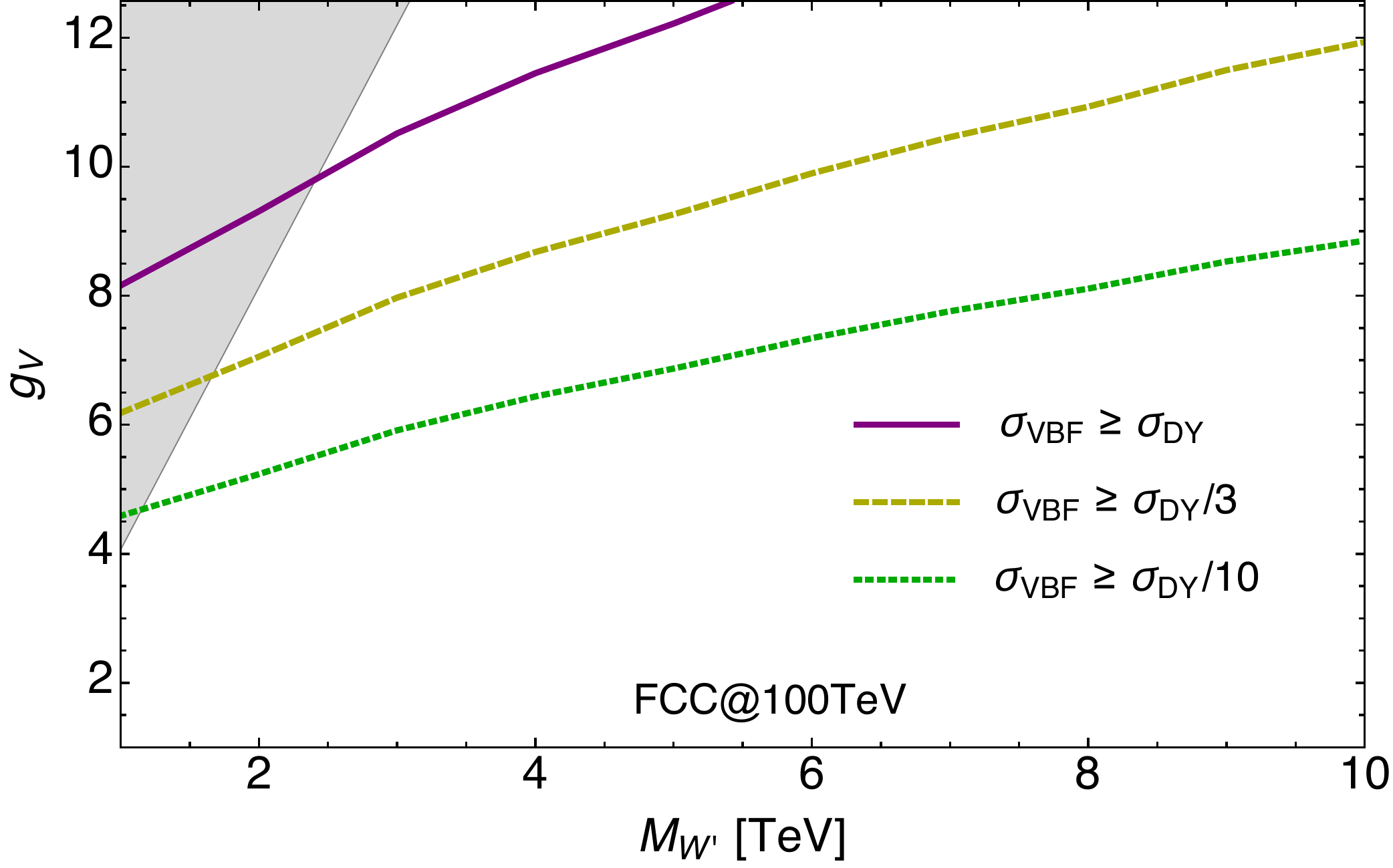} 
\caption{{\it Upper:} cross section for the $W^{'}$ VBF production at the LHC-14 (dashed curve) and at a futuristic 100 TeV pp collider (thick curve) for a coupling $g_V=4$. Cross sections scale as $g^2_V$ with the coupling. We have applied a 30 GeV cut on the jet $p_T$ and a rapidity acceptance $|\eta_j|<5$. {\it Lower left:} contours of different ratios of the VBF over DY $W^{'}$ production cross sections on the ($M_{W^{'}}$, $g_V$) parameter space at the LHC-14; {\it Lower right:} same quantity at a 100 TeV collider. The shaded areas in the upper-left corner of the parameter space correspond to values $g_V v/M_{V} > 1$ which are indicative of a theoretically excluded region (where $v/f\gtrsim 1$) in MCHM.}
\label{fig:xsec}
\end{figure}

%The above arguments are clearly shown in Fig.~\ref{fig:xsec}. The upper panel shows the cross section for the VBF production of a $W^{'}$ resonance ($W^{'+}+W^{'-}$) at the 14 TeV LHC and at 100 TeV FCC, for a fixed coupling $g_V=4$. It is evident that the VBF production takes considerable advantage of the increase in the collider center-of-mass-energy. 
%The two lower plots in Fig.~\ref{fig:xsec} show contour regions in the mass-coupling parameter space with different values of the ratio between the VBF and the DY production cross sections for a $W^{'}$ resonance. We see that VBF production gets a cross section of the same order of magnitude of the DY for $g_V \gtrsim 6$. In this large-coupling regime, VBF becomes therefore a dominant production mechanism. Further, one can take advantage of the unique topology of the VBF mechanism. The VBF signal is characterized by the presence of two forward-backward final jets that permit a clear distinction of the signal from the background even in the case of broad vector resonances, a regime which is instead difficult to explore via the DY production \cite{Pappadopulo:2014qza}.

We have performed a search analysis, based on Monte Carlo simulations, of the VBF $W^{'} \to tb$ signal depicted in Fig.~\ref{fig:feyn2} at a 100 TeV collider. Signal and background events have been simulated at LO with MadGraph 5\cite{Alwall:2011uj} and passed to PYTHIA \cite{Cacciari:2011ma} for showering and hadronization. We have also applied a smearing to the jet energy in order to mimic detector effects \cite{Kulchitsky:2000gg}. The main backgrounds include the $WWbb$, which is mainly made of $t\bar{t}$ events with a minor contribution from single-top $Wt$ events, the $Wbb$+jets and the t-channel single top $tb$+jets. This latter, which has a t-channel topology similar to the signal, represents the dominant background after applying our selection.
\begin{figure}
\centering
\includegraphics[width=0.5\textwidth]{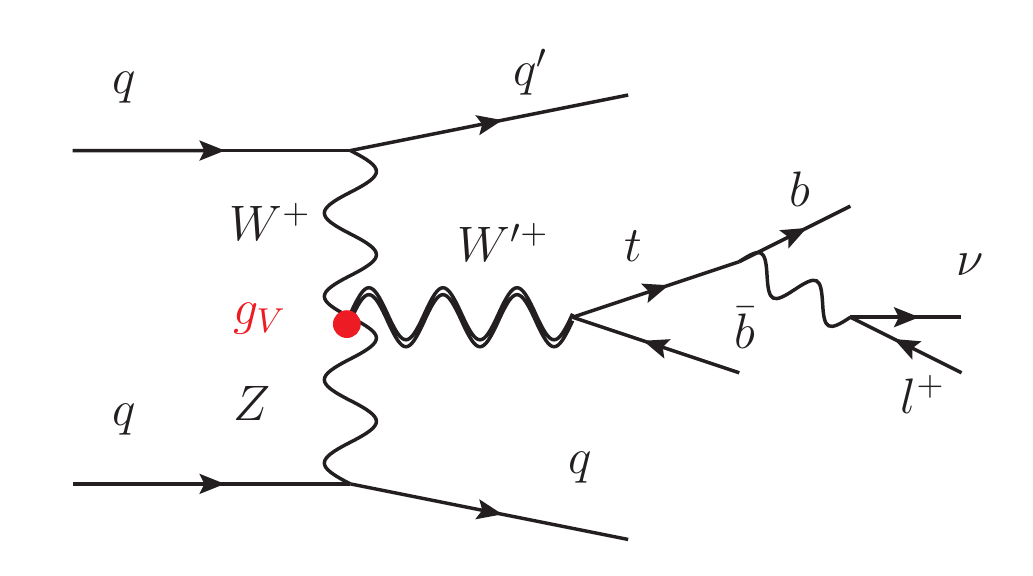} 
\caption{Feynman diagram for the VBF $W^{'} \to tb$ process. Both the $W^{'+}$ and the $W^{'-}$ processes are considered in the analysis. }
\label{fig:feyn2}
\end{figure}
We focused on the final state:
\begin{equation}
e/ \mu + n_{jet}\ \text{jets} , \qquad  n_{jet} \geq 4 \ (2\ \text{b-tag}).
\label{eq:final-state}
\end{equation}
\noindent
and we applied the following isolation criteria and $p_T$ acceptance cuts on the lepton and jets:
\begin{align}\label{eq:accept}
\begin{split}
& p_T\ j > 30 \ \text{GeV}\ , \ p_T\  l > 40 \ \text{GeV}\ , \ \Delta R(l-j)> 0.2 \ , \ |\eta_j|<5, 6  
\end{split}
\end{align}
We explored a region at high $W^{'}$ masses, where the top is boosted and, as a consequence, the lepton tends to be harder and at a lower $R$ separation from the b-jet, which also comes from the top decay. The relatively hard acceptance cut on the lepton $p_T$ has been chosen in order to obtain a better distinction from the b-jet \cite{ATLAS:2014ffa}. 

We have employed a simple search strategy which relies on the main characteristics of the signal: the distinctive VBF topology with the two final forward-backward jets emitted at high rapidity and with a large $\eta$ separation (see Fig.~\ref{fig:eta_jets_SB}) and the presence of a heavy resonance which leads to hard final states. \\
We have thus imposed a first cut on $HT_2$, defined as the scalar sum on the $p_T$ of the leading and second-leading jet, $HT_2 >800$ GeV, which already reduces significantly the background and we have then imposed a forward-backward jet tagging, by requiring that at least one signal jet had $\eta>2.5$ and at least one jet $\eta<-2.5$ and that the forward-backward jets had a rapidity separation $|\Delta\eta \ _{ FJ,BJ}|>8$. \\
We found that at a 100 TeV collider, the signal is really boosted and, for a significant fraction  of the events, the two final forward-backward jets have a rapidity larger than 6, as shown in Fig.~\ref{fig:eta_jets_SB}. We thus point out that it would be advantageous to extend the rapidity acceptance of a future $pp$ collider to the forward region up to values $\sim$6. \\
The subsequent steps of the analysis consist on a simple reconstruction procedure of the top and the bottom in the final state which allow the $W^{'}$ resonance reconstruction. We have thus imposed a bound on the reconstructed $W^{'}$ invariant mass, $m_{W^{'}}$, and on the $p_T$ of the top and of the bottom: 
%\begin{table}[t]
%\begin{center}
%{\small
\begin{equation}
\label{tab:cut-final}
\begin{tabular}[]{c|ccccc}
% ($M_{W^{'}}$ (TeV), $g_V$) & (2, 4) & (3, 4) & (4, 4\&8) & (5, 8) & (6, 12) \\
$M_{W^{'}}$ (TeV) & 2 & 3 & 4 & 5 & 6 \\
 \hline
  && && & \\[-0.3cm]
$m_{W^{'}}\ >$  (TeV) & 1.5 & 2.5 & 3.5 & 4.0 & 5.0 \\
$p_{T} \ b, t \ >$  (TeV) & 0.75 & 0.9 & 1.5 & 1.5 & 1.5
\end{tabular}
\end{equation}
%}
%\caption{ \small .
%\label{tab:cut-final}}
%\end{center}
%\end{table}

\begin{figure}
\centering
\includegraphics[width=0.6\textwidth]{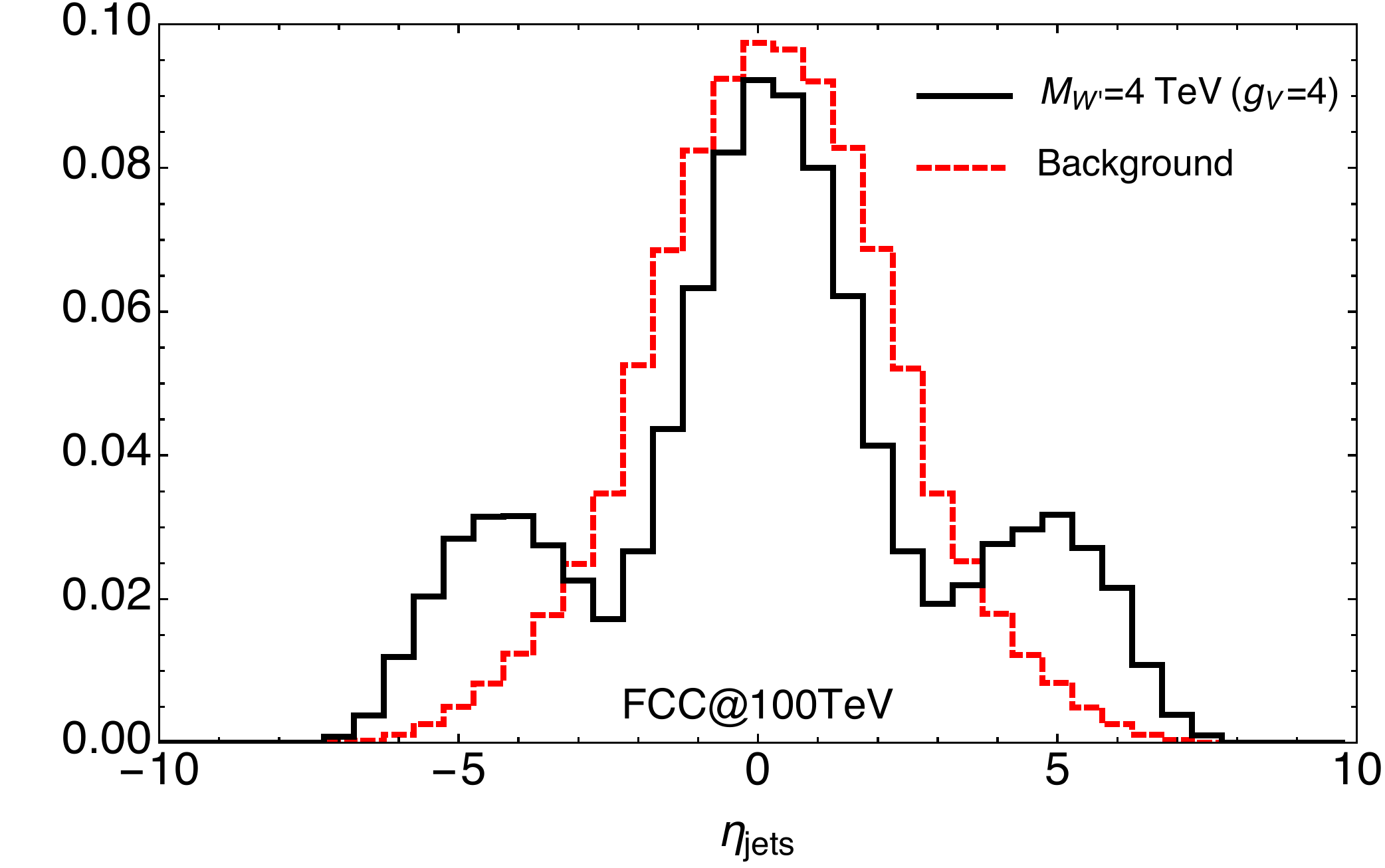} 
\caption{Normalized rapidity distribution of all of the final jets which have passed the acceptance requirements in eq. (\ref{eq:accept}), with the exception of the $|\eta_j|$ restriction, for the total background (red dashed curve) and the signal with $m_{W^{'}}=4$ TeV, $g_V=4$ (black curve) at a 100 TeV collider. }
\label{fig:eta_jets_SB}
\end{figure}
\begin{table}[]
\begin{center}
%{\small
\begin{tabular}[]{c|cccc}
%\multicolumn{1}{c|}{}&\multicolumn{1}{c}{}&\multicolumn{1}{c}{ } &\multicolumn{1}{c}{ } &\multicolumn{1}{c}{}  &\multicolumn{1}{c}{} &\multicolumn{1}{|c}{}  \\
\multicolumn{1}{c|}{{\sf 100 TeV}} & \multicolumn{2}{c}{signal} &  \multicolumn{2}{c}{bckg} \\
  & $|\eta_j|<5$ & $|\eta_j|<6$ & $|\eta_j|<5$ & $|\eta_j|<6$ \\
 \hline
% && && \\
 ($M_{W^{'}}$ (TeV), $g_V$) & & & &\\
%&&&&  \\
(2, 4) & 0.56 & 1.1 & 70 & 100\\ 
(3, 4) & 0.13 & 0.25 & 31 & 45 \\
(4, 4) & 0.022 & 0.042 & 4.8 & 7.2 \\
(4, 8) & 0.082 & 0.15 & 4.8 & 7.2 \\
(5, 8) & 0.028 & 0.051 & 3.6 & 4.9 \\
(6, 12) & 0.013 & 0.022 & 1.4 & 1.8 \\
\hline  
\end{tabular}
%}
\caption{Signal and background cross sections, in fb, at a 100 TeV collider after the complete selection.
\label{tab:final-100}}
\end{center}
\end{table}
The final results of our selection are shown on Table \ref{tab:final-100}. We used these results to extract the discovery/exclusion reach on the $W^{'}$ (mass, coupling) parameter space of a 100 TeV collider.
\begin{figure}
\centering
\includegraphics[scale=0.36]{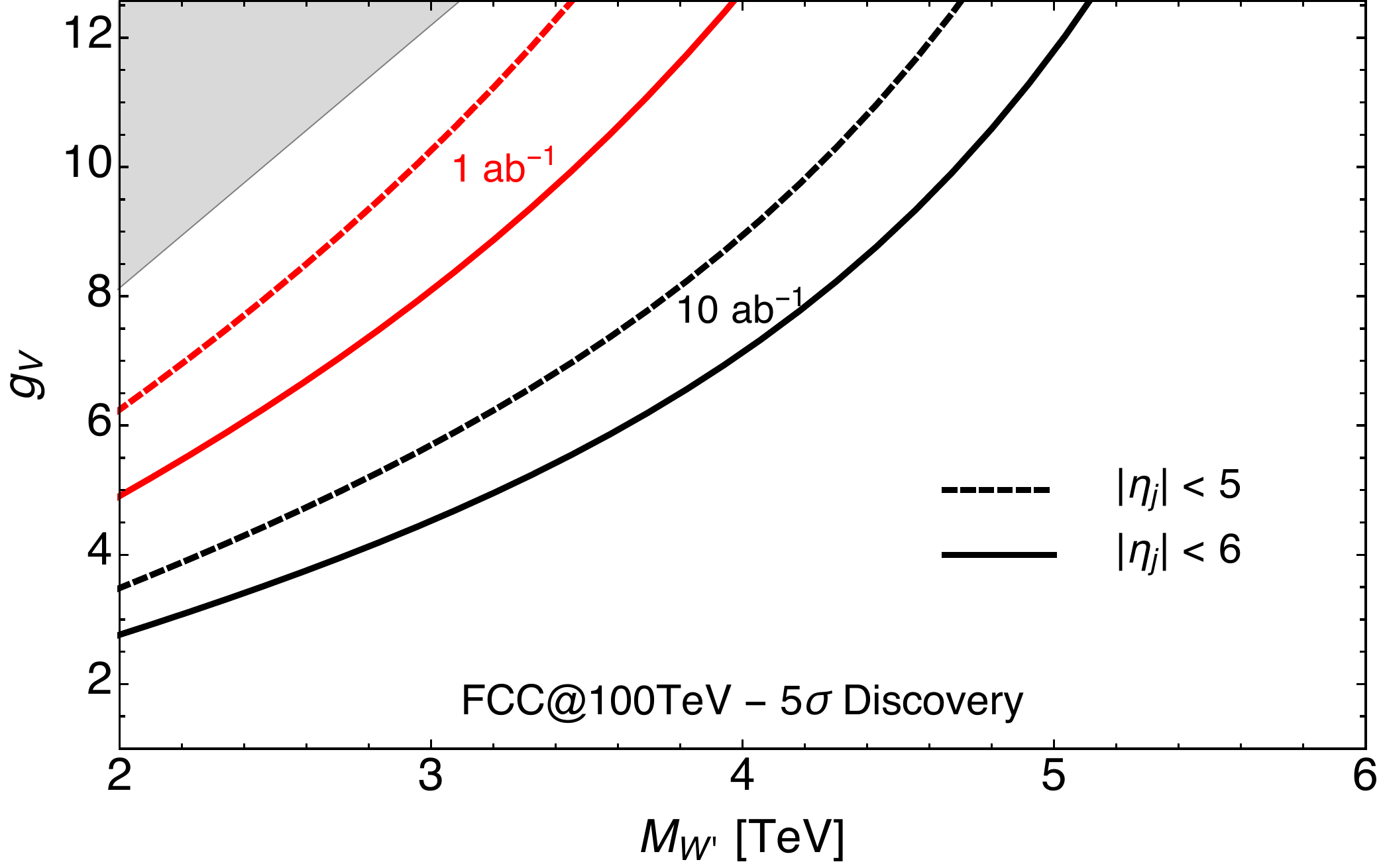} \hspace{5mm}
\includegraphics[scale=0.36]{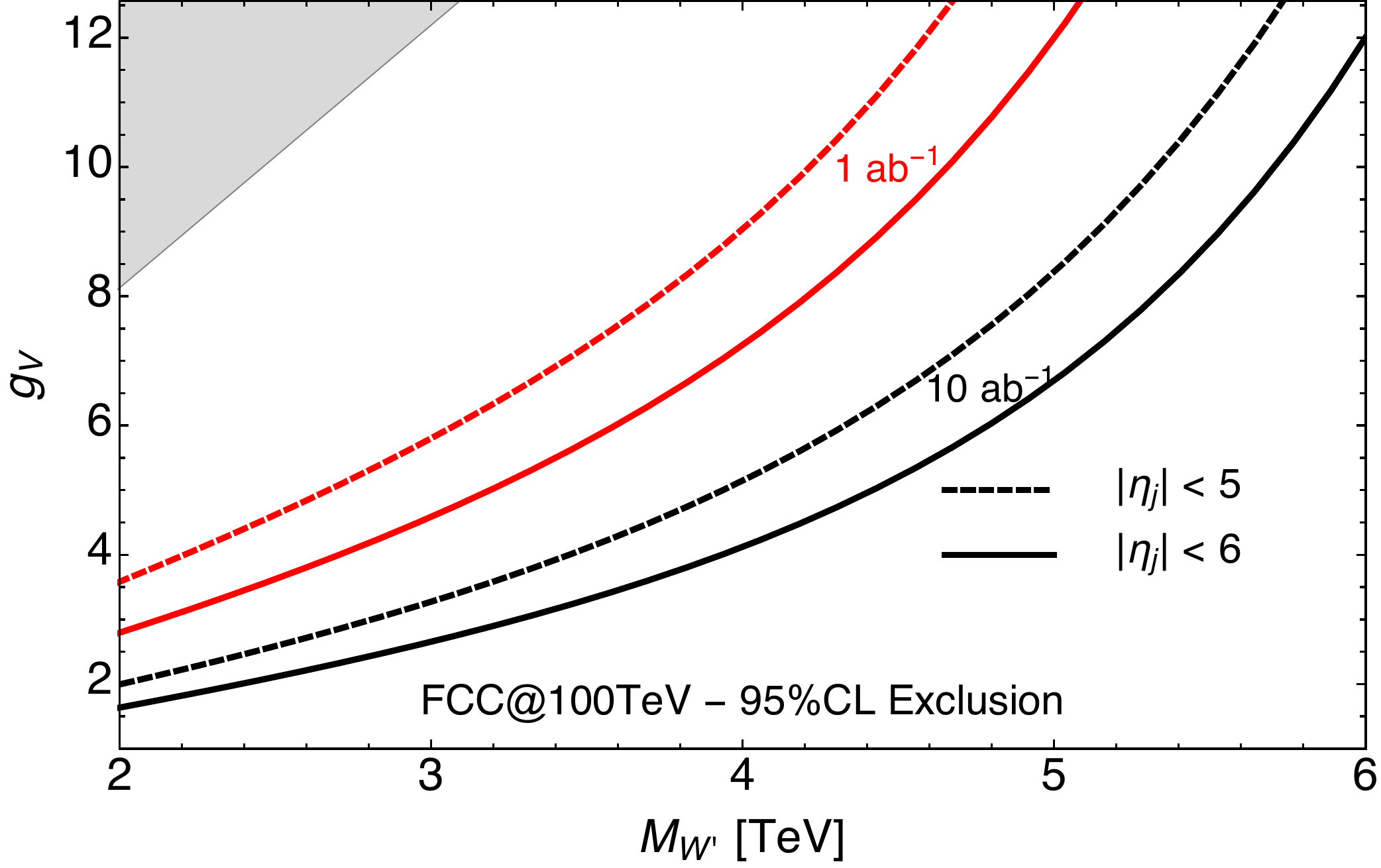} 
\caption{\small The 100 TeV pp collider reach, with 1 and 10 ab$^{-1}$, on a $W^{'}$ produced via VBF in the $tb$ channel. {\it Left:} 5$\sigma$ discovery potential. {\it Right:} 95\% CL exclusion reach. The continuous (dotted) curves are obtained for a jet rapidity acceptance $|\eta_j|<$ 6 (5). The shaded areas in the upper-left corner of the parameter space correspond to values $g_V v/M_{V} > 1$ which are indicative of a theoretically excluded region (where $v/f\gtrsim 1$) in MCHM. }
\label{fig:reach-100}
\end{figure}
We found that while the 14 TeV LHC can access only a small portion of the MCHM parameter space (the high-luminosity LHC, with 3 ab$^{-1}$ can exclude a $W^{'}$ vector resonance up to about 2.1 TeV),
a future 100 TeV $pp$ collider has a much wider sensitivity. The left plot in Fig.~\ref{fig:reach-100} shows that at a 100 TeV collider a 5$\sigma$ discovery is achieved for a $W^{'}$ in the VBF channel with masses up to 5.1 (4) TeV with 10 (1) ab$^{-1}$ of integrated luminosity in the large $g_V$ coupling region. The exclusion potential of a 100 TeV collider, as shown in the lower plot of Fig.~\ref{fig:reach-100}, extends up to $W^{'}$ masses of 6.1 (5.1) TeV with 10 (1) ab$^{-1}$. These values refer to a jet-rapidity acceptance $|\eta_j|<6$. As we anticipated, we found that the reach of a 100 TeV collider is significantly enhanced, by about a 10\% in the $W^{'}$ mass reach, if the rapidity acceptance on the jets can be increased from 5, the present LHC rapidity coverage, up to 6. 

In conclusion, we have shown that a future 100 TeV $pp$ collider offers the possibility to test, using the VBF channel, a wide range of a vector resonance mass and coupling parameter space and, most importantly, that the reach of a 100 TeV collider in VBF extends up to the more strongly-coupled regime of composite Higgs models, which is not within the reach of LHC.\\
Motivated by MCHM with partially composite 3rd generation quarks, we have focused on the $W^{'} \to tb$ channel and have outlined a simple signal-selection strategy which mainly relies on the distinctive VBF topology. Several improvements and extensions to our analysis are possible. Firstly, one could further exploit the boosted nature of the final states to apply more refined top-reconstruction techniques (as for example explained in Section \ref{Auerbach}). It could be also interesting, in this case, to consider a different final state, for example a totally hadronic final state, which would lead to a larger signal cross section compared to the semileptonic channel considered in our analysis. A natural extension of our study is then to consider different $W^{'}$ decay channels. In the case, for example, where the top degree of compositeness is relatively small ($s_L\lesssim 0.5$) the dominant decay channels become $W^{'}\to WZ/Wh$. Finally, in our analysis we have neglected the possible contributions of vector-like quarks, which could lead to spectacular new signatures \cite{Vignaroli:2014bpa} at a 100 TeV collider.  

%\subsubsection{Resonances in $VH$ final states ---- To be removed}

%\subsubsection{Resonances in other di-boson channels ($Z\gamma,W\gamma,HH,\ldots$) --- To be removed}

\subsubsection{Photon Cascade Decay of the Warped Graviton}\label{CYChen}
The warped Randall-Sundrum models, where Standard Model gauge fields~\cite{Davoudiasl:1999tf,Pomarol:1999ad} and fermions~\cite{Grossman:1999ra} are allowed to propagate in the  5-dimensional bulk, provides an excellent framework to address both the Planck-weak and flavor hierarchy problems of the SM~\cite{Gherghetta:2000qt,Huber:2000ie,Huber:2003tu}. 
In this class of models, the Kaluza-Klein (KK) gravitons (as well as gauge and fermion KK states) mostly interact with heavy SM fields such as the top quark, Higgs, and longitudinal modes of $Z/W$ gauge bosons because all the particles involved in the relevant couplings are localized near the TeV/IR brane.
Hence, once produced, they give rise to interesting collider signatures, and various aspects of collider phenomenology (mostly in the context of the LHC) have been studied in, for example, in Refs.~\cite{Davoudiasl:2000wi,Agashe:2006hk,Lillie:2007yh,Fitzpatrick:2007qr,Agashe:2007zd,Lillie:2007ve,Djouadi:2007eg,Agashe:2007ki,Antipin:2007pi,Agashe:2008jb,Davoudiasl:2009cd,Chen:2014oha}.

At the same time, future colliders beyond the LHC could not only offer discovery opportunities for the KK particles but also allow their precision studies. 
Given this situation, it is opportune and interesting to examine potential signals from KK particles in more detail and explore other non-conventional decay modes especially if they can provide independent information on the underlying model parameters. 
With this goal in mind, we investigate here a novel signature coming from the cascade decay of KK gravitons, i.e., those with final states having other KK particles. 
We particularly focus on the ``photon cascade'' decay mode, for which the original study was presented in Ref.~\cite{Agashe:2014wba},
\begin{eqnarray}
p p \rightarrow G_1 \rightarrow \gamma_1 \gamma\,, \label{eq:channel}
\end{eqnarray}
where $G_1$ and $\gamma_1$ denote the first KK states of the graviton and photon, respectively. 
This channel could potentially be a good cross-check of the conventional search channels accompanying two heavy SM states. 
Moreover, in the post-discovery and precision study phase, the specific dependence of the rate on the volume factor for the process in eq.~\eqref{eq:channel} -- which differs from that in conventional ones -- may enable us to extract the underlying model parameters {\it separately} in conjunction with information from other channels.
In this context, we emphasize that, in general, this new decay mode is {\it complementary} to the ones that were previously investigated.

%%%%%%%%%%%%%%%%%%%%%%%%%%%%%%%% 
%%%%%%%%%%%%%%%%%%%%%%%%%%%%%%%%
%%%%%%%%%%%%%%%%%%%%%%%%%%%%%%%%
%\subsection{The model \label{sec:model}}
%%%%%%%%%%%%%%%%%%%%%%%%%%%%%%%%
%%%%%%%%%%%%%%%%%%%%%%%%%%%%%%%%
%%%%%%%%%%%%%%%%%%%%%%%%%%%%%%%%
A rough sketch of particle profiles in the model under consideration is given as follows: 1) {\it all} KK states are localized near the TeV/IR brane, 2) left-handed top and bottom quarks are either localized near TeV brane or have a (roughly) flat profile; for concreteness we choose this latter option, with the $t_R$ localized near the TeV brane, 3) SM photon and gluon have flat profiles, and 4) light SM fermions are localized near the UV brane (to suppress their couplings with the Higgs).
We then find the following features of the KK graviton coupling, in particular, for their production and decay: i) the couplings to $t_R/h/W_L/Z_L$ are the largest, i.e., $\mathcal{O}(1)$, ii) the coupling to the SM gluon, relevant for production, is suppressed by a ``volume factor'' $\sim 1/(k\pi R)$, and iii) the coupling of $\gamma_1$ with $\gamma$ is ``in-between'' the previous two. 
Here, $k$ and $R$ are the curvature scale and the compactification radius of the RS background, respectively; the Planck-weak hierarchy is generated for $k R \simeq 11$. 

The couplings relevant to vertices $G_1gg$ and $G_1\gamma_1\gamma$ have been studied in Ref.~\cite{Davoudiasl:2000wi}, and it turns out that they are proportional to $1/(k\pi R)$ and $1/\sqrt{k\pi R}$. 
Therefore, the total rate for $gg\rightarrow G_1 \rightarrow WW,\,ZZ$ etc., goes like $\sim 1/(k\pi R)^2$, whereas that for $gg\rightarrow G_1 \rightarrow \gamma_1\gamma$ scales with $\sim 1/(k\pi R)^3$. 
Roughly speaking, we then anticipate that the ratio between conventional channels and photon cascade channel should scale as $k\pi R$, enabling us to extract information on the 5D parameter $kR$. 
Hence, a measurement of the process proposed here after potential discovery made in the conventional heavy SM channels can shed light on the parameters of warped models. 

%\subsection{Signal, background, and cuts}
The signal process that we consider is $pp \rightarrow G_1 \rightarrow \gamma_1\gamma$ followed by $\gamma_1 \rightarrow W^+W^- \rightarrow (jj)\ell\nu$.
We adopted the KK photon couplings and model parameter values in Ref.~\cite{Agashe:2007ki}, which motivated the choice BR$(\gamma_1 \rightarrow W^+W^-)\simeq 0.44$ as a typical value. 
Basically, the signal process is characterized by a three-step cascade decay of $G_1$ into $\gamma$, (hadronic) $W$, and $\ell$ ($e$ or $\mu$) along with an (invisible) neutrino:
\begin{equation}
G_1\rightarrow \gamma \gamma_1 \rightarrow \gamma W^{\pm} W^{\mp} \rightarrow \gamma W^{\pm}(\rightarrow jj) \ell^{\mp} \nu
\end{equation}
Since all visible particles (including the hadronic $W$) are fully distinguishable, many distinctive kinematic features can be easily applied without any combinatorial issues. 
We first expect a hard photon due to a sizable mass gap between $G_1$ and $\gamma_1$. 
A large mass hierarchy between $\gamma_1$ and $W$ again enables us to have hard jets, lepton, and missing transverse momentum. 
In particular, the $W$ is so significantly boosted that the two jets are highly collimated, demanding us to employ the boosted object techniques to tag merged two-prong $W$ jets while suppressing the single-prong QCD jet background.   

On top of the cuts motivated by the hardness of final state objects, the two invariant masses defined by $G_1$ and $\gamma_1$ and the associated invariant mass cuts play a crucial role in suppressing backgrounds. 
In order to reconstruct them, we first obtain the energy-momentum of the invisible neutrino using the missing transverse momentum constraint and $\nu/W$ mass-shell conditions. 
Although there arise two solutions, we consider the event of interest as accepted in our analysis if either of the resulting solutions pass the two invariant mass window cuts:
\begin{eqnarray}
m_1^{\gamma}-\Gamma_1^{\gamma}< &m_{\ell jj \nu}& < m_1^{\gamma}+\Gamma_1^{\gamma}\,,  \label{eq:masswindowkka}\\
 m_1^{G}-\Gamma_1^{G}< &m_{\gamma \ell jj \nu}& < m_1^G+2\Gamma_1^G\,, \label{eq:modifiedmasswindowkkg}
\end{eqnarray}
where $\Gamma_1^G$ and $\Gamma_1^{\gamma}$ are the widths of $G_1$ and $\gamma_1$, respectively. 
Here the asymmetric form of the latter criteria considers the possibility of skewed Breit-Wigner distributions for KK gravitons having a large $c\, (\equiv k/\bar{M}_P)$ parameter. More quantitatively, it turns out that the KK graviton with a large $c$ value (e.g., $\sim 2$) has a particle width larger than $\sim 20\%$ of its mass. 
Moreover, the $G_1gg$ coupling emerges from a dim-5 operator (in turn, due to its spin-2 nature) which grows with energy, and therefore, larger $m_{\gamma \ell jj \nu}$ values are preferred.
In particular, at higher energy colliders such as a 100 TeV collider, the KK graviton mass of a few TeV is in the regime of low $x$, so that we expect that the associated invariant mass distribution is not significantly affected by the gluon parton distribution function, i.e., the skewness becomes manifest. 
The left panel of Fig.~\ref{fig:fig1} demonstrates the shift of the peak position for three different KK photon masses with the $c$ parameter fixed to 2. We observe that the peak position is shifted by about half $\Gamma_1^G$ in all cases, comparing with relevant theory expectations denoted by the black dashed vertical lines.  

\begin{figure}[t]
\centering
\includegraphics[width=7.5cm]{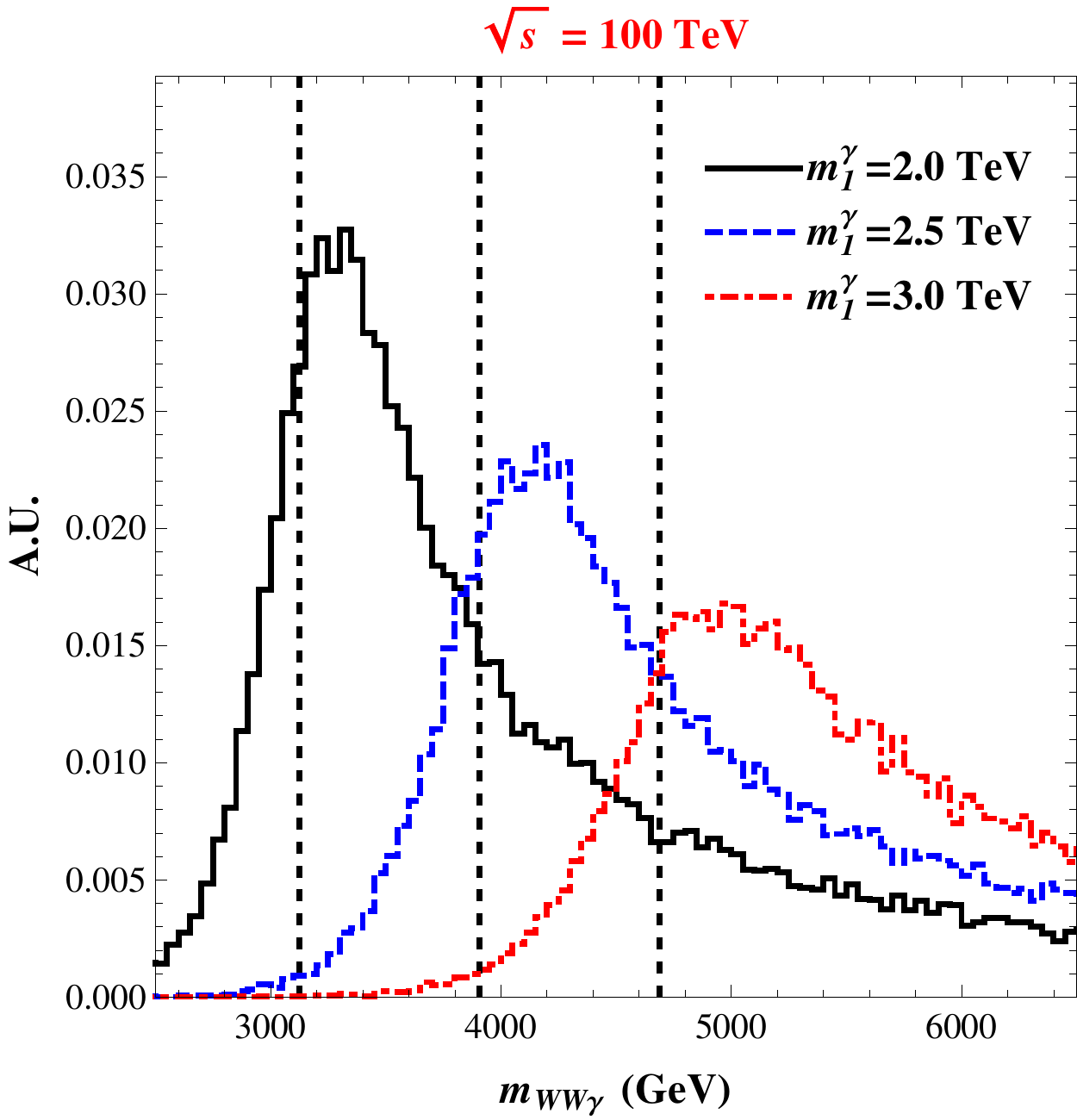} \hspace{0.2cm}
\includegraphics[width=7.5cm]{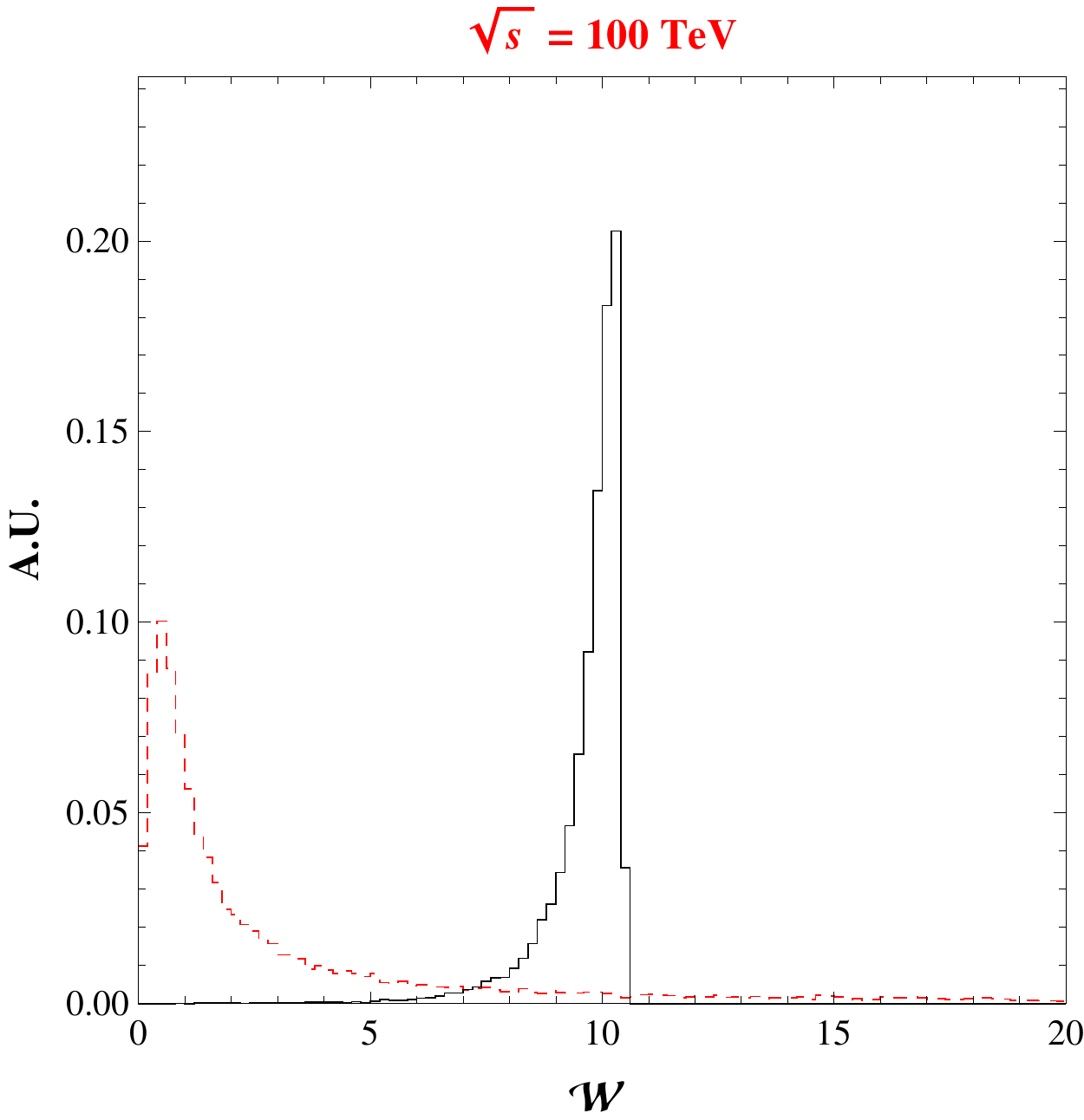}
\caption{\label{fig:fig1} Left panel: unit-normalized $m_{WW\gamma}$ distributions for three different $m_{\gamma_1}$ values with $c=2$. $\sqrt{s}$ is set to be 100 TeV. The vertical dashed lines indicate the relevant theory expectations. Right panel: unit-normalized distributions of the $\mathcal{W}$ variable for the dominant background $Wj\gamma$ (black solid histogram) and the signal (red dashed histogram) events at $\sqrt{s}=100$ TeV.}
\end{figure}

When it comes to the selection process in regard to $m_{\ell jj \nu}$ and $m_{\gamma \ell jj \nu}$, we introduce a new {\it weighted} measure $\mathcal{W}$ defined as
\begin{equation}
{\cal W}=
\frac{|m_{\ell jj \nu}-m_1^{\gamma}|}{\Gamma_1^{\gamma}}+\frac{|m_{\gamma \ell jj \nu}-(m_1^G+0.5\Gamma_1^G)|}{1.5\Gamma_1^G}\;,
\end{equation}
in order to capture the events where one of the invariant
mass windows is marginally not satisfied while the other is satisfied. The right panel of Fig.~\ref{fig:fig1} shows the performance of the $\mathcal{W}$ measure. The signal events typically give a small $\mathcal{W}$, whereas background events are peaked at a larger value of $\mathcal{W}$. This is because it is rather unlikely for the reconstructed $m_{G_1}$ and $m_{\gamma_1}$ in background events to be within both invariant mass windows simultaneously.

Given that the collider signature of our signal process is characterized by $\gamma \ell jj + \met$, several SM processes should be taken into consideration as potential backgrounds. Obviously, the irreducible SM background is from $WW\gamma$ production. Due to the existence of merged $jj$ in the signal, we expect that the most important reducible background comes from $Wj\gamma$, which contains a (single-prong) QCD jet. It turns out that this reducible background dominates over the irreducible one from $WW\gamma$. Several other backgrounds such as $WWj$ (by a photon fake), $ZZ\gamma$, and $Wjj\gamma$ can be considered, but they can be sufficiently suppressed by a set of cuts that we impose.  

%\subsection{Results/discovery reach}
We next discuss the discovery opportunity of the KK graviton based on the cuts listed in Table~\ref{tab:cutflow100TeV}. 
To this end, we take a couple of representative study points (SPs) at $\sqrt{s}=100$ TeV: 1) SP1 with $m_1^\gamma=2.5$ TeV and $c=2$ and 2) SP2 with $m_1^\gamma=3$ TeV and $c=2$.
As mentioned before, the dominant SM background is $Wj\gamma$. 
For the distributions of the transverse momenta of the leading jet $P_T^j$ and the photon $P_T^\gamma$,
the background peaks in the low $P_T$ region while the signal events tend to have larger values of $P_T$ above 150 GeV and 600 GeV 
for the jets and photon, respectively.
Therefore, we find that $P_T^j > 150$ GeV and $P_T^\gamma > 600$ GeV are very powerful in reducing the background.
Furthermore, the invariant mass constructed with the two $W$'s and the photon can provide a powerful handle for the signal-background separation. 
As discussed above, we implement this observation as the $\mathcal{W}$ measure and confirm its performance (see the right panel of Fig.~\ref{fig:fig1}).   

\begin{table}[t]
\centering
\begin{tabular}{c|c  c|c c}
 & SP1 & SP2 & $WW\gamma$ & $Wj\gamma$ \\
 \hline \hline
No cut                          & 0.4    & 0.13 &  --   & --\\
Basic cuts                       & 0.35   & 0.12 & (391) & (1.68$\times10^5$) \\
$p_T^{\gamma}>600$              & 0.31   & 0.11 & 1.81  & 132.0 \\
$p_T^{j/\ell}>150$               & 0.26   & 0.10 & 0.28  & 42.5 \\
$|\eta^{\textnormal{all}}|<2.0$  & 0.21   & 0.08 & 0.19  & 29.6 \\
$E_T^{\textnormal{miss}}>150$    & 0.20   & 0.077 & 0.10  & 13.1\\
$\Delta R_{jj}<0.4$              & 0.19   & 0.077 & 0.09  & -- \\
$60<m_{jj}<100$                  & 0.19   & 0.077 & 0.09  & -- \\
\hline
${\cal W}<0.9$(SP1)                    & 0.03    &--     & 0.0025& 0.29\\
${\cal W}<2.0$(SP2)                   & --      & 0.014 &  0.0055&1.19    \\
\hline \hline
$\mathcal{L}$ (ab$^{-1}$)       & 3     &  3     & 3   & 3 \\
%Number of events                & 390 & 90    &    & 7.5 & 870 \\
\hline
Number of events (SP1)           & 90    &--     & 7.5  & 870\\
Number of events (SP2)           & --      & 42  &  16.5 &3570    \\
\hline
$S/\sqrt{B}$                    & $5.0\sigma$  & $1.1\sigma$ & -- & --
\end{tabular}
\caption{\label{tab:cutflow100TeV} Signal and background cross sections in fb according to a sequence of cuts for two study points, SP1: $m_1^{\gamma}=2.5$ TeV with $c=2$  and SP2: $m_1^{\gamma}=3$ TeV with $c=2$, and two dominant SM backgrounds at a $pp$ collider of $\sqrt{s}=100$ TeV. 
To evaluate the background cross sections as leading order (in parentheses) without introducing any possible divergence, we require basic selection cuts such as $p_T^j>20$ GeV, $p_T^{\gamma/\ell}>10$ GeV, $|\eta^j|<5$, $|\eta^{\gamma/\ell}|<2.5$, and $\Delta R_{jj/j\gamma}>0.01$. All momenta and masses are in the unit of GeV.}
\end{table}

We make our parton-level Monte Carlo simulation for both signal and background processes, using \texttt{MG5\_aMC@NLO}~\cite{Alwall:2014hca} and \texttt{CalcHEP3.4}~\cite{Belyaev:2012qa} using the parton distribution functions \texttt{NNPDF23}~\cite{Ball:2012cx}. To implement the warped hierarchy/flavor model under consideration appropriately, we first take existing model files in Ref.~\cite{CP3}. 
As the model files do not contain the vertex of $G_1\gamma_1 \gamma$, we add the relevant vertex structure based on the corresponding one encoded in $G_1\gamma\gamma$ as previously explained. In addition, various decay modes of $\gamma_1$ are written by modifying the existing vertices in the model files. 
We obtain the total number of signal (defined as $S$) and background (defined as $B$) events to calculate the significance $S/\sqrt{B}$ as follows:
\begin{eqnarray}
 S&=&\epsilon_W \times N_S, \\
 B&=&\epsilon_W \times N_{WW\gamma}+2\times (1-\epsilon_j)\times N_{Wj\gamma},
\end{eqnarray}
where $N_S$, $N_{WW\gamma}$, and $N_{Wj\gamma}$ are numbers of events after all cuts, i.e., the numbers in the second and third last rows in Table~\ref{tab:cutflow100TeV}, for the signal (either SP1 or SP2), $WW\gamma$, and $Wj\gamma$ backgrounds, respectively. 
Here the tagging efficiency $\epsilon_W$ (rejection rate $\epsilon_j$) for a two-prong $W$-jet (single-prong QCD jet) is set to be 0.5 (0.95)~\cite{CMS:2014joa}.
For a more conservative analysis, we include a factor of two for the $Wj\gamma$ background to account for the next-to-leading order corrections.

Table~\ref{tab:cutflow100TeV} shows signal (SP1 and SP2) and background ($WW\gamma$ and $Wj\gamma$) cross sections in fb according to a set of cuts at a $pp$ collider of $\sqrt{s}=100$ TeV. 
We find that $m_1^{\gamma}=2.5$~TeV (or equivalently, $m^G_1 \sim 3.7$~TeV) can allow
a $5\sigma$ discovery of our photon cascade decay signal with an integrated luminosity of 3~ab$^{-1}$.  
This scale of the KK graviton mass is in a fairly good agreement with precision electroweak constraints~\cite{Carena:2006bn,Carena:2007ua} as well as the current bounds inferred from a null observation of KK gluons~\cite{Chatrchyan:2013lca,TheATLAScollaboration:2013kha}.

%\subsection{Conclusions and outlook}
To summarize, we have studied an unconventional search channel of the KK graviton featured by a novel ``cascade decay'' into a photon and an on-shell KK photon $\gamma_1$ which subsequently decays into a semi-leptonic $W$ pair. 
The highly energetic photon due to the mass gap between $G_1$ and $\gamma_1$ provides a distinct and elementary final state signature which can be detected efficiently. 
Although the photon coupling is suppressed by a 5D ``volume factor'', the strong coupling between $G_1$ and $\gamma_1$ renders this mode merely semi-suppressed. 
Consequently, we pointed out that the different dependence of the total rate on the volume factor from that of conventional search channels could enable us to determine the underlying model parameters. 
We found that the discovery reach of the KK graviton at future colliders such as a 100 TeV collider would be roughly $m_1^G=4$ TeV at an integrated luminosity of 3 ab$^{-1}$. 
Finally, we emphasize that the proposed search strategy with regard to $\gamma \ell jj+\met$ signature can be straightforwardly applied to other BSM models containing  processes yielding the same final state. 

\subsubsection{Seesaw Models and Resonances with Cascade Decays Involving RH Neutrinos}\label{BuphalDev}
%\begin{itemize}
%\item P.S.~Bhupal Dev, D.~Kim, R.~Mohapatra,, {\it Disambiguating Seesaw Models using Invariant Mass Variables at Hadron Colliders} {\bf contribution received}
%\item J.~Gluza, G.~Bambhaniya, J.~Chakrabortty, T.~Jeli\'nski, M.~Kordiaczy\'nska, R.~Szafron, {\it In quest of Right-Handed Currents [RHC]} {\bf contribution received}
%\end{itemize}
A widely discussed paradigm for neutrino masses is the so-called  type-I seesaw mechanism~\cite{Minkowski:1977sc,Mohapatra:1979ia,Yanagida:1979as,GellMann:1980vs} which postulates the existence of heavy right-handed (RH) neutrinos $N$ with Majorana masses. The mass scale of the RH neutrinos, synonymous with the seesaw scale, is {\em a priori} unknown, and its determination would play a big role in vindicating the seesaw mechanism as the new physics responsible for neutrino mass generation. In a bottom-up approach, the seesaw scale could be anywhere ranging from the left-handed (LH) neutrino mass scale of sub-eV all the way up to the grand unification theory (GUT) scale. However, there are arguments based on naturalness of the Higgs mass which suggest the seesaw scale to be  below $10^7$ GeV or so~\cite{Vissani:1997ys,Clarke:2015hta}. It is therefore of interest to focus on the seesaw scale being in the multi-TeV range which can be accessed at the current and foreseeable future collider energies. In particular, hadron colliders can probe TeV-scale seesaw through the ``smoking gun'' lepton number violating (LNV) signal of same-sign dilepton plus dijet final states: $pp\to W^*\to N\ell^\pm \to \ell^\pm \ell^\pm jj$~\cite{Keung:1983uu} and other related processes, such as the collinear-enhanced $t$-channel photon exchange processes~\cite{Datta:1993nm, Dev:2013wba, Alva:2014gxa}. In addition,  there are many kinds of complementary low energy searches for rare processes, such as neutrinoless double beta decay ($0\nu\beta\beta$)~\cite{Rodejohann:2011mu}, lepton flavor violation (LFV)~\cite{deGouvea:2013zba}, anomalous Higgs decays~\cite{BhupalDev:2012zg, Cely:2012bz, Maiezza:2015lza, Dermisek:2015vra} and so on which are sensitive to TeV-scale models of neutrino mass. It is important to emphasize that the collider probe of the seesaw is truly complementary to the low-energy searches of LNV and LFV at the intensity frontier. For a  recent review on the collider aspects of TeV-scale seesaw, see e.g., Ref.~\cite{Deppisch:2015qwa}. 

In the simplest seesaw extension of the SM, i.e. with the minimal addition of the heavy Majorana neutrinos while keeping the SM gauge group unchanged, there are two key aspects that can be tested experimentally, namely,  the Majorana mass $M_N$ of the mostly sterile neutrinos and their mixing $V_{\ell N}$ with the active neutrinos. In the traditional ``vanilla" seesaw mechanism~\cite{Minkowski:1977sc, Mohapatra:1979ia, Yanagida:1979as, GellMann:1980vs}, the left-right neutrino mixing is suppressed by the light neutrino mass $M_\nu \lesssim 0.1~{\rm eV}$:   
\begin{align}
V_{\ell N} \ \simeq \ \sqrt{\frac{M_\nu}{M_N}} \ \lesssim \ 10^{-6} \sqrt{\frac{100~{\rm GeV}}{M_N}}\; .
\label{canon}
\end{align}
Thus for a TeV-scale seesaw, the experimental effects of the light-heavy neutrino mixing are expected to be too small, unless the RH neutrinos have additional interactions, e.g. when they are charged under a $U(1)$ or $SU(2)$ gauge group. There exists a class of low-scale Type-I seesaw scenarios~\cite{Pilaftsis:1991ug, Buchmuller:1991ce, Gluza:2002vs, Kersten:2007vk,  Xing:2009in, Gavela:2009cd, He:2009ua,  Adhikari:2010yt, Ibarra:2010xw, Mitra:2011qr}, where $V_{\ell N}$ can be sizable due to specific textures of the Dirac and Majorana mass matrices in the seesaw formula $M_\nu\simeq -M_DM_N^{-1}M_D^T$. However, the constraints of small neutrino masses usually suppress the LNV $\ell^\pm \ell^\pm jj$ signals~\cite{Kersten:2007vk, Ibarra:2010xw, Lopez-Pavon:2015cga} in these models. 

Another natural realization of a low-scale seesaw scenario with large light-heavy neutrino mixing is the inverse seesaw model~\cite{Mohapatra:1986aw, Mohapatra:1986bd}. In this case, the magnitude of the neutrino mass becomes decoupled from the heavy neutrino mass, thus allowing for a large mixing 
\begin{align}
\label{eq:thetainvseesaw}
		V_{\ell N} \simeq \ \sqrt{\frac{M_\nu}{\mu_S}} \ \approx \ 10^{-2}\sqrt{\frac{1~\text{keV}}{\mu_S}} \; ,
\end{align}
where $\mu_S$ is the small LNV parameter in the theory, whose smallness is ``technically natural'', i.e. in the limit of $\mu_{S}\to  0$,
lepton   number  symmetry   is  restored   and  the  LH  neutrinos
 are massless  to all  orders in  perturbation theory, as in the SM. 

As for the LNV signature at colliders, in a natural seesaw scenario with approximate lepton number conservation, the LNV amplitude for the on-shell production of heavy neutrinos at average four-momentum squared $\bar{s}=(M_{N_1}^2+M_{N_2}^2)/2$ can be written as 
\begin{align}
{\cal A}_{\rm LNV}(\bar{s}) \ = \ -V_{\ell N}^2\frac{2\Delta M_N}{\Delta M_N^2+\Gamma_N^2}+{\cal O}\left(\frac{\Delta M_N}{M_N}\right) \; ,
\label{ampLNV}
\end{align}
for $\Delta M_N\lesssim \Gamma_N$, i.e. for small mass difference $\Delta M_N=|M_{N_1}-M_{N_2}|$ between the heavy neutrinos compared to their average decay width $\Gamma_N\equiv (\Gamma_{N_1}+\Gamma_{N_2})/2$. Thus, the LNV amplitude in \eqref{ampLNV} will be suppressed by the small mass splitting, except for the case $\Delta M_N\simeq \Gamma_N$ when it can be resonantly enhanced~\cite{Bray:2007ru}. In general, whether the dilepton signal can be of same-sign or mostly of opposite-sign depends on how degenerate the RH neutrinos are and to what extent they satisfy the coherence condition~\cite{Akhmedov:2007fk}. For seesaw models with suppressed same-sign dilepton signal, one can use the opposite-sign dilepton signal~\cite{Khachatryan:2014dka} and rely on the specific kinematic features to distill the signal from the huge SM background. Another option is to use the trilepton channel: $pp\to W^*\to N\ell^\pm \to \ell^\pm \ell^\mp \ell^\pm + \slashed{E}_T$~\cite{delAguila:2008cj, delAguila:2008hw, Chen:2011hc, Das:2012ze, Das:2014jxa, Bambhaniya:2014kga}, which has a relatively smaller SM background. 

The current direct search limits using the same-sign dilepton channel with 20 fb$^{-1}$ data at $\sqrt s=8$ TeV LHC~\cite{Khachatryan:2015gha, Aad:2015xaa} range from $|V_{\ell N}|^2 \lesssim 10^{-2}-1$ for $M_N=100-500$ GeV for $\ell=e,\mu$. These limits could be improved by roughly an order of magnitude and extended for heavy neutrino masses up to a TeV or so with the Run-II phase of the LHC~\cite{Deppisch:2015qwa} or with a future lepton collider~\cite{Antusch:2015mia, Banerjee:2015gca}. On the other hand, with the currently
allowed mixing, a $5\sigma$ discovery can be made for a TeV-scale heavy Majorana neutrino at a 100 TeV collider with 1 ${\rm ab}^{-1}$ of integrated luminosity~\cite{Alva:2014gxa}. We should note here that the $W\gamma$ vector boson fusion processes~\cite{Dev:2013wba, Alva:2014gxa, Deppisch:2015qwa} become increasingly important at these energies and must be taken into account, along with the usual Drell-Yan production mechanism so far considered in analyzing the LHC data.

On the theory front, a natural framework which could provide a TeV-scale renormalizable theory of the seesaw mechanism is the Left-Right (L-R) Symmetric extension of the SM (LRSM)~\cite{Pati:1974yy, Mohapatra:1974hk, Mohapatra:1974gc, Senjanovic:1975rk}, see also Section 4.5 of Volume 2 of this report. The two essential ingredients of seesaw, i.e., the existence of the RH neutrinos (and exactly three of them) and the seesaw scale,  emerge naturally in LRSM -- the former as the parity gauge partners of the LH neutrinos and the latter as the scale of parity restoration. There also exist examples~\cite{Dev:2013oxa} where the small neutrino masses via type-I seesaw at TeV-scale can arise without excessive fine tuning of the LRSM parameters. In addition, the discovery of the RH gauge bosons below 10 TeV could falsify the popular mechanism of leptogenesis as a viable explanation of the observed matter-antimatter asymmetry in our Universe~\cite{Frere:2008ct, Dev:2014iva, Dev:2015vra, Dhuria:2015cfa}. There are therefore considerable theoretical motivations to search for TeV-scale L-R seesaw signatures at the LHC and future colliders. It is worth emphasizing that in the LRSM, the Majorana nature of the RH neutrinos inevitably leads to the LNV signature of $\ell^\pm \ell^\pm jj$~\cite{Keung:1983uu, Ferrari:2000sp, Nemevsek:2011hz, Chen:2011hc, Chakrabortty:2012pp, Das:2012ii, AguilarSaavedra:2012gf, Chen:2013fna, Rizzo:2014xma, Ng:2015hba, Dev:2015kca}, irrespective of the light-heavy neutrino mixing parameter $V_{\ell N}$. A RH charged gauge boson mass up to $\sim 5.5$ TeV with 300 fb$^{-1}$ of data at the 14 TeV LHC~\cite{Ferrari:2000sp} or up to $\sim 32$ TeV with 1 ab$^{-1}$ of data at a 100 TeV collider~\cite{Rizzo:2014xma, Dev:2015kca} can be probed using the same-sign dilepton channel. 

\begin{figure}[t!]
\centering
\includegraphics[width=6cm]{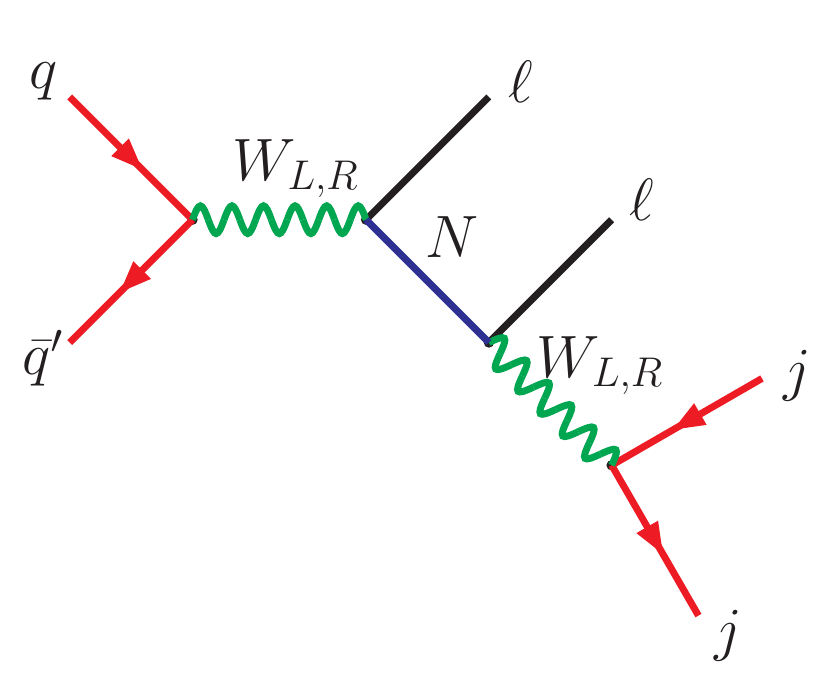}
\caption{Feynman diagram for the ``smoking gun'' collider signal of seesaw in the LRSM.}
\label{fig1}
\end{figure}
For $M_{W_R}>M_N>M_W$, there are {\em four} different sources in the LRSM for the origin of the $\ell \ell jj$ signal at the LHC~\cite{Chen:2013fna, Dev:2013vba} (see Figure~\ref{fig1}):
\begin{eqnarray}
LL:&& \; pp\  \rightarrow \ W_L^* \ \rightarrow \ \ell N \ \rightarrow \ \ell \ell W_L \ \rightarrow \ \ell \ell j j \, ,\label{eq:LL}\\
RR:&& \; pp \ \rightarrow \ W_R \ \rightarrow \ \ell N \ \rightarrow \ \ell \ell W_R^* \ \rightarrow \ \ell \ell j j \, ,\label{eq:RR}\\
RL:&& \; pp \ \rightarrow \ W_R \ \rightarrow \ \ell N \ \rightarrow \ \ell \ell W_L \ \rightarrow \ \ell \ell j j \, ,\label{eq:RL}\\
LR:&& \; pp \ \rightarrow \ W_L^* \ \rightarrow \ \ell N \ \rightarrow \ \ell \ell W_R^* \  \rightarrow \  \ell \ell j j \, ,\label{eq:LR}
\end{eqnarray}
where the first ($LL$) mode is the only one that arises in the SM seesaw via $s$-channel exchange of the SM $W$-boson, whereas all the four modes can arise in L-R models. These signals are uniquely suited to probe the Majorana and Dirac flavor structure of the neutrino seesaw and are therefore an important probe of the detailed nature of the seesaw mechanism. To this end, it is important to disambiguate the different mechanisms \eqref{eq:LL}-\eqref{eq:LR} in case of a positive collider signal in future. This can be done systematically~\cite{Dev:2015kca} by determining the kinematic endpoints of different invariant mass distributions, e.g., $m_{\ell\ell}$, $m_{\ell jj}$, $m_{\ell \ell j}$ etc., irrespectively of the dynamical details. This general kinematic strategy is equally applicable to both same and opposite-sign dilepton signals. In this sense, its efficacy is not just limited to the type-I seesaw models, but also to many of its variants, such as the inverse~\cite{Mohapatra:1986aw, Mohapatra:1986bd}, linear~\cite{Malinsky:2005bi} and generalized~\cite{Gavela:2009cd, Dev:2012sg, Dev:2015pga} seesaw models, which typically predict a dominant opposite-sign dilepton signal. Some of these variants might indeed be relevant in the potential discovery of parity restoration, if the recent observations from both CMS~\cite{Khachatryan:2014dka} and ATLAS~\cite{Aad:2015xaa} indicating a paucity of $\ell^\pm \ell^\pm jj$ events is confirmed in future collider data. 

%%%JG

%\subsubsection{Resonances with cascade decays involving RH neutrinos}
%\begin{itemize}
%\item J.~Gluza, G.~Bambhaniya, J.~Chakrabortty, T.~Jeli\'nski, M.~Kordiaczy\'nska, R.~Szafron, {\it In quest of Right-Handed Currents [RHC]} {\bf contribution received}
%\end{itemize}
 
Let us consider pure right-handed current (RHC) signals for the process of eq.~\eqref{eq:RR}. Here we  try to be as model independent as possible
and  we assume and discuss general left and right gauge couplings $g_L,g_R$. We are testing effects of a heavy particle sector neglecting small heavy neutrino and $W_2^\pm$ mixings with corresponding light SM states.
In this way pure effects coming from right-handed sector on the $pp \to lljj$ process are discussed \cite{Gluza:2015goa}.  
It is also shown when  mixings of heavy neutrino states can be factored out.

RHC Lagrangian includes general couplings $g_L,g_R$, which are important for gauge couplings unification \cite{Deppisch:2015cua,Dev:2015pga}

\begin{eqnarray}
 \mathcal{L}_L +\mathcal{L}_R =\frac{{g_L}}{\sqrt{2}}
\bar{\nu}_a\gamma^{\mu}P_L(U_{PMNS})_{aj}l_jW_{1\mu}^{+}
+
\frac{{g_R}}{\sqrt{2}}
\overline{N}_a\gamma^{\mu}P_R(K_{R})_{aj}l_jW_{2\mu}^{+} +\mathrm{h.c.} \;\;\;\;\label{lagr2}
\end{eqnarray}
\\
$\mathcal{L}_L$ describes the SM physics of charged currents. It includes the neutrino mixing matrix 
$U_{PMNS}$, responsible for neutrino oscillations phenomena. $\mathcal{L}_R$ is responsible for non-standard effects connected with heavy neutrinos 
$N_a$ and right-handed currents mediated by an additional heavy charged gauge boson $W_2$. $K_R$ defines a mixing matrix between flavour and massive heavy neutrino states. We assume it unitary, for a discussion, see ref.~\cite{Gluza:2015goa}.
 
The main RHC Feynman diagram which gives the $pp \to lljj$ process 
comes through two $W_2$ gauge bosons and heavy neutrinos $N_a$ in intermediate states, i.e. $pp {\to} W_2 \to N_a l  \to ll W_2 \to lljj $. 
For literature, see refs.~\cite{Keung:1983uu,delAguila:2007em,delAguila:2008cj,Maiezza:2010ic,Nemevsek:2011hz,Das:2012ii,Dev:2013oxa,Chen:2013fna,Vasquez:2014mxa,Deppisch:2014qpa,Heikinheimo:2014tba,Deppisch:2014zta,Aguilar-Saavedra:2014ola,Chakrabortty:2012pp,Ng:2015hba,Dobrescu:2015qna,Dev:2015pga,Coloma:2015una,Gluza:2015goa,Bandyopadhyay:2015fka}. This signal mimics the signature of neutrinoless double beta decay when the two leptons are same-sign electrons.
 
 To account lepton number violation, which might come if Majorana neutrinos are involved, we use the following notation:
 \begin{eqnarray}
 \sigma_{ij}^{\pm\pm}=\sigma(pp\to l_i^\pm l_j^\pm jj),\\
 \sigma_{ij}^{\pm\mp}=\sigma(pp\to l_i^\pm l_j^\mp jj).
 \end{eqnarray}
 We collectively denote all these cross-sections by $\sigma_{ij}$. 
 The process depends on the gauge couplings $g_L$ and $g_R$, the quark mixing matrices $U_{CKM}^{L,R}$, which can be chosen of the same form \cite{Senjanovic:2014pva,Senjanovic:2015yea}, the Parton Distribution Functions (PDFs) $f_{\alpha}(x,Q^2)$ and the heavy neutrino mixing matrix $(K_R)_{aj}$, see eq.~\eqref{lagr2}. The mass scales which are important for the process are: $M_{W_2}$, $M_{N_a}$ and $\sqrt{s}$. 
 It can be shown, that quite generally we have  $\Gamma(W_2)/M_{W_2}$, $\Gamma(N_a)/M_{N_a}\ll1$ and the NWA can be used to compute $\sigma_{ij}$ \cite{Gluza:2015goa,dirac}. When masses of heavy neutrinos are degenerate then dependence on the mixing matrix elements $(K_R)_{aj}$ can be factorized from the whole expression in the following way:
\begin{eqnarray}
\sigma_{ij}^{\pm\pm}&=&\widehat{\sigma}^{\pm\pm}\left|\sum_a(K_R^\dag)_{ia}(K_R^*)_{aj}\right|^2,\\
\sigma_{ij}^{\pm\mp}&=&\widehat{\sigma}^{\pm\mp}\left|\sum_a(K_R^\dag)_{ia}(K_R)_{aj}\right|^2,
\end{eqnarray}
where $\widehat{\sigma}^{\pm\pm}$, $\widehat{\sigma}^{\pm\mp}$ are ``bare'' cross-sections calculated for $(K_R)_{aj}=\delta_{aj}$, and
\begin{eqnarray}
\widehat{\sigma}^{\pm\pm}&=&\sigma(pp\to W_2^\pm)\times\mathrm{BR}(W_2^\pm\to N_1l_1^\pm)\mathrm{BR}(N_1\to l_1^\pm jj),\label{sighata1} \\
\widehat{\sigma}^{+-}&=&[\sigma(pp\to W_2^{+})+\sigma(pp\to W_2^{-})]\times\mathrm{BR}(W_2^{+}\to N_1l_1^{+})\mathrm{BR}(N_1\to l_1^{-}jj).\label{sighata2}
\end{eqnarray}
\begin{figure}[t!]
\begin{center}
{\includegraphics[scale=0.6]{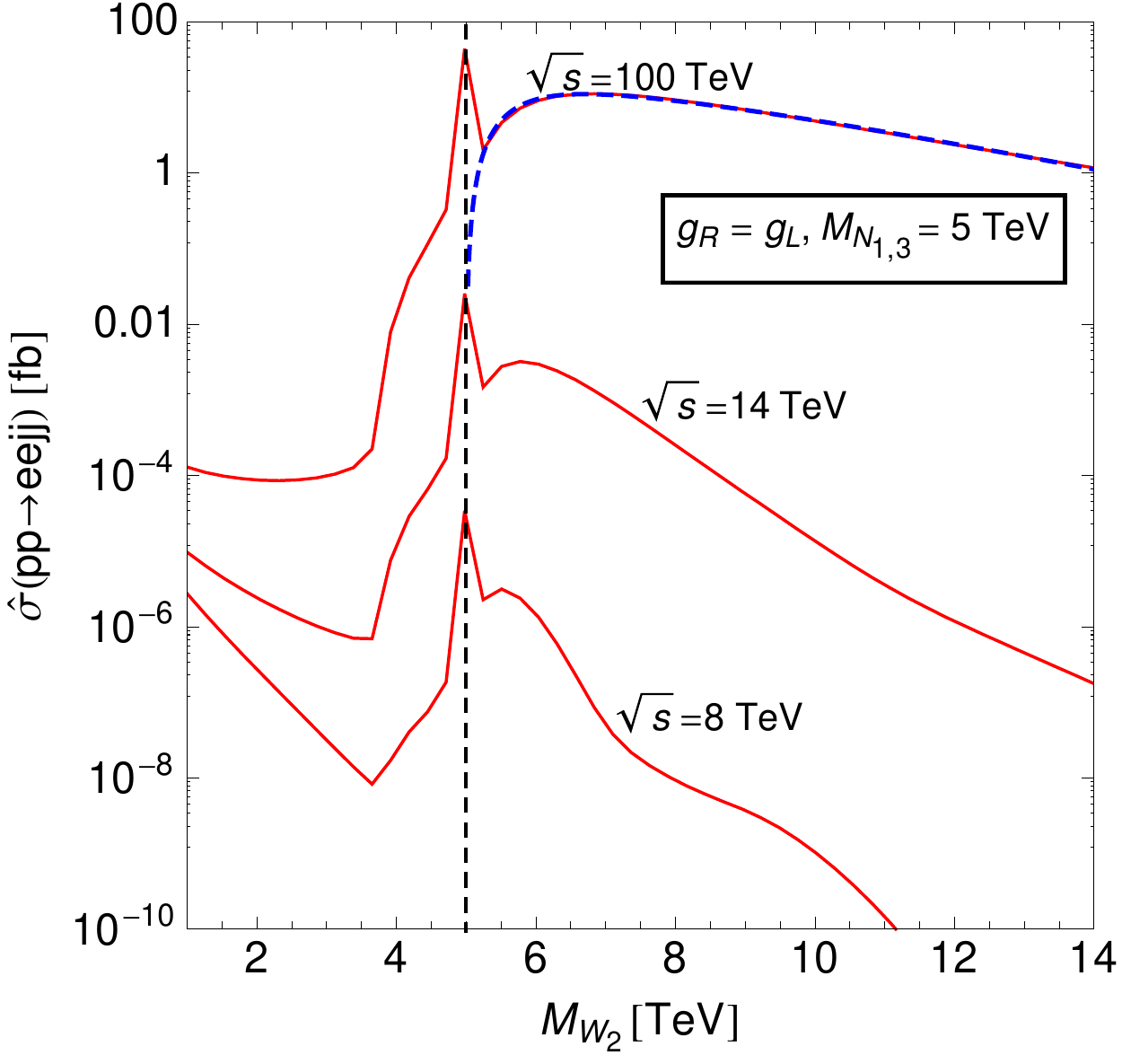}}\hspace{3mm}
{\includegraphics[scale=0.6]{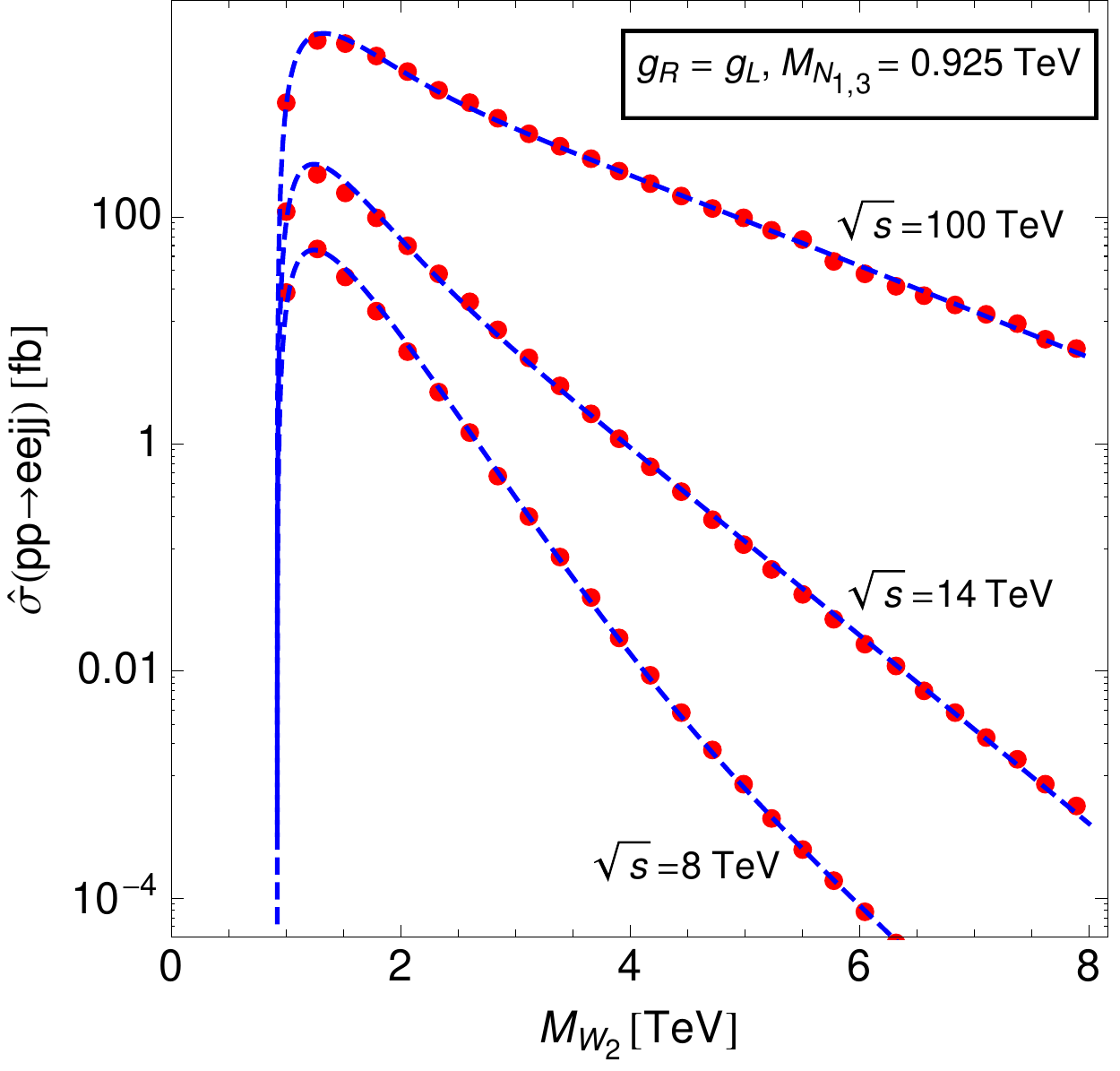}}
\end{center}
\caption{Examples of the neutrino mixing $K_R$ independent analysis. {\it Left:} The dependence of the total cross-section $\widehat{\sigma}_{ee}=\widehat{\sigma}^{+-}_{ee}+\widehat{\sigma}^{++}_{ee}+\widehat{\sigma}^{--}_{ee}$ on $M_{W_2}$ derived with the help of the \textsc{MadGraph} for $\sqrt{s}=8$, $14$, $100\,\mathrm{TeV}$. Heavy neutrino masses are $M_{N_{1,3}}=5\,\mathrm{TeV}$ and $M_{N_2}=10\,\mathrm{TeV}$, while $g_L=g_R$.  NWA is valid only in the region in which $M_{W_2}>M_{N_{1,3}}$ and $M_{W_2}$ is neither close to $M_{N_a}$ nor to $\sqrt{s}$ (where the distance is measured in $\Gamma(W_2)$ units). 
{\it Right:} An example of fitting analytical formula \eqref{sigP} (blue dashed curves) to the numerical results (red dots) obtained for $\sqrt{s}=8$, $14$ and $100\,\mathrm{TeV}$, $M_{N_{1,3}}=0.925\,\mathrm{TeV}$, $M_{N_2}=10\,\mathrm{TeV}$ with the help of the MadGraph 5 \cite{Alwall:2011uj}. Values of the fitted parameters $a$, $b$, $c$ and $d$ are given in the main text, see eqs.~\eqref{abcd8}-\eqref{abcd100}.
}\label{figgrgl}
\end{figure}
These can be written as
\begin{equation}
\widehat{\sigma}=\sum_{\alpha\beta}\int\limits_{M_{N_1}^2/s}^1 dx\int\limits_{
M_{N_1}^2/xs}^{1} dy f_{\alpha}(y,Q^2)f_{\beta}(x,Q^2)\widehat{\sigma}_{\alpha 
\beta}(xys),
\end{equation}
where $f_{\alpha}(x,Q^2)$ are PDFs of partons $\alpha$, while $Q$ is a characteristic scale of partonic process. Finally, $\widehat{\sigma}_{\alpha\beta}$ stands for partonic cross-section. Using the NWA one can write the ``bare'' cross sections in the following form \cite{dirac}:
\begin{eqnarray}\label{sigNWA}
\widehat{\sigma}&=&
\frac{g_R^2\pi}{18s}\frac{F_W(x_1)}{[18+\sum\limits_{b}F_W(x_b)]}
\sum_{\alpha\beta}\Phi_{\alpha\beta}\left(\frac{M_{W_2}^2}{s},M_{W_2}^2\right).
\end{eqnarray}
Here, we have used differential parton-parton luminosities $\Phi_{\alpha\beta}(\tau)$, see e.g. \cite{Campbell:2006wx}, defined as:
\begin{equation}\label{Phi}
%\frac{dL_{\alpha\beta}}{d\tau}
\Phi_{\alpha\beta}(\tau,Q^2)=\frac{1}{1+\delta_{\alpha\beta}}\int_{\tau}^1\frac{dx}{x}f_\alpha\left(x,Q^2\right)f_\beta\left(\frac{\tau}{x},Q^2\right)
\end{equation}
and
\begin{eqnarray}
F_W(x)&=&(2-3x+x^3)\theta(1-x).
%  F(x)&=&\frac{12}{x}\left[1-\frac{x}{2}-\frac{x^2}{6}+\frac{1-x}{x}\ln(1-x)\right].
\end{eqnarray}

Let us shortly comment on the form of eq.~\eqref{sigNWA}. First, the typical asymptotic $\sim1/s$ is clearly visible. Second, note that $\widehat{\sigma}\sim g_R^2$ while one would rather expect $\widehat{\sigma}\sim g_R^8$ from counting powers of gauge couplings entering matrix element corresponding to the process $pp \to l_il_jjj$. That difference is no longer surprising when one recalls that dominant contributions to the cross-section come from configurations in which $W_2$ and $N_a$ are nearly on-shell, what precisely corresponds to NWA. Finally, let us also remark that the simple consequence of $\widehat{\sigma}\sim g_R^2$ is that the cross-section scales like $(g_R/g_L)^2g_L^2$.

It turns out that one can estimate the total cross-section $\widehat{\sigma}_{ee}=\widehat{\sigma}^{+-}_{ee}+\widehat{\sigma}^{++}_{ee}+\widehat{\sigma}^{--}_{ee}$ under investigation using naive approximation: 
\begin{equation}\label{sigP}
\widehat{\sigma}_{ee}=
%\frac{g_L^2\pi}{18s}
\frac{F_W(x_1)}{1+\frac{1}{18}\sum_{b}F_W(x_b)}P\left(\mu\right),
\end{equation}
where $\mu=M_{W_2}/(1\,\mathrm{TeV})$ while $P(\mu)=a(e^{-b\mu}+ce^{-d\mu})$. For example, in the scenario in which $M_{N_{1,3}}=0.925\,\mathrm{TeV}$, 
%$M_{N_2}=10\,\mathrm{TeV}$ and  $N_{2}$ does not mix with $N_{1,3}$
the values of fitted parameters $a$, $b$, $c$ and $d$ are (see Fig.~\ref{figgrgl}):
\begin{eqnarray}
(8\,\mathrm{TeV})&&\quad a=0.18\times10^5\,\mathrm{fb},\quad b=3.62,\quad c=0.002,\quad\mathrm{and}\quad d=2.17\label{abcd8},\\
(14\,\mathrm{TeV})&&\quad a=1.32\times10^5\,\mathrm{fb},\quad b=3.97,\quad c=0.016,\quad\mathrm{and}\quad d=1.92\label{abcd14},\\
(100\,\mathrm{TeV})&&\quad a=5.40\times10^5\,\mathrm{fb},\quad b=3.04,\quad c=0.020,\quad\mathrm{and}\quad d=0.94\label{abcd100}.
\end{eqnarray}

Heavy neutrino masses are taken as in ref.~\cite{Gluza:2015goa}, though here the cross sections in Fig.~\ref{figgrgl}  is general and independent of the $K_R$ mixing matrix parametrization. 
For $M_{N}=10\,\mathrm{TeV}$ the cross section is already very small, at the $\sim$ ab level in a whole range of considered $M_{W_2}$ masses. 

%Let us note that a more systematic approach to extracting analytical dependence of $\widehat{\sigma}$ on $M_{W_2}$ is also possible. When one uses the following analytic parametrization of PDFs \cite{Harland-Lang:2014zoa}: 
%\begin{eqnarray}\label{PDF}
%&&xf_\alpha(x,Q_0^2)=
%A_{\alpha}x^{\delta_{\alpha}}(1-x)^{\eta_{\alpha}}\left[\sum_{i=1}^4a_{\alpha}^iT_i\left(1-2\sqrt{x}\right)\right],
%\end{eqnarray}
%where $T_i(y)$ are Chebyshev polynomials, then the luminosities \eqref{Phi} can be explicitly expressed in terms of Gamma function $\Gamma(z)$ and hypergeometric function ${}_2F_1(a,b,c;z)$. Hence $\widehat{\sigma}$ can be expressed as analytical function of $M_{W_2}^2/s$. Of course, to this end one has to know the values of $A_\alpha$, $\delta_{\alpha}$, etc. at $Q=M_{W_2}$ \cite{dirac}. 
 
%In this report we have focused on the signal estimation and analytical estimations. 
From this analysis we see that the $pp \to lljj$ process signal coming from right-handed currents (RHC) is about two order of magnitudes larger at a 100 TeV collider compared to the LHC at 14 TeV and the process can be parametrized in a simple way for $M_{W_2}\geq M_{N_i}$ independently of the heavy neutrino mixing scenarios. The background for the process depends on charges of dileptons. For the same $l^\pm l^\pm$ (LNV) and opposite $l^\pm l^\mp$  (LNC) signal cases see, e.g., the discussions in refs.~\cite{Nemevsek:2011hz,Han:2012vk,Ruiz:2015gsa,Dev:2015kca}. Further studies are needed to assess the potential for extracting signal from background at a 100 TeV collider.

%%%%%%%%%%%%%%%%%%%%%%%%%%%%%%%%%%%%%%%%%%%%%%%%%%%%%%

%%%%%%%%%%%%%%%%%%%%%%%%%%%%%%%%%%%%%%%%%%%%%%%%%%%%%%

%\subsection{Resonances: production in association with forward SM particles}
%
%\subsubsection{Associated production with a forward top or bottom}
%
%\subsubsection{Associated production with a forward light jet}
%
%\subsubsection{Excited quarks final states (multileptons, l+gamma, llqq)}

\subsection{New Fermionic Resonances}

\subsubsection{Seesaw Leptons at Future Hadron Collider Experiments}\label{Ruiz}
%\begin{itemize}
%\item R.~Ruiz, C.~Weiland, {\it Seesaw Leptons at Future Hadron Collider Experiments} {\bf contribution received}
%\end{itemize}
%\begin{figure}[!th]
%\subfigure[]{	\includegraphics[width=0.48\textwidth,height=6cm]{Contributions/Ruiz_et_al_included/typeISeesaw_MultiChannelProduction_100TeV.pdf}	\label{fig:seesawSingletXSec}	}
%\subfigure[]{	\includegraphics[width=0.48\textwidth,height=6cm]{Contributions/Ruiz_et_al_included/seesawLeptons_Production_100TeV.pdf}	\label{fig:seesawTripletXSec}	}
%\caption{100 TeV production rates of (a) singlet~\cite{Degrande:2016aje} and (b) triplet pair~\cite{ArkaniHamed:2015vfh}
%Seesaw leptons as a function of mass.}
%\label{fig:seesawXSec}
%\end{figure}
\begin{figure}[!t]
\begin{center}
\includegraphics[scale=0.5]{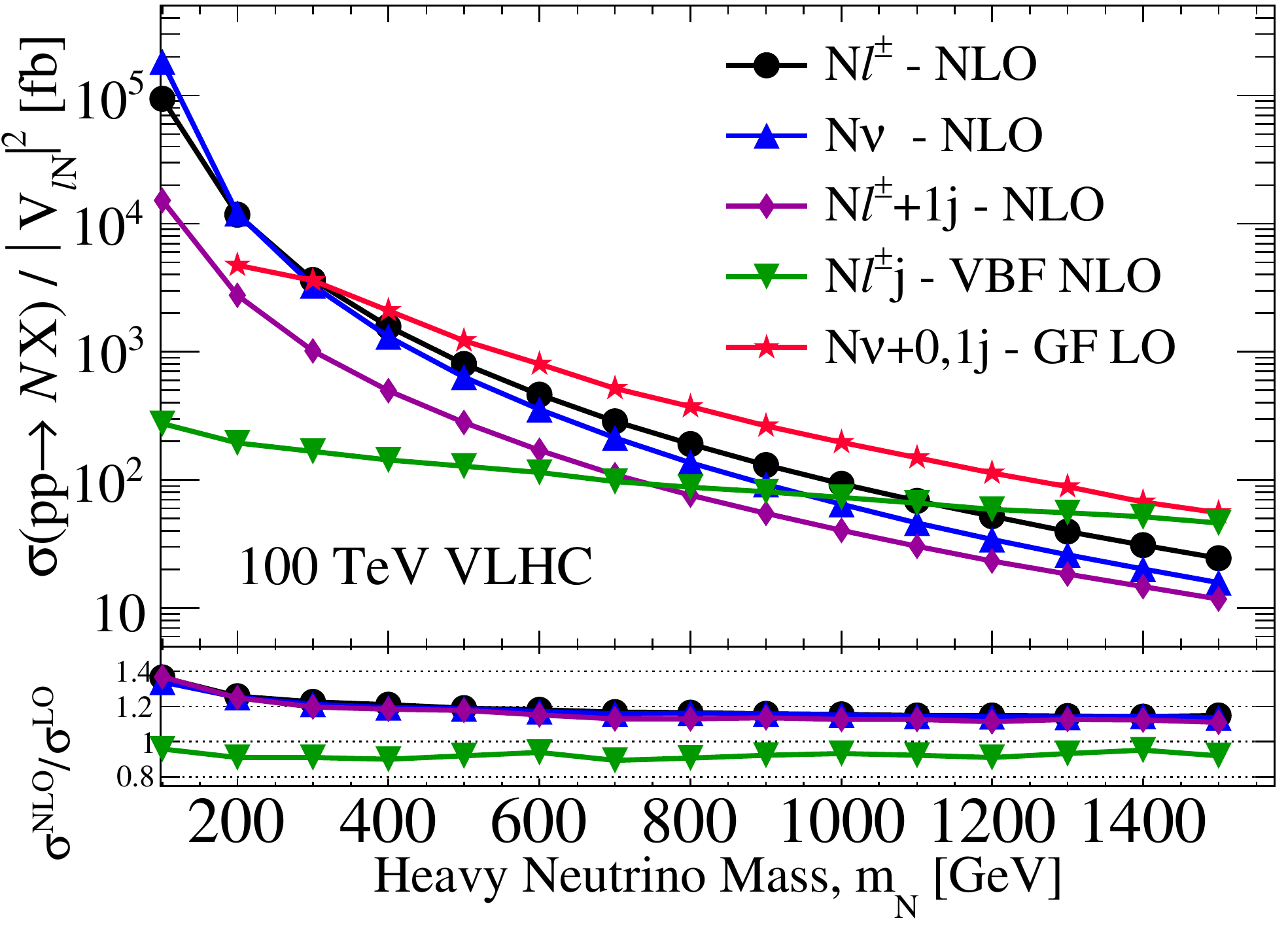}
%\hspace{3mm}
%\includegraphics[width=0.48\textwidth,height=6cm]{Contributions/Ruiz_et_al_included/seesawLeptons_Production_100TeV.pdf}
\caption{100 TeV pair production rates of a singlet seesaw lepton \cite{Degrande:2016aje} as a function of mass.}
\label{fig:seesawXSec}
\end{center}
\end{figure}
%\section{Heavy Seesaw Lepton Production at 100 TeV}
As we already stressed in Section \ref{BuphalDev}, collider tests of neutrino mass-generating offer a high degree of complementarity to low energy probes like neutrinoless double-beta decay, and precision lepton experiments.
In particular, low energy realizations of fermionic Seesaw mechanisms predict 
EW- and TeV-scale SU$(2)_L$ singlets $(N)$ and triplets $(T^\pm,T^0)$ that couple to gauge bosons through mixing  (singlet) 
or directly via gauge quantum numbers (triplet).

If kinematically accessible, these particles can be resonantly produced in hadron collisions through a variety of mechanisms.
Fig.~\ref{fig:seesawXSec} shows LO and NLO production rates of a singlet neutrino for $m_N > M_W$~\cite{Degrande:2016aje}. 
While Drell-Yan (DY) largely dominates at 14 TeV, the situation is qualitatively different at 100 TeV, where the gluon fusion (GF) process
\begin{equation}
 g ~g ~\rightarrow ~N ~\overset{(-)}{\nu_\ell}
\end{equation}
is the leading production mode for $m_N\lesssim 1.5$ TeV~\cite{Degrande:2016aje}. Beyond this, vector boson fusion (VBF)
\begin{equation}
 q ~\gamma ~	\overset{W\gamma~\text{Fusion}}{\longrightarrow} ~N ~\ell^\pm ~q'
\end{equation}
is dominant.
For $m_N\approx 1$ TeV, the GF, DY, and VBF mechanisms all share cross sections of the order of 100 fb.
In Figure \ref{fig:Seesaw} we have already shown the NLO triplet pair production rates for 
both the charged current and neutral current Drell-Yan processes~\cite{Ruiz:2015zca,ArkaniHamed:2015vfh}
\begin{equation}
q ~\overline{q'} ~\rightarrow W^* ~\rightarrow ~T^0 ~T^\pm \quad\text{and}\quad
q ~\overline{q}  ~\rightarrow \gamma ~\rightarrow ~T^+ ~T^-.
\end{equation}
Compared to the 14 TeV LHC, the reach of a 100 TeV $pp$ collider grows considerably from $\sigma_{14\text{ TeV}}= 1$ ab for triplet masses $m_T \approx 2.5$ TeV to $\sigma_{100\text{ TeV}}= 1$ ab for $m_T \approx 10-11$ TeV.

We now briefly summarize preliminary discovery potential of Seesaw leptons at 100 TeV.
We note that model-independent benchmark searches are rather robust since it is straightforward 
to reinterpret collider results for a particular neutrino flavor model.

%\subsection{Type I Seesaw Discovery Potential}
\begin{figure}[!t]
\includegraphics[width=0.48\textwidth,height=6.2cm]{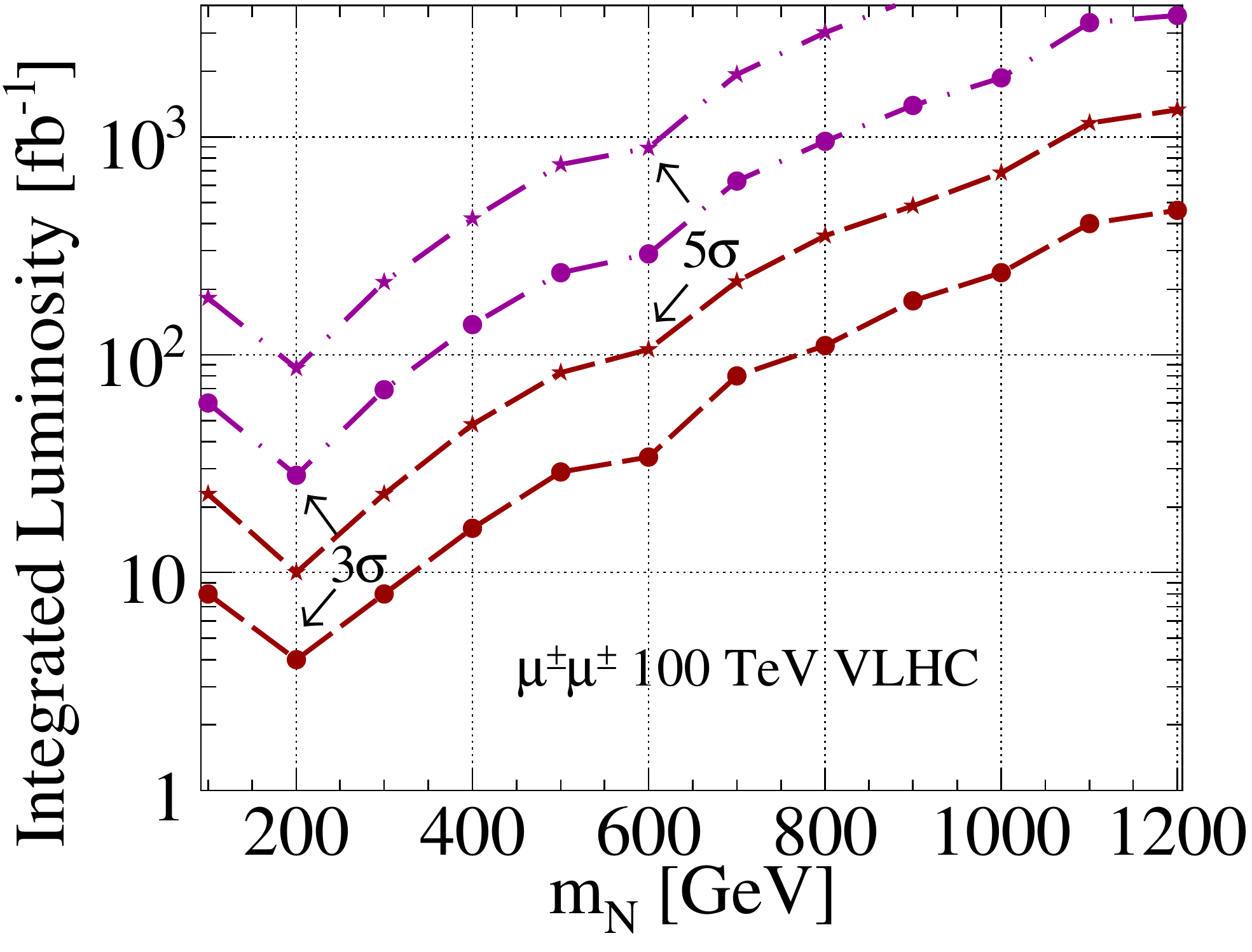}\hspace{4mm}
\includegraphics[width=0.48\textwidth,height=6.2cm]{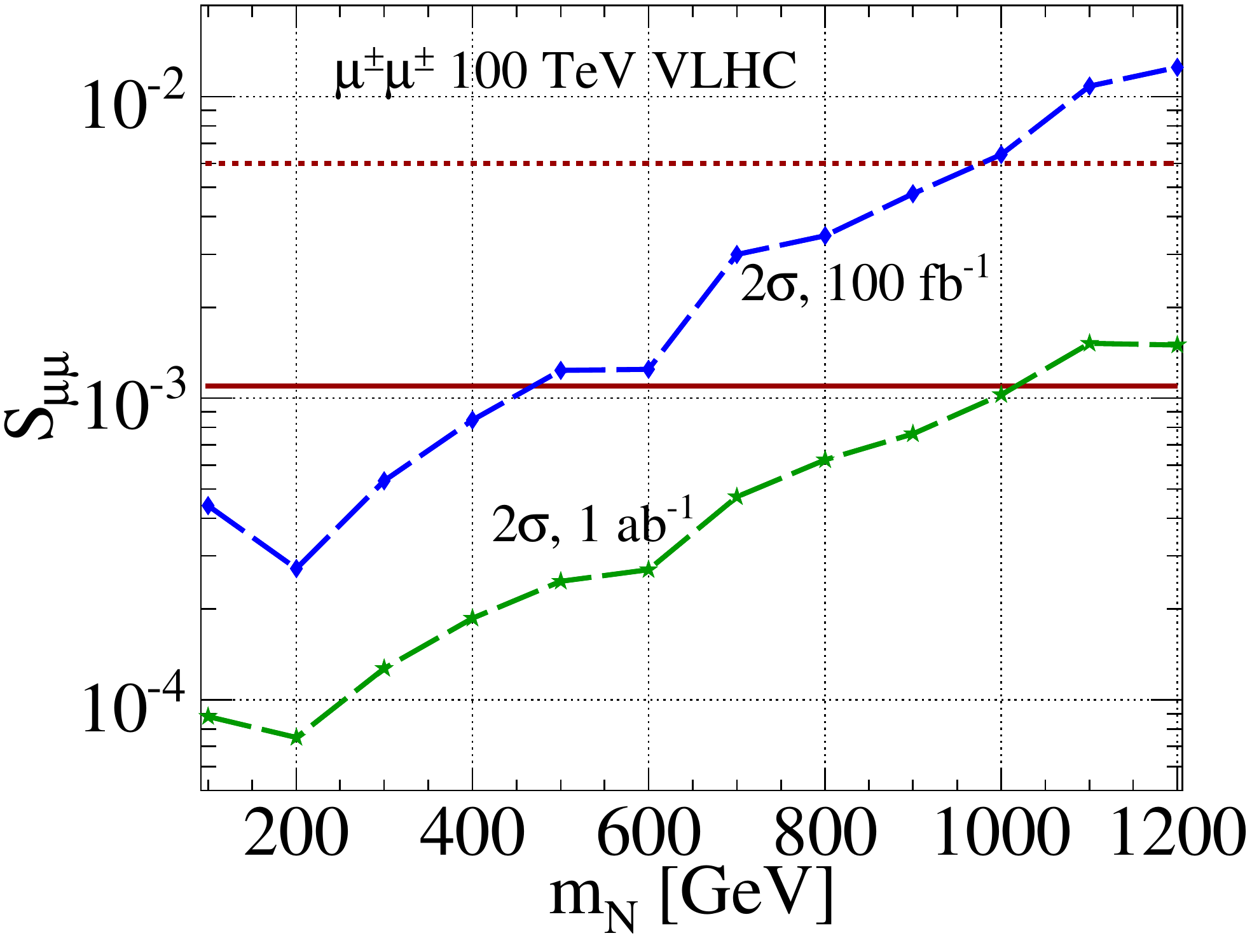}
\caption{{\it Left:} required luminosity for $3~(5) \sigma$ evidence (discovery) at a 100 TeV collider as a function of heavy neutrino mass $m_N$ in the $\mu^\pm\mu^\pm$ final state assuming optimistic (dash) and pessimistic (dash-dot) mixing scenarios \cite{Alva:2014gxa}; {\it Right:} sensitivity to $N-\mu$ mixing at a 100 TeV collider \cite{Alva:2014gxa}. Both plots are in the Type-I Seesaw model.}
\label{fig:SeesawIDiscoveryPotential}
\end{figure}

A key prediction of Type I-based scenarios is the existence of lepton number violating interactions, $N\rightarrow  \ell^\pm W^\mp$, 
which implies the same-sign leptons collider signature~\cite{Keung:1983uu} already discussed in Section \ref{BuphalDev}
\begin{equation}
 q ~\overline{q'} \rightarrow  ~N ~\ell^\pm_1 \rightarrow ~\ell^\pm_1 ~\ell^\pm_2 ~ W^\mp  \rightarrow ~\ell^\pm_1 ~\ell^\pm_2 ~j ~j.
 \label{LNVfinalstate}
\end{equation}
The largeness of the $N\ell^\pm$ VBF production cross section relative to the charged current DY process
offers considerable gain to inclusive searches for heavy Majorana neutrinos.
Assuming currently allowed mixing, a $5\sigma$ discovery can be made via the $\mu^\pm\mu^\pm$ channel for {$m_N = 1070$} GeV at 100 TeV after 1 ab$^{-1}$;
conversely, $N-\mu$ mixing as small as $S_{\mu\mu}\approx\vert V_{\mu N}\vert^2\lesssim 8\times10^{-5}$ may be probed~\cite{Alva:2014gxa}.
In the same-sign muon final state, the left panel of Fig.~\ref{fig:SeesawIDiscoveryPotential} shows the required luminosity for $3~(5) \sigma$ evidence (discovery) 
as a function of $m_N$ assuming an optimistic (dash) and pessimistic (dash-dot) mixing scenario;
in the right panel of Fig.~\ref{fig:SeesawIDiscoveryPotential} the sensitivity to mixing between heavy neutrinos and muons flavor states is shown.
\begin{figure}[!t]
\includegraphics[width=0.48\textwidth,height=6.2cm]{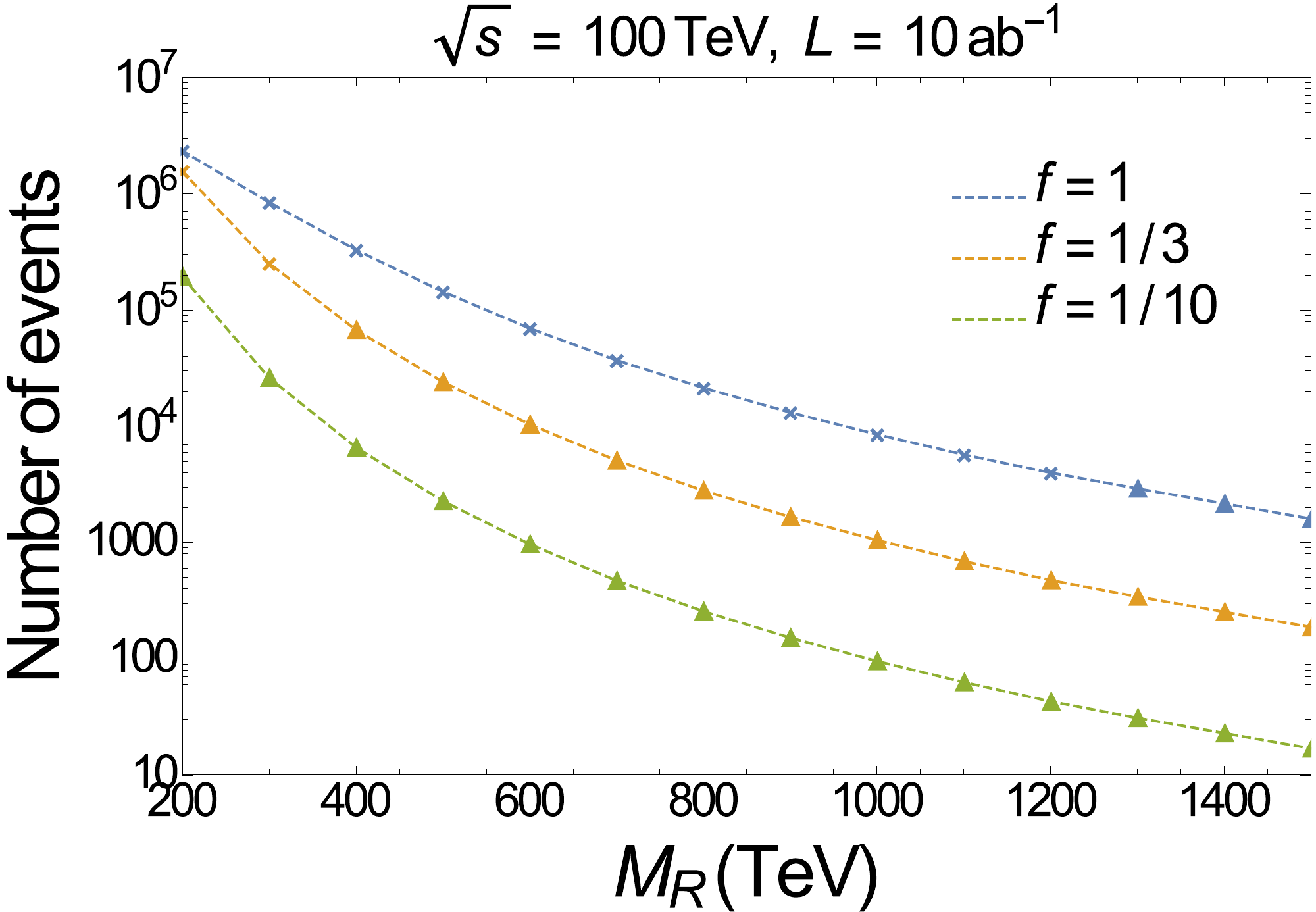}\hspace{6mm}
\includegraphics[width=0.48\textwidth,height=6.2cm]{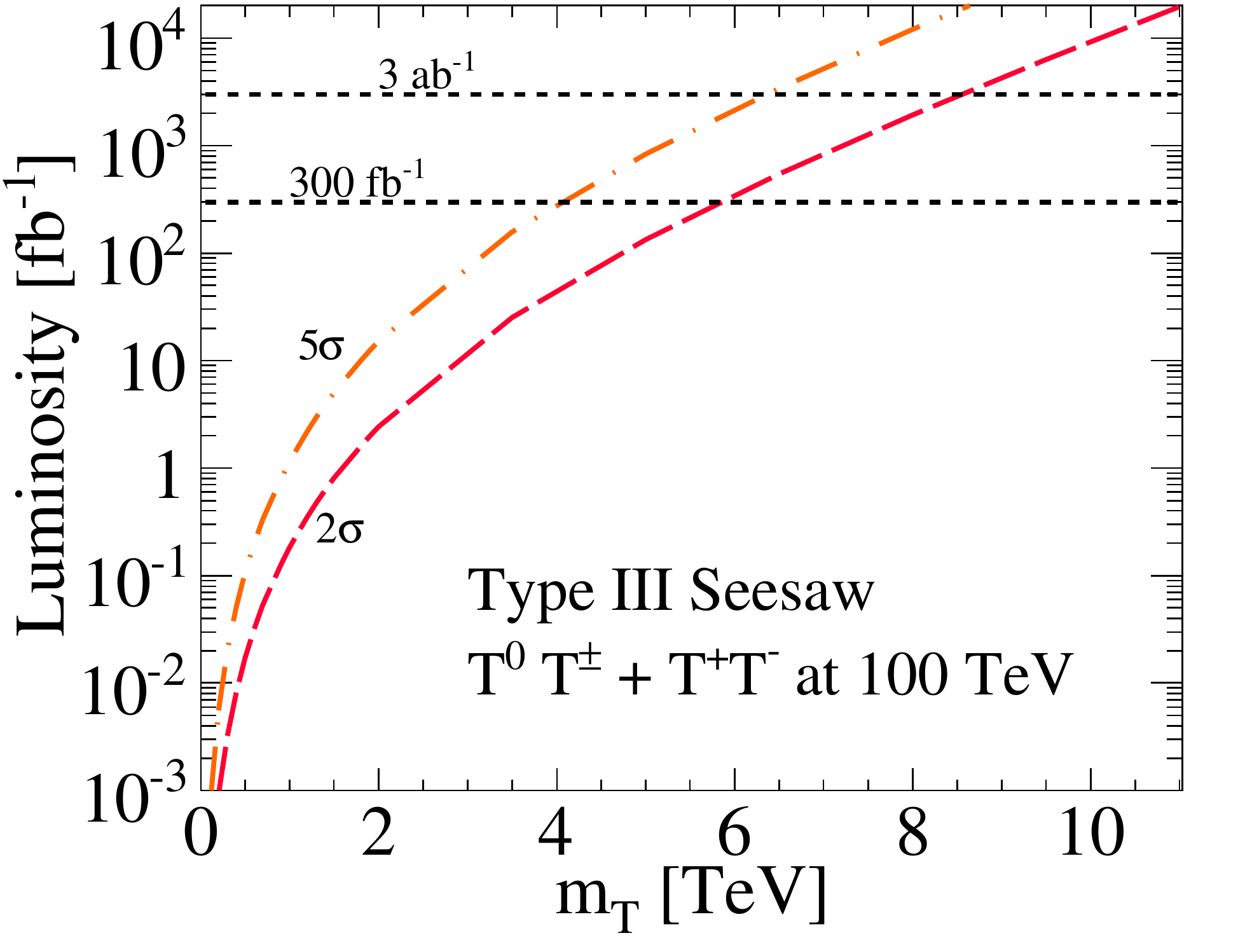}
\caption{
{\it Left:} in ISS, number of $\mu^\pm \tau^\mp jj$ events at a 100 TeV collider as a function of the seesaw scale $M_R$ for representative  scaling factors f~\cite{Arganda:2015ija}. Triangles respect constraints from LFV radiative decays while crosses do not respect the $\tau \rightarrow \mu \gamma$ BaBar upper limit~\cite{Aubert:2009ag}. {\it Right:} in Type III Seesaw, required luminosity for $2~(5) \sigma$ sensitivity (discovery) at a 100 TeV collider versus heavy lepton mass $m_T$ with fully reconstructible final states~\cite{Ruiz:2015zca}.}
\label{fig:SeesawDiscoveryPotentical}
\end{figure}
The Inverse Seesaw relies on an approximately conserved lepton number symmetry to suppress the light neutrino masses and lower the seesaw scale while
keeping large Yukawa couplings. As a consequence, heavy neutrinos form pseudo-Dirac pairs and lepton number violating processes
such as the one in eq.~\eqref{LNVfinalstate} are suppressed. 
In ref.~\cite{Arganda:2015ija}, a search relying on lepton flavor violating (LFV) final states
\begin{equation}
 q ~\overline{q'} \rightarrow  ~N ~\ell^\pm_1 \rightarrow ~\ell^\pm_1 ~\ell^\mp_2 ~ W^\mp  \rightarrow ~\ell^\pm_1 ~\ell^\mp_2 ~j ~j\,,
 \label{LFVfinalstate}
\end{equation}
was proposed.
The left panel of Fig.~\ref{fig:SeesawDiscoveryPotentical} shows the number of expected events for inclusive $\mu^\pm\tau^\mp jj$ production. Similar numbers would be expected for the $e^\pm\tau^\mp$ analogue, while experimental limits on $\mu\rightarrow e \gamma$~\cite{Adam:2013mnn} severely limit the $e^\pm\mu^\mp jj$ event rate. This assumes only the lightest pseudo-Dirac pair is kinematically accessible, and uses the $\mu_X$-parametrization~\cite{Arganda:2014dta} with a neutrino Yukawa coupling
\begin{equation}
 Y_{\nu}=f \left(\begin{array}{ccc}
-1&1&0\\
1&1&0.9\\
1&1&1
\end{array}\right)\,.
\label{Ytmmax1}
\end{equation}
$M_R$ is defined as in ref.~\cite{Arganda:2015ija}
and can be interpreted as the seesaw scale. The mass of the heavy neutrinos is equal to $M_R$ up to corrections proportional
to $Y_\nu v/M_{R}$, explaining the difference at low $M_R$.

%\subsection{Type III Seesaw Discovery Potential}
For masses well above the EW scale, 
triplet fermions preferentially decay to the Higgs and longitudinal polarizations of the $W$ and $Z$ bosons, a manifestation of the Goldstone Equivalence Theorem.
For EW boson decays to jets or charged lepton pairs, 
heavy lepton pairs can decay into fully reconstructible final-states with four jets and two high-$p_T$ leptons that scale like $p_T^\ell \sim m_T/2$:
\begin{eqnarray}
 T^0T^\pm	\rightarrow ~\ell\ell' + WZ/Wh		&\rightarrow& 
 ~\ell\ell' ~+~ 4j ~/~  2j+2b\,,
 \\
 T^+T^- 	\rightarrow ~\ell\ell' + ZZ/Zh/hh 	&\rightarrow& 
 ~\ell\ell' ~+~ 4j ~/~  2j+2b ~/~ 4b\,.
\end{eqnarray}
Assuming a nominal detector acceptance and efficiency of $\mathcal{A}=0.75$, at 100 TeV and after 10 fb$^{-1}$, 
a $5\sigma$ discovery can be achieved for $m_T\approx1.4-1.6$ TeV~\cite{Ruiz:2015zca}.
Taking instead $\mathcal{A}=1.0$, 
The right panel of Fig.~\ref{fig:SeesawDiscoveryPotentical} shows the discovery potential of the combined charged current and neutral current processes. After 3 ab$^{-1}$, there is $5~(2)\sigma$ discovery (sensitivity) up to $m_T \approx 6~(8)$ TeV.

\subsubsection{Fermionic Top Partners in Composite Higgs Models}\label{Wulzer}
%\begin{itemize}
%\item O.~Matsedonskyi, G.~Panico, A.~Wulzer, {\it Top partners at 100 TeV} {\bf (see What Next document)}
%\end{itemize}
An $100$~TeV collider can probe models with a terrific amount of Electro-Weak fine tuning. Even if none of these models had to be discovered, the result will be extremely informative as it will strongly disfavour (or exclude) a Natural origin of the Electro-Weak scale, pushing us towards the investigation of alternatives. We illustrate this point by estimating the reach, in terms of exclusions, for vector-like coloured fermions with a sizeable coupling to third-generation quarks, the so-called ``top partners''. Top partners are a common prediction of composite Higgs models in which the partial compositeness paradigm is assumed for the generation of fermion masses (see, e.g., refs.~\cite{2014arXiv1401.2457B,Panico:2015jxa} for a review). In these models, their mass $M$ is directly related to the amount of fine-tuning $\Delta$ according to the approximate formula
\begin{equation}
\label{tuningTP}
\Delta\sim \left(\frac{M}{500~{\textrm{GeV}}}\right)^2\,.
\end{equation}

\begin{figure}[t]
\begin{center}
\includegraphics[width=7.8cm]{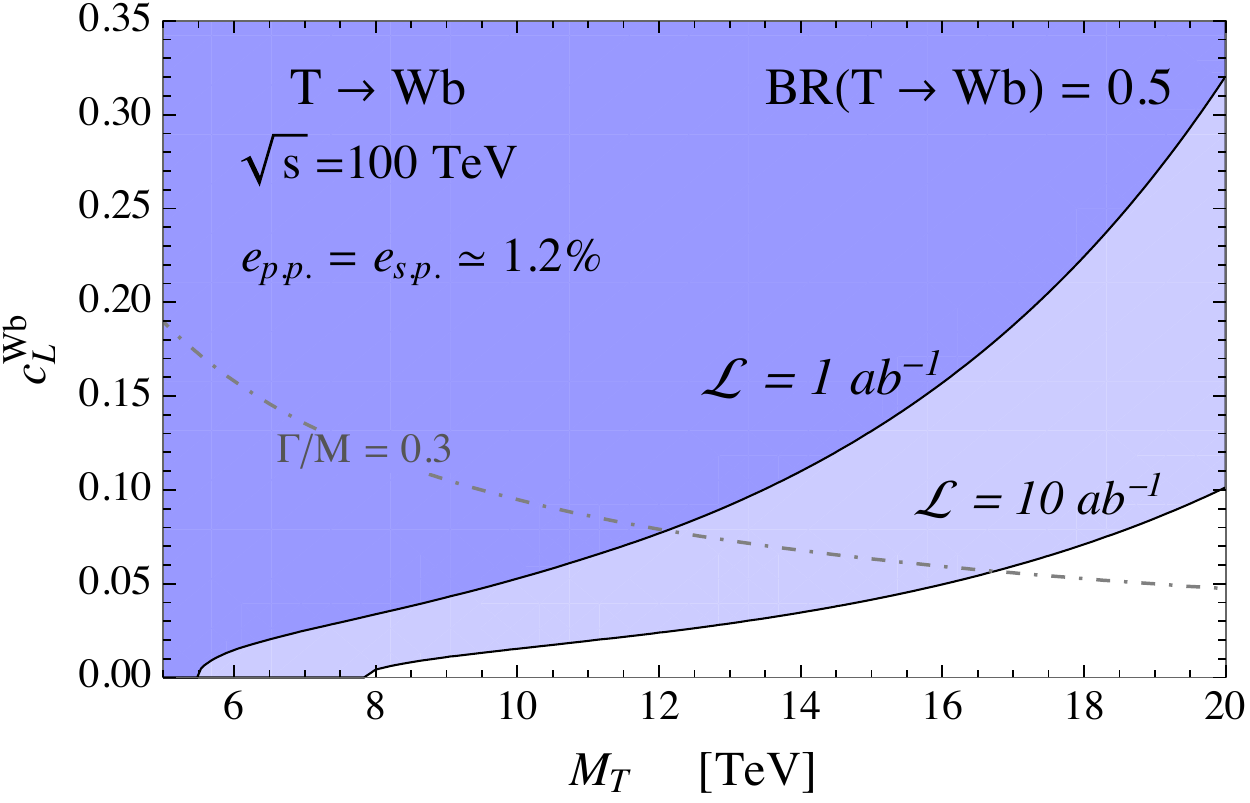}
\includegraphics[width=7.8cm]{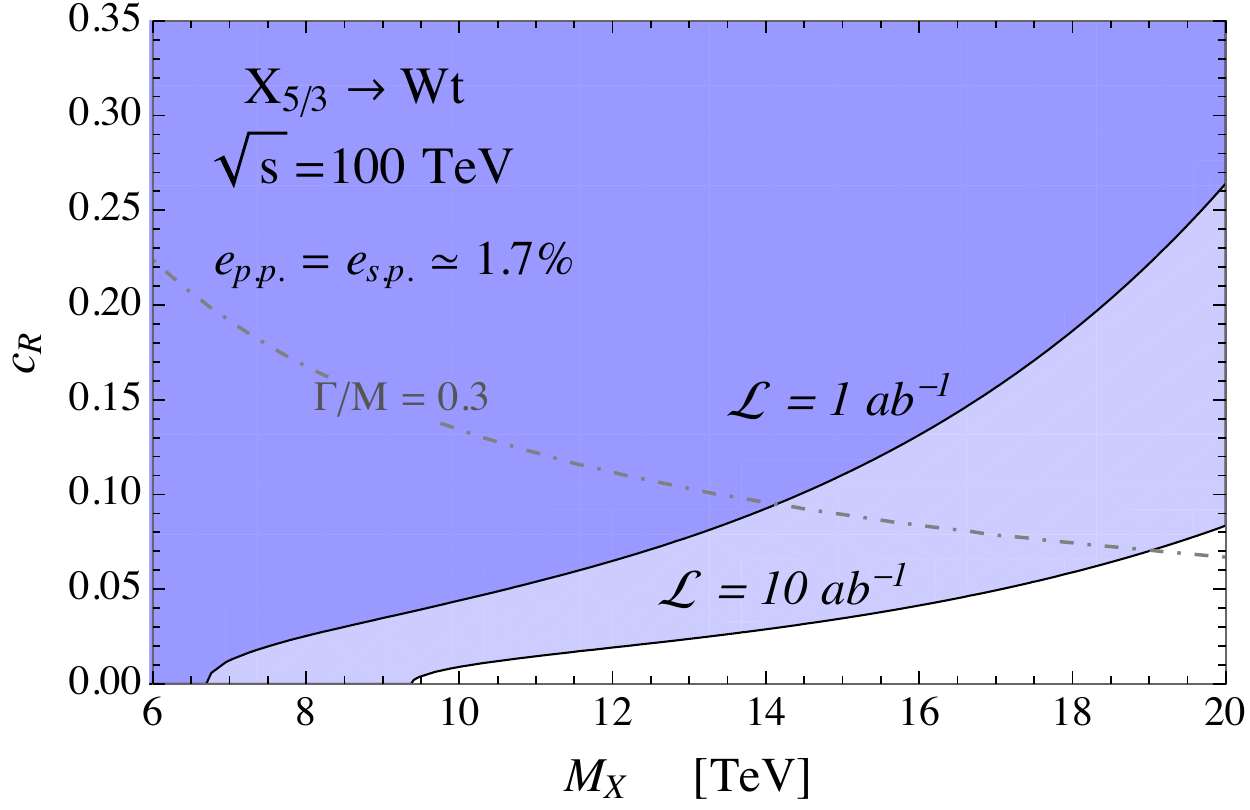}
\caption{{\it Left:} Exclusion reach for a top partner $T$ of electric charge $2/3$; {\it Right:} same plot for an $X_{5/3}$ of charge $5/3$. The plots are obtained by assuming that future searches at $100$~TeV will be sensitive to the same number of signal events as the current $8$~TeV ones. Namely, excluded signal yields $S_{exc} \simeq 25$ and $S_{exc} \simeq 10$ are assumed for the $T$ and the $X_{5/3}$. Signal selection efficiencies are also extracted from $8$~TeV results. In the case of  the single production mode, for which no dedicated searches are currently available, the efficiency ($e_{s.p.}$) is taken equal to the pair production one for simplicity. Further details can be found in ref.~\cite{Matsedonskyi:2014mna}.
}\label{exc}
\end{center}
\end{figure}

Top partners are coloured, thus they are unmistakably produced in pair by QCD interactions. They are also endowed with a sizeable coupling to third generation quarks and SM vector bosons or Higgs. The latter coupling is responsible for their decay, but also for their single production in association with a forward jet and a third generation quark. Exclusion contours are displayed in Fig.~\ref{exc}, in the plane defined by the top partner mass and its single production coupling. Top partners of electric charge $2/3$ (and \mbox{BR$(Wb)=0.5$}, which is typical for a SM singlet) and $5/3$ are shown, respectively, in the left and right plots. The results are based on a rough extrapolation based on current LHC Run-I limits and details are reported in the figure caption and, more extensively, in ref.~\cite{Matsedonskyi:2014mna}. A partial confirmation of the validity of the extrapolation, based on $100$~TeV simulations, can be found in ref.~\cite{Andreazza:2015bja}.

The result is that top partner masses of around $9$~TeV can be excluded at an $100$~TeV collider with $10$~ab$^{-1}$ luminosity in a completely model-independent way (i.e., for vanishing single-production coupling). According to eq.~(\ref{tuningTP}), this corresponds to $\Delta\sim300$. For composite Higgs models that cannot be excluded at the LHC, namely for $\xi=v^2/f^2=0.05$ or $\xi=0.01$, single production couplings of order $c\simeq 0.2$ or $c\simeq 0.1$ are expected \cite{Matsedonskyi:2014mna} and the reach considerably increases.

\subsubsection{Exotic Quarks in Twin Higgs Models: Displaced Decays in Association with a Prompt $t\bar{t}$ Pair}\label{Salvioni}
%\begin{itemize}
%\item H.-C.~Cheng, S.~Jung, E.~Salvioni, Y. Tsai, {\it Exotic Quarks in Twin Higgs Models: Displaced decays in association with a prompt $t\bar{t}$ pair} {\bf Contribution received}
%\end{itemize}
Models of ``neutral naturalness'', where the top partners do not carry SM color, provide a new set of signatures at hadron collider experiments. Since the low-energy connection between the SM and the sector that stabilizes the weak scale is feeble, the decays of the lightest BSM particles into the SM typically have macroscopic lifetimes, leading to displaced signatures. These challenging signals, combined with the low production rates of uncolored particles, imply that it is not unconceivable that a neutrally natural theory with a fine-tuning of $O(10)\%$ may completely escape detection at the LHC. However, a general feature of this class of models is that they only solve the little hierarchy problem, and thus require UV completion at a relatively low scale of at most $\sim 10$ TeV. The particles belonging to the UV theory would likely become accessible at a future 100 TeV collider, thanks to the high partonic energies available, allowing the future collider to probe an entirely new set of experimental signals. In this subsection, following ref.~\cite{Cheng:2015buv}, we begin the identification of the signatures of UV completions of neutral naturalness, by considering its prime example, the Twin Higgs \cite{Chacko:2005pe}, as benchmark model.

In the Twin Higgs, all the new particles lighter than about a TeV are complete SM singlets. However, the model requires to be extended below $\sim 10$ TeV, to remove residual logarithmic divergences. In non-supersymmetric UV completions, new exotic fermions charged under both the SM and the twin gauge symmetries must accompany the top quark. Their masses are expected to be in the $1$-$10$ TeV range. Some of these new fermions carry SM color, and would therefore be pair produced with large rates at a 100 TeV collider. Once produced, each of these ``exotic quarks'' decays into a SM top quark plus twin particles. Some of the twin particles can decay back to the SM with long lifetimes, giving rise to spectacular displaced vertices in combination with the prompt $t\bar{t}$ pair. Therefore, the signatures we consider are, labeling the exotic quarks as $\mathcal{T},\mathcal{B}$,
\begin{align}\label{exoquark signal}
pp\;\to&\; (\mathcal{T}\to t \hat{Z})(\overline{\mathcal{T}}\to \bar{t}\hat{Z})\; \to\; t\bar{t} \;+\; \mathrm{twin\;hadrons}\,,\qquad \mathrm{twin\;hadron}\;\to\; \mathrm{displaced\;signal}\,, \nonumber \\
pp\;\to&\; (\mathcal{B}\to t \hat{W})(\overline{\mathcal{B}}\to \bar{t}\hat{W})\; \to\; t\bar{t} \;+\; \mathrm{twin\;leptons}\,,\qquad \mathrm{twin\;lepton}\;\to\; \mathrm{displaced\;signal}\,,
\end{align}
where $\hat{Z},\,\hat{W}$ are the twin gauge bosons (we denote all the twin particles with a hat). 

We consider the ``fraternal'' version of the Twin Higgs, inspired by naturalness, where only the third generation SM fermions acquire a twin partner \cite{Craig:2015pha}. Therefore the twin hadrons in eq.~\eqref{exoquark signal} are produced by the $\hat{Z}\to \hat{b}\bar{\hat{b}}$ decay, followed by twin hadronization. Depending on the parameters, the long-lived twin hadron can be either a $CP$-even scalar meson $\hat{\chi}_{b0}$, a vector meson $\hat{\Upsilon}$ or a glueball $\hat{G}_{0^{++}}$. The $\hat{\chi}_{b0}$ and $\hat{G}_{0^{++}}$ decay through mixing with the $125$ GeV Higgs and therefore primarily into $b\bar{b}$, whereas the $\hat{\Upsilon}$ decays via photon kinetic mixing and thus almost democratically into all the SM electrically charged particles. The corresponding proper decay lengths are \cite{Cheng:2015buv,Craig:2015pha}
\begin{eqnarray} 
c\tau_{\hat{\chi}_{b0}}&\simeq& 3.8\,\text{cm}\left(\frac{m_b}{m_{\hat{b}}}\right)^2\left(\frac{f}{1\;\mathrm{TeV}}\right)^4\left(\frac{5\,\text{GeV}}{\Lambda}\right)^5 , \nonumber\\
c\tau_{\hat{\Upsilon}} &\simeq& 1.3\,\text{cm}\, \left(\frac{m_{\hat{A}}}{100\,\text{GeV}}\right)^4\left(\frac{10^{-3}}{\epsilon}\right)^2\left(\frac{5\,\text{GeV}}{\Lambda}\right)^5 , \label{decaylengths} \\
c\tau_{\hat{G}_{0^{++}}}&\simeq& 1\,\text{cm}\left(\frac{5\;\mathrm{GeV}}{\Lambda}\right)^7\left(\frac{f}{1\;\mathrm{TeV}}\right)^4, \nonumber
\end{eqnarray}
where the benchmark value of the $Z_2$-breaking scale $f$ ensures that Higgs coupling deviations are too small to be detected at the LHC. The choice of twin-QCD confinement scale $\Lambda$ is instead motivated by naturalness arguments \cite{Craig:2015pha}. For $\epsilon$, which determines the size of the kinetic mixing between the twin photon (with mass $m_{\hat{A}}$) and SM photon, we use the value naturally generated by exotic quark loops. In our study we kept most of the above parameters fixed to the benchmark values of eq.~\eqref{decaylengths}, except for the mass of the $\hat{b}$ quark. As a consequence, the typical decay lengths that we consider are as follows: the proper lifetime of $\hat{\chi}_{b0}$ approximately varies from $1$ cm to $10$ m, whereas the $\hat{\Upsilon}$ and $\hat{G}_{0^{++}}$ have centimeter-scale decays.

For the hadronic $(b\bar{b})$ displaced decay search, the detector is modeled, following the ATLAS searches of refs.~\cite{Aad:2015uaa,Aad:2015asa}, as the sum of two annuli with radii $1< r <28$ cm and $200 < r < 750$ cm, representing the inner detector (ID) and hadronic calorimeter plus muon spectrometer, respectively, with an efficiency for displaced vertex (DV) identification equal to a constant $10\%$. In the search for $\hat{\Upsilon}$ decays we concentrate on dimuon DVs in the ID, which is modeled, following the CMS search of ref.~\cite{CMS:2014hka}, as an annulus with radii $1< r <50$ cm and efficiency for DV identification equal to a constant $50\%$. In addition, simple cuts are applied directly on the twin hadrons, to roughly reproduce reasonable experimental requirements. 

The combination of the prompt $t\bar{t}$ and displaced signal ensures a straightforward triggering, and is expected to remove completely the SM background. Then, assuming no events are observed, a signal hypothesis can be excluded at $95\%$ if it would predict more than $3$ events. The projected reach of the displaced twin hadron + $t\bar{t}$ search at a 100 TeV collider is shown in Fig.~\ref{fig:twinhadron}, assuming an integrated luminosity of $1$ ab$^{-1}$. The sensitivity extends up to $m_{\mathcal{T}}\sim 11$ TeV. For comparison, we also show the estimated reach obtained from the search for direct stop production \cite{Gershtein:2013iqa}, which is sensitive to the exotic quarks if all the produced twin particles leave the detector as missing energy. We find that the potential of the search for $t\bar{t}$+displaced signals can be significantly superior. Estimates for luminosities different from 1 ab$^{-1}$ can be obtained by assuming that the signal rate scales with the partonic luminosities. This is a reasonable first approximation, because at large $m_{\mathcal{T}}$ the exotic quark branching ratios are approximately independent of the mass, and the variation of the typical twin hadron boost factor gives subdominant effects.  
\begin{figure}
\begin{center}
\includegraphics[width=11cm]{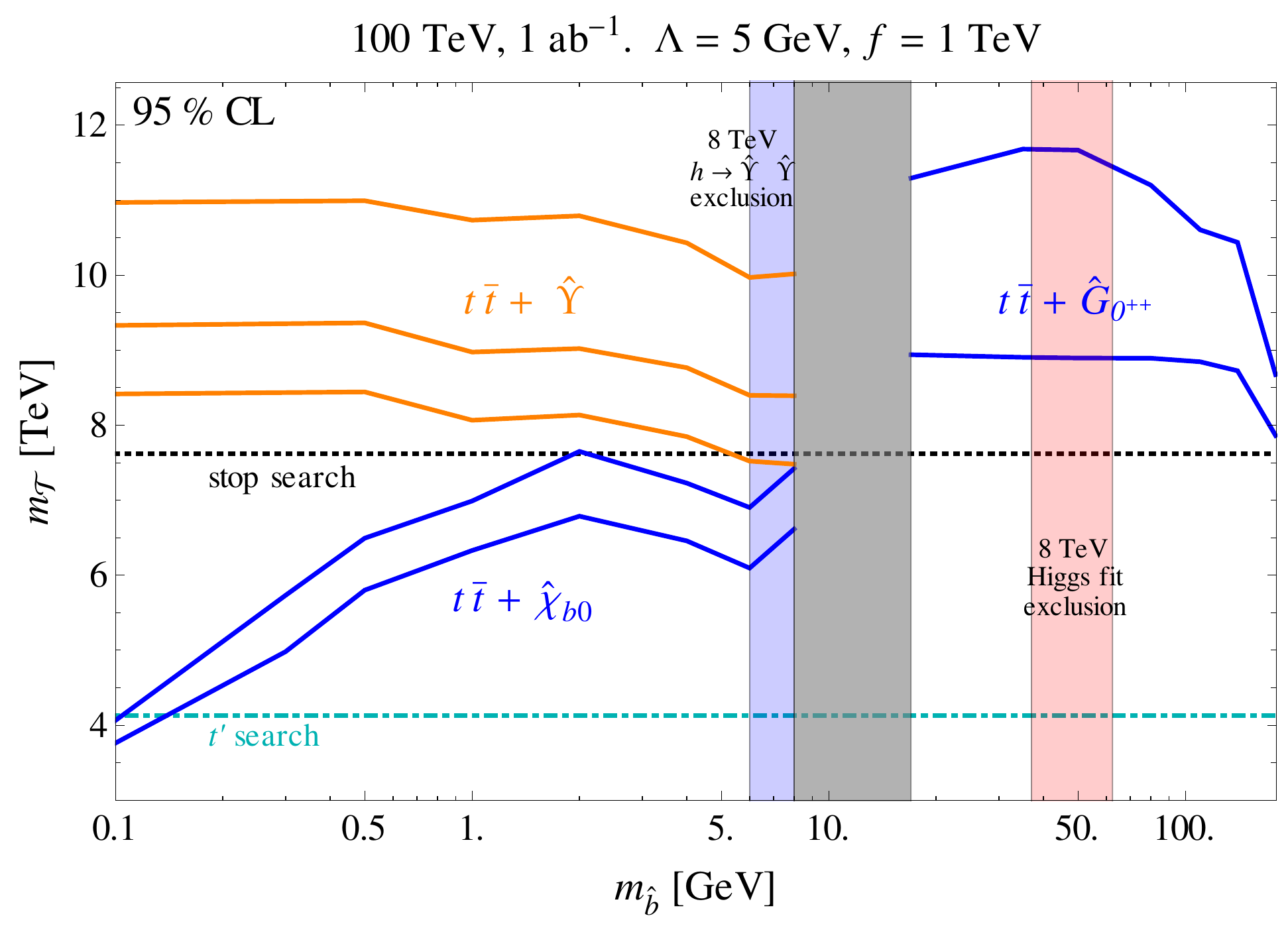}
%\vspace{2.5mm}
\caption{Projected bounds on the mass of the exotic quark $\mathcal{T}$ from the twin hadron displaced signal at a 100 TeV collider. The orange and blue curves correspond to the $t\bar{t}+$displaced twin hadron signals. %The different curves for each signal illustrate the uncertainty in the estimate of the twin hadron fractions. %Twin bottomonium (glueball) production dominates for light (heavy) $\hat{b}$, whereas in the gray shaded region both types of twin hadrons appear. 
Also shown are the bounds from top partner (dot-dashed light blue lines) and stop (dotted black lines) searches. %The region shaded in blue (red) is already excluded by the $8$ TeV LHC searches for Higgs displaced decays (global Higgs fit).
}
\label{fig:twinhadron} 
\end{center}
\end{figure}

The twin leptons are produced in the decay of the twin $W$ boson, $\hat{W}\to\hat{\ell}\hat{\ell}$ (since we assume twin electromagnetism is broken, for our purposes the distinction between twin tau and twin neutrino is irrelevant, so we simply denote all twin leptons by $\hat{\ell}$). The twin leptons can mix with the SM neutrinos and thus effectively behave as sterile neutrinos, decaying into either three SM leptons or one SM lepton and a pair of quarks. As a consequence, both hadronic and leptonic displaced decay searches are relevant. The proper decay length is given by
\begin{equation}
c\tau_{\hat{\ell}}=\,10\,\mathrm{cm}\,\left(\frac{10^{-3}}{\sin\theta_\nu}\right)^2\left(\frac{m_{\hat{\ell}}}{6\;\mathrm{GeV}}\right)^5.
\end{equation} 
where $\sin\theta_\nu$ controls the $\hat{\ell}$-$\nu$ mixing. In the range of parameters we consider, the lifetime varies from $1$ cm to $10$ m. The simulation of the twin lepton signals is described in detail in ref.~\cite{Cheng:2015buv}. Here we only observe that the twin leptons are very boosted, and thus their decay products are very collimated, with typical angular separation of $\Delta R\sim O(0.01)$. Therefore searches for lepton jets \cite{Aad:2014yea} play an important role.

The projected reach of the displaced twin lepton + $t\bar{t}$ search at a 100 TeV collider with $1$ ab$^{-1}$ is shown in Fig.~\ref{fig:twinlepton}, again under the assumption of zero SM background. The sensitivity extends up to $m_{\mathcal{B}} \sim 11$ TeV, with hadronic displaced signals more promising due to the larger branching ratio of $\hat{\ell}$ into quarks. To compare the reach of the twin hadron and twin lepton signals, it is useful to recall that $m_{\mathcal{T}} \simeq (m_\mathcal{B}^2 + y_t^2 f^2 /2)^{1/2}$, therefore the $\mathcal{T}$ and $\mathcal{B}$ are approximately degenerate for masses much larger than $1$ TeV. 

\begin{figure}[t] 
\includegraphics[width=0.48\textwidth]{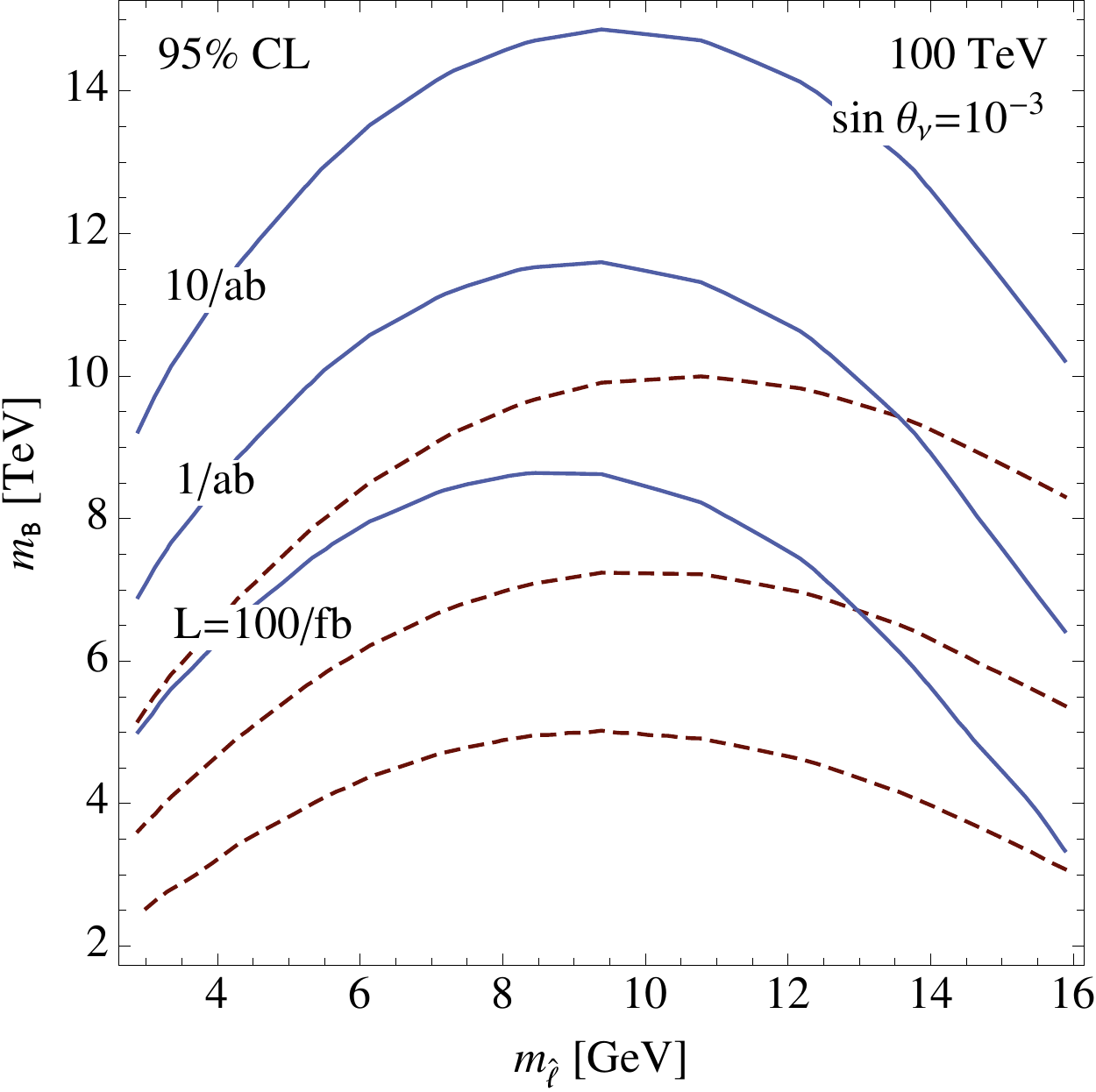}\hspace{3mm}
\includegraphics[width=0.48\textwidth]{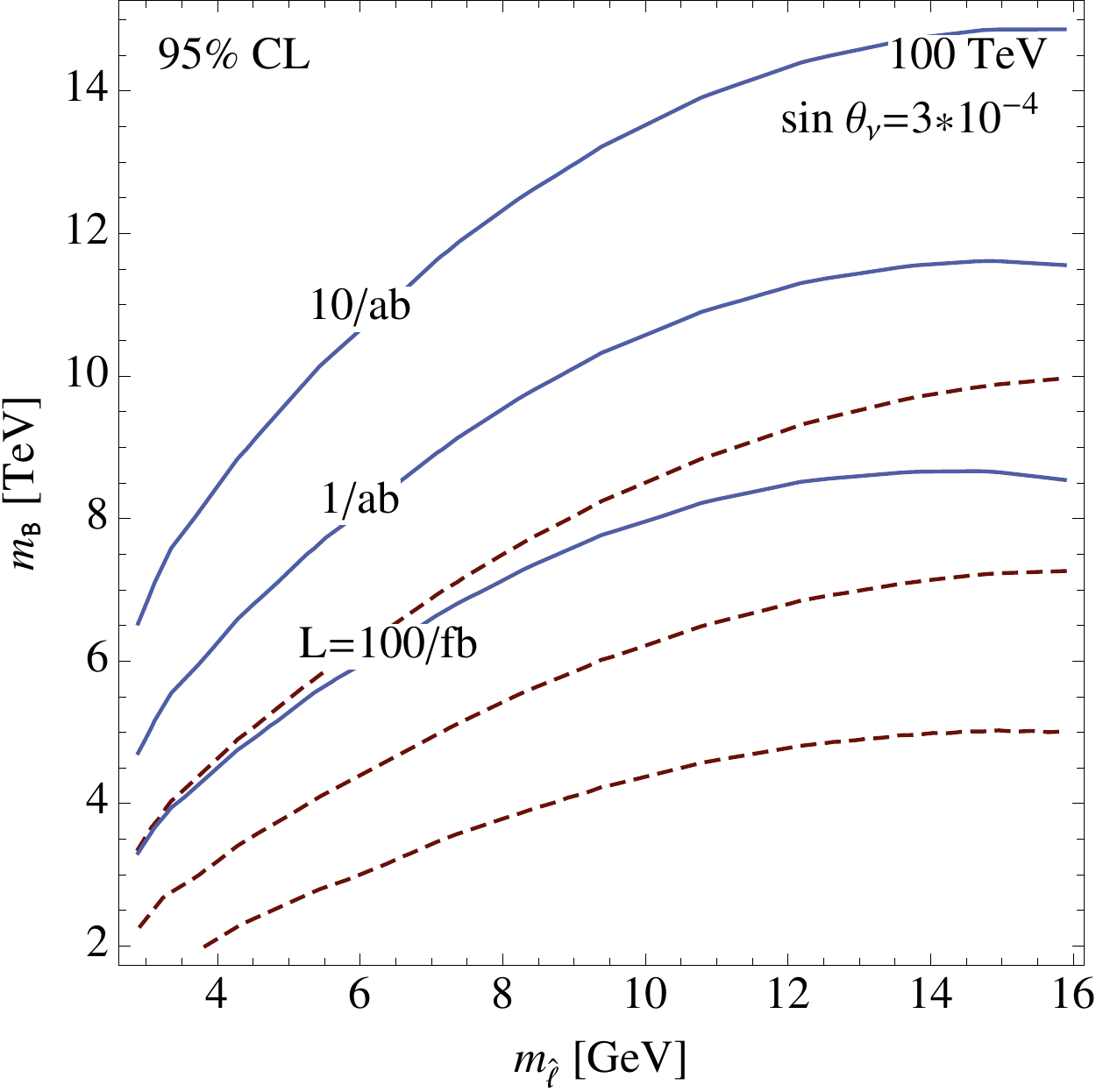}
\caption{Projected bounds on the mass of the exotic quark $\mathcal{B}$ from the twin lepton displaced signal at a 100 TeV collider. The blue solid lines correspond to hadronic displaced searches and the brown dashed lines to leptonic displaced searches. {\it Left:} $\sin\theta_{\nu}=10^{-3}$; {\it Right:} $\sin\theta_{\nu}=3\cdot 10^{-4}$.}
\label{fig:twinlepton}
\end{figure}

To summarize, we presented a first study of displaced decays produced in association with a prompt $t\bar{t}$ pair at a future 100 TeV collider. Both hadronic and leptonic displaced signals were considered. The presence of the decay products of the tops guarantees triggering, and is expected to remove completely the SM background. This signature can arise, for example, from exotic quarks that appear in several UV completions of Twin Higgs models. We estimated that a 100 TeV collider with $1$ ab$^{-1}$ will be able to probe these new fermions up to masses of $\sim 11$ TeV, thus providing a strong test of the most motivated region of parameters.

%\subsubsection{Four jet final states from pairs of colored resonances}

%\subsubsection{Final states with large multiplicities}

%\subsubsection{Pair production of uncolored particles through Higgs portal}
\subsubsection{Probing Naturalness Model-Independently at a 100 TeV Collider}\label{Curtin}
%\begin{itemize}
%\item D.~Curtin, P.~Saraswat, {\it Probing Naturalness Model-Independently at a 100 TeV collider} {\bf contribution received}
%\end{itemize}
One of the primary goals of the current and future collider program is to search for new physics associated with the stabilization of the electroweak scale. In symmetry-based solutions to the hierarchy problem, the large quantum corrections to the Higgs scalar from the top quark must be canceled by ``top partners,'' new states with couplings related by symmetry to that of the top quark. In traditional theories such as minimal supersymmetric or composite Higgs models, these top partners carry SM color charge like the top quark and are copiously produced at hadron colliders. In this broad class of models, searches for new colored particles directly probe electroweak naturalness. Placing the top partner mass beyond the LHC reach of $\sim 1 {\ \rm TeV}$ already implies a tuning in the Higgs mass of at least a few percent; searches for colored particles at a 100 TeV collider would probe even more parameter space, with a null result worsening the tuning.   

\begin{figure}[t]
\begin{center}
\includegraphics[height=7cm]{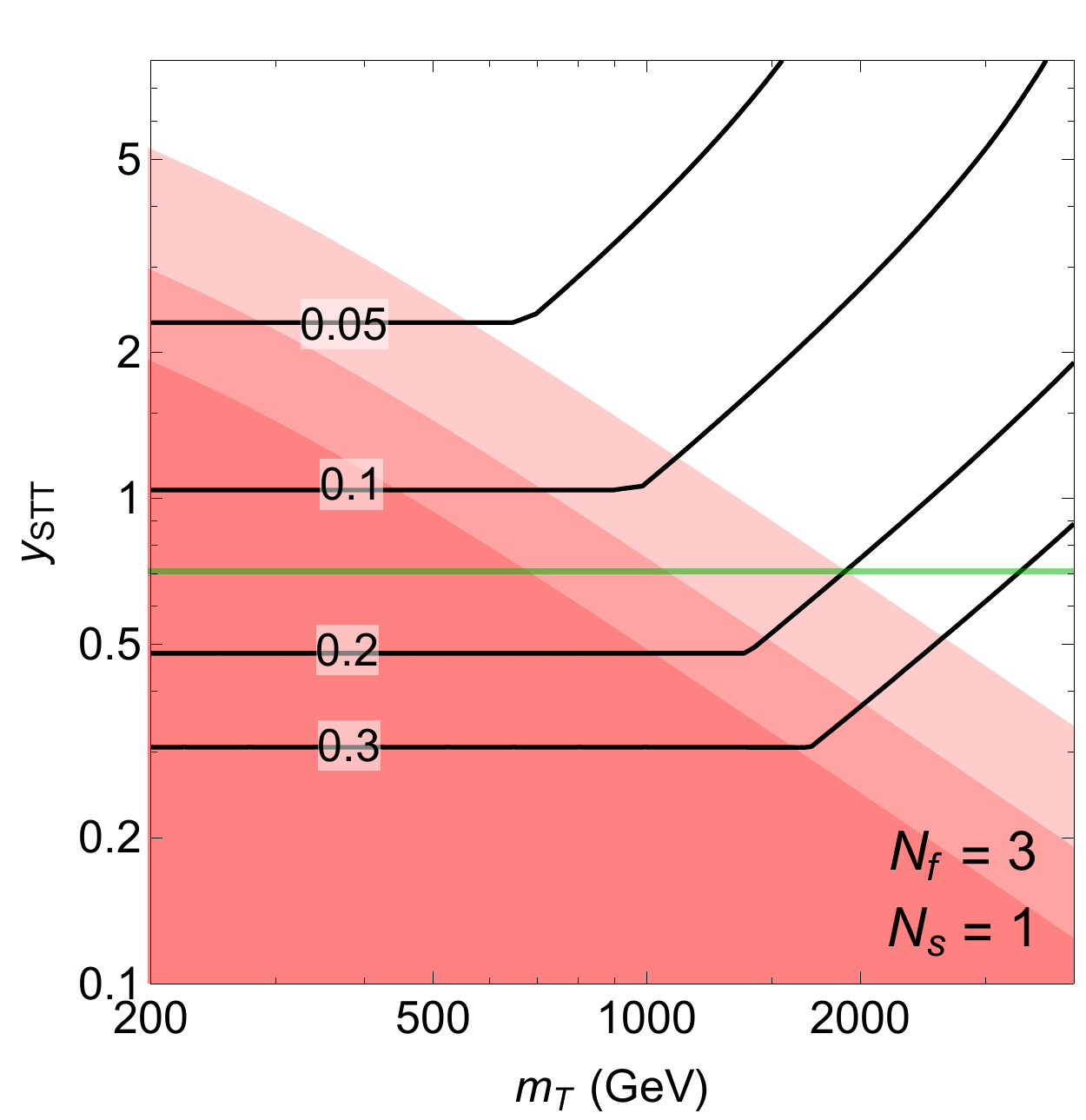}
\end{center}
\caption{
Model-independent sensitivity estimate of future lepton and hadron colliders for theories of Neutral Naturalness, with three fermionic top partners of mass $m_T$, and a single scalar mediator coupling to the top partners with Yukawa coupling $y_{STT}$.
$m_T \gtrsim 500 {\ \rm GeV}$ leads to tuning worse than 10\% due to incomplete cancellation of the top loop. 
\emph{Green horizontal line:} prediction for $y_{STT}$ in the Twin Higgs model.
\emph{Red shading:} reach of future lepton colliders from precision Higgs coupling measurements.
\emph{Black contours:} additional tuning due to the scalar mediator that can be probed by a 100 TeV collider via direct searches, assuming a mass reach for UV completions up to $\sim 20 {\ \rm TeV}$. Requiring both top partner and mediator tunings to be better than 10\% requires either high-scale direct production signals at 100 TeV, or Higgs coupling deviations at lepton colliders, or both.
See text and \cite{Curtin:2015bka} for details.
}
\label{f.scalarmediator}
\end{figure}

However, al already mentioned earlier in this report, it is possible to formulate theories of ``neutral naturalness'' in which the top partners do not in fact carry SM color charge. This can be achieved if the symmetry which protects the Higgs is discrete and does not commute with SM color. Known examples include the folded SUSY~\cite{Burdman:2006tz} and Quirky Little Higgs~\cite{Cai:2008au} models in which the top partners carry electroweak charge but not color, and the Twin Higgs~\cite{Chacko:2005pe} in which the top partners are completely neutral under the SM gauge forces. In these models therefore searches for new colored or even electroweak charged particles do not directly constrain naturalness. While specific neutral naturalness models proposed thus far predict various other new physics signatures (see e.g. \cite{Craig:2015xla, Curtin:2015fna,Chacko:2015fbc, Curtin:2015fna,Cheng:2015buv, Burdman:2015oej}), it is not obvious whether or not future collider experiments can \emph{robustly} constrain neutral naturalness-- i.e., is there a ``no-lose theorem'' indicating that the physics behind electroweak naturalness will be observable?

In ref.~\cite{Curtin:2015bka}, this question was addressed by exploring neutral top partner models using a bottom-up, effective field theory (EFT) approach. Model-independently, a 100 TeV hadron collider would be able to directly probe the interaction of neutral top partners with the Higgs to a modest degree. The top partners can be directly pair-produced through off-shell Higgs bosons, however because these neutral partners may be invisible to detectors this is a challenging signal, requiring the identification of initial state radiation or forward jets recoiling off of missing energy~\cite{Curtin:2014jma, Craig:2014lda}. In typical neutral naturalness models, such direct searches can probe masses up to $\sim 300 {\ \rm GeV}$~\cite{Craig:2014lda,Curtin:2015bka}. Additionally, measurements of double Higgs production at a 100 TeV~\cite{Barr:2014sga, Azatov:2015oxa, He:2015spf} may be able to probe the loop-induced corrections to the triple Higgs coupling for similar top partner masses.

However, the true power of a 100 TeV collider for probing neutral naturalness is its unprecedented reach in energy. As argued in ref.~\cite{Curtin:2015bka}, a common feature of all neutral top partner EFTs is the need for a UV completion of the top partner dynamics~\cite{Batra:2008jy,Barbieri:2015lqa, Low:2015nqa, Geller:2014kta,
Craig:2013fga, Craig:2014fka, Chang:2006ra}, typically at a scale about a loop factor above the top partner mass, e.g. $\sim 10 {\ \rm TeV}$. This defines a new energy scale at which further new states must appear, likely carrying SM color charge. (In the case of known Twin Higgs theories this is the scale at which colored squarks or resonances appear.) As demonstrated explicitly in ref.~\cite{Cheng:2015buv}, a 100 TeV collider can allow access to these high mass states which are a necessary component of neutral naturalness. The resulting constraints on theory are complementary to those from precision measurements of Higgs properties, which can be achieved at future lepton colliders.

Applying this reasoning to EFT scenarios for Neutral Naturalness, ref.~\cite{Curtin:2015bka}  found model-independent arguments regarding the extent to which naturalness would be tested by results from future colliders. To cover the full range of possible neutral naturalness scenarios,  all perturbative neutral top partner structures had to be classified, including those which have yet to be realized in a top-down theory. These include scalar top partners with direct couplings to the Higgs, as well as fermionic top partners which must couple to the Higgs via some additional mediator states. In all cases the top partner EFTs must be completed by further new physics at some UV scale $\Lambda$, e.g. to regulate the divergences of the new scalar particles. Reference \cite{Curtin:2015bka} makes the assumption that at or below the scale $\Lambda$ new colored particles appear, as is realized in all neutral naturalness models known so far~\cite{Batra:2008jy,Barbieri:2015lqa, Low:2015nqa, Geller:2014kta,
Craig:2013fga, Craig:2014fka, Chang:2006ra}, so that new states will be discovered if $\Lambda \lesssim 10 - 20 {\ \rm TeV}$. The constraints on these models from both direct probes of the top partner interactions as well as the reach of a 100 TeV collider for new states at $\Lambda$ was analyzed. In all cases, it was found that untuned models always lead to observable new physics at future colliders, as long as the top partner sector had a similar number of degrees of freedom as the top quark sector.

\begin{figure}
\begin{center}
\includegraphics[width=11cm]{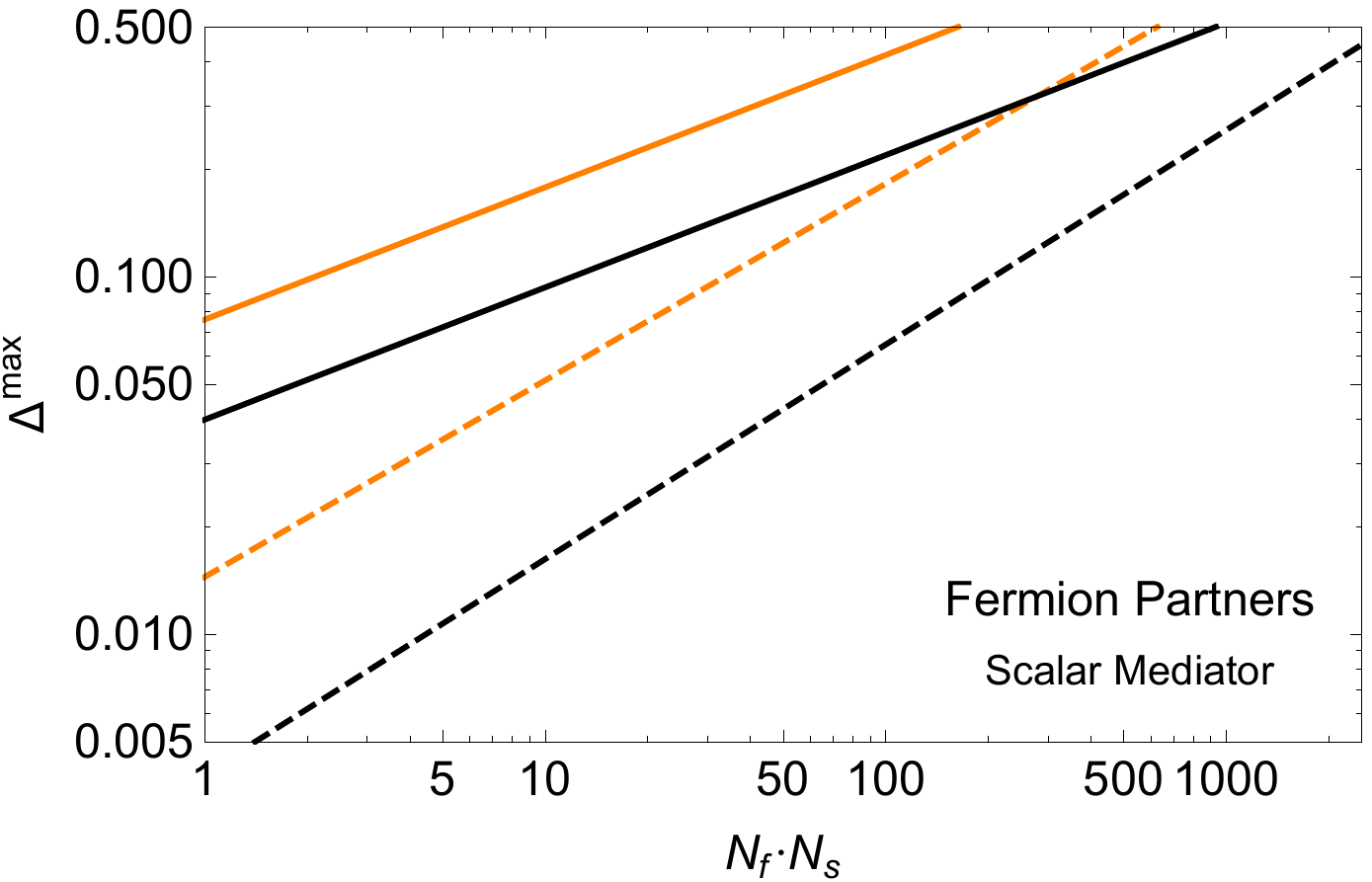}
\end{center}
\caption{
For theories of Neutral Naturalness with $N_f$ fermionic top partners and $N_s$ scalar mediators, $\Delta^\mathrm{max}$ is the ``unavoidable tuning price'' the theory has to pay in order to avoid detection at both future lepton and hadron colliders. The black (orange) curves assume that the 100 TeV collider can probe UV completions at $\sim$ 20 (10) TeV. Dashed curves: combines independent tunings multiplicatively, i.e. $\Delta = \prod_i \Delta_i$. Solid curves: only considers the most severe of several independent tunings, i.e. $\Delta = \mathrm{Min}\{\Delta_i\}$. 
Natural theories where the number of top partners is not large have to produce signals at the 100 TeV collider, future lepton colliders, or both.
See text and \cite{Curtin:2015bka} for details.
}
\label{f.scalarmediator2}
\end{figure}

A representative example of these methods is provided by the case of fermionic top partners $T$ interacting through one or more scalar mediators $S$, as shown below. 
\begin{center}
\begin{tabular}{m{4cm}m{2cm}m{4cm}}
\includegraphics[width=3cm]{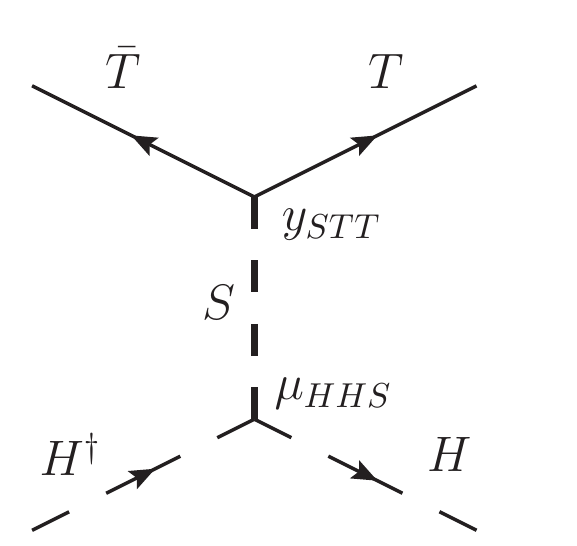} &
$\longrightarrow$ &
\includegraphics[width=3cm]{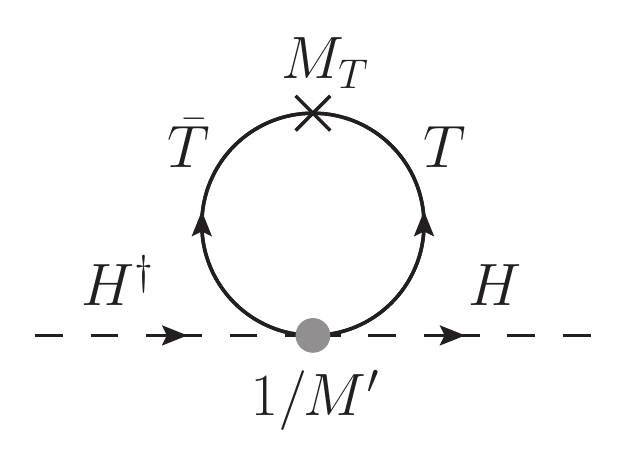}
\end{tabular}
\end{center}
Integrating out the heavy scalar(s) leads to the $|H|^2 \bar T T$ effective interaction, which cancels the quadratically divergent top loop at low scales. 
 SUSY Twin Higgs theories~\cite{Chang:2006ra, Craig:2013fga} are top-down models falling into this category, though the EFT approach applies more generally. 
 In this class of models the Higgs mixes with the singlet scalar mediator, producing deviations in the effective Higgs coupling which can be probed most effectively at lepton colliders, see red shading in Fig.~\ref{f.scalarmediator}. 
In the Twin Higgs model, this allows TeV-scale top partners to be detected. On the other hand, light top partners could escape detection in the fully model-independent case if  $y_{STT}$ is large. In order for this to be natural, the UV completion scale would have to be quite low. Since the 100 TeV collider can probe UV completions up to $\sim 10 {\ \rm TeV}$, these natural theories will lead to direct production of new states. 
The black contours in Fig.~\ref{f.scalarmediator} show the \emph{additional} tuning suffered \emph{only due to the singlet mediator}, if the UV completion scale is high enough to avoid direct production of new states at a 100 TeV collider. This effectively probes large values of $y_{STT}$ in natural theories inaccessible to lepton colliders. Combined with the requirement that top partners not be much heavier than 500 GeV to avoid incomplete cancellation of the top loop, it leads to the conclusion that every natural theory with SM-charged states at the UV completion scale can be discovered. 

This argument generalizes beyond the canonical case of  three top partners ($N_f = 3$) and one scalar mediator ($N_s = 1$). More generally, one can define an \emph{unavoidable tuning price} that a theory of Neutral Naturalness has to pay in order to avoid \emph{both} Higgs coupling deviations at lepton colliders \emph{and} direct production of new states at 100 TeV. This is shown as a function of $(N_f \ \cdot \ N_s)$ in Fig.~\ref{f.scalarmediator2}. Natural theories where the number of top partners is not large have to produce signals at the 100 TeV collider, future lepton colliders, or both.

Different tuning arguments have to be constructed for different scenarios of Neutral Naturalness (e.g. scalar top partners), but the principle of the argument remains the same, as do the conclusions.  
Combining the results for all top partner scenarios, ref.~\cite{Curtin:2015bka} finds that neutral naturalness models are generically observable at future colliders unless they are tuned at the $\sim 10\%$ level or worse. Within the abovementioned assumptions this provides a ``no-lose theorem'' for future colliders as model-independent probes of naturalness. Avoiding this result by violating requires very exotic model-building, such as extremely large top partner multiplicity, or solutions to the hierarchy problem not based on symmetry at all (e.g.~\cite{Dienes:2001se, Graham:2015cka, Dong:2015gya}). Results from future colliders will therefore provide a qualitative advance in our understanding of the origin of the electroweak scale beyond what can be achieved at the LHC.

\subsection{Non-Resonant Signatures}

\subsubsection{Measuring Top Couplings via $tW/tZ$ Scattering}
%\begin{itemize}
%\item J.A.~Dror, M.~Farina, E.~Salvioni, J.~Serra, {\it Strong $tW$ scattering} {\bf contribution received}
%\end{itemize}
Although the top quark was discovered more than twenty years ago, some of its properties are still poorly known. In particular, only recently the couplings of the top to the electroweak $Z$ gauge boson have been directly probed, in $t \bar t Z$ production at the LHC \cite{Baur:2004uw}, though with uncertainties that are currently several times the SM values, while projected sensitivities at Run-II are barely below $100\%$ \cite{Rontsch:2014cca}. 
The lack of experimental precision is due to the complicated environment in hadronic machines, aggravated by the relatively high mass thresholds.
However, in ref.~\cite{Dror:2015nkp} a different approach to probe the properties of the top was put forward that takes advantage of the high energies accessible at hadronic machines: 
certain scattering amplitudes, such as $t W \to t W$, grow quadratically with momenta whenever the electroweak couplings of the top deviate from their SM predictions. 
Such a behaviour is reminiscent of $WW$ scattering when the Higgs couplings to the electroweak gauge bosons depart from the SM \cite{Bellazzini:2012tv}, and it is a genuine signal of models where the top quark, along with the Higgs, is part of a strongly interacting sector \cite{Pomarol:2008bh}.%
\footnote{Indeed, its large mass indicates that the top quark is a key player in composite Higgs scenarios, and crucial BSM particles such as the top-partners \cite{DeSimone:2012ul} could potentially be exchanged in $t W$ scattering.}

As shown in Fig.~\ref{fig:tW}, $t W$ scattering participates in the process $pp \to t \bar t W j$, giving rise to a clean same-sign leptons signature. 
A machine such as a hadron collider at $100$ TeV would significantly profit from the enhanced sensitivity to non-standard top couplings at high energies present in this channel, thanks to the large momenta carried by the initial state partons. This is true already at the inclusive level. The dominant background for such a search is expected to come from QCD production of $pp \to t \bar{t}W$+0(1)\,jets, which arises at $O(g_s^{2(3)}g_w)$ and has a cross section $\sigma_{\mathrm{QCD}} \approx 25$\,pb. The signal arises at $O(g_s g_w^{3})$, with a cross section $\sigma_\mathrm{EW} \approx 4$\,pb (cross sections computed at LO with MadGraph5 \cite{Alloul:2013bka} and a custom FeynRules \cite{Alwall:2014hca} model). These numbers should be compared with the QCD and EW cross sections at the 13 TeV LHC, of $\approx 0.7$\,pb and $\approx 0.06$\,pb, respectively. 
Nevertheless, the potential improvement in sensitivity can be best seen by studying the unique kinematical features of the final state particles.

Let us be specific and focus on the $Z$ coupling to the right-handed top quark, 
\begin{equation} 
c_R \, g_{Z t_R t_R} \bar t_R \gamma_\mu t_R Z^\mu \, ,
\end{equation} 
 where $g_{Z t_R t_R} = - {2 \over 3}(g s_w^2/c_w)$ and $c_R = 1$ in the SM. 
The effect on this coupling from heavy new physics can be effectively parametrised by the dimension-6 operator \cite{Dror:2015nkp}
\begin{equation}
\frac{i \bar c_R}{v^{2}} H^{\dagger} \overleftrightarrow{D_{\mu}} H \bar{t}_{R}\gamma^{\mu}t_{R} \, ,
\label{cRoperator}
\end{equation}
and gives rise to a deviation from the SM, $c_R - 1 = {3 \over 4} \bar c_R/s_w^2$, of an expected size $\bar c_R \sim g_*^2 v^2/\Lambda^2$, where $\Lambda$ is the mass of the resonance that has been integrated out, and $g_*$ its coupling to the top quark. 
Such a non-standard coupling makes the scattering amplitude $t W \to t W$ grow with energy. 
The leading divergence is given by
\begin{equation}
\mathcal{M} = - \frac{g^2}{2m_W^2} \sqrt{\hat{s}(\hat{s}+\hat{t})} \, \bar c_R + O(\sqrt{\hat s}) \ .
\label{cRamplitude}
\end{equation}
The high energy behaviour of this amplitude has been explicitly shown in ref.~\cite{Dror:2015nkp}. 

\begin{figure}[t!]
\begin{center}
\includegraphics[width=0.4\textwidth]{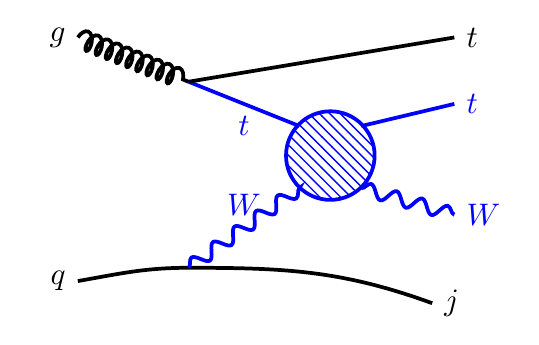}
\end{center}\
\caption{Feynman diagram for the $t W \rightarrow t W $ scattering in $pp$ collisions. Anomalous top couplings lead to the final $tW$ pair having large invariant masses, providing a unique handle to identify the signal.}
\label{fig:tW}
\end{figure}

Here we directly focus on the effects that such a high energy growth has on the kinematical variables associated with $t \bar t W j$ production. 
In particular, for a sizeable $\bar c_R$ the particles that participate in the strong scattering, the $W$ and either one of the two tops (the other is a spectator), will have larger invariant masses than in the SM. 
\begin{figure}[t]
\begin{center}
\includegraphics[width=0.6\textwidth]{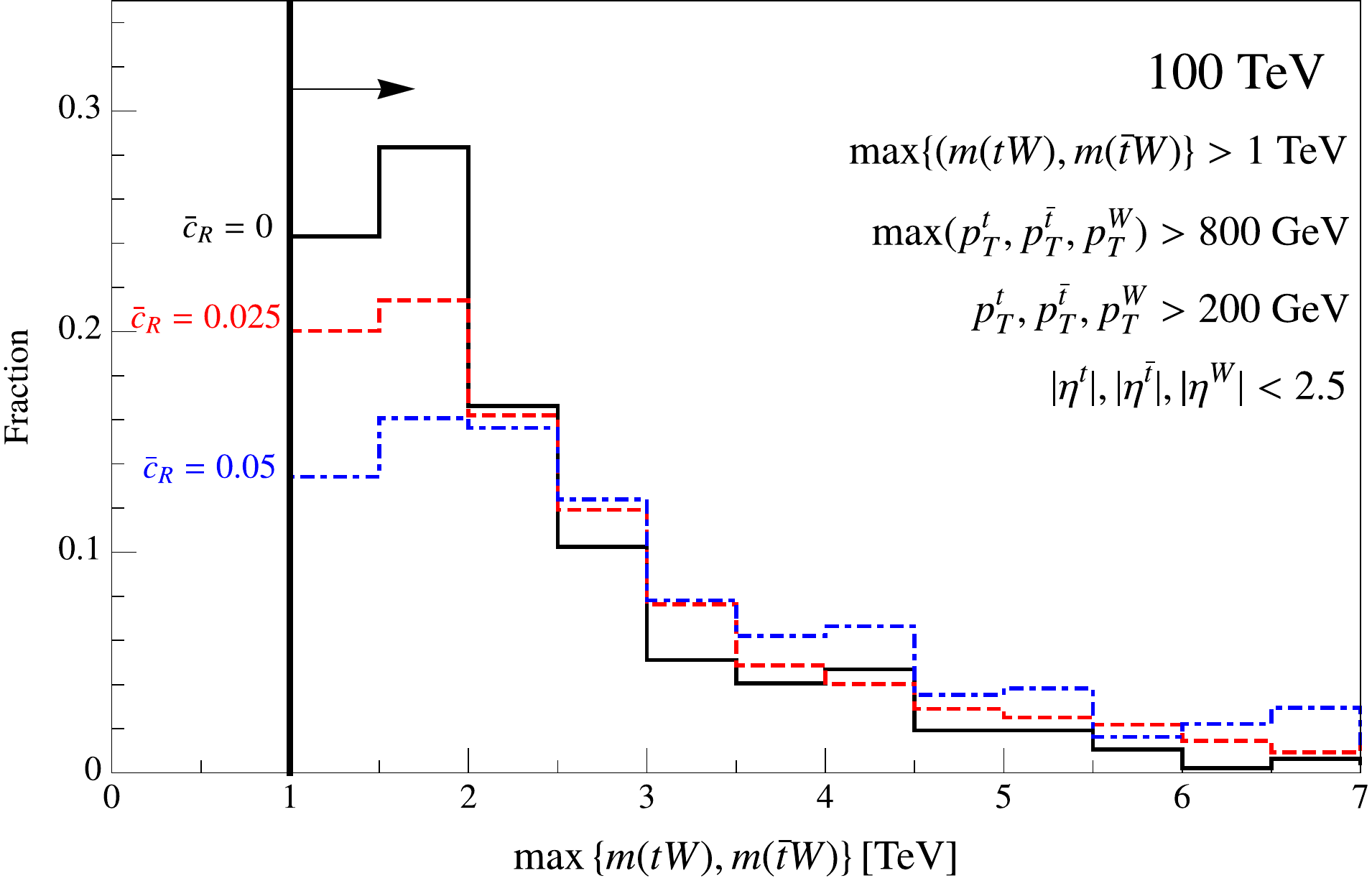}
\end{center}
\caption{Invariant mass distributions for the $ t \bar{t} W j $ electroweak production at a $ 100 \text{ TeV} $ collider. We applied some benchmark cuts (inset top-right) on the tops and the $W$, specifically on the transverse momentum ($ p _T $), pseudorapidity ($ \eta $), and invariant mass ($ m  $).}
\label{fig:mdistribution}
\end{figure}
This is depicted in Fig.~\ref{fig:mdistribution}, where we show the (normalized) distribution of events in a 100 TeV collider as a function of the maximum invariant mass between the pairs $t W$ and $\bar t W$, for the set of cuts shown in the legend.%
\footnote{One should be aware that at a 100 TeV collider and for large invariant masses there could be large logarithms arising from the collinear singularity of the gluon splitting. These have been partly tamed by cutting on the $p _T$ of the tops.} 
The events in the presence of anomalous $Z t_R t_R$ couplings are typically harder than in the SM ($\bar c_R = 0$). The power of a 100 TeV collider in performing this type of ``precision'' probes of the top couplings is apparent once we notice that the values of $\bar c_R$ used for the distributions are an order of magnitude smaller than those that the LHC will be able to probe after $300 \, \textrm{fb}^{-1}$ of integrated luminosity ($\bar c_R \approx 0.3$ \cite{Dror:2015nkp}).
Awaiting for a detailed study, the improvement in sensitivity can be estimated by assuming that the a 100 TeV collider will be able to measure cross sections with absolute uncertainties at the same level as at the LHC (a sensible assumption given $\mathcal{L} = 10 \, \textrm{ab}^{-1}$), but for energies a factor $\sqrt{s_{\textrm{100 TeV}}/s_{\textrm{13 TeV}}} = 100/13 \approx 8$ larger. 
Recalling that the new physics effects we are interested in grow as $ \bar c_R \hat s$ (see eq.~(\ref{cRamplitude})), we can then expect to probe at a 100 TeV collider values of $\bar c_R$ at the per cent to per mille level (similar conclusions hold for the couplings of the left-handed top). 

It is conceivable then that through a careful study of $pp \to t \bar t W j$ production, a 100 TeV collider would be able to greatly improve our sensitivity to new physics modifying the top-$Z$ couplings. Furthermore, as explained in ref.~\cite{Dror:2015nkp} this is not the only process which shows a strong high energy behaviour in the presence of non-standard top couplings. One prominent example is $ t Z \rightarrow t h $ scattering, identified through $t\bar{t}h$+jets production, which would also constitute an important (and complementary) probe of the nature of the top-Higgs sector using a 100 TeV collider.

\subsubsection{Running Electroweak Couplings as a Probe of New Physics}
%List of contributions:
%\begin{itemize}
%\item D.S.M.~Alves, J.~Galloway, J.~Ruderman, J.R.~Walsh, {\it Running Electroweak Couplings as a Probe of New Physics} {\bf contribution received}
%\end{itemize}
In this report has clearly emerged how a future 100 TeV collider can improve in the 
%The search for new physics would enjoy a dramatic improvement at a hadron collider operating with center of mass energy $\sqrt s = 100 \, {\rm TeV}$.  The improvement for 
production of heavy states. However, we will now argue that there are also novel opportunities for precision studies that could uncover new light (relative to $\sqrt s$) states indirectly.  In particular, any new states interacting with gauge bosons will impact how the associated gauge couplings evolve with energy, thereby providing a model-independent handle on their existence provided sufficiently clean channels involving these couplings can be identified and studied experimentally.  Such a possibility has been demonstrated for electroweak (EW) processes \cite{Alves:2014cda} to be discussed below, with similar applications possible in the colored sector of the theory \cite{Kaplan:2008pt, Becciolini:2014lya}.
Analogous possibilities are also familiar from precision studies at LEP, where constraints on new {\it heavy} fields could be applied through accurate determination of the $Z$ boson properties via the modification of gauge boson propagators by new states.  

Many theories extending the SM introduce several new states coupling to weak gauge bosons, potentially making them promising cases for such indirect tests.  Moreover, the model-independence of this setup amounts to an insensitivity to how these new states may decay, thereby opening the possibility of inferring the presence of new physics that may be difficult to discover directly due to reduction of conventional handles (as is the case with reduced missing energy in supersymmetry (SUSY) for compressed spectra \cite{LeCompte:2011cn, Alwall:2008ve} or in models of Stealth SUSY \cite{Fan:2011yu}) or due to increased backgrounds.  The evolution of EW gauge couplings is fully determined within the SM: the coupling $\alpha_1$ grows in the UV while $\alpha_2$ decreases as shown in Fig.~\ref{fig:WinoSleptons}.  
%%%%%%%%%%%%%%%%%%%%%
\begin{figure}[ht]
\begin{center}
\includegraphics[width=0.5\textwidth]{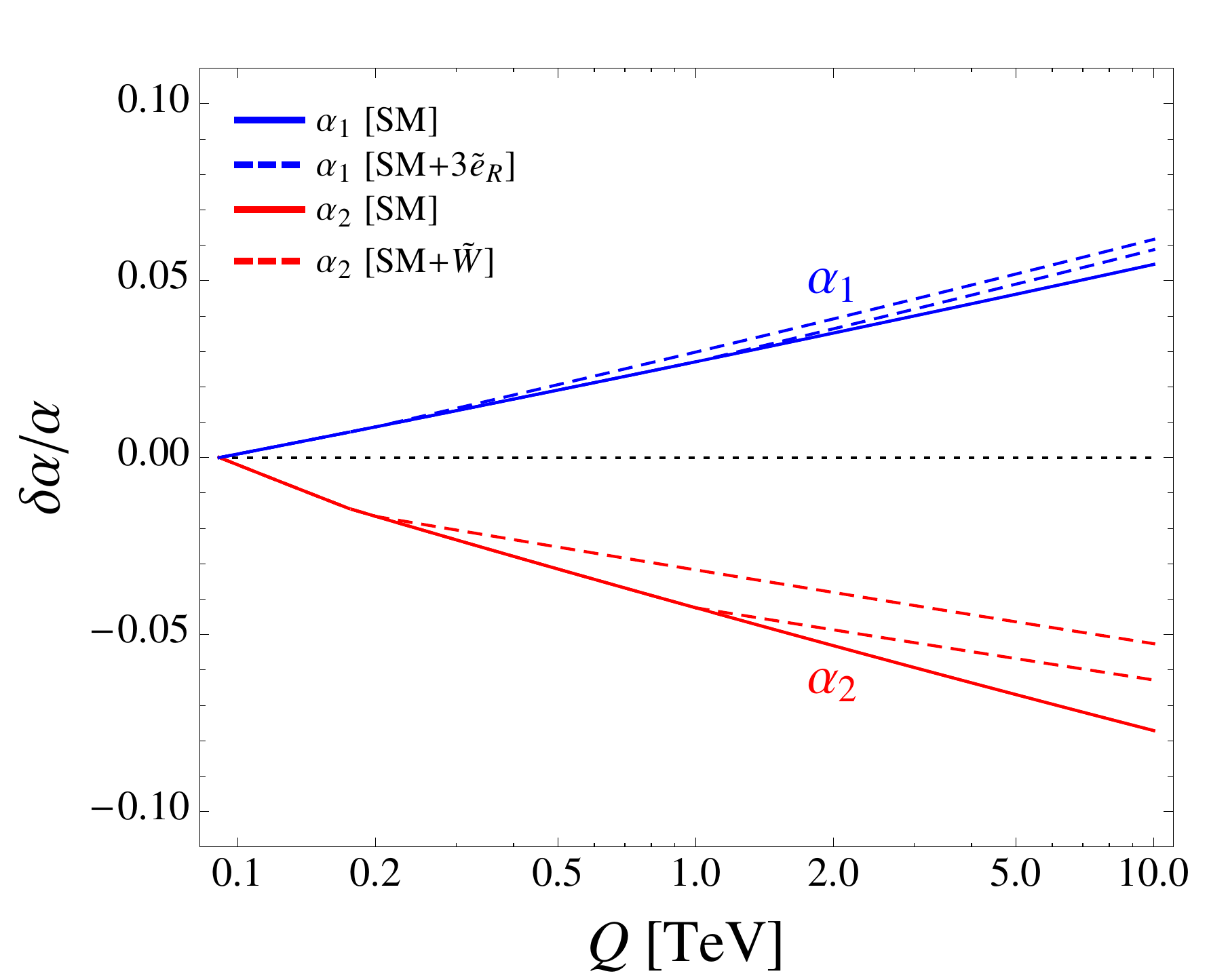}
\caption{\small {Evolution of the two EW gauge couplings within the SM (solid lines) and in the presence of new states of the MSSM (dashed lines).  Shown are the contributions to running of $\alpha_2$ from the presence of a triplet fermion (wino), and to the running of $\alpha_1$ in the presence of three $SU(2)_L$ singlet scalars of hypercharge $1$ (right-handed sleptons); each are shown assuming the new states entering at either $200 \, {\rm GeV}$ or $1 \, {\rm TeV}$.  Figure from \cite{Alves:2014cda}.}}
\label{fig:WinoSleptons}
\end{center}
\end{figure}
%%%%%%%%%%%%%%%%%%%%%
At energies above the mass of any charged particles extending the SM, however, this behavior can change: new fields contribute to the beta functions at scales above their masses, such that
the asymptotic freedom of $\alpha_2$ may no longer persist.  Indeed in the minimal supersymmetric Standard Model (MSSM) the sign of $\alpha_2$'s beta function is flipped once all superpartners are included, such that above that threshold the coupling will increase in the UV.  Even the qualitative running behavior can thus serve as a consistency check of the SM itself, or as an indirect probe of new fields if they exist.  As indicated in Fig.~\ref{fig:WinoSleptons}, however, deviations in the running coupling are typically of order $1\%$ after a decade of running in the presence of an isolated new state of the MSSM.  As such, percent-level experimental precision is needed in order to assess cases in which beta functions are modified by a single field at a given threshold.  

Experimental sensitivity to the running of $\alpha_{1,2}$ relies on minimizing uncertainties in the process under examination.  Statistical uncertainties are minimized by identifying a process whose cross-section remains sizable at high energy; theoretical uncertainties are minimized for processes that are well determined theoretically; and experimental uncertainties are minimized for processes that are sufficiently clean.  Drell-Yan (DY) processes proceeding through neutral and charged currents satisfy these three criteria.  At hadron colliders, both $\alpha_1$ and $\alpha_2$ can thus be sensitively probed with the (neutral current) process $pp \to Z^*/\gamma^* \to \ell^+ \ell^-$, while $\alpha_2$ is constrained with the (charged current) process $pp \to W^{\pm *} \to \ell^\pm \nu$ \cite{Rainwater:2007qa}.  Modifications to the running of $\alpha_{1,2}$ may be observed in the shape of the dilepton invariant mass spectrum in the neutral current case, while in the charged current case the shape of the transverse mass spectrum is modified:
\begin{eqnarray}
\frac{\textrm{d}\sigma}{\textrm{d} M_{\ell \ell}} (pp \to Z^*/\gamma^* \to \ell^+\ell^-) &\equiv& \frac{\textrm{d}\sigma^{Z/\gamma}}{\textrm{d} M_{\ell \ell}} \bigl(\alpha_{1,2} (Q= M_{\ell \ell} )\bigr) \\
\frac{\textrm{d}\sigma}{\textrm{d} M_T} (pp \to W^* \to \ell\nu) &\equiv& \int_{M_T}^{\infty} \textrm{d} M_{\ell \nu} \frac{\textrm{d}\sigma^{W^\pm}}{\textrm{d} M_T \textrm{d} M_{\ell \nu}} \bigl(\alpha_2 (Q=M_{\ell \nu}) \bigr) \, .
\end{eqnarray}
Thus both effects rely on the fact that the couplings are evaluated at a scale, $Q$, corresponding to the invariant mass of the final state.

The main uncertainties impacting the precision with which final state distributions of DY processes can be used to constrain running couplings are statistical, theoretical (scale and PDF), and experimental.  Statistical uncertainties are sufficiently small, for an integrated luminosity of $3 \, {\rm ab}^{-1}$ at $100 \, {\rm TeV}$, assuming final states with $M_{\ell \ell}, M_{\ell \nu} \lesssim 3 \, {\rm TeV}$.  At these energies, theoretical uncertainties entering through scale and PDF are $\lesssim 1-2 \%$, determined using the generators {\tt DYNNLO} and {\tt FEWZ} \cite{Catani:2007vq, Catani:2009sm, Melnikov:2006kv, Gavin:2010az, Gavin:2012sy, Li:2012wna} with the NNPDF2.3 PDF set \cite{Ball:2012cx}.  Finally experimental uncertainties are assumed to be similar to those of the LHC, where neutral current DY measurements at $7$ and $8 \, {\rm TeV}$ indicate uncorrelated uncertainties again at the level of $1-2\%$ \cite{Aad:2013iua, Chatrchyan:2013tia, CMS:2014jea}.  Treating these uncertainties accordingly, the significance with which a $100 \, {\rm TeV}$ collider is indirectly sensitive to typical SUSY states is as shown in Fig.~\ref{fig:SU2multiplets}.  Shown is also a comparison to how well the $14\, {\rm TeV}$ LHC can perform analogous measurements assuming the running of $\alpha_2$ is as in the MSSM.  
%%%%%%%%%%%%%%%%%%%%%
\begin{figure}[ht]
\begin{center}
\includegraphics[width=0.9\textwidth]{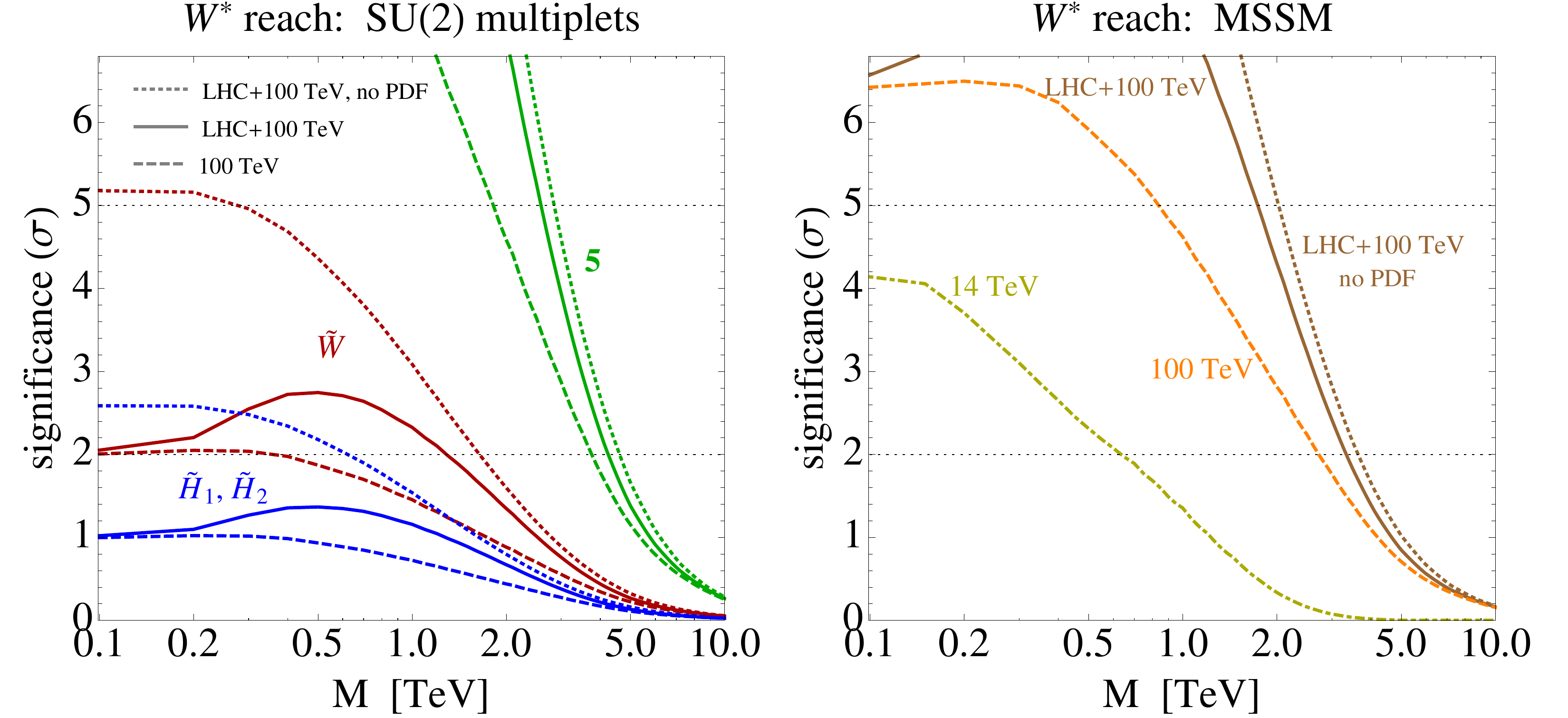}
\caption{\small {Sensitivity to new states contributing to the running of $SU(2)_L$.  {\it Left}: Sensitivity at $100 \, {\rm TeV}$ and its combination with the LHC to Higgsinos, wino, and a {\bf 5} of $SU(2)_L$.  {\it Right}: Sensitivity of the same machine(s) to the entire MSSM entering the running of $\alpha_2$.  Figures from \cite{Alves:2014cda}.}}
\label{fig:SU2multiplets}
\end{center}
\end{figure}
%%%%%%%%%%%%%%%%%%%%%

A general treatment can be carried out by comparing sensitivity to states of a mass, $M$, with contributions $\Delta b_{1,2}$ to the two EW gauge couplings.  At the leading log level, this parameter space depends only on the representation (charge) of the new states; dependence on the spin of the new states enters through their finite contribution to gauge boson propagation, which must be accounted for only in higher order matching.  Thus working at leading log level, results are as shown in Fig.~\ref{fig:model_indep} for current and future runs of the LHC, together with comparisons to what can be gained at $100 \, {\rm TeV}$ and to what is learned through precision studies at LEP where new physics effects could be observed through the presence of higher dimension operators that may be generated upon integrating out heavy states.
%%%%%%%%%%%%%%%%%%%%%
\begin{figure}[t]
\begin{center}
\includegraphics[width=0.9\textwidth]{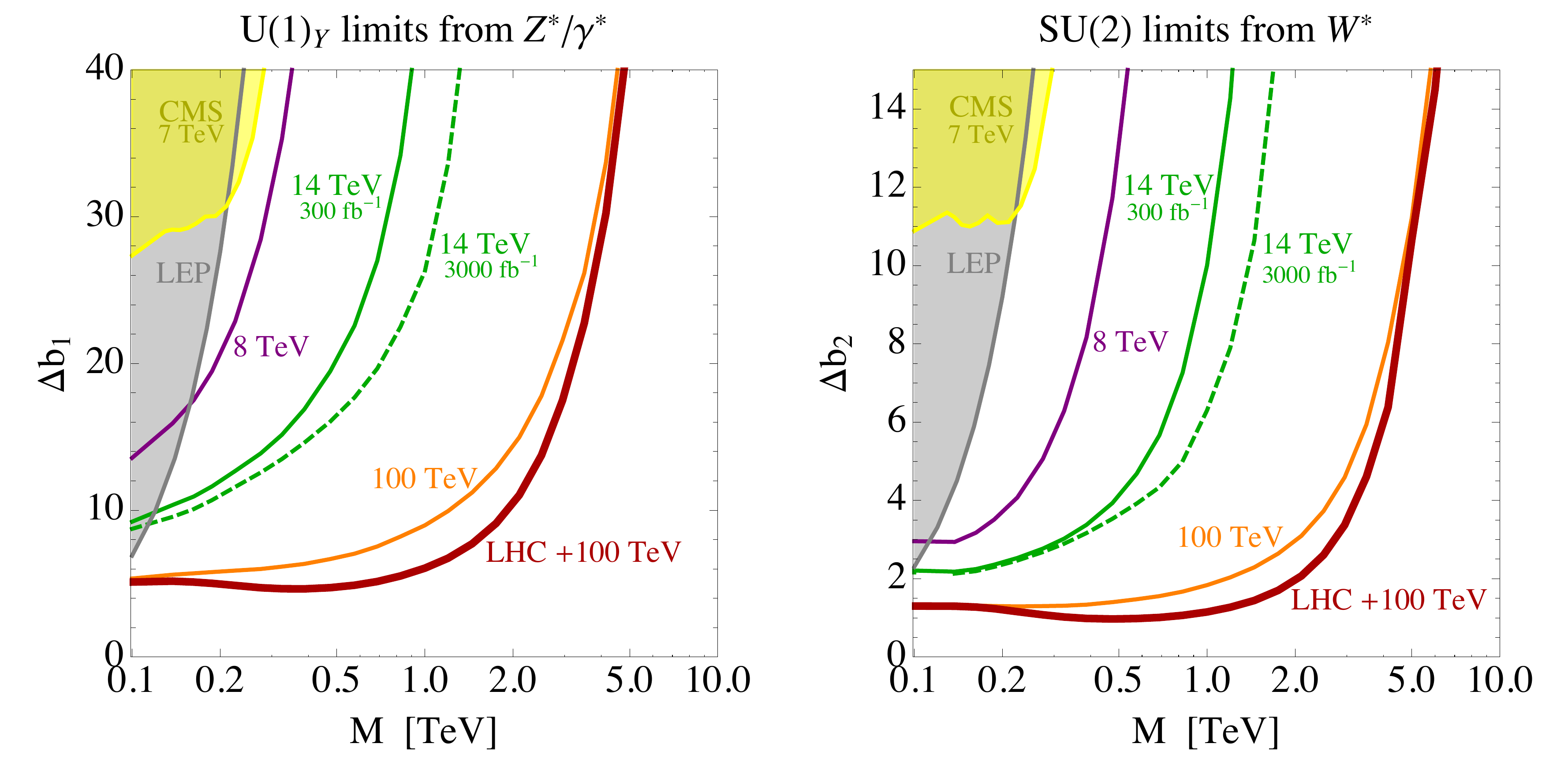}
\caption{\small {Sensitivity of past, current, and future colliders to generic new states of mass $M$ contributing to beta functions of the EW gauge group.  {\it Left}: results for states charged under hypercharge, determined through neutral current DY.  {\it Right}: results from $SU(2)$ representations using charged current DY.}}
\label{fig:model_indep}
\end{center}
\end{figure}
%%%%%%%%%%%%%%%%%%%%%

The effect of systematic uncertainties on the sensitivity of hadron colliders in extracting running coupling information is a crucial consideration of this program.  Fig.~\ref{fig:reachBands} 
%%%%%%%%%%%%%%%%%%%%%
\begin{figure}[t]
\begin{center}
\includegraphics[width=0.9\textwidth]{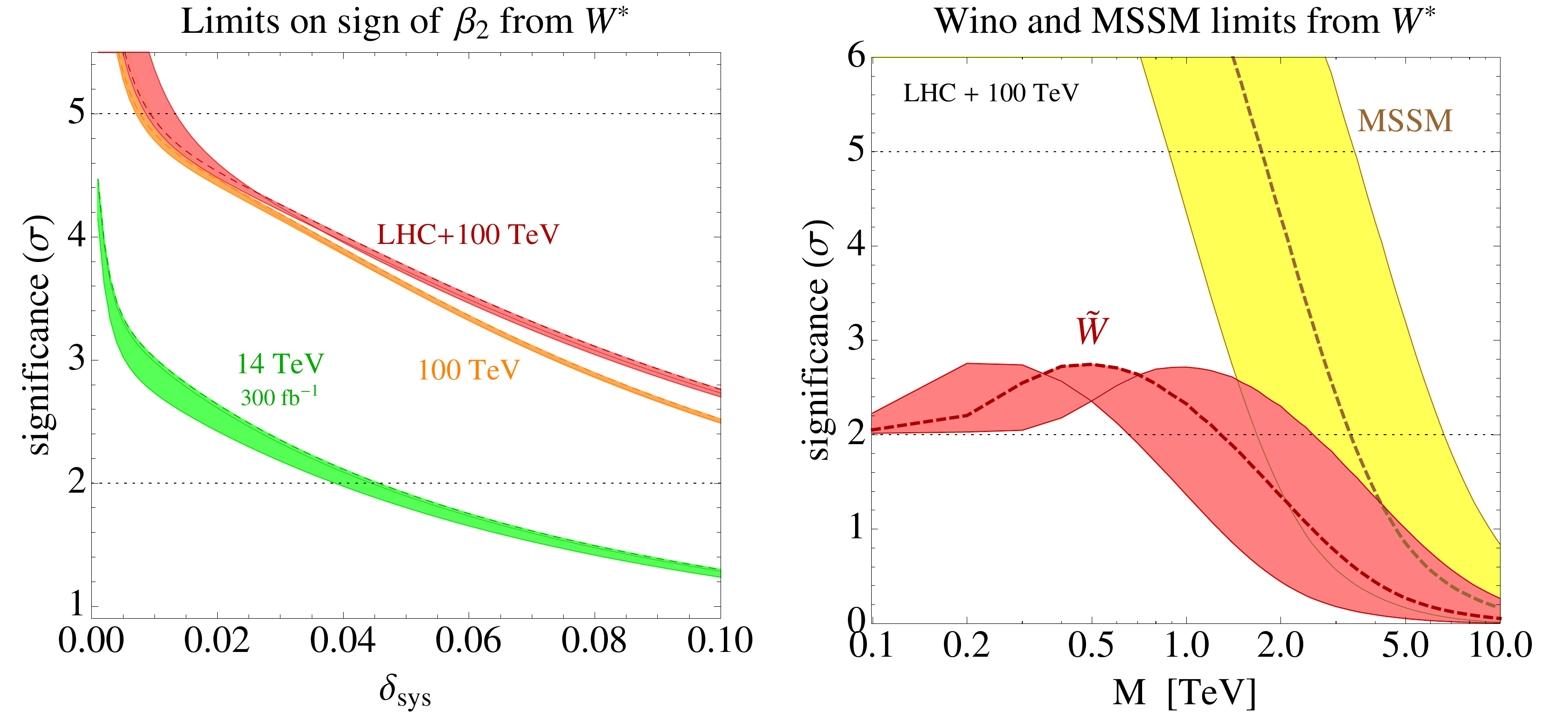}
\caption{\small {Impact of varying systematic uncertainties.  {\it Left}: Variation in sensitivity to the sign of $\alpha_2$'s beta function with respect to the overall size of uncorrelated systematic uncertainties. {\it Right}: Variation in sensitivity to contributions to running $\alpha_2$ from a wino and the entire MSSM entering at mass $M$, with respect to varying scale choice within a factor of two of $M_{\ell \nu}$.  Figures from \cite{Alves:2014cda}.}}
\label{fig:reachBands}
\end{center}
\end{figure}
%%%%%%%%%%%%%%%%%%%%%
shows how the reach of the LHC and of a $100 \, {\rm TeV}$ machine respond to varying these uncertainties, taking sensitivity to the sign of $\beta_2$ and to the presence of a wino or the full MSSM as examples.  A change in sign of $\beta_2$, as would be obtained with the MSSM contributing to the running, can be probed at the $3\sigma$ level between LHC and a future $100\, {\rm TeV}$ collider even as systematic uncertainties approach the $10\%$ level.  The effect of varying scale choice, taking $M_{\ell \nu}/2 \leq Q \leq 2 M_{\ell \nu}$ as the scale at which the EW couplings are evaluated, is shown for the case of a wino or the full MSSM contributing to $\beta_2$: this has the effect of varying the pertinent thresholds that can be constrained within about a factor of four. This analysis is all carried out at leading log order in the EW couplings, and thus the uncertainty band coming from scale choice may be significantly reduced by carrying out a higher order calculation of these processes.  Taking the central scale choice as the fiducial value, a $100 \, {\rm TeV}$ machine can thus provide indirect sensitivity to the presence of a wino up to masses $\approx 1.5 \, {\rm TeV}$ and of the MSSM up to $\approx 3.5 \, {\rm TeV}$.

%
%\subsubsection{High mass $t\bar{t}$}
%List of contributions:
%\begin{itemize}
%\item C.~Helsens, J.~Ferrando, D.~Miller, M.~Mangano, B.~Fuks, J.A.~Aguilar-Saavedra, {\it Boosted tops as a window to new physics} {\bf in SM  section ?}
%\end{itemize}

%\subsubsection{Vector boson fusion ($WW$ scattering)}

%\subsection{Other signatures}
%List of contributions:
%\begin{itemize}
%\item J.~Gluza, G.~Bambhaniya, J.~Chakrabortty, T.~Jeli\'nski, M.~Kordiaczy\'nska, R.~Szafron, {\it In quest of Right-Handed Currents [RHC]} {\bf contribution received}
%\end{itemize}
%
%\subsubsection{Displaced signatures}
%
%\subsubsection{Dark sector signatures}

%- ttbar resonances\\
%- LFV resonance searches: emu, etau, mutau\\
%- WÕ->lnu\\
%- WÕ->top+b-jet\\
%- excited lepton final states (multileptons, l+gamma, llqq)\\
%- excited quarks: gamma+jet and other boson+jet searches\\
%- generic multi-lepton searches\\
%- VLQs\\
%- leptoquarks\\
%Other stuff included in the exotics program at LHC
%- heavy neutrinos\\
%- black hole signatures: multijets, lepton+jet(s),...\\
%- Unconventional signatures: displaced vertices, dark sector (dark Z / photons: lepton-jets), kinked tracks, HIPs, multi-charged particles, magnetic monopoles, ...\\

%We instead aim at analysing different new physics signatures,

\newpage
\section*{Acknowledgements} 
\label{sec:ack}
P.~A.~is supported by NSF grant PHY-1216270.  
\\
T.~C.~is supported by an LHC Theory Initiative Postdoctoral Fellowship, under the National Science Foundation grant PHY-0969510.   
\\
R.~T.~D.~acknowledges support from the Institute for Advanced Study Marvin L. Goldberger Membership and DOE grant de-sc0009988.
\\
S.~E.~and B.~Z.~acknowledge partial support from the U.S. Department of Energy, Office of Science, under grant DE-SC0007859. 
\\
J.~E.~thanks the European Research Council for support via the Advanced Investigator Grant 26732, and from the STFC via Grant ST/L000326/1. 
\\
J.~L.~Feng thanks the Technion Center for Particle Physics for hospitality. The work of J.~L.~Feng is supported in part by U.S. National Science Foundation Grant No. PHY--1316792, United States-Israel Binational Science Foundation Grant No. 2010221, a Guggenheim Foundation grant, and Simons Investigator Award 376204.  
\\
The research of S.~G.~at Perimeter Institute is supported by the Government of Canada through Industry Canada and by the Province of Ontario through the Ministry of Economic Development and Innovation.  
\\
M.~R.~is supported by the NSF Grant PHY-1415548.  
\\
C.~B.~V.~is supported by National Science Foundation Grant No. PHY-1315155 and the Maryland Center for Fundamental Physics. 
\\
J.~W.~is supported in part by the Department of Energy under grant DE-SC0007859.  W.X. is supported by the U.S. Department of Energy (DOE) under cooperative research agreement DE-SC-00012567.
\\
The work of G.~P.~has been partly supported by the Catalan ICREA Academia Program, and grants FPA2014-55613-P, 2014-SGR-1450 and Severo Ochoa excellence program SO-2012-0234.
\\
The work of K.A.O. is supported in part by DOE grant DE-SC0011842 at the University of Minnesota. 
\\
The work of N.~V.~has been partially funded by the Danish National Research Foundation, grant number DNRF90.
\\
The work of K.~M.~has been supported by the National Science Foundation under Grant No. PHY-0854889.
\\
The work of T.~R.~was supported by the Department of Energy, Contract DE-AC02-76SF00515.
\\
The work of R.~N.~M.~is supported in part by the US National Science Foundation Grant No. PHY-1315155.
\\
D.~K.~is supported by the Department of Energy under Grant DE-SC0010296.
\\
%The work of J.~P., J.~L., and S~.C.~has been supported by 
Argonne National Laboratory's work was supported by the U.S. Department of Energy, Office of Science, Office of Basic Energy Sciences, under contract DE-AC02-06CH11357.
\\
D.~C.~is supported by National Science Foundation grant No. PHY-1315155 and the Maryland Center for Fundamental Physics.
\\
H.-C.~C. and E.~S.~were supported in part by the US Department of Energy grant DE-SC-000999. S.~J.~was partly supported by the US Department of Energy under contract DE-AC02-76SF00515. Y.~T.~was supported in part by the US National Science Foundation Grant PHY-1315155, and by the Maryland Center for Fundamental Physics.
\\
The work of A.~K.~was supported by the Fermi National Accelerator Laboratory and by a Department of Energy grant to Duke University. Fermilab is operated by Fermi Research Alliance, LLC, under Contract No. DE-AC02-07CH11359 with the United States Department of Energy.
\\
The work of Y.~S, S.~I.~ and S.~T.~ supported by BSF Grant No.~2010221,  by the ICORE Program of Planning and Budgeting Committee, and by ISF Grant Nos.~1937/12 (Y.~S.~ and S.~I.~) and 1787/11 (S.~T.~).
\\
The work of R.~R.~is supported by Science and Technology Facilities Council (STFC).
\\
The work of C.~W.~is supported by the European Research Council under the European UnionÕs Seventh Framework Programme (FP/2007-2013) / ERC Grant NuMass agreement n. [617143] and partial support from the EU Grant No. FP7 ITN INVISIBLES (Marie Curie Actions, Grant No. PITN-GA-2011-289442).
\\
D.~S.~M.~A.~is supported by the NSF under grants PHY-0947827 and PHY-1316753, and by the LHC Theory Initiative NSF-PHY-0969510.
\\
A.~W.~acknowledges the MIUR- FIRB grant RBFR12H1MW, the ERC Advanced Grant no.267985 (DaMeSyFla) and the UNIPD-PRAT grant  CPDA153129.
\\
J.~K.~, A.~T.~and F.~Y.~acknowledge support from the Cluster of Excellence {\it Precision Physics, Fundamental Interactions and Structure of Matter} (PRISMA -- EXC 1098).
\\
The work of C.-Y.~C is supported by NSERC, Canada. Research at the Perimeter Institute is supported in part by the Government of Canada through NSERC and by the Province of Ontario through MEDT. 
\\
%M.~C., F.~S. and M.~T. acknowledge the hospitality of the Institut d'Astrophysique de Paris ({\sc Iap}).
M.~C., F.~S. and M.~T. acknowledge ERC under the EU Seventh Framework Programme (FP7/2007-2013) Starting Grant (agreement n.\ 278234 --- `{\it NewDark}' project),{\it Erc} Advanced Grant 267117 (`{\it Dark}') hosted by Universit\'e Pierre \& Marie Curie - Paris 6, FNRS {\it Anr} under contract {\it Anr} 2010 {\it Blanc} 041301,Centro de Excelencia Severo Ochoa Programme SEV-2012-0249, MINECO through a Severo Ochoa fellowship with Program SEV-2012-0249,FPA2015-65929-P and Consolider MultiDark CSD2009-00064, and the hospitality of the Institut d'Astrophysique de Paris ({\sc Iap}).
\\
The work of J.~D. is supported by the NSF through grant PHY-1316222 and by the NSERC Grant PGSD3-438393-2013. M.~F. is supported by the DOE Grant DE-SC0003883. 
%JS is grateful to R.~Torre for very interesting discussions.
\\
The work of M.L.~M. is supported by the European Reasearch
Council advanced grant 291377 {\it ``LHCtheory'': Theoretical predictions and
analyses of LHC physics: advancing the precision frontier}. 
\\
G.~N.~is supported by the Swiss National Science Foundation (SNF) under grant 200020-155935.
\\
The work of J.~C.~has been supported by the Department of Science \& Technology, Government of INDIA under the Grant Agreement number IFA12-PH-34 (INSPIRE Faculty Award).
\\
The work of J.~G. and T.~J.~has been supported by the Polish National Center of Science (NCN) under the Grant Agreement number DEC-2013/11/B/ST2/04023 
\\
The work of R.~S. has been supported by the Natural Sciences and Engineering Research Council of Canada (NSERC).
\\
The work of P.S.B.D. is supported by the DFG grant RO 2516/5-1.

\newpage

\bibliographystyle{report}
\bibliography{report}

\providecommand{\href}[2]{#2}\begingroup\raggedright\begin{thebibliography}{100}

\bibitem{benedikt}
{\em {M. Benedikt, ``FCC study overview and status'', talk at the FCC week
  2016, Rome, 11-15 April 2016}\/},
  \url{http://indico.cern.ch/event/438866/contributions/1085016/}.

\bibitem{cepc_website}
{CEPC project website}.
\newblock \url{http://cepc.ihep.ac.cn}.

\bibitem{ArkaniHamed:2015vfh}
N.~Arkani-Hamed, T.~Han, M.~Mangano, and L.-T. Wang, {\em {Physics
  Opportunities of a 100 TeV Proton-Proton Collider}\/},
\href{http://arxiv.org/abs/1511.06495}{{\tt arXiv:1511.06495 [hep-ph]}}.
%%CITATION = ARXIV:1511.06495;%%.

\bibitem{Eichten:1984eu}
E.~Eichten, I.~Hinchliffe, K.~D. Lane, and C.~Quigg, {\em {Super Collider
  Physics}\/},  \href{http://dx.doi.org/10.1103/RevModPhys.56.579,
  10.1103/RevModPhys.58.1065}{Rev. Mod. Phys. {\bf 56} (1984)  579--707}.
[Addendum: Rev. Mod. Phys.58,1065(1986)].
%%CITATION = RMPHA,56,579;%%.

\bibitem{Hinchliffe:2015qma}
I.~Hinchliffe, A.~Kotwal, M.~L. Mangano, C.~Quigg, and L.-T. Wang, {\em
  {Luminosity goals for a 100-TeV pp collider}\/},
\href{http://arxiv.org/abs/1504.06108}{{\tt arXiv:1504.06108 [hep-ph]}}.
%%CITATION = ARXIV:1504.06108;%%.

\bibitem{SM-report}
M.~Mangano, G.~Zanderighi, et al., {\em Physics at a 100 TeV pp collider:
  Standard Model processes\/},  CERN-TH-2016-112, 2016.

\bibitem{Higgs-report}
R.~Contino, D.~Curtin, A.~Katz, M.~L. Mangano, G.~Panico, M.~J. Ramsey-Musolf,
  G.~Zanderighi, et al., {\em Physics at a 100 TeV pp collider: Higgs and EW
  symmetry breaking studies\/},  CERN-TH-2016-113, 2016.

\bibitem{Dainese:2016gch}
A.~Dainese et al., {\em {Heavy ions at the Future Circular Collider}\/},
\href{http://arxiv.org/abs/1605.01389}{{\tt arXiv:1605.01389 [hep-ph]}}.
%%CITATION = ARXIV:1605.01389;%%.

\bibitem{inj-report}
B.~Goddard, G.~Isidori, F.~Teubert, et al., {\em Physics opportunities with the
  FCC-hh injectors\/},  to appear, 2016.

\bibitem{Gomez-Ceballos:2013zzn}
{TLEP Design Study Working Group Collaboration}, M.~Bicer et al., {\em {First
  Look at the Physics Case of TLEP}\/},
  \href{http://dx.doi.org/10.1007/JHEP01(2014)164}{JHEP {\bf 01} (2014)  164},
\href{http://arxiv.org/abs/1308.6176}{{\tt arXiv:1308.6176 [hep-ex]}}.
%%CITATION = ARXIV:1308.6176;%%.

\bibitem{AbelleiraFernandez:2012cc}
{LHeC Study Group Collaboration}, J.~L. Abelleira~Fernandez et al., {\em {A
  Large Hadron Electron Collider at CERN: Report on the Physics and Design
  Concepts for Machine and Detector}\/},
  \href{http://dx.doi.org/10.1088/0954-3899/39/7/075001}{J. Phys. {\bf G39}
  (2012)  075001},
\href{http://arxiv.org/abs/1206.2913}{{\tt arXiv:1206.2913 [physics.acc-ph]}}.
%%CITATION = ARXIV:1206.2913;%%.

\bibitem{Cacciari:2014gra}
M.~Cacciari, G.~P. Salam, and G.~Soyez, {\em {SoftKiller, a particle-level
  pileup removal method}\/},
  \href{http://dx.doi.org/10.1140/epjc/s10052-015-3267-2}{Eur. Phys. J. {\bf
  C75} (2015) no.~2, 59},
\href{http://arxiv.org/abs/1407.0408}{{\tt arXiv:1407.0408 [hep-ph]}}.
%%CITATION = ARXIV:1407.0408;%%.

\bibitem{Berta:2014eza}
P.~Berta, M.~Spousta, D.~W. Miller, and R.~Leitner, {\em {Particle-level pileup
  subtraction for jets and jet shapes}\/},
  \href{http://dx.doi.org/10.1007/JHEP06(2014)092}{JHEP {\bf 06} (2014)  092},
\href{http://arxiv.org/abs/1403.3108}{{\tt arXiv:1403.3108 [hep-ex]}}.
%%CITATION = ARXIV:1403.3108;%%.

\bibitem{Bertolini:2014bba}
D.~Bertolini, P.~Harris, M.~Low, and N.~Tran, {\em {Pileup Per Particle
  Identification}\/},  \href{http://dx.doi.org/10.1007/JHEP10(2014)059}{JHEP
  {\bf 10} (2014)  59},
\href{http://arxiv.org/abs/1407.6013}{{\tt arXiv:1407.6013 [hep-ph]}}.
%%CITATION = ARXIV:1407.6013;%%.

\bibitem{Schaetzel:2013vka}
S.~Schaetzel and M.~Spannowsky, {\em {Tagging highly boosted top quarks}\/},
  \href{http://dx.doi.org/10.1103/PhysRevD.89.014007}{Phys. Rev. {\bf D89}
  (2014) no.~1, 014007},
\href{http://arxiv.org/abs/1308.0540}{{\tt arXiv:1308.0540 [hep-ph]}}.
%%CITATION = ARXIV:1308.0540;%%.

\bibitem{Larkoski:2015yqa}
A.~J. Larkoski, F.~Maltoni, and M.~Selvaggi, {\em {Tracking down hyper-boosted
  top quarks}\/},  \href{http://dx.doi.org/10.1007/JHEP06(2015)032}{JHEP {\bf
  06} (2015)  032},
\href{http://arxiv.org/abs/1503.03347}{{\tt arXiv:1503.03347 [hep-ph]}}.
%%CITATION = ARXIV:1503.03347;%%.

\bibitem{Spannowsky:2015eba}
M.~Spannowsky and M.~Stoll, {\em {Tracking New Physics at the LHC and
  beyond}\/},  \href{http://dx.doi.org/10.1103/PhysRevD.92.054033}{Phys. Rev.
  {\bf D92} (2015) no.~5, 054033},
\href{http://arxiv.org/abs/1505.01921}{{\tt arXiv:1505.01921 [hep-ph]}}.
%%CITATION = ARXIV:1505.01921;%%.

\bibitem{Bressler:2015uma}
S.~Bressler, T.~Flacke, Y.~Kats, S.~J. Lee, and G.~Perez, {\em {Hadronic
  Calorimeter Shower Size: Challenges and Opportunities for Jet Substructure in
  the Superboosted Regime}\/},
\href{http://arxiv.org/abs/1506.02656}{{\tt arXiv:1506.02656 [hep-ph]}}.
%%CITATION = ARXIV:1506.02656;%%.

\bibitem{Aad:2008zzm}
{ATLAS Collaboration}, G.~Aad et al., {\em {The ATLAS Experiment at the CERN
  Large Hadron Collider}\/},
\href{http://dx.doi.org/10.1088/1748-0221/3/08/S08003}{JINST {\bf 3} (2008)
  S08003}.
%%CITATION = JINST,3,S08003;%%.

\bibitem{Volkov:1973ix}
D.~V. Volkov and V.~P. Akulov, {\em {Is the Neutrino a Goldstone Particle?}\/},
\href{http://dx.doi.org/10.1016/0370-2693(73)90490-5}{Phys. Lett. {\bf B46}
  (1973)  109--110}.
%%CITATION = PHLTA,B46,109;%%.

\bibitem{Wess:1974tw}
J.~Wess and B.~Zumino, {\em {Supergauge Transformations in Four-Dimensions}\/},
\href{http://dx.doi.org/10.1016/0550-3213(74)90355-1}{Nucl. Phys. {\bf B70}
  (1974)  39--50}.
%%CITATION = NUPHA,B70,39;%%.

\bibitem{Freedman:1976xh}
D.~Z. Freedman, P.~van Nieuwenhuizen, and S.~Ferrara, {\em {Progress Toward a
  Theory of Supergravity}\/},
\href{http://dx.doi.org/10.1103/PhysRevD.13.3214}{Phys. Rev. {\bf D13} (1976)
  3214--3218}.
%%CITATION = PHRVA,D13,3214;%%.

\bibitem{Deser:1976eh}
S.~Deser and B.~Zumino, {\em {Consistent Supergravity}\/},
\href{http://dx.doi.org/10.1016/0370-2693(76)90089-7}{Phys. Lett. {\bf B62}
  (1976)  335}.
%%CITATION = PHLTA,B62,335;%%.

\bibitem{Ellis:1983ew}
J.~R. Ellis, J.~S. Hagelin, D.~V. Nanopoulos, K.~A. Olive, and M.~Srednicki,
  {\em {Supersymmetric Relics from the Big Bang}\/},
\href{http://dx.doi.org/10.1016/0550-3213(84)90461-9}{Nucl. Phys. {\bf B238}
  (1984)  453--476}.
%%CITATION = NUPHA,B238,453;%%.

\bibitem{Goldberg:1983nd}
H.~Goldberg, {\em {Constraint on the Photino Mass from Cosmology}\/},
\href{http://dx.doi.org/10.1103/PhysRevLett.50.1419}{Phys.Rev.Lett. {\bf 50}
  (1983)  1419}.
%%CITATION = PRLTA,50,1419;%%.

\bibitem{Hagelin:1984wv}
J.~S. Hagelin, G.~L. Kane, and S.~Raby, {\em {Perhaps Scalar Neutrinos Are the
  Lightest Supersymmetric Partners}\/},
\href{http://dx.doi.org/10.1016/0550-3213(84)90064-6}{Nucl. Phys. {\bf B241}
  (1984)  638}.
%%CITATION = NUPHA,B241,638;%%.

\bibitem{Ibanez:1983kw}
L.~E. Ibanez, {\em {The Scalar Neutrinos as the Lightest Supersymmetric
  Particles and Cosmology}\/},
\href{http://dx.doi.org/10.1016/0370-2693(84)90221-1}{Phys. Lett. {\bf B137}
  (1984)  160}.
%%CITATION = PHLTA,B137,160;%%.

\bibitem{Falk:1994es}
T.~Falk, K.~A. Olive, and M.~Srednicki, {\em {Heavy sneutrinos as dark
  matter}\/},  \href{http://dx.doi.org/10.1016/0370-2693(94)90639-4}{Phys.
  Lett. {\bf B339} (1994)  248--251},
\href{http://arxiv.org/abs/hep-ph/9409270}{{\tt arXiv:hep-ph/9409270
  [hep-ph]}}.
%%CITATION = HEP-PH/9409270;%%.

\bibitem{Dimopoulos:1981yj}
S.~Dimopoulos, S.~Raby, and F.~Wilczek, {\em {Supersymmetry and the Scale of
  Unification}\/},
\href{http://dx.doi.org/10.1103/PhysRevD.24.1681}{Phys. Rev. {\bf D24} (1981)
  1681--1683}.
%%CITATION = PHRVA,D24,1681;%%.

\bibitem{Dimopoulos:1981zb}
S.~Dimopoulos and H.~Georgi, {\em {Softly Broken Supersymmetry and SU(5)}\/},
\href{http://dx.doi.org/10.1016/0550-3213(81)90522-8}{Nucl.Phys. {\bf B193}
  (1981)  150}.
%%CITATION = NUPHA,B193,150;%%.

\bibitem{Martin:1997ns}
S.~P. Martin, {\em {A Supersymmetry primer}\/},
\href{http://arxiv.org/abs/hep-ph/9709356}{{\tt arXiv:hep-ph/9709356
  [hep-ph]}}.
%%CITATION = HEP-PH/9709356;%%.

\bibitem{Georgi:1974sy}
H.~Georgi and S.~Glashow, {\em {Unity of All Elementary Particle Forces}\/},
\href{http://dx.doi.org/10.1103/PhysRevLett.32.438}{Phys.Rev.Lett. {\bf 32}
  (1974)  438--441}.
%%CITATION = PRLTA,32,438;%%.

\bibitem{Arvanitaki:2012ps}
A.~Arvanitaki, N.~Craig, S.~Dimopoulos, and G.~Villadoro, {\em {Mini-Split}\/},
   \href{http://dx.doi.org/10.1007/JHEP02(2013)126}{JHEP {\bf 02} (2013)  126},
\href{http://arxiv.org/abs/1210.0555}{{\tt arXiv:1210.0555 [hep-ph]}}.
%%CITATION = ARXIV:1210.0555;%%.

\bibitem{Wells:2004di}
J.~D. Wells, {\em {Pev-Scale Supersymmetry}\/},
  \href{http://dx.doi.org/10.1103/PhysRevD.71.015013}{Phys. Rev. {\bf D71}
  (2005)  015013},
\href{http://arxiv.org/abs/hep-ph/0411041}{{\tt arXiv:hep-ph/0411041
  [hep-ph]}}.
%%CITATION = HEP-PH/0411041;%%.

\bibitem{ArkaniHamed:2004fb}
N.~Arkani-Hamed and S.~Dimopoulos, {\em {Supersymmetric unification without low
  energy supersymmetry and signatures for fine-tuning at the LHC}\/},
  \href{http://dx.doi.org/10.1088/1126-6708/2005/06/073}{JHEP {\bf 0506} (2005)
   073},
\href{http://arxiv.org/abs/hep-th/0405159}{{\tt arXiv:hep-th/0405159
  [hep-th]}}.
%%CITATION = HEP-TH/0405159;%%.

\bibitem{Giudice:2004tc}
G.~Giudice and A.~Romanino, {\em {Split supersymmetry}\/},
  \href{http://dx.doi.org/10.1016/j.nuclphysb.2004.11.048}{Nucl.Phys. {\bf
  B699} (2004)  65--89},
\href{http://arxiv.org/abs/hep-ph/0406088}{{\tt arXiv:hep-ph/0406088
  [hep-ph]}}.
%%CITATION = HEP-PH/0406088;%%.

\bibitem{Chanowitz:1977ye}
M.~S. Chanowitz, J.~R. Ellis, and M.~K. Gaillard, {\em {The Price of Natural
  Flavor Conservation in Neutral Weak Interactions}\/},
\href{http://dx.doi.org/10.1016/0550-3213(77)90057-8}{Nucl. Phys. {\bf B128}
  (1977)  506--536}.
%%CITATION = NUPHA,B128,506;%%.

\bibitem{Nanopoulos:1978hh}
D.~V. Nanopoulos and D.~A. Ross, {\em {Limits on the Number of Flavors in Grand
  Unified Theories from Higher Order Corrections to Fermion Masses}\/},
\href{http://dx.doi.org/10.1016/0550-3213(79)90507-8}{Nucl. Phys. {\bf B157}
  (1979)  273--284}.
%%CITATION = NUPHA,B157,273;%%.

\bibitem{Georgi:1979df}
H.~Georgi and C.~Jarlskog, {\em {A New Lepton - Quark Mass Relation in a
  Unified Theory}\/},
\href{http://dx.doi.org/10.1016/0370-2693(79)90842-6}{Phys. Lett. {\bf B86}
  (1979)  297--300}.
%%CITATION = PHLTA,B86,297;%%.

\bibitem{Antusch:2009gu}
S.~Antusch and M.~Spinrath, {\em {New GUT predictions for quark and lepton mass
  ratios confronted with phenomenology}\/},
  \href{http://dx.doi.org/10.1103/PhysRevD.79.095004}{Phys. Rev. {\bf D79}
  (2009)  095004},
\href{http://arxiv.org/abs/0902.4644}{{\tt arXiv:0902.4644 [hep-ph]}}.
%%CITATION = ARXIV:0902.4644;%%.

\bibitem{Antusch:2013rxa}
S.~Antusch, S.~F. King, and M.~Spinrath, {\em {GUT predictions for quark-lepton
  Yukawa coupling ratios with messenger masses from non-singlets}\/},
  \href{http://dx.doi.org/10.1103/PhysRevD.89.055027}{Phys. Rev. {\bf D89}
  (2014) no.~5, 055027},
\href{http://arxiv.org/abs/1311.0877}{{\tt arXiv:1311.0877 [hep-ph]}}.
%%CITATION = ARXIV:1311.0877;%%.

\bibitem{Hempfling:1993kv}
R.~Hempfling, {\em {Yukawa coupling unification with supersymmetric threshold
  corrections}\/},
\href{http://dx.doi.org/10.1103/PhysRevD.49.6168}{Phys. Rev. {\bf D49} (1994)
  6168--6172}.
%%CITATION = PHRVA,D49,6168;%%.

\bibitem{Hall:1993gn}
L.~J. Hall, R.~Rattazzi, and U.~Sarid, {\em {The Top quark mass in
  supersymmetric SO(10) unification}\/},
  \href{http://dx.doi.org/10.1103/PhysRevD.50.7048}{Phys. Rev. {\bf D50} (1994)
   7048--7065},
\href{http://arxiv.org/abs/hep-ph/9306309}{{\tt arXiv:hep-ph/9306309
  [hep-ph]}}.
%%CITATION = HEP-PH/9306309;%%.

\bibitem{Carena:1994bv}
M.~Carena, M.~Olechowski, S.~Pokorski, and C.~E.~M. Wagner, {\em {Electroweak
  symmetry breaking and bottom - top Yukawa unification}\/},
  \href{http://dx.doi.org/10.1016/0550-3213(94)90313-1}{Nucl. Phys. {\bf B426}
  (1994)  269--300},
\href{http://arxiv.org/abs/hep-ph/9402253}{{\tt arXiv:hep-ph/9402253
  [hep-ph]}}.
%%CITATION = HEP-PH/9402253;%%.

\bibitem{Blazek:1995nv}
T.~Blazek, S.~Raby, and S.~Pokorski, {\em {Finite supersymmetric threshold
  corrections to CKM matrix elements in the large tan Beta regime}\/},
  \href{http://dx.doi.org/10.1103/PhysRevD.52.4151}{Phys. Rev. {\bf D52} (1995)
   4151--4158},
\href{http://arxiv.org/abs/hep-ph/9504364}{{\tt arXiv:hep-ph/9504364
  [hep-ph]}}.
%%CITATION = HEP-PH/9504364;%%.

\bibitem{Antusch:2008tf}
S.~Antusch and M.~Spinrath, {\em {Quark and lepton masses at the GUT scale
  including SUSY threshold corrections}\/},
  \href{http://dx.doi.org/10.1103/PhysRevD.78.075020}{Phys. Rev. {\bf D78}
  (2008)  075020},
\href{http://arxiv.org/abs/0804.0717}{{\tt arXiv:0804.0717 [hep-ph]}}.
%%CITATION = ARXIV:0804.0717;%%.

\bibitem{Antusch:2015nwi}
S.~Antusch and C.~Sluka, {\em {Predicting the Sparticle Spectrum from GUTs via
  SUSY Threshold Corrections with SusyTC}\/},
\href{http://arxiv.org/abs/1512.06727}{{\tt arXiv:1512.06727 [hep-ph]}}.
%%CITATION = ARXIV:1512.06727;%%.

\bibitem{Antusch:2016nak}
S.~Antusch and C.~Sluka, {\em {Testable SUSY spectra from GUTs at a 100 TeV pp
  collider}\/},
\href{http://arxiv.org/abs/1604.00212}{{\tt arXiv:1604.00212 [hep-ph]}}.
%%CITATION = ARXIV:1604.00212;%%.

\bibitem{Ellis:1990nz}
J.~R. Ellis, G.~Ridolfi, and F.~Zwirner, {\em {Radiative corrections to the
  masses of supersymmetric Higgs bosons}\/},
\href{http://dx.doi.org/10.1016/0370-2693(91)90863-L}{Phys. Lett. {\bf B257}
  (1991)  83--91}.
%%CITATION = PHLTA,B257,83;%%.

\bibitem{Ellis:1991zd}
J.~R. Ellis, G.~Ridolfi, and F.~Zwirner, {\em {On radiative corrections to
  supersymmetric Higgs boson masses and their implications for LEP
  searches}\/},
\href{http://dx.doi.org/10.1016/0370-2693(91)90626-2}{Phys. Lett. {\bf B262}
  (1991)  477--484}.
%%CITATION = PHLTA,B262,477;%%.

\bibitem{Bagnaschi:2014rsa}
E.~Bagnaschi, G.~F. Giudice, P.~Slavich, and A.~Strumia, {\em {Higgs Mass and
  Unnatural Supersymmetry}\/},
  \href{http://dx.doi.org/10.1007/JHEP09(2014)092}{JHEP {\bf 09} (2014)  092},
\href{http://arxiv.org/abs/1407.4081}{{\tt arXiv:1407.4081 [hep-ph]}}.
%%CITATION = ARXIV:1407.4081;%%.

\bibitem{Hall:2011aa}
L.~J. Hall, D.~Pinner, and J.~T. Ruderman, {\em {A Natural SUSY Higgs Near 126
  GeV}\/},  \href{http://dx.doi.org/10.1007/JHEP04(2012)131}{JHEP {\bf 04}
  (2012)  131},
\href{http://arxiv.org/abs/1112.2703}{{\tt arXiv:1112.2703 [hep-ph]}}.
%%CITATION = ARXIV:1112.2703;%%.

\bibitem{ArkaniHamed:2012gw}
N.~Arkani-Hamed, A.~Gupta, D.~E. Kaplan, N.~Weiner, and T.~Zorawski, {\em
  {Simply Unnatural Supersymmetry}\/},
\href{http://arxiv.org/abs/1212.6971}{{\tt arXiv:1212.6971 [hep-ph]}}.
%%CITATION = ARXIV:1212.6971;%%.

\bibitem{Ibanez:1982fr}
L.~E. Ibanez and G.~G. Ross, {\em {SU(2)-L x U(1) Symmetry Breaking as a
  Radiative Effect of Supersymmetry Breaking in Guts}\/},
\href{http://dx.doi.org/10.1016/0370-2693(82)91239-4}{Phys. Lett. {\bf B110}
  (1982)  215--220}.
%%CITATION = PHLTA,B110,215;%%.

\bibitem{Inoue:1982pi}
K.~Inoue, A.~Kakuto, H.~Komatsu, and S.~Takeshita, {\em {Aspects of Grand
  Unified Models with Softly Broken Supersymmetry}\/},
  \href{http://dx.doi.org/10.1143/PTP.68.927}{Prog. Theor. Phys. {\bf 68}
  (1982)  927}.
[Erratum: Prog. Theor. Phys.70,330(1983)].
%%CITATION = PTPKA,68,927;%%.

\bibitem{Ibanez:1982ee}
L.~E. Ibanez, {\em {Locally Supersymmetric SU(5) Grand Unification}\/},
\href{http://dx.doi.org/10.1016/0370-2693(82)90604-9}{Phys. Lett. {\bf B118}
  (1982)  73--78}.
%%CITATION = PHLTA,B118,73;%%.

\bibitem{Ellis:1983bp}
J.~R. Ellis, J.~S. Hagelin, D.~V. Nanopoulos, and K.~Tamvakis, {\em {Weak
  Symmetry Breaking by Radiative Corrections in Broken Supergravity}\/},
\href{http://dx.doi.org/10.1016/0370-2693(83)91283-2}{Phys. Lett. {\bf B125}
  (1983)  275}.
%%CITATION = PHLTA,B125,275;%%.

\bibitem{AlvarezGaume:1983gj}
L.~Alvarez-Gaume, J.~Polchinski, and M.~B. Wise, {\em {Minimal Low-Energy
  Supergravity}\/},
\href{http://dx.doi.org/10.1016/0550-3213(83)90591-6}{Nucl. Phys. {\bf B221}
  (1983)  495}.
%%CITATION = NUPHA,B221,495;%%.

\bibitem{Hooft:1979bh}
G.~'t~Hooft, {\em {Naturalness, chiral symmetry, and spontaneous chiral
  symmetry breaking}\/},
NATO Sci. Ser. B {\bf 59} (1980)  135.
%%CITATION = PRINT-80-0083 (UTRECHT);%%.

\bibitem{Susskind:1978ms}
L.~Susskind, {\em {Dynamics of Spontaneous Symmetry Breaking in the
  Weinberg-Salam Theory}\/},
\href{http://dx.doi.org/10.1103/PhysRevD.20.2619}{Phys.Rev. {\bf D20} (1979)
  2619--2625}.
%%CITATION = PHRVA,D20,2619;%%.

\bibitem{Ellis:1986yg}
J.~R. Ellis, K.~Enqvist, D.~V. Nanopoulos, and F.~Zwirner, {\em {Observables in
  Low-Energy Superstring Models}\/},
\href{http://dx.doi.org/10.1142/S0217732386000105}{Mod. Phys. Lett. {\bf A1}
  (1986)  57}.
%%CITATION = MPLAE,A1,57;%%.

\bibitem{Barbieri:1987fn}
R.~Barbieri and G.~Giudice, {\em {Upper Bounds on Supersymmetric Particle
  Masses}\/},
\href{http://dx.doi.org/10.1016/0550-3213(88)90171-X}{Nucl.Phys. {\bf B306}
  (1988)  63}.
%%CITATION = NUPHA,B306,63;%%.

\bibitem{Craig:2013cxa}
N.~Craig, {\em {The State of Supersymmetry after Run I of the LHC}\/},
\href{http://arxiv.org/abs/1309.0528}{{\tt arXiv:1309.0528 [hep-ph]}}.
%%CITATION = ARXIV:1309.0528;%%.

\bibitem{Batell:2015fma}
B.~Batell, G.~F. Giudice, and M.~McCullough, {\em {Natural Heavy
  Supersymmetry}\/},
\href{http://arxiv.org/abs/1509.00834}{{\tt arXiv:1509.00834 [hep-ph]}}.
%%CITATION = ARXIV:1509.00834;%%.

\bibitem{Graham:2015cka}
P.~W. Graham, D.~E. Kaplan, and S.~Rajendran, {\em {Cosmological Relaxation of
  the Electroweak Scale}\/},
\href{http://arxiv.org/abs/1504.07551}{{\tt arXiv:1504.07551 [hep-ph]}}.
%%CITATION = ARXIV:1504.07551;%%.

\bibitem{Feng:1999zg}
J.~L. Feng, K.~T. Matchev, and T.~Moroi, {\em {Focus points and naturalness in
  supersymmetry}\/},  \href{http://dx.doi.org/10.1103/PhysRevD.61.075005}{Phys.
  Rev. {\bf D61} (2000)  075005},
\href{http://arxiv.org/abs/hep-ph/9909334}{{\tt arXiv:hep-ph/9909334
  [hep-ph]}}.
%%CITATION = HEP-PH/9909334;%%.

\bibitem{Baer:2012up}
H.~Baer, V.~Barger, P.~Huang, A.~Mustafayev, and X.~Tata, {\em {Radiative
  natural SUSY with a 125 GeV Higgs boson}\/},
  \href{http://dx.doi.org/10.1103/PhysRevLett.109.161802}{Phys. Rev. Lett. {\bf
  109} (2012)  161802},
\href{http://arxiv.org/abs/1207.3343}{{\tt arXiv:1207.3343 [hep-ph]}}.
%%CITATION = ARXIV:1207.3343;%%.

\bibitem{LeCompte:2011cn}
T.~J. LeCompte and S.~P. Martin, {\em {Large Hadron Collider reach for
  supersymmetric models with compressed mass spectra}\/},
  \href{http://dx.doi.org/10.1103/PhysRevD.84.015004}{Phys. Rev. {\bf D84}
  (2011)  015004},
\href{http://arxiv.org/abs/1105.4304}{{\tt arXiv:1105.4304 [hep-ph]}}.
%%CITATION = ARXIV:1105.4304;%%.

\bibitem{Fan:2011yu}
J.~Fan, M.~Reece, and J.~T. Ruderman, {\em {Stealth Supersymmetry}\/},
  \href{http://dx.doi.org/10.1007/JHEP11(2011)012}{JHEP {\bf 11} (2011)  012},
\href{http://arxiv.org/abs/1105.5135}{{\tt arXiv:1105.5135 [hep-ph]}}.
%%CITATION = ARXIV:1105.5135;%%.

\bibitem{Fan:2012jf}
J.~Fan, M.~Reece, and J.~T. Ruderman, {\em {A Stealth Supersymmetry
  Sampler}\/},  \href{http://dx.doi.org/10.1007/JHEP07(2012)196}{JHEP {\bf 07}
  (2012)  196},
\href{http://arxiv.org/abs/1201.4875}{{\tt arXiv:1201.4875 [hep-ph]}}.
%%CITATION = ARXIV:1201.4875;%%.

\bibitem{Barbier:2004ez}
R.~Barbier et al., {\em {R-parity violating supersymmetry}\/},
  \href{http://dx.doi.org/10.1016/j.physrep.2005.08.006}{Phys. Rept. {\bf 420}
  (2005)  1--202},
\href{http://arxiv.org/abs/hep-ph/0406039}{{\tt arXiv:hep-ph/0406039
  [hep-ph]}}.
%%CITATION = HEP-PH/0406039;%%.

\bibitem{Csaki:2011ge}
C.~Csaki, Y.~Grossman, and B.~Heidenreich, {\em {MFV SUSY: A Natural Theory for
  R-Parity Violation}\/},
  \href{http://dx.doi.org/10.1103/PhysRevD.85.095009}{Phys. Rev. {\bf D85}
  (2012)  095009},
\href{http://arxiv.org/abs/1111.1239}{{\tt arXiv:1111.1239 [hep-ph]}}.
%%CITATION = ARXIV:1111.1239;%%.

\bibitem{Papucci:2011wy}
M.~Papucci, J.~T. Ruderman, and A.~Weiler, {\em {Natural SUSY Endures}\/},
  \href{http://dx.doi.org/10.1007/JHEP09(2012)035}{JHEP {\bf 1209} (2012)
  035},
\href{http://arxiv.org/abs/1110.6926}{{\tt arXiv:1110.6926 [hep-ph]}}.
%%CITATION = ARXIV:1110.6926;%%.

\bibitem{Dimopoulos:1995mi}
S.~Dimopoulos and G.~Giudice, {\em {Naturalness constraints in supersymmetric
  theories with nonuniversal soft terms}\/},
  \href{http://dx.doi.org/10.1016/0370-2693(95)00961-J}{Phys.Lett. {\bf B357}
  (1995)  573--578},
\href{http://arxiv.org/abs/hep-ph/9507282}{{\tt arXiv:hep-ph/9507282
  [hep-ph]}}.
%%CITATION = HEP-PH/9507282;%%.

\bibitem{Cohen:1996vb}
A.~G. Cohen, D.~Kaplan, and A.~Nelson, {\em {The More minimal supersymmetric
  standard model}\/},
  \href{http://dx.doi.org/10.1016/S0370-2693(96)01183-5}{Phys.Lett. {\bf B388}
  (1996)  588--598},
\href{http://arxiv.org/abs/hep-ph/9607394}{{\tt arXiv:hep-ph/9607394
  [hep-ph]}}.
%%CITATION = HEP-PH/9607394;%%.

\bibitem{Fox:2002bu}
P.~J. Fox, A.~E. Nelson, and N.~Weiner, {\em {Dirac gaugino masses and
  supersoft supersymmetry breaking}\/},
  \href{http://dx.doi.org/10.1088/1126-6708/2002/08/035}{JHEP {\bf 08} (2002)
  035},
\href{http://arxiv.org/abs/hep-ph/0206096}{{\tt arXiv:hep-ph/0206096
  [hep-ph]}}.
%%CITATION = HEP-PH/0206096;%%.

\bibitem{Kribs:2012gx}
G.~D. Kribs and A.~Martin, {\em {Supersoft Supersymmetry is Super-Safe}\/},
  \href{http://dx.doi.org/10.1103/PhysRevD.85.115014}{Phys. Rev. {\bf D85}
  (2012)  115014},
\href{http://arxiv.org/abs/1203.4821}{{\tt arXiv:1203.4821 [hep-ph]}}.
%%CITATION = ARXIV:1203.4821;%%.

\bibitem{Cohen:2013xda}
T.~Cohen, T.~Golling, M.~Hance, A.~Henrichs, K.~Howe, J.~Loyal, S.~Padhi, and
  J.~G. Wacker, {\em {SUSY Simplified Models at 14, 33, and 100 TeV Proton
  Colliders}\/},  \href{http://dx.doi.org/10.1007/JHEP04(2014)117}{JHEP {\bf
  04} (2014)  117},
\href{http://arxiv.org/abs/1311.6480}{{\tt arXiv:1311.6480 [hep-ph]}}.
%%CITATION = ARXIV:1311.6480;%%.

\bibitem{Jung:2013zya}
S.~Jung and J.~D. Wells, {\em {Gaugino physics of split supersymmetry spectra
  at the LHC and future proton colliders}\/},
  \href{http://dx.doi.org/10.1103/PhysRevD.89.075004}{Phys. Rev. {\bf D89}
  (2014) no.~7, 075004},
\href{http://arxiv.org/abs/1312.1802}{{\tt arXiv:1312.1802 [hep-ph]}}.
%%CITATION = ARXIV:1312.1802;%%.

\bibitem{Low:2014cba}
M.~Low and L.-T. Wang, {\em {Neutralino dark matter at 14 TeV and 100 TeV}\/},
  \href{http://dx.doi.org/10.1007/JHEP08(2014)161}{JHEP {\bf 08} (2014)  161},
\href{http://arxiv.org/abs/1404.0682}{{\tt arXiv:1404.0682 [hep-ph]}}.
%%CITATION = ARXIV:1404.0682;%%.

\bibitem{Cohen:2014hxa}
T.~Cohen, R.~T. D'Agnolo, M.~Hance, H.~K. Lou, and J.~G. Wacker, {\em {Boosting
  Stop Searches with a 100 TeV Proton Collider}\/},
  \href{http://dx.doi.org/10.1007/JHEP11(2014)021}{JHEP {\bf 1411} (2014)
  021},
\href{http://arxiv.org/abs/1406.4512}{{\tt arXiv:1406.4512 [hep-ph]}}.
%%CITATION = ARXIV:1406.4512;%%.

\bibitem{Ellis:2014kla}
S.~A.~R. Ellis, G.~L. Kane, and B.~Zheng, {\em {Superpartners at LHC and Future
  Colliders: Predictions from Constrained Compactified M-Theory}\/},
\href{http://arxiv.org/abs/1408.1961}{{\tt arXiv:1408.1961 [hep-ph]}}.
%%CITATION = ARXIV:1408.1961;%%.

\bibitem{Acharya:2014pua}
B.~S. Acharya, K.~Bozek, C.~Pongkitivanichkul, and K.~Sakurai, {\em {Prospects
  for observing charginos and neutralinos at a 100 TeV proton-proton
  collider}\/},
\href{http://arxiv.org/abs/1410.1532}{{\tt arXiv:1410.1532 [hep-ph]}}.
%%CITATION = ARXIV:1410.1532;%%.

\bibitem{Gori:2014oua}
S.~Gori, S.~Jung, L.-T. Wang, and J.~D. Wells, {\em {Prospects for
  Electroweakino Discovery at a 100 TeV Hadron Collider}\/},
  \href{http://dx.doi.org/10.1007/JHEP12(2014)108}{JHEP {\bf 1412} (2014)
  108},
\href{http://arxiv.org/abs/1410.6287}{{\tt arXiv:1410.6287 [hep-ph]}}.
%%CITATION = ARXIV:1410.6287;%%.

\bibitem{diCortona:2014yua}
G.~G. di~Cortona, {\em {Hunting electroweakinos at future hadron colliders and
  direct detection experiments}\/},
\href{http://arxiv.org/abs/1412.5952}{{\tt arXiv:1412.5952 [hep-ph]}}.
%%CITATION = ARXIV:1412.5952;%%.

\bibitem{Berlin:2015aba}
A.~Berlin, T.~Lin, M.~Low, and L.-T. Wang, {\em {Neutralinos in Vector Boson
  Fusion at High Energy Colliders}\/},
  \href{http://dx.doi.org/10.1103/PhysRevD.91.115002}{Phys. Rev. {\bf D91}
  (2015) no.~11, 115002},
\href{http://arxiv.org/abs/1502.05044}{{\tt arXiv:1502.05044 [hep-ph]}}.
%%CITATION = ARXIV:1502.05044;%%.

\bibitem{Beauchesne:2015jra}
H.~Beauchesne, K.~Earl, and T.~Gr\'{e}goire, {\em {LHC constraints on
  Mini-Split anomaly and gauge mediation and prospects for LHC 14 and a future
  100 TeV pp collider}\/},
  \href{http://dx.doi.org/10.1007/JHEP08(2015)117}{JHEP {\bf 08} (2015)  117},
\href{http://arxiv.org/abs/1503.03099}{{\tt arXiv:1503.03099 [hep-ph]}}.
%%CITATION = ARXIV:1503.03099;%%.

\bibitem{Beenakker:1996ch}
W.~Beenakker, R.~Hopker, M.~Spira, and P.~Zerwas, {\em {Squark and gluino
  production at hadron colliders}\/},
  \href{http://dx.doi.org/10.1016/S0550-3213(97)00084-9}{Nucl.Phys. {\bf B492}
  (1997)  51--103},
\href{http://arxiv.org/abs/hep-ph/9610490}{{\tt arXiv:hep-ph/9610490
  [hep-ph]}}.
%%CITATION = HEP-PH/9610490;%%.

\bibitem{Beenakker:1997ut}
W.~Beenakker, M.~Kr\"{a}mer, T.~Plehn, M.~Spira, and P.~Zerwas, {\em {Stop
  production at hadron colliders}\/},
  \href{http://dx.doi.org/10.1016/S0550-3213(98)00014-5}{Nucl.Phys. {\bf B515}
  (1998)  3--14},
\href{http://arxiv.org/abs/hep-ph/9710451}{{\tt arXiv:hep-ph/9710451
  [hep-ph]}}.
%%CITATION = HEP-PH/9710451;%%.

\bibitem{Beenakker:2009ha}
W.~Beenakker, S.~Brensing, M.~Kr\"{a}mer, A.~Kulesza, E.~Laenen, and
  I.~Niessen, {\em {Soft-gluon resummation for squark and gluino
  hadroproduction}\/},
  \href{http://dx.doi.org/10.1088/1126-6708/2009/12/041}{JHEP {\bf 12} (2009)
  041},
\href{http://arxiv.org/abs/0909.4418}{{\tt arXiv:0909.4418 [hep-ph]}}.
%%CITATION = ARXIV:0909.4418;%%.

\bibitem{Beenakker:2010nq}
W.~Beenakker, S.~Brensing, M.~Kr\"{a}mer, A.~Kulesza, E.~Laenen, and
  I.~Niessen, {\em {Supersymmetric top and bottom squark production at hadron
  colliders}\/},  \href{http://dx.doi.org/10.1007/JHEP08(2010)098}{JHEP {\bf
  08} (2010)  098},
\href{http://arxiv.org/abs/1006.4771}{{\tt arXiv:1006.4771 [hep-ph]}}.
%%CITATION = ARXIV:1006.4771;%%.

\bibitem{Borschensky:2014cia}
C.~Borschensky, M.~Kr\"{a}mer, A.~Kulesza, M.~Mangano, S.~Padhi, et al., {\em
  {Squark and gluino production cross sections in pp collisions at $\sqrt{s}$ =
  13, 14, 33 and 100 TeV}\/},
\href{http://arxiv.org/abs/1407.5066}{{\tt arXiv:1407.5066 [hep-ph]}}.
%%CITATION = ARXIV:1407.5066;%%.

\bibitem{Nadolsky:2008zw}
P.~M. Nadolsky, H.-L. Lai, Q.-H. Cao, J.~Huston, J.~Pumplin, et al., {\em
  {Implications of CTEQ global analysis for collider observables}\/},
  \href{http://dx.doi.org/10.1103/PhysRevD.78.013004}{Phys.Rev. {\bf D78}
  (2008)  013004},
\href{http://arxiv.org/abs/0802.0007}{{\tt arXiv:0802.0007 [hep-ph]}}.
%%CITATION = ARXIV:0802.0007;%%.

\bibitem{Martin:2009iq}
A.~Martin, W.~Stirling, R.~Thorne, and G.~Watt, {\em {Parton distributions for
  the LHC}\/},
  \href{http://dx.doi.org/10.1140/epjc/s10052-009-1072-5}{Eur.Phys.J. {\bf C63}
  (2009)  189--285},
\href{http://arxiv.org/abs/0901.0002}{{\tt arXiv:0901.0002 [hep-ph]}}.
%%CITATION = ARXIV:0901.0002;%%.

\bibitem{Bornhauser:2007bf}
S.~Bornhauser, M.~Drees, H.~K. Dreiner, and J.~S. Kim, {\em {Electroweak
  contributions to squark pair production at the LHC}\/},
  \href{http://dx.doi.org/10.1103/PhysRevD.76.095020}{Phys. Rev. {\bf D76}
  (2007)  095020},
\href{http://arxiv.org/abs/0709.2544}{{\tt arXiv:0709.2544 [hep-ph]}}.
%%CITATION = ARXIV:0709.2544;%%.

\bibitem{Hollik:2008yi}
W.~Hollik and E.~Mirabella, {\em {Squark anti-squark pair production at the
  LHC: The Electroweak contribution}\/},
  \href{http://dx.doi.org/10.1088/1126-6708/2008/12/087}{JHEP {\bf 12} (2008)
  087},
\href{http://arxiv.org/abs/0806.1433}{{\tt arXiv:0806.1433 [hep-ph]}}.
%%CITATION = ARXIV:0806.1433;%%.

\bibitem{Hollik:2008vm}
W.~Hollik, E.~Mirabella, and M.~K. Trenkel, {\em {Electroweak contributions to
  squark-gluino production at the LHC}\/},
  \href{http://dx.doi.org/10.1088/1126-6708/2009/02/002}{JHEP {\bf 02} (2009)
  002},
\href{http://arxiv.org/abs/0810.1044}{{\tt arXiv:0810.1044 [hep-ph]}}.
%%CITATION = ARXIV:0810.1044;%%.

\bibitem{Mirabella:2009ap}
E.~Mirabella, {\em {NLO electroweak contributions to gluino pair production at
  hadron colliders}\/},
  \href{http://dx.doi.org/10.1088/1126-6708/2009/12/012}{JHEP {\bf 12} (2009)
  012},
\href{http://arxiv.org/abs/0908.3318}{{\tt arXiv:0908.3318 [hep-ph]}}.
%%CITATION = ARXIV:0908.3318;%%.

\bibitem{Germer:2010vn}
J.~Germer, W.~Hollik, E.~Mirabella, and M.~K. Trenkel, {\em {Hadronic
  production of squark-squark pairs: The electroweak contributions}\/},
  \href{http://dx.doi.org/10.1007/JHEP08(2010)023}{JHEP {\bf 08} (2010)  023},
\href{http://arxiv.org/abs/1004.2621}{{\tt arXiv:1004.2621 [hep-ph]}}.
%%CITATION = ARXIV:1004.2621;%%.

\bibitem{Germer:2011an}
J.~Germer, W.~Hollik, and E.~Mirabella, {\em {Hadronic production of
  bottom-squark pairs with electroweak contributions}\/},
  \href{http://dx.doi.org/10.1007/JHEP05(2011)068}{JHEP {\bf 05} (2011)  068},
\href{http://arxiv.org/abs/1103.1258}{{\tt arXiv:1103.1258 [hep-ph]}}.
%%CITATION = ARXIV:1103.1258;%%.

\bibitem{Germer:2014jpa}
J.~Germer, W.~Hollik, J.~M. Lindert, and E.~Mirabella, {\em {Top-squark pair
  production at the LHC: a complete analysis at next-to-leading order}\/},
  \href{http://dx.doi.org/10.1007/JHEP09(2014)022}{JHEP {\bf 09} (2014)  022},
\href{http://arxiv.org/abs/1404.5572}{{\tt arXiv:1404.5572 [hep-ph]}}.
%%CITATION = ARXIV:1404.5572;%%.

\bibitem{Hollik:2015lha}
W.~Hollik, J.~M. Lindert, E.~Mirabella, and D.~Pagani, {\em {Electroweak
  corrections to squark-antisquark production at the LHC}\/},
  \href{http://dx.doi.org/10.1007/JHEP08(2015)099}{JHEP {\bf 08} (2015)  099},
\href{http://arxiv.org/abs/1506.01052}{{\tt arXiv:1506.01052 [hep-ph]}}.
%%CITATION = ARXIV:1506.01052;%%.

\bibitem{Ball:2013hta}
{NNPDF Collaboration}, R.~D. Ball, V.~Bertone, S.~Carrazza, L.~Del~Debbio,
  S.~Forte, A.~Guffanti, N.~P. Hartland, and J.~Rojo, {\em {Parton
  distributions with QED corrections}\/},
  \href{http://dx.doi.org/10.1016/j.nuclphysb.2013.10.010}{Nucl. Phys. {\bf
  B877} (2013)  290--320},
\href{http://arxiv.org/abs/1308.0598}{{\tt arXiv:1308.0598 [hep-ph]}}.
%%CITATION = ARXIV:1308.0598;%%.

\bibitem{Alwall:2014hca}
J.~Alwall, R.~Frederix, S.~Frixione, V.~Hirschi, F.~Maltoni, O.~Mattelaer,
  H.~S. Shao, T.~Stelzer, P.~Torrielli, and M.~Zaro, {\em {The automated
  computation of tree-level and next-to-leading order differential cross
  sections, and their matching to parton shower simulations}\/},
  \href{http://dx.doi.org/10.1007/JHEP07(2014)079}{JHEP {\bf 07} (2014)  079},
\href{http://arxiv.org/abs/1405.0301}{{\tt arXiv:1405.0301 [hep-ph]}}.
%%CITATION = ARXIV:1405.0301;%%.

\bibitem{Khachatryan:2015wza}
{CMS Collaboration}, V.~Khachatryan et al., {\em {Searches for third-generation
  squark production in fully hadronic final states in proton-proton collisions
  at $ \sqrt{s} = 8$ TeV}\/},
  \href{http://dx.doi.org/10.1007/JHEP06(2015)116}{JHEP {\bf 06} (2015)  116},
\href{http://arxiv.org/abs/1503.08037}{{\tt arXiv:1503.08037 [hep-ex]}}.
%%CITATION = ARXIV:1503.08037;%%.

\bibitem{Aad:2015pfx}
{ATLAS Collaboration}, G.~Aad et al., {\em {ATLAS Run 1 searches for direct
  pair production of third-generation squarks at the Large Hadron Collider}\/},
   \href{http://dx.doi.org/10.1140/epjc/s10052-015-3726-9}{Eur. Phys. J. {\bf
  C75} (2015) no.~10, 510},
\href{http://arxiv.org/abs/1506.08616}{{\tt arXiv:1506.08616 [hep-ex]}}.
%%CITATION = ARXIV:1506.08616;%%.

\bibitem{Chatrchyan:2008aa}
{CMS Collaboration}, S.~Chatrchyan et al., {\em {The CMS experiment at the CERN
  LHC}\/},
\href{http://dx.doi.org/10.1088/1748-0221/3/08/S08004}{JINST {\bf 3} (2008)
  S08004}.
%%CITATION = JINST,3,S08004;%%.

\bibitem{Plehn:2010st}
T.~Plehn, M.~Spannowsky, M.~Takeuchi, and D.~Zerwas, {\em {Stop Reconstruction
  with Tagged Tops}\/},  \href{http://dx.doi.org/10.1007/JHEP10(2010)078}{JHEP
  {\bf 1010} (2010)  078},
\href{http://arxiv.org/abs/1006.2833}{{\tt arXiv:1006.2833 [hep-ph]}}.
%%CITATION = ARXIV:1006.2833;%%.

\bibitem{Kaplan:2008ie}
D.~E. Kaplan, K.~Rehermann, M.~D. Schwartz, and B.~Tweedie, {\em {Top Tagging:
  A Method for Identifying Boosted Hadronically Decaying Top Quarks}\/},
  \href{http://dx.doi.org/10.1103/PhysRevLett.101.142001}{Phys.Rev.Lett. {\bf
  101} (2008)  142001},
\href{http://arxiv.org/abs/0806.0848}{{\tt arXiv:0806.0848 [hep-ph]}}.
%%CITATION = ARXIV:0806.0848;%%.

\bibitem{Acosta:2005zd}
{CDF Collaboration}, D.~Acosta et al., {\em {Measurement of the $t\bar{t}$
  production cross section in $p\bar{p}$ collisions at $\sqrt{s} = 1.96$ TeV
  using lepton plus jets events with semileptonic B decays to muons}\/},
  \href{http://dx.doi.org/10.1103/PhysRevD.72.032002}{Phys.Rev. {\bf D72}
  (2005)  032002},
\href{http://arxiv.org/abs/hep-ex/0506001}{{\tt arXiv:hep-ex/0506001
  [hep-ex]}}.
%%CITATION = HEP-EX/0506001;%%.

\bibitem{Abazov:2004bv}
{D0 Collaboration}, V.~Abazov et al., {\em {A Search for anomalous heavy-flavor
  quark production in association with $W$ bosons}\/},
  \href{http://dx.doi.org/10.1103/PhysRevLett.94.152002}{Phys.Rev.Lett. {\bf
  94} (2005)  152002},
\href{http://arxiv.org/abs/hep-ex/0411084}{{\tt arXiv:hep-ex/0411084
  [hep-ex]}}.
%%CITATION = HEP-EX/0411084;%%.

\bibitem{Abulencia:2005qa}
{CDF Collaboration}, A.~Abulencia et al., {\em {Search for anomalous
  semileptonic decay of heavy flavor hadrons produced in association with a W
  boson at CDF II}\/},
  \href{http://dx.doi.org/10.1103/PhysRevD.73.051101}{Phys.Rev. {\bf D73}
  (2006)  051101},
\href{http://arxiv.org/abs/hep-ex/0512065}{{\tt arXiv:hep-ex/0512065
  [hep-ex]}}.
%%CITATION = HEP-EX/0512065;%%.

\bibitem{Aaltonen:2007dm}
{CDF Collaboration}, T.~Aaltonen et al., {\em {First measurement of the
  production of a $W$ boson in association with a single charm quark in $p
  \bar{p}$ collisions at $\sqrt{s}$ = 1.96-TeV}\/},
  \href{http://dx.doi.org/10.1103/PhysRevLett.100.091803}{Phys.Rev.Lett. {\bf
  100} (2008)  091803},
\href{http://arxiv.org/abs/0711.2901}{{\tt arXiv:0711.2901 [hep-ex]}}.
%%CITATION = ARXIV:0711.2901;%%.

\bibitem{Aaltonen:2009ad}
{CDF Collaboration}, T.~Aaltonen et al., {\em {Measurement of the $t\bar{t}$
  Production Cross Section in 2 fb$^{-1}$ of $p\bar{p}$ Collisions at
  $\sqrt{s}=1.96$ TeV Using Lepton Plus Jets Events with Soft Muon
  b-Tagging}\/},  \href{http://dx.doi.org/10.1103/PhysRevD.79.052007}{Phys.Rev.
  {\bf D79} (2009)  052007},
\href{http://arxiv.org/abs/0901.4142}{{\tt arXiv:0901.4142 [hep-ex]}}.
%%CITATION = ARXIV:0901.4142;%%.

\bibitem{Aaltonen:2010se}
{CDF Collaboration}, T.~Aaltonen et al., {\em {Measurement of the $t\bar{t}$
  Production Cross Section in $p\bar{p}$ Collisions at sqrt(s)=1.96 TeV using
  Soft Electron b-Tagging}\/},
  \href{http://dx.doi.org/10.1103/PhysRevD.81.092002}{Phys.Rev. {\bf D81}
  (2010)  092002},
\href{http://arxiv.org/abs/1002.3783}{{\tt arXiv:1002.3783 [hep-ex]}}.
%%CITATION = ARXIV:1002.3783;%%.

\bibitem{Aad:2009wy}
{ATLAS Collaboration}, G.~Aad et al., {\em {Expected Performance of the ATLAS
  Experiment - Detector, Trigger and Physics}\/},
\href{http://arxiv.org/abs/0901.0512}{{\tt arXiv:0901.0512 [hep-ex]}}.
%%CITATION = ARXIV:0901.0512;%%.

\bibitem{ATLAS:2010afa}
{ATLAS Collaboration}, {\em {Soft muon tagging and Dstar/mu correlations in 7
  TeV collisions with ATLAS}\/},   ATLAS-CONF-2010-100, CERN, Geneva, Dec,
  2010.
\newblock \url{https://cds.cern.ch/record/1316469}.

\bibitem{Chatrchyan:2012jua}
{CMS Collaboration}, S.~Chatrchyan et al., {\em {Identification of b-quark jets
  with the CMS experiment}\/},
  \href{http://dx.doi.org/10.1088/1748-0221/8/04/P04013}{JINST {\bf 8} (2013)
  P04013},
\href{http://arxiv.org/abs/1211.4462}{{\tt arXiv:1211.4462 [hep-ex]}}.
%%CITATION = ARXIV:1211.4462;%%.

\bibitem{ATLAS-CONF-2013-068}
{\em {Search for pair-produced top squarks decaying into a charm quark and the
  lightest neutralinos with 20.3 fb$^{-1}$ of $pp$ collisions at
  $\sqrt{s}=8~$TeV with the ATLAS detector at the LHC}\/},
  ATLAS-CONF-2013-068, CERN, Geneva, Jul, 2013.
\newblock \url{https://cds.cern.ch/record/1562880}.

\bibitem{Avetisyan:2013onh}
A.~Avetisyan et al., {\em {Methods and Results for Standard Model Event
  Generation at $\sqrt{s}$ = 14 TeV, 33 TeV and 100 TeV Proton Colliders (A
  Snowmass Whitepaper)}\/},  in {\em {Community Summer Study 2013: Snowmass on
  the Mississippi (CSS2013) Minneapolis, MN, USA, July 29-August 6, 2013}}.
\newblock 2013.
\newblock \href{http://arxiv.org/abs/1308.1636}{{\tt arXiv:1308.1636
  [hep-ex]}}.
\newblock
\url{http://lss.fnal.gov/archive/test-fn/0000/fermilab-fn-0965-t.pdf}.
\newblock
%%CITATION = ARXIV:1308.1636;%%.

\bibitem{Alwall:2011uj}
J.~Alwall, M.~Herquet, F.~Maltoni, O.~Mattelaer, and T.~Stelzer, {\em {MadGraph
  5 : Going Beyond}\/},  \href{http://dx.doi.org/10.1007/JHEP06(2011)128}{JHEP
  {\bf 1106} (2011)  128},
\href{http://arxiv.org/abs/1106.0522}{{\tt arXiv:1106.0522 [hep-ph]}}.
%%CITATION = ARXIV:1106.0522;%%.

\bibitem{Sjostrand:2006za}
T.~Sjostrand, S.~Mrenna, and P.~Z. Skands, {\em {PYTHIA 6.4 Physics and
  Manual}\/},  \href{http://dx.doi.org/10.1088/1126-6708/2006/05/026}{JHEP {\bf
  0605} (2006)  026},
\href{http://arxiv.org/abs/hep-ph/0603175}{{\tt arXiv:hep-ph/0603175
  [hep-ph]}}.
%%CITATION = HEP-PH/0603175;%%.

\bibitem{deFavereau:2013fsa}
{DELPHES 3 Collaboration}, J.~de~Favereau, C.~Delaere, P.~Demin, A.~Giammanco,
  V.~Lema{\^\i}tre, A.~Mertens, and M.~Selvaggi, {\em {DELPHES 3, A modular
  framework for fast simulation of a generic collider experiment}\/},
  \href{http://dx.doi.org/10.1007/JHEP02(2014)057}{JHEP {\bf 02} (2014)  057},
\href{http://arxiv.org/abs/1307.6346}{{\tt arXiv:1307.6346 [hep-ex]}}.
%%CITATION = ARXIV:1307.6346;%%.

\bibitem{Anderson:2013kxz}
J.~Anderson, A.~Avetisyan, R.~Brock, S.~Chekanov, T.~Cohen, et al., {\em
  {Snowmass Energy Frontier Simulations}\/},  in {\em {Community Summer Study
  2013: Snowmass on the Mississippi (CSS2013) Minneapolis, MN, USA, July
  29-August 6, 2013}}.
\newblock 2013.
\newblock
\href{http://arxiv.org/abs/1309.1057}{{\tt arXiv:1309.1057 [hep-ex]}}.
\newblock
%%CITATION = ARXIV:1309.1057;%%.

\bibitem{Cacciari:2008gp}
M.~Cacciari, G.~P. Salam, and G.~Soyez, {\em {The anti-$k_t$ jet clustering
  algorithm}\/},  \href{http://dx.doi.org/10.1088/1126-6708/2008/04/063}{JHEP
  {\bf 04} (2008)  063},
\href{http://arxiv.org/abs/0802.1189}{{\tt arXiv:0802.1189 [hep-ph]}}.
%%CITATION = 0802.1189;%%.

\bibitem{Moneta:2010pm}
L.~Moneta, K.~Belasco, K.~S. Cranmer, S.~Kreiss, A.~Lazzaro, et al., {\em {The
  RooStats Project}\/},  PoS {\bf ACAT2010} (2010)  057,
\href{http://arxiv.org/abs/1009.1003}{{\tt arXiv:1009.1003 [physics.data-an]}}.
%%CITATION = ARXIV:1009.1003;%%.

\bibitem{Plehn:2012pr}
T.~Plehn, M.~Spannowsky, and M.~Takeuchi, {\em {Stop searches in 2012}\/},
  \href{http://dx.doi.org/10.1007/JHEP08(2012)091}{JHEP {\bf 08} (2012)  091},
\href{http://arxiv.org/abs/1205.2696}{{\tt arXiv:1205.2696 [hep-ph]}}.
%%CITATION = ARXIV:1205.2696;%%.

\bibitem{Kaplan:2012gd}
D.~E. Kaplan, K.~Rehermann, and D.~Stolarski, {\em {Searching for Direct Stop
  Production in Hadronic Top Data at the LHC}\/},
  \href{http://dx.doi.org/10.1007/JHEP07(2012)119}{JHEP {\bf 07} (2012)  119},
\href{http://arxiv.org/abs/1205.5816}{{\tt arXiv:1205.5816 [hep-ph]}}.
%%CITATION = ARXIV:1205.5816;%%.

\bibitem{Dutta:2012kx}
B.~Dutta, T.~Kamon, N.~Kolev, K.~Sinha, and K.~Wang, {\em {Searching for Top
  Squarks at the LHC in Fully Hadronic Final State}\/},
  \href{http://dx.doi.org/10.1103/PhysRevD.86.075004}{Phys. Rev. {\bf D86}
  (2012)  075004},
\href{http://arxiv.org/abs/1207.1873}{{\tt arXiv:1207.1873 [hep-ph]}}.
%%CITATION = ARXIV:1207.1873;%%.

\bibitem{Buckley:2013lpa}
M.~R. Buckley, T.~Plehn, and M.~Takeuchi, {\em {Buckets of Tops}\/},
  \href{http://dx.doi.org/10.1007/JHEP08(2013)086}{JHEP {\bf 08} (2013)  086},
\href{http://arxiv.org/abs/1302.6238}{{\tt arXiv:1302.6238 [hep-ph]}}.
%%CITATION = ARXIV:1302.6238;%%.

\bibitem{Stolarski:2013msa}
D.~Stolarski, {\em {Reach in All Hadronic Stop Decays: A Snowmass White
  Paper}\/},  in {\em {Community Summer Study 2013: Snowmass on the Mississippi
  (CSS2013) Minneapolis, MN, USA, July 29-August 6, 2013}}.
\newblock 2013.
\newblock \href{http://arxiv.org/abs/1309.1514}{{\tt arXiv:1309.1514
  [hep-ph]}}.
\newblock
\url{http://inspirehep.net/record/1253108/files/arXiv:1309.1514.pdf}.
\newblock
%%CITATION = ARXIV:1309.1514;%%.

\bibitem{Chatrchyan:2013sza}
{CMS Collaboration}, S.~Chatrchyan et al., {\em {Interpretation of Searches for
  Supersymmetry with simplified Models}\/},
  \href{http://dx.doi.org/10.1103/PhysRevD.88.052017}{Phys. Rev. {\bf D88}
  (2013) no.~5, 052017},
\href{http://arxiv.org/abs/1301.2175}{{\tt arXiv:1301.2175 [hep-ex]}}.
%%CITATION = ARXIV:1301.2175;%%.

\bibitem{Thaler:2008ju}
J.~Thaler and L.-T. Wang, {\em {Strategies to Identify Boosted Tops}\/},
  \href{http://dx.doi.org/10.1088/1126-6708/2008/07/092}{JHEP {\bf 0807} (2008)
   092},
\href{http://arxiv.org/abs/0806.0023}{{\tt arXiv:0806.0023 [hep-ph]}}.
%%CITATION = ARXIV:0806.0023;%%.

\bibitem{Almeida:2008yp}
L.~G. Almeida, S.~J. Lee, G.~Perez, G.~F. Sterman, I.~Sung, and J.~Virzi, {\em
  {Substructure of high-$p_T$ Jets at the LHC}\/},
  \href{http://dx.doi.org/10.1103/PhysRevD.79.074017}{Phys. Rev. {\bf D79}
  (2009)  074017},
\href{http://arxiv.org/abs/0807.0234}{{\tt arXiv:0807.0234 [hep-ph]}}.
%%CITATION = ARXIV:0807.0234;%%.

\bibitem{Thaler:2011gf}
J.~Thaler and K.~Van~Tilburg, {\em {Maximizing Boosted Top Identification by
  Minimizing N-subjettiness}\/},
  \href{http://dx.doi.org/10.1007/JHEP02(2012)093}{JHEP {\bf 02} (2012)  093},
\href{http://arxiv.org/abs/1108.2701}{{\tt arXiv:1108.2701 [hep-ph]}}.
%%CITATION = ARXIV:1108.2701;%%.

\bibitem{Soper:2012pb}
D.~E. Soper and M.~Spannowsky, {\em {Finding top quarks with shower
  deconstruction}\/},
  \href{http://dx.doi.org/10.1103/PhysRevD.87.054012}{Phys. Rev. {\bf D87}
  (2013)  054012},
\href{http://arxiv.org/abs/1211.3140}{{\tt arXiv:1211.3140 [hep-ph]}}.
%%CITATION = ARXIV:1211.3140;%%.

\bibitem{Chatrchyan:2012ku}
{{CMS Collaboration}}, S.~Chatrchyan et al., {\em {Search for anomalous t t-bar
  production in the highly-boosted all-hadronic final state}\/},
  \href{http://dx.doi.org/10.1007/JHEP09(2012)029}{JHEP {\bf 1209} (2012)
  029},
\href{http://arxiv.org/abs/1204.2488}{{\tt arXiv:1204.2488 [hep-ex]}}.
%%CITATION = ARXIV:1204.2488;%%.

\bibitem{Aad:2012raa}
{ATLAS Collaboration}, G.~Aad et al., {\em {Search for resonances decaying into
  top-quark pairs using fully hadronic decays in $pp$ collisions with ATLAS at
  $\sqrt{s}=7$ TeV}\/},  \href{http://dx.doi.org/10.1007/JHEP01(2013)116}{JHEP
  {\bf 01} (2013)  116},
\href{http://arxiv.org/abs/1211.2202}{{\tt arXiv:1211.2202 [hep-ex]}}.
%%CITATION = ARXIV:1211.2202;%%.

\bibitem{CMS-PAS-BTV-11-001}
{CMS Collaboration}, {\em {Performance of the b-jet identification in CMS}\/},
   CMS-PAS-BTV-11-001, CERN, Geneva, 2011.
\newblock \url{https://cds.cern.ch/record/1366061}.

\bibitem{ATLAS-CONF-2011-102}
{ATLAS Collaboration}, {\em {Commissioning of the ATLAS high-performance
  b-tagging algorithms in the 7 TeV collision data}\/},   ATLAS-CONF-2011-102,
  CERN, Geneva, Jul, 2011.
\newblock \url{https://cds.cern.ch/record/1369219}.

\bibitem{Beenakker:1996ed}
W.~Beenakker, R.~Hopker, and M.~Spira, {\em {PROSPINO: A Program for the
  production of supersymmetric particles in next-to-leading order QCD}\/},
\href{http://arxiv.org/abs/hep-ph/9611232}{{\tt arXiv:hep-ph/9611232
  [hep-ph]}}.
%%CITATION = HEP-PH/9611232;%%.

\bibitem{ATL-PHYS-PUB-2013-002}
{\em {Searches for Supersymmetry at the high luminosity LHC with the ATLAS
  Detector}\/},   ATL-PHYS-PUB-2013-002, CERN, Geneva, Feb, 2013.
\newblock \url{https://cds.cern.ch/record/1512933}.

\bibitem{ATLAS-CONF-2012-147}
{ATLAS Collaboration}, {\em {Search for New Phenomena in Monojet plus Missing
  Transverse Momentum Final States using 10fb$^{-1}$ of pp Collisions at
  $\sqrt{s}$=8 TeV with the ATLAS detector at the LHC}\/},
  ATLAS-CONF-2012-147, CERN, Geneva, Nov, 2012.
\newblock \url{https://cds.cern.ch/record/1493486}.

\bibitem{CMS-PAS-EXO-12-048}
{CMS Collaboration}, {\em {Search for new physics in monojet events in pp
  collisions at sqrt(s)= 8 TeV}\/},   CMS-PAS-EXO-12-048, CERN, Geneva, 2013.
\newblock \url{https://cds.cern.ch/record/1525585}.

\bibitem{Chatrchyan:2012paa}
{CMS Collaboration}, S.~Chatrchyan et al., {\em {Search for new physics in
  events with same-sign dileptons and $b$ jets in $pp$ collisions at
  $\sqrt{s}=8$ TeV}\/},  \href{http://dx.doi.org/10.1007/JHEP03(2013)037,
  10.1007/JHEP07(2013)041}{JHEP {\bf 1303} (2013)  037},
\href{http://arxiv.org/abs/1212.6194}{{\tt arXiv:1212.6194 [hep-ex]}}.
%%CITATION = ARXIV:1212.6194;%%.

\bibitem{Lester:1999tx}
C.~G. Lester and D.~J. Summers, {\em {Measuring masses of semiinvisibly
  decaying particles pair produced at hadron colliders}\/},
  \href{http://dx.doi.org/10.1016/S0370-2693(99)00945-4}{Phys. Lett. {\bf B463}
  (1999)  99--103},
\href{http://arxiv.org/abs/hep-ph/9906349}{{\tt arXiv:hep-ph/9906349
  [hep-ph]}}.
%%CITATION = HEP-PH/9906349;%%.

\bibitem{Barr:2003rg}
A.~Barr, C.~Lester, and P.~Stephens, {\em {$m_{T2}$: The Truth behind the
  glamour}\/},  \href{http://dx.doi.org/10.1088/0954-3899/29/10/304}{J.Phys.
  {\bf G29} (2003)  2343--2363},
\href{http://arxiv.org/abs/hep-ph/0304226}{{\tt arXiv:hep-ph/0304226
  [hep-ph]}}.
%%CITATION = HEP-PH/0304226;%%.

\bibitem{Burns:2008va}
M.~Burns, K.~Kong, K.~T. Matchev, and M.~Park, {\em {Using Subsystem MT2 for
  Complete Mass Determinations in Decay Chains with Missing Energy at Hadron
  Colliders}\/},  \href{http://dx.doi.org/10.1088/1126-6708/2009/03/143}{JHEP
  {\bf 0903} (2009)  143},
\href{http://arxiv.org/abs/0810.5576}{{\tt arXiv:0810.5576 [hep-ph]}}.
%%CITATION = ARXIV:0810.5576;%%.

\bibitem{ATLAS-CONF-2013-047}
{ATLAS Collaboration}, G.~Aad et al., {\em {Search for squarks and gluinos with
  the ATLAS detector in final states with jets and missing transverse momentum
  using $\sqrt{s}=8$ TeV proton--proton collision data}\/},
  \href{http://dx.doi.org/10.1007/JHEP09(2014)176}{JHEP {\bf 09} (2014)  176},
\href{http://arxiv.org/abs/1405.7875}{{\tt arXiv:1405.7875 [hep-ex]}}.
%%CITATION = ARXIV:1405.7875;%%.

\bibitem{Ellis:2015xba}
S.~A.~R. Ellis and B.~Zheng, {\em {Reaching for squarks and gauginos at a 100
  TeV p-p collider}\/},
  \href{http://dx.doi.org/10.1103/PhysRevD.92.075034}{Phys. Rev. {\bf D92}
  (2015) no.~7, 075034},
\href{http://arxiv.org/abs/1506.02644}{{\tt arXiv:1506.02644 [hep-ph]}}.
%%CITATION = ARXIV:1506.02644;%%.

\bibitem{Randall:1998uk}
L.~Randall and R.~Sundrum, {\em {Out of this world supersymmetry breaking}\/},
  \href{http://dx.doi.org/10.1016/S0550-3213(99)00359-4}{Nucl. Phys. {\bf B557}
  (1999)  79--118},
\href{http://arxiv.org/abs/hep-th/9810155}{{\tt arXiv:hep-th/9810155
  [hep-th]}}.
%%CITATION = HEP-TH/9810155;%%.

\bibitem{Giudice:1998xp}
G.~F. Giudice, M.~A. Luty, H.~Murayama, and R.~Rattazzi, {\em {Gaugino mass
  without singlets}\/},
  \href{http://dx.doi.org/10.1088/1126-6708/1998/12/027}{JHEP {\bf 12} (1998)
  027},
\href{http://arxiv.org/abs/hep-ph/9810442}{{\tt arXiv:hep-ph/9810442
  [hep-ph]}}.
%%CITATION = HEP-PH/9810442;%%.

\bibitem{Wells:2003tf}
J.~D. Wells, {\em {Implications of Supersymmetry Breaking with a Little
  Hierarchy Between Gauginos and Scalars}\/},  in {\em {11Th International
  Conference on Supersymmetry and the Unification of Fundamental Interactions
  (Susy 2003) Tucson, Arizona, June 5-10, 2003}}.
\newblock 2003.
\newblock
\href{http://arxiv.org/abs/hep-ph/0306127}{{\tt arXiv:hep-ph/0306127
  [hep-ph]}}.
\newblock
%%CITATION = HEP-PH/0306127;%%.

\bibitem{Acharya:2007rc}
B.~S. Acharya, K.~Bobkov, G.~L. Kane, P.~Kumar, and J.~Shao, {\em {Explaining
  the Electroweak Scale and Stabilizing Moduli in M Theory}\/},
  \href{http://dx.doi.org/10.1103/PhysRevD.76.126010}{Phys. Rev. {\bf D76}
  (2007)  126010},
\href{http://arxiv.org/abs/hep-th/0701034}{{\tt arXiv:hep-th/0701034
  [hep-th]}}.
%%CITATION = HEP-TH/0701034;%%.

\bibitem{Profumo:2004wk}
S.~Profumo and C.~E. Yaguna, {\em {Gluino coannihilations and heavy bino dark
  matter}\/},  \href{http://dx.doi.org/10.1103/PhysRevD.69.115009}{Phys. Rev.
  {\bf D69} (2004)  115009},
\href{http://arxiv.org/abs/hep-ph/0402208}{{\tt arXiv:hep-ph/0402208
  [hep-ph]}}.
%%CITATION = HEP-PH/0402208;%%.

\bibitem{Ellis:2015vaa}
J.~Ellis, F.~Luo, and K.~A. Olive, {\em {Gluino Coannihilation Revisited}\/},
  \href{http://dx.doi.org/10.1007/JHEP09(2015)127}{JHEP {\bf 09} (2015)  127},
\href{http://arxiv.org/abs/1503.07142}{{\tt arXiv:1503.07142 [hep-ph]}}.
%%CITATION = ARXIV:1503.07142;%%.

\bibitem{Kribs:2013oda}
G.~D. Kribs and A.~Martin, {\em {Dirac Gauginos in Supersymmetry -- Suppressed
  Jets + MET Signals: A Snowmass Whitepaper}\/},
\href{http://arxiv.org/abs/1308.3468}{{\tt arXiv:1308.3468 [hep-ph]}}.
%%CITATION = ARXIV:1308.3468;%%.

\bibitem{Khachatryan:2015vra}
{CMS Collaboration}, V.~Khachatryan et al., {\em {Searches for Supersymmetry
  using the M$_{T2}$ Variable in Hadronic Events Produced in pp Collisions at 8
  TeV}\/},  \href{http://dx.doi.org/10.1007/JHEP05(2015)078}{JHEP {\bf 05}
  (2015)  078},
\href{http://arxiv.org/abs/1502.04358}{{\tt arXiv:1502.04358 [hep-ex]}}.
%%CITATION = ARXIV:1502.04358;%%.

\bibitem{Aad:2015iea}
{ATLAS Collaboration}, G.~Aad et al., {\em {Summary of the searches for squarks
  and gluinos using $\sqrt{s}=8 $ TeV pp collisions with the ATLAS experiment
  at the LHC}\/},  \href{http://dx.doi.org/10.1007/JHEP10(2015)054}{JHEP {\bf
  10} (2015)  054},
\href{http://arxiv.org/abs/1507.05525}{{\tt arXiv:1507.05525 [hep-ex]}}.
%%CITATION = ARXIV:1507.05525;%%.

\bibitem{Jung:2014bda}
S.~Jung, {\em {Resolving the existence of Higgsinos in the LHC inverse
  problem}\/},  \href{http://dx.doi.org/10.1007/JHEP06(2014)111}{JHEP {\bf 06}
  (2014)  111},
\href{http://arxiv.org/abs/1404.2691}{{\tt arXiv:1404.2691 [hep-ph]}}.
%%CITATION = ARXIV:1404.2691;%%.

\bibitem{Hisano:2006nn}
J.~Hisano, S.~Matsumoto, M.~Nagai, O.~Saito, and M.~Senami, {\em
  {Non-perturbative effect on thermal relic abundance of dark matter}\/},
  \href{http://dx.doi.org/10.1016/j.physletb.2007.01.012}{Phys. Lett. {\bf
  B646} (2007)  34--38},
\href{http://arxiv.org/abs/hep-ph/0610249}{{\tt arXiv:hep-ph/0610249
  [hep-ph]}}.
%%CITATION = HEP-PH/0610249;%%.

\bibitem{Cohen:2013ama}
T.~Cohen, M.~Lisanti, A.~Pierce, and T.~R. Slatyer, {\em {Wino Dark Matter
  Under Siege}\/},  \href{http://dx.doi.org/10.1088/1475-7516/2013/10/061}{JCAP
  {\bf 1310} (2013)  061},
\href{http://arxiv.org/abs/1307.4082}{{\tt arXiv:1307.4082}}.
%%CITATION = ARXIV:1307.4082;%%.

\bibitem{Fan:2013faa}
J.~Fan and M.~Reece, {\em {In Wino Veritas? Indirect Searches Shed Light on
  Neutralino Dark Matter}\/},
  \href{http://dx.doi.org/10.1007/JHEP10(2013)124}{JHEP {\bf 1310} (2013)
  124},
\href{http://arxiv.org/abs/1307.4400}{{\tt arXiv:1307.4400 [hep-ph]}}.
%%CITATION = ARXIV:1307.4400;%%.

\bibitem{Baer:2012ts}
H.~Baer, V.~Barger, A.~Lessa, W.~Sreethawong, and X.~Tata, {\em {$Wh$ plus
  missing-$E_T$ signature from gaugino pair production at the LHC}\/},
  \href{http://dx.doi.org/10.1103/PhysRevD.85.055022}{Phys. Rev. {\bf D85}
  (2012)  055022},
\href{http://arxiv.org/abs/1201.2949}{{\tt arXiv:1201.2949 [hep-ph]}}.
%%CITATION = ARXIV:1201.2949;%%.

\bibitem{Howe:2012xe}
K.~Howe and P.~Saraswat, {\em {Excess Higgs Production in Neutralino
  Decays}\/},  \href{http://dx.doi.org/10.1007/JHEP10(2012)065}{JHEP {\bf 10}
  (2012)  065},
\href{http://arxiv.org/abs/1208.1542}{{\tt arXiv:1208.1542 [hep-ph]}}.
%%CITATION = ARXIV:1208.1542;%%.

\bibitem{Arbey:2012fa}
A.~Arbey, M.~Battaglia, and F.~Mahmoudi, {\em {Higgs Production in Neutralino
  Decays in the MSSM - The LHC and a Future $e^{+}e^-$ Collider}\/},
  \href{http://dx.doi.org/10.1140/epjc/s10052-015-3316-x}{Eur. Phys. J. {\bf
  C75} (2015) no.~3, 108},
\href{http://arxiv.org/abs/1212.6865}{{\tt arXiv:1212.6865 [hep-ph]}}.
%%CITATION = ARXIV:1212.6865;%%.

\bibitem{Griest:1990kh}
K.~Griest and D.~Seckel, {\em {Three exceptions in the calculation of relic
  abundances}\/},
\href{http://dx.doi.org/10.1103/PhysRevD.43.3191}{Phys. Rev. {\bf D43} (1991)
  3191--3203}.
%%CITATION = PHRVA,D43,3191;%%.

\bibitem{Ellis:1998kh}
J.~R. Ellis, T.~Falk, and K.~A. Olive, {\em {Neutralino - Stau coannihilation
  and the cosmological upper limit on the mass of the lightest supersymmetric
  particle}\/},  \href{http://dx.doi.org/10.1016/S0370-2693(98)01392-6}{Phys.
  Lett. {\bf B444} (1998)  367--372},
\href{http://arxiv.org/abs/hep-ph/9810360}{{\tt arXiv:hep-ph/9810360
  [hep-ph]}}.
%%CITATION = HEP-PH/9810360;%%.

\bibitem{Citron:2012fg}
M.~Citron, J.~Ellis, F.~Luo, J.~Marrouche, K.~A. Olive, and K.~J. de~Vries,
  {\em {End of the CMSSM coannihilation strip is nigh}\/},
  \href{http://dx.doi.org/10.1103/PhysRevD.87.036012}{Phys. Rev. {\bf D87}
  (2013) no.~3, 036012},
\href{http://arxiv.org/abs/1212.2886}{{\tt arXiv:1212.2886 [hep-ph]}}.
%%CITATION = ARXIV:1212.2886;%%.

\bibitem{Desai:2014uha}
N.~Desai, J.~Ellis, F.~Luo, and J.~Marrouche, {\em {Closing in on the Tip of
  the CMSSM Stau Coannihilation Strip}\/},
  \href{http://dx.doi.org/10.1103/PhysRevD.90.055031}{Phys. Rev. {\bf D90}
  (2014) no.~5, 055031},
\href{http://arxiv.org/abs/1404.5061}{{\tt arXiv:1404.5061 [hep-ph]}}.
%%CITATION = ARXIV:1404.5061;%%.

\bibitem{Feng:2003xh}
J.~L. Feng, A.~Rajaraman, and F.~Takayama, {\em {Superweakly interacting
  massive particles}\/},
  \href{http://dx.doi.org/10.1103/PhysRevLett.91.011302}{Phys. Rev. Lett. {\bf
  91} (2003)  011302},
\href{http://arxiv.org/abs/hep-ph/0302215}{{\tt arXiv:hep-ph/0302215
  [hep-ph]}}.
%%CITATION = HEP-PH/0302215;%%.

\bibitem{Feng:2003uy}
J.~L. Feng, A.~Rajaraman, and F.~Takayama, {\em {SuperWIMP dark matter signals
  from the early universe}\/},
  \href{http://dx.doi.org/10.1103/PhysRevD.68.063504}{Phys. Rev. {\bf D68}
  (2003)  063504},
\href{http://arxiv.org/abs/hep-ph/0306024}{{\tt arXiv:hep-ph/0306024
  [hep-ph]}}.
%%CITATION = HEP-PH/0306024;%%.

\bibitem{Ellis:1984eq}
J.~R. Ellis, J.~E. Kim, and D.~V. Nanopoulos, {\em {Cosmological Gravitino
  Regeneration and Decay}\/},
\href{http://dx.doi.org/10.1016/0370-2693(84)90334-4}{Phys. Lett. {\bf B145}
  (1984)  181}.
%%CITATION = PHLTA,B145,181;%%.

\bibitem{Moroi:1993mb}
T.~Moroi, H.~Murayama, and M.~Yamaguchi, {\em {Cosmological constraints on the
  light stable gravitino}\/},
\href{http://dx.doi.org/10.1016/0370-2693(93)91434-O}{Phys. Lett. {\bf B303}
  (1993)  289--294}.
%%CITATION = PHLTA,B303,289;%%.

\bibitem{Ellis:2003dn}
J.~R. Ellis, K.~A. Olive, Y.~Santoso, and V.~C. Spanos, {\em {Gravitino dark
  matter in the CMSSM}\/},
  \href{http://dx.doi.org/10.1016/j.physletb.2004.03.021}{Phys. Lett. {\bf
  B588} (2004)  7--16},
\href{http://arxiv.org/abs/hep-ph/0312262}{{\tt arXiv:hep-ph/0312262
  [hep-ph]}}.
%%CITATION = HEP-PH/0312262;%%.

\bibitem{Feng:2015wqa}
J.~L. Feng, S.~Iwamoto, Y.~Shadmi, and S.~Tarem, {\em {Long-Lived Sleptons at
  the LHC and a 100 TeV Proton Collider}\/},
  \href{http://dx.doi.org/10.1007/JHEP12(2015)166}{JHEP {\bf 12} (2015)  166},
\href{http://arxiv.org/abs/1505.02996}{{\tt arXiv:1505.02996 [hep-ph]}}.
%%CITATION = ARXIV:1505.02996;%%.

\bibitem{Groom:2001kq}
D.~E. Groom, N.~V. Mokhov, and S.~I. Striganov, {\em {Muon stopping power and
  range tables 10-MeV to 100-TeV}\/},
\href{http://dx.doi.org/10.1006/adnd.2001.0861}{Atom. Data Nucl. Data Tabl.
  {\bf 78} (2001)  183--356}.
%%CITATION = ADNDA,78,183;%%.

\bibitem{Chatrchyan:2013oca}
{CMS Collaboration}, S.~Chatrchyan et al., {\em {Searches for long-lived
  charged particles in pp collisions at $\sqrt{s}$=7 and 8 TeV}\/},
  \href{http://dx.doi.org/10.1007/JHEP07(2013)122}{JHEP {\bf 07} (2013)  122},
\href{http://arxiv.org/abs/1305.0491}{{\tt arXiv:1305.0491 [hep-ex]}}.
%%CITATION = ARXIV:1305.0491;%%.

\bibitem{ATLAS:2014fka}
{ATLAS Collaboration}, G.~Aad et al., {\em {Searches for heavy long-lived
  charged particles with the ATLAS detector in proton-proton collisions at $
  \sqrt{s}=8 $ TeV}\/},  \href{http://dx.doi.org/10.1007/JHEP01(2015)068}{JHEP
  {\bf 01} (2015)  068},
\href{http://arxiv.org/abs/1411.6795}{{\tt arXiv:1411.6795 [hep-ex]}}.
%%CITATION = ARXIV:1411.6795;%%.

\bibitem{Avetisyan:2013dta}
A.~Avetisyan, S.~Bhattacharya, M.~Narain, S.~Padhi, J.~Hirschauer, et al., {\em
  {Snowmass Energy Frontier Simulations using the Open Science Grid (A Snowmass
  2013 whitepaper)}\/},  in {\em {Community Summer Study 2013: Snowmass on the
  Mississippi (CSS2013) Minneapolis, MN, USA, July 29-August 6, 2013}}.
\newblock 2013.
\newblock
\href{http://arxiv.org/abs/1308.0843}{{\tt arXiv:1308.0843 [hep-ex]}}.
\newblock
%%CITATION = ARXIV:1308.0843;%%.

\bibitem{GEANT4}
{GEANT4 Collaboration}, S.~Agostinelli et al., {\em {GEANT4: A Simulation
  toolkit}\/},
\href{http://dx.doi.org/10.1016/S0168-9002(03)01368-8}{Nucl.Instrum.Meth. {\bf
  A506} (2003)  250--303}.
%%CITATION = NUIMA,A506,250;%%.

\bibitem{Salvucci:2012np}
A.~Salvucci, {\em {Measurement of muon momentum resolution of the ATLAS
  detector}\/},  \href{http://dx.doi.org/10.1051/epjconf/20122812039}{EPJ Web
  Conf. {\bf 28} (2012)  12039},
\href{http://arxiv.org/abs/1201.4704}{{\tt arXiv:1201.4704 [physics.ins-det]}}.
%%CITATION = ARXIV:1201.4704;%%.

\bibitem{Aad:2014yja}
{ATLAS Collaboration}, G.~Aad et al., {\em {Search For Higgs Boson Pair
  Production in the $\gamma\gamma b\bar{b}$ Final State using $pp$ Collision
  Data at $\sqrt{s}=8$ TeV from the ATLAS Detector}\/},
\href{http://arxiv.org/abs/1406.5053}{{\tt arXiv:1406.5053 [hep-ex]}}.
%%CITATION = ARXIV:1406.5053;%%.

\bibitem{Aad:2015xja}
{ATLAS Collaboration}, G.~Aad et al., {\em {Searches for Higgs boson pair
  production in the $hh\to bb\tau\tau, \gamma\gamma WW^*, \gamma\gamma bb,
  bbbb$ channels with the ATLAS detector}\/},
  \href{http://dx.doi.org/10.1103/PhysRevD.92.092004}{Phys. Rev. {\bf D92}
  (2015)  092004},
\href{http://arxiv.org/abs/1509.04670}{{\tt arXiv:1509.04670 [hep-ex]}}.
%%CITATION = ARXIV:1509.04670;%%.

\bibitem{CMS-PAS-HIG-13-032}
{CMS Collaboration}, {\em {Search for resonant HH production in 2gamma+2b
  channel}\/},   CMS-PAS-HIG-13-032, CERN, Geneva, 2014.
\newblock \url{https://cds.cern.ch/record/1697512}.

\bibitem{CMS-PAS-HIG-14-013}
{CMS Collaboration}, {\em {Search for di-Higgs resonances decaying to 4 bottom
  quarks}\/},   CMS-PAS-HIG-14-013, CERN, Geneva, 2014.
\newblock \url{https://cds.cern.ch/record/1748425}.

\bibitem{Aad:2015uka}
{ATLAS Collaboration}, G.~Aad et al., {\em {Search for Higgs boson pair
  production in the $b\bar{b}b\bar{b}$ final state from pp collisions at
  $\sqrt{s} = 8$ TeVwith the ATLAS detector}\/},
  \href{http://dx.doi.org/10.1140/epjc/s10052-015-3628-x}{Eur. Phys. J. {\bf
  C75} (2015) no.~9, 412},
\href{http://arxiv.org/abs/1506.00285}{{\tt arXiv:1506.00285 [hep-ex]}}.
%%CITATION = ARXIV:1506.00285;%%.

\bibitem{CMS-PAS-HIG-13-025}
{CMS Collaboration}, {\em {2HDM scenario, H to hh and A to Zh}\/},
  CMS-PAS-HIG-13-025, CERN, Geneva, 2013.
\newblock \url{https://cds.cern.ch/record/1637776}.

\bibitem{ATL-PHYS-PUB-2014-019}
{\em {Prospects for measuring Higgs pair production in the channel
  $H(\rightarrow\gamma\gamma)H(\rightarrow b\overline{b}) $ using the ATLAS
  detector at the HL-LHC}\/},   ATL-PHYS-PUB-2014-019, CERN, Geneva, Oct, 2014.
\newblock \url{https://cds.cern.ch/record/1956733}.

\bibitem{CMS:2015nat}
{CMS Collaboration}, {\em {Higgs pair production at the High Luminosity
  LHC}\/},   CMS-PAS-FTR-15-002, CERN, Geneva, 2015.
\newblock \url{http://cds.cern.ch/record/2063038}.

\bibitem{Barr:2014sga}
A.~J. Barr, M.~J. Dolan, C.~Englert, D.~E. Ferreira~de Lima, and M.~Spannowsky,
  {\em {Higgs Self-Coupling Measurements at a 100 TeV Hadron Collider}\/},
  \href{http://dx.doi.org/10.1007/JHEP02(2015)016}{JHEP {\bf 02} (2015)  016},
\href{http://arxiv.org/abs/1412.7154}{{\tt arXiv:1412.7154 [hep-ph]}}.
%%CITATION = ARXIV:1412.7154;%%.

\bibitem{Azatov:2015oxa}
A.~Azatov, R.~Contino, G.~Panico, and M.~Son, {\em {Effective field theory
  analysis of double Higgs boson production via gluon fusion}\/},
  \href{http://dx.doi.org/10.1103/PhysRevD.92.035001}{Phys. Rev. {\bf D92}
  (2015) no.~3, 035001},
\href{http://arxiv.org/abs/1502.00539}{{\tt arXiv:1502.00539 [hep-ph]}}.
%%CITATION = ARXIV:1502.00539;%%.

\bibitem{Li:2015yia}
Q.~Li, Z.~Li, Q.-S. Yan, and X.~Zhao, {\em {Probe Higgs boson pair production
  via the $3\ell 2j+\not{E}$ mode}\/},
  \href{http://dx.doi.org/10.1103/PhysRevD.92.014015}{Phys. Rev. {\bf D92}
  (2015) no.~1, 014015},
\href{http://arxiv.org/abs/1503.07611}{{\tt arXiv:1503.07611 [hep-ph]}}.
%%CITATION = ARXIV:1503.07611;%%.

\bibitem{Papaefstathiou:2015iba}
A.~Papaefstathiou, {\em {Discovering Higgs boson pair production through rare
  final states at a 100 TeV collider}\/},
  \href{http://dx.doi.org/10.1103/PhysRevD.91.113016}{Phys. Rev. {\bf D91}
  (2015) no.~11, 113016},
\href{http://arxiv.org/abs/1504.04621}{{\tt arXiv:1504.04621 [hep-ph]}}.
%%CITATION = ARXIV:1504.04621;%%.

\bibitem{Brust:2012uf}
C.~Brust, A.~Katz, and R.~Sundrum, {\em {SUSY Stops at a Bump}\/},
  \href{http://dx.doi.org/10.1007/JHEP08(2012)059}{JHEP {\bf 08} (2012)  059},
\href{http://arxiv.org/abs/1206.2353}{{\tt arXiv:1206.2353 [hep-ph]}}.
%%CITATION = ARXIV:1206.2353;%%.

\bibitem{Evans:2012bf}
J.~A. Evans and Y.~Kats, {\em {LHC Coverage of RPV MSSM with Light Stops}\/},
  \href{http://dx.doi.org/10.1007/JHEP04(2013)028}{JHEP {\bf 04} (2013)  028},
\href{http://arxiv.org/abs/1209.0764}{{\tt arXiv:1209.0764 [hep-ph]}}.
%%CITATION = ARXIV:1209.0764;%%.

\bibitem{Bai:2013xla}
Y.~Bai, A.~Katz, and B.~Tweedie, {\em {Pulling Out All the Stops: Searching for
  RPV SUSY with Stop-Jets}\/},
  \href{http://dx.doi.org/10.1007/JHEP01(2014)040}{JHEP {\bf 01} (2014)  040},
\href{http://arxiv.org/abs/1309.6631}{{\tt arXiv:1309.6631 [hep-ph]}}.
%%CITATION = ARXIV:1309.6631;%%.

\bibitem{Csaki:2012fh}
C.~Csaki, L.~Randall, and J.~Terning, {\em {Light Stops from Seiberg
  Duality}\/},  \href{http://dx.doi.org/10.1103/PhysRevD.86.075009}{Phys.Rev.
  {\bf D86} (2012)  075009},
\href{http://arxiv.org/abs/1201.1293}{{\tt arXiv:1201.1293 [hep-ph]}}.
%%CITATION = ARXIV:1201.1293;%%.

\bibitem{Han:2012fw}
Z.~Han, A.~Katz, D.~Krohn, and M.~Reece, {\em {(Light) Stop Signs}\/},
  \href{http://dx.doi.org/10.1007/JHEP08(2012)083}{JHEP {\bf 08} (2012)  083},
\href{http://arxiv.org/abs/1205.5808}{{\tt arXiv:1205.5808 [hep-ph]}}.
%%CITATION = ARXIV:1205.5808;%%.

\bibitem{Kilic:2012kw}
C.~Kilic and B.~Tweedie, {\em {Cornering Light Stops with Dileptonic mT2}\/},
  \href{http://dx.doi.org/10.1007/JHEP04(2013)110}{JHEP {\bf 04} (2013)  110},
\href{http://arxiv.org/abs/1211.6106}{{\tt arXiv:1211.6106 [hep-ph]}}.
%%CITATION = ARXIV:1211.6106;%%.

\bibitem{Czakon:2014fka}
M.~Czakon, A.~Mitov, M.~Papucci, J.~T. Ruderman, and A.~Weiler, {\em {Closing
  the stop gap}\/},
  \href{http://dx.doi.org/10.1103/PhysRevLett.113.201803}{Phys. Rev. Lett. {\bf
  113} (2014) no.~20, 201803},
\href{http://arxiv.org/abs/1407.1043}{{\tt arXiv:1407.1043 [hep-ph]}}.
%%CITATION = ARXIV:1407.1043;%%.

\bibitem{LeCompte:2011fh}
T.~J. LeCompte and S.~P. Martin, {\em {Compressed supersymmetry after 1/fb at
  the Large Hadron Collider}\/},
  \href{http://dx.doi.org/10.1103/PhysRevD.85.035023}{Phys. Rev. {\bf D85}
  (2012)  035023},
\href{http://arxiv.org/abs/1111.6897}{{\tt arXiv:1111.6897 [hep-ph]}}.
%%CITATION = ARXIV:1111.6897;%%.

\bibitem{Dreiner:2012gx}
H.~K. Dreiner, M.~Kr\"{a}mer, and J.~Tattersall, {\em {How low can SUSY go?
  Matching, monojets and compressed spectra}\/},
  \href{http://dx.doi.org/10.1209/0295-5075/99/61001}{Europhys. Lett. {\bf 99}
  (2012)  61001},
\href{http://arxiv.org/abs/1207.1613}{{\tt arXiv:1207.1613 [hep-ph]}}.
%%CITATION = ARXIV:1207.1613;%%.

\bibitem{Bhattacherjee:2012mz}
B.~Bhattacherjee and K.~Ghosh, {\em {Degenerate SUSY search at the 8 TeV
  LHC}\/},
\href{http://arxiv.org/abs/1207.6289}{{\tt arXiv:1207.6289 [hep-ph]}}.
%%CITATION = ARXIV:1207.6289;%%.

\bibitem{Drees:2012dd}
M.~Drees, M.~Hanussek, and J.~S. Kim, {\em {Light Stop Searches at the LHC with
  Monojet Events}\/},
  \href{http://dx.doi.org/10.1103/PhysRevD.86.035024}{Phys. Rev. {\bf D86}
  (2012)  035024},
\href{http://arxiv.org/abs/1201.5714}{{\tt arXiv:1201.5714 [hep-ph]}}.
%%CITATION = ARXIV:1201.5714;%%.

\bibitem{Belanger:2012mk}
G.~Belanger, M.~Heikinheimo, and V.~Sanz, {\em {Model-Independent Bounds on
  Squarks from Monophoton Searches}\/},
  \href{http://dx.doi.org/10.1007/JHEP08(2012)151}{JHEP {\bf 08} (2012)  151},
\href{http://arxiv.org/abs/1205.1463}{{\tt arXiv:1205.1463 [hep-ph]}}.
%%CITATION = ARXIV:1205.1463;%%.

\bibitem{Alves:2012ft}
D.~S.~M. Alves, M.~R. Buckley, P.~J. Fox, J.~D. Lykken, and C.-T. Yu, {\em
  {Stops and $\not E_T$: The shape of things to come}\/},
  \href{http://dx.doi.org/10.1103/PhysRevD.87.035016}{Phys. Rev. {\bf D87}
  (2013) no.~3, 035016},
\href{http://arxiv.org/abs/1205.5805}{{\tt arXiv:1205.5805 [hep-ph]}}.
%%CITATION = ARXIV:1205.5805;%%.

\bibitem{Krizka:2012ah}
K.~Krizka, A.~Kumar, and D.~E. Morrissey, {\em {Very Light Scalar Top Quarks at
  the LHC}\/},  \href{http://dx.doi.org/10.1103/PhysRevD.87.095016}{Phys. Rev.
  {\bf D87} (2013) no.~9, 095016},
\href{http://arxiv.org/abs/1212.4856}{{\tt arXiv:1212.4856 [hep-ph]}}.
%%CITATION = ARXIV:1212.4856;%%.

\bibitem{Carena:2012gp}
M.~Carena, S.~Gori, N.~R. Shah, C.~E.~M. Wagner, and L.-T. Wang, {\em {Light
  Stau Phenomenology and the Higgs $\gamma\gamma$ Rate}\/},
  \href{http://dx.doi.org/10.1007/JHEP07(2012)175}{JHEP {\bf 07} (2012)  175},
\href{http://arxiv.org/abs/1205.5842}{{\tt arXiv:1205.5842 [hep-ph]}}.
%%CITATION = ARXIV:1205.5842;%%.

\bibitem{Carena:2013iba}
M.~Carena, S.~Gori, N.~R. Shah, C.~E.~M. Wagner, and L.-T. Wang, {\em {Light
  Stops, Light Staus and the 125 GeV Higgs}\/},
  \href{http://dx.doi.org/10.1007/JHEP08(2013)087}{JHEP {\bf 08} (2013)  087},
\href{http://arxiv.org/abs/1303.4414}{{\tt arXiv:1303.4414 [hep-ph]}}.
%%CITATION = ARXIV:1303.4414;%%.

\bibitem{Batell:2015koa}
B.~Batell, M.~McCullough, D.~Stolarski, and C.~B. Verhaaren, {\em {Putting a
  Stop to di-Higgs Modifications}\/},
  \href{http://dx.doi.org/10.1007/JHEP09(2015)216}{JHEP {\bf 09} (2015)  216},
\href{http://arxiv.org/abs/1508.01208}{{\tt arXiv:1508.01208 [hep-ph]}}.
%%CITATION = ARXIV:1508.01208;%%.

\bibitem{Plehn:1996wb}
T.~Plehn, M.~Spira, and P.~M. Zerwas, {\em {Pair production of neutral Higgs
  particles in gluon-gluon collisions}\/},
  \href{http://dx.doi.org/10.1016/0550-3213(96)00418-X}{Nucl. Phys. {\bf B479}
  (1996)  46--64}, \href{http://arxiv.org/abs/hep-ph/9603205}{{\tt
  arXiv:hep-ph/9603205 [hep-ph]}}.
[Erratum: Nucl. Phys.B531,655(1998)].
%%CITATION = HEP-PH/9603205;%%.

\bibitem{Djouadi:1999rca}
A.~Djouadi, W.~Kilian, M.~Muhlleitner, and P.~M. Zerwas, {\em {Production of
  neutral Higgs boson pairs at LHC}\/},
  \href{http://dx.doi.org/10.1007/s100529900083}{Eur. Phys. J. {\bf C10} (1999)
   45--49},
\href{http://arxiv.org/abs/hep-ph/9904287}{{\tt arXiv:hep-ph/9904287
  [hep-ph]}}.
%%CITATION = HEP-PH/9904287;%%.

\bibitem{Belyaev:1999mx}
A.~Belyaev, M.~Drees, O.~J.~P. Eboli, J.~K. Mizukoshi, and S.~F. Novaes, {\em
  {Supersymmetric Higgs pair production at hadron colliders}\/},
  \href{http://dx.doi.org/10.1103/PhysRevD.60.075008}{Phys. Rev. {\bf D60}
  (1999)  075008},
\href{http://arxiv.org/abs/hep-ph/9905266}{{\tt arXiv:hep-ph/9905266
  [hep-ph]}}.
%%CITATION = HEP-PH/9905266;%%.

\bibitem{BarrientosBendezu:2001di}
A.~A. Barrientos~Bendezu and B.~A. Kniehl, {\em {Pair production of neutral
  Higgs bosons at the CERN large hadron collider}\/},
  \href{http://dx.doi.org/10.1103/PhysRevD.64.035006}{Phys. Rev. {\bf D64}
  (2001)  035006},
\href{http://arxiv.org/abs/hep-ph/0103018}{{\tt arXiv:hep-ph/0103018
  [hep-ph]}}.
%%CITATION = HEP-PH/0103018;%%.

\bibitem{Kane:1993td}
G.~L. Kane, C.~F. Kolda, L.~Roszkowski, and J.~D. Wells, {\em {Study of
  constrained minimal supersymmetry}\/},
  \href{http://dx.doi.org/10.1103/PhysRevD.49.6173}{Phys. Rev. {\bf D49} (1994)
   6173--6210},
\href{http://arxiv.org/abs/hep-ph/9312272}{{\tt arXiv:hep-ph/9312272
  [hep-ph]}}.
%%CITATION = HEP-PH/9312272;%%.

\bibitem{Ellis:1996xu}
J.~R. Ellis, T.~Falk, K.~A. Olive, and M.~Schmitt, {\em {Supersymmetric dark
  matter in the light of LEP-1.5}\/},
  \href{http://dx.doi.org/10.1016/0370-2693(96)01130-6}{Phys. Lett. {\bf B388}
  (1996)  97--105},
\href{http://arxiv.org/abs/hep-ph/9607292}{{\tt arXiv:hep-ph/9607292
  [hep-ph]}}.
%%CITATION = HEP-PH/9607292;%%.

\bibitem{Ellis:1997wva}
J.~R. Ellis, T.~Falk, K.~A. Olive, and M.~Schmitt, {\em {Constraints on
  neutralino dark matter from LEP-2 and cosmology}\/},
  \href{http://dx.doi.org/10.1016/S0370-2693(97)01122-2}{Phys. Lett. {\bf B413}
  (1997)  355--364},
\href{http://arxiv.org/abs/hep-ph/9705444}{{\tt arXiv:hep-ph/9705444
  [hep-ph]}}.
%%CITATION = HEP-PH/9705444;%%.

\bibitem{Ellis:1998jk}
J.~R. Ellis, T.~Falk, G.~Ganis, K.~A. Olive, and M.~Schmitt, {\em {Charginos
  and neutralinos in the light of radiative corrections: Sealing the fate of
  Higgsino dark matter}\/},
  \href{http://dx.doi.org/10.1103/PhysRevD.58.095002}{Phys. Rev. {\bf D58}
  (1998)  095002},
\href{http://arxiv.org/abs/hep-ph/9801445}{{\tt arXiv:hep-ph/9801445
  [hep-ph]}}.
%%CITATION = HEP-PH/9801445;%%.

\bibitem{Barger:1997kb}
V.~D. Barger and C.~Kao, {\em {Relic density of neutralino dark matter in
  supergravity models}\/},
  \href{http://dx.doi.org/10.1103/PhysRevD.57.3131}{Phys. Rev. {\bf D57} (1998)
   3131--3139},
\href{http://arxiv.org/abs/hep-ph/9704403}{{\tt arXiv:hep-ph/9704403
  [hep-ph]}}.
%%CITATION = HEP-PH/9704403;%%.

\bibitem{Ellis:2000we}
J.~R. Ellis, T.~Falk, G.~Ganis, and K.~A. Olive, {\em {Supersymmetric dark
  matter in the light of LEP and the Tevatron collider}\/},
  \href{http://dx.doi.org/10.1103/PhysRevD.62.075010}{Phys. Rev. {\bf D62}
  (2000)  075010},
\href{http://arxiv.org/abs/hep-ph/0004169}{{\tt arXiv:hep-ph/0004169
  [hep-ph]}}.
%%CITATION = HEP-PH/0004169;%%.

\bibitem{Roszkowski:2001sb}
L.~Roszkowski, R.~Ruiz~de Austri, and T.~Nihei, {\em {New cosmological and
  experimental constraints on the CMSSM}\/},
  \href{http://dx.doi.org/10.1088/1126-6708/2001/08/024}{JHEP {\bf 08} (2001)
  024},
\href{http://arxiv.org/abs/hep-ph/0106334}{{\tt arXiv:hep-ph/0106334
  [hep-ph]}}.
%%CITATION = HEP-PH/0106334;%%.

\bibitem{Djouadi:2001yk}
A.~Djouadi, M.~Drees, and J.~L. Kneur, {\em {Constraints on the minimal
  supergravity model and prospects for SUSY particle production at future
  linear $e^{+} e^{-}$ colliders}\/},
  \href{http://dx.doi.org/10.1088/1126-6708/2001/08/055}{JHEP {\bf 08} (2001)
  055},
\href{http://arxiv.org/abs/hep-ph/0107316}{{\tt arXiv:hep-ph/0107316
  [hep-ph]}}.
%%CITATION = HEP-PH/0107316;%%.

\bibitem{Chattopadhyay:2001va}
U.~Chattopadhyay, A.~Corsetti, and P.~Nath, {\em {Supersymmetric dark matter
  and Yukawa unification}\/},
  \href{http://dx.doi.org/10.1103/PhysRevD.66.035003}{Phys. Rev. {\bf D66}
  (2002)  035003},
\href{http://arxiv.org/abs/hep-ph/0201001}{{\tt arXiv:hep-ph/0201001
  [hep-ph]}}.
%%CITATION = HEP-PH/0201001;%%.

\bibitem{Ellis:2002rp}
J.~R. Ellis, K.~A. Olive, and Y.~Santoso, {\em {Constraining supersymmetry}\/},
   \href{http://dx.doi.org/10.1088/1367-2630/4/1/332}{New J. Phys. {\bf 4}
  (2002)  32},
\href{http://arxiv.org/abs/hep-ph/0202110}{{\tt arXiv:hep-ph/0202110
  [hep-ph]}}.
%%CITATION = HEP-PH/0202110;%%.

\bibitem{Buchmueller:2015uqa}
O.~Buchmueller, M.~Citron, J.~Ellis, S.~Guha, J.~Marrouche, K.~A. Olive,
  K.~de~Vries, and J.~Zheng, {\em {Collider Interplay for Supersymmetry, Higgs
  and Dark Matter}\/},
  \href{http://dx.doi.org/10.1140/epjc/s10052-015-3675-3}{Eur. Phys. J. {\bf
  C75} (2015) no.~10, 469},
\href{http://arxiv.org/abs/1505.04702}{{\tt arXiv:1505.04702 [hep-ph]}}.
%%CITATION = ARXIV:1505.04702;%%.

\bibitem{Boehm:1999bj}
C.~Boehm, A.~Djouadi, and M.~Drees, {\em {Light scalar top quarks and
  supersymmetric dark matter}\/},
  \href{http://dx.doi.org/10.1103/PhysRevD.62.035012}{Phys. Rev. {\bf D62}
  (2000)  035012},
\href{http://arxiv.org/abs/hep-ph/9911496}{{\tt arXiv:hep-ph/9911496
  [hep-ph]}}.
%%CITATION = HEP-PH/9911496;%%.

\bibitem{Ellis:2001nx}
J.~R. Ellis, K.~A. Olive, and Y.~Santoso, {\em {Calculations of neutralino stop
  coannihilation in the CMSSM}\/},
  \href{http://dx.doi.org/10.1016/S0927-6505(02)00151-2}{Astropart. Phys. {\bf
  18} (2003)  395--432},
\href{http://arxiv.org/abs/hep-ph/0112113}{{\tt arXiv:hep-ph/0112113
  [hep-ph]}}.
%%CITATION = HEP-PH/0112113;%%.

\bibitem{Edsjo:2003us}
J.~Edsjo, M.~Schelke, P.~Ullio, and P.~Gondolo, {\em {Accurate relic densities
  with neutralino, chargino and sfermion coannihilations in mSUGRA}\/},
  \href{http://dx.doi.org/10.1088/1475-7516/2003/04/001}{JCAP {\bf 0304} (2003)
   001},
\href{http://arxiv.org/abs/hep-ph/0301106}{{\tt arXiv:hep-ph/0301106
  [hep-ph]}}.
%%CITATION = HEP-PH/0301106;%%.

\bibitem{DiazCruz:2007fc}
J.~L. Diaz-Cruz, J.~R. Ellis, K.~A. Olive, and Y.~Santoso, {\em {On the
  Feasibility of a Stop NLSP in Gravitino Dark Matter Scenarios}\/},
  \href{http://dx.doi.org/10.1088/1126-6708/2007/05/003}{JHEP {\bf 05} (2007)
  003},
\href{http://arxiv.org/abs/hep-ph/0701229}{{\tt arXiv:hep-ph/0701229
  [HEP-PH]}}.
%%CITATION = HEP-PH/0701229;%%.

\bibitem{Gogoladze:2011be}
I.~Gogoladze, S.~Raza, and Q.~Shafi, {\em {Light stop from b{\^O}{\o}à¥à¤tau
  Yukawa unification}\/},
  \href{http://dx.doi.org/10.1016/j.physletb.2011.11.026}{Phys. Lett. {\bf
  B706} (2012)  345--349},
\href{http://arxiv.org/abs/1104.3566}{{\tt arXiv:1104.3566 [hep-ph]}}.
%%CITATION = ARXIV:1104.3566;%%.

\bibitem{Ajaib:2011hs}
M.~A. Ajaib, T.~Li, and Q.~Shafi, {\em {Stop-Neutralino Coannihilation in the
  Light of LHC}\/},  \href{http://dx.doi.org/10.1103/PhysRevD.85.055021}{Phys.
  Rev. {\bf D85} (2012)  055021},
\href{http://arxiv.org/abs/1111.4467}{{\tt arXiv:1111.4467 [hep-ph]}}.
%%CITATION = ARXIV:1111.4467;%%.

\bibitem{eoz}
J.~Ellis, K.~A. Olive, and J.~Zheng, {\em {The Extent of the Stop
  Coannihilation Strip}\/},
  \href{http://dx.doi.org/10.1140/epjc/s10052-014-2947-7}{Eur. Phys. J. {\bf
  C74} (2014)  2947},
\href{http://arxiv.org/abs/1404.5571}{{\tt arXiv:1404.5571 [hep-ph]}}.
%%CITATION = ARXIV:1404.5571;%%.

\bibitem{Feng:1999mn}
J.~L. Feng, K.~T. Matchev, and T.~Moroi, {\em {Multi - TeV scalars are natural
  in minimal supergravity}\/},
  \href{http://dx.doi.org/10.1103/PhysRevLett.84.2322}{Phys. Rev. Lett. {\bf
  84} (2000)  2322--2325},
\href{http://arxiv.org/abs/hep-ph/9908309}{{\tt arXiv:hep-ph/9908309
  [hep-ph]}}.
%%CITATION = HEP-PH/9908309;%%.

\bibitem{Feng:2000gh}
J.~L. Feng, K.~T. Matchev, and F.~Wilczek, {\em {Neutralino dark matter in
  focus point supersymmetry}\/},
  \href{http://dx.doi.org/10.1016/S0370-2693(00)00512-8}{Phys. Lett. {\bf B482}
  (2000)  388--399},
\href{http://arxiv.org/abs/hep-ph/0004043}{{\tt arXiv:hep-ph/0004043
  [hep-ph]}}.
%%CITATION = HEP-PH/0004043;%%.

\bibitem{Baer:2005ky}
H.~Baer, T.~Krupovnickas, S.~Profumo, and P.~Ullio, {\em {Model independent
  approach to focus point supersymmetry: From dark matter to collider
  searches}\/},  \href{http://dx.doi.org/10.1088/1126-6708/2005/10/020}{JHEP
  {\bf 10} (2005)  020},
\href{http://arxiv.org/abs/hep-ph/0507282}{{\tt arXiv:hep-ph/0507282
  [hep-ph]}}.
%%CITATION = HEP-PH/0507282;%%.

\bibitem{Feng:2011aa}
J.~L. Feng, K.~T. Matchev, and D.~Sanford, {\em {Focus Point Supersymmetry
  Redux}\/},  \href{http://dx.doi.org/10.1103/PhysRevD.85.075007}{Phys. Rev.
  {\bf D85} (2012)  075007},
\href{http://arxiv.org/abs/1112.3021}{{\tt arXiv:1112.3021 [hep-ph]}}.
%%CITATION = ARXIV:1112.3021;%%.

\bibitem{Draper:2013cka}
P.~Draper, J.~L. Feng, P.~Kant, S.~Profumo, and D.~Sanford, {\em {Dark Matter
  Detection in Focus Point Supersymmetry}\/},
  \href{http://dx.doi.org/10.1103/PhysRevD.88.015025}{Phys. Rev. {\bf D88}
  (2013) no.~1, 015025},
\href{http://arxiv.org/abs/1304.1159}{{\tt arXiv:1304.1159 [hep-ph]}}.
%%CITATION = ARXIV:1304.1159;%%.

\bibitem{Hinchliffe:2001bz}
I.~Hinchliffe and F.~E. Paige, {\em {High mass SUSY Models at LHC and VLHC.
  Part 1.}\/},
eConf {\bf C010630} (2001)  E401.
%%CITATION = ECONF,C010630,E401;%%.

\bibitem{Vega:2015fna}
J.~P. Vega and G.~Villadoro, {\em {Susyhd: Higgs Mass Determination in
  Supersymmetry}\/},  \href{http://dx.doi.org/10.1007/JHEP07(2015)159}{JHEP
  {\bf 07} (2015)  159},
\href{http://arxiv.org/abs/1504.05200}{{\tt arXiv:1504.05200 [hep-ph]}}.
%%CITATION = ARXIV:1504.05200;%%.

\bibitem{Barbieri:1990ja}
R.~Barbieri, M.~Frigeni, and F.~Caravaglios, {\em {The Supersymmetric Higgs for
  heavy superpartners}\/},
\href{http://dx.doi.org/10.1016/0370-2693(91)91226-L}{Phys. Lett. {\bf B258}
  (1991)  167--170}.
%%CITATION = PHLTA,B258,167;%%.

\bibitem{Haber:1990aw}
H.~E. Haber and R.~Hempfling, {\em {Can the mass of the lightest Higgs boson of
  the minimal supersymmetric model be larger than m(Z)?}\/},
\href{http://dx.doi.org/10.1103/PhysRevLett.66.1815}{Phys. Rev. Lett. {\bf 66}
  (1991)  1815--1818}.
%%CITATION = PRLTA,66,1815;%%.

\bibitem{Casas:1994us}
J.~A. Casas, J.~R. Espinosa, M.~Quiros, and A.~Riotto, {\em {The Lightest Higgs
  boson mass in the minimal supersymmetric standard model}\/},
  \href{http://dx.doi.org/10.1016/0550-3213(94)00508-C}{Nucl. Phys. {\bf B436}
  (1995)  3--29}, \href{http://arxiv.org/abs/hep-ph/9407389}{{\tt
  arXiv:hep-ph/9407389 [hep-ph]}}.
[Erratum: Nucl. Phys.B439,466(1995)].
%%CITATION = HEP-PH/9407389;%%.

\bibitem{Carena:1995bx}
M.~Carena, J.~R. Espinosa, M.~Quiros, and C.~E.~M. Wagner, {\em {Analytical
  expressions for radiatively corrected Higgs masses and couplings in the
  MSSM}\/},  \href{http://dx.doi.org/10.1016/0370-2693(95)00694-G}{Phys. Lett.
  {\bf B355} (1995)  209--221},
\href{http://arxiv.org/abs/hep-ph/9504316}{{\tt arXiv:hep-ph/9504316
  [hep-ph]}}.
%%CITATION = HEP-PH/9504316;%%.

\bibitem{Draper:2016pys}
P.~Draper and H.~Rzehak, {\em {A Review of Higgs Mass Calculations in
  Supersymmetric Models}\/},
\href{http://arxiv.org/abs/1601.01890}{{\tt arXiv:1601.01890 [hep-ph]}}.
%%CITATION = ARXIV:1601.01890;%%.

\bibitem{Fan:2011jc}
J.~Fan, D.~Krohn, P.~Mosteiro, A.~M. Thalapillil, and L.-T. Wang, {\em {Heavy
  Squarks at the LHC}\/},
  \href{http://dx.doi.org/10.1007/JHEP03(2011)077}{JHEP {\bf 03} (2011)  077},
\href{http://arxiv.org/abs/1102.0302}{{\tt arXiv:1102.0302 [hep-ph]}}.
%%CITATION = ARXIV:1102.0302;%%.

\bibitem{Sato:2012xf}
R.~Sato, S.~Shirai, and K.~Tobioka, {\em {Gluino Decay as a Probe of High Scale
  Supersymmetry Breaking}\/},
  \href{http://dx.doi.org/10.1007/JHEP11(2012)041}{JHEP {\bf 11} (2012)  041},
\href{http://arxiv.org/abs/1207.3608}{{\tt arXiv:1207.3608 [hep-ph]}}.
%%CITATION = ARXIV:1207.3608;%%.

\bibitem{Toharia:2005gm}
M.~Toharia and J.~D. Wells, {\em {Gluino decays with heavier scalar
  superpartners}\/},
  \href{http://dx.doi.org/10.1088/1126-6708/2006/02/015}{JHEP {\bf 02} (2006)
  015},
\href{http://arxiv.org/abs/hep-ph/0503175}{{\tt arXiv:hep-ph/0503175
  [hep-ph]}}.
%%CITATION = HEP-PH/0503175;%%.

\bibitem{Gambino:2005eh}
P.~Gambino, G.~F. Giudice, and P.~Slavich, {\em {Gluino decays in split
  supersymmetry}\/},
  \href{http://dx.doi.org/10.1016/j.nuclphysb.2005.08.011}{Nucl. Phys. {\bf
  B726} (2005)  35--52},
\href{http://arxiv.org/abs/hep-ph/0506214}{{\tt arXiv:hep-ph/0506214
  [hep-ph]}}.
%%CITATION = HEP-PH/0506214;%%.

\bibitem{Choi:2005ge}
K.~Choi, A.~Falkowski, H.~P. Nilles, and M.~Olechowski, {\em {Soft
  supersymmetry breaking in KKLT flux compactification}\/},
  \href{http://dx.doi.org/10.1016/j.nuclphysb.2005.04.032}{Nucl. Phys. {\bf
  B718} (2005)  113--133},
\href{http://arxiv.org/abs/hep-th/0503216}{{\tt arXiv:hep-th/0503216
  [hep-th]}}.
%%CITATION = HEP-TH/0503216;%%.

\bibitem{Choi:2005hd}
K.~Choi, K.~S. Jeong, T.~Kobayashi, and K.-i. Okumura, {\em {Little SUSY
  hierarchy in mixed modulus-anomaly mediation}\/},
  \href{http://dx.doi.org/10.1016/j.physletb.2005.11.078}{Phys. Lett. {\bf
  B633} (2006)  355--361},
\href{http://arxiv.org/abs/hep-ph/0508029}{{\tt arXiv:hep-ph/0508029
  [hep-ph]}}.
%%CITATION = HEP-PH/0508029;%%.

\bibitem{Acharya:2008zi}
B.~S. Acharya, K.~Bobkov, G.~L. Kane, J.~Shao, and P.~Kumar, {\em {The
  G(2)-MSSM: An M Theory motivated model of Particle Physics}\/},
  \href{http://dx.doi.org/10.1103/PhysRevD.78.065038}{Phys. Rev. {\bf D78}
  (2008)  065038},
\href{http://arxiv.org/abs/0801.0478}{{\tt arXiv:0801.0478 [hep-ph]}}.
%%CITATION = ARXIV:0801.0478;%%.

\bibitem{Coughlan:1983ci}
G.~D. Coughlan, W.~Fischler, E.~W. Kolb, S.~Raby, and G.~G. Ross, {\em
  {Cosmological Problems for the Polonyi Potential}\/},
\href{http://dx.doi.org/10.1016/0370-2693(83)91091-2}{Phys. Lett. {\bf B131}
  (1983)  59}.
%%CITATION = PHLTA,B131,59;%%.

\bibitem{Ellis:1986zt}
J.~R. Ellis, D.~V. Nanopoulos, and M.~Quiros, {\em {On the Axion, Dilaton,
  Polonyi, Gravitino and Shadow Matter Problems in Supergravity and Superstring
  Models}\/},
\href{http://dx.doi.org/10.1016/0370-2693(86)90736-7}{Phys. Lett. {\bf B174}
  (1986)  176}.
%%CITATION = PHLTA,B174,176;%%.

\bibitem{deCarlos:1993wie}
B.~de~Carlos, J.~A. Casas, F.~Quevedo, and E.~Roulet, {\em {Model independent
  properties and cosmological implications of the dilaton and moduli sectors of
  4-d strings}\/},  \href{http://dx.doi.org/10.1016/0370-2693(93)91538-X}{Phys.
  Lett. {\bf B318} (1993)  447--456},
\href{http://arxiv.org/abs/hep-ph/9308325}{{\tt arXiv:hep-ph/9308325
  [hep-ph]}}.
%%CITATION = HEP-PH/9308325;%%.

\bibitem{Moroi:1995fs}
T.~Moroi, {\em {Effects of the Gravitino on the Inflationary Universe}}.
\newblock PhD thesis, Tohoku U., 1995.
\newblock
\href{http://arxiv.org/abs/hep-ph/9503210}{{\tt arXiv:hep-ph/9503210
  [hep-ph]}}.
\newblock
%%CITATION = HEP-PH/9503210;%%.

\bibitem{Kane:2015jia}
G.~Kane, K.~Sinha, and S.~Watson, {\em {Cosmological Moduli and the
  Post-Inflationary Universe: A Critical Review}\/},
  \href{http://dx.doi.org/10.1142/S0218271815300220}{Int. J. Mod. Phys. {\bf
  D24} (2015) no.~08, 1530022},
\href{http://arxiv.org/abs/1502.07746}{{\tt arXiv:1502.07746 [hep-th]}}.
%%CITATION = ARXIV:1502.07746;%%.

\bibitem{Altmannshofer:2013lfa}
W.~Altmannshofer, R.~Harnik, and J.~Zupan, {\em {Low Energy Probes of PeV Scale
  Sfermions}\/},  \href{http://dx.doi.org/10.1007/JHEP11(2013)202}{JHEP {\bf
  11} (2013)  202},
\href{http://arxiv.org/abs/1308.3653}{{\tt arXiv:1308.3653 [hep-ph]}}.
%%CITATION = ARXIV:1308.3653;%%.

\bibitem{Baumgart:2014jya}
M.~Baumgart, D.~Stolarski, and T.~Zorawski, {\em {Split supersymmetry radiates
  flavor}\/},  \href{http://dx.doi.org/10.1103/PhysRevD.90.055001}{Phys. Rev.
  {\bf D90} (2014) no.~5, 055001},
\href{http://arxiv.org/abs/1403.6118}{{\tt arXiv:1403.6118 [hep-ph]}}.
%%CITATION = ARXIV:1403.6118;%%.

\bibitem{Blumenhagen:2009gk}
R.~Blumenhagen, J.~P. Conlon, S.~Krippendorf, S.~Moster, and F.~Quevedo, {\em
  {SUSY Breaking in Local String/F-Theory Models}\/},
  \href{http://dx.doi.org/10.1088/1126-6708/2009/09/007}{JHEP {\bf 09} (2009)
  007},
\href{http://arxiv.org/abs/0906.3297}{{\tt arXiv:0906.3297 [hep-th]}}.
%%CITATION = ARXIV:0906.3297;%%.

\bibitem{Aparicio:2014wxa}
L.~Aparicio, M.~Cicoli, S.~Krippendorf, A.~Maharana, F.~Muia, and F.~Quevedo,
  {\em {Sequestered De Sitter String Scenarios: Soft-Terms}\/},
  \href{http://dx.doi.org/10.1007/JHEP11(2014)071}{JHEP {\bf 11} (2014)  071},
\href{http://arxiv.org/abs/1409.1931}{{\tt arXiv:1409.1931 [hep-th]}}.
%%CITATION = ARXIV:1409.1931;%%.

\bibitem{Reece:2015qbf}
M.~Reece and W.~Xue, {\em {SUSY's Ladder: Reframing Sequestering at Large
  Volume}\/},
\href{http://arxiv.org/abs/1512.04941}{{\tt arXiv:1512.04941 [hep-ph]}}.
%%CITATION = ARXIV:1512.04941;%%.

\bibitem{Cremmer:1983bf}
E.~Cremmer, S.~Ferrara, C.~Kounnas, and D.~V. Nanopoulos, {\em {Naturally
  Vanishing Cosmological Constant in N=1 Supergravity}\/},
\href{http://dx.doi.org/10.1016/0370-2693(83)90106-5}{Phys.Lett. {\bf B133}
  (1983)  61}.
%%CITATION = PHLTA,B133,61;%%.

\bibitem{Ellis:1983sf}
J.~R. Ellis, A.~Lahanas, D.~V. Nanopoulos, and K.~Tamvakis, {\em {No-Scale
  Supersymmetric Standard Model}\/},
\href{http://dx.doi.org/10.1016/0370-2693(84)91378-9}{Phys.Lett. {\bf B134}
  (1984)  429}.
%%CITATION = PHLTA,B134,429;%%.

\bibitem{Balasubramanian:2005zx}
V.~Balasubramanian, P.~Berglund, J.~P. Conlon, and F.~Quevedo, {\em
  {Systematics of Moduli Stabilisation in Calabi-Yau Flux
  Compactifications}\/},
  \href{http://dx.doi.org/10.1088/1126-6708/2005/03/007}{JHEP {\bf 03} (2005)
  007},
\href{http://arxiv.org/abs/hep-th/0502058}{{\tt arXiv:hep-th/0502058
  [hep-th]}}.
%%CITATION = HEP-TH/0502058;%%.

\bibitem{Conlon:2005ki}
J.~P. Conlon, F.~Quevedo, and K.~Suruliz, {\em {Large-Volume Flux
  Compactifications: Moduli Spectrum and D3/D7 Soft Supersymmetry Breaking}\/},
   \href{http://dx.doi.org/10.1088/1126-6708/2005/08/007}{JHEP {\bf 08} (2005)
  007},
\href{http://arxiv.org/abs/hep-th/0505076}{{\tt arXiv:hep-th/0505076
  [hep-th]}}.
%%CITATION = HEP-TH/0505076;%%.

\bibitem{Han:2013kza}
T.~Han, S.~Padhi, and S.~Su, {\em {Electroweakinos in the Light of the Higgs
  Boson}\/},  \href{http://dx.doi.org/10.1103/PhysRevD.88.115010}{Phys. Rev.
  {\bf D88} (2013) no.~11, 115010},
\href{http://arxiv.org/abs/1309.5966}{{\tt arXiv:1309.5966 [hep-ph]}}.
%%CITATION = ARXIV:1309.5966;%%.

\bibitem{ATLAS:2013hta}
{ATLAS Collaboration}, {\em {Physics at a High-Luminosity LHC with ATLAS}\/},
  in {\em {Community Summer Study 2013: Snowmass on the Mississippi (CSS2013)
  Minneapolis, MN, USA, July 29-August 6, 2013}}.
\newblock 2013.
\newblock \href{http://arxiv.org/abs/1307.7292}{{\tt arXiv:1307.7292
  [hep-ex]}}.
\newblock
\url{http://inspirehep.net/record/1245017/files/arXiv:1307.7292.pdf}.
\newblock
%%CITATION = ARXIV:1307.7292;%%.

\bibitem{Peskin:2013xra}
M.~E. Peskin, {\em {Estimation of LHC and ILC Capabilities for Precision Higgs
  Boson Coupling Measurements}\/},  in {\em {Community Summer Study 2013:
  Snowmass on the Mississippi (CSS2013) Minneapolis, MN, USA, July 29-August 6,
  2013}}.
\newblock 2013.
\newblock \href{http://arxiv.org/abs/1312.4974}{{\tt arXiv:1312.4974
  [hep-ph]}}.
\newblock
\url{http://www.slac.stanford.edu/econf/C1307292/docs/submittedArxivFiles/1312.4974.pdf}.
\newblock
%%CITATION = ARXIV:1312.4974;%%.

\bibitem{Sjostrand:2014zea}
T.~Sjà¥«strand, S.~Ask, J.~R. Christiansen, R.~Corke, N.~Desai, P.~Ilten,
  S.~Mrenna, S.~Prestel, C.~O. Rasmussen, and P.~Z. Skands, {\em {An
  Introduction to PYTHIA 8.2}\/},
  \href{http://dx.doi.org/10.1016/j.cpc.2015.01.024}{Comput. Phys. Commun. {\bf
  191} (2015)  159--177},
\href{http://arxiv.org/abs/1410.3012}{{\tt arXiv:1410.3012 [hep-ph]}}.
%%CITATION = ARXIV:1410.3012;%%.

\bibitem{Djouadi:2006bz}
A.~Djouadi, M.~M. Muhlleitner, and M.~Spira, {\em {Decays of supersymmetric
  particles: The Program SUSY-HIT (SUspect-SdecaY-Hdecay-InTerface)}\/},  Acta
  Phys. Polon. {\bf B38} (2007)  635--644,
\href{http://arxiv.org/abs/hep-ph/0609292}{{\tt arXiv:hep-ph/0609292
  [hep-ph]}}.
%%CITATION = HEP-PH/0609292;%%.

\bibitem{Cacciari:2005hq}
M.~Cacciari and G.~P. Salam, {\em {Dispelling the $N^{3}$ myth for the $k_t$
  jet-finder}\/},
  \href{http://dx.doi.org/10.1016/j.physletb.2006.08.037}{Phys. Lett. {\bf
  B641} (2006)  57--61},
\href{http://arxiv.org/abs/hep-ph/0512210}{{\tt arXiv:hep-ph/0512210
  [hep-ph]}}.
%%CITATION = HEP-PH/0512210;%%.

\bibitem{Cacciari:2011ma}
M.~Cacciari, G.~P. Salam, and G.~Soyez, {\em {FastJet User Manual}\/},
  \href{http://dx.doi.org/10.1140/epjc/s10052-012-1896-2}{Eur.Phys.J. {\bf C72}
  (2012)  1896},
\href{http://arxiv.org/abs/1111.6097}{{\tt arXiv:1111.6097 [hep-ph]}}.
%%CITATION = ARXIV:1111.6097;%%.

\bibitem{Thaler:2010tr}
J.~Thaler and K.~Van~Tilburg, {\em {Identifying Boosted Objects with
  N-subjettiness}\/},  \href{http://dx.doi.org/10.1007/JHEP03(2011)015}{JHEP
  {\bf 03} (2011)  015},
\href{http://arxiv.org/abs/1011.2268}{{\tt arXiv:1011.2268 [hep-ph]}}.
%%CITATION = ARXIV:1011.2268;%%.

\bibitem{Larkoski:2014uqa}
A.~J. Larkoski, D.~Neill, and J.~Thaler, {\em {Jet Shapes with the Broadening
  Axis}\/},  \href{http://dx.doi.org/10.1007/JHEP04(2014)017}{JHEP {\bf 04}
  (2014)  017},
\href{http://arxiv.org/abs/1401.2158}{{\tt arXiv:1401.2158 [hep-ph]}}.
%%CITATION = ARXIV:1401.2158;%%.

\bibitem{comesoon}
P.~Agrawal, J.~Fan, M.~Reece, and W.~Xue, {\em To appear soon in 2016\/}, .

\bibitem{ATL-PHYS-PUB-2013-011}
{ATLAS Collaboration}, {\em {Prospects for benchmark Supersymmetry searches at
  the high luminosity LHC with the ATLAS Detector}\/},   ATL-PHYS-PUB-2013-011,
  CERN, Geneva, Sep, 2013.
\newblock \url{https://cds.cern.ch/record/1604505}.

\bibitem{ATL-PHYS-PUB-2014-010}
{ATLAS Collaboration}, {\em {Search for Supersymmetry at the high luminosity
  LHC with the ATLAS experiment}\/},   ATL-PHYS-PUB-2014-010, CERN, Geneva,
  Jul, 2014.
\newblock \url{https://cds.cern.ch/record/1735031}.

\bibitem{CMS-PAS-SUS-14-012}
{CMS Collaboration}, {\em {Supersymmetry discovery potential in future LHC and
  HL-LHC running with the CMS detector}\/},   CMS-PAS-SUS-14-012, CERN, Geneva,
  2015.
\newblock \url{https://cds.cern.ch/record/1981344}.

\bibitem{CMS-PAS-FTR-13-014}
{CMS Collaboration}, {\em {Study of the Discovery Reach in Searches for
  Supersymmetry at CMS with 3000/fb}\/},   CMS-PAS-FTR-13-014, CERN, Geneva,
  2013.
\newblock \url{https://cds.cern.ch/record/1607141}.

\bibitem{ColliderReach}
G.~Salam and A.~Weiler, {\em Collider Reach Tool\/},
  \url{http://collider-reach.web.cern.ch}.

\bibitem{Ade:2015xua}
{Planck Collaboration}, P.~A.~R. Ade et al., {\em {Planck 2015 results. XIII.
  Cosmological parameters}\/},
\href{http://arxiv.org/abs/1502.01589}{{\tt arXiv:1502.01589 [astro-ph.CO]}}.
%%CITATION = ARXIV:1502.01589;%%.

\bibitem{Gelmini:2015zpa}
G.~B. Gelmini, {\em {TASI 2014 Lectures: The Hunt for Dark Matter}\/},  in {\em
  {Theoretical Advanced Study Institute in Elementary Particle Physics:
  Journeys Through the Precision Frontier: Amplitudes for Colliders (TASI 2014)
  Boulder, Colorado, June 2-27, 2014}}.
\newblock 2015.
\newblock \href{http://arxiv.org/abs/1502.01320}{{\tt arXiv:1502.01320
  [hep-ph]}}.
\newblock
\url{http://inspirehep.net/record/1342951/files/arXiv:1502.01320.pdf}.
\newblock
%%CITATION = ARXIV:1502.01320;%%.

\bibitem{Hu:2000ke}
W.~Hu, R.~Barkana, and A.~Gruzinov, {\em {Cold and fuzzy dark matter}\/},
  \href{http://dx.doi.org/10.1103/PhysRevLett.85.1158}{Phys. Rev. Lett. {\bf
  85} (2000)  1158--1161},
\href{http://arxiv.org/abs/astro-ph/0003365}{{\tt arXiv:astro-ph/0003365
  [astro-ph]}}.
%%CITATION = ASTRO-PH/0003365;%%.

\bibitem{Griest:2013aaa}
K.~Griest, A.~M. Cieplak, and M.~J. Lehner, {\em {Experimental Limits on
  Primordial Black Hole Dark Matter from the First 2 yr of Kepler Data}\/},
  \href{http://dx.doi.org/10.1088/0004-637X/786/2/158}{Astrophys. J. {\bf 786}
  (2014) no.~2, 158},
\href{http://arxiv.org/abs/1307.5798}{{\tt arXiv:1307.5798 [astro-ph.CO]}}.
%%CITATION = ARXIV:1307.5798;%%.

\bibitem{Alcock:1998fx}
{MACHO, EROS Collaboration}, C.~Alcock et al., {\em {EROS and MACHO combined
  limits on planetary mass dark matter in the galactic halo}\/},
  \href{http://dx.doi.org/10.1086/311355}{Astrophys. J. {\bf 499} (1998)  L9},
\href{http://arxiv.org/abs/astro-ph/9803082}{{\tt arXiv:astro-ph/9803082
  [astro-ph]}}.
%%CITATION = ASTRO-PH/9803082;%%.

\bibitem{Yoo:2003fr}
J.~Yoo, J.~Chaname, and A.~Gould, {\em {The end of the MACHO era: limits on
  halo dark matter from stellar halo wide binaries}\/},
  \href{http://dx.doi.org/10.1086/380562}{Astrophys. J. {\bf 601} (2004)
  311--318},
\href{http://arxiv.org/abs/astro-ph/0307437}{{\tt arXiv:astro-ph/0307437
  [astro-ph]}}.
%%CITATION = ASTRO-PH/0307437;%%.

\bibitem{Kolb:1990vq}
E.~W. Kolb and M.~S. Turner, {\em {The Early Universe}\/},
Front.Phys. {\bf 69} (1990)  1--547.
%%CITATION = FRPHA,69,1;%%.

\bibitem{Griest:1989wd}
K.~Griest and M.~Kamionkowski, {\em {Unitarity Limits on the Mass and Radius of
  Dark Matter Particles}\/},
\href{http://dx.doi.org/10.1103/PhysRevLett.64.615}{Phys. Rev. Lett. {\bf 64}
  (1990)  615}.
%%CITATION = PRLTA,64,615;%%.

\bibitem{Blum:2014dca}
K.~Blum, Y.~Cui, and M.~Kamionkowski, {\em {An Ultimate Target for Dark Matter
  Searches}\/},  \href{http://dx.doi.org/10.1103/PhysRevD.92.023528}{Phys. Rev.
  {\bf D92} (2015) no.~2, 023528},
\href{http://arxiv.org/abs/1412.3463}{{\tt arXiv:1412.3463 [hep-ph]}}.
%%CITATION = ARXIV:1412.3463;%%.

\bibitem{Petraki:2013wwa}
K.~Petraki and R.~R. Volkas, {\em {Review of asymmetric dark matter}\/},
  \href{http://dx.doi.org/10.1142/S0217751X13300287}{Int. J. Mod. Phys. {\bf
  A28} (2013)  1330028},
\href{http://arxiv.org/abs/1305.4939}{{\tt arXiv:1305.4939 [hep-ph]}}.
%%CITATION = ARXIV:1305.4939;%%.

\bibitem{Zurek:2013wia}
K.~M. Zurek, {\em {Asymmetric Dark Matter: Theories, Signatures, and
  Constraints}\/},
  \href{http://dx.doi.org/10.1016/j.physrep.2013.12.001}{Phys. Rept. {\bf 537}
  (2014)  91--121},
\href{http://arxiv.org/abs/1308.0338}{{\tt arXiv:1308.0338 [hep-ph]}}.
%%CITATION = ARXIV:1308.0338;%%.

\bibitem{Cushman:2013zza}
P.~Cushman et al., {\em {Working Group Report: WIMP Dark Matter Direct
  Detection}\/},  in {\em {Community Summer Study 2013: Snowmass on the
  Mississippi (CSS2013) Minneapolis, MN, USA, July 29-August 6, 2013}}.
\newblock 2013.
\newblock \href{http://arxiv.org/abs/1310.8327}{{\tt arXiv:1310.8327
  [hep-ex]}}.
\newblock
\url{http://inspirehep.net/record/1262767/files/arXiv:1310.8327.pdf}.
\newblock
%%CITATION = ARXIV:1310.8327;%%.

\bibitem{Grothaus:2014hja}
P.~Grothaus, M.~Fairbairn, and J.~Monroe, {\em {Directional Dark Matter
  Detection Beyond the Neutrino Bound}\/},
  \href{http://dx.doi.org/10.1103/PhysRevD.90.055018}{Phys. Rev. {\bf D90}
  (2014) no.~5, 055018},
\href{http://arxiv.org/abs/1406.5047}{{\tt arXiv:1406.5047 [hep-ph]}}.
%%CITATION = ARXIV:1406.5047;%%.

\bibitem{Schumann:2015cpa}
{Schumann, Marc and Baudis, Laura and B{\"{u}}tikofer, Lukas and Kish,
  Alexander and Selvi, Marco}, {\em {Dark matter sensitivity of multi-ton
  liquid xenon detectors}\/},
  \href{http://dx.doi.org/10.1088/1475-7516/2015/10/016}{JCAP {\bf 1510} (2015)
  no.~10, 016},
\href{http://arxiv.org/abs/1506.08309}{{\tt arXiv:1506.08309
  [physics.ins-det]}}.
%%CITATION = ARXIV:1506.08309;%%.

\bibitem{Akerib:2015cja}
{LZ Collaboration}, D.~S. Akerib et al., {\em {LUX-ZEPLIN (LZ) Conceptual
  Design Report}\/},
\href{http://arxiv.org/abs/1509.02910}{{\tt arXiv:1509.02910
  [physics.ins-det]}}.
%%CITATION = ARXIV:1509.02910;%%.

\bibitem{Aalseth:2015mba}
C.~E. Aalseth et al., {\em {The DarkSide Multiton Detector for the Direct Dark
  Matter Search}\/},
\href{http://dx.doi.org/10.1155/2015/541362}{Adv. High Energy Phys. {\bf 2015}
  (2015)  541362}.
%%CITATION = 00642,2015,541362;%%.

\bibitem{Aprile:2013doa}
{XENON100 Collaboration}, E.~Aprile et al., {\em {Limits on spin-dependent
  WIMP-nucleon cross sections from 225 live days of XENON100 data}\/},
  \href{http://dx.doi.org/10.1103/PhysRevLett.111.021301}{Phys. Rev. Lett. {\bf
  111} (2013) no.~2, 021301},
\href{http://arxiv.org/abs/1301.6620}{{\tt arXiv:1301.6620 [astro-ph.CO]}}.
%%CITATION = ARXIV:1301.6620;%%.

\bibitem{Akerib:2016lao}
{LUX Collaboration}, D.~S. Akerib et al., {\em {First spin-dependent
  WIMP-nucleon cross section limits from the LUX experiment}\/},
\href{http://arxiv.org/abs/1602.03489}{{\tt arXiv:1602.03489 [hep-ex]}}.
%%CITATION = ARXIV:1602.03489;%%.

\bibitem{ChangJin:550}
C.~Jin, {\em Dark Matter Particle Explorer: The First Chinese Cosmic Ray and
  Hard gamma-ray Detector in Space\/},  Chinese journal of space science (2014)
   .

\bibitem{Dokuchaev:2015ghx}
V.~I. Dokuchaev and {\relax Yu}.~N. Eroshenko, {\em {Physical laboratory at the
  center of the Galaxy}\/},
  \href{http://dx.doi.org/10.3367/UFNe.0185.201508c.0829}{Phys. Usp. {\bf 58}
  (2015)  772--784},
\href{http://arxiv.org/abs/1512.02943}{{\tt arXiv:1512.02943 [astro-ph.HE]}}.
%%CITATION = ARXIV:1512.02943;%%.

\bibitem{cta}
{CTA Consortium Collaboration}, {Actis, M. and Agnetta, G. and Aharonian, F.
  and Akhperjanian, A. and Aleksi\'{c}, J. and Aliu, E. and Allan, D. and
  Allekotte, I. and Antico, F. and Antonelli, L.~A. and others}, {\em {Design
  concepts for the Cherenkov Telescope Array CTA: An advanced facility for
  ground-based high-energy gamma-ray astronomy}\/},
  \href{http://dx.doi.org/10.1007/s10686-011-9247-0}{Exper. Astron. {\bf 32}
  (2011)  193--316},
\href{http://arxiv.org/abs/1008.3703}{{\tt arXiv:1008.3703 [astro-ph.IM]}}.
%%CITATION = ARXIV:1008.3703;%%.

\bibitem{Abramowski:2013ax}
{H.E.S.S. Collaboration}, A.~Abramowski et al., {\em {Search for
  Photon-Linelike Signatures from Dark Matter Annihilations with H.E.S.S.}\/},
  \href{http://dx.doi.org/10.1103/PhysRevLett.110.041301}{Phys.Rev.Lett. {\bf
  110} (2013)  041301},
\href{http://arxiv.org/abs/1301.1173}{{\tt arXiv:1301.1173 [astro-ph.HE]}}.
%%CITATION = ARXIV:1301.1173;%%.

\bibitem{Ackermann:2015lka}
{Fermi-LAT Collaboration}, M.~Ackermann et al., {\em {Updated search for
  spectral lines from Galactic dark matter interactions with pass 8 data from
  the Fermi Large Area Telescope}\/},
  \href{http://dx.doi.org/10.1103/PhysRevD.91.122002}{Phys. Rev. {\bf D91}
  (2015) no.~12, 122002},
\href{http://arxiv.org/abs/1506.00013}{{\tt arXiv:1506.00013 [astro-ph.HE]}}.
%%CITATION = ARXIV:1506.00013;%%.

\bibitem{Giesen:2015ufa}
G.~Giesen, M.~Boudaud, Y.~GÃ©nolini, V.~Poulin, M.~Cirelli, P.~Salati, and
  P.~D. Serpico, {\em {AMS-02 antiprotons, at last! Secondary astrophysical
  component and immediate implications for Dark Matter}\/},
  \href{http://dx.doi.org/10.1088/1475-7516/2015/09/023,
  10.1088/1475-7516/2015/9/023}{JCAP {\bf 1509} (2015) no.~09, 023},
\href{http://arxiv.org/abs/1504.04276}{{\tt arXiv:1504.04276 [astro-ph.HE]}}.
%%CITATION = ARXIV:1504.04276;%%.

\bibitem{AMSTalk}
A.~Kounine, {\em {Talk at AMS Days at CERN}\/}, .

\bibitem{Drlica-Wagner:2015xua}
{DES, Fermi-LAT Collaboration}, A.~Drlica-Wagner et al., {\em {Search for
  Gamma-Ray Emission from DES Dwarf Spheroidal Galaxy Candidates with Fermi-LAT
  Data}\/},  \href{http://dx.doi.org/10.1088/2041-8205/809/1/L4}{Astrophys. J.
  {\bf 809} (2015) no.~1, L4},
\href{http://arxiv.org/abs/1503.02632}{{\tt arXiv:1503.02632 [astro-ph.HE]}}.
%%CITATION = ARXIV:1503.02632;%%.

\bibitem{HESS:2015cda}
{HESS Collaboration}, A.~Abramowski et al., {\em {Constraints on an
  Annihilation Signal from a Core of Constant Dark Matter Density around the
  Milky Way Center with H.E.S.S.}\/},
  \href{http://dx.doi.org/10.1103/PhysRevLett.114.081301}{Phys. Rev. Lett. {\bf
  114} (2015) no.~8, 081301},
\href{http://arxiv.org/abs/1502.03244}{{\tt arXiv:1502.03244 [astro-ph.HE]}}.
%%CITATION = ARXIV:1502.03244;%%.

\bibitem{D'Ambrosio:2002ex}
G.~D'Ambrosio, G.~F. Giudice, G.~Isidori, and A.~Strumia, {\em {Minimal flavor
  violation: An Effective field theory approach}\/},
  \href{http://dx.doi.org/10.1016/S0550-3213(02)00836-2}{Nucl. Phys. {\bf B645}
  (2002)  155--187},
\href{http://arxiv.org/abs/hep-ph/0207036}{{\tt arXiv:hep-ph/0207036
  [hep-ph]}}.
%%CITATION = HEP-PH/0207036;%%.

\bibitem{Abdallah:2015ter}
J.~Abdallah et al., {\em {Simplified Models for Dark Matter Searches at the
  LHC}\/},  \href{http://dx.doi.org/10.1016/j.dark.2015.08.001}{Phys. Dark
  Univ. {\bf 9-10} (2015)  8--23},
\href{http://arxiv.org/abs/1506.03116}{{\tt arXiv:1506.03116 [hep-ph]}}.
%%CITATION = ARXIV:1506.03116;%%.

\bibitem{Backovic:2013dpa}
M.~Backovic, K.~Kong, and M.~McCaskey, {\em {MadDM v.1.0: Computation of Dark
  Matter Relic Abundance Using MadGraph5}\/},
  \href{http://dx.doi.org/10.1016/j.dark.2014.04.001}{Physics of the Dark
  Universe {\bf 5-6} (2014)  18--28},
\href{http://arxiv.org/abs/1308.4955}{{\tt arXiv:1308.4955 [hep-ph]}}.
%%CITATION = ARXIV:1308.4955;%%.

\bibitem{Chala:2015ama}
M.~Chala, F.~Kahlhoefer, M.~McCullough, G.~Nardini, and K.~Schmidt-Hoberg, {\em
  {Constraining Dark Sectors with Monojets and Dijets}\/},
  \href{http://dx.doi.org/10.1007/JHEP07(2015)089}{JHEP {\bf 07} (2015)  089},
\href{http://arxiv.org/abs/1503.05916}{{\tt arXiv:1503.05916 [hep-ph]}}.
%%CITATION = ARXIV:1503.05916;%%.

\bibitem{CERN-LHCC-2015-020}
{ATLAS Collaboration}, {\em {ATLAS Phase-II Upgrade Scoping Document}\/},
  CERN-LHCC-2015-020. LHCC-G-166, CERN, Geneva, Sep, 2015.
\newblock \url{https://cds.cern.ch/record/2055248}.

\bibitem{Butler:2020886}
J.~Butler, D.~Contardo, M.~Klute, J.~Mans, and L.~Silvestris, {\em {Technical
  Proposal for the Phase-II Upgrade of the CMS Detector}\/},
  CERN-LHCC-2015-010. LHCC-P-008, CERN, Geneva, Jun, 2015.
\newblock \url{http://cds.cern.ch/record/2020886}.

\bibitem{TheFermi-LAT:2015kwa}
{Fermi-LAT Collaboration}, M.~Ajello et al., {\em {Fermi-LAT Observations of
  High-Energy Gamma-Ray Emission Toward the Galactic Center}\/},
  \href{http://dx.doi.org/10.3847/0004-637X/819/1/44}{Astrophys. J. {\bf 819}
  (2016) no.~1, 44},
\href{http://arxiv.org/abs/1511.02938}{{\tt arXiv:1511.02938 [astro-ph.HE]}}.
%%CITATION = ARXIV:1511.02938;%%.

\bibitem{Hooper:2010mq}
D.~Hooper and L.~Goodenough, {\em {Dark Matter Annihilation in The Galactic
  Center As Seen by the Fermi Gamma Ray Space Telescope}\/},
  \href{http://dx.doi.org/10.1016/j.physletb.2011.02.029}{Phys. Lett. {\bf
  B697} (2011)  412--428},
\href{http://arxiv.org/abs/1010.2752}{{\tt arXiv:1010.2752 [hep-ph]}}.
%%CITATION = ARXIV:1010.2752;%%.

\bibitem{Gordon:2013vta}
C.~Gordon and O.~Macias, {\em {Dark Matter and Pulsar Model Constraints from
  Galactic Center Fermi-LAT Gamma Ray Observations}\/},
  \href{http://dx.doi.org/10.1103/PhysRevD.88.083521,
  10.1103/PhysRevD.89.049901}{Phys. Rev. {\bf D88} (2013) no.~8, 083521},
  \href{http://arxiv.org/abs/1306.5725}{{\tt arXiv:1306.5725 [astro-ph.HE]}}.
[Erratum: Phys. Rev.D89,no.4,049901(2014)].
%%CITATION = ARXIV:1306.5725;%%.

\bibitem{Daylan:2014rsa}
T.~Daylan, D.~P. Finkbeiner, D.~Hooper, T.~Linden, S.~K.~N. Portillo, N.~L.
  Rodd, and T.~R. Slatyer, {\em {The characterization of the gamma-ray signal
  from the central Milky Way: A case for annihilating dark matter}\/},
  \href{http://dx.doi.org/10.1016/j.dark.2015.12.005}{Phys. Dark Univ. {\bf 12}
  (2016)  1--23},
\href{http://arxiv.org/abs/1402.6703}{{\tt arXiv:1402.6703 [astro-ph.HE]}}.
%%CITATION = ARXIV:1402.6703;%%.

\bibitem{Calore:2014xka}
F.~Calore, I.~Cholis, and C.~Weniger, {\em {Background model systematics for
  the Fermi GeV excess}\/},
  \href{http://dx.doi.org/10.1088/1475-7516/2015/03/038}{JCAP {\bf 1503} (2015)
   038},
\href{http://arxiv.org/abs/1409.0042}{{\tt arXiv:1409.0042 [astro-ph.CO]}}.
%%CITATION = ARXIV:1409.0042;%%.

\bibitem{Agrawal:2014oha}
P.~Agrawal, B.~Batell, P.~J. Fox, and R.~Harnik, {\em {WIMPs at the Galactic
  Center}\/},  \href{http://dx.doi.org/10.1088/1475-7516/2015/05/011}{JCAP {\bf
  1505} (2015)  011},
\href{http://arxiv.org/abs/1411.2592}{{\tt arXiv:1411.2592 [hep-ph]}}.
%%CITATION = ARXIV:1411.2592;%%.

\bibitem{CMS-EXO-12-055}
{CMS Collaboration}, {\em {Search for New Physics in the V/jet + MET final
  state}\/},   CMS-PAS-EXO-12-055, CERN, Geneva, 2015.
\newblock \url{https://cds.cern.ch/record/2036044}.

\bibitem{ATLAS-CONF-2015-081}
{ATLAS Collaboration}, {\em {Search for resonances decaying to photon pairs in
  3.2 fb$^{-1}$ of $pp$ collisions at $\sqrt{s}$ = 13 TeV with the ATLAS
  detector}\/},   ATLAS-CONF-2015-081, CERN, Geneva, Dec, 2015.
\newblock \url{https://cds.cern.ch/record/2114853}.

\bibitem{CMS-PAS-EXO-15-004}
{CMS Collaboration}, {\em {Search for new physics in high mass diphoton events
  in proton-proton collisions at 13TeV}\/},   CMS-PAS-EXO-15-004, CERN, Geneva,
  2015.
\newblock \url{https://cds.cern.ch/record/2114808}.

\bibitem{Backovic:2015fnp}
M.~Backovic, A.~Mariotti, and D.~Redigolo, {\em {Di-photon excess illuminates
  Dark Matter}\/},
\href{http://arxiv.org/abs/1512.04917}{{\tt arXiv:1512.04917 [hep-ph]}}.
%%CITATION = ARXIV:1512.04917;%%.

\bibitem{Knapen:2015dap}
S.~Knapen, T.~Melia, M.~Papucci, and K.~Zurek, {\em {Rays of light from the
  LHC}\/},
\href{http://arxiv.org/abs/1512.04928}{{\tt arXiv:1512.04928 [hep-ph]}}.
%%CITATION = ARXIV:1512.04928;%%.

\bibitem{Han:2015cty}
C.~Han, H.~M. Lee, M.~Park, and V.~Sanz, {\em {The diphoton resonance as a
  gravity mediator of dark matter}\/},
\href{http://arxiv.org/abs/1512.06376}{{\tt arXiv:1512.06376 [hep-ph]}}.
%%CITATION = ARXIV:1512.06376;%%.

\bibitem{Bi:2015uqd}
X.-J. Bi, Q.-F. Xiang, P.-F. Yin, and Z.-H. Yu, {\em {The 750 GeV diphoton
  excess at the LHC and dark matter constraints}\/},
\href{http://arxiv.org/abs/1512.06787}{{\tt arXiv:1512.06787 [hep-ph]}}.
%%CITATION = ARXIV:1512.06787;%%.

\bibitem{Bhattacharya:2016lyg}
S.~Bhattacharya, S.~Patra, N.~Sahoo, and N.~Sahu, {\em {750 GeV Di-photon
  excess at CERN LHC from a dark sector assisted scalar decay}\/},
\href{http://arxiv.org/abs/1601.01569}{{\tt arXiv:1601.01569 [hep-ph]}}.
%%CITATION = ARXIV:1601.01569;%%.

\bibitem{D'Eramo:2016mgv}
F.~D'Eramo, J.~de~Vries, and P.~Panci, {\em {A 750 GeV Portal: LHC
  Phenomenology and Dark Matter Candidates}\/},
\href{http://arxiv.org/abs/1601.01571}{{\tt arXiv:1601.01571 [hep-ph]}}.
%%CITATION = ARXIV:1601.01571;%%.

\bibitem{Mambrini:2015wyu}
Y.~Mambrini, G.~Arcadi, and A.~Djouadi, {\em {The LHC diphoton resonance and
  dark matter}\/},
\href{http://arxiv.org/abs/1512.04913}{{\tt arXiv:1512.04913 [hep-ph]}}.
%%CITATION = ARXIV:1512.04913;%%.

\bibitem{Franceschini:2015kwy}
R.~Franceschini, G.~F. Giudice, J.~F. Kamenik, M.~McCullough, A.~Pomarol,
  R.~Rattazzi, M.~Redi, F.~Riva, A.~Strumia, and R.~Torre, {\em {What is the
  $\gamma \gamma$ resonance at 750 GeV?}\/},
  \href{http://dx.doi.org/10.1007/JHEP03(2016)144}{JHEP {\bf 03} (2016)  144},
\href{http://arxiv.org/abs/1512.04933}{{\tt arXiv:1512.04933 [hep-ph]}}.
%%CITATION = ARXIV:1512.04933;%%.

\bibitem{Adriani:2008zr}
{PAMELA Collaboration}, O.~Adriani et al., {\em {An anomalous positron
  abundance in cosmic rays with energies 1.5-100 GeV}\/},
  \href{http://dx.doi.org/10.1038/nature07942}{Nature {\bf 458} (2009)
  607--609},
\href{http://arxiv.org/abs/0810.4995}{{\tt arXiv:0810.4995 [astro-ph]}}.
%%CITATION = ARXIV:0810.4995;%%.

\bibitem{Accardo:2014lma}
{AMS Collaboration}, L.~Accardo et al., {\em {High Statistics Measurement of
  the Positron Fraction in Primary Cosmic Rays of 0.5--500 GeV with the Alpha
  Magnetic Spectrometer on the International Space Station}\/},
\href{http://dx.doi.org/10.1103/PhysRevLett.113.121101}{Phys. Rev. Lett. {\bf
  113} (2014)  121101}.
%%CITATION = PRLTA,113,121101;%%.

\bibitem{Bernabei:2010mq}
{DAMA, LIBRA Collaboration}, R.~Bernabei et al., {\em {New results from
  DAMA/LIBRA}\/},  \href{http://dx.doi.org/10.1140/epjc/s10052-010-1303-9}{Eur.
  Phys. J. {\bf C67} (2010)  39--49},
\href{http://arxiv.org/abs/1002.1028}{{\tt arXiv:1002.1028 [astro-ph.GA]}}.
%%CITATION = ARXIV:1002.1028;%%.

\bibitem{Cirelli:2005uq}
M.~Cirelli, N.~Fornengo, and A.~Strumia, {\em {Minimal dark matter}\/},
  \href{http://dx.doi.org/10.1016/j.nuclphysb.2006.07.012}{Nucl.Phys. {\bf
  B753} (2006)  178--194},
\href{http://arxiv.org/abs/hep-ph/0512090}{{\tt arXiv:hep-ph/0512090
  [hep-ph]}}.
%%CITATION = HEP-PH/0512090;%%.

\bibitem{Mahbubani:2005pt}
R.~Mahbubani and L.~Senatore, {\em {The Minimal model for dark matter and
  unification}\/},  \href{http://dx.doi.org/10.1103/PhysRevD.73.043510}{Phys.
  Rev. {\bf D73} (2006)  043510},
\href{http://arxiv.org/abs/hep-ph/0510064}{{\tt arXiv:hep-ph/0510064
  [hep-ph]}}.
%%CITATION = HEP-PH/0510064;%%.

\bibitem{Cohen:2011ec}
T.~Cohen, J.~Kearney, A.~Pierce, and D.~Tucker-Smith, {\em {Singlet-Doublet
  Dark Matter}\/},  \href{http://dx.doi.org/10.1103/PhysRevD.85.075003}{Phys.
  Rev. {\bf D85} (2012)  075003},
\href{http://arxiv.org/abs/1109.2604}{{\tt arXiv:1109.2604 [hep-ph]}}.
%%CITATION = ARXIV:1109.2604;%%.

\bibitem{Joglekar:2012vc}
A.~Joglekar, P.~Schwaller, and C.~E.~M. Wagner, {\em {Dark Matter and Enhanced
  Higgs to Di-photon Rate from Vector-like Leptons}\/},
  \href{http://dx.doi.org/10.1007/JHEP12(2012)064}{JHEP {\bf 12} (2012)  064},
\href{http://arxiv.org/abs/1207.4235}{{\tt arXiv:1207.4235 [hep-ph]}}.
%%CITATION = ARXIV:1207.4235;%%.

\bibitem{Cheung:2013dua}
C.~Cheung and D.~Sanford, {\em {Simplified Models of Mixed Dark Matter}\/},
  \href{http://dx.doi.org/10.1088/1475-7516/2014/02/011}{JCAP {\bf 1402} (2014)
   011},
\href{http://arxiv.org/abs/1311.5896}{{\tt arXiv:1311.5896 [hep-ph]}}.
%%CITATION = ARXIV:1311.5896;%%.

\bibitem{Calibbi:2015nha}
L.~Calibbi, A.~Mariotti, and P.~Tziveloglou, {\em {Singlet-Doublet Model: Dark
  matter searches and LHC constraints}\/},
  \href{http://dx.doi.org/10.1007/JHEP10(2015)116}{JHEP {\bf 10} (2015)  116},
\href{http://arxiv.org/abs/1505.03867}{{\tt arXiv:1505.03867 [hep-ph]}}.
%%CITATION = ARXIV:1505.03867;%%.

\bibitem{Cynolter:2015sua}
G.~Cynolter, J.~Kovács, and E.~Lendvai, {\em {Doublet-singlet model and
  unitarity}\/},  \href{http://dx.doi.org/10.1142/S0217732316500139}{Mod. Phys.
  Lett. {\bf A31} (2015) no.~01, 1650013},
\href{http://arxiv.org/abs/1509.05323}{{\tt arXiv:1509.05323 [hep-ph]}}.
%%CITATION = ARXIV:1509.05323;%%.

\bibitem{Tait:2016qbg}
T.~M.~P. Tait and Z.-H. Yu, {\em {Triplet-Quadruplet Dark Matter}\/},
\href{http://arxiv.org/abs/1601.01354}{{\tt arXiv:1601.01354 [hep-ph]}}.
%%CITATION = ARXIV:1601.01354;%%.

\bibitem{Thomas:1998wy}
S.~D. Thomas and J.~D. Wells, {\em {Phenomenology of Massive Vectorlike Doublet
  Leptons}\/},
  \href{http://dx.doi.org/10.1103/PhysRevLett.81.34}{Phys.Rev.Lett. {\bf 81}
  (1998)  34--37},
\href{http://arxiv.org/abs/hep-ph/9804359}{{\tt arXiv:hep-ph/9804359
  [hep-ph]}}.
%%CITATION = HEP-PH/9804359;%%.

\bibitem{Schwaller:2013baa}
P.~Schwaller and J.~Zurita, {\em {Compressed electroweakino spectra at the
  LHC}\/},  \href{http://dx.doi.org/10.1007/JHEP03(2014)060}{JHEP {\bf 03}
  (2014)  060},
\href{http://arxiv.org/abs/1312.7350}{{\tt arXiv:1312.7350 [hep-ph]}}.
%%CITATION = ARXIV:1312.7350;%%.

\bibitem{Ibe:2012sx}
M.~Ibe, S.~Matsumoto, and R.~Sato, {\em {Mass Splitting between Charged and
  Neutral Winos at Two-Loop Level}\/},
  \href{http://dx.doi.org/10.1016/j.physletb.2013.03.015}{Phys.Lett. {\bf B721}
  (2013)  252--260},
\href{http://arxiv.org/abs/1212.5989}{{\tt arXiv:1212.5989 [hep-ph]}}.
%%CITATION = ARXIV:1212.5989;%%.

\bibitem{Aad:2013yna}
{ATLAS Collaboration}, G.~Aad et al., {\em {Search for charginos nearly mass
  degenerate with the lightest neutralino based on a disappearing-track
  signature in pp collisions at $\sqrt{s}$=8 TeV with the ATLAS detector}\/},
  \href{http://dx.doi.org/10.1103/PhysRevD.88.112006}{Phys.Rev. {\bf D88}
  (2013) no.~11, 112006},
\href{http://arxiv.org/abs/1310.3675}{{\tt arXiv:1310.3675 [hep-ex]}}.
%%CITATION = ARXIV:1310.3675;%%.

\bibitem{CMS:2014gxa}
{CMS Collaboration}, V.~Khachatryan et al., {\em {Search for disappearing
  tracks in proton-proton collisions at $ \sqrt{s}=8 $ TeV}\/},
  \href{http://dx.doi.org/10.1007/JHEP01(2015)096}{JHEP {\bf 01} (2015)  096},
\href{http://arxiv.org/abs/1411.6006}{{\tt arXiv:1411.6006 [hep-ex]}}.
%%CITATION = ARXIV:1411.6006;%%.

\bibitem{Cirelli:2014dsa}
M.~Cirelli, F.~Sala, and M.~Taoso, {\em {Wino-like Minimal Dark Matter and
  future colliders}\/},  \href{http://dx.doi.org/10.1007/JHEP01(2015)041,
  10.1007/JHEP10(2014)033}{JHEP {\bf 1410} (2014)  033},
\href{http://arxiv.org/abs/1407.7058}{{\tt arXiv:1407.7058 [hep-ph]}}.
%%CITATION = ARXIV:1407.7058;%%.

\bibitem{Kotwal:2015rba}
A.~V. Kotwal, S.~Chekanov, and M.~Low, {\em {Double Higgs Boson Production in
  the 4$\tau$ Channel from Resonances in Longitudinal Vector Boson Scattering
  at a 100 TeV Collider}\/},
  \href{http://dx.doi.org/10.1103/PhysRevD.91.114018}{Phys. Rev. {\bf D91}
  (2015)  114018},
\href{http://arxiv.org/abs/1504.08042}{{\tt arXiv:1504.08042 [hep-ph]}}.
%%CITATION = ARXIV:1504.08042;%%.

\bibitem{Hill:2013hoa}
R.~J. Hill and M.~P. Solon, {\em {WIMP-nucleon scattering with heavy WIMP
  effective theory}\/},
  \href{http://dx.doi.org/10.1103/PhysRevLett.112.211602}{Phys.Rev.Lett. {\bf
  112} (2014)  211602},
\href{http://arxiv.org/abs/1309.4092}{{\tt arXiv:1309.4092 [hep-ph]}}.
%%CITATION = ARXIV:1309.4092;%%.

\bibitem{Cirelli:2009uv}
M.~Cirelli and A.~Strumia, {\em {Minimal Dark Matter: Model and results}\/},
  \href{http://dx.doi.org/10.1088/1367-2630/11/10/105005}{New J.Phys. {\bf 11}
  (2009)  105005},
\href{http://arxiv.org/abs/0903.3381}{{\tt arXiv:0903.3381 [hep-ph]}}.
%%CITATION = ARXIV:0903.3381;%%.

\bibitem{Hall:2011jd}
L.~J. Hall and Y.~Nomura, {\em {Spread Supersymmetry}\/},
  \href{http://dx.doi.org/10.1007/JHEP01(2012)082}{JHEP {\bf 1201} (2012)
  082},
\href{http://arxiv.org/abs/1111.4519}{{\tt arXiv:1111.4519 [hep-ph]}}.
%%CITATION = ARXIV:1111.4519;%%.

\bibitem{Hall:2012zp}
L.~J. Hall, Y.~Nomura, and S.~Shirai, {\em {Spread Supersymmetry with Wino LSP:
  Gluino and Dark Matter Signals}\/},
  \href{http://dx.doi.org/10.1007/JHEP01(2013)036}{JHEP {\bf 1301} (2013)
  036},
\href{http://arxiv.org/abs/1210.2395}{{\tt arXiv:1210.2395 [hep-ph]}}.
%%CITATION = ARXIV:1210.2395;%%.

\bibitem{Hall:2013eko}
L.~J. Hall and Y.~Nomura, {\em {Grand Unification and Intermediate Scale
  Supersymmetry}\/},  \href{http://dx.doi.org/10.1007/JHEP02(2014)129}{JHEP
  {\bf 1402} (2014)  129},
\href{http://arxiv.org/abs/1312.6695}{{\tt arXiv:1312.6695 [hep-ph]}}.
%%CITATION = ARXIV:1312.6695;%%.

\bibitem{Hall:2014vga}
L.~J. Hall, Y.~Nomura, and S.~Shirai, {\em {Grand Unification, Axion, and
  Inflation in Intermediate Scale Supersymmetry}\/},
  \href{http://dx.doi.org/10.1007/JHEP06(2014)137}{JHEP {\bf 06} (2014)  137},
\href{http://arxiv.org/abs/1403.8138}{{\tt arXiv:1403.8138 [hep-ph]}}.
%%CITATION = ARXIV:1403.8138;%%.

\bibitem{Chao:2012mx}
W.~Chao, M.~Gonderinger, and M.~J. Ramsey-Musolf, {\em {Higgs Vacuum Stability,
  Neutrino Mass, and Dark Matter}\/},
  \href{http://dx.doi.org/10.1103/PhysRevD.86.113017}{Phys.Rev. {\bf D86}
  (2012)  113017},
\href{http://arxiv.org/abs/1210.0491}{{\tt arXiv:1210.0491 [hep-ph]}}.
%%CITATION = ARXIV:1210.0491;%%.

\bibitem{Frigerio:2009wf}
M.~Frigerio and T.~Hambye, {\em {Dark matter stability and unification without
  supersymmetry}\/},
  \href{http://dx.doi.org/10.1103/PhysRevD.81.075002}{Phys.Rev. {\bf D81}
  (2010)  075002},
\href{http://arxiv.org/abs/0912.1545}{{\tt arXiv:0912.1545 [hep-ph]}}.
%%CITATION = ARXIV:0912.1545;%%.

\bibitem{Farina:2013mla}
M.~Farina, D.~Pappadopulo, and A.~Strumia, {\em {A modified naturalness
  principle and its experimental tests}\/},
  \href{http://dx.doi.org/10.1007/JHEP08(2013)022}{JHEP {\bf 1308} (2013)
  022},
\href{http://arxiv.org/abs/1303.7244v3}{{\tt arXiv:1303.7244v3 [hep-ph]}}.
%%CITATION = ARXIV:1303.7244;%%.

\bibitem{Hisano:2015rsa}
J.~Hisano, K.~Ishiwata, and N.~Nagata, {\em {QCD Effects on Direct Detection of
  Wino Dark Matter}\/},  \href{http://dx.doi.org/10.1007/JHEP06(2015)097}{JHEP
  {\bf 06} (2015)  097},
\href{http://arxiv.org/abs/1504.00915}{{\tt arXiv:1504.00915 [hep-ph]}}.
%%CITATION = ARXIV:1504.00915;%%.

\bibitem{Cirelli:2016boh}
{Cirelli, Marco and Panci, Paolo and Sala, Filippo and Taoso, Marco}, {\em work
  in progress\/}, .

\bibitem{Hryczuk:2014hpa}
A.~Hryczuk, I.~Cholis, R.~Iengo, M.~Tavakoli, and P.~Ullio, {\em {Indirect
  Detection Analysis: Wino Dark Matter Case Study}\/},
  \href{http://dx.doi.org/10.1088/1475-7516/2014/07/031}{JCAP {\bf 1407} (2014)
   031},
\href{http://arxiv.org/abs/1401.6212}{{\tt arXiv:1401.6212 [astro-ph.HE]}}.
%%CITATION = ARXIV:1401.6212;%%.

\bibitem{Feng:1999fu}
J.~L. Feng, T.~Moroi, L.~Randall, M.~Strassler, and S.~Su, {\em {Discovering
  supersymmetry at the Tevatron in wino LSP scenarios}\/},
  \href{http://dx.doi.org/10.1103/PhysRevLett.83.1731}{Phys.Rev.Lett. {\bf 83}
  (1999)  1731--1734},
\href{http://arxiv.org/abs/hep-ph/9904250}{{\tt arXiv:hep-ph/9904250
  [hep-ph]}}.
%%CITATION = HEP-PH/9904250;%%.

\bibitem{Barbieri:2004qk}
R.~Barbieri, A.~Pomarol, R.~Rattazzi, and A.~Strumia, {\em {Electroweak
  symmetry breaking after LEP-1 and LEP-2}\/},
  \href{http://dx.doi.org/10.1016/j.nuclphysb.2004.10.014}{Nucl. Phys. {\bf
  B703} (2004)  127--146},
\href{http://arxiv.org/abs/hep-ph/0405040}{{\tt arXiv:hep-ph/0405040
  [hep-ph]}}.
%%CITATION = HEP-PH/0405040;%%.

\bibitem{Fan:2014vta}
J.~Fan, M.~Reece, and L.-T. Wang, {\em {Possible Futures of Electroweak
  Precision: ILC, FCC-ee, and CEPC}\/},
\href{http://arxiv.org/abs/1411.1054}{{\tt arXiv:1411.1054 [hep-ph]}}.
%%CITATION = ARXIV:1411.1054;%%.

\bibitem{Ackermann:2015zua}
{Fermi-LAT Collaboration}, M.~Ackermann et al., {\em {Searching for Dark Matter
  Annihilation from Milky Way Dwarf Spheroidal Galaxies with Six Years of
  Fermi-LAT Data}\/},
\href{http://arxiv.org/abs/1503.02641}{{\tt arXiv:1503.02641 [astro-ph.HE]}}.
%%CITATION = ARXIV:1503.02641;%%.

\bibitem{Bonnivard:2015xpq}
V.~Bonnivard et al., {\em {Dark matter annihilation and decay in dwarf
  spheroidal galaxies: The classical and ultrafaint dSphs}\/},
  \href{http://dx.doi.org/10.1093/mnras/stv1601}{Mon. Not. Roy. Astron. Soc.
  {\bf 453} (2015) no.~1, 849--867},
\href{http://arxiv.org/abs/1504.02048}{{\tt arXiv:1504.02048 [astro-ph.HE]}}.
%%CITATION = ARXIV:1504.02048;%%.

\bibitem{Consortium:2010bc}
{CTA Consortium Collaboration}, M.~Actis et al., {\em {Design concepts for the
  Cherenkov Telescope Array CTA: An advanced facility for ground-based
  high-energy gamma-ray astronomy}\/},
  \href{http://dx.doi.org/10.1007/s10686-011-9247-0}{Exper. Astron. {\bf 32}
  (2011)  193--316},
\href{http://arxiv.org/abs/1008.3703}{{\tt arXiv:1008.3703 [astro-ph.IM]}}.
%%CITATION = ARXIV:1008.3703;%%.

\bibitem{Salati:2002md}
P.~Salati, {\em {Quintessence and the relic density of neutralinos}\/},
  \href{http://dx.doi.org/10.1016/j.physletb.2003.07.073}{Phys. Lett. {\bf
  B571} (2003)  121--131},
\href{http://arxiv.org/abs/astro-ph/0207396}{{\tt arXiv:astro-ph/0207396
  [astro-ph]}}.
%%CITATION = ASTRO-PH/0207396;%%.

\bibitem{Boucenna:2015haa}
S.~M. Boucenna, M.~B. Krauss, and E.~Nardi, {\em {Minimal Asymmetric Dark
  Matter}\/},  \href{http://dx.doi.org/10.1016/j.physletb.2015.06.080}{Phys.
  Lett. {\bf B748} (2015)  191--198},
\href{http://arxiv.org/abs/1503.01119}{{\tt arXiv:1503.01119 [hep-ph]}}.
%%CITATION = ARXIV:1503.01119;%%.

\bibitem{Fabbrichesi:2015bta}
M.~Fabbrichesi and A.~Urbano, {\em {Natural minimal dark matter}\/},
\href{http://arxiv.org/abs/1510.03861}{{\tt arXiv:1510.03861 [hep-ph]}}.
%%CITATION = ARXIV:1510.03861;%%.

\bibitem{Aoki:2015nza}
M.~Aoki, T.~Toma, and A.~Vicente, {\em {Non-thermal Production of Minimal Dark
  Matter via Right-handed Neutrino Decay}\/},
  \href{http://dx.doi.org/10.1088/1475-7516/2015/09/063}{JCAP {\bf 1509} (2015)
  no.~09, 063},
\href{http://arxiv.org/abs/1507.01591}{{\tt arXiv:1507.01591 [hep-ph]}}.
%%CITATION = ARXIV:1507.01591;%%.

\bibitem{Junnarkar:2013ac}
P.~Junnarkar and A.~Walker-Loud, {\em {Scalar strange content of the nucleon
  from lattice QCD}\/},
  \href{http://dx.doi.org/10.1103/PhysRevD.87.114510}{Phys. Rev. {\bf D87}
  (2013)  114510},
\href{http://arxiv.org/abs/1301.1114}{{\tt arXiv:1301.1114 [hep-lat]}}.
%%CITATION = ARXIV:1301.1114;%%.

\bibitem{Cirelli:2007xd}
M.~Cirelli, A.~Strumia, and M.~Tamburini, {\em {Cosmology and Astrophysics of
  Minimal Dark Matter}\/},
  \href{http://dx.doi.org/10.1016/j.nuclphysb.2007.07.023}{Nucl. Phys. {\bf
  B787} (2007)  152--175},
\href{http://arxiv.org/abs/0706.4071}{{\tt arXiv:0706.4071 [hep-ph]}}.
%%CITATION = ARXIV:0706.4071;%%.

\bibitem{DiLuzio:2015oha}
L.~Di~Luzio, R.~GrÃ¶ber, J.~F. Kamenik, and M.~Nardecchia, {\em {Accidental
  matter at the LHC}\/},  \href{http://dx.doi.org/10.1007/JHEP07(2015)074}{JHEP
  {\bf 07} (2015)  074},
\href{http://arxiv.org/abs/1504.00359}{{\tt arXiv:1504.00359 [hep-ph]}}.
%%CITATION = ARXIV:1504.00359;%%.

\bibitem{Cirelli:2015bda}
M.~Cirelli, T.~Hambye, P.~Panci, F.~Sala, and M.~Taoso, {\em {Gamma ray tests
  of Minimal Dark Matter}\/},
  \href{http://dx.doi.org/10.1088/1475-7516/2015/10/026}{JCAP {\bf 1510} (2015)
  no.~10, 026},
\href{http://arxiv.org/abs/1507.05519}{{\tt arXiv:1507.05519 [hep-ph]}}.
%%CITATION = ARXIV:1507.05519;%%.

\bibitem{Garcia-Cely:2015dda}
C.~Garcia-Cely, A.~Ibarra, A.~S. Lamperstorfer, and M.~H.~G. Tytgat, {\em
  {Gamma-rays from Heavy Minimal Dark Matter}\/},
\href{http://arxiv.org/abs/1507.05536}{{\tt arXiv:1507.05536 [hep-ph]}}.
%%CITATION = ARXIV:1507.05536;%%.

\bibitem{Ostdiek:2015aga}
B.~Ostdiek, {\em {Constraining the minimal dark matter fiveplet with LHC
  searches}\/},  \href{http://dx.doi.org/10.1103/PhysRevD.92.055008}{Phys. Rev.
  {\bf D92} (2015) no.~5, 055008},
\href{http://arxiv.org/abs/1506.03445}{{\tt arXiv:1506.03445 [hep-ph]}}.
%%CITATION = ARXIV:1506.03445;%%.

\bibitem{Bramante:2015una}
J.~Bramante, N.~Desai, P.~Fox, A.~Martin, B.~Ostdiek, and T.~Plehn, {\em
  {Towards the Final Word on Neutralino Dark Matter}\/},
\href{http://arxiv.org/abs/1510.03460}{{\tt arXiv:1510.03460 [hep-ph]}}.
%%CITATION = ARXIV:1510.03460;%%.

\bibitem{Alloul:2013bka}
A.~Alloul, N.~D. Christensen, C.~Degrande, C.~Duhr, and B.~Fuks, {\em
  {FeynRules 2.0 - A complete toolbox for tree-level phenomenology}\/},
  \href{http://dx.doi.org/10.1016/j.cpc.2014.04.012}{Comput. Phys. Commun. {\bf
  185} (2014)  2250--2300},
\href{http://arxiv.org/abs/1310.1921}{{\tt arXiv:1310.1921 [hep-ph]}}.
%%CITATION = ARXIV:1310.1921;%%.

\bibitem{Bramante:2014tba}
J.~Bramante, P.~J. Fox, A.~Martin, B.~Ostdiek, T.~Plehn, T.~Schell, and
  M.~Takeuchi, {\em {The Relic neutralino surface at a 100 TeV collider}\/},
  \href{http://dx.doi.org/10.1103/PhysRevD.91.054015}{Phys. Rev. {\bf D91}
  (2015)  054015},
\href{http://arxiv.org/abs/1412.4789}{{\tt arXiv:1412.4789 [hep-ph]}}.
%%CITATION = ARXIV:1412.4789;%%.

\bibitem{lepii}
{LEP2 SUSY Working Group Collaboration}, {\em {LEPSUSYWG, ALEPH, DELPHI, L3 and
  OPAL experiments}\/}, .

\bibitem{Hryczuk:2011vi}
A.~Hryczuk and R.~Iengo, {\em {The one-loop and Sommerfeld electroweak
  corrections to the Wino dark matter annihilation}\/},
  \href{http://dx.doi.org/10.1007/JHEP01(2012)163,
  10.1007/JHEP06(2012)137}{JHEP {\bf 1201} (2012)  163},
\href{http://arxiv.org/abs/1111.2916}{{\tt arXiv:1111.2916 [hep-ph]}}.
%%CITATION = ARXIV:1111.2916;%%.

\bibitem{Bechtle:2012zk}
P.~Bechtle, T.~Bringmann, K.~Desch, H.~Dreiner, M.~Hamer, et al., {\em
  {Constrained Supersymmetry after two years of LHC data: a global view with
  Fittino}\/},  \href{http://dx.doi.org/10.1007/JHEP06(2012)098}{JHEP {\bf
  1206} (2012)  098},
\href{http://arxiv.org/abs/1204.4199}{{\tt arXiv:1204.4199 [hep-ph]}}.
%%CITATION = ARXIV:1204.4199;%%.

\bibitem{Belanger:2013oya}
G.~Belanger, F.~Boudjema, A.~Pukhov, and A.~Semenov, {\em {micrOMEGAs$_3$: A
  program for calculating dark matter observables}\/},
  \href{http://dx.doi.org/10.1016/j.cpc.2013.10.016}{Comput. Phys. Commun. {\bf
  185} (2014)  960--985},
\href{http://arxiv.org/abs/1305.0237}{{\tt arXiv:1305.0237 [hep-ph]}}.
%%CITATION = ARXIV:1305.0237;%%.

\bibitem{Djouadi:2002ze}
A.~Djouadi, J.-L. Kneur, and G.~Moultaka, {\em {SuSpect: A Fortran code for the
  supersymmetric and Higgs particle spectrum in the MSSM}\/},
  \href{http://dx.doi.org/10.1016/j.cpc.2006.11.009}{Comput.Phys.Commun. {\bf
  176} (2007)  426--455},
\href{http://arxiv.org/abs/hep-ph/0211331}{{\tt arXiv:hep-ph/0211331
  [hep-ph]}}.
%%CITATION = HEP-PH/0211331;%%.

\bibitem{Mangano:2006rw}
M.~L. Mangano, M.~Moretti, F.~Piccinini, and M.~Treccani, {\em {Matching matrix
  elements and shower evolution for top-quark production in hadronic
  collisions}\/},  \href{http://dx.doi.org/10.1088/1126-6708/2007/01/013}{JHEP
  {\bf 01} (2007)  013},
\href{http://arxiv.org/abs/hep-ph/0611129}{{\tt arXiv:hep-ph/0611129
  [hep-ph]}}.
%%CITATION = HEP-PH/0611129;%%.

\bibitem{Han:2014kaa}
Z.~Han, G.~D. Kribs, A.~Martin, and A.~Menon, {\em {Hunting Quasi-Degenerate
  Higgsinos}\/},  \href{http://dx.doi.org/10.1103/PhysRevD.89.075007}{Phys.Rev.
  {\bf D89} (2014)  075007},
\href{http://arxiv.org/abs/1401.1235}{{\tt arXiv:1401.1235 [hep-ph]}}.
%%CITATION = ARXIV:1401.1235;%%.

\bibitem{Bramante:2014dza}
J.~Bramante, A.~Delgado, F.~Elahi, A.~Martin, and B.~Ostdiek, {\em {Catching
  sparks from well-forged neutralinos}\/},
  \href{http://dx.doi.org/10.1103/PhysRevD.90.095008}{Phys.Rev. {\bf D90}
  (2014) no.~9, 095008},
\href{http://arxiv.org/abs/1408.6530}{{\tt arXiv:1408.6530 [hep-ph]}}.
%%CITATION = ARXIV:1408.6530;%%.

\bibitem{Han:2014xoa}
C.~Han, L.~Wu, J.~M. Yang, M.~Zhang, and Y.~Zhang, {\em {New approach for
  detecting a compressed bino/wino at the LHC}\/},
  \href{http://dx.doi.org/10.1103/PhysRevD.91.055030}{Phys. Rev. {\bf D91}
  (2015)  055030},
\href{http://arxiv.org/abs/1409.4533}{{\tt arXiv:1409.4533 [hep-ph]}}.
%%CITATION = ARXIV:1409.4533;%%.

\bibitem{Baer:2014kya}
H.~Baer, A.~Mustafayev, and X.~Tata, {\em {Monojet plus soft dilepton signal
  from light higgsino pair production at LHC14}\/},
\href{http://arxiv.org/abs/1409.7058}{{\tt arXiv:1409.7058 [hep-ph]}}.
%%CITATION = ARXIV:1409.7058;%%.

\bibitem{Han:2015lma}
C.~Han, D.~Kim, S.~Munir, and M.~Park, {\em {Accessing the core of naturalness,
  nearly degenerate higgsinos, at the LHC}\/},
  \href{http://dx.doi.org/10.1007/JHEP04(2015)132}{JHEP {\bf 04} (2015)  132},
\href{http://arxiv.org/abs/1502.03734}{{\tt arXiv:1502.03734 [hep-ph]}}.
%%CITATION = ARXIV:1502.03734;%%.

\bibitem{Han:2015lha}
C.~Han and M.~Park, {\em {Revealing the jet substructure in a compressed
  spectrum}\/},
\href{http://arxiv.org/abs/1507.07729}{{\tt arXiv:1507.07729 [hep-ph]}}.
%%CITATION = ARXIV:1507.07729;%%.

\bibitem{Bhattacharyya:2009cc}
N.~Bhattacharyya and A.~Datta, {\em {Tracking down the elusive charginos /
  neutralinos through tau leptons at the Large Hadron Collider}\/},
  \href{http://dx.doi.org/10.1103/PhysRevD.80.055016}{Phys.Rev. {\bf D80}
  (2009)  055016},
\href{http://arxiv.org/abs/0906.1460}{{\tt arXiv:0906.1460 [hep-ph]}}.
%%CITATION = ARXIV:0906.1460;%%.

\bibitem{Giudice:2010wb}
G.~F. Giudice, T.~Han, K.~Wang, and L.-T. Wang, {\em {Nearly Degenerate
  Gauginos and Dark Matter at the LHC}\/},
  \href{http://dx.doi.org/10.1103/PhysRevD.81.115011}{Phys.Rev. {\bf D81}
  (2010)  115011},
\href{http://arxiv.org/abs/1004.4902}{{\tt arXiv:1004.4902 [hep-ph]}}.
%%CITATION = ARXIV:1004.4902;%%.

\bibitem{Calibbi:2013poa}
L.~Calibbi, J.~M. Lindert, T.~Ota, and Y.~Takanishi, {\em {Cornering light
  Neutralino Dark Matter at the LHC}\/},
  \href{http://dx.doi.org/10.1007/JHEP10(2013)132}{JHEP {\bf 1310} (2013)
  132},
\href{http://arxiv.org/abs/1307.4119}{{\tt arXiv:1307.4119}}.
%%CITATION = ARXIV:1307.4119;%%.

\bibitem{Chakraborti:2015mra}
M.~Chakraborti, U.~Chattopadhyay, A.~Choudhury, A.~Datta, and S.~Poddar, {\em
  {Reduced LHC constraints for higgsino-like heavier electroweakinos}\/},
\href{http://arxiv.org/abs/1507.01395}{{\tt arXiv:1507.01395 [hep-ph]}}.
%%CITATION = ARXIV:1507.01395;%%.

\bibitem{Izaguirre:2015zva}
E.~Izaguirre, G.~Krnjaic, and B.~Shuve, {\em {Discovering Inelastic
  Thermal-Relic Dark Matter at Colliders}\/},
\href{http://arxiv.org/abs/1508.03050}{{\tt arXiv:1508.03050 [hep-ph]}}.
%%CITATION = ARXIV:1508.03050;%%.

\bibitem{Carr:2015hta}
{CTA Consortium Collaboration}, J.~Carr et al., {\em {Prospects for Indirect
  Dark Matter Searches with the Cherenkov Telescope Array (CTA)}\/},  in {\em
  {Proceedings, 34th International Cosmic Ray Conference (ICRC 2015)}}.
\newblock 2015.
\newblock \href{http://arxiv.org/abs/1508.06128}{{\tt arXiv:1508.06128
  [astro-ph.HE]}}.
\newblock
\url{http://inspirehep.net/record/1389681/files/arXiv:1508.06128.pdf}.
\newblock
%%CITATION = ARXIV:1508.06128;%%.

\bibitem{Akerib:2013tjd}
{LUX Collaboration}, D.~Akerib et al., {\em {First results from the LUX dark
  matter experiment at the Sanford Underground Research Facility}\/},
  \href{http://dx.doi.org/10.1103/PhysRevLett.112.091303}{Phys.Rev.Lett. {\bf
  112} (2014)  091303},
\href{http://arxiv.org/abs/1310.8214}{{\tt arXiv:1310.8214 [astro-ph.CO]}}.
%%CITATION = ARXIV:1310.8214;%%.

\bibitem{Silveira:1985rk}
V.~Silveira and A.~Zee, {\em {SCALAR PHANTOMS}\/},
\href{http://dx.doi.org/10.1016/0370-2693(85)90624-0}{Phys.Lett. {\bf B161}
  (1985)  136}.
%%CITATION = PHLTA,B161,136;%%.

\bibitem{McDonald:1993ex}
J.~McDonald, {\em {Gauge singlet scalars as cold dark matter}\/},
  \href{http://dx.doi.org/10.1103/PhysRevD.50.3637}{Phys.Rev. {\bf D50} (1994)
  3637--3649},
\href{http://arxiv.org/abs/hep-ph/0702143}{{\tt arXiv:hep-ph/0702143
  [HEP-PH]}}.
%%CITATION = HEP-PH/0702143;%%.

\bibitem{Burgess:2000yq}
C.~Burgess, M.~Pospelov, and T.~ter Veldhuis, {\em {The Minimal model of
  nonbaryonic dark matter: A Singlet scalar}\/},
  \href{http://dx.doi.org/10.1016/S0550-3213(01)00513-2}{Nucl.Phys. {\bf B619}
  (2001)  709--728},
\href{http://arxiv.org/abs/hep-ph/0011335}{{\tt arXiv:hep-ph/0011335
  [hep-ph]}}.
%%CITATION = HEP-PH/0011335;%%.

\bibitem{Patt:2006fw}
B.~Patt and F.~Wilczek, {\em {Higgs-field portal into hidden sectors}\/},
\href{http://arxiv.org/abs/hep-ph/0605188}{{\tt arXiv:hep-ph/0605188
  [hep-ph]}}.
%%CITATION = HEP-PH/0605188;%%.

\bibitem{Barger:2007im}
V.~Barger, P.~Langacker, M.~McCaskey, M.~J. Ramsey-Musolf, and G.~Shaughnessy,
  {\em {LHC Phenomenology of an Extended Standard Model with a Real Scalar
  Singlet}\/},  \href{http://dx.doi.org/10.1103/PhysRevD.77.035005}{Phys.Rev.
  {\bf D77} (2008)  035005},
\href{http://arxiv.org/abs/0706.4311}{{\tt arXiv:0706.4311 [hep-ph]}}.
%%CITATION = ARXIV:0706.4311;%%.

\bibitem{Davoudiasl:2004be}
H.~Davoudiasl, R.~Kitano, T.~Li, and H.~Murayama, {\em {The New minimal
  standard model}\/},
  \href{http://dx.doi.org/10.1016/j.physletb.2005.01.026}{Phys.Lett. {\bf B609}
  (2005)  117--123},
\href{http://arxiv.org/abs/hep-ph/0405097}{{\tt arXiv:hep-ph/0405097
  [hep-ph]}}.
%%CITATION = HEP-PH/0405097;%%.

\bibitem{Eboli:2000ze}
O.~J.~P. Eboli and D.~Zeppenfeld, {\em {Observing an invisible Higgs boson}\/},
   \href{http://dx.doi.org/10.1016/S0370-2693(00)01213-2}{Phys. Lett. {\bf
  B495} (2000)  147--154},
\href{http://arxiv.org/abs/hep-ph/0009158}{{\tt arXiv:hep-ph/0009158
  [hep-ph]}}.
%%CITATION = HEP-PH/0009158;%%.

\bibitem{Low:2011kp}
I.~Low, P.~Schwaller, G.~Shaughnessy, and C.~E.~M. Wagner, {\em {The dark side
  of the Higgs boson}\/},
  \href{http://dx.doi.org/10.1103/PhysRevD.85.015009}{Phys. Rev. {\bf D85}
  (2012)  015009},
\href{http://arxiv.org/abs/1110.4405}{{\tt arXiv:1110.4405 [hep-ph]}}.
%%CITATION = ARXIV:1110.4405;%%.

\bibitem{Djouadi:2011aa}
A.~Djouadi, O.~Lebedev, Y.~Mambrini, and J.~Quevillon, {\em {Implications of
  LHC searches for Higgs--portal dark matter}\/},
  \href{http://dx.doi.org/10.1016/j.physletb.2012.01.062}{Phys.Lett. {\bf B709}
  (2012)  65--69},
\href{http://arxiv.org/abs/1112.3299}{{\tt arXiv:1112.3299 [hep-ph]}}.
%%CITATION = ARXIV:1112.3299;%%.

\bibitem{Englert:2011us}
C.~Englert, J.~Jaeckel, E.~Re, and M.~Spannowsky, {\em {Evasive Higgs Maneuvers
  at the LHC}\/},
  \href{http://dx.doi.org/10.1103/PhysRevD.85.035008}{Phys.Rev. {\bf D85}
  (2012)  035008},
\href{http://arxiv.org/abs/1111.1719}{{\tt arXiv:1111.1719 [hep-ph]}}.
%%CITATION = ARXIV:1111.1719;%%.

\bibitem{Djouadi:2012zc}
A.~Djouadi, A.~Falkowski, Y.~Mambrini, and J.~Quevillon, {\em {Direct Detection
  of Higgs-Portal Dark Matter at the LHC}\/},
  \href{http://dx.doi.org/10.1140/epjc/s10052-013-2455-1}{Eur.Phys.J. {\bf C73}
  (2013)  2455},
\href{http://arxiv.org/abs/1205.3169}{{\tt arXiv:1205.3169 [hep-ph]}}.
%%CITATION = ARXIV:1205.3169;%%.

\bibitem{Englert:2013gz}
C.~Englert, J.~Jaeckel, V.~Khoze, and M.~Spannowsky, {\em {Emergence of the
  Electroweak Scale through the Higgs Portal}\/},
  \href{http://dx.doi.org/10.1007/JHEP04(2013)060}{JHEP {\bf 1304} (2013)
  060},
\href{http://arxiv.org/abs/1301.4224}{{\tt arXiv:1301.4224 [hep-ph]}}.
%%CITATION = ARXIV:1301.4224;%%.

\bibitem{Aad:2014iia}
{ATLAS Collaboration}, G.~Aad et al., {\em {Search for Invisible Decays of a
  Higgs Boson Produced in Association with a Z Boson in ATLAS}\/},
  \href{http://dx.doi.org/10.1103/PhysRevLett.112.201802}{Phys.Rev.Lett. {\bf
  112} (2014)  201802},
\href{http://arxiv.org/abs/1402.3244}{{\tt arXiv:1402.3244 [hep-ex]}}.
%%CITATION = ARXIV:1402.3244;%%.

\bibitem{Chatrchyan:2014tja}
{CMS Collaboration}, S.~Chatrchyan et al., {\em {Search for invisible decays of
  Higgs bosons in the vector boson fusion and associated ZH production
  modes}\/},  \href{http://dx.doi.org/10.1140/epjc/s10052-014-2980-6}{Eur.
  Phys. J. {\bf C74} (2014)  2980},
\href{http://arxiv.org/abs/1404.1344}{{\tt arXiv:1404.1344 [hep-ex]}}.
%%CITATION = ARXIV:1404.1344;%%.

\bibitem{Bernaciak:2014pna}
C.~Bernaciak, T.~Plehn, P.~Schichtel, and J.~Tattersall, {\em {Spying an
  invisible Higgs boson}\/},
  \href{http://dx.doi.org/10.1103/PhysRevD.91.035024}{Phys. Rev. {\bf D91}
  (2015)  035024},
\href{http://arxiv.org/abs/1411.7699}{{\tt arXiv:1411.7699 [hep-ph]}}.
%%CITATION = ARXIV:1411.7699;%%.

\bibitem{Sjostrand:2007gs}
T.~Sjostrand, S.~Mrenna, and P.~Z. Skands, {\em {A Brief Introduction to PYTHIA
  8.1}\/},
  \href{http://dx.doi.org/10.1016/j.cpc.2008.01.036}{Comput.Phys.Commun. {\bf
  178} (2008)  852--867},
\href{http://arxiv.org/abs/0710.3820}{{\tt arXiv:0710.3820 [hep-ph]}}.
%%CITATION = ARXIV:0710.3820;%%.

\bibitem{Craig:2014lda}
N.~Craig, H.~K. Lou, M.~McCullough, and A.~Thalapillil, {\em {The Higgs Portal
  Above Threshold}\/},
\href{http://arxiv.org/abs/1412.0258}{{\tt arXiv:1412.0258 [hep-ph]}}.
%%CITATION = ARXIV:1412.0258;%%.

\bibitem{Malik:2014ggr}
S.~A. Malik et al., {\em {Interplay and Characterization of Dark Matter
  Searches at Colliders and in Direct Detection Experiments}\/},
  \href{http://dx.doi.org/10.1016/j.dark.2015.03.003}{Phys. Dark Univ. {\bf
  9-10} (2015)  51--58},
\href{http://arxiv.org/abs/1409.4075}{{\tt arXiv:1409.4075 [hep-ex]}}.
%%CITATION = ARXIV:1409.4075;%%.

\bibitem{Abercrombie:2015wmb}
D.~Abercrombie et al., {\em {Dark Matter Benchmark Models for Early LHC Run-2
  Searches: Report of the ATLAS/CMS Dark Matter Forum}\/},
\href{http://arxiv.org/abs/1507.00966}{{\tt arXiv:1507.00966 [hep-ex]}}.
%%CITATION = ARXIV:1507.00966;%%.

\bibitem{Buchmueller:2013dya}
O.~Buchmueller, M.~J. Dolan, and C.~McCabe, {\em {Beyond Effective Field Theory
  for Dark Matter Searches at the LHC}\/},
  \href{http://dx.doi.org/10.1007/JHEP01(2014)025}{JHEP {\bf 01} (2014)  025},
\href{http://arxiv.org/abs/1308.6799}{{\tt arXiv:1308.6799 [hep-ph]}}.
%%CITATION = ARXIV:1308.6799;%%.

\bibitem{Buchmueller:2014yoa}
O.~Buchmueller, M.~J. Dolan, S.~A. Malik, and C.~McCabe, {\em {Characterising
  dark matter searches at colliders and direct detection experiments: Vector
  mediators}\/},  \href{http://dx.doi.org/10.1007/JHEP01(2015)037}{JHEP {\bf
  01} (2015)  037},
\href{http://arxiv.org/abs/1407.8257}{{\tt arXiv:1407.8257 [hep-ph]}}.
%%CITATION = ARXIV:1407.8257;%%.

\bibitem{DiFranzo:2013vra}
A.~DiFranzo, K.~I. Nagao, A.~Rajaraman, and T.~M.~P. Tait, {\em {Simplified
  Models for Dark Matter Interacting with Quarks}\/},
  \href{http://dx.doi.org/10.1007/JHEP11(2013)014}{JHEP {\bf 1311} (2013)
  014},
\href{http://arxiv.org/abs/1308.2679}{{\tt arXiv:1308.2679 [hep-ph]}}.
%%CITATION = ARXIV:1308.2679;%%.

\bibitem{Harris:2014hga}
P.~Harris, V.~V. Khoze, M.~Spannowsky, and C.~Williams, {\em {Constraining Dark
  Sectors at Colliders: Beyond the Effective Theory Approach}\/},
\href{http://arxiv.org/abs/1411.0535}{{\tt arXiv:1411.0535 [hep-ph]}}.
%%CITATION = ARXIV:1411.0535;%%.

\bibitem{Buckley:2014fba}
M.~R. Buckley, D.~Feld, and D.~Goncalves, {\em {Scalar Simplified Models for
  Dark Matter}\/},
\href{http://arxiv.org/abs/1410.6497}{{\tt arXiv:1410.6497 [hep-ph]}}.
%%CITATION = ARXIV:1410.6497;%%.

\bibitem{Haisch:2015ioa}
U.~Haisch and E.~Re, {\em {Simplified dark matter top-quark interactions at the
  LHC}\/},  \href{http://dx.doi.org/10.1007/JHEP06(2015)078}{JHEP {\bf 06}
  (2015)  078},
\href{http://arxiv.org/abs/1503.00691}{{\tt arXiv:1503.00691 [hep-ph]}}.
%%CITATION = ARXIV:1503.00691;%%.

\bibitem{Khoze:2015sra}
V.~V. Khoze, G.~Ro, and M.~Spannowsky, {\em {Spectroscopy of scalar mediators
  to dark matter at the LHC and at 100 TeV}\/},
  \href{http://dx.doi.org/10.1103/PhysRevD.92.075006}{Phys. Rev. {\bf D92}
  (2015) no.~7, 075006},
\href{http://arxiv.org/abs/1505.03019}{{\tt arXiv:1505.03019 [hep-ph]}}.
%%CITATION = ARXIV:1505.03019;%%.

\bibitem{Harris:2015kda}
P.~Harris, V.~V. Khoze, M.~Spannowsky, and C.~Williams, {\em {Closing up on
  Dark Sectors at Colliders: from 14 to 100 TeV}\/},
\href{http://arxiv.org/abs/1509.02904}{{\tt arXiv:1509.02904 [hep-ph]}}.
%%CITATION = ARXIV:1509.02904;%%.

\bibitem{Abazov:2003gp}
{D0 Collaboration}, V.~M. Abazov et al., {\em {Search for large extra
  dimensions in the monojet + missing $E_T$ channel at D\O}\/},
  \href{http://dx.doi.org/10.1103/PhysRevLett.90.251802}{Phys. Rev. Lett. {\bf
  90} (2003)  251802},
\href{http://arxiv.org/abs/hep-ex/0302014}{{\tt arXiv:hep-ex/0302014
  [hep-ex]}}.
%%CITATION = HEP-EX/0302014;%%.

\bibitem{Aaltonen:2012jb}
{CDF Collaboration}, T.~Aaltonen et al., {\em {A Search for dark matter in
  events with one jet and missing transverse energy in $p\bar{p}$ collisions at
  $\sqrt{s} = 1.96$ TeV}\/},
  \href{http://dx.doi.org/10.1103/PhysRevLett.108.211804}{Phys. Rev. Lett. {\bf
  108} (2012)  211804},
\href{http://arxiv.org/abs/1203.0742}{{\tt arXiv:1203.0742 [hep-ex]}}.
%%CITATION = ARXIV:1203.0742;%%.

\bibitem{Chatrchyan:2012me}
{CMS Collaboration}, S.~Chatrchyan et al., {\em {Search for dark matter and
  large extra dimensions in monojet events in $pp$ collisions at $\sqrt{s}=7$
  TeV}\/},  \href{http://dx.doi.org/10.1007/JHEP09(2012)094}{JHEP {\bf 09}
  (2012)  094},
\href{http://arxiv.org/abs/1206.5663}{{\tt arXiv:1206.5663 [hep-ex]}}.
%%CITATION = ARXIV:1206.5663;%%.

\bibitem{ATLAS-CONF-2012-084}
{ATLAS Collaboration}, {\em {Search for dark matter candidates and large extra
  dimensions in events with a jet and missing transverse momentum with the
  ATLAS detector}\/},   ATLAS-CONF-2012-084, CERN, Geneva, Jul, 2012.
\newblock \url{https://cds.cern.ch/record/1460396}.

\bibitem{Khachatryan:2014rra}
{CMS Collaboration}, V.~Khachatryan et al., {\em {Search for dark matter, extra
  dimensions, and unparticles in monojet events in proton-proton collisions at
  $\sqrt{s} = 8$ TeV}\/},
  \href{http://dx.doi.org/10.1140/epjc/s10052-015-3451-4}{Eur. Phys. J. {\bf
  C75} (2015) no.~5, 235},
\href{http://arxiv.org/abs/1408.3583}{{\tt arXiv:1408.3583 [hep-ex]}}.
%%CITATION = ARXIV:1408.3583;%%.

\bibitem{Diehl:2014dda}
{ATLAS Collaboration}, E.~Diehl, {\em {The search for dark matter using
  monojets and monophotons with the ATLAS detector}\/},  AIP Conf.Proc. (2014)
  no.~1604, 324, .
  \url{http://scitation.aip.org/content/aip/proceeding/aipcp/10.1063/1.4883448}.

\bibitem{Feng:2005gj}
J.~L. Feng, S.~Su, and F.~Takayama, {\em {Lower limit on dark matter production
  at the large hadron collider}\/},
  \href{http://dx.doi.org/10.1103/PhysRevLett.96.151802}{Phys. Rev. Lett. {\bf
  96} (2006)  151802},
\href{http://arxiv.org/abs/hep-ph/0503117}{{\tt arXiv:hep-ph/0503117
  [hep-ph]}}.
%%CITATION = HEP-PH/0503117;%%.

\bibitem{Cao:2009uw}
Q.-H. Cao, C.-R. Chen, C.~S. Li, and H.~Zhang, {\em {Effective Dark Matter
  Model: Relic density, CDMS II, Fermi LAT and LHC}\/},
  \href{http://dx.doi.org/10.1007/JHEP08(2011)018}{JHEP {\bf 08} (2011)  018},
\href{http://arxiv.org/abs/0912.4511}{{\tt arXiv:0912.4511 [hep-ph]}}.
%%CITATION = ARXIV:0912.4511;%%.

\bibitem{Beltran:2010ww}
M.~Beltran, D.~Hooper, E.~W. Kolb, Z.~A.~C. Krusberg, and T.~M.~P. Tait, {\em
  {Maverick dark matter at colliders}\/},
  \href{http://dx.doi.org/10.1007/JHEP09(2010)037}{JHEP {\bf 09} (2010)  037},
\href{http://arxiv.org/abs/1002.4137}{{\tt arXiv:1002.4137 [hep-ph]}}.
%%CITATION = ARXIV:1002.4137;%%.

\bibitem{Goodman:2010yf}
J.~Goodman, M.~Ibe, A.~Rajaraman, W.~Shepherd, T.~M.~P. Tait, and H.-B. Yu,
  {\em {Constraints on Light Majorana dark Matter from Colliders}\/},
  \href{http://dx.doi.org/10.1016/j.physletb.2010.11.009}{Phys. Lett. {\bf
  B695} (2011)  185--188},
\href{http://arxiv.org/abs/1005.1286}{{\tt arXiv:1005.1286 [hep-ph]}}.
%%CITATION = ARXIV:1005.1286;%%.

\bibitem{Goodman:2010ku}
J.~Goodman, M.~Ibe, A.~Rajaraman, W.~Shepherd, T.~M.~P. Tait, and H.-B. Yu,
  {\em {Constraints on Dark Matter from Colliders}\/},
  \href{http://dx.doi.org/10.1103/PhysRevD.82.116010}{Phys. Rev. {\bf D82}
  (2010)  116010},
\href{http://arxiv.org/abs/1008.1783}{{\tt arXiv:1008.1783 [hep-ph]}}.
%%CITATION = ARXIV:1008.1783;%%.

\bibitem{Fox:2011pm}
P.~J. Fox, R.~Harnik, J.~Kopp, and Y.~Tsai, {\em {Missing Energy Signatures of
  Dark Matter at the LHC}\/},
  \href{http://dx.doi.org/10.1103/PhysRevD.85.056011}{Phys.Rev. {\bf D85}
  (2012)  056011},
\href{http://arxiv.org/abs/1109.4398}{{\tt arXiv:1109.4398 [hep-ph]}}.
%%CITATION = ARXIV:1109.4398;%%.

\bibitem{Haisch:2012kf}
U.~Haisch, F.~Kahlhoefer, and J.~Unwin, {\em {The impact of heavy-quark loops
  on LHC dark matter searches}\/},
  \href{http://dx.doi.org/10.1007/JHEP07(2013)125}{JHEP {\bf 07} (2013)  125},
\href{http://arxiv.org/abs/1208.4605}{{\tt arXiv:1208.4605 [hep-ph]}}.
%%CITATION = ARXIV:1208.4605;%%.

\bibitem{Branco:2011iw}
G.~C. Branco, P.~M. Ferreira, L.~Lavoura, M.~N. Rebelo, M.~Sher, and J.~P.
  Silva, {\em {Theory and phenomenology of two-Higgs-doublet models}\/},
  \href{http://dx.doi.org/10.1016/j.physrep.2012.02.002}{Phys. Rept. {\bf 516}
  (2012)  1--102},
\href{http://arxiv.org/abs/1106.0034}{{\tt arXiv:1106.0034 [hep-ph]}}.
%%CITATION = ARXIV:1106.0034;%%.

\bibitem{Schabinger:2005ei}
R.~Schabinger and J.~D. Wells, {\em {A Minimal spontaneously broken hidden
  sector and its impact on Higgs boson physics at the large hadron
  collider}\/},  \href{http://dx.doi.org/10.1103/PhysRevD.72.093007}{Phys. Rev.
  {\bf D72} (2005)  093007},
\href{http://arxiv.org/abs/hep-ph/0509209}{{\tt arXiv:hep-ph/0509209
  [hep-ph]}}.
%%CITATION = HEP-PH/0509209;%%.

\bibitem{Englert:2011yb}
C.~Englert, T.~Plehn, D.~Zerwas, and P.~M. Zerwas, {\em {Exploring the Higgs
  portal}\/},  \href{http://dx.doi.org/10.1016/j.physletb.2011.08.002}{Phys.
  Lett. {\bf B703} (2011)  298--305},
\href{http://arxiv.org/abs/1106.3097}{{\tt arXiv:1106.3097 [hep-ph]}}.
%%CITATION = ARXIV:1106.3097;%%.

\bibitem{Hambye:2013sna}
T.~Hambye and A.~Strumia, {\em {Dynamical generation of the weak and Dark
  Matter scale}\/},  \href{http://dx.doi.org/10.1103/PhysRevD.88.055022}{Phys.
  Rev. {\bf D88} (2013)  055022},
\href{http://arxiv.org/abs/1306.2329}{{\tt arXiv:1306.2329 [hep-ph]}}.
%%CITATION = ARXIV:1306.2329;%%.

\bibitem{Carone:2013wla}
C.~D. Carone and R.~Ramos, {\em {Classical scale-invariance, the electroweak
  scale and vector dark matter}\/},
  \href{http://dx.doi.org/10.1103/PhysRevD.88.055020}{Phys. Rev. {\bf D88}
  (2013)  055020},
\href{http://arxiv.org/abs/1307.8428}{{\tt arXiv:1307.8428 [hep-ph]}}.
%%CITATION = ARXIV:1307.8428;%%.

\bibitem{Khoze:2014xha}
V.~V. Khoze, C.~McCabe, and G.~Ro, {\em {Higgs vacuum stability from the dark
  matter portal}\/},  \href{http://dx.doi.org/10.1007/JHEP08(2014)026}{JHEP
  {\bf 08} (2014)  026},
\href{http://arxiv.org/abs/1403.4953}{{\tt arXiv:1403.4953 [hep-ph]}}.
%%CITATION = ARXIV:1403.4953;%%.

\bibitem{Carena:2004xs}
M.~Carena, A.~Daleo, B.~A. Dobrescu, and T.~M.~P. Tait, {\em {$Z^\prime$ gauge
  bosons at the Tevatron}\/},
  \href{http://dx.doi.org/10.1103/PhysRevD.70.093009}{Phys. Rev. {\bf D70}
  (2004)  093009},
\href{http://arxiv.org/abs/hep-ph/0408098}{{\tt arXiv:hep-ph/0408098
  [hep-ph]}}.
%%CITATION = HEP-PH/0408098;%%.

\bibitem{Fox:2012ru}
P.~J. Fox and C.~Williams, {\em {Next-to-Leading Order Predictions for Dark
  Matter Production at Hadron Colliders}\/},
  \href{http://dx.doi.org/10.1103/PhysRevD.87.054030}{Phys. Rev. {\bf D87}
  (2013) no.~5, 054030},
\href{http://arxiv.org/abs/1211.6390}{{\tt arXiv:1211.6390 [hep-ph]}}.
%%CITATION = ARXIV:1211.6390;%%.

\bibitem{MCFMweb}
R.~K.~E. J.~M.~Campbell and C.~Williams, {\em {MCFM website}\/},
http://mcfm.fnal.gov  .
%%CITATION = ARXIV:1911.6390;%%.

\bibitem{Arnold:2008rz}
K.~Arnold et al., {\em {VBFNLO: A Parton level Monte Carlo for processes with
  electroweak bosons}\/},
  \href{http://dx.doi.org/10.1016/j.cpc.2009.03.006}{Comput. Phys. Commun. {\bf
  180} (2009)  1661--1670},
\href{http://arxiv.org/abs/0811.4559}{{\tt arXiv:0811.4559 [hep-ph]}}.
%%CITATION = ARXIV:0811.4559;%%.

\bibitem{Arnold:2011wj}
K.~Arnold et al., {\em {VBFNLO: A Parton Level Monte Carlo for Processes with
  Electroweak Bosons -- Manual for Version 2.5.0}\/},
\href{http://arxiv.org/abs/1107.4038}{{\tt arXiv:1107.4038 [hep-ph]}}.
%%CITATION = ARXIV:1107.4038;%%.

\bibitem{Baglio:2014uba}
J.~Baglio et al., {\em {Release Note - VBFNLO 2.7.0}\/},
\href{http://arxiv.org/abs/1404.3940}{{\tt arXiv:1404.3940 [hep-ph]}}.
%%CITATION = ARXIV:1404.3940;%%.

\bibitem{Ball:2014uwa}
{NNPDF Collaboration}, R.~D. Ball et al., {\em {Parton distributions for the
  LHC Run II}\/},  \href{http://dx.doi.org/10.1007/JHEP04(2015)040}{JHEP {\bf
  04} (2015)  040},
\href{http://arxiv.org/abs/1410.8849}{{\tt arXiv:1410.8849 [hep-ph]}}.
%%CITATION = ARXIV:1410.8849;%%.

\bibitem{Read:2002hq}
A.~L. Read, {\em {Presentation of search results: The CL(s) technique}\/},
  \href{http://dx.doi.org/10.1088/0954-3899/28/10/313}{J. Phys. {\bf G28}
  (2002)  2693--2704}.
[,11(2002)].
%%CITATION = JPAGA,G28,2693;%%.

\bibitem{Cowan:2010js}
G.~Cowan, K.~Cranmer, E.~Gross, and O.~Vitells, {\em {Asymptotic formulae for
  likelihood-based tests of new physics}\/},
  \href{http://dx.doi.org/10.1140/epjc/s10052-011-1554-0,
  10.1140/epjc/s10052-013-2501-z}{Eur. Phys. J. {\bf C71} (2011)  1554},
  \href{http://arxiv.org/abs/1007.1727}{{\tt arXiv:1007.1727
  [physics.data-an]}}.
[Erratum: Eur. Phys. J.C73,2501(2013)].
%%CITATION = ARXIV:1007.1727;%%.

\bibitem{ATLAS:1502664}
{ATLAS Collaboration}, {\em {Letter of Intent for the Phase-II Upgrade of the
  ATLAS Experiment}\/},   CERN-LHCC-2012-022. LHCC-I-023, CERN, Geneva, Dec,
  2012.
\newblock \url{https://cds.cern.ch/record/1502664}.
\newblock Draft version for comments.

\bibitem{Khachatryan:2014gga}
{CMS Collaboration}, V.~Khachatryan et al., {\em {Performance of the CMS
  missing transverse momentum reconstruction in pp data at $\sqrt{s}$ = 8
  TeV}\/},  \href{http://dx.doi.org/10.1088/1748-0221/10/02/P02006}{JINST {\bf
  10} (2015) no.~02, P02006},
\href{http://arxiv.org/abs/1411.0511}{{\tt arXiv:1411.0511 [physics.ins-det]}}.
%%CITATION = ARXIV:1411.0511;%%.

\bibitem{Cheng:2012qr}
H.-Y. Cheng and C.-W. Chiang, {\em {Revisiting Scalar and Pseudoscalar
  Couplings with Nucleons}\/},
  \href{http://dx.doi.org/10.1007/JHEP07(2012)009}{JHEP {\bf 07} (2012)  009},
\href{http://arxiv.org/abs/1202.1292}{{\tt arXiv:1202.1292 [hep-ph]}}.
%%CITATION = ARXIV:1202.1292;%%.

\bibitem{Kurylov:2003ra}
A.~Kurylov and M.~Kamionkowski, {\em {Generalized analysis of weakly
  interacting massive particle searches}\/},
  \href{http://dx.doi.org/10.1103/PhysRevD.69.063503}{Phys. Rev. {\bf D69}
  (2004)  063503},
\href{http://arxiv.org/abs/hep-ph/0307185}{{\tt arXiv:hep-ph/0307185
  [hep-ph]}}.
%%CITATION = HEP-PH/0307185;%%.

\bibitem{Hisano:2010ct}
J.~Hisano, K.~Ishiwata, and N.~Nagata, {\em {Gluon contribution to the dark
  matter direct detection}\/},
  \href{http://dx.doi.org/10.1103/PhysRevD.82.115007}{Phys. Rev. {\bf D82}
  (2010)  115007},
\href{http://arxiv.org/abs/1007.2601}{{\tt arXiv:1007.2601 [hep-ph]}}.
%%CITATION = ARXIV:1007.2601;%%.

\bibitem{Cheung:2013pfa}
K.~Cheung, C.-T. Lu, P.-Y. Tseng, and T.-C. Yuan, {\em {Collider Constraints on
  the Dark Matter Interpretation of the CDMS II Results}\/},
\href{http://arxiv.org/abs/1308.0067}{{\tt arXiv:1308.0067 [hep-ph]}}.
%%CITATION = ARXIV:1308.0067;%%.

\bibitem{Hoferichter:2015dsa}
M.~Hoferichter, J.~Ruiz~de Elvira, B.~Kubis, and U.-G. Mei{\ss}ner, {\em
  {High-Precision Determination of the Pion-Nucleon $\sigma$ Term from
  Roy-Steiner Equations}\/},
  \href{http://dx.doi.org/10.1103/PhysRevLett.115.092301}{Phys. Rev. Lett. {\bf
  115} (2015)  092301},
\href{http://arxiv.org/abs/1506.04142}{{\tt arXiv:1506.04142 [hep-ph]}}.
%%CITATION = ARXIV:1506.04142;%%.

\bibitem{Crivellin:2013ipa}
A.~Crivellin, M.~Hoferichter, and M.~Procura, {\em {Accurate evaluation of
  hadronic uncertainties in spin-independent WIMP-nucleon scattering:
  Disentangling two- and three-flavor effects}\/},
  \href{http://dx.doi.org/10.1103/PhysRevD.89.054021}{Phys. Rev. {\bf D89}
  (2014)  054021},
\href{http://arxiv.org/abs/1312.4951}{{\tt arXiv:1312.4951 [hep-ph]}}.
%%CITATION = ARXIV:1312.4951;%%.

\bibitem{Ackermann:2011wa}
{Fermi-LAT Collaboration}, M.~Ackermann et al., {\em {Constraining Dark Matter
  Models from a Combined Analysis of Milky Way Satellites with the Fermi Large
  Area Telescope}\/},
  \href{http://dx.doi.org/10.1103/PhysRevLett.107.241302}{Phys. Rev. Lett. {\bf
  107} (2011)  241302},
\href{http://arxiv.org/abs/1108.3546}{{\tt arXiv:1108.3546 [astro-ph.HE]}}.
%%CITATION = ARXIV:1108.3546;%%.

\bibitem{Abdo:2010ex}
{Fermi-LAT Collaboration}, A.~A. Abdo et al., {\em {Observations of Milky Way
  Dwarf Spheroidal galaxies with the Fermi-LAT detector and constraints on Dark
  Matter models}\/},
  \href{http://dx.doi.org/10.1088/0004-637X/712/1/147}{Astrophys. J. {\bf 712}
  (2010)  147--158},
\href{http://arxiv.org/abs/1001.4531}{{\tt arXiv:1001.4531 [astro-ph.CO]}}.
%%CITATION = ARXIV:1001.4531;%%.

\bibitem{Zheng:2010js}
J.-M. Zheng, Z.-H. Yu, J.-W. Shao, X.-J. Bi, Z.~Li, and H.-H. Zhang, {\em
  {Constraining the interaction strength between dark matter and visible
  matter: I. fermionic dark matter}\/},
  \href{http://dx.doi.org/10.1016/j.nuclphysb.2011.09.009}{Nucl. Phys. {\bf
  B854} (2012)  350--374},
\href{http://arxiv.org/abs/1012.2022}{{\tt arXiv:1012.2022 [hep-ph]}}.
%%CITATION = ARXIV:1012.2022;%%.

\bibitem{Boehm:2014hva}
C.~Boehm, M.~J. Dolan, C.~McCabe, M.~Spannowsky, and C.~J. Wallace, {\em
  {Extended gamma-ray emission from Coy Dark Matter}\/},
  \href{http://dx.doi.org/10.1088/1475-7516/2014/05/009}{JCAP {\bf 1405} (2014)
   009},
\href{http://arxiv.org/abs/1401.6458}{{\tt arXiv:1401.6458 [hep-ph]}}.
%%CITATION = ARXIV:1401.6458;%%.

\bibitem{Pree:2016hwc}
T.~d. Pree, K.~Hahn, P.~Harris, and C.~Roskas, {\em {Cosmological constraints
  on Dark Matter models for collider searches}\/},
\href{http://arxiv.org/abs/1603.08525}{{\tt arXiv:1603.08525 [hep-ph]}}.
%%CITATION = ARXIV:1603.08525;%%.

\bibitem{Frandsen:2012rk}
M.~T. Frandsen, F.~Kahlhoefer, A.~Preston, S.~Sarkar, and K.~Schmidt-Hoberg,
  {\em {LHC and Tevatron Bounds on the Dark Matter Direct Detection
  Cross-Section for Vector Mediators}\/},
  \href{http://dx.doi.org/10.1007/JHEP07(2012)123}{JHEP {\bf 07} (2012)  123},
\href{http://arxiv.org/abs/1204.3839}{{\tt arXiv:1204.3839 [hep-ph]}}.
%%CITATION = ARXIV:1204.3839;%%.

\bibitem{Fairbairn:2014aqa}
M.~Fairbairn and J.~Heal, {\em {Complementarity of dark matter searches at
  resonance}\/},  \href{http://dx.doi.org/10.1103/PhysRevD.90.115019}{Phys.
  Rev. {\bf D90} (2014) no.~11, 115019},
\href{http://arxiv.org/abs/1406.3288}{{\tt arXiv:1406.3288 [hep-ph]}}.
%%CITATION = ARXIV:1406.3288;%%.

\bibitem{Kahlhoefer:2015bea}
F.~Kahlhoefer, K.~Schmidt-Hoberg, T.~Schwetz, and S.~Vogl, {\em {Implications
  of unitarity and gauge invariance for simplified dark matter models}\/},
\href{http://arxiv.org/abs/1510.02110}{{\tt arXiv:1510.02110 [hep-ph]}}.
%%CITATION = ARXIV:1510.02110;%%.

\bibitem{Xiang:2015lfa}
Q.-F. Xiang, X.-J. Bi, P.-F. Yin, and Z.-H. Yu, {\em {Searches for dark matter
  signals in simplified models at future hadron colliders}\/},
  \href{http://dx.doi.org/10.1103/PhysRevD.91.095020}{Phys. Rev. {\bf D91}
  (2015)  095020},
\href{http://arxiv.org/abs/1503.02931}{{\tt arXiv:1503.02931 [hep-ph]}}.
%%CITATION = ARXIV:1503.02931;%%.

\bibitem{Aad:2015zva}
{ATLAS Collaboration}, G.~Aad et al., {\em {Search for new phenomena in final
  states with an energetic jet and large missing transverse momentum in pp
  collisions at $\sqrt{s}=$8 TeV with the ATLAS detector}\/},
  \href{http://dx.doi.org/10.1140/epjc/s10052-015-3517-3,
  10.1140/epjc/s10052-015-3639-7}{Eur. Phys. J. {\bf C75} (2015) no.~7, 299},
  \href{http://arxiv.org/abs/1502.01518}{{\tt arXiv:1502.01518 [hep-ex]}}.
[Erratum: Eur. Phys. J.C75,no.9,408(2015)].
%%CITATION = ARXIV:1502.01518;%%.

\bibitem{Monojets}
{\em {Sensitivity to WIMP Dark Matter in the Final States Containing Jets and
  Missing Transverse Momentum with the ATLAS Detector at 14 TeV LHC}\/},
  ATL-PHYS-PUB-2014-007, CERN, Geneva, Jun, 2014.
\newblock \url{https://cds.cern.ch/record/1708859}.

\bibitem{CMS:kxa}
{CMS Collaboration}, {\em {Search for Narrow Resonances using the Dijet Mass
  Spectrum with 19.6 fb$^{-1}$ of pp Collisions at $\sqrt{s}$=8 TeV}\/},
  CMS-PAS-EXO-12-059, CERN, Geneva, 2013.
\newblock \url{https://cds.cern.ch/record/1519066}.

\bibitem{Yu:2013wta}
F.~Yu, {\em {Di-jet resonances at future hadron colliders: A Snowmass
  whitepaper}\/},
\href{http://arxiv.org/abs/1308.1077}{{\tt arXiv:1308.1077 [hep-ph]}}.
%%CITATION = ARXIV:1308.1077;%%.

\bibitem{Conte:2014zja}
E.~Conte, B.~Dumont, B.~Fuks, and C.~Wymant, {\em {Designing and recasting LHC
  analyses with MadAnalysis 5}\/},
  \href{http://dx.doi.org/10.1140/epjc/s10052-014-3103-0}{Eur. Phys. J. {\bf
  C74} (2014) no.~10, 3103},
\href{http://arxiv.org/abs/1405.3982}{{\tt arXiv:1405.3982 [hep-ph]}}.
%%CITATION = ARXIV:1405.3982;%%.

\bibitem{Belanger:2014vza}
G.~Bélanger, F.~Boudjema, A.~Pukhov, and A.~Semenov, {\em {micrOMEGAs4.1: two
  dark matter candidates}\/},
  \href{http://dx.doi.org/10.1016/j.cpc.2015.03.003}{Comput. Phys. Commun. {\bf
  192} (2015)  322--329},
\href{http://arxiv.org/abs/1407.6129}{{\tt arXiv:1407.6129 [hep-ph]}}.
%%CITATION = ARXIV:1407.6129;%%.

\bibitem{Alves:2014cda}
D.~S.~M. Alves, J.~Galloway, J.~T. Ruderman, and J.~R. Walsh, {\em {Running
  Electroweak Couplings as a Probe of New Physics}\/},
  \href{http://dx.doi.org/10.1007/JHEP02(2015)007}{JHEP {\bf 02} (2015)  007},
\href{http://arxiv.org/abs/1410.6810}{{\tt arXiv:1410.6810 [hep-ph]}}.
%%CITATION = ARXIV:1410.6810;%%.

\bibitem{Curtin:2013fra}
D.~Curtin, R.~Essig, S.~Gori, P.~Jaiswal, A.~Katz, et al., {\em {Exotic Decays
  of the 125 GeV Higgs Boson}\/},
\href{http://arxiv.org/abs/1312.4992}{{\tt arXiv:1312.4992 [hep-ph]}}.
%%CITATION = ARXIV:1312.4992;%%.

\bibitem{Holdom:1985ag}
B.~Holdom, {\em {Two U(1)'s and Epsilon Charge Shifts}\/},
\href{http://dx.doi.org/10.1016/0370-2693(86)91377-8}{Phys.Lett. {\bf B166}
  (1986)  196}.
%%CITATION = PHLTA,B166,196;%%.

\bibitem{Galison:1983pa}
P.~Galison and A.~Manohar, {\em {Two Z's or Not Two Z's?}\/},
\href{http://dx.doi.org/10.1016/0370-2693(84)91161-4}{Phys.Lett. {\bf B136}
  (1984)  279}.
%%CITATION = PHLTA,B136,279;%%.

\bibitem{Dienes:1996zr}
K.~R. Dienes, C.~F. Kolda, and J.~March-Russell, {\em {Kinetic mixing and the
  supersymmetric gauge hierarchy}\/},
  \href{http://dx.doi.org/10.1016/S0550-3213(97)00173-9}{Nucl.Phys. {\bf B492}
  (1997)  104--118},
\href{http://arxiv.org/abs/hep-ph/9610479}{{\tt arXiv:hep-ph/9610479
  [hep-ph]}}.
%%CITATION = HEP-PH/9610479;%%.

\bibitem{Curtin:2014cca}
D.~Curtin, R.~Essig, S.~Gori, and J.~Shelton, {\em {Illuminating Dark Photons
  with High-Energy Colliders}\/},
  \href{http://dx.doi.org/10.1007/JHEP02(2015)157}{JHEP {\bf 1502} (2015)
  157},
\href{http://arxiv.org/abs/1412.0018}{{\tt arXiv:1412.0018 [hep-ph]}}.
%%CITATION = ARXIV:1412.0018;%%.

\bibitem{Bjorken:2009mm}
J.~D. Bjorken, R.~Essig, P.~Schuster, and N.~Toro, {\em {New Fixed-Target
  Experiments to Search for Dark Gauge Forces}\/},
\href{http://dx.doi.org/10.1103/PhysRevD.80.075018}{Phys. Rev. {\bf D80} (2009)
   075018}.
%%CITATION = 0906.0580;%%.

\bibitem{Batell:2009yf}
B.~Batell, M.~Pospelov, and A.~Ritz, {\em {Probing a Secluded U(1) at
  B-factories}\/},
\href{http://dx.doi.org/10.1103/PhysRevD.79.115008}{Phys. Rev. {\bf D79} (2009)
   115008}.
%%CITATION = 0903.0363;%%.

\bibitem{Essig:2009nc}
R.~Essig, P.~Schuster, and N.~Toro, {\em {Probing Dark Forces and Light Hidden
  Sectors at Low-Energy e+e- Colliders}\/},
  \href{http://dx.doi.org/10.1103/PhysRevD.80.015003}{Phys. Rev. {\bf D80}
  (2009)  015003},
\href{http://arxiv.org/abs/0903.3941}{{\tt arXiv:0903.3941 [hep-ph]}}.
%%CITATION = 0903.3941;%%.

\bibitem{Freytsis:2009bh}
M.~Freytsis, G.~Ovanesyan, and J.~Thaler, {\em {Dark Force Detection in Low
  Energy e-p Collisions}\/},
  \href{http://dx.doi.org/10.1007/JHEP01(2010)111}{JHEP {\bf 1001} (2010)
  111},
\href{http://arxiv.org/abs/0909.2862}{{\tt arXiv:0909.2862 [hep-ph]}}.
%%CITATION = ARXIV:0909.2862;%%.

\bibitem{Essig:2010xa}
R.~Essig, P.~Schuster, N.~Toro, and B.~Wojtsekhowski, {\em {An Electron Fixed
  Target Experiment to Search for a New Vector Boson A' Decaying to e+e-}\/},
  \href{http://dx.doi.org/10.1007/JHEP02(2011)009}{JHEP {\bf 1102} (2011)
  009},
\href{http://arxiv.org/abs/1001.2557}{{\tt arXiv:1001.2557 [hep-ph]}}.
%%CITATION = ARXIV:1001.2557;%%.

\bibitem{Blumlein:2011mv}
J.~Blumlein and J.~Brunner, {\em {New Exclusion Limits for Dark Gauge Forces
  from Beam-Dump Data}\/},
  \href{http://dx.doi.org/10.1016/j.physletb.2011.05.046}{Phys.Lett. {\bf B701}
  (2011)  155--159},
\href{http://arxiv.org/abs/1104.2747}{{\tt arXiv:1104.2747 [hep-ex]}}.
%%CITATION = ARXIV:1104.2747;%%.

\bibitem{Andreas:2012mt}
S.~Andreas, C.~Niebuhr, and A.~Ringwald, {\em {New Limits on Hidden Photons
  from Past Electron Beam Dumps}\/},
  \href{http://dx.doi.org/10.1103/PhysRevD.86.095019}{Phys.Rev. {\bf D86}
  (2012)  095019},
\href{http://arxiv.org/abs/1209.6083}{{\tt arXiv:1209.6083 [hep-ph]}}.
%%CITATION = ARXIV:1209.6083;%%.

\bibitem{Pospelov:2008zw}
M.~Pospelov, {\em {Secluded U(1) below the weak scale}\/},
  \href{http://dx.doi.org/10.1103/PhysRevD.80.095002}{Phys.Rev. {\bf D80}
  (2009)  095002},
\href{http://arxiv.org/abs/0811.1030}{{\tt arXiv:0811.1030 [hep-ph]}}.
%%CITATION = ARXIV:0811.1030;%%.

\bibitem{Reece:2009un}
M.~Reece and L.-T. Wang, {\em {Searching for the light dark gauge boson in
  GeV-scale experiments}\/},
\href{http://dx.doi.org/10.1088/1126-6708/2009/07/051}{JHEP {\bf 07} (2009)
  051}.
%%CITATION = 0904.1743;%%.

\bibitem{Aubert:2009cp}
{BaBar Collaboration}, B.~Aubert et al., {\em {Search for Dimuon Decays of a
  Light Scalar Boson in Radiative Transitions $\Upsilon \rightarrow \gamma
  A^0$}\/},
  \href{http://dx.doi.org/10.1103/PhysRevLett.103.081803}{Phys.Rev.Lett. {\bf
  103} (2009)  081803},
\href{http://arxiv.org/abs/0905.4539}{{\tt arXiv:0905.4539 [hep-ex]}}.
%%CITATION = ARXIV:0905.4539;%%.

\bibitem{Hook:2010tw}
A.~Hook, E.~Izaguirre, and J.~G. Wacker, {\em {Model Independent Bounds on
  Kinetic Mixing}\/},  \href{http://dx.doi.org/10.1155/2011/859762}{Adv.High
  Energy Phys. {\bf 2011} (2011)  859762},
\href{http://arxiv.org/abs/1006.0973}{{\tt arXiv:1006.0973 [hep-ph]}}.
%%CITATION = ARXIV:1006.0973;%%.

\bibitem{Bjorken:1988as}
J.~D. Bjorken et al., {\em {Search for Neutral Metastable Penetrating Particles
  Produced in the SLAC Beam Dump}\/},
\href{http://dx.doi.org/10.1103/PhysRevD.38.3375}{Phys. Rev. {\bf D38} (1988)
  3375}.
%%CITATION = PHRVA,D38,3375;%%.

\bibitem{Riordan:1987aw}
E.~M. Riordan et al., {\em {A Search for Short Lived Axions in an Electron Beam
  Dump Experiment}\/},
\href{http://dx.doi.org/10.1103/PhysRevLett.59.755}{Phys. Rev. Lett. {\bf 59}
  (1987)  755}.
%%CITATION = PRLTA,59,755;%%.

\bibitem{Bross:1989mp}
A.~Bross et al., {\em {A Search for Shortlived Particles Produced in an
  Electron Beam Dump}\/},
\href{http://dx.doi.org/10.1103/PhysRevLett.67.2942}{Phys. Rev. Lett. {\bf 67}
  (1991)  2942--2945}.
%%CITATION = PRLTA,67,2942;%%.

\bibitem{Babusci:2012cr}
{KLOE-2 Collaboration}, D.~Babusci et al., {\em {Limit on the production of a
  light vector gauge boson in phi meson decays with the KLOE detector}\/},
  \href{http://dx.doi.org/10.1016/j.physletb.2013.01.067}{Phys.Lett. {\bf B720}
  (2013)  111--115},
\href{http://arxiv.org/abs/1210.3927}{{\tt arXiv:1210.3927 [hep-ex]}}.
%%CITATION = ARXIV:1210.3927;%%.

\bibitem{Archilli:2011zc}
F.~Archilli, D.~Babusci, D.~Badoni, I.~Balwierz, G.~Bencivenni, et al., {\em
  {Search for a vector gauge boson in phi meson decays with the KLOE
  detector}\/},
  \href{http://dx.doi.org/10.1016/j.physletb.2011.11.033}{Phys.Lett. {\bf B706}
  (2012)  251--255},
\href{http://arxiv.org/abs/1110.0411}{{\tt arXiv:1110.0411 [hep-ex]}}.
%%CITATION = ARXIV:1110.0411;%%.

\bibitem{Abrahamyan:2011gv}
{APEX Collaboration}, S.~Abrahamyan et al., {\em {Search for a new gauge boson
  in the $A'$ Experiment (APEX)}\/},
  \href{http://dx.doi.org/10.1103/PhysRevLett.107.191804}{Phys. Rev. Lett. {\bf
  107} (2011)  191804},
\href{http://arxiv.org/abs/1108.2750}{{\tt arXiv:1108.2750 [hep-ex]}}.
%%CITATION = 1108.2750;%%.

\bibitem{Merkel:2011ze}
{A1 Collaboration}, H.~Merkel et al., {\em {Search for Light Gauge Bosons of
  the Dark Sector at the Mainz Microtron}\/},
Phys. Rev. Lett. {\bf 106} (2011)  251802.
%%CITATION = 1101.4091;%%.

\bibitem{Dent:2012mx}
J.~B. Dent, F.~Ferrer, and L.~M. Krauss, {\em {Constraints on Light Hidden
  Sector Gauge Bosons from Supernova Cooling}\/},
\href{http://arxiv.org/abs/1201.2683}{{\tt arXiv:1201.2683 [astro-ph.CO]}}.
%%CITATION = ARXIV:1201.2683;%%.

\bibitem{Davoudiasl:2012ig}
H.~Davoudiasl, H.-S. Lee, and W.~J. Marciano, {\em {Dark Side of Higgs Diphoton
  Decays and Muon g-2}\/},
  \href{http://dx.doi.org/10.1103/PhysRevD.86.095009}{Phys.Rev. {\bf D86}
  (2012)  095009},
\href{http://arxiv.org/abs/1208.2973}{{\tt arXiv:1208.2973 [hep-ph]}}.
%%CITATION = ARXIV:1208.2973;%%.

\bibitem{Davoudiasl:2012ag}
H.~Davoudiasl, H.-S. Lee, and W.~J. Marciano, {\em {'Dark' Z implications for
  Parity Violation, Rare Meson Decays, and Higgs Physics}\/},
  \href{http://dx.doi.org/10.1103/PhysRevD.85.115019}{Phys.Rev. {\bf D85}
  (2012)  115019},
\href{http://arxiv.org/abs/1203.2947}{{\tt arXiv:1203.2947 [hep-ph]}}.
%\%CITATION = ARXIV:1203.2947;%%.

\bibitem{Davoudiasl:2013aya}
H.~Davoudiasl, H.-S. Lee, I.~Lewis, and W.~J. Marciano, {\em {Higgs Decays as a
  Window into the Dark Sector}\/},
\href{http://arxiv.org/abs/1304.4935}{{\tt arXiv:1304.4935 [hep-ph]}}.
%\%CITATION = ARXIV:1304.4935;%%.

\bibitem{Endo:2012hp}
M.~Endo, K.~Hamaguchi, and G.~Mishima, {\em {Constraints on Hidden Photon
  Models from Electron g-2 and Hydrogen Spectroscopy}\/},
  \href{http://dx.doi.org/10.1103/PhysRevD.86.095029}{Phys.Rev. {\bf D86}
  (2012)  095029},
\href{http://arxiv.org/abs/1209.2558}{{\tt arXiv:1209.2558 [hep-ph]}}.
%%CITATION = ARXIV:1209.2558;%%.

\bibitem{Balewski:2013oza}
J.~Balewski, J.~Bernauer, W.~Bertozzi, J.~Bessuille, B.~Buck, et al., {\em
  {DarkLight: A Search for Dark Forces at the Jefferson Laboratory
  Free-Electron Laser Facility}\/},
\href{http://arxiv.org/abs/1307.4432}{{\tt arXiv:1307.4432}}.
%%CITATION = ARXIV:1307.4432;%%.

\bibitem{Adlarson:2013eza}
{WASA-at-COSY Collaboration}, P.~Adlarson et al., {\em {Search for a dark
  photon in the $\pi^0 \to e^+e^-\gamma$ decay}\/},
  \href{http://dx.doi.org/10.1016/j.physletb.2013.08.055}{Phys.Lett. {\bf B726}
  (2013)  187--193},
\href{http://arxiv.org/abs/1304.0671}{{\tt arXiv:1304.0671 [hep-ex]}}.
%%CITATION = ARXIV:1304.0671;%%.

\bibitem{Agakishiev:2013fwl}
{HADES Collaboration}, G.~Agakishiev et al., {\em {Searching a Dark Photon with
  HADES}\/},
  \href{http://dx.doi.org/10.1016/j.physletb.2014.02.035}{Phys.Lett. {\bf B731}
  (2014)  265--271},
\href{http://arxiv.org/abs/1311.0216}{{\tt arXiv:1311.0216 [hep-ex]}}.
%%CITATION = ARXIV:1311.0216;%%.

\bibitem{Blumlein:2013cua}
J.~Blümlein and J.~Brunner, {\em {New Exclusion Limits on Dark Gauge Forces
  from Proton Bremsstrahlung in Beam-Dump Data}\/},
  \href{http://dx.doi.org/10.1016/j.physletb.2014.02.029}{Phys.Lett. {\bf B731}
  (2014)  320--326},
\href{http://arxiv.org/abs/1311.3870}{{\tt arXiv:1311.3870 [hep-ph]}}.
%%CITATION = ARXIV:1311.3870;%%.

\bibitem{Andreas:2013lya}
S.~Andreas, S.~Donskov, P.~Crivelli, A.~Gardikiotis, S.~Gninenko, et al., {\em
  {Proposal for an Experiment to Search for Light Dark Matter at the SPS}\/},
\href{http://arxiv.org/abs/1312.3309}{{\tt arXiv:1312.3309 [hep-ex]}}.
%%CITATION = ARXIV:1312.3309;%%.

\bibitem{Battaglieri:2014hga}
M.~Battaglieri, S.~Boyarinov, S.~Bueltmann, V.~Burkert, A.~Celentano, et al.,
  {\em {The Heavy Photon Search Test Detector}\/},
\href{http://arxiv.org/abs/1406.6115}{{\tt arXiv:1406.6115 [physics.ins-det]}}.
%%CITATION = ARXIV:1406.6115;%%.

\bibitem{Merkel:2014avp}
H.~Merkel, P.~Achenbach, C.~A. Gayoso, T.~Beranek, J.~Bericic, et al., {\em
  {Search for light massive gauge bosons as an explanation of the $(g-2)_\mu$
  anomaly at MAMI}\/},
\href{http://arxiv.org/abs/1404.5502}{{\tt arXiv:1404.5502 [hep-ex]}}.
%%CITATION = ARXIV:1404.5502;%%.

\bibitem{Lees:2014xha}
{BaBar Collaboration}, J.~Lees et al., {\em {Search for a dark photon in e+e-
  collisions at BABAR}\/},
\href{http://arxiv.org/abs/1406.2980}{{\tt arXiv:1406.2980 [hep-ex]}}.
%%CITATION = ARXIV:1406.2980;%%.

\bibitem{Adare:2014ega}
A.~Adare, S.~Afanasiev, C.~Aidala, N.~Ajitanand, Y.~Akiba, et al., {\em
  {Closing the Door for Dark Photons as the Explanation for the Muon g-2
  Anomaly}\/},
\href{http://arxiv.org/abs/1409.0851}{{\tt arXiv:1409.0851 [nucl-ex]}}.
%%CITATION = ARXIV:1409.0851;%%.

\bibitem{Kazanas:2014mca}
D.~Kazanas, R.~N. Mohapatra, S.~Nussinov, V.~Teplitz, and Y.~Zhang, {\em
  {Supernova Bounds on the Dark Photon Using its Electromagnetic Decay}\/},
\href{http://arxiv.org/abs/1410.0221}{{\tt arXiv:1410.0221 [hep-ph]}}.
%%CITATION = ARXIV:1410.0221;%%.

\bibitem{Echenard:2014lma}
B.~Echenard, R.~Essig, and Y.-M. Zhong, {\em {Projections for Dark Photon
  Searches at Mu3e}\/},
\href{http://arxiv.org/abs/1411.1770}{{\tt arXiv:1411.1770 [hep-ph]}}.
%%CITATION = ARXIV:1411.1770;%%.

\bibitem{Gorbunov:2014wqa}
D.~Gorbunov, A.~Makarov, and I.~Timiryasov, {\em {Decaying light particles on
  board the SHiP (I): Signal rate estimates for hidden photons}\/},
\href{http://arxiv.org/abs/1411.4007}{{\tt arXiv:1411.4007 [hep-ph]}}.
%%CITATION = ARXIV:1411.4007;%%.

\bibitem{NA482}
E.~Goudzovski, {\em {Search for the dark photon in $\pi^0$ decays by NA48/2 at
  CERN , MesonNet workshop, LNF, Frascati, Sept.~2014}\/}, .

\bibitem{Gopalakrishna:2008dv}
S.~Gopalakrishna, S.~Jung, and J.~D. Wells, {\em {Higgs boson decays to four
  fermions through an abelian hidden sector}\/},
  \href{http://dx.doi.org/10.1103/PhysRevD.78.055002}{Phys.Rev. {\bf D78}
  (2008)  055002},
\href{http://arxiv.org/abs/0801.3456}{{\tt arXiv:0801.3456 [hep-ph]}}.
%%CITATION = ARXIV:0801.3456;%%.

\bibitem{Chang:2013lfa}
C.-F. Chang, E.~Ma, and T.-C. Yuan, {\em {Multilepton Higgs Decays through the
  Dark Portal}\/},
\href{http://arxiv.org/abs/1308.6071}{{\tt arXiv:1308.6071 [hep-ph]}}.
%%CITATION = ARXIV:1308.6071;%%.

\bibitem{Falkowski:2014ffa}
A.~Falkowski and R.~Vega-Morales, {\em {Exotic Higgs decays in the golden
  channel}\/},
\href{http://arxiv.org/abs/1405.1095}{{\tt arXiv:1405.1095 [hep-ph]}}.
%%CITATION = ARXIV:1405.1095;%%.

\bibitem{Cline:2014dwa}
J.~M. Cline, G.~Dupuis, Z.~Liu, and W.~Xue, {\em {The windows for kinetically
  mixed Z'-mediated dark matter and the galactic center gamma ray excess}\/},
\href{http://arxiv.org/abs/1405.7691}{{\tt arXiv:1405.7691 [hep-ph]}}.
%%CITATION = ARXIV:1405.7691;%%.

\bibitem{Hoenig:2014dsa}
I.~Hoenig, G.~Samach, and D.~Tucker-Smith, {\em {Searching for dilepton
  resonances below the Z mass at the LHC}\/},
\href{http://arxiv.org/abs/1408.1075}{{\tt arXiv:1408.1075 [hep-ph]}}.
%%CITATION = ARXIV:1408.1075;%%.

\bibitem{ArkaniHamed:2008qp}
N.~Arkani-Hamed and N.~Weiner, {\em {LHC Signals for a SuperUnified Theory of
  Dark Matter}\/},  \href{http://dx.doi.org/10.1088/1126-6708/2008/12/104}{JHEP
  {\bf 0812} (2008)  104},
\href{http://arxiv.org/abs/0810.0714}{{\tt arXiv:0810.0714 [hep-ph]}}.
%%CITATION = ARXIV:0810.0714;%%.

\bibitem{Cheung:2009fk}
C.~Cheung, J.~T. Ruderman, L.-T. Wang, and I.~Yavin, {\em {Kinetic Mixing as
  the Origin of Light Dark Scales}\/},
  \href{http://arxiv.org/abs/0902.3246}{{\tt 0902.3246}}.
  \url{http://arxiv.org/abs/0902.3246}.

\bibitem{Baumgart:2009tn}
M.~Baumgart, C.~Cheung, J.~T. Ruderman, L.-T. Wang, and I.~Yavin, {\em
  {Non-Abelian Dark Sectors and Their Collider Signatures}\/},
  \href{http://dx.doi.org/10.1088/1126-6708/2009/04/014}{JHEP {\bf 0904} (2009)
   014},
\href{http://arxiv.org/abs/0901.0283}{{\tt arXiv:0901.0283 [hep-ph]}}.
%%CITATION = ARXIV:0901.0283;%%.

\bibitem{Morrissey:2009ur}
D.~E. Morrissey, D.~Poland, and K.~M. Zurek, {\em {Abelian Hidden Sectors at a
  GeV}\/},  \href{http://dx.doi.org/10.1088/1126-6708/2009/07/050}{JHEP {\bf
  0907} (2009)  050},
\href{http://arxiv.org/abs/0904.2567}{{\tt arXiv:0904.2567 [hep-ph]}}.
%%CITATION = ARXIV:0904.2567;%%.

\bibitem{CMS:xwa}
{CMS Collaboration}, {\em {Properties of the Higgs-like boson in the decay H to
  ZZ to 4l in pp collisions at sqrt s =7 and 8 TeV}\/},   CMS-PAS-HIG-13-002,
  CERN, Geneva, 2013.
\newblock \url{https://cds.cern.ch/record/1523767}.

\bibitem{ATLAS:2013gma}
{ATLAS Collaboration}, {\em {Measurement of the total ZZ production cross
  section in proton-proton collisions $\sqrt{s}$ = 8 TeV in 20 fb$^{-1}$ with
  the ATLAS detector}\/},   ATLAS-CONF-2013-020, CERN, Geneva, Mar, 2013.
\newblock \url{https://cds.cern.ch/record/1525555}.

\bibitem{CMS:2013lea}
{CMS Collaboration}, {\em {Search for a non-standard-model Higgs boson decaying
  to a pair of new light bosons in four-muon final states}\/},
  CMS-PAS-HIG-13-010, CERN, Geneva, 2013.
\newblock \url{https://cds.cern.ch/record/1563546}.

\bibitem{Craig:2015pha}
N.~Craig, A.~Katz, M.~Strassler, and R.~Sundrum, {\em {Naturalness in the Dark
  at the LHC}\/},
\href{http://arxiv.org/abs/1501.05310}{{\tt arXiv:1501.05310 [hep-ph]}}.
%%CITATION = ARXIV:1501.05310;%%.

\bibitem{Curtin:2015fna}
D.~Curtin and C.~B. Verhaaren, {\em {Discovering Uncolored Naturalness in
  Exotic Higgs Decays}\/},
\href{http://arxiv.org/abs/1506.06141}{{\tt arXiv:1506.06141 [hep-ph]}}.
%%CITATION = ARXIV:1506.06141;%%.

\bibitem{Strassler:2006im}
M.~J. Strassler and K.~M. Zurek, {\em {Echoes of a hidden valley at hadron
  colliders}\/},
  \href{http://dx.doi.org/10.1016/j.physletb.2007.06.055}{Phys.Lett. {\bf B651}
  (2007)  374--379},
\href{http://arxiv.org/abs/hep-ph/0604261}{{\tt arXiv:hep-ph/0604261
  [hep-ph]}}.
%%CITATION = HEP-PH/0604261;%%.

\bibitem{Strassler:2006ri}
M.~J. Strassler and K.~M. Zurek, {\em {Discovering the Higgs through
  highly-displaced vertices}\/},
  \href{http://dx.doi.org/10.1016/j.physletb.2008.02.008}{Phys.Lett. {\bf B661}
  (2008)  263--267},
\href{http://arxiv.org/abs/hep-ph/0605193}{{\tt arXiv:hep-ph/0605193
  [hep-ph]}}.
%%CITATION = HEP-PH/0605193;%%.

\bibitem{Strassler:2006qa}
M.~J. Strassler, {\em {Possible effects of a hidden valley on supersymmetric
  phenomenology}\/},
\href{http://arxiv.org/abs/hep-ph/0607160}{{\tt arXiv:hep-ph/0607160
  [hep-ph]}}.
%%CITATION = HEP-PH/0607160;%%.

\bibitem{Han:2007ae}
T.~Han, Z.~Si, K.~M. Zurek, and M.~J. Strassler, {\em {Phenomenology of hidden
  valleys at hadron colliders}\/},
  \href{http://dx.doi.org/10.1088/1126-6708/2008/07/008}{JHEP {\bf 0807} (2008)
   008},
\href{http://arxiv.org/abs/0712.2041}{{\tt arXiv:0712.2041 [hep-ph]}}.
%%CITATION = ARXIV:0712.2041;%%.

\bibitem{CMS:2014hka}
{CMS Collaboration}, V.~Khachatryan et al., {\em {Search for long-lived
  particles that decay into final states containing two electrons or two muons
  in proton-proton collisions at $\sqrt{s} =$ 8 TeV}\/},
  \href{http://dx.doi.org/10.1103/PhysRevD.91.052012}{Phys. Rev. {\bf D91}
  (2015) no.~5, 052012},
\href{http://arxiv.org/abs/1411.6977}{{\tt arXiv:1411.6977 [hep-ex]}}.
%%CITATION = ARXIV:1411.6977;%%.

\bibitem{Aad:2014yea}
{ATLAS Collaboration}, G.~Aad et al., {\em {Search for long-lived neutral
  particles decaying into lepton jets in proton-proton collisions at
  $\sqrt{s}=8$ TeV with the ATLAS detector}\/},
  \href{http://dx.doi.org/10.1007/JHEP11(2014)088}{JHEP {\bf 11} (2014)  088},
\href{http://arxiv.org/abs/1409.0746}{{\tt arXiv:1409.0746 [hep-ex]}}.
%%CITATION = ARXIV:1409.0746;%%.

\bibitem{deSimone:2014pda}
A.~De~Simone, G.~F. Giudice, and A.~Strumia, {\em {Benchmarks for Dark Matter
  Searches at the LHC}\/},
  \href{http://dx.doi.org/10.1007/JHEP06(2014)081}{JHEP {\bf 1406} (2014)
  081},
\href{http://arxiv.org/abs/1402.6287}{{\tt arXiv:1402.6287 [hep-ph]}}.
%%CITATION = ARXIV:1402.6287;%%.

\bibitem{Harigaya:2014dwa}
K.~Harigaya, K.~Kaneta, and S.~Matsumoto, {\em {Gaugino coannihilations}\/},
  \href{http://dx.doi.org/10.1103/PhysRevD.89.115021}{Phys.Rev. {\bf D89}
  (2014)  115021},
\href{http://arxiv.org/abs/1403.0715}{{\tt arXiv:1403.0715 [hep-ph]}}.
%%CITATION = ARXIV:1403.0715;%%.

\bibitem{Arbey:2013iza}
A.~Arbey, M.~Battaglia, and F.~Mahmoudi, {\em {Combining monojet,
  supersymmetry, and dark matter searches}\/},
  \href{http://dx.doi.org/10.1103/PhysRevD.89.077701}{Phys. Rev. {\bf D89}
  (2014) no.~7, 077701},
\href{http://arxiv.org/abs/1311.7641}{{\tt arXiv:1311.7641 [hep-ph]}}.
%%CITATION = ARXIV:1311.7641;%%.

\bibitem{Cahill-Rowley:2014twa}
M.~Cahill-Rowley, J.~L. Hewett, A.~Ismail, and T.~G. Rizzo, {\em {Lessons and
  prospects from the pMSSM after LHC Run I}\/},
  \href{http://dx.doi.org/10.1103/PhysRevD.91.055002}{Phys. Rev. {\bf D91}
  (2015) no.~5, 055002},
\href{http://arxiv.org/abs/1407.4130}{{\tt arXiv:1407.4130 [hep-ph]}}.
%%CITATION = ARXIV:1407.4130;%%.

\bibitem{Aad:2015baa}
{ATLAS Collaboration}, G.~Aad et al., {\em {Summary of the ATLAS experiment's
  sensitivity to supersymmetry after LHC Run 1 - interpreted in the
  phenomenological MSSM}\/},
  \href{http://dx.doi.org/10.1007/JHEP10(2015)134}{JHEP {\bf 10} (2015)  134},
\href{http://arxiv.org/abs/1508.06608}{{\tt arXiv:1508.06608 [hep-ex]}}.
%%CITATION = ARXIV:1508.06608;%%.

\bibitem{CMS:2015ebf}
{CMS Collaboration}, {\em {Phenomenological MSSM interpretation of CMS results
  at $\sqrt{s}=$ 7 and 8 TeV}\/},   CMS-PAS-SUS-15-010, CERN, Geneva, 2015.
\newblock \url{https://cds.cern.ch/record/2063744}.

\bibitem{Arvey:2015nra}
A.~Arbey, M.~Battaglia, L.~Covi, J.~Hasenkamp, and F.~Mahmoudi, {\em {LHC
  constraints on Gravitino Dark Matter}\/},
  \href{http://dx.doi.org/10.1103/PhysRevD.92.115008}{Phys. Rev. {\bf D92}
  (2015) no.~11, 115008},
\href{http://arxiv.org/abs/1505.04595}{{\tt arXiv:1505.04595 [hep-ph]}}.
%%CITATION = ARXIV:1505.04595;%%.

\bibitem{Servant:2013uwa}
G.~Servant and S.~Tulin, {\em {Baryogenesis and Dark Matter through a Higgs
  Asymmetry}\/},  \href{http://dx.doi.org/10.1103/PhysRevLett.111.151601}{Phys.
  Rev. Lett. {\bf 111} (2013) no.~15, 151601},
\href{http://arxiv.org/abs/1304.3464}{{\tt arXiv:1304.3464 [hep-ph]}}.
%%CITATION = ARXIV:1304.3464;%%.

\bibitem{Buckley:2011kk}
M.~R. Buckley, {\em {Asymmetric Dark Matter and Effective Operators}\/},
  \href{http://dx.doi.org/10.1103/PhysRevD.84.043510}{Phys. Rev. {\bf D84}
  (2011)  043510},
\href{http://arxiv.org/abs/1104.1429}{{\tt arXiv:1104.1429 [hep-ph]}}.
%%CITATION = ARXIV:1104.1429;%%.

\bibitem{MarchRussell:2012hi}
J.~March-Russell, J.~Unwin, and S.~M. West, {\em {Closing in on Asymmetric Dark
  Matter I: Model independent limits for interactions with quarks}\/},
  \href{http://dx.doi.org/10.1007/JHEP08(2012)029}{JHEP {\bf 08} (2012)  029},
\href{http://arxiv.org/abs/1203.4854}{{\tt arXiv:1203.4854 [hep-ph]}}.
%%CITATION = ARXIV:1203.4854;%%.

\bibitem{Lee:1956qn}
T.~D. Lee and C.-N. Yang, {\em {Question of Parity Conservation in Weak
  Interactions}\/},
\href{http://dx.doi.org/10.1103/PhysRev.104.254}{Phys. Rev. {\bf 104} (1956)
  254--258}.
%%CITATION = PHRVA,104,254;%%.

\bibitem{Kobzarev:1966qya}
I.~{\relax Yu}. Kobzarev, L.~B. Okun, and I.~{\relax Ya}. Pomeranchuk, {\em {On
  the possibility of experimental observation of mirror particles}\/},  Sov. J.
  Nucl. Phys. {\bf 3} (1966) no.~6, 837--841.
[Yad. Fiz.3,1154(1966)].
%%CITATION = SJNCA,3,837;%%.

\bibitem{Foot:1991bp}
R.~Foot, H.~Lew, and R.~R. Volkas, {\em {A Model with fundamental improper
  space-time symmetries}\/},
\href{http://dx.doi.org/10.1016/0370-2693(91)91013-L}{Phys. Lett. {\bf B272}
  (1991)  67--70}.
%%CITATION = PHLTA,B272,67;%%.

\bibitem{Berezhiani:1995am}
Z.~G. Berezhiani, A.~D. Dolgov, and R.~N. Mohapatra, {\em {Asymmetric
  inflationary reheating and the nature of mirror universe}\/},
  \href{http://dx.doi.org/10.1016/0370-2693(96)00219-5}{Phys. Lett. {\bf B375}
  (1996)  26--36},
\href{http://arxiv.org/abs/hep-ph/9511221}{{\tt arXiv:hep-ph/9511221
  [hep-ph]}}.
%%CITATION = HEP-PH/9511221;%%.

\bibitem{Kribs:2009fy}
G.~D. Kribs, T.~S. Roy, J.~Terning, and K.~M. Zurek, {\em {Quirky Composite
  Dark Matter}\/},  \href{http://dx.doi.org/10.1103/PhysRevD.81.095001}{Phys.
  Rev. {\bf D81} (2010)  095001},
\href{http://arxiv.org/abs/0909.2034}{{\tt arXiv:0909.2034 [hep-ph]}}.
%%CITATION = ARXIV:0909.2034;%%.

\bibitem{Blennow:2010qp}
M.~Blennow, B.~Dasgupta, E.~Fernandez-Martinez, and N.~Rius, {\em {Aidnogenesis
  via Leptogenesis and Dark Sphalerons}\/},
  \href{http://dx.doi.org/10.1007/JHEP03(2011)014}{JHEP {\bf 03} (2011)  014},
\href{http://arxiv.org/abs/1009.3159}{{\tt arXiv:1009.3159 [hep-ph]}}.
%%CITATION = ARXIV:1009.3159;%%.

\bibitem{Frandsen:2011kt}
M.~T. Frandsen, S.~Sarkar, and K.~Schmidt-Hoberg, {\em {Light asymmetric dark
  matter from new strong dynamics}\/},
  \href{http://dx.doi.org/10.1103/PhysRevD.84.051703}{Phys. Rev. {\bf D84}
  (2011)  051703},
\href{http://arxiv.org/abs/1103.4350}{{\tt arXiv:1103.4350 [hep-ph]}}.
%%CITATION = ARXIV:1103.4350;%%.

\bibitem{Bai:2013xga}
Y.~Bai and P.~Schwaller, {\em {Scale of dark QCD}\/},
  \href{http://dx.doi.org/10.1103/PhysRevD.89.063522}{Phys. Rev. {\bf D89}
  (2014) no.~6, 063522},
\href{http://arxiv.org/abs/1306.4676}{{\tt arXiv:1306.4676 [hep-ph]}}.
%%CITATION = ARXIV:1306.4676;%%.

\bibitem{Strassler:2008fv}
M.~J. Strassler, {\em {On the Phenomenology of Hidden Valleys with Heavy
  Flavor}\/},
\href{http://arxiv.org/abs/0806.2385}{{\tt arXiv:0806.2385 [hep-ph]}}.
%%CITATION = ARXIV:0806.2385;%%.

\bibitem{Schwaller:2015gea}
P.~Schwaller, D.~Stolarski, and A.~Weiler, {\em {Emerging Jets}\/},
  \href{http://dx.doi.org/10.1007/JHEP05(2015)059}{JHEP {\bf 1505} (2015)
  059},
\href{http://arxiv.org/abs/1502.05409}{{\tt arXiv:1502.05409 [hep-ph]}}.
%%CITATION = ARXIV:1502.05409;%%.

\bibitem{Cohen:2015toa}
T.~Cohen, M.~Lisanti, and H.~K. Lou, {\em {Semi-visible Jets: Dark Matter
  Undercover at the LHC}\/},
\href{http://arxiv.org/abs/1503.00009}{{\tt arXiv:1503.00009 [hep-ph]}}.
%%CITATION = ARXIV:1503.00009;%%.

\bibitem{Spergel:1999mh}
D.~N. Spergel and P.~J. Steinhardt, {\em {Observational evidence for
  selfinteracting cold dark matter}\/},
  \href{http://dx.doi.org/10.1103/PhysRevLett.84.3760}{Phys.Rev.Lett. {\bf 84}
  (2000)  3760--3763},
\href{http://arxiv.org/abs/astro-ph/9909386}{{\tt arXiv:astro-ph/9909386
  [astro-ph]}}.
%%CITATION = ASTRO-PH/9909386;%%.

\bibitem{Tulin:2013teo}
S.~Tulin, H.-B. Yu, and K.~M. Zurek, {\em {Beyond Collisionless Dark Matter:
  Particle Physics Dynamics for Dark Matter Halo Structure}\/},
  \href{http://dx.doi.org/10.1103/PhysRevD.87.115007}{Phys.Rev. {\bf D87}
  (2013) no.~11, 115007},
\href{http://arxiv.org/abs/1302.3898}{{\tt arXiv:1302.3898 [hep-ph]}}.
%%CITATION = ARXIV:1302.3898;%%.

\bibitem{Vogelsberger:2014kha}
M.~Vogelsberger, S.~Genel, V.~Springel, P.~Torrey, D.~Sijacki, et al., {\em
  {Properties of galaxies reproduced by a hydrodynamic simulation}\/},
  \href{http://dx.doi.org/10.1038/nature13316}{Nature {\bf 509} (2014)
  177--182},
\href{http://arxiv.org/abs/1405.1418}{{\tt arXiv:1405.1418 [astro-ph.CO]}}.
%%CITATION = ARXIV:1405.1418;%%.

\bibitem{Sawala:2014xka}
T.~Sawala, C.~S. Frenk, A.~Fattahi, J.~F. Navarro, R.~G. Bower, et al., {\em
  {Local Group galaxies emerge from the dark}\/},
\href{http://arxiv.org/abs/1412.2748}{{\tt arXiv:1412.2748 [astro-ph.GA]}}.
%%CITATION = ARXIV:1412.2748;%%.

\bibitem{Hisano:2003ec}
J.~Hisano, S.~Matsumoto, and M.~M. Nojiri, {\em {Explosive dark matter
  annihilation}\/},
  \href{http://dx.doi.org/10.1103/PhysRevLett.92.031303}{Phys. Rev. Lett. {\bf
  92} (2004)  031303},
\href{http://arxiv.org/abs/hep-ph/0307216}{{\tt arXiv:hep-ph/0307216
  [hep-ph]}}.
%%CITATION = HEP-PH/0307216;%%.

\bibitem{Cirelli:2008pk}
M.~Cirelli, M.~Kadastik, M.~Raidal, and A.~Strumia, {\em {Model-independent
  implications of the e+-, anti-proton cosmic ray spectra on properties of Dark
  Matter}\/},  \href{http://dx.doi.org/10.1016/j.nuclphysb.2013.05.002,
  10.1016/j.nuclphysb.2008.11.031}{Nucl.Phys. {\bf B813} (2009)  1--21},
\href{http://arxiv.org/abs/0809.2409}{{\tt arXiv:0809.2409 [hep-ph]}}.
%%CITATION = ARXIV:0809.2409;%%.

\bibitem{ArkaniHamed:2008qn}
N.~Arkani-Hamed, D.~P. Finkbeiner, T.~R. Slatyer, and N.~Weiner, {\em {A Theory
  of Dark Matter}\/},
  \href{http://dx.doi.org/10.1103/PhysRevD.79.015014}{Phys.Rev. {\bf D79}
  (2009)  015014},
\href{http://arxiv.org/abs/0810.0713}{{\tt arXiv:0810.0713 [hep-ph]}}.
%%CITATION = ARXIV:0810.0713;%%.

\bibitem{Shepherd:2009sa}
W.~Shepherd, T.~M.~P. Tait, and G.~Zaharijas, {\em {Bound states of weakly
  interacting dark matter}\/},
  \href{http://dx.doi.org/10.1103/PhysRevD.79.055022}{Phys. Rev. {\bf D79}
  (2009)  055022},
\href{http://arxiv.org/abs/0901.2125}{{\tt arXiv:0901.2125 [hep-ph]}}.
%%CITATION = ARXIV:0901.2125;%%.

\bibitem{Altmannshofer:2014cla}
W.~Altmannshofer, P.~J. Fox, R.~Harnik, G.~D. Kribs, and N.~Raj, {\em {Dark
  Matter Signals in Dilepton Production at Hadron Colliders}\/},
\href{http://arxiv.org/abs/1411.6743}{{\tt arXiv:1411.6743 [hep-ph]}}.
%%CITATION = ARXIV:1411.6743;%%.

\bibitem{Buschmann:2015awa}
M.~Buschmann, J.~Kopp, J.~Liu, and P.~A.~N. Machado, {\em {Lepton Jets from
  Radiating Dark Matter}\/},
  \href{http://dx.doi.org/10.1007/JHEP07(2015)045}{JHEP {\bf 07} (2015)  045},
\href{http://arxiv.org/abs/1505.07459}{{\tt arXiv:1505.07459 [hep-ph]}}.
%%CITATION = ARXIV:1505.07459;%%.

\bibitem{Carloni:2010tw}
L.~Carloni and T.~Sjostrand, {\em {Visible Effects of Invisible Hidden Valley
  Radiation}\/},  \href{http://dx.doi.org/10.1007/JHEP09(2010)105}{JHEP {\bf
  1009} (2010)  105},
\href{http://arxiv.org/abs/1006.2911}{{\tt arXiv:1006.2911 [hep-ph]}}.
%%CITATION = ARXIV:1006.2911;%%.

\bibitem{Carloni:2011kk}
L.~Carloni, J.~Rathsman, and T.~Sjostrand, {\em {Discerning Secluded Sector
  gauge structures}\/},  \href{http://dx.doi.org/10.1007/JHEP04(2011)091}{JHEP
  {\bf 1104} (2011)  091},
\href{http://arxiv.org/abs/1102.3795}{{\tt arXiv:1102.3795 [hep-ph]}}.
%%CITATION = ARXIV:1102.3795;%%.

\bibitem{Cheung:2009su}
C.~Cheung, J.~T. Ruderman, L.-T. Wang, and I.~Yavin, {\em {Lepton Jets in
  (Supersymmetric) Electroweak Processes}\/},
  \href{http://dx.doi.org/10.1007/JHEP04(2010)116}{JHEP {\bf 1004} (2010)
  116},
\href{http://arxiv.org/abs/0909.0290}{{\tt arXiv:0909.0290 [hep-ph]}}.
%%CITATION = ARXIV:0909.0290;%%.

\bibitem{Katz:2009qq}
A.~Katz and R.~Sundrum, {\em {Breaking the Dark Force}\/},
  \href{http://dx.doi.org/10.1088/1126-6708/2009/06/003}{JHEP {\bf 0906} (2009)
   003},
\href{http://arxiv.org/abs/0902.3271}{{\tt arXiv:0902.3271 [hep-ph]}}.
%%CITATION = ARXIV:0902.3271;%%.

\bibitem{Bai:2009it}
Y.~Bai and Z.~Han, {\em {Measuring the Dark Force at the LHC}\/},
  \href{http://dx.doi.org/10.1103/PhysRevLett.103.051801}{Phys.Rev.Lett. {\bf
  103} (2009)  051801},
\href{http://arxiv.org/abs/0902.0006}{{\tt arXiv:0902.0006 [hep-ph]}}.
%%CITATION = ARXIV:0902.0006;%%.

\bibitem{Chan:2011aa}
Y.~F. Chan, M.~Low, D.~E. Morrissey, and A.~P. Spray, {\em {LHC Signatures of a
  Minimal Supersymmetric Hidden Valley}\/},
  \href{http://dx.doi.org/10.1007/JHEP05(2012)155}{JHEP {\bf 1205} (2012)
  155},
\href{http://arxiv.org/abs/1112.2705}{{\tt arXiv:1112.2705 [hep-ph]}}.
%%CITATION = ARXIV:1112.2705;%%.

\bibitem{Falkowski:2010gv}
A.~Falkowski, J.~T. Ruderman, T.~Volansky, and J.~Zupan, {\em {Discovering
  Higgs Decays to Lepton Jets at Hadron Colliders}\/},
  \href{http://dx.doi.org/10.1103/PhysRevLett.105.241801}{Phys.Rev.Lett. {\bf
  105} (2010)  241801},
\href{http://arxiv.org/abs/1007.3496}{{\tt arXiv:1007.3496 [hep-ph]}}.
%%CITATION = ARXIV:1007.3496;%%.

\bibitem{Gupta:2015lfa}
A.~Gupta, R.~Primulando, and P.~Saraswat, {\em {A New Probe of Dark Sector
  Dynamics at the LHC}\/},
\href{http://arxiv.org/abs/1504.01385}{{\tt arXiv:1504.01385 [hep-ph]}}.
%%CITATION = ARXIV:1504.01385;%%.

\bibitem{Autran:2015mfa}
M.~Autran, K.~Bauer, T.~Lin, and D.~Whiteson, {\em {Mono-Z': searches for dark
  matter in events with a resonance and missing transverse energy}\/},
\href{http://arxiv.org/abs/1504.01386}{{\tt arXiv:1504.01386 [hep-ph]}}.
%%CITATION = ARXIV:1504.01386;%%.

\bibitem{Aad:2012qua}
{ATLAS Collaboration}, G.~Aad et al., {\em {A Search for Prompt Lepton-Jets in
  $pp$ Collisions at $\sqrt{s}=7$ TeV with the ATLAS Detector}\/},
  \href{http://dx.doi.org/10.1016/j.physletb.2013.01.034}{Phys.Lett. {\bf B719}
  (2013)  299--317},
\href{http://arxiv.org/abs/1212.5409}{{\tt arXiv:1212.5409 [hep-ex]}}.
%%CITATION = ARXIV:1212.5409;%%.

\bibitem{Hamaguchi:2004df}
K.~Hamaguchi, Y.~Kuno, T.~Nakaya, and M.~M. Nojiri, {\em {A Study of late
  decaying charged particles at future colliders}\/},
  \href{http://dx.doi.org/10.1103/PhysRevD.70.115007}{Phys. Rev. {\bf D70}
  (2004)  115007},
\href{http://arxiv.org/abs/hep-ph/0409248}{{\tt arXiv:hep-ph/0409248
  [hep-ph]}}.
%%CITATION = HEP-PH/0409248;%%.

\bibitem{Feng:2004yi}
J.~L. Feng and B.~T. Smith, {\em {Slepton trapping at the large hadron and
  international linear colliders}\/},
  \href{http://dx.doi.org/10.1103/PhysRevD.71.015004,
  10.1103/PhysRevD.71.019904}{Phys. Rev. {\bf D71} (2005)  015004},
  \href{http://arxiv.org/abs/hep-ph/0409278}{{\tt arXiv:hep-ph/0409278
  [hep-ph]}}.
[Erratum: Phys. Rev.D71,019904(2005)].
%%CITATION = HEP-PH/0409278;%%.

\bibitem{Buchmuller:2007ui}
W.~Buchmuller, L.~Covi, K.~Hamaguchi, A.~Ibarra, and T.~Yanagida, {\em
  {Gravitino Dark Matter in R-Parity Breaking Vacua}\/},
  \href{http://dx.doi.org/10.1088/1126-6708/2007/03/037}{JHEP {\bf 03} (2007)
  037},
\href{http://arxiv.org/abs/hep-ph/0702184}{{\tt arXiv:hep-ph/0702184
  [HEP-PH]}}.
%%CITATION = HEP-PH/0702184;%%.

\bibitem{Arcadi:2014tsa}
G.~Arcadi, L.~Covi, and F.~Dradi, {\em {LHC prospects for minimal decaying Dark
  Matter}\/},  \href{http://dx.doi.org/10.1088/1475-7516/2014/10/063}{JCAP {\bf
  1410} (2014) no.~10, 063},
\href{http://arxiv.org/abs/1408.1005}{{\tt arXiv:1408.1005 [hep-ph]}}.
%%CITATION = ARXIV:1408.1005;%%.

\bibitem{Han:2010rf}
T.~Han, I.~Lewis, and Z.~Liu, {\em {Colored Resonant Signals at the LHC:
  Largest Rate and Simplest Topology}\/},
  \href{http://dx.doi.org/10.1007/JHEP12(2010)085}{JHEP {\bf 1012} (2010)
  085},
\href{http://arxiv.org/abs/1010.4309}{{\tt arXiv:1010.4309 [hep-ph]}}.
%%CITATION = ARXIV:1010.4309;%%.

\bibitem{Langacker:2008yv}
P.~Langacker, {\em {The Physics of Heavy $Z^\prime$ Gauge Bosons}\/},
  \href{http://dx.doi.org/10.1103/RevModPhys.81.1199}{Rev.Mod.Phys. {\bf 81}
  (2009)  1199--1228},
\href{http://arxiv.org/abs/0801.1345}{{\tt arXiv:0801.1345 [hep-ph]}}.
%%CITATION = ARXIV:0801.1345;%%.

\bibitem{Agashe:2014kda}
{Particle Data Group Collaboration}, K.~A. Olive et al., {\em {Review of
  Particle Physics}\/},
\href{http://dx.doi.org/10.1088/1674-1137/38/9/090001}{Chin. Phys. {\bf C38}
  (2014)  090001}.
%%CITATION = CHPHD,C38,090001;%%.

\bibitem{Cai:2008ss}
H.~Cai, H.-C. Cheng, and J.~Terning, {\em {A Spin-1 Top Quark Superpartner}\/},
   \href{http://dx.doi.org/10.1103/PhysRevLett.101.171805}{Phys. Rev. Lett.
  {\bf 101} (2008)  171805},
\href{http://arxiv.org/abs/0806.0386}{{\tt arXiv:0806.0386 [hep-ph]}}.
%%CITATION = ARXIV:0806.0386;%%.

\bibitem{Chen:2012uw}
C.-Y. Chen, A.~Freitas, T.~Han, and K.~S.~M. Lee, {\em {New Physics from the
  Top at the LHC}\/},  \href{http://dx.doi.org/10.1007/JHEP11(2012)124}{JHEP
  {\bf 11} (2012)  124},
\href{http://arxiv.org/abs/1207.4794}{{\tt arXiv:1207.4794 [hep-ph]}}.
%%CITATION = ARXIV:1207.4794;%%.

\bibitem{Chen:2014haa}
C.-Y. Chen, A.~Freitas, T.~Han, and K.~S.~M. Lee, {\em {Heavy Color-Octet
  Particles at the LHC}\/},
  \href{http://dx.doi.org/10.1007/JHEP05(2015)135}{JHEP {\bf 05} (2015)  135},
\href{http://arxiv.org/abs/1410.8113}{{\tt arXiv:1410.8113 [hep-ph]}}.
%%CITATION = ARXIV:1410.8113;%%.

\bibitem{Foot:1988aq}
R.~Foot, H.~Lew, X.~G. He, and G.~C. Joshi, {\em {Seesaw Neutrino Masses
  Induced by a Triplet of Leptons}\/},
\href{http://dx.doi.org/10.1007/BF01415558}{Z. Phys. {\bf C44} (1989)  441}.
%%CITATION = ZEPYA,C44,441;%%.

\bibitem{Arhrib:2009mz}
A.~Arhrib, B.~Bajc, D.~K. Ghosh, T.~Han, G.-Y. Huang, I.~Puljak, and
  G.~Senjanovic, {\em {Collider Signatures for Heavy Lepton Triplet in Type
  I+III Seesaw}\/},  \href{http://dx.doi.org/10.1103/PhysRevD.82.053004}{Phys.
  Rev. {\bf D82} (2010)  053004},
\href{http://arxiv.org/abs/0904.2390}{{\tt arXiv:0904.2390 [hep-ph]}}.
%%CITATION = ARXIV:0904.2390;%%.

\bibitem{Li:2009mw}
T.~Li and X.-G. He, {\em {Neutrino Masses and Heavy Triplet Leptons at the LHC:
  Testability of Type III Seesaw}\/},
  \href{http://dx.doi.org/10.1103/PhysRevD.80.093003}{Phys. Rev. {\bf D80}
  (2009)  093003},
\href{http://arxiv.org/abs/0907.4193}{{\tt arXiv:0907.4193 [hep-ph]}}.
%%CITATION = ARXIV:0907.4193;%%.

\bibitem{Rizzo:2014xma}
T.~G. Rizzo, {\em {Exploring new gauge bosons at a 100 TeV collider}\/},
  \href{http://dx.doi.org/10.1103/PhysRevD.89.095022}{Phys. Rev. {\bf D89}
  (2014) no.~9, 095022},
\href{http://arxiv.org/abs/1403.5465}{{\tt arXiv:1403.5465 [hep-ph]}}.
%%CITATION = ARXIV:1403.5465;%%.

\bibitem{Baur:1987ga}
U.~Baur, I.~Hinchliffe, and D.~Zeppenfeld, {\em {Excited quark production at
  hadron colliders}\/},
\href{http://dx.doi.org/10.1142/S0217751X87000661}{Int. J. Mod. Phys. {\bf A2}
  (1987)  1285}.
%%CITATION = IMPAE,A2,1285;%%.

\bibitem{Baur:1989kv}
U.~Baur, M.~Spira, and P.~M. Zerwas, {\em {Excited quark and lepton production
  at hadron colliders}\/},
\href{http://dx.doi.org/10.1103/PhysRevD.42.815}{Phys. Rev. {\bf D42} (1990)
  815--825}.
%%CITATION = PHRVA,D42,815;%%.

\bibitem{Kretzer:2003it}
S.~Kretzer, H.~L. Lai, F.~I. Olness, and W.~K. Tung, {\em {CTEQ6 parton
  distributions with heavy quark mass effects}\/},
  \href{http://dx.doi.org/10.1103/PhysRevD.69.114005}{Phys. Rev. {\bf D69}
  (2004)  114005},
\href{http://arxiv.org/abs/hep-ph/0307022}{{\tt arXiv:hep-ph/0307022
  [hep-ph]}}.
%%CITATION = HEP-PH/0307022;%%.

\bibitem{Pumplin:2002vw}
J.~Pumplin, D.~Stump, J.~Huston, H.~Lai, P.~M. Nadolsky, et al., {\em {New
  generation of parton distributions with uncertainties from global QCD
  analysis}\/},  \href{http://dx.doi.org/10.1088/1126-6708/2002/07/012}{JHEP
  {\bf 0207} (2002)  012},
\href{http://arxiv.org/abs/hep-ph/0201195}{{\tt arXiv:hep-ph/0201195
  [hep-ph]}}.
%%CITATION = HEP-PH/0201195;%%.

\bibitem{Lai:2010vv}
H.-L. Lai, M.~Guzzi, J.~Huston, Z.~Li, P.~M. Nadolsky, J.~Pumplin, and C.~P.
  Yuan, {\em {New parton distributions for collider physics}\/},
  \href{http://dx.doi.org/10.1103/PhysRevD.82.074024}{Phys. Rev. {\bf D82}
  (2010)  074024},
\href{http://arxiv.org/abs/1007.2241}{{\tt arXiv:1007.2241 [hep-ph]}}.
%%CITATION = ARXIV:1007.2241;%%.

\bibitem{Sacrifice}
{\em Sacrifice 0.9.1 for Pythia 8\/},
  \url{https://agile.hepforge.org/trac/wiki/Sacrifice}, 2013.
\newblock [Online; accessed 07-March-2016].

\bibitem{Conte:2012fm}
E.~Conte, B.~Fuks, and G.~Serret, {\em {MadAnalysis 5, A User-Friendly
  Framework for Collider Phenomenology}\/},
  \href{http://dx.doi.org/10.1016/j.cpc.2012.09.009}{Comput. Phys. Commun. {\bf
  184} (2013)  222--256},
\href{http://arxiv.org/abs/1206.1599}{{\tt arXiv:1206.1599 [hep-ph]}}.
%%CITATION = ARXIV:1206.1599;%%.

\bibitem{Brun:1997pa}
R.~Brun and F.~Rademakers, {\em {ROOT: An object oriented data analysis
  framework}\/},
\href{http://dx.doi.org/10.1016/S0168-9002(97)00048-X}{Nucl. Instrum. Meth.
  {\bf A389} (1997)  81--86}.
%%CITATION = NUIMA,A389,81;%%.

\bibitem{Aad2016}
{ATLAS Collaboration}, G.~Aad et al., {\em {Search for new phenomena in dijet
  mass and angular distributions from $pp$ collisions at $\sqrt{s}=$ 13 TeV
  with the ATLAS detector}\/},
  \href{http://dx.doi.org/10.1016/j.physletb.2016.01.032}{Phys. Lett. {\bf
  B754} (2016)  302--322},
\href{http://arxiv.org/abs/1512.01530}{{\tt arXiv:1512.01530 [hep-ex]}}.
%%CITATION = ARXIV:1512.01530;%%.

\bibitem{Apanasevich:2013cta}
L.~Apanasevich, S.~Upadhyay, N.~Varelas, D.~Whiteson, and F.~Yu, {\em
  {Sensitivity of potential future $pp$ colliders to quark compositeness}\/},
\href{http://arxiv.org/abs/1307.7149}{{\tt arXiv:1307.7149 [hep-ex]}}.
%%CITATION = ARXIV:1307.7149;%%.

\bibitem{Kong:2013xta}
K.~Kong and F.~Yu, {\em {Discovery potential of Kaluza-Klein gluons at hadron
  colliders: A Snowmass whitepaper}\/},
\href{http://arxiv.org/abs/1308.1078}{{\tt arXiv:1308.1078 [hep-ph]}}.
%%CITATION = ARXIV:1308.1078;%%.

\bibitem{Agashe:2013kxa}
K.~Agashe, M.~Bauer, F.~Goertz, S.~J. Lee, L.~Vecchi, et al., {\em
  {Constraining RS Models by Future Flavor and Collider Measurements: A
  Snowmass Whitepaper}\/},
\href{http://arxiv.org/abs/1310.1070}{{\tt arXiv:1310.1070 [hep-ph]}}.
%%CITATION = ARXIV:1310.1070;%%.

\bibitem{Dobrescu:2013cmh}
B.~A. Dobrescu and F.~Yu, {\em {Coupling-mass mapping of dijet peak
  searches}\/},  \href{http://dx.doi.org/10.1103/PhysRevD.88.035021,
  10.1103/PhysRevD.90.079901}{Phys.Rev. {\bf D88} (2013) no.~3, 035021},
\href{http://arxiv.org/abs/1306.2629}{{\tt arXiv:1306.2629 [hep-ph]}}.
%%CITATION = ARXIV:1306.2629;%%.

\bibitem{Mangano:2002ea}
M.~L. Mangano, M.~Moretti, F.~Piccinini, R.~Pittau, and A.~D. Polosa, {\em
  {ALPGEN, a generator for hard multiparton processes in hadronic
  collisions}\/},  \href{http://dx.doi.org/10.1088/1126-6708/2003/07/001}{JHEP
  {\bf 0307} (2003)  001},
\href{http://arxiv.org/abs/hep-ph/0206293}{{\tt arXiv:hep-ph/0206293
  [hep-ph]}}.
%%CITATION = HEP-PH/0206293;%%.

\bibitem{Khachatryan:2015sja}
{CMS Collaboration}, V.~Khachatryan et al., {\em {Search for resonances and
  quantum black holes using dijet mass spectra in proton-proton collisions at
  $\sqrt{s} =$ 8 TeV}\/},
  \href{http://dx.doi.org/10.1103/PhysRevD.91.052009}{Phys. Rev. {\bf D91}
  (2015) no.~5, 052009},
\href{http://arxiv.org/abs/1501.04198}{{\tt arXiv:1501.04198 [hep-ex]}}.
%%CITATION = ARXIV:1501.04198;%%.

\bibitem{Auerbach:2014xua}
B.~Auerbach, S.~Chekanov, J.~Love, J.~Proudfoot, and A.~V. Kotwal, {\em
  {Sensitivity to new high-mass states decaying to $t\bar{t}$ at a 100 TeV
  collider}\/},  \href{http://dx.doi.org/10.1103/PhysRevD.91.034014}{Phys. Rev.
  {\bf D91} (2015) no.~3, 034014},
\href{http://arxiv.org/abs/1412.5951}{{\tt arXiv:1412.5951 [hep-ph]}}.
%%CITATION = ARXIV:1412.5951;%%.

\bibitem{PhysRevLett.83.3370}
L.~Randall and R.~Sundrum, {\em {A Large mass hierarchy from a small extra
  dimension}\/},  \href{http://dx.doi.org/10.1103/PhysRevLett.83.3370}{Phys.
  Rev. Lett. {\bf 83} (1999)  3370--3373},
\href{http://arxiv.org/abs/hep-ph/9905221}{{\tt arXiv:hep-ph/9905221
  [hep-ph]}}.
%%CITATION = HEP-PH/9905221;%%.

\bibitem{Lillie:2007yh}
B.~Lillie, L.~Randall, and L.-T. Wang, {\em {The Bulk RS KK-gluon at the
  LHC}\/},  \href{http://dx.doi.org/10.1088/1126-6708/2007/09/074}{JHEP {\bf
  09} (2007)  074},
\href{http://arxiv.org/abs/hep-ph/0701166}{{\tt arXiv:hep-ph/0701166
  [hep-ph]}}.
%%CITATION = HEP-PH/0701166;%%.

\bibitem{Chekanov:2014fga}
S.~Chekanov, {\em {HepSim: a repository with predictions for high-energy
  physics experiments}\/},  Advances in High Energy Physics {\bf 2015} (2015)
  136093. Available as \url{http://atlaswww.hep.anl.gov/hepsim/}.

\bibitem{Ellis:2009su}
S.~D. Ellis, C.~K. Vermilion, and J.~R. Walsh, {\em {Techniques for improved
  heavy particle searches with jet substructure}\/},
  \href{http://dx.doi.org/10.1103/PhysRevD.80.051501}{Phys. Rev. {\bf D80}
  (2009)  051501},
\href{http://arxiv.org/abs/0903.5081}{{\tt arXiv:0903.5081 [hep-ph]}}.
%%CITATION = 0903.5081;%%.

\bibitem{Butterworth:2002tt}
J.~Butterworth, B.~Cox, and J.~R. Forshaw, {\em {$W W$ scattering at the CERN
  LHC}\/},  \href{http://dx.doi.org/10.1103/PhysRevD.65.096014}{Phys.Rev. {\bf
  D65} (2002)  096014},
\href{http://arxiv.org/abs/hep-ph/0201098}{{\tt arXiv:hep-ph/0201098
  [hep-ph]}}.
%%CITATION = HEP-PH/0201098;%%.

\bibitem{Chekanov:2010vc}
S.~Chekanov and J.~Proudfoot, {\em {Searches for TeV-scale particles at the LHC
  using jet shapes}\/},
  \href{http://dx.doi.org/10.1103/PhysRevD.81.114038}{Phys. Rev. {\bf D81}
  (2010)  114038},
\href{http://arxiv.org/abs/1002.3982}{{\tt arXiv:1002.3982 [hep-ph]}}.
%%CITATION = 1002.3982;%%.

\bibitem{Junk:1999kv}
T.~Junk, {\em {Confidence level computation for combining searches with small
  statistics}\/},
  \href{http://dx.doi.org/10.1016/S0168-9002(99)00498-2}{Nucl.~Instrum.~Meth.
  {\bf A~434} (1999)  435--443},
\href{http://arxiv.org/abs/hep-ex/9902006}{{\tt arXiv:hep-ex/9902006
  [hep-ex]}}.
%%CITATION = HEP-EX/9902006;%%.

\bibitem{Pappadopulo:2014qza}
D.~Pappadopulo, A.~Thamm, R.~Torre, and A.~Wulzer, {\em {Heavy Vector Triplets:
  Bridging Theory and Data}\/},
  \href{http://dx.doi.org/10.1007/JHEP09(2014)060}{JHEP {\bf 09} (2014)  060},
\href{http://arxiv.org/abs/1402.4431}{{\tt arXiv:1402.4431 [hep-ph]}}.
%%CITATION = ARXIV:1402.4431;%%.

\bibitem{Thamm:2015zwa}
A.~Thamm, R.~Torre, and A.~Wulzer, {\em {Future tests of Higgs compositeness:
  direct vs indirect}\/},
  \href{http://dx.doi.org/10.1007/JHEP07(2015)100}{JHEP {\bf 07} (2015)  100},
\href{http://arxiv.org/abs/1502.01701}{{\tt arXiv:1502.01701 [hep-ph]}}.
%%CITATION = ARXIV:1502.01701;%%.

\bibitem{Agashe:2004ib}
K.~Agashe, R.~Contino, and A.~Pomarol, {\em The {M}inimal {C}omposite {H}iggs
  {M}odel\/},  \href{http://dx.doi.org/10.1016/j.nuclphysb.2005.04.035}{Nucl.
  Phys. {\bf B 719} (2005)  165--187}, \href{hep-ph/0412089}{{\tt 0412089}}.
  [\href{http://inspirebeta.net/record/666275}{Inspire}].

\bibitem{2014arXiv1401.2457B}
B.~Bellazzini, C.~Csaki, and J.~Serra, {\em {Composite Higgses}\/},
  \href{http://dx.doi.org/10.1140/epjc/s10052-014-2766-x}{Eur. Phys. J. {\bf C
  74} (2014)  2766}, \href{hep-ph/1401.2457}{{\tt 1401.2457}}.
  [\href{http://inspirehep.net/record/1276832}{Inspire}].

\bibitem{Panico:2015jxa}
G.~Panico and A.~Wulzer, {\em {The Composite Nambu-Goldstone Higgs}\/},
  \href{http://dx.doi.org/10.1007/978-3-319-22617-0}{Lect. Notes Phys. {\bf
  913} (2016)  1--316}, \href{hep-ph/1506.01961}{{\tt 1506.01961}}.
  [\href{http://inspirehep.net/record/1374915}{Inspire}].

\bibitem{DeSimone:2012ul}
A.~{De Simone}, O.~Matsedonskyi, R.~Rattazzi, and A.~Wulzer, {\em {A First Top
  Partner's Hunter Guide}\/},
  \href{http://dx.doi.org/10.1007/JHEP04(2013)004}{JHEP {\bf 04} (2013)  004},
  \href{hep-ph/1211.5663}{{\tt 1211.5663}}.
  [\href{http://inspirehep.net/record/1203860}{Inspire}].

\bibitem{CMS-PAS-EXO-12-061}
{CMS Collaboration}, {\em {Search for Resonances in the Dilepton Mass
  Distribution in pp Collisions at sqrt(s) = 8 TeV}\/},   CMS-PAS-EXO-12-061,
  CERN, Geneva, 2013.
\newblock \url{https://cds.cern.ch/record/1519132}.

\bibitem{Khachatryan:2014xja}
{CMS Collaboration}, V.~Khachatryan et al., {\em {Search for new resonances
  decaying via WZ to leptons in proton-proton collisions at $\sqrt s =$ 8
  TeV}\/},  \href{http://dx.doi.org/10.1016/j.physletb.2014.11.026}{Phys. Lett.
  {\bf B740} (2015)  83--104},
\href{http://arxiv.org/abs/1407.3476}{{\tt arXiv:1407.3476 [hep-ex]}}.
%%CITATION = ARXIV:1407.3476;%%.

\bibitem{ATLAScollaboration:2014eb}
{ATLAS Collaboration}, G.~Aad et al., {\em {Search for high-mass dilepton
  resonances in $pp$ collisions at $\sqrt{s} = 8$ TeV with the ATLAS
  detector}\/},  \href{http://dx.doi.org/10.1103/PhysRevD.90.052005}{Phys. Rev.
  {\bf D 90} (2014)  052005}, \href{hep-ex/1405.4123}{{\tt 1405.4123}}.
  [\href{http://inspirehep.net/record/1296830}{Inspire}].

\bibitem{ATLAScollaboration:2014uc}
{ATLAS Collaboration}, G.~Aad et al., {\em {Search for $WZ$ resonances in the
  fully leptonic channel using pp collisions at $\sqrt{s} = 8$ TeV with the
  ATLAS detector}\/},
  \href{http://dx.doi.org/10.1016/j.physletb.2014.08.039}{Phys. Lett. {\bf B
  737} (2014)  223--243}, \href{hep-ex/1406.4456}{{\tt 1406.4456}}.
  [\href{http://inspirehep.net/record/1300821}{Inspire}].

\bibitem{CMS-NOTE-2012/006}
{CMS Collaboration}, {\em {CMS at the High-Energy Frontier. Contribution to the
  Update of the European Strategy for Particle Physics}\/},
  CMS-NOTE-2012-006. CERN-CMS-NOTE-2012-006, CERN, Geneva, Oct, 2012.
\newblock \url{https://cds.cern.ch/record/1494600}.

\bibitem{ATL-PHYS-PUB-2013-014}
{ATLAS Collaboration}, {\em {Projections for measurements of Higgs boson cross
  sections, branching ratios and coupling parameters with the ATLAS detector at
  a HL-LHC}\/},   ATL-PHYS-PUB-2013-014, CERN, Geneva, Oct, 2013.
\newblock \url{https://cds.cern.ch/record/1611186}.

\bibitem{Dawson:2013bba}
S.~Dawson, A.~Gritsan, H.~Logan, J.~Qian, C.~Tully, et al., {\em {Working Group
  Report: Higgs Boson}\/},
\href{http://arxiv.org/abs/1310.8361}{{\tt arXiv:1310.8361 [hep-ex]}}.
%%CITATION = ARXIV:1310.8361;%%.

\bibitem{Froggatt:1978nt}
C.~Froggatt and H.~B. Nielsen, {\em {Hierarchy of Quark Masses, Cabibbo Angles
  and CP Violation}\/},
\href{http://dx.doi.org/10.1016/0550-3213(79)90316-X}{Nucl.Phys. {\bf B147}
  (1979)  277}.
%%CITATION = NUPHA,B147,277;%%.

\bibitem{Leurer:1992wg}
M.~Leurer, Y.~Nir, and N.~Seiberg, {\em {Mass matrix models}\/},
  \href{http://dx.doi.org/10.1016/0550-3213(93)90112-3}{Nucl. Phys. {\bf B398}
  (1993)  319--342},
\href{http://arxiv.org/abs/hep-ph/9212278}{{\tt arXiv:hep-ph/9212278
  [hep-ph]}}.
%%CITATION = HEP-PH/9212278;%%.

\bibitem{Leurer:1993gy}
M.~Leurer, Y.~Nir, and N.~Seiberg, {\em {Mass matrix models: The Sequel}\/},
  \href{http://dx.doi.org/10.1016/0550-3213(94)90074-4}{Nucl. Phys. {\bf B420}
  (1994)  468--504},
\href{http://arxiv.org/abs/hep-ph/9310320}{{\tt arXiv:hep-ph/9310320
  [hep-ph]}}.
%%CITATION = HEP-PH/9310320;%%.

\bibitem{Huitu:2016pwk}
K.~Huitu, V.~Keus, N.~Koivunen, and O.~Lebedev, {\em {Higgs--Flavon Mixing and
  $h \rightarrow \mu\tau$}\/},
\href{http://arxiv.org/abs/1603.06614}{{\tt arXiv:1603.06614 [hep-ph]}}.
%%CITATION = ARXIV:1603.06614;%%.

\bibitem{Georgi:1972hy}
H.~Georgi and S.~L. Glashow, {\em {Attempts to calculate the electron mass}\/},
\href{http://dx.doi.org/10.1103/PhysRevD.7.2457}{Phys. Rev. {\bf D7} (1973)
  2457--2463}.
%%CITATION = PHRVA,D7,2457;%%.

\bibitem{Kaplan:1991dc}
D.~B. Kaplan, {\em {Flavor at SSC energies: A New mechanism for dynamically
  generated fermion masses}\/},
\href{http://dx.doi.org/10.1016/S0550-3213(05)80021-5}{Nucl. Phys. {\bf B365}
  (1991)  259--278}.
%%CITATION = NUPHA,B365,259;%%.

\bibitem{Gherghetta:2000qt}
T.~Gherghetta and A.~Pomarol, {\em {Bulk fields and supersymmetry in a slice of
  AdS}\/},  \href{http://dx.doi.org/10.1016/S0550-3213(00)00392-8}{Nucl. Phys.
  {\bf B586} (2000)  141--162},
\href{http://arxiv.org/abs/hep-ph/0003129}{{\tt arXiv:hep-ph/0003129
  [hep-ph]}}.
%%CITATION = HEP-PH/0003129;%%.

\bibitem{Grossman:1999ra}
Y.~Grossman and M.~Neubert, {\em {Neutrino masses and mixings in
  nonfactorizable geometry}\/},
  \href{http://dx.doi.org/10.1016/S0370-2693(00)00054-X}{Phys. Lett. {\bf B474}
  (2000)  361--371},
\href{http://arxiv.org/abs/hep-ph/9912408}{{\tt arXiv:hep-ph/9912408
  [hep-ph]}}.
%%CITATION = HEP-PH/9912408;%%.

\bibitem{Blanke:2008zb}
M.~Blanke, A.~J. Buras, B.~Duling, S.~Gori, and A.~Weiler, {\em {$\Delta$ F=2
  Observables and Fine-Tuning in a Warped Extra Dimension with Custodial
  Protection}\/},  \href{http://dx.doi.org/10.1088/1126-6708/2009/03/001}{JHEP
  {\bf 03} (2009)  001},
\href{http://arxiv.org/abs/0809.1073}{{\tt arXiv:0809.1073 [hep-ph]}}.
%%CITATION = ARXIV:0809.1073;%%.

\bibitem{Casagrande:2008hr}
S.~Casagrande, F.~Goertz, U.~Haisch, M.~Neubert, and T.~Pfoh, {\em {Flavor
  Physics in the Randall-Sundrum Model: I. Theoretical Setup and Electroweak
  Precision Tests}\/},
  \href{http://dx.doi.org/10.1088/1126-6708/2008/10/094}{JHEP {\bf 10} (2008)
  094},
\href{http://arxiv.org/abs/0807.4937}{{\tt arXiv:0807.4937 [hep-ph]}}.
%%CITATION = ARXIV:0807.4937;%%.

\bibitem{Bauer:2009cf}
M.~Bauer, S.~Casagrande, U.~Haisch, and M.~Neubert, {\em {Flavor Physics in the
  Randall-Sundrum Model: II. Tree-Level Weak-Interaction Processes}\/},
  \href{http://dx.doi.org/10.1007/JHEP09(2010)017}{JHEP {\bf 09} (2010)  017},
\href{http://arxiv.org/abs/0912.1625}{{\tt arXiv:0912.1625 [hep-ph]}}.
%%CITATION = ARXIV:0912.1625;%%.

\bibitem{Bauer:2015fxa}
M.~Bauer, M.~Carena, and K.~Gemmler, {\em {Flavor from the Electroweak
  Scale}\/},  \href{http://dx.doi.org/10.1007/JHEP11(2015)016}{JHEP {\bf 11}
  (2015)  016},
\href{http://arxiv.org/abs/1506.01719}{{\tt arXiv:1506.01719 [hep-ph]}}.
%%CITATION = ARXIV:1506.01719;%%.

\bibitem{Bauer:2015kzy}
M.~Bauer, M.~Carena, and K.~Gemmler, {\em {Creating the Fermion Mass
  Hierarchies with Multiple Higgs Bosons}\/},
\href{http://arxiv.org/abs/1512.03458}{{\tt arXiv:1512.03458 [hep-ph]}}.
%%CITATION = ARXIV:1512.03458;%%.

\bibitem{Calibbi:2015sfa}
L.~Calibbi, A.~Crivellin, and B.~Zaldà¤¥à¤var, {\em {Flavor portal to dark
  matter}\/},  \href{http://dx.doi.org/10.1103/PhysRevD.92.016004}{Phys. Rev.
  {\bf D92} (2015) no.~1, 016004},
\href{http://arxiv.org/abs/1501.07268}{{\tt arXiv:1501.07268 [hep-ph]}}.
%%CITATION = ARXIV:1501.07268;%%.

\bibitem{Bauer:2016rxs}
M.~Bauer, T.~Schell, and T.~Plehn, {\em {Hunting the Flavon}\/},
\href{http://arxiv.org/abs/1603.06950}{{\tt arXiv:1603.06950 [hep-ph]}}.
%%CITATION = ARXIV:1603.06950;%%.

\bibitem{Charles:2013aka}
J.~Charles, S.~Descotes-Genon, Z.~Ligeti, S.~Monteil, M.~Papucci, and
  K.~Trabelsi, {\em {Future sensitivity to new physics in $B_d, B_s$, and K
  mixings}\/},  \href{http://dx.doi.org/10.1103/PhysRevD.89.033016}{Phys. Rev.
  {\bf D89} (2014) no.~3, 033016},
\href{http://arxiv.org/abs/1309.2293}{{\tt arXiv:1309.2293 [hep-ph]}}.
%%CITATION = ARXIV:1309.2293;%%.

\bibitem{Chatrchyan:2013bka}
{CMS Collaboration}, S.~Chatrchyan et al., {\em {Measurement of the B(s) to mu+
  mu- branching fraction and search for B0 to mu+ mu- with the CMS
  Experiment}\/},
  \href{http://dx.doi.org/10.1103/PhysRevLett.111.101804}{Phys. Rev. Lett. {\bf
  111} (2013)  101804},
\href{http://arxiv.org/abs/1307.5025}{{\tt arXiv:1307.5025 [hep-ex]}}.
%%CITATION = ARXIV:1307.5025;%%.

\bibitem{Aaij:2013aka}
{LHCb Collaboration}, R.~Aaij et al., {\em {Measurement of the $B^0_s \to \mu^+
  \mu^-$ branching fraction and search for $B^0 \to \mu^+ \mu^-$ decays at the
  LHCb experiment}\/},
  \href{http://dx.doi.org/10.1103/PhysRevLett.111.101805}{Phys. Rev. Lett. {\bf
  111} (2013)  101805},
\href{http://arxiv.org/abs/1307.5024}{{\tt arXiv:1307.5024 [hep-ex]}}.
%%CITATION = ARXIV:1307.5024;%%.

\bibitem{CMS:2014xfa}
{LHCb, CMS Collaboration}, V.~Khachatryan et al., {\em {Observation of the rare
  $B^0_s\to\mu^+\mu^-$ decay from the combined analysis of CMS and LHCb
  data}\/},  \href{http://dx.doi.org/10.1038/nature14474}{Nature {\bf 522}
  (2015)  68--72},
\href{http://arxiv.org/abs/1411.4413}{{\tt arXiv:1411.4413 [hep-ex]}}.
%%CITATION = ARXIV:1411.4413;%%.

\bibitem{Baldini:2013ke}
A.~M. Baldini et al., {\em {MEG Upgrade Proposal}\/},
\href{http://arxiv.org/abs/1301.7225}{{\tt arXiv:1301.7225 [physics.ins-det]}}.
%%CITATION = ARXIV:1301.7225;%%.

\bibitem{Natori:2014yba}
{DeeMe Collaboration}, H.~Natori, {\em {DeeMe experiment - An experimental
  search for a mu-e conversion reaction at J-PARC MLF}\/},
\href{http://dx.doi.org/10.1016/j.nuclphysbps.2014.02.010}{Nucl. Phys. Proc.
  Suppl. {\bf 248-250} (2014)  52--57}.
%%CITATION = NUPHZ,248-250,52;%%.

\bibitem{Kuno:2013mha}
{COMET Collaboration}, Y.~Kuno, {\em {A search for muon-to-electron conversion
  at J-PARC: The COMET experiment}\/},
\href{http://dx.doi.org/10.1093/ptep/pts089}{PTEP {\bf 2013} (2013)  022C01}.
%%CITATION = INSPIRE-1223754;%%.

\bibitem{Abrams:2012er}
{Mu2e Collaboration}, R.~J. Abrams et al., {\em {Mu2e Conceptual Design
  Report}\/},
\href{http://arxiv.org/abs/1211.7019}{{\tt arXiv:1211.7019 [physics.ins-det]}}.
%%CITATION = ARXIV:1211.7019;%%.

\bibitem{Kiehn:2015eta}
{Mu3e Collaboration}, M.~Kiehn, {\em {The Mu3e Experiment - Introduction and
  Current Status}\/},
PoS {\bf NUFACT2014} (2015)  088.
%%CITATION = POSCI,NUFACT2014,088;%%.

\bibitem{fastjet}
M.~Cacciari, G.~Salam, and G.~Soyez, {\em {{FastJet. A C++ library for the
  $k_T$ algorithm}}\/}, .

\bibitem{Schmidt:2012az}
B.~Schmidt and M.~Steinhauser, {\em {CRunDec: a C++ package for running and
  decoupling of the strong coupling and quark masses}\/},
  \href{http://dx.doi.org/10.1016/j.cpc.2012.03.023}{Comput. Phys. Commun. {\bf
  183} (2012)  1845--1848},
\href{http://arxiv.org/abs/1201.6149}{{\tt arXiv:1201.6149 [hep-ph]}}.
%%CITATION = ARXIV:1201.6149;%%.

\bibitem{Alves:2006df}
A.~Alves, O.~Eboli, and T.~Plehn, {\em {It's a gluino}\/},
  \href{http://dx.doi.org/10.1103/PhysRevD.74.095010}{Phys. Rev. {\bf D74}
  (2006)  095010},
\href{http://arxiv.org/abs/hep-ph/0605067}{{\tt arXiv:hep-ph/0605067
  [hep-ph]}}.
%%CITATION = HEP-PH/0605067;%%.

\bibitem{ATL-PHYS-PUB-2015-040}
{\em {A new tagger for the charge identification of b-jets}\/},
  ATL-PHYS-PUB-2015-040, CERN, Geneva, Sep, 2015.
\newblock \url{https://cds.cern.ch/record/2048132}.

\bibitem{Mohan:2015doa}
K.~Mohan and N.~Vignaroli, {\em {Vector resonances in weak-boson-fusion at
  future pp colliders}\/},
  \href{http://dx.doi.org/10.1007/JHEP10(2015)031}{JHEP {\bf 10} (2015)  031},
\href{http://arxiv.org/abs/1507.03940}{{\tt arXiv:1507.03940 [hep-ph]}}.
%%CITATION = ARXIV:1507.03940;%%.

\bibitem{Agashe:2004rs}
K.~Agashe, R.~Contino, and A.~Pomarol, {\em {The Minimal composite Higgs
  model}\/},  \href{http://dx.doi.org/10.1016/j.nuclphysb.2005.04.035}{Nucl.
  Phys. {\bf B719} (2005)  165--187},
\href{http://arxiv.org/abs/hep-ph/0412089}{{\tt arXiv:hep-ph/0412089
  [hep-ph]}}.
%%CITATION = HEP-PH/0412089;%%.

\bibitem{Contino:2006nn}
R.~Contino, T.~Kramer, M.~Son, and R.~Sundrum, {\em {Warped/composite
  phenomenology simplified}\/},
  \href{http://dx.doi.org/10.1088/1126-6708/2007/05/074}{JHEP {\bf 05} (2007)
  074},
\href{http://arxiv.org/abs/hep-ph/0612180}{{\tt arXiv:hep-ph/0612180
  [hep-ph]}}.
%%CITATION = HEP-PH/0612180;%%.

\bibitem{Vignaroli:2014bpa}
N.~Vignaroli, {\em {New Wà¥{\"A}à¥à¤ signals at the LHC}\/},
  \href{http://dx.doi.org/10.1103/PhysRevD.89.095027}{Phys. Rev. {\bf D89}
  (2014) no.~9, 095027},
\href{http://arxiv.org/abs/1404.5558}{{\tt arXiv:1404.5558 [hep-ph]}}.
%%CITATION = ARXIV:1404.5558;%%.

\bibitem{Kulchitsky:2000gg}
Y.~A. Kulchitsky, M.~V. Kuzmin, J.~A. Budagov, V.~B. Vinogradov, and M.~Nessi,
  {\em {Hadron energy reconstruction for the ATLAS barrel prototype combined
  calorimeter in the framework of the nonparametrical method}\/},
\href{http://arxiv.org/abs/hep-ex/0004009}{{\tt arXiv:hep-ex/0004009
  [hep-ex]}}.
%%CITATION = HEP-EX/0004009;%%.

\bibitem{ATLAS:2014ffa}
{ATLAS Collaboration}, {\em {Estimation of non-prompt and fake lepton
  backgrounds in final states with top quarks produced in proton-proton
  collisions at $\sqrt{s}=8$~TeV with the ATLAS detector}\/},
  ATLAS-CONF-2014-058, CERN, Geneva, Oct, 2014.
\newblock \url{https://cds.cern.ch/record/1951336}.

\bibitem{Davoudiasl:1999tf}
H.~Davoudiasl, J.~L. Hewett, and T.~G. Rizzo, {\em {Bulk gauge fields in the
  Randall-Sundrum model}\/},
  \href{http://dx.doi.org/10.1016/S0370-2693(99)01430-6}{Phys. Lett. {\bf B473}
  (2000)  43--49},
\href{http://arxiv.org/abs/hep-ph/9911262}{{\tt arXiv:hep-ph/9911262
  [hep-ph]}}.
%%CITATION = HEP-PH/9911262;%%.

\bibitem{Pomarol:1999ad}
A.~Pomarol, {\em {Gauge bosons in a five-dimensional theory with localized
  gravity}\/},  \href{http://dx.doi.org/10.1016/S0370-2693(00)00737-1}{Phys.
  Lett. {\bf B486} (2000)  153--157},
\href{http://arxiv.org/abs/hep-ph/9911294}{{\tt arXiv:hep-ph/9911294
  [hep-ph]}}.
%%CITATION = HEP-PH/9911294;%%.

\bibitem{Huber:2000ie}
S.~J. Huber and Q.~Shafi, {\em {Fermion masses, mixings and proton decay in a
  Randall-Sundrum model}\/},
  \href{http://dx.doi.org/10.1016/S0370-2693(00)01399-X}{Phys. Lett. {\bf B498}
  (2001)  256--262},
\href{http://arxiv.org/abs/hep-ph/0010195}{{\tt arXiv:hep-ph/0010195
  [hep-ph]}}.
%%CITATION = HEP-PH/0010195;%%.

\bibitem{Huber:2003tu}
S.~J. Huber, {\em {Flavor violation and warped geometry}\/},
  \href{http://dx.doi.org/10.1016/S0550-3213(03)00502-9}{Nucl. Phys. {\bf B666}
  (2003)  269--288},
\href{http://arxiv.org/abs/hep-ph/0303183}{{\tt arXiv:hep-ph/0303183
  [hep-ph]}}.
%%CITATION = HEP-PH/0303183;%%.

\bibitem{Davoudiasl:2000wi}
H.~Davoudiasl, J.~L. Hewett, and T.~G. Rizzo, {\em {Experimental probes of
  localized gravity: On and off the wall}\/},
  \href{http://dx.doi.org/10.1103/PhysRevD.63.075004}{Phys. Rev. {\bf D63}
  (2001)  075004},
\href{http://arxiv.org/abs/hep-ph/0006041}{{\tt arXiv:hep-ph/0006041
  [hep-ph]}}.
%%CITATION = HEP-PH/0006041;%%.

\bibitem{Agashe:2006hk}
K.~Agashe, A.~Belyaev, T.~Krupovnickas, G.~Perez, and J.~Virzi, {\em {LHC
  Signals from Warped Extra Dimensions}\/},
  \href{http://dx.doi.org/10.1103/PhysRevD.77.015003}{Phys. Rev. {\bf D77}
  (2008)  015003},
\href{http://arxiv.org/abs/hep-ph/0612015}{{\tt arXiv:hep-ph/0612015
  [hep-ph]}}.
%%CITATION = HEP-PH/0612015;%%.

\bibitem{Fitzpatrick:2007qr}
A.~L. Fitzpatrick, J.~Kaplan, L.~Randall, and L.-T. Wang, {\em {Searching for
  the Kaluza-Klein Graviton in Bulk RS Models}\/},
  \href{http://dx.doi.org/10.1088/1126-6708/2007/09/013}{JHEP {\bf 09} (2007)
  013},
\href{http://arxiv.org/abs/hep-ph/0701150}{{\tt arXiv:hep-ph/0701150}}.
%%CITATION = HEP-PH/0701150;%%.

\bibitem{Agashe:2007zd}
K.~Agashe, H.~Davoudiasl, G.~Perez, and A.~Soni, {\em {Warped Gravitons at the
  LHC and Beyond}\/},
  \href{http://dx.doi.org/10.1103/PhysRevD.76.036006}{Phys. Rev. {\bf D76}
  (2007)  036006},
\href{http://arxiv.org/abs/hep-ph/0701186}{{\tt arXiv:hep-ph/0701186
  [hep-ph]}}.
%%CITATION = HEP-PH/0701186;%%.

\bibitem{Lillie:2007ve}
B.~Lillie, J.~Shu, and T.~M.~P. Tait, {\em {Kaluza-Klein Gluons as a Diagnostic
  of Warped Models}\/},
  \href{http://dx.doi.org/10.1103/PhysRevD.76.115016}{Phys. Rev. {\bf D76}
  (2007)  115016},
\href{http://arxiv.org/abs/0706.3960}{{\tt arXiv:0706.3960 [hep-ph]}}.
%%CITATION = ARXIV:0706.3960;%%.

\bibitem{Djouadi:2007eg}
A.~Djouadi, G.~Moreau, and R.~K. Singh, {\em {Kaluza-Klein excitations of gauge
  bosons at the LHC}\/},
  \href{http://dx.doi.org/10.1016/j.nuclphysb.2007.12.024}{Nucl. Phys. {\bf
  B797} (2008)  1--26},
\href{http://arxiv.org/abs/0706.4191}{{\tt arXiv:0706.4191 [hep-ph]}}.
%%CITATION = ARXIV:0706.4191;%%.

\bibitem{Agashe:2007ki}
K.~Agashe, H.~Davoudiasl, S.~Gopalakrishna, T.~Han, G.-Y. Huang, G.~Perez,
  Z.-G. Si, and A.~Soni, {\em {LHC Signals for Warped Electroweak Neutral Gauge
  Bosons}\/},  \href{http://dx.doi.org/10.1103/PhysRevD.76.115015}{Phys. Rev.
  {\bf D76} (2007)  115015},
\href{http://arxiv.org/abs/0709.0007}{{\tt arXiv:0709.0007 [hep-ph]}}.
%%CITATION = ARXIV:0709.0007;%%.

\bibitem{Antipin:2007pi}
O.~Antipin, D.~Atwood, and A.~Soni, {\em {Search for RS gravitons via W(L)W(L)
  decays}\/},  \href{http://dx.doi.org/10.1016/j.physletb.2008.07.009}{Phys.
  Lett. {\bf B666} (2008)  155--161},
\href{http://arxiv.org/abs/0711.3175}{{\tt arXiv:0711.3175 [hep-ph]}}.
%%CITATION = ARXIV:0711.3175;%%.

\bibitem{Agashe:2008jb}
K.~Agashe, S.~Gopalakrishna, T.~Han, G.-Y. Huang, and A.~Soni, {\em {LHC
  Signals for Warped Electroweak Charged Gauge Bosons}\/},
  \href{http://dx.doi.org/10.1103/PhysRevD.80.075007}{Phys. Rev. {\bf D80}
  (2009)  075007},
\href{http://arxiv.org/abs/0810.1497}{{\tt arXiv:0810.1497 [hep-ph]}}.
%%CITATION = ARXIV:0810.1497;%%.

\bibitem{Davoudiasl:2009cd}
H.~Davoudiasl, S.~Gopalakrishna, E.~Ponton, and J.~Santiago, {\em {Warped
  5-Dimensional Models: Phenomenological Status and Experimental Prospects}\/},
   \href{http://dx.doi.org/10.1088/1367-2630/12/7/075011}{New J. Phys. {\bf 12}
  (2010)  075011},
\href{http://arxiv.org/abs/0908.1968}{{\tt arXiv:0908.1968 [hep-ph]}}.
%%CITATION = ARXIV:0908.1968;%%.

\bibitem{Chen:2014oha}
C.-Y. Chen, H.~Davoudiasl, and D.~Kim, {\em {Z with missing energy as a warped
  graviton signal at hadron colliders}\/},
  \href{http://dx.doi.org/10.1103/PhysRevD.89.096007}{Phys. Rev. {\bf D89}
  (2014) no.~9, 096007},
\href{http://arxiv.org/abs/1403.3399}{{\tt arXiv:1403.3399 [hep-ph]}}.
%%CITATION = ARXIV:1403.3399;%%.

\bibitem{Agashe:2014wba}
K.~Agashe, C.-Y. Chen, H.~Davoudiasl, and D.~Kim, {\em {Photon cascade decay of
  the warped graviton at LHC14 and a 100 TeV hadron collider}\/},
  \href{http://dx.doi.org/10.1103/PhysRevD.91.076002}{Phys. Rev. {\bf D91}
  (2015) no.~7, 076002},
\href{http://arxiv.org/abs/1412.6215}{{\tt arXiv:1412.6215 [hep-ph]}}.
%%CITATION = ARXIV:1412.6215;%%.

\bibitem{Belyaev:2012qa}
A.~Belyaev, N.~D. Christensen, and A.~Pukhov, {\em {CalcHEP 3.4 for collider
  physics within and beyond the Standard Model}\/},
  \href{http://dx.doi.org/10.1016/j.cpc.2013.01.014}{Comput.Phys.Commun. {\bf
  184} (2013)  1729--1769},
\href{http://arxiv.org/abs/1207.6082}{{\tt arXiv:1207.6082 [hep-ph]}}.
%%CITATION = ARXIV:1207.6082;%%.

\bibitem{Ball:2012cx}
R.~D. Ball et al., {\em {Parton distributions with LHC data}\/},
  \href{http://dx.doi.org/10.1016/j.nuclphysb.2012.10.003}{Nucl. Phys. {\bf
  B867} (2013)  244--289},
\href{http://arxiv.org/abs/1207.1303}{{\tt arXiv:1207.1303 [hep-ph]}}.
%%CITATION = ARXIV:1207.1303;%%.

\bibitem{CP3}
O.~Antipin and T.~Hapola. \url{http://cp3-origins.dk/research/units/ed-tools}.

\bibitem{CMS:2014joa}
{CMS Collaboration}, {\em {V Tagging Observables and Correlations}\/},
  CMS-PAS-JME-14-002, CERN, Geneva, 2014.
\newblock \url{https://cds.cern.ch/record/1754913}.

\bibitem{Carena:2006bn}
M.~Carena, E.~Ponton, J.~Santiago, and C.~E.~M. Wagner, {\em {Light Kaluza
  Klein States in Randall-Sundrum Models with Custodial SU(2)}\/},
  \href{http://dx.doi.org/10.1016/j.nuclphysb.2006.10.012}{Nucl. Phys. {\bf
  B759} (2006)  202--227},
\href{http://arxiv.org/abs/hep-ph/0607106}{{\tt arXiv:hep-ph/0607106
  [hep-ph]}}.
%%CITATION = HEP-PH/0607106;%%.

\bibitem{Carena:2007ua}
M.~Carena, E.~Ponton, J.~Santiago, and C.~E.~M. Wagner, {\em {Electroweak
  constraints on warped models with custodial symmetry}\/},
  \href{http://dx.doi.org/10.1103/PhysRevD.76.035006}{Phys. Rev. {\bf D76}
  (2007)  035006},
\href{http://arxiv.org/abs/hep-ph/0701055}{{\tt arXiv:hep-ph/0701055
  [hep-ph]}}.
%%CITATION = HEP-PH/0701055;%%.

\bibitem{Chatrchyan:2013lca}
{CMS Collaboration}, S.~Chatrchyan et al., {\em {Searches for new physics using
  the $t\bar{t}$ invariant mass distribution in $pp$ collisions at $\sqrt{s}$=8
  TeV}\/},  \href{http://dx.doi.org/10.1103/PhysRevLett.111.211804,
  10.1103/PhysRevLett.112.119903}{Phys. Rev. Lett. {\bf 111} (2013) no.~21,
  211804}, \href{http://arxiv.org/abs/1309.2030}{{\tt arXiv:1309.2030
  [hep-ex]}}.
[Erratum: Phys. Rev. Lett.112,no.11,119903(2014)].
%%CITATION = ARXIV:1309.2030;%%.

\bibitem{TheATLAScollaboration:2013kha}
{\em {A search for $t\bar{t}$ resonances in the lepton plus jets final state
  with ATLAS using 14 fb$^{-1}$ of pp collisions at $\sqrt{s}=8$ TeV}\/},
  ATLAS-CONF-2013-052, CERN, Geneva, May, 2013.
\newblock \url{https://cds.cern.ch/record/1547568}.
\newblock Not published in the proceedings.

\bibitem{Minkowski:1977sc}
P.~Minkowski, {\em {$\mu \to e\gamma$ at a Rate of One Out of $10^{9}$ Muon
  Decays?}\/},
\href{http://dx.doi.org/10.1016/0370-2693(77)90435-X}{Phys. Lett. {\bf B67}
  (1977)  421--428}.
%%CITATION = PHLTA,B67,421;%%.

\bibitem{Mohapatra:1979ia}
R.~N. Mohapatra and G.~Senjanovic, {\em {Neutrino Mass and Spontaneous Parity
  Violation}\/},
\href{http://dx.doi.org/10.1103/PhysRevLett.44.912}{Phys. Rev. Lett. {\bf 44}
  (1980)  912}.
%%CITATION = PRLTA,44,912;%%.

\bibitem{Yanagida:1979as}
T.~Yanagida, {\em {Horizontal symmetry and masses of neutrinos}\/},  Conf.
  Proc. {\bf C7902131} (1979)  95--99.
[Conf. Proc.C7902131,95(1979)].
%%CITATION = CONFP,C7902131,95;%%.

\bibitem{GellMann:1980vs}
M.~Gell-Mann, P.~Ramond, and R.~Slansky, {\em {Complex Spinors and Unified
  Theories}\/},  Conf. Proc. {\bf C790927} (1979)  315--321,
\href{http://arxiv.org/abs/1306.4669}{{\tt arXiv:1306.4669 [hep-th]}}.
%%CITATION = ARXIV:1306.4669;%%.

\bibitem{Vissani:1997ys}
F.~Vissani, {\em {Do experiments suggest a hierarchy problem?}\/},
  \href{http://dx.doi.org/10.1103/PhysRevD.57.7027}{Phys. Rev. {\bf D57} (1998)
   7027--7030},
\href{http://arxiv.org/abs/hep-ph/9709409}{{\tt arXiv:hep-ph/9709409
  [hep-ph]}}.
%%CITATION = HEP-PH/9709409;%%.

\bibitem{Clarke:2015hta}
J.~D. Clarke, R.~Foot, and R.~R. Volkas, {\em {Natural leptogenesis and
  neutrino masses with two Higgs doublets}\/},
  \href{http://dx.doi.org/10.1103/PhysRevD.92.033006}{Phys. Rev. {\bf D92}
  (2015) no.~3, 033006},
\href{http://arxiv.org/abs/1505.05744}{{\tt arXiv:1505.05744 [hep-ph]}}.
%%CITATION = ARXIV:1505.05744;%%.

\bibitem{Keung:1983uu}
W.-Y. Keung and G.~Senjanovic, {\em {Majorana Neutrinos and the Production of
  the Right-handed Charged Gauge Boson}\/},
\href{http://dx.doi.org/10.1103/PhysRevLett.50.1427}{Phys. Rev. Lett. {\bf 50}
  (1983)  1427}.
%%CITATION = PRLTA,50,1427;%%.

\bibitem{Datta:1993nm}
A.~Datta, M.~Guchait, and A.~Pilaftsis, {\em {Probing lepton number violation
  via majorana neutrinos at hadron supercolliders}\/},
  \href{http://dx.doi.org/10.1103/PhysRevD.50.3195}{Phys. Rev. {\bf D50} (1994)
   3195--3203},
\href{http://arxiv.org/abs/hep-ph/9311257}{{\tt arXiv:hep-ph/9311257
  [hep-ph]}}.
%%CITATION = HEP-PH/9311257;%%.

\bibitem{Dev:2013wba}
P.~S.~B. Dev, A.~Pilaftsis, and U.-k. Yang, {\em {New Production Mechanism for
  Heavy Neutrinos at the LHC}\/},
  \href{http://dx.doi.org/10.1103/PhysRevLett.112.081801}{Phys. Rev. Lett. {\bf
  112} (2014) no.~8, 081801},
\href{http://arxiv.org/abs/1308.2209}{{\tt arXiv:1308.2209 [hep-ph]}}.
%%CITATION = ARXIV:1308.2209;%%.

\bibitem{Alva:2014gxa}
D.~Alva, T.~Han, and R.~Ruiz, {\em {Heavy Majorana neutrinos from $W\gamma$
  fusion at hadron colliders}\/},
  \href{http://dx.doi.org/10.1007/JHEP02(2015)072}{JHEP {\bf 02} (2015)  072},
\href{http://arxiv.org/abs/1411.7305}{{\tt arXiv:1411.7305 [hep-ph]}}.
%%CITATION = ARXIV:1411.7305;%%.

\bibitem{Rodejohann:2011mu}
W.~Rodejohann, {\em {Neutrino-less Double Beta Decay and Particle Physics}\/},
  \href{http://dx.doi.org/10.1142/S0218301311020186}{Int. J. Mod. Phys. {\bf
  E20} (2011)  1833--1930},
\href{http://arxiv.org/abs/1106.1334}{{\tt arXiv:1106.1334 [hep-ph]}}.
%%CITATION = ARXIV:1106.1334;%%.

\bibitem{deGouvea:2013zba}
A.~de~Gouvea and P.~Vogel, {\em {Lepton Flavor and Number Conservation, and
  Physics Beyond the Standard Model}\/},
  \href{http://dx.doi.org/10.1016/j.ppnp.2013.03.006}{Prog. Part. Nucl. Phys.
  {\bf 71} (2013)  75--92},
\href{http://arxiv.org/abs/1303.4097}{{\tt arXiv:1303.4097 [hep-ph]}}.
%%CITATION = ARXIV:1303.4097;%%.

\bibitem{BhupalDev:2012zg}
P.~S. Bhupal~Dev, R.~Franceschini, and R.~N. Mohapatra, {\em {Bounds on TeV
  Seesaw Models from LHC Higgs Data}\/},
  \href{http://dx.doi.org/10.1103/PhysRevD.86.093010}{Phys. Rev. {\bf D86}
  (2012)  093010},
\href{http://arxiv.org/abs/1207.2756}{{\tt arXiv:1207.2756 [hep-ph]}}.
%%CITATION = ARXIV:1207.2756;%%.

\bibitem{Cely:2012bz}
C.~G. Cely, A.~Ibarra, E.~Molinaro, and S.~T. Petcov, {\em {Higgs Decays in the
  Low Scale Type I See-Saw Model}\/},
  \href{http://dx.doi.org/10.1016/j.physletb.2012.11.026}{Phys. Lett. {\bf
  B718} (2013)  957--964},
\href{http://arxiv.org/abs/1208.3654}{{\tt arXiv:1208.3654 [hep-ph]}}.
%%CITATION = ARXIV:1208.3654;%%.

\bibitem{Maiezza:2015lza}
A.~Maiezza, M.~Nemev\v{s}ek, and F.~Nesti, {\em {Lepton Number Violation in
  Higgs Decay at LHC}\/},
  \href{http://dx.doi.org/10.1103/PhysRevLett.115.081802}{Phys. Rev. Lett. {\bf
  115} (2015)  081802},
\href{http://arxiv.org/abs/1503.06834}{{\tt arXiv:1503.06834 [hep-ph]}}.
%%CITATION = ARXIV:1503.06834;%%.

\bibitem{Dermisek:2015vra}
R.~Dermisek, E.~Lunghi, and S.~Shin, {\em {Contributions of flavor violating
  couplings of a Higgs boson to $pp \to WW$}\/},
  \href{http://dx.doi.org/10.1007/JHEP08(2015)126}{JHEP {\bf 08} (2015)  126},
\href{http://arxiv.org/abs/1503.08829}{{\tt arXiv:1503.08829 [hep-ph]}}.
%%CITATION = ARXIV:1503.08829;%%.

\bibitem{Deppisch:2015qwa}
F.~F. Deppisch, P.~S. Bhupal~Dev, and A.~Pilaftsis, {\em {Neutrinos and
  Collider Physics}\/},
  \href{http://dx.doi.org/10.1088/1367-2630/17/7/075019}{New J. Phys. {\bf 17}
  (2015) no.~7, 075019},
\href{http://arxiv.org/abs/1502.06541}{{\tt arXiv:1502.06541 [hep-ph]}}.
%%CITATION = ARXIV:1502.06541;%%.

\bibitem{Pilaftsis:1991ug}
A.~Pilaftsis, {\em {Radiatively induced neutrino masses and large Higgs
  neutrino couplings in the standard model with Majorana fields}\/},
  \href{http://dx.doi.org/10.1007/BF01482590}{Z. Phys. {\bf C55} (1992)
  275--282},
\href{http://arxiv.org/abs/hep-ph/9901206}{{\tt arXiv:hep-ph/9901206
  [hep-ph]}}.
%%CITATION = HEP-PH/9901206;%%.

\bibitem{Buchmuller:1991ce}
W.~Buchmuller, C.~Greub, and P.~Minkowski, {\em {Neutrino masses, neutral
  vector bosons and the scale of B-L breaking}\/},
\href{http://dx.doi.org/10.1016/0370-2693(91)90952-M}{Phys. Lett. {\bf B267}
  (1991)  395--399}.
%%CITATION = PHLTA,B267,395;%%.

\bibitem{Gluza:2002vs}
J.~Gluza, {\em {On teraelectronvolt Majorana neutrinos}\/},  Acta Phys. Polon.
  {\bf B33} (2002)  1735--1746,
\href{http://arxiv.org/abs/hep-ph/0201002}{{\tt arXiv:hep-ph/0201002
  [hep-ph]}}.
%%CITATION = HEP-PH/0201002;%%.

\bibitem{Kersten:2007vk}
J.~Kersten and A.~{\relax Yu}. Smirnov, {\em {Right-Handed Neutrinos at CERN
  LHC and the Mechanism of Neutrino Mass Generation}\/},
  \href{http://dx.doi.org/10.1103/PhysRevD.76.073005}{Phys. Rev. {\bf D76}
  (2007)  073005},
\href{http://arxiv.org/abs/0705.3221}{{\tt arXiv:0705.3221 [hep-ph]}}.
%%CITATION = ARXIV:0705.3221;%%.

\bibitem{Xing:2009in}
Z.-z. Xing, {\em {Naturalness and Testability of TeV Seesaw Mechanisms}\/},
  \href{http://dx.doi.org/10.1143/PTPS.180.112}{Prog. Theor. Phys. Suppl. {\bf
  180} (2009)  112--127},
\href{http://arxiv.org/abs/0905.3903}{{\tt arXiv:0905.3903 [hep-ph]}}.
%%CITATION = ARXIV:0905.3903;%%.

\bibitem{Gavela:2009cd}
M.~B. Gavela, T.~Hambye, D.~Hernandez, and P.~Hernandez, {\em {Minimal Flavour
  Seesaw Models}\/},
  \href{http://dx.doi.org/10.1088/1126-6708/2009/09/038}{JHEP {\bf 09} (2009)
  038},
\href{http://arxiv.org/abs/0906.1461}{{\tt arXiv:0906.1461 [hep-ph]}}.
%%CITATION = ARXIV:0906.1461;%%.

\bibitem{He:2009ua}
X.-G. He, S.~Oh, J.~Tandean, and C.-C. Wen, {\em {Large Mixing of Light and
  Heavy Neutrinos in Seesaw Models and the LHC}\/},
  \href{http://dx.doi.org/10.1103/PhysRevD.80.073012}{Phys. Rev. {\bf D80}
  (2009)  073012},
\href{http://arxiv.org/abs/0907.1607}{{\tt arXiv:0907.1607 [hep-ph]}}.
%%CITATION = ARXIV:0907.1607;%%.

\bibitem{Adhikari:2010yt}
R.~Adhikari and A.~Raychaudhuri, {\em {Light neutrinos from massless texture
  and below TeV seesaw scale}\/},
  \href{http://dx.doi.org/10.1103/PhysRevD.84.033002}{Phys. Rev. {\bf D84}
  (2011)  033002},
\href{http://arxiv.org/abs/1004.5111}{{\tt arXiv:1004.5111 [hep-ph]}}.
%%CITATION = ARXIV:1004.5111;%%.

\bibitem{Ibarra:2010xw}
A.~Ibarra, E.~Molinaro, and S.~T. Petcov, {\em {TeV Scale See-Saw Mechanisms of
  Neutrino Mass Generation, the Majorana Nature of the Heavy Singlet Neutrinos
  and $(\beta\beta)_{0\nu}$-Decay}\/},
  \href{http://dx.doi.org/10.1007/JHEP09(2010)108}{JHEP {\bf 09} (2010)  108},
\href{http://arxiv.org/abs/1007.2378}{{\tt arXiv:1007.2378 [hep-ph]}}.
%%CITATION = ARXIV:1007.2378;%%.

\bibitem{Mitra:2011qr}
M.~Mitra, G.~Senjanovic, and F.~Vissani, {\em {Neutrinoless Double Beta Decay
  and Heavy Sterile Neutrinos}\/},
  \href{http://dx.doi.org/10.1016/j.nuclphysb.2011.10.035}{Nucl. Phys. {\bf
  B856} (2012)  26--73},
\href{http://arxiv.org/abs/1108.0004}{{\tt arXiv:1108.0004 [hep-ph]}}.
%%CITATION = ARXIV:1108.0004;%%.

\bibitem{Lopez-Pavon:2015cga}
J.~Lopez-Pavon, E.~Molinaro, and S.~T. Petcov, {\em {Radiative Corrections to
  Light Neutrino Masses in Low Scale Type I Seesaw Scenarios and Neutrinoless
  Double Beta Decay}\/},  \href{http://dx.doi.org/10.1007/JHEP11(2015)030}{JHEP
  {\bf 11} (2015)  030},
\href{http://arxiv.org/abs/1506.05296}{{\tt arXiv:1506.05296 [hep-ph]}}.
%%CITATION = ARXIV:1506.05296;%%.

\bibitem{Mohapatra:1986aw}
R.~N. Mohapatra, {\em {Mechanism for Understanding Small Neutrino Mass in
  Superstring Theories}\/},
\href{http://dx.doi.org/10.1103/PhysRevLett.56.561}{Phys. Rev. Lett. {\bf 56}
  (1986)  561--563}.
%%CITATION = PRLTA,56,561;%%.

\bibitem{Mohapatra:1986bd}
R.~N. Mohapatra and J.~W.~F. Valle, {\em {Neutrino Mass and Baryon Number
  Nonconservation in Superstring Models}\/},
\href{http://dx.doi.org/10.1103/PhysRevD.34.1642}{Phys. Rev. {\bf D34} (1986)
  1642}.
%%CITATION = PHRVA,D34,1642;%%.

\bibitem{Bray:2007ru}
S.~Bray, J.~S. Lee, and A.~Pilaftsis, {\em {Resonant CP violation due to heavy
  neutrinos at the LHC}\/},
  \href{http://dx.doi.org/10.1016/j.nuclphysb.2007.07.002}{Nucl. Phys. {\bf
  B786} (2007)  95--118},
\href{http://arxiv.org/abs/hep-ph/0702294}{{\tt arXiv:hep-ph/0702294
  [HEP-PH]}}.
%%CITATION = HEP-PH/0702294;%%.

\bibitem{Akhmedov:2007fk}
E.~K. Akhmedov, {\em {Do charged leptons oscillate?}\/},
  \href{http://dx.doi.org/10.1088/1126-6708/2007/09/116}{JHEP {\bf 09} (2007)
  116},
\href{http://arxiv.org/abs/0706.1216}{{\tt arXiv:0706.1216 [hep-ph]}}.
%%CITATION = ARXIV:0706.1216;%%.

\bibitem{Khachatryan:2014dka}
{CMS Collaboration}, V.~Khachatryan et al., {\em {Search for heavy neutrinos
  and $\mathrm {W}$ bosons with right-handed couplings in proton-proton
  collisions at $\sqrt{s} = 8\,\text {TeV} $}\/},
  \href{http://dx.doi.org/10.1140/epjc/s10052-014-3149-z}{Eur. Phys. J. {\bf
  C74} (2014) no.~11, 3149},
\href{http://arxiv.org/abs/1407.3683}{{\tt arXiv:1407.3683 [hep-ex]}}.
%%CITATION = ARXIV:1407.3683;%%.

\bibitem{delAguila:2008cj}
F.~del Aguila and J.~A. Aguilar-Saavedra, {\em {Distinguishing seesaw models at
  LHC with multi-lepton signals}\/},
  \href{http://dx.doi.org/10.1016/j.nuclphysb.2008.12.029}{Nucl. Phys. {\bf
  B813} (2009)  22--90},
\href{http://arxiv.org/abs/0808.2468}{{\tt arXiv:0808.2468 [hep-ph]}}.
%%CITATION = ARXIV:0808.2468;%%.

\bibitem{delAguila:2008hw}
F.~del Aguila and J.~A. Aguilar-Saavedra, {\em {Electroweak scale seesaw and
  heavy Dirac neutrino signals at LHC}\/},
  \href{http://dx.doi.org/10.1016/j.physletb.2009.01.010}{Phys. Lett. {\bf
  B672} (2009)  158--165},
\href{http://arxiv.org/abs/0809.2096}{{\tt arXiv:0809.2096 [hep-ph]}}.
%%CITATION = ARXIV:0809.2096;%%.

\bibitem{Chen:2011hc}
C.-Y. Chen and P.~S.~B. Dev, {\em {Multi-Lepton Collider Signatures of Heavy
  Dirac and Majorana Neutrinos}\/},
  \href{http://dx.doi.org/10.1103/PhysRevD.85.093018}{Phys. Rev. {\bf D85}
  (2012)  093018},
\href{http://arxiv.org/abs/1112.6419}{{\tt arXiv:1112.6419 [hep-ph]}}.
%%CITATION = ARXIV:1112.6419;%%.

\bibitem{Das:2012ze}
A.~Das and N.~Okada, {\em {Inverse seesaw neutrino signatures at the LHC and
  ILC}\/},  \href{http://dx.doi.org/10.1103/PhysRevD.88.113001}{Phys. Rev. {\bf
  D88} (2013)  113001},
\href{http://arxiv.org/abs/1207.3734}{{\tt arXiv:1207.3734 [hep-ph]}}.
%%CITATION = ARXIV:1207.3734;%%.

\bibitem{Das:2014jxa}
A.~Das, P.~S. Bhupal~Dev, and N.~Okada, {\em {Direct bounds on electroweak
  scale pseudo-Dirac neutrinos from $\sqrt s=8$ TeV LHC data}\/},
  \href{http://dx.doi.org/10.1016/j.physletb.2014.06.058}{Phys. Lett. {\bf
  B735} (2014)  364--370},
\href{http://arxiv.org/abs/1405.0177}{{\tt arXiv:1405.0177 [hep-ph]}}.
%%CITATION = ARXIV:1405.0177;%%.

\bibitem{Bambhaniya:2014kga}
G.~Bambhaniya, S.~Goswami, S.~Khan, P.~Konar, and T.~Mondal, {\em {Looking for
  hints of a reconstructible seesaw model at the Large Hadron Collider}\/},
  \href{http://dx.doi.org/10.1103/PhysRevD.91.075007}{Phys. Rev. {\bf D91}
  (2015)  075007},
\href{http://arxiv.org/abs/1410.5687}{{\tt arXiv:1410.5687 [hep-ph]}}.
%%CITATION = ARXIV:1410.5687;%%.

\bibitem{Khachatryan:2015gha}
{CMS Collaboration}, V.~Khachatryan et al., {\em {Search for heavy Majorana
  neutrinos in $\mu^\pm \mu^\pm+$ jets events in proton-proton collisions at
  $\sqrt{s}$ = 8 TeV}\/},
  \href{http://dx.doi.org/10.1016/j.physletb.2015.06.070}{Phys. Lett. {\bf
  B748} (2015)  144--166},
\href{http://arxiv.org/abs/1501.05566}{{\tt arXiv:1501.05566 [hep-ex]}}.
%%CITATION = ARXIV:1501.05566;%%.

\bibitem{Aad:2015xaa}
{ATLAS Collaboration}, G.~Aad et al., {\em {Search for heavy Majorana neutrinos
  with the ATLAS detector in pp collisions at $ \sqrt{s}=8 $ TeV}\/},
  \href{http://dx.doi.org/10.1007/JHEP07(2015)162}{JHEP {\bf 07} (2015)  162},
\href{http://arxiv.org/abs/1506.06020}{{\tt arXiv:1506.06020 [hep-ex]}}.
%%CITATION = ARXIV:1506.06020;%%.

\bibitem{Antusch:2015mia}
S.~Antusch and O.~Fischer, {\em {Testing sterile neutrino extensions of the
  Standard Model at future lepton colliders}\/},
  \href{http://dx.doi.org/10.1007/JHEP05(2015)053}{JHEP {\bf 05} (2015)  053},
\href{http://arxiv.org/abs/1502.05915}{{\tt arXiv:1502.05915 [hep-ph]}}.
%%CITATION = ARXIV:1502.05915;%%.

\bibitem{Banerjee:2015gca}
S.~Banerjee, P.~S.~B. Dev, A.~Ibarra, T.~Mandal, and M.~Mitra, {\em {Prospects
  of Heavy Neutrino Searches at Future Lepton Colliders}\/},
  \href{http://dx.doi.org/10.1103/PhysRevD.92.075002}{Phys. Rev. {\bf D92}
  (2015)  075002},
\href{http://arxiv.org/abs/1503.05491}{{\tt arXiv:1503.05491 [hep-ph]}}.
%%CITATION = ARXIV:1503.05491;%%.

\bibitem{Pati:1974yy}
J.~C. Pati and A.~Salam, {\em {Lepton Number as the Fourth Color}\/},
  \href{http://dx.doi.org/10.1103/PhysRevD.10.275,
  10.1103/PhysRevD.11.703.2}{Phys. Rev. {\bf D10} (1974)  275--289}.
[Erratum: Phys. Rev.D11,703(1975)].
%%CITATION = PHRVA,D10,275;%%.

\bibitem{Mohapatra:1974hk}
R.~N. Mohapatra and J.~C. Pati, {\em {Left-Right Gauge Symmetry and an
  Isoconjugate Model of CP Violation}\/},
\href{http://dx.doi.org/10.1103/PhysRevD.11.566}{Phys. Rev. {\bf D11} (1975)
  566--571}.
%%CITATION = PHRVA,D11,566;%%.

\bibitem{Mohapatra:1974gc}
R.~N. Mohapatra and J.~C. Pati, {\em {A Natural Left-Right Symmetry}\/},
\href{http://dx.doi.org/10.1103/PhysRevD.11.2558}{Phys. Rev. {\bf D11} (1975)
  2558}.
%%CITATION = PHRVA,D11,2558;%%.

\bibitem{Senjanovic:1975rk}
G.~Senjanovic and R.~N. Mohapatra, {\em {Exact Left-Right Symmetry and
  Spontaneous Violation of Parity}\/},
\href{http://dx.doi.org/10.1103/PhysRevD.12.1502}{Phys. Rev. {\bf D12} (1975)
  1502}.
%%CITATION = PHRVA,D12,1502;%%.

\bibitem{Dev:2013oxa}
P.~S.~B. Dev, C.-H. Lee, and R.~Mohapatra, {\em {Natural TeV-Scale Left-Right
  Seesaw for Neutrinos and Experimental Tests}\/},
\href{http://arxiv.org/abs/1309.0774}{{\tt arXiv:1309.0774 [hep-ph]}}.
%%CITATION = ARXIV:1309.0774;%%.

\bibitem{Frere:2008ct}
J.-M. Frere, T.~Hambye, and G.~Vertongen, {\em {Is leptogenesis falsifiable at
  LHC?}\/},  \href{http://dx.doi.org/10.1088/1126-6708/2009/01/051}{JHEP {\bf
  01} (2009)  051},
\href{http://arxiv.org/abs/0806.0841}{{\tt arXiv:0806.0841 [hep-ph]}}.
%%CITATION = ARXIV:0806.0841;%%.

\bibitem{Dev:2014iva}
P.~S.~B. Dev, C.-H. Lee, and R.~N. Mohapatra, {\em {Leptogenesis Constraints on
  the Mass of Right-handed Gauge Bosons}\/},
  \href{http://dx.doi.org/10.1103/PhysRevD.90.095012}{Phys. Rev. {\bf D90}
  (2014) no.~9, 095012},
\href{http://arxiv.org/abs/1408.2820}{{\tt arXiv:1408.2820 [hep-ph]}}.
%%CITATION = ARXIV:1408.2820;%%.

\bibitem{Dev:2015vra}
P.~S. Bhupal~Dev, C.-H. Lee, and R.~N. Mohapatra, {\em {TeV Scale Lepton Number
  Violation and Baryogenesis}\/},
  \href{http://dx.doi.org/10.1088/1742-6596/631/1/012007}{J. Phys. Conf. Ser.
  {\bf 631} (2015) no.~1, 012007},
\href{http://arxiv.org/abs/1503.04970}{{\tt arXiv:1503.04970 [hep-ph]}}.
%%CITATION = ARXIV:1503.04970;%%.

\bibitem{Dhuria:2015cfa}
M.~Dhuria, C.~Hati, R.~Rangarajan, and U.~Sarkar, {\em {Falsifying leptogenesis
  for a TeV scale $W^{\pm}_{R}$ at the LHC}\/},
  \href{http://dx.doi.org/10.1103/PhysRevD.92.031701}{Phys. Rev. {\bf D92}
  (2015) no.~3, 031701},
\href{http://arxiv.org/abs/1503.07198}{{\tt arXiv:1503.07198 [hep-ph]}}.
%%CITATION = ARXIV:1503.07198;%%.

\bibitem{Ferrari:2000sp}
A.~Ferrari, J.~Collot, M.-L. Andrieux, B.~Belhorma, P.~de~Saintignon, J.-Y.
  Hostachy, P.~Martin, and M.~Wielers, {\em {Sensitivity study for new gauge
  bosons and right-handed Majorana neutrinos in $p p$ collisions at $s$ =
  14-TeV}\/},
\href{http://dx.doi.org/10.1103/PhysRevD.62.013001}{Phys. Rev. {\bf D62} (2000)
   013001}.
%%CITATION = PHRVA,D62,013001;%%.

\bibitem{Nemevsek:2011hz}
M.~Nemevsek, F.~Nesti, G.~Senjanovic, and Y.~Zhang, {\em {First Limits on
  Left-Right Symmetry Scale from LHC Data}\/},
  \href{http://dx.doi.org/10.1103/PhysRevD.83.115014}{Phys. Rev. {\bf D83}
  (2011)  115014},
\href{http://arxiv.org/abs/1103.1627}{{\tt arXiv:1103.1627 [hep-ph]}}.
%%CITATION = ARXIV:1103.1627;%%.

\bibitem{Chakrabortty:2012pp}
J.~Chakrabortty, J.~Gluza, R.~Sevillano, and R.~Szafron, {\em {Left-Right
  Symmetry at LHC and Precise 1-Loop Low Energy Data}\/},
  \href{http://dx.doi.org/10.1007/JHEP07(2012)038}{JHEP {\bf 07} (2012)  038},
\href{http://arxiv.org/abs/1204.0736}{{\tt arXiv:1204.0736 [hep-ph]}}.
%%CITATION = ARXIV:1204.0736;%%.

\bibitem{Das:2012ii}
S.~P. Das, F.~F. Deppisch, O.~Kittel, and J.~W.~F. Valle, {\em {Heavy Neutrinos
  and Lepton Flavour Violation in Left-Right Symmetric Models at the LHC}\/},
  \href{http://dx.doi.org/10.1103/PhysRevD.86.055006}{Phys. Rev. {\bf D86}
  (2012)  055006},
\href{http://arxiv.org/abs/1206.0256}{{\tt arXiv:1206.0256 [hep-ph]}}.
%%CITATION = ARXIV:1206.0256;%%.

\bibitem{AguilarSaavedra:2012gf}
J.~A. Aguilar-Saavedra and F.~R. Joaquim, {\em {Measuring heavy neutrino
  couplings at the LHC}\/},
  \href{http://dx.doi.org/10.1103/PhysRevD.86.073005}{Phys. Rev. {\bf D86}
  (2012)  073005},
\href{http://arxiv.org/abs/1207.4193}{{\tt arXiv:1207.4193 [hep-ph]}}.
%%CITATION = ARXIV:1207.4193;%%.

\bibitem{Chen:2013fna}
C.-Y. Chen, P.~S.~B. Dev, and R.~N. Mohapatra, {\em {Probing Heavy-Light
  Neutrino Mixing in Left-Right Seesaw Models at the LHC}\/},
  \href{http://dx.doi.org/10.1103/PhysRevD.88.033014}{Phys. Rev. {\bf D88}
  (2013)  033014},
\href{http://arxiv.org/abs/1306.2342}{{\tt arXiv:1306.2342 [hep-ph]}}.
%%CITATION = ARXIV:1306.2342;%%.

\bibitem{Ng:2015hba}
J.~N. Ng, A.~de~la Puente, and B.~W.-P. Pan, {\em {Search for Heavy
  Right-Handed Neutrinos at the LHC and Beyond in the Same-Sign Leptons Final
  State}\/},
\href{http://arxiv.org/abs/1505.01934}{{\tt arXiv:1505.01934 [hep-ph]}}.
%%CITATION = ARXIV:1505.01934;%%.

\bibitem{Dev:2015kca}
P.~S.~B. Dev, D.~Kim, and R.~N. Mohapatra, {\em {Disambiguating Seesaw Models
  using Invariant Mass Variables at Hadron Colliders}\/},
\href{http://arxiv.org/abs/1510.04328}{{\tt arXiv:1510.04328 [hep-ph]}}.
%%CITATION = ARXIV:1510.04328;%%.

\bibitem{Dev:2013vba}
P.~S.~B. Dev and R.~N. Mohapatra, {\em {Probing TeV Left-Right Seesaw at Energy
  and Intensity Frontiers: a Snowmass White Paper}\/},  in {\em {Community
  Summer Study 2013: Snowmass on the Mississippi (CSS2013) Minneapolis, MN,
  USA, July 29-August 6, 2013}}.
\newblock 2013.
\newblock \href{http://arxiv.org/abs/1308.2151}{{\tt arXiv:1308.2151
  [hep-ph]}}.
\newblock
\url{http://inspirehep.net/record/1247267/files/arXiv:1308.2151.pdf}.
\newblock
%%CITATION = ARXIV:1308.2151;%%.

\bibitem{Malinsky:2005bi}
M.~Malinsky, J.~C. Romao, and J.~W.~F. Valle, {\em {Novel supersymmetric SO(10)
  seesaw mechanism}\/},
  \href{http://dx.doi.org/10.1103/PhysRevLett.95.161801}{Phys. Rev. Lett. {\bf
  95} (2005)  161801},
\href{http://arxiv.org/abs/hep-ph/0506296}{{\tt arXiv:hep-ph/0506296
  [hep-ph]}}.
%%CITATION = HEP-PH/0506296;%%.

\bibitem{Dev:2012sg}
P.~S.~B. Dev and A.~Pilaftsis, {\em {Minimal Radiative Neutrino Mass Mechanism
  for Inverse Seesaw Models}\/},
  \href{http://dx.doi.org/10.1103/PhysRevD.86.113001}{Phys. Rev. {\bf D86}
  (2012)  113001},
\href{http://arxiv.org/abs/1209.4051}{{\tt arXiv:1209.4051 [hep-ph]}}.
%%CITATION = ARXIV:1209.4051;%%.

\bibitem{Dev:2015pga}
P.~S. Bhupal~Dev and R.~N. Mohapatra, {\em {Unified explanation of the $eejj$,
  diboson and dijet resonances at the LHC}\/},
  \href{http://dx.doi.org/10.1103/PhysRevLett.115.181803}{Phys. Rev. Lett. {\bf
  115} (2015) no.~18, 181803},
\href{http://arxiv.org/abs/1508.02277}{{\tt arXiv:1508.02277 [hep-ph]}}.
%%CITATION = ARXIV:1508.02277;%%.

\bibitem{Gluza:2015goa}
J.~Gluza and T.~Jeli\'nski, {\em {Heavy neutrinos and the $pp \to lljj$ CMS
  data}\/},  \href{http://dx.doi.org/10.1016/j.physletb.2015.06.077}{Phys.
  Lett. {\bf B748} (2015)  125--131},
\href{http://arxiv.org/abs/1504.05568}{{\tt arXiv:1504.05568 [hep-ph]}}.
%%CITATION = ARXIV:1504.05568;%%.

\bibitem{Deppisch:2015cua}
F.~F. Deppisch, L.~Graf, S.~Kulkarni, S.~Patra, W.~Rodejohann, N.~Sahu, and
  U.~Sarkar, {\em {Reconciling the 2 TeV Excesses at the LHC in a Linear Seesaw
  Left-Right Model}\/},
\href{http://arxiv.org/abs/1508.05940}{{\tt arXiv:1508.05940 [hep-ph]}}.
%%CITATION = ARXIV:1508.05940;%%.

\bibitem{delAguila:2007em}
F.~del Aguila, J.~A. Aguilar-Saavedra, and R.~Pittau, {\em {Heavy neutrino
  signals at large hadron colliders}\/},
  \href{http://dx.doi.org/10.1088/1126-6708/2007/10/047}{JHEP {\bf 10} (2007)
  047},
\href{http://arxiv.org/abs/hep-ph/0703261}{{\tt arXiv:hep-ph/0703261
  [hep-ph]}}.
%%CITATION = HEP-PH/0703261;%%.

\bibitem{Maiezza:2010ic}
A.~Maiezza, M.~Nemevsek, F.~Nesti, and G.~Senjanovic, {\em {Left-Right Symmetry
  at LHC}\/},  \href{http://dx.doi.org/10.1103/PhysRevD.82.055022}{Phys.Rev.
  {\bf D82} (2010)  055022},
\href{http://arxiv.org/abs/1005.5160}{{\tt arXiv:1005.5160 [hep-ph]}}.
%%CITATION = ARXIV:1005.5160;%%.

\bibitem{Vasquez:2014mxa}
J.~C. Vasquez, {\em {Right-handed lepton mixings at the LHC}\/},
\href{http://arxiv.org/abs/1411.5824}{{\tt arXiv:1411.5824 [hep-ph]}}.
%%CITATION = ARXIV:1411.5824;%%.

\bibitem{Deppisch:2014qpa}
F.~F. Deppisch, T.~E. Gonzalo, S.~Patra, N.~Sahu, and U.~Sarkar, {\em {Signal
  of Right-Handed Charged Gauge Bosons at the LHC?}\/},
  \href{http://dx.doi.org/10.1103/PhysRevD.90.053014}{Phys. Rev. {\bf D90}
  (2014) no.~5, 053014},
\href{http://arxiv.org/abs/1407.5384}{{\tt arXiv:1407.5384 [hep-ph]}}.
%%CITATION = ARXIV:1407.5384;%%.

\bibitem{Heikinheimo:2014tba}
M.~Heikinheimo, M.~Raidal, and C.~Spethmann, {\em {Testing Right-Handed
  Currents at the LHC}\/},
  \href{http://dx.doi.org/10.1140/epjc/s10052-014-3107-9}{Eur. Phys. J. {\bf
  C74} (2014) no.~10, 3107},
\href{http://arxiv.org/abs/1407.6908}{{\tt arXiv:1407.6908 [hep-ph]}}.
%%CITATION = ARXIV:1407.6908;%%.

\bibitem{Deppisch:2014zta}
F.~F. Deppisch, T.~E. Gonzalo, S.~Patra, N.~Sahu, and U.~Sarkar, {\em {Double
  beta decay, lepton flavor violation, and collider signatures of left-right
  symmetric models with spontaneous $D$-parity breaking}\/},
  \href{http://dx.doi.org/10.1103/PhysRevD.91.015018}{Phys. Rev. {\bf D91}
  (2015) no.~1, 015018},
\href{http://arxiv.org/abs/1410.6427}{{\tt arXiv:1410.6427 [hep-ph]}}.
%%CITATION = ARXIV:1410.6427;%%.

\bibitem{Aguilar-Saavedra:2014ola}
J.~A. Aguilar-Saavedra and F.~R. Joaquim, {\em {Closer look at the possible CMS
  signal of a new gauge boson}\/},
  \href{http://dx.doi.org/10.1103/PhysRevD.90.115010}{Phys. Rev. {\bf D90}
  (2014) no.~11, 115010},
\href{http://arxiv.org/abs/1408.2456}{{\tt arXiv:1408.2456 [hep-ph]}}.
%%CITATION = ARXIV:1408.2456;%%.

\bibitem{Dobrescu:2015qna}
B.~A. Dobrescu and Z.~Liu, {\em {A W' Boson near 2 TeV: Predictions for Run 2
  of the LHC}\/},
\href{http://arxiv.org/abs/1506.06736}{{\tt arXiv:1506.06736 [hep-ph]}}.
%%CITATION = ARXIV:1506.06736;%%.

\bibitem{Coloma:2015una}
P.~Coloma, B.~A. Dobrescu, and J.~Lopez-Pavon, {\em {Right-handed neutrinos and
  the 2 TeV $W'$ boson}\/},
\href{http://arxiv.org/abs/1508.04129}{{\tt arXiv:1508.04129 [hep-ph]}}.
%%CITATION = ARXIV:1508.04129;%%.

\bibitem{Bandyopadhyay:2015fka}
T.~Bandyopadhyay, B.~Brahmachari, and A.~Raychaudhuri, {\em {Implications of
  the CMS search for $W_R$ on Grand Unification}\/},
\href{http://arxiv.org/abs/1509.03232}{{\tt arXiv:1509.03232 [hep-ph]}}.
%%CITATION = ARXIV:1509.03232;%%.

\bibitem{Senjanovic:2014pva}
G.~Senjanovi{\'c} and V.~Tello, {\em {Right Handed Quark Mixing in Left-Right
  Symmetric Theory}\/},
  \href{http://dx.doi.org/10.1103/PhysRevLett.114.071801}{Phys. Rev. Lett. {\bf
  114} (2015) no.~7, 071801},
\href{http://arxiv.org/abs/1408.3835}{{\tt arXiv:1408.3835 [hep-ph]}}.
%%CITATION = ARXIV:1408.3835;%%.

\bibitem{Senjanovic:2015yea}
G.~Senjanovi<E6> and V.~Tello, {\em {Restoration of Parity and the Right-Handed
  Analog of the CKM Matrix}\/},
\href{http://arxiv.org/abs/1502.05704}{{\tt arXiv:1502.05704 [hep-ph]}}.
%%CITATION = ARXIV:1502.05704;%%.

\bibitem{dirac}
J.~Gluza, T.~Jelinski, and R.~Szafron, {\em {Lepton Number Violation and
  `Diracness' of massive neutrinos composed of Majorana states}\/},  2016.
\newblock \href{http://arxiv.org/abs/1604.01388}{{\tt arXiv:1604.01388
  [hep-ph]}}.

\bibitem{Campbell:2006wx}
J.~M. Campbell, J.~W. Huston, and W.~J. Stirling, {\em {Hard Interactions of
  Quarks and Gluons: A Primer for LHC Physics}\/},
  \href{http://dx.doi.org/10.1088/0034-4885/70/1/R02}{Rept. Prog. Phys. {\bf
  70} (2007)  89},
\href{http://arxiv.org/abs/hep-ph/0611148}{{\tt arXiv:hep-ph/0611148
  [hep-ph]}}.
%%CITATION = HEP-PH/0611148;%%.

\bibitem{Han:2012vk}
T.~Han, I.~Lewis, R.~Ruiz, and Z.-g. Si, {\em {Lepton Number Violation and
  $W^\prime$ Chiral Couplings at the LHC}\/},
  \href{http://dx.doi.org/10.1103/PhysRevD.87.035011,
  10.1103/PhysRevD.87.039906}{Phys. Rev. {\bf D87} (2013) no.~3, 035011},
  \href{http://arxiv.org/abs/1211.6447}{{\tt arXiv:1211.6447 [hep-ph]}}.
[Erratum: Phys. Rev.D87,no.3,039906(2013)].
%%CITATION = ARXIV:1211.6447;%%.

\bibitem{Ruiz:2015gsa}
R.~E. Ruiz, {\em {Hadron Collider Tests of Neutrino Mass-Generating
  Mechanisms}}.
\newblock PhD thesis, Pittsburgh U., 2015.
\newblock \href{http://arxiv.org/abs/1509.06375}{{\tt arXiv:1509.06375
  [hep-ph]}}.
\newblock
\url{http://inspirehep.net/record/1394386/files/arXiv:1509.06375.pdf}.
\newblock
%%CITATION = ARXIV:1509.06375;%%.

\bibitem{Degrande:2016aje}
C.~Degrande, O.~Mattelaer, R.~Ruiz, and J.~Turner, {\em {Fully-Automated
  Precision Predictions for Heavy Neutrino Production Mechanisms at Hadron
  Colliders}\/},
\href{http://arxiv.org/abs/1602.06957}{{\tt arXiv:1602.06957 [hep-ph]}}.
%%CITATION = ARXIV:1602.06957;%%.

\bibitem{Ruiz:2015zca}
R.~Ruiz, {\em {QCD Corrections to Pair Production of Type III Seesaw Leptons at
  Hadron Colliders}\/},  \href{http://dx.doi.org/10.1007/JHEP12(2015)165}{JHEP
  {\bf 12} (2015)  165},
\href{http://arxiv.org/abs/1509.05416}{{\tt arXiv:1509.05416 [hep-ph]}}.
%%CITATION = ARXIV:1509.05416;%%.

\bibitem{Arganda:2015ija}
E.~Arganda, M.~J. Herrero, X.~Marcano, and C.~Weiland, {\em {Exotic
  à¥à¤à¤°jj events from heavy ISS neutrinos at the LHC}\/},
  \href{http://dx.doi.org/10.1016/j.physletb.2015.11.013}{Phys. Lett. {\bf
  B752} (2016)  46--50},
\href{http://arxiv.org/abs/1508.05074}{{\tt arXiv:1508.05074 [hep-ph]}}.
%%CITATION = ARXIV:1508.05074;%%.

\bibitem{Aubert:2009ag}
{BaBar Collaboration}, B.~Aubert et al., {\em {Searches for Lepton Flavor
  Violation in the Decays $\tau^{\pm}\rightarrow e^{\pm}\gamma$ and $\tau^{\pm}
  \rightarrow \mu^{\pm}\gamma$}\/},
  \href{http://dx.doi.org/10.1103/PhysRevLett.104.021802}{Phys.Rev.Lett. {\bf
  104} (2010)  021802},
\href{http://arxiv.org/abs/0908.2381}{{\tt arXiv:0908.2381 [hep-ex]}}.
%%CITATION = ARXIV:0908.2381;%%.

\bibitem{Adam:2013mnn}
{MEG Collaboration}, J.~Adam et al., {\em {New constraint on the existence of
  the $\mu^+ \to e^+\gamma$ decay}\/},
  \href{http://dx.doi.org/10.1103/PhysRevLett.110.201801}{Phys.Rev.Lett. {\bf
  110} (2013)  201801},
\href{http://arxiv.org/abs/1303.0754}{{\tt arXiv:1303.0754 [hep-ex]}}.
%%CITATION = ARXIV:1303.0754;%%.

\bibitem{Arganda:2014dta}
E.~Arganda, M.~J. Herrero, X.~Marcano, and C.~Weiland, {\em {Imprints of
  massive inverse seesaw model neutrinos in lepton flavor violating Higgs boson
  decays}\/},  \href{http://dx.doi.org/10.1103/PhysRevD.91.015001}{Phys. Rev.
  {\bf D91} (2015) no.~1, 015001},
\href{http://arxiv.org/abs/1405.4300}{{\tt arXiv:1405.4300 [hep-ph]}}.
%%CITATION = ARXIV:1405.4300;%%.

\bibitem{Matsedonskyi:2014mna}
O.~Matsedonskyi, G.~Panico, and A.~Wulzer, {\em {On the Interpretation of Top
  Partners Searches}\/},  \href{http://dx.doi.org/10.1007/JHEP12(2014)097}{JHEP
  {\bf 12} (2014)  097},
\href{http://arxiv.org/abs/1409.0100}{{\tt arXiv:1409.0100 [hep-ph]}}.
%%CITATION = ARXIV:1409.0100;%%.

\bibitem{Andreazza:2015bja}
A.~Andreazza et al., {\em {What Next: White Paper of the INFN-CSN1}\/},
Frascati Phys. Ser. {\bf 60} (2015)  1--302.
%%CITATION = 00309,60,1;%%.

\bibitem{Cheng:2015buv}
H.-C. Cheng, S.~Jung, E.~Salvioni, and Y.~Tsai, {\em {Exotic Quarks in Twin
  Higgs Models}\/},
\href{http://arxiv.org/abs/1512.02647}{{\tt arXiv:1512.02647 [hep-ph]}}.
%%CITATION = ARXIV:1512.02647;%%.

\bibitem{Chacko:2005pe}
Z.~Chacko, H.-S. Goh, and R.~Harnik, {\em {The Twin Higgs: Natural electroweak
  breaking from mirror symmetry}\/},
  \href{http://dx.doi.org/10.1103/PhysRevLett.96.231802}{Phys.Rev.Lett. {\bf
  96} (2006)  231802},
\href{http://arxiv.org/abs/hep-ph/0506256}{{\tt arXiv:hep-ph/0506256
  [hep-ph]}}.
%%CITATION = HEP-PH/0506256;%%.

\bibitem{Aad:2015uaa}
{ATLAS Collaboration}, G.~Aad et al., {\em {Search for long-lived, weakly
  interacting particles that decay to displaced hadronic jets in proton-proton
  collisions at $\sqrt{s}=8$ TeV with the ATLAS detector}\/},
\href{http://arxiv.org/abs/1504.03634}{{\tt arXiv:1504.03634 [hep-ex]}}.
%%CITATION = ARXIV:1504.03634;%%.

\bibitem{Aad:2015asa}
{ATLAS Collaboration}, G.~Aad et al., {\em {Search for pair-produced long-lived
  neutral particles decaying in the ATLAS hadronic calorimeter in $pp$
  collisions at $\sqrt{s}$ = 8 TeV}\/},
  \href{http://dx.doi.org/10.1016/j.physletb.2015.02.015}{Phys.Lett. {\bf B743}
  (2015)  15--34},
\href{http://arxiv.org/abs/1501.04020}{{\tt arXiv:1501.04020 [hep-ex]}}.
%%CITATION = ARXIV:1501.04020;%%.

\bibitem{Gershtein:2013iqa}
Y.~Gershtein, M.~Luty, M.~Narain, L.~T. Wang, D.~Whiteson, et al., {\em {New
  Particles Working Group Report of the Snowmass 2013 Community Summer
  Study}\/},
\href{http://arxiv.org/abs/1311.0299}{{\tt arXiv:1311.0299 [hep-ex]}}.
%%CITATION = ARXIV:1311.0299;%%.

\bibitem{Curtin:2015bka}
D.~Curtin and P.~Saraswat, {\em {Towards a No-Lose Theorem for Naturalness}\/},
\href{http://arxiv.org/abs/1509.04284}{{\tt arXiv:1509.04284 [hep-ph]}}.
%%CITATION = ARXIV:1509.04284;%%.

\bibitem{Burdman:2006tz}
G.~Burdman, Z.~Chacko, H.-S. Goh, and R.~Harnik, {\em {Folded supersymmetry and
  the LEP paradox}\/},
  \href{http://dx.doi.org/10.1088/1126-6708/2007/02/009}{JHEP {\bf 0702} (2007)
   009},
\href{http://arxiv.org/abs/hep-ph/0609152}{{\tt arXiv:hep-ph/0609152
  [hep-ph]}}.
%%CITATION = HEP-PH/0609152;%%.

\bibitem{Cai:2008au}
H.~Cai, H.-C. Cheng, and J.~Terning, {\em {A Quirky Little Higgs Model}\/},
  \href{http://dx.doi.org/10.1088/1126-6708/2009/05/045}{JHEP {\bf 0905} (2009)
   045},
\href{http://arxiv.org/abs/0812.0843}{{\tt arXiv:0812.0843 [hep-ph]}}.
%%CITATION = ARXIV:0812.0843;%%.

\bibitem{Craig:2015xla}
N.~Craig and A.~Katz, {\em {The Fraternal WIMP Miracle}\/},
\href{http://arxiv.org/abs/1505.07113}{{\tt arXiv:1505.07113 [hep-ph]}}.
%%CITATION = ARXIV:1505.07113;%%.

\bibitem{Chacko:2015fbc}
Z.~Chacko, D.~Curtin, and C.~B. Verhaaren, {\em {A Quirky Probe of Neutral
  Naturalness}\/},
\href{http://arxiv.org/abs/1512.05782}{{\tt arXiv:1512.05782 [hep-ph]}}.
%%CITATION = ARXIV:1512.05782;%%.

\bibitem{Burdman:2015oej}
G.~Burdman and R.~T. D'Agnolo, {\em {Scalar Leptons in Folded
  Supersymmetry}\/},
\href{http://arxiv.org/abs/1512.00040}{{\tt arXiv:1512.00040 [hep-ph]}}.
%%CITATION = ARXIV:1512.00040;%%.

\bibitem{Curtin:2014jma}
D.~Curtin, P.~Meade, and C.-T. Yu, {\em {Testing Electroweak Baryogenesis with
  Future Colliders}\/},  \href{http://dx.doi.org/10.1007/JHEP11(2014)127}{JHEP
  {\bf 1411} (2014)  127},
\href{http://arxiv.org/abs/1409.0005}{{\tt arXiv:1409.0005 [hep-ph]}}.
%%CITATION = ARXIV:1409.0005;%%.

\bibitem{He:2015spf}
H.-J. He, J.~Ren, and W.~Yao, {\em {Probing New Physics of Cubic Higgs
  Interaction via Higgs Pair Production at Hadron Colliders}\/},
\href{http://arxiv.org/abs/1506.03302}{{\tt arXiv:1506.03302 [hep-ph]}}.
%%CITATION = ARXIV:1506.03302;%%.

\bibitem{Batra:2008jy}
P.~Batra and Z.~Chacko, {\em {A Composite Twin Higgs Model}\/},
  \href{http://dx.doi.org/10.1103/PhysRevD.79.095012}{Phys.Rev. {\bf D79}
  (2009)  095012},
\href{http://arxiv.org/abs/0811.0394}{{\tt arXiv:0811.0394 [hep-ph]}}.
%%CITATION = ARXIV:0811.0394;%%.

\bibitem{Barbieri:2015lqa}
R.~Barbieri, D.~Greco, R.~Rattazzi, and A.~Wulzer, {\em {The Composite Twin
  Higgs scenario}\/},
\href{http://arxiv.org/abs/1501.07803}{{\tt arXiv:1501.07803 [hep-ph]}}.
%%CITATION = ARXIV:1501.07803;%%.

\bibitem{Low:2015nqa}
M.~Low, A.~Tesi, and L.-T. Wang, {\em {Twin Higgs mechanism and a composite
  Higgs boson}\/},
  \href{http://dx.doi.org/10.1103/PhysRevD.91.095012}{Phys.Rev. {\bf D91}
  (2015) no.~9, 095012},
\href{http://arxiv.org/abs/1501.07890}{{\tt arXiv:1501.07890 [hep-ph]}}.
%%CITATION = ARXIV:1501.07890;%%.

\bibitem{Geller:2014kta}
M.~Geller and O.~Telem, {\em {A Holographic Twin Higgs Model}\/},
  \href{http://dx.doi.org/10.1103/PhysRevLett.114.191801}{Phys.Rev.Lett. {\bf
  114} (2015) no.~19, 191801},
\href{http://arxiv.org/abs/1411.2974}{{\tt arXiv:1411.2974 [hep-ph]}}.
%%CITATION = ARXIV:1411.2974;%%.

\bibitem{Craig:2013fga}
N.~Craig and K.~Howe, {\em {Doubling down on naturalness with a supersymmetric
  twin Higgs}\/},  \href{http://dx.doi.org/10.1007/JHEP03(2014)140}{JHEP {\bf
  1403} (2014)  140},
\href{http://arxiv.org/abs/1312.1341}{{\tt arXiv:1312.1341 [hep-ph]}}.
%%CITATION = ARXIV:1312.1341;%%.

\bibitem{Craig:2014fka}
N.~Craig and H.~K. Lou, {\em {Scherk-Schwarz Supersymmetry Breaking in 4D}\/},
  \href{http://dx.doi.org/10.1007/JHEP12(2014)184}{JHEP {\bf 1412} (2014)
  184},
\href{http://arxiv.org/abs/1406.4880}{{\tt arXiv:1406.4880 [hep-ph]}}.
%%CITATION = ARXIV:1406.4880;%%.

\bibitem{Chang:2006ra}
S.~Chang, L.~J. Hall, and N.~Weiner, {\em {A Supersymmetric twin Higgs}\/},
  \href{http://dx.doi.org/10.1103/PhysRevD.75.035009}{Phys. Rev. {\bf D75}
  (2007)  035009},
\href{http://arxiv.org/abs/hep-ph/0604076}{{\tt arXiv:hep-ph/0604076
  [hep-ph]}}.
%%CITATION = HEP-PH/0604076;%%.

\bibitem{Dienes:2001se}
K.~R. Dienes, {\em {Solving the hierarchy problem without supersymmetry or
  extra dimensions: An Alternative approach}\/},
  \href{http://dx.doi.org/10.1016/S0550-3213(01)00344-3}{Nucl. Phys. {\bf B611}
  (2001)  146--178},
\href{http://arxiv.org/abs/hep-ph/0104274}{{\tt arXiv:hep-ph/0104274
  [hep-ph]}}.
%%CITATION = HEP-PH/0104274;%%.

\bibitem{Dong:2015gya}
X.~Dong, D.~Z. Freedman, and Y.~Zhao, {\em {AdS/CFT and the Little Hierarchy
  Problem}\/},
\href{http://arxiv.org/abs/1510.01741}{{\tt arXiv:1510.01741 [hep-th]}}.
%%CITATION = ARXIV:1510.01741;%%.

\bibitem{Baur:2004uw}
U.~Baur, A.~Juste, L.~H. Orr, and D.~Rainwater, {\em {Probing electroweak top
  quark couplings at hadron colliders}\/},
  \href{http://dx.doi.org/10.1103/PhysRevD.71.054013}{Phys. Rev. {\bf D71}
  (2005)  054013},
\href{http://arxiv.org/abs/hep-ph/0412021}{{\tt arXiv:hep-ph/0412021
  [hep-ph]}}.
%%CITATION = HEP-PH/0412021;%%.

\bibitem{Rontsch:2014cca}
R.~R{\"{o}}ntsch and M.~Schulze, {\em {Constraining couplings of top quarks to
  the Z boson in {$t\overline{t} $} + Z production at the LHC}\/},
  \href{http://dx.doi.org/10.1007/JHEP09(2015)132,
  10.1007/JHEP07(2014)091}{JHEP {\bf 07} (2014)  091},
  \href{http://arxiv.org/abs/1404.1005}{{\tt arXiv:1404.1005 [hep-ph]}}.
[Erratum: JHEP09,132(2015)].
%%CITATION = ARXIV:1404.1005;%%.

\bibitem{Dror:2015nkp}
J.~A. Dror, M.~Farina, E.~Salvioni, and J.~Serra, {\em {Strong tW Scattering at
  the LHC}\/},  \href{http://dx.doi.org/10.1007/JHEP01(2016)071}{JHEP {\bf 01}
  (2016)  071},
\href{http://arxiv.org/abs/1511.03674}{{\tt arXiv:1511.03674 [hep-ph]}}.
%%CITATION = ARXIV:1511.03674;%%.

\bibitem{Bellazzini:2012tv}
B.~Bellazzini, C.~Csaki, J.~Hubisz, J.~Serra, and J.~Terning, {\em {Composite
  Higgs Sketch}\/},  \href{http://dx.doi.org/10.1007/JHEP11(2012)003}{JHEP {\bf
  11} (2012)  003}, \href{hep-ph/1205.4032}{{\tt 1205.4032}}.
  [\href{http://inspirehep.net/record/1115304}{Inspire}].

\bibitem{Pomarol:2008bh}
A.~Pomarol and J.~Serra, {\em {Top Quark Compositeness: Feasibility and
  Implications}\/},  \href{http://dx.doi.org/10.1103/PhysRevD.78.074026}{Phys.
  Rev. {\bf D78} (2008)  074026},
\href{http://arxiv.org/abs/0806.3247}{{\tt arXiv:0806.3247 [hep-ph]}}.
%%CITATION = ARXIV:0806.3247;%%.

\bibitem{Kaplan:2008pt}
D.~E. Kaplan and M.~D. Schwartz, {\em {Constraining Light Colored Particles
  with Event Shapes}\/},
  \href{http://dx.doi.org/10.1103/PhysRevLett.101.022002}{Phys. Rev. Lett. {\bf
  101} (2008)  022002},
\href{http://arxiv.org/abs/0804.2477}{{\tt arXiv:0804.2477 [hep-ph]}}.
%%CITATION = ARXIV:0804.2477;%%.

\bibitem{Becciolini:2014lya}
D.~Becciolini, M.~Gillioz, M.~Nardecchia, F.~Sannino, and M.~Spannowsky, {\em
  {Constraining new colored matter from the ratio of 3 to 2 jets cross sections
  at the LHC}\/},  \href{http://dx.doi.org/10.1103/PhysRevD.91.015010,
  10.1103/PhysRevD.92.079905}{Phys. Rev. {\bf D91} (2015) no.~1, 015010},
  \href{http://arxiv.org/abs/1403.7411}{{\tt arXiv:1403.7411 [hep-ph]}}.
[Addendum: Phys. Rev.D92,no.7,079905(2015)].
%%CITATION = ARXIV:1403.7411;%%.

\bibitem{Alwall:2008ve}
J.~Alwall, M.-P. Le, M.~Lisanti, and J.~G. Wacker, {\em {Searching for Directly
  Decaying Gluinos at the Tevatron}\/},
  \href{http://dx.doi.org/10.1016/j.physletb.2008.06.065}{Phys. Lett. {\bf
  B666} (2008)  34--37},
\href{http://arxiv.org/abs/0803.0019}{{\tt arXiv:0803.0019 [hep-ph]}}.
%%CITATION = ARXIV:0803.0019;%%.

\bibitem{Rainwater:2007qa}
D.~Rainwater and T.~M.~P. Tait, {\em {Testing Grand Unification at the
  (S)LHC}\/},  \href{http://dx.doi.org/10.1103/PhysRevD.75.115014}{Phys. Rev.
  {\bf D75} (2007)  115014},
\href{http://arxiv.org/abs/hep-ph/0701093}{{\tt arXiv:hep-ph/0701093
  [hep-ph]}}.
%%CITATION = HEP-PH/0701093;%%.

\bibitem{Catani:2007vq}
S.~Catani and M.~Grazzini, {\em {An NNLO subtraction formalism in hadron
  collisions and its application to Higgs boson production at the LHC}\/},
  \href{http://dx.doi.org/10.1103/PhysRevLett.98.222002}{Phys. Rev. Lett. {\bf
  98} (2007)  222002},
\href{http://arxiv.org/abs/hep-ph/0703012}{{\tt arXiv:hep-ph/0703012
  [hep-ph]}}.
%%CITATION = HEP-PH/0703012;%%.

\bibitem{Catani:2009sm}
S.~Catani, L.~Cieri, G.~Ferrera, D.~de~Florian, and M.~Grazzini, {\em {Vector
  boson production at hadron colliders: a fully exclusive QCD calculation at
  NNLO}\/},  \href{http://dx.doi.org/10.1103/PhysRevLett.103.082001}{Phys. Rev.
  Lett. {\bf 103} (2009)  082001},
\href{http://arxiv.org/abs/0903.2120}{{\tt arXiv:0903.2120 [hep-ph]}}.
%%CITATION = ARXIV:0903.2120;%%.

\bibitem{Melnikov:2006kv}
K.~Melnikov and F.~Petriello, {\em {Electroweak gauge boson production at
  hadron colliders through O(alpha(s)**2)}\/},
  \href{http://dx.doi.org/10.1103/PhysRevD.74.114017}{Phys. Rev. {\bf D74}
  (2006)  114017},
\href{http://arxiv.org/abs/hep-ph/0609070}{{\tt arXiv:hep-ph/0609070
  [hep-ph]}}.
%%CITATION = HEP-PH/0609070;%%.

\bibitem{Gavin:2010az}
R.~Gavin, Y.~Li, F.~Petriello, and S.~Quackenbush, {\em {FEWZ 2.0: A code for
  hadronic Z production at next-to-next-to-leading order}\/},
  \href{http://dx.doi.org/10.1016/j.cpc.2011.06.008}{Comput. Phys. Commun. {\bf
  182} (2011)  2388--2403},
\href{http://arxiv.org/abs/1011.3540}{{\tt arXiv:1011.3540 [hep-ph]}}.
%%CITATION = ARXIV:1011.3540;%%.

\bibitem{Gavin:2012sy}
R.~Gavin, Y.~Li, F.~Petriello, and S.~Quackenbush, {\em {W Physics at the LHC
  with FEWZ 2.1}\/},
  \href{http://dx.doi.org/10.1016/j.cpc.2012.09.005}{Comput. Phys. Commun. {\bf
  184} (2013)  208--214},
\href{http://arxiv.org/abs/1201.5896}{{\tt arXiv:1201.5896 [hep-ph]}}.
%%CITATION = ARXIV:1201.5896;%%.

\bibitem{Li:2012wna}
Y.~Li and F.~Petriello, {\em {Combining QCD and electroweak corrections to
  dilepton production in FEWZ}\/},
  \href{http://dx.doi.org/10.1103/PhysRevD.86.094034}{Phys. Rev. {\bf D86}
  (2012)  094034},
\href{http://arxiv.org/abs/1208.5967}{{\tt arXiv:1208.5967 [hep-ph]}}.
%%CITATION = ARXIV:1208.5967;%%.

\bibitem{Aad:2013iua}
{ATLAS Collaboration}, G.~Aad et al., {\em {Measurement of the high-mass
  Drell--Yan differential cross-section in pp collisions at sqrt(s)=7 TeV with
  the ATLAS detector}\/},
  \href{http://dx.doi.org/10.1016/j.physletb.2013.07.049}{Phys. Lett. {\bf
  B725} (2013)  223--242},
\href{http://arxiv.org/abs/1305.4192}{{\tt arXiv:1305.4192 [hep-ex]}}.
%%CITATION = ARXIV:1305.4192;%%.

\bibitem{Chatrchyan:2013tia}
{CMS Collaboration}, S.~Chatrchyan et al., {\em {Measurement of the
  differential and double-differential Drell-Yan cross sections in
  proton-proton collisions at $\sqrt{s} =$ 7 TeV}\/},
  \href{http://dx.doi.org/10.1007/JHEP12(2013)030}{JHEP {\bf 12} (2013)  030},
\href{http://arxiv.org/abs/1310.7291}{{\tt arXiv:1310.7291 [hep-ex]}}.
%%CITATION = ARXIV:1310.7291;%%.

\bibitem{CMS:2014jea}
{CMS Collaboration}, V.~Khachatryan et al., {\em {Measurements of differential
  and double-differential Drell-Yan cross sections in proton-proton collisions
  at 8 TeV}\/},  \href{http://dx.doi.org/10.1140/epjc/s10052-015-3364-2}{Eur.
  Phys. J. {\bf C75} (2015) no.~4, 147},
\href{http://arxiv.org/abs/1412.1115}{{\tt arXiv:1412.1115 [hep-ex]}}.
%%CITATION = ARXIV:1412.1115;%%.

\end{thebibliography}\endgroup
%\bibliography{report,02-DM/DM_bib0,02-DM/DM_experimentalBounds/DM_experimentalBounds,02-DM/DM_higgsportal/portalbib,02-DM/DM_SM_portal_1/fcc-neutralinos,02-DM/DM_SM_portal_2/Wino_FCChh_CST,02-DM/DM_SM_portal_3/100TeVFiveplet,02-DM/DM_SM_portal_4/serelic,02-DM/DM_BSM_portal_1/HKSW_100,02-DM/DM_BSM_portal_2/refs,02-DM/DM_BSM_portal_3/zdark,02-DM/DM_nonwimp/nonwimp,02-DM/DM_nonwimp_radiating/radiating-dm-100-TeV}

%% Run the following command to merge bib files into 'report_test.bib':
% bibtool -s -d -x report.aux report.bib 02-DM/DM_bib0.bib 02-DM/DM_experimentalBounds/DM_experimentalBounds.bib 02-DM/DM_higgsportal/portalbib.bib 02-DM/DM_SM_portal_1/fcc-neutralinos.bib 02-DM/DM_SM_portal_2/Wino_FCChh_CST.bib 02-DM/DM_SM_portal_3/100TeVFiveplet.bib 02-DM/DM_SM_portal_4/serelic.bib 02-DM/DM_BSM_portal_1/HKSW_100.bib 02-DM/DM_BSM_portal_2/refs.bib 02-DM/DM_BSM_portal_3/zdark.bib 02-DM/DM_nonwimp/nonwimp.bib 02-DM/DM_nonwimp_radiating/radiating-dm-100-TeV.bib > report_test.bib

%% Then comment out the \bibliography line above, uncomment the line below, and re-run pdflatex/bibtex.
%\bibliography{report_test}

\end{document}